%% file: nlm09.tex
\documentclass[11pt]{cernrepdbg}
\usepackage{graphicx}
\usepackage{epsfig}
\usepackage{color}
\usepackage{cite}
\usepackage{xspace}
\usepackage{dcolumn}
\usepackage{bm}
\usepackage{axodraw,rotating}
\usepackage{latexsym}
\usepackage{pstricks}
\usepackage{amsmath}
\usepackage[T1]{fontenc}
\usepackage{amssymb}
\usepackage{amsfonts}
\usepackage{slashed}
\usepackage{amsbsy}
\usepackage{units,booktabs}
\usepackage{cite}
\usepackage{wrapfig,rotating}
\usepackage{graphicx}
\usepackage{epsfig}

\makeatletter
\renewcommand\thesubsubsection{\thesubsection.\@arabic\c@subsubsection}
\renewcommand*\l@section[2]{%
  \ifnum \c@tocdepth >\z@
    \addpenalty\@secpenalty
    \addvspace{0.9em \@plus\p@}%
    \setlength\@tempdima{1.5em}%
    \begingroup
      \parindent \z@ \rightskip \@pnumwidth
      \parfillskip -\@pnumwidth
      \leavevmode \bfseries
      \advance\leftskip\@tempdima
      \hskip -\leftskip
      #1\nobreak\hfil \nobreak\hb@xt@\@pnumwidth{\hss #2}\par
    \endgroup
  \fi}
\makeatother

\begin{document}

\bibliographystyle{lesHouches}

\title{\centering{THE SM AND NLO MULTILEG WORKING GROUP:\\
          \textbf{Summary Report}}}

\author{\underline{Convenors}:
T.~Binoth$^{1}$,
G.~Dissertori$^{2}$,
J.~Huston$^{3}$,  
R.~Pittau$^{4}$
\\
\underline{Contributing authors}:
J.~R.~Andersen$^{5}$, 
J.~Archibald$^{6}$,
S.~Badger$^{7}$,
R.~D.~Ball$^{1}$,
G.~Bevilacqua$^{8}$,
I.~Bierenbaum$^{9}$,
T.~Binoth$^{1}$,
F.~Boudjema$^{10}$, 
R.~Boughezal$^{11}$,
A.~Bredenstein$^{12}$, 
R.~Britto$^{13}$,
M.~Campanelli$^{14}$, 
J.~Campbell$^{15}$, 
L.~Carminati$^{16,17}$,
G.~Chachamis$^{18}$,
V.~Ciulli$^{19}$,
G.~Cullen$^{1}$,
M.~Czakon$^{20}$,
L.~Del~Debbio$^{1}$, 
A.~Denner$^{18}$, 
G.~Dissertori$^{2}$,
S.~Dittmaier$^{21}$,
S.~Forte$^{16,17}$, 
R.~Frederix$^{11}$,
S.~Frixione$^{5,22,23}$,
E.~Gardi$^{1}$,
M.~V.~Garzelli$^{4,16}$,
S.~Gascon-Shotkin$^{24}$,
T.~Gehrmann$^{11}$, 
A.~Gehrmann--De Ridder$^{25}$, 
W.~Giele$^{15}$, 
T.~Gleisberg$^{26}$, 
E.~W.~N.~Glover$^{6}$, 
N.~Greiner$^{11}$,
A.~Guffanti$^{21}$,
J.-Ph.~Guillet$^{10}$,
A.~van Hameren$^{27}$,
G.~Heinrich$^{6}$, 
S.~H{\"o}che$^{11}$, 
M.~Huber$^{28}$,
J.~Huston$^{3}$,  
M.~Jaquier$^{11}$, 
S.~Kallweit$^{18}$, 
S.~Karg$^{20}$,
N.~Kauer$^{29}$,
F.~Krauss$^{6}$,
J.~I.~Latorre$^{30}$, 
A.~Lazopoulos$^{25}$,
P.~Lenzi$^{19}$, 
G.~Luisoni$^{11}$, 
R.~Mackeprang$^{31}$, 
L.~Magnea$^{5,32}$,
D.~Ma\^{\i}tre$^{6}$,
D.~Majumder$^{33}$,
I.~Malamos$^{34}$,
F.~Maltoni$^{35}$, 
K.~Mazumdar$^{33}$,
P.~Nadolsky$^{36}$, 
P.~Nason$^{37}$, 
C.~Oleari$^{37}$, 
F.~Olness$^{36}$,  
C.~G.~Papadopoulos$^{8}$,
G.~Passarino$^{32}$,
E.~Pilon$^{10}$,
R.~Pittau$^{4}$,
S.~Pozzorini$^{5}$,
T.~Reiter$^{38}$,
J.~Reuter$^{21}$,
M.~Rodgers$^{6}$,
G.~Rodrigo$^{9}$,
J.~Rojo$^{16,17}$, 
G.~Sanguinetti$^{10}$,  
F.-P. Schilling$^{39}$, 
M.~Schumacher$^{21}$, 
S.~Schumann$^{40}$,
R.~Schwienhorst$^{3}$,
P.~Skands$^{15}$, 
H.~Stenzel$^{41}$,
F.~St\"ockli$^{5}$,
R.~Thorne$^{14,42}$,
M.~Ubiali$^{1,35}$, 
P.~Uwer$^{43}$,
A.~Vicini$^{16,17}$,
M.~Warsinsky$^{21}$,
G.~Watt$^{5}$,
J.~Weng$^{2}$,
I.~Wigmore$^{1}$,
S.~Weinzierl$^{44}$,
J.~Winter$^{15}$, 
M.~Worek$^{45}$,
G.~Zanderighi$^{46}$
\\
\mbox{} }
\institute{\centering{\small
$^{1}$ The University of Edinburgh,
School of Physics and Astronomy,
Edinburgh EH9\,3JZ,
UK\\
$^{2}$ Institute for Particle Physics, ETH Zurich, CH-8093 Zurich, Switzerland\\
$^{3}$ Michigan State University, East Lansing, Michigan 48824, USA\\
$^{4}$ Departamento de F\'{i}sica Te\'orica y del Cosmos,
Centro Andaluz de F\'{i}sica de Part\'{i}culas Elementales (CAFPE),
Universidad de Granada, E-18071 Granada, Spain \\
$^{5}$ PH TH, CERN, CH-1211 Geneva, Switzerland \\
$^{6}$ Institute for Particle Physics Phenomenology, 
        University of Durham, Durham, DH1 3LE, UK\\
$^{7}$ Deutsches Elektronensynchrotron DESY,
Platanenallee 6, D-15738 Zeuthen, Germany\\
$^{8}$ Institute of Nuclear Physics, NCSR Demokritos, 
GR-15310 Athens, Greece \\
$^{9}$ Instituto de F\'{i}sica Corpuscular, CSIC-Universitat de 
Val\`{e}ncia, Apartado de Correos 22085, E-46071 Valencia, Spain\\
$^{10}$ LAPTH, Universite de Savoie, 
CNRS, BP. 110, 74941 Annecy-le-Vieux, France\\
$^{11}$ Institut f\"ur Theoretische Physik,
Universit\"at Z\"urich, CH-8057 Z\"urich, Switzerland\\
$^{12}$ High Energy Accelerator Research Organization (KEK),
Tsukuba, Ibaraki 305-0801, Japan\\
$^{13}$ Institut de Physique Th\'eorique, Orme des Merisiers, 
CEA/Saclay, 91191 Gif-sur-Yvette Cedex, France\\
$^{14}$ Universtiy College London, Gower Street, WC1E 6BT, London, UK\\
$^{15}$ Theory Dept., Fermi National 
Accelerator Laboratory, Batavia (IL), USA \\
$^{16}$ INFN Milano, via Celoria 16, I-20133 Milano, Italy\\ 
$^{17}$ Universit\`a di Milano, Dipartimento di Fisica, via Celoria 16, 
I-20133 Milano, Italy\\ 
$^{18}$ Paul Scherrer Institut, W\"urenlingen und Villigen,
CH-5232 Villigen PSI, Switzerland\\
$^{19}$ Universit\`a di Firenze \& INFN, via Sansone 1, 50019 Sesto F.no, Firenze, Italy\\
$^{20}$ Institut f\"ur 
Theoretische Physik E, RWTH Aachen University D-52056 Aachen,
Germany\\
$^{21}$ Albert-Ludwigs Universit\"at Freiburg, Physikalisches Institut, Hermann-Herder-Str.~3, 79104 Freiburg im Breisgau, Germany \\
$^{22}$ On leave of absence from INFN, Sez. di Genova, Italy \\
$^{23}$ ITPP, EPFL, CH-1015 Lausanne, Switzerland \\
$^{24}$ Universite Claude Bernard Lyon-I, Institut de Physique Nucleaire 
de Lyon (IPNL), 4 rue Enrico Fermi, F-69622 Villeurbanne, CEDEX, France\\
$^{25}$ Institute for Theoretical Physics, ETH, CH-8093 Zurich, Switzerland\\
$^{26}$ SLAC National Accelerator Laboratory, Stanford University, Stanford, CA 
94309, USA \\
$^{27}$ Institute of Nuclear Physics, Polish Academy 
of Sciences, PL-31342 Cracow, Poland \\
$^{28}$ Max-Planck-Institut f\"ur Physik (Werner-Heisenberg-Institut),
  D-80805 M\"unchen, Germany \\
$^{29}$ Department of Physics,
Royal Holloway,
University of London,
Egham TW20\,0EX, UK\\
$\vphantom{^{29}}$
School of Physics and Astronomy,
University of Southampton,
Southampton SO17\,1BJ, UK\\
$^{30}$ Departament d'Estructura i Constituents de la Mat\`eria, 
Universitat de Barcelona, Diagonal 647, E-08028 Barcelona, Spain\\
$^{31}$ PH ADT, CERN, CH-1211 Geneva 23, Switzerland\\
$^{32}$ Dipartimento di Fisica Teorica, Universit\`a di Torino, Italy,
           INFN, Sezione di Torino, Italy\\
$^{33}$ Tata Institute of Fundamental Research, Homi Bhabha Road, Mumbai 400005\\
$^{34}$ Radboud Universiteit Nijmegen,
Department of Theoretical High Energy Physics,
Institute for Mathematics, Astrophysics and Particle Physics,
6525 AJ Nijmegen, the Netherlands\\
$^{35}$5 Center for Particle Physics and Phenomenology (CP3), Universit\'e Catholique de Louvain, Chemin du Cyclotron, B-1348 Louvain-la-Neuve, Belgium\\
$^{36}$ Dept. of Physics, Southern Methodist University, Dallas (TX), 
75275-0175, USA \\
$^{37}$ INFN, Sezione di Milano-Bicocca, Piazza della Scienza 3, 
I-20126 Milano, Italy\\
$^{38}$ Nikhef,
Science Park 105,
1098\,XG Amsterdam,
The Netherlands\\
$^{39}$ Inst. f. Exp. Kernphysik, Karlsruhe Institute of Technology (KIT), Karlsruhe, Germany \\
$^{40}$ Institut f\"ur Theoretische Physik, Universit\"at
Heidelberg, D-69120, Heidelberg, Germany \\
$^{41}$ II. Physikalisches Institut, Justus-Liebig Universit\"at Giessen, 
D-35392 Giessen, Germany \\     
$^{42}$ Associate of IPPP, Durham\\
$^{43}$ Inst. f. Physik, Humboldt-Universit\"at zu Berlin, Berlin, Germany \\
$^{44}$ Institut f{\"u}r Physik, Universit{\"a}t Mainz, D - 55099 Mainz, 
Germany \\
$^{45}$ Fachbereich C Physik, 
Bergische Universit\"at Wuppertal, D-42097 Wuppertal, Germany \\
$^{46}$ Rudolf Peierls Centre for Theoretical Physics, 1 Keble Road,
  University of Oxford, UK\\
}}
 
\maketitle

 \begin{center}
   \textit{Report of the SM and NLO Multileg  Working Group 
     for the Workshop ``Physics at TeV
     Colliders'', Les Houches, France, 8--26 June, 2009.  }
\end{center}

\newpage
  
\setcounter{tocdepth}{1}
\tableofcontents
\setcounter{footnote}{0}


\section[Introduction]{INTRODUCTION}
{\graphicspath{{introduction/}}
\input{introduction/les_houches_intro.tex}

\clearpage
\part[NLO TECHNIQUES, STANDARDIZATION, AUTOMATION]{NLO TECHNIQUES, STANDARDIZATION, AUTOMATION}


\section[Recent advances in analytic computations for one-loop amplitudes]
{RECENT ADVANCES IN ANALYTIC COMPUTATIONS FOR ONE-LOOP AMPLITUDES%
\protect\footnote{Contributed by: S. Badger and R. Britto.}}
{\graphicspath{{badger/}}
\input{badger/badger.tex}}


\section[A generic implementation of D-dimensional unitarity]
{A GENERIC IMPLEMENTATION OF D-DIMENSIONAL UNITARITY%
\protect\footnote{Contributed by: A. Lazopoulos.}}
{\graphicspath{{lazopoulos/}}
\input{lazopoulos/lazopoulos.tex}}


\section[Analytical calculation of the rational part of 1-loop
amplitudes: the R2 contribution]
{ANALYTICAL CALCULATION OF THE RATIONAL PART OF 1-LOOP AMPLITUDES: 
THE R2 CONTRIBUTION%
\protect\footnote{Contributed by: M.V. Garzelli and I. Malamos.}}
{\graphicspath{{garzelli/}}
\input{garzelli/garzelli.tex}}


\section[Recent Developments of GOLEM]
{RECENT DEVELOPMENTS OF GOLEM
\protect\footnote{Contributed by: 
T. Binoth,
G. Cullen,
N. Greiner,
A. Guffanti,
J.-Ph. Guillet,
G. Heinrich,
S. Karg,
N. Kauer,
T. Reiter,
J. Reuter,
M. Rodgers and
I. Wigmore.
}}
{\graphicspath{{reiter/}}
\input{reiter/reiter.tex}}


\section[Common Ntuple Output format for NLO Calculations]
{COMMON NTUPLE OUTPUT FORMAT FOR NLO CALCULATIONS
\protect\footnote{Contributed by: 
J. Campbell, J. Huston, P. Nadolsky, F.-P. Schilling, 
P. Uwer and J. Weng.
}}
{\graphicspath{{schilling/}}
\input{schilling/schilling.tex}}


\section[First steps towards a duality relation at two loops]
{FIRST STEPS TOWARDS A DUALITY RELATION AT TWO LOOPS
\protect\footnote{Contributed by: I. Bierenbaum and G. Rodrigo.
}}
{\graphicspath{{rodrigo/}}
\input{rodrigo/rodrigo.tex}}

\clearpage
\part[NEW HIGH ORDER CALCULATIONS, WISHLIST]{NEW HIGH ORDER CALCULATIONS, 
WISHLIST}


\section[A NLO study of $t\bar{t}H \to t\bar{t}b\bar{b}$ 
signal versus $t\bar{t}b\bar{b}$ background]
{A NLO STUDY OF $t\bar{t}H \to t\bar{t}b\bar{b}$ 
SIGNAL VERSUS $t\bar{t}b\bar{b}$ BACKGROUND%
\protect\footnote{Contributed by: 
G. Bevilacqua, M. Czakon, M.V. Garzelli, A. van Hameren, 
C.G. Papadopoulos, R. Pittau and M. Worek.
}}
{\graphicspath{{pittau/}}
\input{pittau/pittau.tex}}


\section[NLO QCD corrections to $\mathrm{\mathbf{t\bar t b\bar b}}$ production at the LHC]
{NLO QCD CORRECTIONS TO $\mathrm{\mathbf{t\bar t b\bar b}}$ PRODUCTION AT THE LHC
\protect\footnote{Contributed by: 
A. Bredenstein, A. Denner, S. Dittmaier and S. Pozzorini.
}}
{\graphicspath{{pozzorini/}}
\input{pozzorini/pozzorini.tex}}


\section[Understanding soft and collinear divergences to all orders]
{UNDERSTANDING SOFT AND COLLINEAR DIVERGENCES TO ALL ORDERS
\protect\footnote{Contributed by: E. Gardi and  L. Magnea.
}}
{\graphicspath{{magnea/}}

\input{magnea/magnea.tex}}


\section[Lessons learned from the NNLO calculation of $e^+ e^- \rightarrow 3 \; \mbox{jets}$]
{LESSONS LEARNED FROM THE NNLO CALCULATION OF $e^+ e^- \rightarrow 3 \; \mbox{jets}$
\protect\footnote{Contributed by: S. Weinzierl.
}}
{\graphicspath{{weinzierl/}}
\input{weinzierl/weinzierl.tex}}

\clearpage
\part[OBSERVABLES]{OBSERVABLES}


\section[Comparison of predictions for inclusive {\boldmath$W$}\/ +
  3 jet production at the LHC between \textbf{\textsc{BlackHat}},
  \textbf{\textsc{Rocket}} and \textbf{\textsc{Sherpa}}]
{COMPARISON OF PREDICTIONS FOR INCLUSIVE {\boldmath$W$}\/ +
  3 JET PRODUCTION AT THE LHC BETWEEN \textbf{\textsc{BlackHat}},
  \textbf{\textsc{Rocket}} AND \textbf{\textsc{Sherpa}}
\protect\footnote{Contributed by:
S. H{\"o}che, J. Huston, D. Ma\^{\i}tre, J. Winter and G. Zanderighi. 
}}
{\graphicspath{{zanderighi/}}
\input{zanderighi/zanderighi.tex}}


\section[Comparison of t-channel $2\rightarrow3$ production at NLO with CompHEP samples
]
{COMPARISON OF T-CHANNEL $2\rightarrow3$ PRODUCTION AT NLO WITH COMPHEP SAMPLES
\protect\footnote{Contributed by:
R.~Schwienhorst, R.~Frederix and F.~ Maltoni.
}}
{\graphicspath{{schwienhorst/}}
\input{schwienhorst/schwienhorst.tex}}


\section[Tuned comparison of NLO QCD corrections to \boldmath{$pp\to ZZ{+}\mathrm{jet}{+}X$} production at hadron colliders]
{TUNED COMPARISON OF NLO QCD CORRECTIONS TO \boldmath{$pp\to ZZ{+}\mathrm{jet}{+}X$} PRODUCTION AT HADRON COLLIDERS
\protect\footnote{Contributed by: 
T. Binoth, T. Gleisberg, S. Karg, N. Kauer, G. Sanguinetti and 
S. Dittmaier, S. Kallweit, P. Uwer. 
}}
{\graphicspath{{karg/}}

\input{karg/karg.tex}}


\section[W pair production: NNLO virtual corrections with full mass dependence]
{W PAIR PRODUCTION: NNLO VIRTUAL CORRECTIONS WITH FULL MASS DEPENDENCE
\protect\footnote{Contributed by: G. Chachamis.
}}
{\graphicspath{{chachamis/}}
\input{chachamis/chachamis.tex}}


\section[A Simple Radiation Pattern in Hard Multi-Jet Events in Association with a Weak Boson]
{A SIMPLE RADIATION PATTERN IN HARD MULTI-JET EVENTS IN ASSOCIATION WITH A WEAK BOSON
\protect\footnote{Contributed by: 
J. R. Andersen, M. Campanelli, J. Campbell, V. Ciulli, J. Huston,
P. Lenzi and  R. Mackeprang.
}}
{\graphicspath{{andersen/}}

\input{andersen/andersen.tex}}


\section[NNLO QCD Effects on
  $\mathrm{H}\rightarrow\mathrm{WW}\rightarrow\ell\nu\ell\nu$ at
  Hadron Colliders]
{NNLO QCD EFFECTS ON
  $\mathrm{H}\rightarrow\mathrm{WW}\rightarrow\ell\nu\ell\nu$ AT
  HADRON COLLIDERS%
\protect\footnote{Contributed by: G. Dissertori and F. St\"ockli.
}}
{\graphicspath{{dissertori/}}
\input{dissertori/dissertori.tex}}


\section[Determination of the strong coupling
constant based on NNLO+NLLA results for hadronic event shapes and a study 
of hadronisation corrections]
{DETERMINATION OF THE STRONG COUPLING
CONSTANT BASED ON NNLO+NLLA RESULTS FOR HADRONIC EVENT SHAPES AND A STUDY 
OF HADRONISATION CORRECTIONS
\protect\footnote{Contributed by: 
G. Dissertori, A. Gehrmann--De Ridder, T. Gehrmann, 
E.W.N. Glover, G. Heinrich, M. Jaquier, G. Luisoni and H. Stenzel.
}}
{\graphicspath{{heinrich/}}
\input{heinrich/heinrich.tex}}


\section[Comparisons of Fixed Order Calculations and Parton Shower Monte Carlo for Higgs Boson Production in Vector Boson Fusion]
{COMPARISONS OF FIXED ORDER CALCULATIONS AND PARTON SHOWER MONTE CARLO FOR HIGGS BOSON PRODUCTION IN VECTOR BOSON FUSION%
\protect\footnote{Contributed by: A. Denner, S. Dittmaier, M. Schumacher and M. Warsinsky.
}}
{\graphicspath{{warsinsky/}}
\input{warsinsky/warsinsky.tex}}


\section[A Study of Smooth Photon Isolation: a Novel Implementation]
{A STUDY OF SMOOTH PHOTON ISOLATION: A NOVEL IMPLEMENTATION
\protect\footnote{Contributed by: L. Carminati, S. Frixione,
S. Gascon-Shotkin, J-P. Guillet, G. Heinrich, J. Huston, K. Mazumdar,D. Majumder and  E. Pilon.
}}
{\graphicspath{{huston/}}

\input{huston/huston.tex}}


\section[On the correlation between $\alpha_s\left( M_Z^2\right)$ and
PDFs within the NNPDF approach]
{ON THE CORRELATION BETWEEN $\alpha_s\left( M_Z^2\right)$ AND
PDFS WITHIN THE NNPDF APPROACH
\protect\footnote{Contributed by: 
R. D. Ball, L. Del Debbio, S. Forte, A. Guffanti, J. I. Latorre, J. Rojo, 
M. Ubiali and A. Vicini.
}}
{\graphicspath{{forte/}}

\input{forte/forte.tex}}


\section[The Les Houches benchmarks for GM-VFN heavy quark schemes in 
deep-inelastic scattering]
{THE LES HOUCHES BENCHMARKS FOR GM-VFN HEAVY QUARK SCHEMES IN 
DEEP-INELASTIC SCATTERING
\protect\footnote{Contributed by:
J. Rojo, S. Forte, J. Huston, P. Nadolsky, P. Nason, F. Olness,  
R. Thorne and G. Watt.
}}
{\graphicspath{{rojo/}}
\input{rojo/rojo.tex}}

\clearpage
\part[HIGGS PHENOMENOLOGY]{HIGGS PHENOMENOLOGY}


\section[Additional Jet Production from Higgs Boson + Dijets through
Gluon Fusion]
{ADDITIONAL JET PRODUCTION FROM HIGGS BOSON + DIJETS THROUGH
GLUON FUSION
\protect\footnote{Contributed by: J. R. Andersen, J. Campbell and S.  H\"oche.
}}
{\graphicspath{{campbell/}}

\input{campbell/campbell.tex}}


\section[NLO Electroweak Corrections to Higgs Boson
Production at Hadron Colliders:
Towards a Full Complex Mass Scheme]
{NLO ELECTROWEAK CORRECTIONS TO HIGGS BOSON
PRODUCTION AT HADRON COLLIDERS:
TOWARDS A FULL COMPLEX MASS SCHEME
\protect\footnote{Contributed by: G. Passarino.
}}
{\graphicspath{{passarino/}}
\input{passarino/passarino.tex}}


\section[QCD-electroweak effects and a new prediction for Higgs production 
in gluon fusion within the SM]
{QCD-ELECTROWEAK EFFECTS AND A NEW PREDICTION FOR HIGGS PRODUCTION 
IN GLUON FUSION WITHIN THE SM
\protect\footnote{Contributed by: R. Boughezal. 
}}
{\graphicspath{{boughezal/}}
\input{boughezal/boughezal.tex}}

\clearpage
\part[MC/NLO INTERFACE]{MC/NLO INTERFACE
\protect\footnote{Contributed by: 
T. Binoth, F. Boudjema, G. Dissertori and A. Lazopoulos, 
A. Denner, S. Dittmaier, R. Frederix, N. Greiner and S. H\"oche, 
W. Giele, P. Skands and J. Winter, T. Gleisberg, J. Archibald, G. Heinrich, 
F. Krauss and D. Ma\^{\i}tre, M. Huber, J. Huston, N. Kauer, F. Maltoni, 
C. Oleari, G. Passarino, R. Pittau, S. Pozzorini, T. Reiter, 
S. Schumann and G. Zanderighi.
}}

\input{binoth/binoth.tex}


\section[Example implementation of an EW MC/OLP interface between {\sc
    SHERPA} and {\sc RADY}]
{EXAMPLE IMPLEMENTATION OF AN EW MC/OLP INTERFACE BETWEEN {\sc
    SHERPA} AND {\sc RADY}
\protect\footnote{Contributed by: 
J. Archibald, S. Dittmaier, F. Krauss and M. Huber.
}}
{\graphicspath{{huber/}}
\input{huber/huber.tex}}


\section[The LHA for Monte Carlo tools and one-loop programs:
an application using \texttt{BlackHat}~and \texttt{Rocket}~with \texttt{MadFKS}]
{THE LHA FOR MONTE CARLO TOOLS AND ONE-LOOP PROGRAMS:
AN APPLICATION USING \texttt{BlackHat}~AND \texttt{Rocket}~WITH \texttt{MadFKS}
\protect\footnote{Contributed by: 
R. Frederix, D. Ma\^itre and G. Zanderighi.
}}
{\graphicspath{{frederix/}}
\input{frederix/frederix.tex}}

\clearpage

\bibliography{nlm09}
\end{document}

%% file: introduction/les_houches_intro.tex


After years of waiting, and after six Les Houches workshops, the era of LHC 
running is finally upon us, albeit at a lower initial center-of-mass energy 
than originally planned. Thus, there has been a great sense of anticipation 
from both the experimental and theoretical communities. The last two years, 
in particular,  have seen great productivity in the area of multi-parton 
calculations at leading order (LO), next-to-leading order  (NLO)  and 
next-to-next-to-leading order (NNLO), and this productivity is reflected 
in the proceedings of the NLM group. Both religions, 
Feynmanians and Unitarians, 
as well as agnostic experimenters, were well-represented in both the 
discussions at Les Houches, and in the contributions to the write-up.

Next-to-leading order (NLO) is the first order at which the normalization, 
and in some cases the shape, of perturbative cross sections can be considered 
reliable~\cite{Campbell:2006wx}. This can be especially true when probing 
extreme kinematic regions, as for example with boosted Higgs searches 
considered in several of the contributions to this writeup. A full 
understanding for both standard model and beyond the standard model physics 
at the LHC requires the development of fast, reliable programs for the 
calculation of multi-parton final states at NLO. 
There have been many advances in 
the development of NLO techniques, standardization and automation for 
such processes and this is 
reflected in the contributions to the  first section of this writeup.

Many calculations have previously been 
performed with the aid of semi-numerical techniques. Such techniques, 
although retaining the desired accuracy, lead to codes which are slow to 
run. Advances in the calculation of compact analytic expressions for 
Higgs + 2 jets (see for example the contribution of S. Badger and R. Britto 
to these proceedings)  have resulted in the development of much faster codes, 
which extend the phenomenology that can be conducted, as well as making the 
code available to the public for the first time.

A prioritized list of NLO cross sections was assembled at Les Houches in 
2005~\cite{Buttar:2006zd} and added to in 2007~\cite{Bern:2008ef}. This list 
includes cross sections which are experimentally important, and which are 
theoretically feasible (if difficult) to calculate. Basically all 
$2\rightarrow3$ cross sections of interest have been calculated, with the 
frontier now extending to $2\rightarrow4$ calculations. Often these 
calculations exist only as private codes. 
That wishlist is shown below in Table 1. Since 2007, two  additional 
calculations have been completed: $t\bar{t}b\bar{b}$ and $W$ + 3 jets, 
reflecting the advance of the NLO technology to $2\rightarrow4$ processes. 
In addition, the cross section for $b\bar{b}b\bar{b}$ has been calculated 
for the $q\bar{q}$ initial state with the $gg$ initial state calculation 
in progress (see the contribution of T. Binoth et al).

Final states of such 
complexity usually lead to multi-scale problems, and the correct choice of 
scales to use can be problematic not only at LO, but 
also at NLO. The size of the higher order corrections and of the residual 
scale dependence at NLO can depend strongly on whether the considered 
cross section is inclusive, or whether a jet veto cut has been 
applied~\footnote{The same considerations apply as well to NNLO cross 
sections. See the contribution of G. Dissertori and F. Stoeckli.}. 

Depending on the process, dramatically different behavior can be observed 
upon the application of a jet veto. There is a trade-off between suppressing 
the NLO cross section and increasing the perturbative uncertainty, with 
application of a jet veto sometimes destroying the cancellation between 
infra-red logs of real and virtual origin, and sometimes just suppressing 
large (and very scale-sensitive) tree-level contributions. So far, there is 
no general rule predicting the type of behavior to be expected, but this is 
an important matter for further investigation.

\begin{table}
  \begin{center}
     \begin{tabular}{|l|l|}
\hline \hline
Process ($V\in\{Z,W,\gamma\}$) & Comments\\
\hline
Calculations completed since Les Houches 2005&\\
\hline
&\\
1. $pp\to VV$jet & $WW$jet completed by
Dittmaier/Kallweit/Uwer~\cite{Dittmaier:2007th,Dittmaier:2009un};\\
 &
Campbell/Ellis/Zanderighi~\cite{Campbell:2007ev}.\\ 
 &
$ZZ$jet completed by \\
&
Binoth/Gleisberg/Karg/Kauer/Sanguinetti~\cite{Binoth:2009wk}\\
2. $pp \to$ Higgs+2jets & NLO QCD to the $gg$ channel \\
& completed by Campbell/Ellis/Zanderighi~\cite{Campbell:2006xx};\\
& NLO QCD+EW to the VBF channel\\
& completed by Ciccolini/Denner/Dittmaier~\cite{Ciccolini:2007jr,Ciccolini:2007ec}\\
3. $pp\to V\,V\,V$ & $ZZZ$ completed 
by Lazopoulos/Melnikov/Petriello~\cite{Lazopoulos:2007ix}
 \\
 & and $WWZ$ by Hankele/Zeppenfeld~\cite{Hankele:2007sb}\\
 & (see also Binoth/Ossola/Papadopoulos/Pittau~\cite{Binoth:2008kt})  \\
&\\
4. $pp\to t\bar{t}\,b\bar{b}$ &  relevant for $t\bar{t}H$ computed by\\
 & Bredenstein/Denner/Dittmaier/Pozzorini~\cite{Bredenstein:2009aj,Bredenstein:2010rs} \\
 & and Bevilacqua/Czakon/Papadopoulos/Pittau/Worek~\cite{Bevilacqua:2009zn} \\
5. $pp \to V$+3jets & calculated by the Blackhat/Sherpa~\cite{Berger:2009ep} \\
 & and Rocket~\cite{Ellis:2009zw} collaborations\\
&\\
 \hline 
Calculations remaining from Les Houches 2005&\\
\hline
&\\
6. $pp\to t\bar{t}$+2jets & relevant for $t\bar{t}H$ computed by  \\
& Bevilacqua/Czakon/Papadopoulos/Worek
~\cite{Bevilacqua:2010ve}
\\ 
7. $pp\to VV\,b\bar{b}$,  & relevant for VBF $\rightarrow H\rightarrow VV$,~$t\bar{t}H$ \\
8. $pp\to VV$+2jets  & relevant for VBF $\rightarrow H\rightarrow VV$ \\
& VBF contributions calculated by \\
& (Bozzi/)J\"ager/Oleari/Zeppenfeld~\cite{Jager:2006zc,Jager:2006cp,Bozzi:2007ur}
\\
\hline
NLO calculations added to list in 2007&\\
\hline
&\\
9. $pp\to b\bar{b}b\bar{b}$ & $q\bar{q}$  channel calculated by Golem collaboration~\cite{Binoth:2010pb} \\
&\\
\hline
NLO calculations added to list in 2009&\\
\hline
&\\
10. $pp \to V$+4 jets & top pair production, various new physics signatures\\
11. $pp \to W b \bar{b}j$ & top, new physics signatures\\
12. $pp \to t\bar{t}t\bar{t}$ & various new physics signatures \\
\hline
Calculations beyond NLO added in 2007&\\
\hline
&\\
13. $gg\to W^*W^*$ ${\cal O}(\alpha^2\alpha_s^3)$& backgrounds to Higgs\\
14. NNLO $pp\to t\bar{t}$ & normalization of a benchmark process\\
15. NNLO to VBF and $Z/\gamma$+jet  & Higgs couplings and SM benchmark\\
&\\
\hline 
Calculations including electroweak effects&\\
\hline
&\\
16. NNLO QCD+NLO EW for $W/Z$ & precision calculation of a SM benchmark\\
&\\
\hline
\hline
\end{tabular}
\end{center}
\caption{The updated experimenter's wishlist for LHC processes 
\label{tab:wishlist}}
\end{table}

From the experimental side, an addition to the above 
wish-list that will be crucial is the determination of the accuracy 
to which each of the calculations needs to be known. This is clearly related 
to the experimental accuracy at which the cross sections can be measured at 
the LHC, and can determine, for example, for what processes it may be 
necessary to calculate electo-weak corrections, in addition to the higher 
order QCD corrections. On the theoretical side, it would also be interesting 
to categorize the impact of a jet veto on the size and stability of each 
of the NLO cross sections.

The technology does exist to carry out a calculation for $W/Z$ production
at NNLO (QCD) and at NLO (EW). This process was placed on the 
wish-list in 2007 
and it is unfortunate that the combined calculation 
has not yet been carried out, 
as this precision benchmark will be very useful and important at the LHC. 

To reach full utility, the codes for any of these 
complex NLO calculations should be made public and/or the authors should generate ROOT ntuples 
providing the parton level event information from which experimentalists 
can assemble any cross sections of interest. Where possible, decays (with spin correlations) should be included. A general format for the 
storage of the output of NLO programs in ROOT ntuples was developed at 
Les Houches (see the contribution from J. Campbell et al). The goal is for 
this to be a semi-official standard. Of course the ultimate goal will be 
the ability to link any NLO calculation to a parton shower Monte Carlo. 
A general framework for this linkage, the Binoth Les Houches Accord was 
developed at this workshop~\cite{Binoth:2010xt} and a first 
example of its useage 
is also included in these proceedings (see the contribution of 
J. Archibald et al).

A measurement of Higgs production in the $t\bar{t}H$ channel is important 
for a precision determination of the Higgs Yukawa couplings; for the 
Higgs decay into $b\bar{b}$, the measurement suffers from a sizeable 
background from $t\bar{t}b\bar{b}$ production. Knowledge of the NLO cross 
sections for both signal and background can allow analysis strategies to be 
developed taking into account differences in the two processes, as for example 
differences in the transverse momentum distributions of the $b$ quarks. 
Vetoing on 
the presence of an extra jet reduces the size of the NLO corrections, 
but  at the possible expense of an increase in the scale dependence of the 
resulting exclusive cross sections.The application of a jet veto has a similar 
effect for the size of the NLO corrections for $t\bar{t}H$ as for 
$t\bar{t}b\bar{b}$, but results in a smaller increase in the scale uncertainty.
The use of $\it traditional $ scales for this process, 
such as $m_t + m_{b\bar{b}}/2$, result in a K-factor close to 2, 
suggesting the presence of large logarithms that may spoil the convergence 
of the perturbative expansion. New choices of scale can reduce the size 
of the higher order corrections. 
Two contributions to these proceedings discuss $t\bar{t}b\bar{b}$ 
calculations (see G. Bevilacqua et al, and A. Bredenstein et al).  

There were three additions to the wishlist: 
\begin{itemize}
\item $V(W,Z)$ + 4 jets
\item $Wb\bar{b}j$ (with massive b)
\item $t\bar{t}t\bar{t}$
\end{itemize}
In addition,there is the need/desire to have $Z$ + 3 jets to accompany the
existing calculation of $W$ + 3 jets.

Experimentalists typically deal with leading order (LO) calculations, 
especially in the context of parton shower Monte Carlos. Some of the 
information from a NLO calculation can be encapsulated in the K-factor, 
the ratio of the NLO to LO cross section for a given process, with the 
caveat that the value of the K-factor depends upon a number of variables, 
including the values of the  renormalization and factorization scales, as 
well as the parton distribution functions (PDFs) used at LO and NLO. In 
addition, the NLO corrections often result in a shape change, so that one 
K-factor is not sufficient to describe the impact of the NLO corrections 
on the LO cross section.  
Even with these caveats, it is still useful to calculate the K-factors for 
interesting processes at the Tevatron and LHC. A K-factor table, 
originally shown in the CHS review article~\cite{Campbell:2006wx} and then 
later expanded in the Les Houches 2007 proceedings~\cite{Bern:2008ef}, is 
shown below. The K-factors are shown for several different choices of scale 
and with the use of either LO or NLO PDFs for the LO calculation. Also shown 
are the K-factors when the CTEQ modified LO PDFs are used~\cite{Lai:2009ne}. 

\begin{table}[h]
\begin{center}
\begin{tabular}{|l|l|l|c|c|c|c|c|c|c|}
\hline
  & \multicolumn{2}{|l|}{Fact. scales} & 
 \multicolumn{3}{|c|}{Tevatron K-factor} &
 \multicolumn{4}{|c|}{LHC K-factor} \\ 
  & \multicolumn{2}{|l|}{\quad} & 
 \multicolumn{3}{|c|}{} & \multicolumn{4}{|c|}{} \\
Process & $\mu_0$ & $\mu_1$ &
 ${\cal K}(\mu_0)$ & ${\cal K}(\mu_1)$ & ${\cal K}^\prime(\mu_0)$ &
 ${\cal K}(\mu_0)$ & ${\cal K}(\mu_1)$ & ${\cal K}^\prime(\mu_0)$ &
 ${\cal K}^{\prime\prime}(\mu_0)$  \\
\hline
&&&&&&&&&\\
$W$        & $m_W$ & $2m_W$	&
 1.33 & 1.31 & 1.21 & 1.15 & 1.05 & 1.15 & 0.95 \\
$W$+1 jet          & $m_W$ & $ p_T^{\rm jet}$ &
 1.42 & 1.20 & 1.43 & 1.21 & 1.32 & 1.42 & 0.99\\
$W$+2 jets & $m_W$ & $ p_T^{\rm jet}$	    &
 1.16 & 0.91 & 1.29 & 0.89 & 0.88 & 1.10 & 0.90\\
$WW$+1 jet                & $m_W$ & $2m_W$    &
 1.19 & 1.37 & 1.26 & 1.33 & 1.40 & 1.42 & 1.10\\
$t{\bar t}$        & $m_t$ & $2m_t$         &
 1.08 & 1.31 & 1.24 & 1.40 & 1.59 & 1.19 & 1.09\\
$t{\bar t}$+1 jet        & $m_t$ & $2m_t$    &
 1.13 & 1.43 & 1.37 & 0.97 & 1.29 & 1.10 & 0.85\\
$b{\bar b}$        & $m_b$ & $2m_b$         &
 1.20 & 1.21 & 2.10 & 0.98 & 0.84 & 2.51 & -- \\
Higgs      & $m_H$ & $ p_T^{\rm jet}$       & 
 2.33 & -- & 2.33 & 1.72 & -- & 2.32 & 1.43 \\
Higgs via VBF      & $m_H$ & $ p_T^{\rm jet}$ &
 1.07 & 0.97 & 1.07 & 1.23 & 1.34 & 0.85 & 0.83  \\
Higgs+1 jet     & $m_H$ & $ p_T^{\rm jet}$ &
 2.02 & -- & 2.13 & 1.47 & -- & 1.90 & 1.33\\
Higgs+2 jets     & $m_H$ & $ p_T^{\rm jet}$
 & -- & -- & -- & 1.15 & -- & -- & 1.13 \\
&&&&&&&&&\\
\hline
\end{tabular}
\caption{\label{tab:K-fact}
K-factors for various processes at the LHC (at 14 TeV) calculated using
a selection of input parameters.
In all cases, for NLO calculations, the CTEQ6M PDF set is used.
For LO calculations, ${\cal K}$ uses the CTEQ6L1 set,
whilst ${\cal K}^\prime$ uses the same PDF set, CTEQ6M,
as at NLO, and ${\cal K}^{\prime\prime}$ uses
the LO-MC (2-loop) PDF set CT09MC2.
For Higgs+1 or 2 jets, a jet cut of $40\ \mathrm{GeV}/c$
and $|\eta|<4.5$ has been applied. 
A cut of $p_{T}^{\mathrm{jet}}>20\ \mathrm{GeV}/c$
has been applied to the $t\bar{t}$+jet process,
and a cut of $p_{T}^{\mathrm{jet}}>50\ \mathrm{GeV}/c$ to the $WW$+jet
process. In the $W$(Higgs)+2 jets process, the jets are separated by $\Delta R>0.4$ 
(with $R_{sep}=1.3$), whilst the vector boson fusion (VBF) 
calculations are performed for a Higgs boson of mass $120$~GeV.
In each case the value of the K-factor is compared
at two often-used scale choices, $\mu_0$ and $\mu_1$.}
\end{center}
\end{table}

Several patterns can be observed in the K-factor table. NLO corrections appear
to be larger for processes in which there is a great deal of color 
annihilation, such as $gg \rightarrow$ Higgs in which two color octet gluons produce a 
color singlet Higgs boson. NLO corrections also tend to decrease as more 
final-state legs are added~\footnote{A rule-of-thumb derived by Lance Dixon 
is that the K-factor is often proportional to the factor 
$C_{i1} + C_{i2}  Ð -  C_{f,max}$, where $C_{i1}$ and $C_{i2}$ are the 
Casimir color factors for the initial state and $C_{f,max}$ is the 
Casimir factor for the biggest color representation that the final state 
can be in. Of course, this is not intended to be a rigorous rule, 
just an illustrative one.}. The K-factors at the LHC are similar to the 
K-factors for the same processes at the Tevatron, but have a tendency to be 
smaller.

The cross section for the production a W boson and 3 jets has recently been 
calculated at NLO~\cite{Berger:2009ep}, \cite{KeithEllis:2009bu}. As expected, 
the scale dependence for this cross section shows a monotonic behavior at LO 
and a greatly reduced scale dependence at NLO.   It can be observed that, 
using a scale of $m_W$, the K-factor at the Tevatron is approximately unity, 
while at the LHC it is less than 0.6.

\begin{figure}[ht]
\includegraphics[width=0.8\columnwidth,height=1.6\textheight,keepaspectratio]
{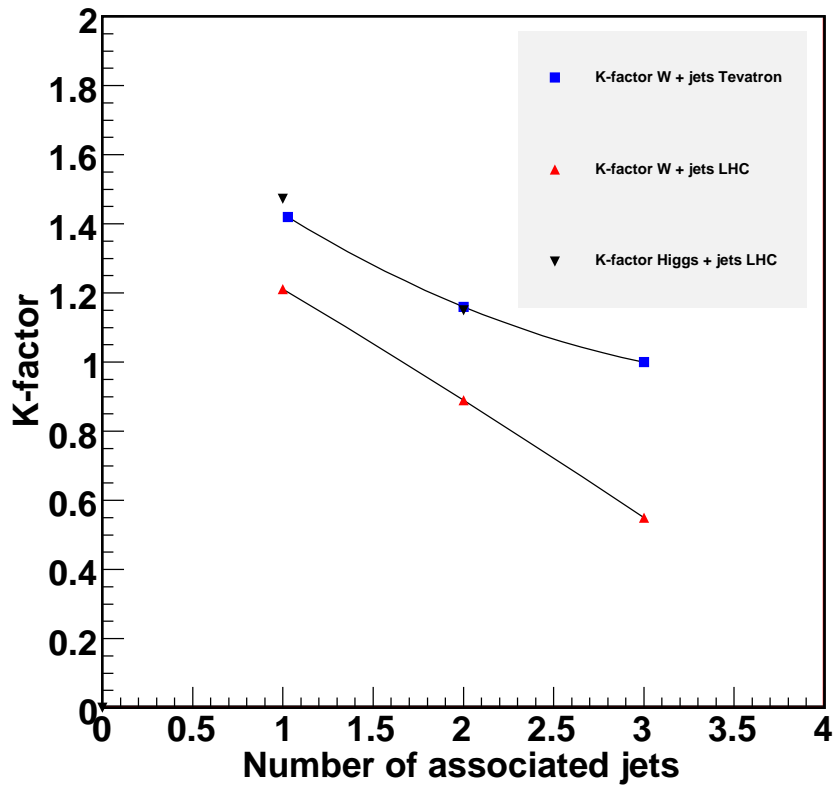}
\caption{The K-factors (NLO/LO) are plotted for $W$ production at the 
Tevatron and LHC and for Higgs production at the LHC as a function of the 
number of accompanying jets. The $k_T$ jet algorithm with a D parameter 
of 0.4 has been used. 
\label{fig:Kfact}}
\end{figure}

The K-factors for W + 1, 2 or 3 jets, at a renormalization/factorization 
scale  of $m_W$, are plotted in Figure \ref{fig:Kfact} 
(along with similar K-factors for Higgs + 1 or 2 jets)~\footnote{For 
these plots, the NLO CTEQ6.6 PDFs~\cite{Nadolsky:2008zw}
have been used with both the LO and NLO matrix elements, in order to 
separate any PDF effects from matrix element effects. If a LO PDF such as 
CTEQ6L1 were used instead, the LO cross sections would shift upwards, but 
the trends would be the same.}~\cite{huston:2010xp}. 
In this plot, a pattern becomes obvious. 
The K-factors appear to decrease linearly as the number of final state 
jets increases, with a similar slope at the Tevatron as at the LHC (but with 
an offset). A similar slope is observed for Higgs  boson+ jets at the LHC. 
To further  understand this pattern (in addition to the color flow argument 
discussed in the previous section), we first have to review jet algorithms 
at LO and NLO. 

At LO, one parton equals one jet. By choosing a jet algorithm with size 
parameter D, we are requiring any two partons to be a distance D or greater 
apart. The matrix elements have $1/\Delta R$ poles, so a larger value of D 
means smaller cross sections. At NLO, there can be two partons in a jet, 
and jets for the first time can have some structure. No $\Delta R$ cut 
is needed since the virtual corrections cancel the collinear singularity 
from the gluon emission (but there are residual logs that can become 
important if the value of D is too small). Increasing the size parameter 
D increases the phase space for including an extra gluon in the jet, 
and thus increases the cross section at NLO (in most cases). The larger 
the number of final state partons, the greater the differences will be 
between the LO and NLO dependence on jet size. The other factors mentioned 
above (such as the color arguments) are also important, but the impact of 
the jet algorithms at LO and NLO will become increasingly important for 
NLO calculations with large numbers of final-state partons. 

It has been observed  that the use of $\it typical$ scales for $W$ + 3 jet 
production at LO can result in substantial shape differences between the 
LO and NLO predictions~\cite{Ellis:2009zw,Berger:2009ep}.
 Conversely, the use of other scales, 
such as the sum of the transverse momentum of all final state objects 
($H_T$), or the use of dynamically generated scales, such as with the 
CKKW or MLM procedures, result in good agreement between the LO and 
NLO predictions. A contribution (see S. Hoche et al) compares the results 
of different scale choices at LO and NLO for this final state 
(as well as being first direct comparison of the Rocket and Blackhat 
calculations). 

An important search channel for Higgs boson production at the LHC is 
Vector Boson Fusion (VBF). The experimental signature consists of the 
presence of two forward-backward tag jets, and the suppression of 
additional hadronic activity in the central rapidity region. A comparison 
was made (in the contribution of A. Denner et al) between NLO predictions 
for this final state and leading order parton shower predictions. 
Differences between the two can partially be taken into account by the
re-weighting of Herwig events, using a weight that depends on the transverse
momentum of the Higgs boson. 

The pattern of gluon radiation for a final state consisting of (1) a  
$W$ boson plus multiple jets (see J. Andersen et al) and (2) a Higgs boson 
plus multiple jets (see J. Andersen et al) was examined in these proceedings, 
with comparison of NLO calculations with LO + parton shower predictions 
and with the predictions of a BFKL Monte Carlo. A universal behaviour is 
observed for a strong correlation between the rapidity span (of the two most 
forward-backward jets) and the jet activity between these two jets. This 
implies that information about jet vetoes in Higgs boson production in 
association with dijets can be extracted from measurements in $W$ boson + 
dijets. 

There was a great deal of discussion at Les Houches, and several resulting 
contributions, on the standardization of parton distribution functions (PDFs) 
and their uncertainties, and a quantitative understanding on reasons for any 
differences among PDFs produced by the major global PDF groups. 
The NNPDF group presented a study on the correlation between $\alpha_s(M_Z)$ 
and the gluon distribution (see R. Ball et al); they found that, at least 
within the NNPDF approach,  the sum in quadrature of PDF and $\alpha_s$ 
uncertainties provides an excellent approximation  of that obtained from an 
exact error propagation treatment. For the case of Higgs boson production 
through $gg$ fusion, similar PDF uncertainties were found for CTEQ6.6 and 
MSTW2008, over a Higgs boson mass range of 100-300 GeV, with slightly larger 
uncertainties from the NNPDF1.2 PDFs. 

A benchmarking study for heavy quark schemes was carried out at and after 
Les Houches (see the contribution of J. Rojo et al). The study compared 
quantitatively different General-Mass Variable Flavour Number  (GM-VFN) 
schemes for the treatment of heavy quarks in deep-inelastic scattering. 
GM-VFN schemes used by the three main global PDF fitting groups were 
compared, and benchmark tables provided for easy comparison with any 
future GM-VFN schemes. 

Although much of the work at Les Houches dealt with the impact of NLO 
corrections on LHC and Tevatron cross sections, many important cross sections
have also been calculated at NNLO, and the extra  order can provide 
additional important information. Higgs searches have been ongoing at the 
Tevatron for many years, with the $gg \rightarrow Higgs(\rightarrow WW)$ 
channel being the most promising in the Higgs mass range of approximately 
twice the $W$ mass. In one contribution, the impact of the NNLO corrections
on the above channel (with the $W$ bosons decaying to leptons) was examined 
(see the contribution of G. Dissertori and F. Stockli). In the contribution
of R. Boughezal, new NNLO predictions for Higgs boson production 
at the Tevatron,taking into account electroweak corrections in addition, 
were summarized. 
The new predictions are typically 4-6\% lower than those previously used
for Tevatron exclusion limits. 
In the contribution of G. Passarino, the complete NLO EW corrections to
the (gg) Higgs production cross section were reviewed. Two schemes for
including higher order EW corrections were discussed and Higgs
pseudo-observables were introduced.

Some of the most 
precise determinations of the strong coupling constant result from 
measurements at $e^+e^-$ colliders, especially LEP. A contribution to Les 
Houches (see G. Dissertori et al) describes a precise determination of the
strong coupling constant based on a NNLO+NNLA analysis of hadronic events 
shapes. 

A  combined theoretical/experimental study of the Frixione photon isolation 
scheme was carried out in the context of the Les Houches workshop. The 
finite size of the calorimeter cells into which fragmentation energy can be 
deposited was taken into account in the Frixione isolation scheme by 
introducing a finite number of nested cones together with the corresponding 
maximal values for the transverse energy allowed inside each of the cones. 
Together with novel techniques for the experimental estimation of the 
amount of underlying event/pileup transverse energy inside the isolation 
cone, this technique will allow for comparisons to theoretical calculations 
where the fragmentation component need not be calculated. 

Finally, we would like to say that this group, and these proceedings, would not have been a 
success without the 
efforts of our colleague and co-convener, Thomas Binoth, who has left us far 
too soon. We would like to dedicate these proceedings to him. 


%% file: badger/badger.tex




\subsection{Introduction}

Key insights of recent years have sparked progress in both analytic and numerical techniques for the computation of multi-particle
one-loop amplitudes. 
Fully analytic computations offer the possibility of extremely fast and accurate evaluation of
NLO cross-sections. Details of the analytic structure of the amplitudes, universal factorisation and cancellation of
unphysical poles play an important role in the development of fully numerical methods. Further interplay between the two
approaches  would be useful to find the most flexible and efficient tools for NLO processes at the LHC.
Achievements of new methods in numerical computation are presented
elsewhere in this report. 
In this section we summarise recent developments in analytic computations.

Most current techniques involve unitarity cuts and generalised cuts, which are evaluated in regions of phase space where loop
propagators are on shell.  The expansion of the amplitude in master integrals with rational coefficients can then be
evaluated algebraically in terms of tree-level quantities.  Unlike in traditional reduction, individual coefficients are
derived independently, based on the  known  analytic structure of the master integrals.  A demonstration of the strength
of analytic methods at one-loop was seen in the full computation of the six-gluon amplitude, whose components have been
helpfully consolidated in \cite{Dunbar:2008zz}.  A recent achievement, which we describe below, is the completion of all helicity amplitudes for
the production of Higgs plus two jets
\cite{Ellis:2005qe,Campbell:2006xx,Badger:2006us,Berger:2006sh,Badger:2007si,Glover:2008ffa,Dixon:2009uk,Badger:2009hw,Badger:2009vh}.

\subsection{INTEGRAL COEFFICIENTS FROM DOUBLE CUTS}

The familiar ``double'' unitarity cuts (two propagators on shell)  yield complete information about large classes of amplitudes.  This cut-constructibility underlies the unitarity method of \cite{Bern:1994zx,Bern:1994cg} for finding coefficients of the master integrals without reducing Feynman integrals.
``Spinor integration'' methods, based on Cauchy's residue theorem applied to the unitarity cut, have recently been used to generate closed-form expressions for the coefficients of scalar master integrals
\cite{Britto:2006fc,Britto:2007tt,Britto:2008vq,Feng:2008ju}.
The first such formulae \cite{Britto:2006fc,Britto:2007tt} were produced for 4-dimensional master integrals in massless theories, starting from tree-level input (the cut integrand) manifesting only physical singularities. From the integrand, the coefficients are obtained through a series of algebraic replacements. 
The formulae have been generalised to $D$-dimensional integrals in \cite{Britto:2008sw} and to scalar masses in \cite{Britto:2008vq}. 

The cut integrand, written analytically as a product of tree-level amplitudes, may be derived in a very compact form using ``MHV diagrams'' \cite{Cachazo:2004kj,Georgiou:2004by,Dixon:2004za,Bern:2004ba,Badger:2004ty,Schwinn:2005pi,Ozeren:2005mp,Boels:2007pj,Schwinn:2008fm} or on-shell recursion relations, particularly in four dimensions with at least two massless particles involved in each tree amplitude \cite{Britto:2004ap,Britto:2005fq,Luo:2005rx,Luo:2005my,Bedford:2005yy,Cachazo:2005ca,Britto:2005dg,Badger:2005zh,Badger:2005jv,Forde:2005ue,Ferrario:2006np,Ozeren:2006ft,Schwinn:2007ee,ArkaniHamed:2008yf,Brandhuber:2008pf,Cheung:2008dn,Drummond:2008cr}.   Extensions to dimensions other than four have been explored in \cite{Quigley:2005cu,Cheung:2009dc,Boels:2009bv,Dennen:2009vk}.  
However, on-shell recursion relations typically feature unphysical singularities, ``spurious poles'', in their individual terms.  In \cite{Feng:2008ju}, the closed form coefficients of \cite{Britto:2008vq} have been generalised to allow any rational functions as cut integrands, in particular with the possible presence of spurious poles.

Current techniques of evaluating unitarity cuts permit numerous variations.  As mentioned above, the cuts may be evaluated by the residue theorem each time, incorporating simplifications depending on the specific forms of the integrands; or the available closed forms for coefficients may be used blindly.  Certainly, there are intermediate and related approaches as well, which are being explored for optimal efficiency, considering also numerical evaluations.  A recent study \cite{Mastrolia:2009dr} frames the double-cut phase space integral in terms of Stokes' theorem, bypassing spinors in favour of a conjugate pair of  complex scalar integration variables.  The cut is evaluated by indefinite integration in one variable followed by Cauchy's residue theorem applied to the conjugate variable.  The cuts of bubble integrals are rational functions, so their coefficients may be extracted algebraically by Hermite polynomial reduction.  It has also been observed that a unitarity cut, viewed as the imaginary part of the loop amplitude, may be interpreted as a Berry phase of the effective momentum space experienced by the two on-shell particles going around the loop \cite{Mastrolia:2009rk}.  The result of the phase-space integration is thus the flux of a 2-form given by the product of the two tree amplitudes on either side of the cut.

\subsection{GENERALISED UNITARITY}

Generalised unitarity has become an essential tool in the computation of one-loop amplitudes over the past two years.
Analytic techniques have focused on generalisations to full QCD amplitudes with arbitrary internal and external masses.

Multiple cuts are well established as an efficient method for the computation of one-loop amplitudes \cite{Bern:1997sc}.
The quadruple cut technique \cite{Britto:2004nc} isolates box coefficients in the one-loop basis, reducing computation
to an algebraic procedure. Forde's Laurent expansion technique \cite{Forde:2007mi} has been widely used in numerical
applications and has also been generalised to the massive case \cite{Kilgore:2007qr}. Further
understanding into the analytic structure has led to the interpretation of the triple cut \cite{BjerrumBohr:2007vu} and double cut
\cite{Mastrolia:2009dr} in terms of Cauchy's and Stokes's Theorem respectively. 

$D$-dimensional cuts with generalised
unitarity have also been applied to analytic computations \cite{Badger:2008cm} using the well known interpretation
of the $D$-dimensional loop integral as a massive vector \cite{Bern:1995db,Giele:2008ve}. In contrast to numerical applications
\cite{Giele:2008ve,Ellis:2008ir}, this allows for a direct computation of the rational contributions without the need to
compute quintuple cuts.

Although the $D$-dimensional cutting method is completely general, in some cases it is preferable to use on-shell
recursion relations for the rational terms \cite{Bern:2005hs}. As long as a suitable analytic continuation can be found which avoids
non-factorising channels, extremely compact analytic forms can be obtained
\cite{Berger:2006ci,Berger:2006vq,Dixon:2009uk,Badger:2007si,Glover:2008ffa}.
Recently combinations of these techniques have been applied in the context of $H+2j$ productions
\cite{Dixon:2009uk,Badger:2009hw,Badger:2009vh} and in preliminary studies of $t\bar{t}$ production
\cite{Badger:2008za}. Since the methods are all completely algebraic, they are particularly suitable for automation with
compact tree-level input taken from on-shell recursion.

For massive one-loop amplitudes, the analytic structure is less understood than in the massless case. In particular, the
addition of wave-function and tadpole contributions introduces complications, as these integrals lack four-dimensional branch cuts in momentum channels.  A recent analysis proposes computing
tadpole coefficients from coefficients of higher-point integrals by introducing an auxiliary, unphysical propagator
\cite{Britto:2009wz}.  The original tadpole integral is then related to an auxiliary integral with two propagators, which can be treated by a conventional double cut.  Relations have been found giving the tadpole coefficients in terms of the bubble coefficients of both the original and auxiliary integrals, and the triangle coefficients of the auxiliary integrals.  The proof of these relations is accomplished with the help of the integrand classification of \cite{Ossola:2006us}.

Single cuts, used in
conjunction with generalised cutting principles, can be an effective method for deriving full QCD amplitudes
\cite{Glover:2008ur}.  A different single-cut method, proposed as an alternative to generalised unitarity cuts, relies on a ``dual'' prescription for the imaginary parts of propagators \cite{Catani:2008xa}.  

\subsection{COMPACT ANALYTIC EXPRESSIONS FOR HIGGS PLUS TWO JETS}

The program of completing the computation of all helicity amplitudes for $H+2j$ production at Hadron colliders as
recently been completed. This allows for a much faster evaluation (about 10 ms for the full colour/helicity sum) of the cross-section previously available from a
semi-numerical computation \cite{Ellis:2005qe,Campbell:2006xx}. A wide variety of the techniques listed above were
employed to ensure a compact analytic form.

The calculation was performed in the large top-mass limit where the Higgs couples to the gluons through an effective
dimension five operator. A complex Higgs field was decomposed into self-dual ($\phi$) and anti-self-dual
($\phi^\dagger$) pieces from which the standard model amplitudes can be constructed from the sum of $\phi$ and parity
symmetric $\phi^\dagger$ amplitudes,
\begin{equation}
	A(H,\{p_k\}) = A(\phi,\{p_k\})+A(\phi^\dagger,\{p_k\}).
\end{equation}
Helicity amplitudes have been calculated using the standard 2-component Weyl spinor representations and written in terms
of spinor products. Results are presented unrenormalised in the four dimensional helicity scheme.

\subsection{Full analytic results}

The full set of amplitudes collects together the work from a number of different groups which we summarise below:

\begin{table}[h!]
	\centering
	\begin{tabular}{|c|c|c|}
	\hline 
	\multicolumn{3}{|c|}{$H\rightarrow gggg$} \\ 
	\hline 
	Helicity & $\phi$ & $\phi^\dagger$ \\
	\hline 
	$----$ & \cite{Badger:2006us} & \cite{Berger:2006sh} \\ 
	\hline 
	$+---$ & \cite{Badger:2009hw} & \cite{Berger:2006sh} \\
	\hline 
	$--++$ & \cite{Badger:2007si} & \cite{Badger:2007si} \\ 
	\hline 
	$-+-+$ & \cite{Glover:2008ffa} & \cite{Glover:2008ffa} \\ 
	\hline 
	\end{tabular}\hspace{1cm}
	\begin{tabular}{|c|c|c|}
	\hline 
	\multicolumn{3}{|c|}{$H\rightarrow \bar{q}qgg$} \\
	\hline 
	Helicity & $\phi$ & $\phi^\dagger$ \\
	\hline 
	$-++-$ & \cite{Dixon:2009uk} & \cite{Dixon:2009uk} \\
	\hline 
	$-+-+$ & \cite{Dixon:2009uk} & \cite{Dixon:2009uk} \\
	\hline 
	$-+--$ & \cite{Badger:2009vh} & \cite{Berger:2006sh} \\
	\hline
	\multicolumn{3}{c}{} \\
	\end{tabular}
	\caption{The set of independent $\phi$ and $\phi^\dagger$ helicity amplitudes contributing to $H+2j$ production
	together with the references where they can be obtained.}
\end{table}

The analytic form of the four quark squared amplitude was presented in the original semi-numerical computation
\cite{Ellis:2005qe}. The helicity amplitudes for this process were computed in reference \cite{Dixon:2009uk}. The results where obtained
using 4-dimensional cutting techniques for the cut-constructible parts. Where applicable on-shell recursion relations
gave a compact representation of the rational terms. For the most complicated NMHV configuration and the ``all-minus"
configuration non-factorising channels in the complex plane were unavoidable and on-shell recursion was not possible. In
these cases extremely compact forms were obtained from Feynman diagrams after all information from unphysical poles in
the cut-constructible part had been accounted for. It was essential to make full use of the universal IR pole structure
in addition to information coming from spurious poles in the logarithmic part.

This calculation relied on some non-trivial relations between terms in the amplitude:
\begin{itemize}
	\item The rational terms in the $\phi gggg$ amplitude obey:
		\begin{eqnarray}
			&&\mathcal{R}\left\{A_{4;1}(\phi;1_g,2_g,3_g,4_g)\right\} = 	
			\left(1-\frac{N_f}{N_c}+\frac{N_s}{N_c}\right)R^{N_p}(\phi;1_g,2_g,3_g,4_g) \nonumber\\&&
			+ 2\left(A_4^{(0)}(\phi,1_g,2_g,3_g,4_g)-A_4^{(0)}(\phi^\dagger,1_g,2_g,3_g,4_g)\right)
		\end{eqnarray}
	\item The rational terms in the $\phi\bar{q}qgg$ amplitude obey:
		\begin{eqnarray}
			&&\mathcal{R}\bigg\{A_{4}^{L}(\phi;1_{\bar{q}},2_q,3_g,4_g)+A_{4}^{R}(\phi;1_{\bar{q}},2_q,3_g,4_g)
			+A_{4}^{f}(\phi;1_{\bar{q}},2_q,3_g,4_g)\bigg\}
			\nonumber\\&&= 
			2\left(A_4^{(0)}(\phi,1_{\bar{q}},2_q,3_g,4_g)-A_4^{(0)}(\phi^\dagger,1_{\bar{q}},2_q,3_g,4_g)\right)
		\end{eqnarray}
	\item The sub-leading colour amplitudes in the $H\bar{q}qgg$ amplitude are completely determined from the leading
		singularities.
\end{itemize}
The identities are strongly reminiscent of cancellations seen in SUSY decompositions of pure QCD amplitudes except that
they are broken by a universal factor proportional to the tree-level $\phi$ and $\phi^\dagger$ amplitudes.

As an example we present the colour ordered amplitude for the most complicated ``NMHV" configuration in the $Hgggg$
channel \cite{Badger:2009hw}. The Feynman diagram representation of this amplitude consists of 739 diagrams with up to
rank 4 tensor pentagon integrals. This leading colour amplitude is sufficient to give the full colour information when
summed over the appropriate permutations, we refer the reader to \cite{Badger:2009hw} for further details. 
\begin{eqnarray}
	A^{(1)}_{4;1}(H,1^+,2^-,3^-,4^-) &=& 
	-A^{(0)}_{4}(H,1^+,2^-,3^-,4^-)\sum_{i=1}^4 \frac{1}{\epsilon^2}\left(\frac{\mu_R^2}{-s_{i,i+1}} \right)^\epsilon 
	\nonumber\\
	&+& F_4(H,1^+,2^-,3^-,4^-) + R_4(H,1^+,2^-,3^-,4^-)
\end{eqnarray}
where
\begin{eqnarray}
	&&A^{(0)}_4(H,1^+,2^-,3^-,4^-) = \nonumber\\
	&& - \frac{m_H^4 \langle 24\rangle ^4}{s_{124} \langle 12\rangle  \langle 14\rangle \langle 2|p_H|3] \langle 4|p_H|3]}
	+\frac{ \langle 4|p_H|1]^3}{s_{123} \langle 4|p_H|3] [12] [23]}
	-\frac{ \langle 2|p_H|1]^3}{s_{134} \langle 2|p_H|3] [14] [34]},
\end{eqnarray}
and
\begin{eqnarray}
&&F_4(H,1^+,2^-,3^-,4^-)= \bigg\{
\frac{1}{4s_{124}}\bigg(\frac{\langle3|p_{H}|1]^4}{\langle3|p_{H}|2]\langle3|p_{H}|4][21][41]}+\frac{\langle24\rangle^4m_{H}^4}{\langle12\rangle\langle14\rangle\langle2|p_{H}|3]\langle4|p_{H}|3]}
	\bigg)W^{(3)}\nonumber\\&&
-\frac{s_{234}^3}{4\langle1|p_{H}|2]\langle1|p_{H}|4][23][34]}W^{(1)}
-\bigg(\frac{\langle2|p_{H}|1]^3}{2s_{134}\langle2|p_{H}|3][34][41]} 
	+\frac{\langle34\rangle^3m_{H}^4}{2s_{134}\langle1|p_{H}|2]\langle3|p_{H}|2]\langle41\rangle}
	\bigg)W^{(2)}\nonumber\\&&
	+2C_{3;H|12|34}^{\rm 3m}(1^+,2^-,3^-,4^-){\rm I_3^{3m}}(m_{H}^2,s_{12},s_{34})
+\left(1-\frac{N_f}{4N_c}\right)\bigg(
\frac{\langle3|p_H|1]^2}{s_{124}[24]^2} {\rm F^{1m}}(s_{12},s_{14};s_{124})\nonumber\\&&
-\frac{4\langle24\rangle\langle3|p_H|1]^2}{s_{124}[42]}\hat{L}_1\left(s_{124},s_{12}\right)
+\frac{4\langle23\rangle\langle4|p_H|1]^2}{s_{123}[32]}\hat{L}_1\left(s_{123},s_{12}\right)
\bigg)-\bigg(1-\frac{N_f}{N_c}+\frac{N_s}{N_c}\bigg)\times\nonumber\\&&\bigg(
\frac{[12][41]\langle3|p_H|2]\langle3|p_H|4]}{2s_{124}[24]^4}{\rm F^{1m}}(s_{12},s_{14};s_{124})
+\left(\frac{2s_{124}\langle34\rangle^2[41]^2}{\langle24\rangle[42]^3}-\frac{\langle24\rangle\langle3|p_H|1]^2}{3s_{124}[42]}\right)\hat{L}_1\left(s_{124},s_{12}\right)\nonumber\\&&
+\frac{2s_{124}\langle24\rangle\langle34\rangle^2[41]^2}{3[42]}\hat{L}_3\left(s_{124},s_{12}\right)
+\frac{\langle34\rangle[41]\left(3s_{124}\langle34\rangle[41]+\langle24\rangle\langle3|p_H|1][42]\right)}{3[42]^2}\hat{L}_2\left(s_{124},s_{12}\right)\nonumber\\&&
+\frac{\langle3|p_H|1](4s_{124}\langle34\rangle[41]+\langle3|p_H|1](2s_{14}+s_{24})}{s_{124}\langle24\rangle[42]^3}\hat{L}_0\left(s_{124},s_{12}\right)
-\frac{2s_{123}\langle23\rangle\langle34\rangle^2[31]^2}{3[32]}\hat{L}_3\left(s_{123},s_{12}\right)\nonumber\\&&
+\frac{\langle23\rangle\langle34\rangle[31]\langle4|p_H|1]}{3[32]}\hat{L}_2\left(s_{123},s_{12}\right)
+\frac{\langle23\rangle\langle4|p_H|1]^2}{3s_{123}[32]}\hat{L}_1\left(s_{123},s_{12}\right)
\bigg)\bigg\}
+\bigg\{(2 \leftrightarrow 4)\bigg\}.
\end{eqnarray}
For convenience we have introduced the following combinations of the finite pieces of one-mass $({\rm F^{1m}})$ and two-mass hard $({\rm F^{2mh}})$ box functions, 
\begin{eqnarray}
W^{(1)}&=&{\rm F^{1m}}(s_{23},s_{34};s_{234})+{\rm F^{2mh}}(s_{41},s_{234};m_H^2,s_{23})+{\rm F^{2mh}}(s_{12},s_{234};s_{34},m_H^2)\nonumber \\
W^{(2)}&=&{\rm F^{1m}}(s_{14},s_{34};s_{134})+{\rm F^{2mh}}(s_{12},s_{134};m_H^2,s_{34})+{\rm F^{2mh}}(s_{23},s_{134};s_{14},m_H^2)\nonumber \\
W^{(3)}&=&{\rm F^{1m}}(s_{12},s_{14};s_{124})+{\rm F^{2mh}}(s_{23},s_{124};m_H^2,s_{14})+{\rm F^{2mh}}(s_{34},s_{124};s_{12},m_H^2)\nonumber .
\end{eqnarray}
The bubble coefficients have been re-arranged into logarithm functions, $L_k=\frac{\log(s/t)}{(s-t)^k}$, which have smooth behaviour in the
various collinear limits,
\begin{eqnarray}
\hat{L}_3(s,t)&=&L_3(s,t)-\frac{1}{2(s-t)^2}\bigg(\frac{1}{s}+\frac{1}{t}\bigg)\quad,\quad
\hat{L}_1(s,t)=L_1(s,t), \nonumber \\
\hat{L}_2(s,t)&=&L_2(s,t)-\frac{1}{2(s-t)}\bigg(\frac{1}{s}+\frac{1}{t}\bigg)\quad,\quad
\hat{L}_0(s,t)=L_0(s,t).
\end{eqnarray}
Representations for the scalar integrals can be found in the literature
\cite{Bern:1993kr,Denner:1991qq,vanHameren:2005ed,Ellis:2007qk}. The three mass triangle coefficient was
obtained from Forde's Laurent expansion procedure \cite{Forde:2007mi},
\begin{eqnarray}
	C_{3;H|12|34}^{\rm 3m}(1^+,2^-,3^-,4^-) = 
	\sum_{\gamma=\gamma_\pm(p_H,p_1+p_2)}
	\frac{-m_\phi^4\langle K^\flat_1 2 \rangle^3\langle34\rangle^3}{2\gamma(\gamma+m_\phi^2)\langle{K^\flat_1}{1}\rangle\langle{K^\flat_1}{3}\rangle\langle{K^\flat_1}{4}\rangle\langle
	12\rangle},
\end{eqnarray}
where $K_1=p_H,K_2=p_1+p_2$, and
\begin{eqnarray}
	K^{\flat,\mu}_1 &=& \gamma\frac{\gamma K_1^{\mu}-K_1^2K_2^{\mu}}{\gamma^2-K_1^2K_2^2},\quad
	K^{\flat,\mu}_2 = \gamma\frac{\gamma K_2^{\mu}-K_2^2K_1^{\mu}}{\gamma^2-K_1^2K_2^2}, 
\nonumber \\
	\gamma_\pm(K_1,K_2) &=& K_1\cdot K_2 \pm \sqrt{ K_1\cdot K_2^2-K_1^2K_2^2}.
\end{eqnarray}
The rational part (which incorporates the rational $A^{(1)}_4(\phi^{\dagger},1^+,2^-,3^-,4^-)$  amplitude derived
in~\cite{Berger:2006sh}) is
\begin{eqnarray}
R_4(H,1^+,2^-,3^-,4^-)=&&\bigg\{\bigg(1-\frac{N_f}{N_c}+\frac{N_s}{N_c}\bigg)\frac{1}{2}\bigg(\frac{\langle 23\rangle\langle34\rangle\langle4|p_{H}|1][31]}{3s_{123}\langle12\rangle[21][32]}
-\frac{\langle3|p_{H}|1]^2}{s_{124}[42]^2}\nonumber\\&&+\frac{\langle24\rangle\langle34\rangle\langle3|p_{H}|1][41]}{3s_{124}s_{12}[42]}
-\frac{[12]^2\langle23\rangle^2}{s_{14}[42]^2}-\frac{\langle24\rangle(s_{23}s_{24}+s_{23}s_{34}+s_{24}s_{34})}{3\langle12\rangle\langle14\rangle[23][34][42]}\nonumber\\&&
+\frac{\langle2|p_{H}|1]\langle4|p_{H}|1]}{3s_{234}[23][34]}-\frac{2[12]\langle23\rangle[31]^2}{3[23]^2[41][34]}  \bigg) \bigg\} + \bigg\{(2\leftrightarrow 4)\bigg\}.
\end{eqnarray}

Further study into the origin of the simplicity in the sub-leading colour amplitudes would be interesting and may shed
light on possible cancellations in other processes \cite{Badger:2008rn}. The full results for all helicity configurations have been made available at
{\tt http://mcfm.fnal.gov}.



%% file: lazopoulos/lazopoulos.tex
%



\subsection{Introduction}
\label{intro}
Until recently, the computational bottleneck in next-to-leading order QCD calculations has been the evaluation of the virtual part of the partonic cross section. The mainstream technology for evaluating one loop integrals corresponding to one loop Feynman diagrams has been to project the contribution of each individual Feynman diagram on a complete basis of scalar master integrals that are known analytically.  There are many generic frameworks implemented that have been used the last years in very demanding calculations (see 
for example \cite{Denner:2006fy,Denner:2005nn,Denner:2002ii}  or the publically available  \cite{Binoth:2008uq}.

Alternative approaches based on unitarity have been employed extensively in the past to recover one loop amplitudes (see e.g. \cite{Bern:1994cg,Bern:1994zx}). Following a generalization of the unitarity idea to quadruple cuts \cite{Britto:2004nc} and a novel approach in reduction that allows one to partially fraction numerically one loop integrals at the integrand level on a per point base \cite{Ossola:2006us, Ossola:2007ax}, the numerical evaluation of one loop amplitudes using tree-level quantities as building blocks appeared to be possible and efficient enough to tackle processes with many final state partons. Since then there have been mainly three approaches developed along the lines of numerical reduction with tree-level objects as building blocks: the Black Hat approach \cite{Berger:2006cz,Berger:2008sj}(based also on the reduction method of \cite{Forde:2007mi}, implemeneted in \cite{Berger:2008sj,Berger:2008ag}), the D-dimensional unitarity approach \cite{Ellis:2007br,Giele:2008ve,Ellis:2008kd,Ellis:2008ir} (implemented in various independent codes~
\cite{Giele:2008bc, Winter:2009kd, Melnikov:2009dn, Ellis:2008qc, Lazopoulos:2008ex} ) 
and the helac-1loop approach\cite{Ossola:2007ax,Ossola:2008xq,Ossola:2007zz,Ossola:2008zza,Ossola:2008zzb,Draggiotis:2009yb,Garzelli:2009is} (implemented in \cite{vanHameren:2009dr}). All three approaches have already delivered  differential cross sections for processes of prime phenomenological importance \cite{Berger:2008sz,Berger:2009zg,Berger:2009ba,Berger:2009ep,Berger:2009xp,Bevilacqua:2009zn,Ossola:2007bb, Melnikov:2009dn, KeithEllis:2009bu, Melnikov:2009wh, Ellis:2009zw}. 

In what follows I report on the progress of a generic implementation of the D-dimensional unitarity approach, emphasizing the particular points where my implementation differs from the published mainstream algorithm. 

\subsection{The main algorithm and variations.}
Within the framework of dimensional regularization, one loop integrals are analytically continued in $D=4-2\epsilon$ dimensions. The particular way one treats this analytic continuation with respect to the degrees of freedom, $D_s$ of the unobserved particles that circulate in the loop defines the regularization scheme used.   The fundamental idea of (generalized) D-dimensional unitarity\footnote{This is not meant to be a complete, or even, a stand alone description of the D-dimensional unitarity algorithm. The reader that is not familiar with the details or the formalism, can find them in \cite{Ellis:2007br,Giele:2008ve,Ellis:2008kd,Ellis:2008ir}} is the observation that the amplitude depends on $D_s$ in a linear way, when the loop is not purely fermionic:

\begin{equation}
A^{D_s}=A_0+D_s A_1
\end{equation}

When the loop is purely fermionic, the colour-ordered amplitude is just proportional to $2^{D_s/2}$:
\begin{equation}
A^{D_s}=2^{D_s/2}A_0
\end{equation}

The strategy is to evaluate $A^{D_s}$ numerically for two integer dimensions $D_s$, extract $A_0$ and $A_1$. The full $D_s$ dependence of $A^{D_s}$ is then recovered and the amplitude can be evaluated in the regularization scheme of preference.

The lowest values of $D_s$ one needs to accommodate  fermions are $D_s=6$ and $D_s=8$. Then 
\begin{equation}
A_0=4A^6-3A^8\;\;\;\;A_1=\frac{A^8-A^6}{2}
\end{equation}

Choosing the FDH scheme ($D_s=4$), the amplitude takes the form
\begin{equation}
A^{FDH} = 2A^6-A^8
\end{equation}

One would now normally set up two OPP systems, to evaluate $A^6$ and $A^8$ using the corresponding residues throughout. Instead, thanks to the linearity of the OPP system, one can evaluate directly\footnote{One can, of course, choose the scheme of ones preference.} $A^{FDH}$. The residues that appear on the left hand side of the OPP equations would now correspond to the difference $2 A^6|_{res} - A^8|{res}$. Potential cancelations between the two contributions, which would have propagated in the coefficients of the OPP system, are now prevented.

The residue of the amplitude with respect to a given multiple cut is recognized \cite{Ellis:2007br} to be a product of tree-level amplitudes sewn together with a sum over polarization states of the particles corresponding to the cut propagators:
\begin{equation}
A^{D_s}|_{res} = \sum_{\lambda_1,\ldots \lambda_N} (\bar{w}^{\lambda_1}_{\mu_1} M^{\mu_1\mu_2} w_{\mu_2}^{\lambda_2})
( \bar{w}_{\mu_2}^{\lambda_2} M^{\mu_2\mu_3}w_{\mu_3}^{\lambda_3})\ldots
(\bar{w}^{\lambda_N}_{\mu_N}M^{\mu_N\mu_1}w_{\mu_1}^{\lambda_1})
\label{original_sum}
\end{equation}
where $w_{\mu_i}^{\lambda^i}$ is the wave function of the cut particle (eg. a spinor for a fermion or a polarization vector for a gluon) corresponding to helicity $\lambda^i$. The tree level amplitudes are obtained via the Berends-Giele recursion relation. This, in effect, means that the current  $J_{12}^{\mu_2}\equiv w^{\lambda_1}_{\mu_1} M^{\mu_1\mu_2}$ is evaluated numerically and then multiplied by the external wave function $\bar{w}_{\mu_2}^{\lambda_2}$ to get the amplitude as a complex number. A rather trivial but vastly simplifying observation is that one can perform the polarization sums in $N-1$ of the cut propagators and use the spin projectors to join the Berends-Giele currents. We then have
\begin{equation}
J_{k,k+1}^{\mu_{k+1}} = \tilde{J}_{k-1,k;\mu_k} M^{\mu_k\mu_{k+1}}\;\;\;\;\;\;\; \tilde{J}_{k-1,k;\mu_k} \equiv J_{k-1,k}^{\nu}D_{\nu,\mu_k}(p_k)
\end{equation}
with $D_{\nu,\mu_k}(p_k)$ the spin projector for the cut particle $k$ carrying (loop) momentum $p_k$. This transforms the multiple sum of eq.~\ref{original_sum} in a single sum over the polarization states of a single cut particle. One can see this schematically as a ring with a single polarization sum, similar (but not identical) to the way the calculation of residues is organized in the approach of \cite{vanHameren:2009dr}. As a side remark, choosing to sum over polarization states of another propagator would provide a non-trivial numerical test for the evaluation of the cut.
 
Regarding the computational burden of evaluating the rational part in extra dimensions,  it is clear, from the representation of the Dirac algebra in $6$ and $8$ dimensions, and the fact that the loop momenta chosen are always restricted to $5$ dimensions, that $A^8|_{res}=A^6|_{res}+\hat{A}^8|_{res}$. In particular, in terms of the rings described above, $\hat{A}^8|_{res}=\sum_{h=5\ldots h_{max}^8} R_{h}$ where  $h^8_{max}=6$ when the cut line is a gluon, and   $h^8_{max}=8$ when the cut line is a fermion. Obviously it is advantageous to cut a gluon line when this is possible.

\subsection{A note on numerical stability}

Since D-dimensional unitarity involves pentuple cuts, it is exposed to a direct criticism regarding the numerical stability of the method, 
not only as far as the calculation of the rational part is concerned, but also in connection to the cut constructible part, which, in other methods, is evaluated
in strictly four dimensions. In particular, pentagon coefficients, carrying inverse Gram determinants up to the fifth power, are potential 
sources of precision loss in phase space points that are close to configurations with co-planar momenta. The detection of such problematic
points can be achieved with either comparing the evaluated pole coefficients (in an $\epsilon$-expansion of the amplitude) with those known 
from analytical formulas, or by checking redundant individual OPP systems for consistency. Both methods are used in the implementation described here.

The percentage of such problematic points depends on the number and type of external particles as well as the particular phase space cuts imposed. The most direct attitude towards this issue (aside from just discarding those points) is to evaluate them in double-double ($32$ digits) or quadruple ($64$ digits) precision using libraries available in the literature. Even though the percentage of problematic points is always less than $5\%$, the exact rate matters, since the penalty in terms of cpu time paid for the increased precision arithmetics can reach a factor of $100$. 

An alternative approach that drastically increases precision for the cut constructible part of the amplitude (without losing information necessary for the rational part or evaluating extra residues) and slightly improves the behavior of the rational part is described in detail in \cite{Lazopoulos:2009zn}. The basic idea is to separate the pentagon contributions throughout the OPP system in another, separately solved system of equations. The remaining system contains the four-dimensional OPP system for the cut-constructible part. Moreover, the factorized pentagon coefficients can be manipulated easily, taking care to avoid numerical cancellations.   

\subsection{Summary}

The present implementation of D-dimensional unitarity is restricted to amplitudes relevant for NLO QCD calculations with massless particles. In terms of color ordered primitives, the full gluonic primitive, primitives with one or more fermion lines and primitives with closed fermion loops are all implemented and checked either by verifying their singularity structure or against published results. The implementation is independent of the number of external particles. Further details about primitives not available in the literature will be given in a forthcoming publication. The numerical stability of the algorithm is enhanced since the effect of large pentagon coefficients is reduced. A number of modifications from the main algorithm that were described above, help to improve the accuracy as well as the computational complexity of the method.



%% file: garzelli/garzelli.tex
\newcommand{\bqa}{\begin{eqnarray}}
\newcommand{\eqa}{\end{eqnarray}}
\newcommand{\bea}{\begin{eqnarray}}
\newcommand{\eea}{\end{eqnarray}}
\def\tld#1{\tilde {#1}}
\def\slh#1{ \rlap/{#1}}
\def\db#1{\bar D_{#1}}
%





The automatic computation of beyond-leading-order amplitudes for
processes involving
multiparticle production in the framework of the 
$\mathrm{SU(3)_C \times SU(2)_W \times U(1)_Y}$ Standard
Model (SM) of interactions,
is one of the goals of the present developments in
Theoretical Physics at High-Energy Colliders. 
The demand for (at least) NLO predictions for partonic processes is an 
established issue~\cite{Binoth:2009fk}, 
since processes with many external legs (jets) are expected at the 
LHC~\cite{Alwall:2007fs} and LO predictions can suffer of uncertainties,
like renormalization and factorization scale dependencies, 
which can increase with the number of external legs.  
Thus, recently, some of the current available Matrix-Element MonteCarlo (MC)
event generators, such as HELAC/PHEGAS~\cite{Cafarella:2007pc} and 
SHERPA~\cite{Gleisberg:2003xi,Gleisberg:2008ta},
used even by the experimental community
in the prediction of inclusive and exclusive observables 
of interest,
have been interfaced with proper codes for the automatic 
evaluation of 1-loop correction contributions~\cite{vanHameren:2009dr,Berger:2009ba}. 
This work has just started,
and, even if it has already allowed to get theoretical 
predictions for most of the specific
processes and sub-processes suggested by the 2007 Les Houches 
wishlist~\cite{Bern:2008ef}, 
has not yet achieved  
the stage of a completely automatic and systematic evaluation. 

Even if interesting results (see e.g.~\cite{Bredenstein:2009aj}) 
have been obtained also in the 
context of the traditional Passarino-Veltman reduction technique~\cite{Passarino:1978jh},  
many of the methods recently developed for 
the evaluation of virtual contributions at NLO~\cite{Berger:2009zb}, 
that have pushed crucial progresses
in this field, are based on Unitarity 
arguments~\cite{Bern:1993mq,Bern:1994fz,Bern:1997sc}.
Due to the unitarity of the S-matrix, it is always possible to
express any 1-loop amplitude, for any given number of external legs,
as a linear combination of 1-loop scalar functions up to 4 external 
legs (boxes, triangles, bubbles and, eventually, tadpoles), i.e.
the cut-constructible (CC) part, plus a residual piece, called the
rational part (R).  Since libraries  for the evaluation of these 
well-known scalar integrals 
exist~\cite{vanHameren:2005ed,Ellis:2007qk}, 
the problem of determining a
1-loop amplitude is reduced to the one of determining the 
coefficients of the combination and of evaluating the residual piece. 
One of the methods developed so far for the determination
of these coefficients is the OPP reduction~\cite{Ossola:2006us}. 
According to it, the coefficients
of the expansion of the amplitude in terms of scalar integrals
are obtained by working at the integrand level in a completely
algebraic way, indipendently of the theory of interactions at hand,
i.e. the method can be applied to any renormalizable gauge theory. 
Both massless and massive particles can be included in the loop.
The method does not require to introduce any additional integer dimension 
beyond the four we are used to consider in special relativity.

As for the residual piece, in the OPP framework, two kinds of R
terms have been recognized, denoted by ${\rm R_1}$ and $\rm{R_2}$, 
respectively~\cite{Ossola:2008xq}. 
The ${\rm R_1}$ terms come essentially from   
the D-dimensional dependence of the denominators appearing
in the integrands of 1-loop amplitudes, when one applies a
dimensional regularization procedure (working in D = $4+\epsilon$ dim.)
to calculate the amplitudes themselves. In fact, in the OPP
reduction procedure aimed to obtain the CC part, the denominators
are treated as 4-dimensional, by neglecting their D-dimensional dependence. 
The $\rm{R_2}$ terms, instead, 
come from the $\epsilon$-dimensional part of the numerators.

In the OPP framework, the ${\rm R_1}$ contribution can be reconstructed at the
same time of the CC part, without needing to know the
analytical structure of the amplitude, i.e. the contributing
Feynman diagrams. In other words, to build the CC part and the ${\rm R_1}$
part of the amplitudes, it is enough to know the numerator of the
integrand in four dimensions as a numerical function of the loop momentum $q$, 
and this numerator $N(q)$ can already include the sum of many contributions, 
corresponding to different loop particle contributions, given a fixed structure
of external legs. 

On the other hand, the numerical knowledge of $N(q)$ is not enough to
extrapolate the $\epsilon$-dependence of the numerator, i.e. the $\rm{R_2}$
terms have to be calculated separately. 
The strategy we have adopted so far to build $\rm{R_2}$ 
is based on the consideration that the divergencies appearing in the 
$\rm{R_2}$ integrands that have to be regularized have a completely 
ultraviolet (UV) nature~\cite{Binoth:2006hk}. Thus, irreducible $\rm{R_2}$ 
contributions up to four external legs are enough to build the total 
$\rm{R_2}$ for processes with any number of external legs. 
Since the number of $\rm{R_2}$ building-blocks is finite, it is possible 
to write them in the form of effective Feynman rules, derived
for each of the theories of particle interactions at hand.
We have first derived these Feynman rules in the framework of
QCD, where the number of involved particles is relatively low, thus
only a few effective vertices exist~\cite{Draggiotis:2009yb}. 
We have provided these rules
both in the 't Hooft-Feynman gauge and in the Four Dimensional
Helicity (FDH) scheme, which
is a particular case of the previous one, corresponding to selecting
only a part of the rational terms ($\epsilon$ is set to 0 before integration).
The rules have been derived analytically by using FORM~\cite{Vermaseren:2000nd,Vermaseren:2008kw}. 
We have then extended the list of effective Feynman rules, by
adding the ones for 1-loop QCD corrections to Electroweak
(EW) amplitudes, the ones for EW corrections to QCD amplitudes 
and, finally, the ones for EW corrections to EW processes~\cite{Garzelli:2009is}. 

We have tested our analytical results, by considering the
fact that $\rm{R_2}$ is not gauge invariant by itself, 
but the sum ${\rm R_1}$+$\rm{R_2}$ has to fulfill this constraint, 
i.e. a given set of Ward identities~\cite{Bardin:1999ak} 
has to be satisfied.
We have derived these Ward identities in the formalism of the
Background Field Method (BFM)~\cite{Denner:1994xt,Denner:1995jd}.
We have derived analytical formulas for ${\rm R_1}$ terms up to 4-external
legs, and we have verified that the sums ${\rm R_1}$+$\rm{R_2}$ satisfy the
Ward identities. This can be considered a non-trivial test, since
the analytical expressions of ${\rm R_1}$ effective vertices
are in general much more complicated,
by involving many terms, with denominators including different
combinations of the momenta of the external particles.
It is worth mentioning that, at the aim
of calculating ${\rm R_1}$ for an arbitrary process, it is not possible to 
apply 
a procedure, based on Feynman rules, 
analogous to the one used to calculate $\rm{R_2}$, since, 
in case of many external legs, one cannot rely on ${\rm R_1}$ contributions 
up to 4-external legs only.
Furthermore, the analytical structure of the ${\rm R_1}$ terms becomes soon
very complicated by increasing the number of external legs, thus,
in general, 
it is easier to proceed with a numerical evaluation of ${\rm R_1}$.

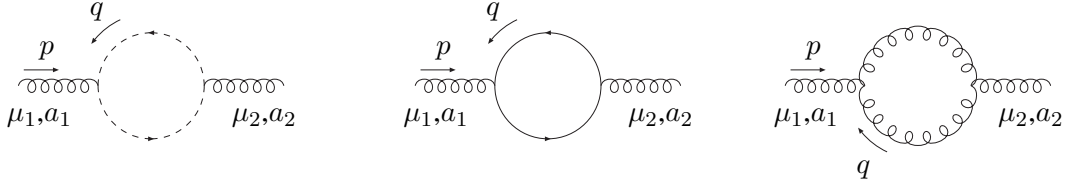
\begin{figure}[t]
\begin{center}
\begin{picture}(300,51)
\SetScale{0.5}
\SetOffset(-70,0)
\DashArrowArc(150,50)(40,0,180){5}
\DashArrowArc(150,50)(40,180,360){5}
\Gluon(50,50)(110,50){5}{5}
\Gluon(190,50)(250,50){5}{5}
\LongArrow(54,62)(79,62)
\LongArrowArc(150,50)(55,115,145)
\Text(33,35)[bl]{$p$}
\Text(52,50)[bl]{$q$}
\Text(33,13)[]{$\mu_1$,$a_1$}
\Text(118,13)[]{$\mu_2$,$a_2$}
\SetOffset(80,0)
\ArrowArc(150,50)(40,0,180)
\ArrowArc(150,50)(40,180,360)
\Gluon(50,50)(110,50){5}{5}
\Gluon(190,50)(250,50){5}{5}
\LongArrow(54,62)(79,62)
\LongArrowArc(150,50)(55,115,145)
\Text(33,35)[bl]{$p$}
\Text(52,50)[bl]{$q$}
\Text(33,13)[]{$\mu_1$,$a_1$}
\Text(118,13)[]{$\mu_2$,$a_2$}
\SetOffset(220,0)
\GlueArc(150,50)(40,0,180){5}{8}
\GlueArc(150,50)(40,180,360){5}{8}
\Gluon(50,50)(110,50){5}{5}
\Gluon(190,50)(250,50){5}{5}
\LongArrow(54,62)(79,62)
\LongArrowArcn(150,50)(55,-115,-145)
\Text(33,35)[bl]{$p$}
\Text(52,-10)[bl]{$q$}
\Text(33,13)[]{$\mu_1$,$a_1$}
\Text(118,13)[]{$\mu_2$,$a_2$}
\end{picture}
\end{center}
\caption{\em Diagrams contributing to the gluon self-energy.}
\label{selfg}
\end{figure}

As an explicit and simple example of the procedure we have adopted
to derive the
${\rm{R_2}}$ effective Feynman rules, we detail the calculation of
${\rm{R_2}}$ coming from the gluon self-energy.
The contributing diagrams 
are drawn in fig.~\ref{selfg}.

As for the ghost loop with 2 external gluons, we can write
the numerator as
\begin{eqnarray}
\bar{N}(\bar{q}) = \frac{g^2}{(2\pi)^4} f^{a_1bc}\,
f^{a_2cb}\,(p+\bar{q})^{\mu_1}
\bar{q}^{\mu_2}
\,.
\end{eqnarray}
In the previous equation, dimensional regularization is assumed, so that
we use a bar to denote objects living
in $D=~4+\epsilon$ dimensions.
Notice that, when a $D$-dimensional index
is contracted with a 4-dimensional (observable) vector
$v_\mu$, the $4$-dimensional part is automatically selected. For example,
\bqa
\label{noeps}
\bar q \cdot v \equiv (q+ {\tld q}) \cdot v\,= q \cdot v\,~~~{\rm and}~~~
\rlap/ {\bar v} \equiv  \bar \gamma_{\bar \mu}\, v^\mu = \rlap /v\, ,
\eqa
where we have used a tilde to represent $\epsilon$-dimensional
quantities.
Since $\mu_1$ and $\mu_2$ are external Lorentz indices, that are
eventually contracted with 4-dimensional external currents, their
$\epsilon$-dimensional component is killed due to
eq.~\ref{noeps}. Therefore, ${\rm{R_2}}= 0$ for this diagram, being
the $\epsilon$-dependent part of the numerator
$\tld{N}(\tld{q}^2,q,\epsilon)= 0$.
With analogous arguments, one easily shows that
ghost loops never contribute to ${\rm{R_2}}$, even in case of 3 or 4
external gluons. This is not only the case of QCD ghosts, but
even the one of EW ghosts that enter EW loops.  
In general, loops of ghost particles give instead a non-vanishing
contribution to ${\rm R_1}$.  

The contribution due to $N_f$ quark loops is given by
the second diagram of fig.~\ref{selfg}, whose numerator reads
\bqa
\bar N(\bar q) &=& -\frac{g^2}{(2 \pi)^4} N_f \frac{\delta_{a_1a_2}}{2}\,
Tr[{\gamma}^{{\mu_1}}(\rlap/{\bar{q}}+m_q) {\gamma}^{{\mu_2}}
(\rlap/{\bar{q}}+\rlap/{p}+m_q)]\,,
\eqa
where the external indices $\mu_1$ and $\mu_2$ have been directly taken
in 4 dimensions.
By anti-commuting ${\gamma}^{{\mu_2}}$ and $\rlap/{\bar{q}}$ 
and using the fact that,
due to Lorentz invariance, odd powers of ${\tld{q}}$ do not contribute, 
one immediately get the result
\bqa
\label{ntilda2}
\tld{N}({\tld{q}}^2)= \frac{g^2}{8 \pi^4}\, N_f\,
\delta_{a_1a_2}\,
g_{\mu_1 \mu_2} {\tld q}^2\,.
\eqa
Eq.~\ref{ntilda2}, integrated with the help of the 
eq.
\bqa
\int d^n\bar{q} \frac{\tld{q}^2}{\db{i}\db{j}}  & = & - \frac{i\pi^2}{2}
\left[m_i^2 + m_j^2 - \frac{(p_i-p_j)^2}{3}\right] 
+ O (\epsilon) \,, 
\eqa
where $\bar{D}_i=(p_i + \bar{q})^2-m_i^2$,
gives the term proportional to $N_f$ in the 2-point effective vertex of
fig.~\ref{effectivevertices}.
\begin{figure}[ht]
\begin{center}
\begin{picture}(300,51)
\SetScale{0.5}
\SetOffset(-50,0)
\LongArrow(47,62)(73,62)
\Text(27,35)[bl]{$p$}
\Text(17,15)[]{$\mu_1$,$a_1$}
\Text(68,15)[]{$\mu_2$,$a_2$}
\Gluon(35,50)(85,50){5}{5}
\Gluon(85,50)(135,50){5}{5}
\GCirc(85,50){6}{0}
\Text(90,25)[l]{$\displaystyle = \frac{i g^2 N_{col}}{48 \pi^2} \,
\delta_{a_1a_2}\, \Bigl[\,\frac{p^2}{2} g_{\mu_1\mu_2}
+\lambda_{HV}\,\Bigl( g_{\mu_1\mu_2} p^2-p_{\mu_1} p_{\mu_2}\Bigr) $}
\Text(170,-10)[l]{$\displaystyle
+\frac{N_f}{N_{col}}\,(p^2-6\, m_q^2)\,g_{\mu_1\mu_2}\Bigr]$}
\end{picture}
\end{center}

\caption{\label{effectivevertices} $\rm{R_2}$ gluon-gluon effective vertex.
$\lambda_{HV}= 1$ in the 't Hooft-Feynman scheme and
$\lambda_{HV}= 0$ in the FDH scheme.
$N_{col}$ is the number of colors and $N_f$ is the number of fermions running
in the quark loop.
}
\end{figure}

The effective rules providing the QCD NLO $\rm{R_2}$ corrections have already
been implemented in a numerical code based on 
tensor reduction~\cite{vanHameren:2009vq}. Furthermore, they have been 
used in a unitary context by the HELAC-NLO system,
in the study of NLO processes 
like $pp \rightarrow t  \bar{t} b \bar{b}$~\cite{Bevilacqua:2009zn} and 
$pp \rightarrow t \bar{t} H$, with the Higgs boson subsequently 
decaying in $b \bar{b}$~\cite{tth}. At the purpose of calculating the total
$\rm{R_2}$ contribution to a physical process,
these Feynman rules have to be considered on the same footing as 
the standard ones. The only constraint in using them
is that in each tree-level like effective Feynman 
diagram contributing to $\rm{R_2}$, one and only one $\rm{R_2}$ effective 
vertex has to be included.

One of the advantages of the incorporation of the Feynman rules we have derived,
in numerical codes for the evaluation of SM amplitudes at NLO,
is the fact that the CPU time needed to compute $\rm{R_2}$  
becomes, in practice, very low. This is important
if one considers that, by using other codes based on Unitarity or
Generalized Unitarity methods, such as Blackhat~\cite{Berger:2008sj} 
and Rocket~\cite{Giele:2008bc}, the
time necessary for the computation of R is comparable or even
longer than the time necessary to derive the CC part of the amplitude.
In particular, it is interesting to observe that while all these methods
retrieve the CC part of the amplitude in more or less similar ways, 
the procedure to build R is very different. In the framework of the 
Generalized Unitarity
techniques~\cite{Giele:2008ve,Ellis:2008ir}, 
for instance, the rational terms are calculated using
the same reduction procedure used to calculate the CC part of the
amplitude, at the price of working in more than 4 integer dimensions.
If, from one hand, this is an elegant solution allowing to treat in
a unified way the CC and R part, on the other, it
requires to work in a number of integer dimensions larger than 4, 
and, thus, to introduce e.g. 
proper spinor representations in more than 4 dimensions 
and so on.
 
We think that in general the R part of the amplitudes deserves
more attention, since at present it is the less understood part of the 1-loop
virtual corrections, and a careful comparison between different methods
to obtain it could help in the attempt of better understanding the 
nature and the origin of the rational terms 
and to improve the computational strategies to calculate them.

\section*{ACKNOWLEDGEMENTS}
We acknowledge the collaboration with R.~Pittau. We thank
C.~Papadopoulos, R.~Kleiss, S.~Dittmaier, F.~Piccinini and 
M.~Moretti for useful discussions.
The work of M.V.G. is supported by the italian INFN. The work of
I.M. is supported by the RTN European Program MRTN-CT-2006-035505
(HEPTOOLS, Tools and Precision Calculations for Physics Discoveries
at Colliders). M.V.G. enjoyed the stimulating atmosphere of the
Les Houches Workshop, where part of this work was carried out.  


%% file: reiter/reiter.tex
\subsection{INTRODUCTION}\label{golem_main:sec:introduction}
The ability to calculate processes with multi-particle final states
beyond leading order will be a necessary requirement to describe
the signals and backgrounds with a precision that allows
to study new physics at the
LHC~\cite{:2008uu,Bern:2008ef,Buttar:2006zd,Campbell:2006wx}.
One of the challenges of next-to-leading
order~(NLO) calculations is the numerically stable evaluation and
integration of the virtual corrections. Both the development of new
unitarity based methods and the improvement of traditional methods
based on Feynman diagrams have led to new results for NLO
predictions, as reported by several groups in these proceedings.
We want to stress the importance of
automatisation for such computations, as it reduces the time spent
on a single calculation.
Moreover, automated NLO programs can be combined 
with existing tools for tree level matrix element generation and 
phase space integration via a standard interface, thus providing 
flexible NLO tools  which can be made publicly~available.

The Golem approach to the calculation of one-loop matrix elements
can be summarised as a Feynman diagrammatic expansion of
helicity amplitudes with a semi-numerical reduction of the tensor
integrals~\cite{Binoth:2005ff,Binoth:2008uq}.
This method produces a fast, analytic representation
for the matrix element and works for processes involving massive and massless
particles. The algorithm for the reduction of the tensor integrals
extracts all infrared divergences analytically in terms of triangles;
its implementation, which is described in
Section~\ref{golem_main:sec:golem95}, switches between the
analytic reduction of tensor integrals and their numeric evaluation;
this is a special feature of the Golem approach
which preserves numerical stability in phase space regions of small
Gram determinants.
Working entirely with Feynman diagrams, we generate the rational terms of an
amplitude at no extra cost together with the cut-constructable~parts.

Section~\ref{golem_main:sec:golem-2.0} describes our current implementation
of the Golem method and in Section~\ref{golem_main:sec:results:qqbbbb} we
present results which have been achieved recently using our~formalism.

\subsection{Reduction of Tensor Integrals with \texttt{golem95}}
\label{golem_main:sec:golem95}
In~\cite{Binoth:2005ff,Binoth:1999sp} we describe an algorithm for the reduction
of one-loop tensor integrals which works for an arbitrary number of
legs, both massive and massless. This algorithm has been implemented
as a Fortran\,90 library, \texttt{golem95},
for integrals with up to six external momenta~\cite{Binoth:2008uq}
and massless~propagators.

The algebraic reduction of higher rank four-point  and
three-point functions to expressions containing only scalar  integrals 
necessarily leads to inverse 
Gram determinants appearing in the coefficients of
those scalar integrals. These determinants can become arbitrarily small and
can therefore hamper a numerically stable evaluation of the one-loop
amplitude. Our algorithm avoids a full reduction in phase space regions
where a Gram determinant becomes small. 
In these cases the tensor integrals, corresponding to integrals with
Feynman parameters in the numerator,  are evaluated by means of numerical
integration. The use of one-dimensional integral representations hereby
guarantees a fast and stable~evaluation.

We have recently extended the library \texttt{golem95} to the case
with internal masses. All infrared divergent integrals have been implemented 
explicitly. For the finite boxes and triangles,
LoopTools~\cite{Hahn:1998yk,Hahn:2000kx,vanOldenborgh:1989wn} 
needs to be linked. 
This ``massive" version of the golem95 library is currently 
in the testing phase and will be available shortly at 
\texttt{lappweb.in2p3.fr/lapth/Golem/golem95.html}.

For integrals with internal masses, the option to evaluate the tensor integrals 
numerically prior to reduction 
in regions where the Gram determinant tends to zero, is not yet 
supported. 
However, one-dimensional integral representations valid for all 
possible kinematic configurations are under~construction.

\subsection{Towards an Automated One-Loop Matrix Element Generator}
\label{golem_main:sec:golem-2.0}
We have implemented the Golem formalism into a one-loop matrix element
generator based on helicity projections of Feynman diagrams. This
program, currently called \texttt{golem\,2{.}0},
has been successfully applied in the calculation of
the process $q\bar{q}\rightarrow b\bar{b}b\bar{b}$, which is described
in Section~\ref{golem_main:sec:results:qqbbbb}. A future version of our
matrix element generator will support the standard interface to Monte-Carlo
tools~\cite{Binoth:2010xt}. Using this interface, a seamless integration
into existing Monte Carlo tools becomes~possible.

The implementation in form of a Python package uses
QGraf~\cite{Nogueira:1991ex} to generate all tree and one-loop
diagrams for a given process. On top of the Standard Model, our package
supports the import of model files in the CompHEP~\cite{Pukhov:1999gg}
table format;
an interface with FeynRules~\cite{Christensen:2008py}
is currently under~development.

The output of the diagram generator is simplified algebraically
using Form~\cite{Vermaseren:2000nd} and the Form library
\texttt{spinney}~\cite{Spinney:not-yet-published} which
adds the required functionality for manipulating helicity spinors.
We use this library at many places in the simplification
process, such as
\begin{itemize}
\item the translation from QGraf to Form. The elements of a spinor chain
   in QGraf are not generated in the order in which the elements appear
   along the fermion line. Instead, the correct order can be restored
   by considering explicit spinor indices as in~$(\gamma^\mu)_{\alpha\beta}$.
\item the application of flipping rules~\cite{Denner:1992vza,Denner:1992me}
   for the correct treatment of fermion number violating interactions
   as in models containing Majorana fermions.
\item carrying out the numerator algebra. We use the 't\ Hooft-Veltman
   scheme and dimension splitting for dealing with an $n$-dimensional
   numerator algebra. The relevant formulae have been worked out and
   implemented in~\texttt{spinney}.
\item the contraction of Lorentz indices by means of Chisholm identities.
   These identities have originally been formulated for Dirac traces
   but have been extended in \texttt{spinney} to the case of
   spinor~chains.
\end{itemize}

\noindent
After the simplification by Form, each diagram of a helicity amplitude is
expressed entirely in terms of spinor products, model parameters and
integral form factors as defined in \texttt{golem95}.
Diagrams containing four-gluon vertices are represented as a sum over
colour subamplitudes.
This representation is in principle suitable for numerical evaluation
but we optimise the expression in several steps to improve speed, size
and numerical stability of the resulting Fortran program.
First of all, our Form code factors out coefficients common to all terms
in the diagram and introduces abbreviations for products of spinor~products.

\begin{figure}[hbt]
\begin{center}
\includegraphics[width=0.30\textwidth]{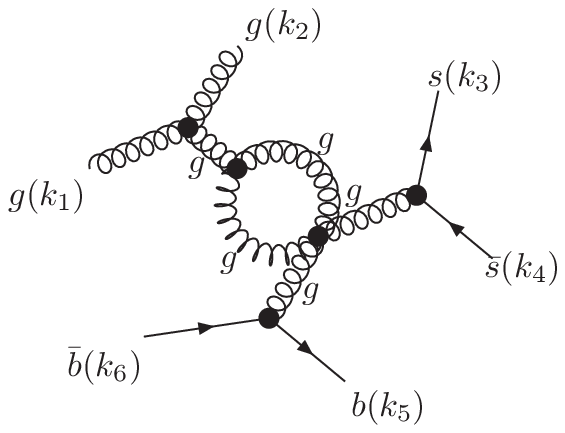}
\includegraphics[width=0.30\textwidth]{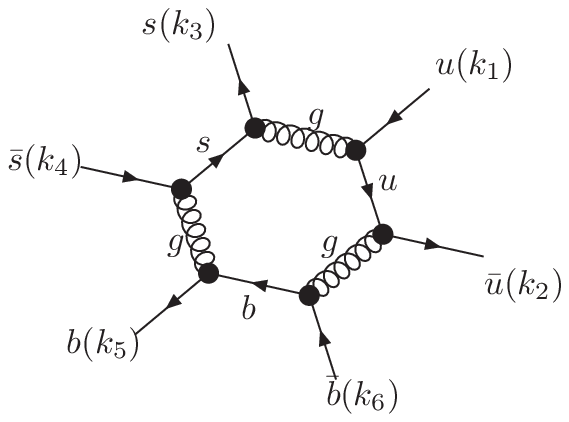}
\includegraphics[width=0.30\textwidth]{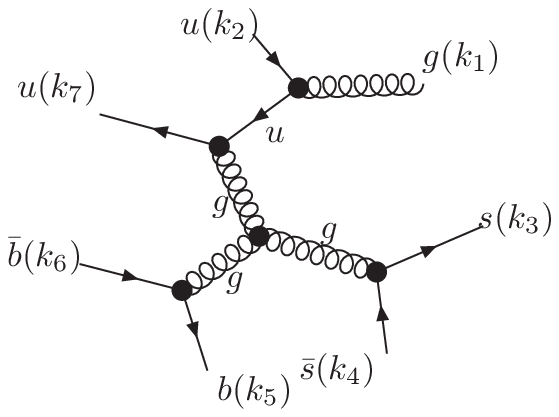}
\end{center}
\caption{Three diagrams contributing to different subprocesses
of $pp\rightarrow s\bar{s}b\bar{b}$ at NLO in QCD. The diagrams
have been drawn automatically by \texttt{golem-2{.}0} using
\LaTeX{} and AxoDraw~\cite{Vermaseren:1994je}.
}\label{golem_main:fig:golem-feyn}
\end{figure}

In the next step we use the optimising code generator
\texttt{haggies}~\cite{Reiter:2009ts} for generating efficient
Fortran\,90 code for each diagram. The program \texttt{haggies}
combines a multivariate Horner scheme~\cite{Ceberio:2003ga} and
common coefficient extraction~\cite{Gopalakrishnan:2009at} with
common subexpression elimination and a linear search strategy
for variable allocation~\cite{Poletto:1999els}. Its built in
type checker allows one to build up expressions from different
built-in and derived data types. The combination of these strategies
optimises an expression with respect to the number of multiplications
and function calls required for its~evaluation. As an example,
we consider the hexagon diagram in Figure~\ref{golem_main:fig:golem-feyn}:
the Form output consists of 535 terms for one specific
helicity amplitude,
requiring 1100 multiplications for its evaluation. The program
generated by \texttt{haggies} evaluates the same expression using
585 multiplications, saving roughly fifty percent of the required
operations. Similarly, we process the sum of the
64 tree diagrams contributing to $qg\rightarrow s\bar{s}b\bar{b}q$
(See Fig.~\ref{golem_main:fig:golem-feyn}, right).
One of the non-vanishing helicity amplitudes
would require 12,279 multiplications and 5,128 additions
before optimisation, whereas
\texttt{haggies} produces a program that evaluates the result with
only 2,166 multiplications and 687 additions, saving more than 80\% of
all arithmetic operations.

The program \texttt{haggies} works independent from the input and
output language and is very flexible in the output format. Its design
incorporates the possibility of writing code for object oriented
programming languages, even if they do not support operator overloading.
The program is therefore well suited for many problems both within and 
outside high energy physics. It is publicly available under
\texttt{http://www.nikhef.nl/\~{}thomasr/download.php}\,.

In the code generated by \texttt{golem-2{.}0},
the one-loop integrals are evaluated by the \texttt{golem95} library.
Its internal, recursive structure uses a cache for storing function
evaluations which are required in different form factors belonging
to the same diagram topology.
We improved the performance of the numerical evaluation of the
one-loop amplitude further by relating diagrams of which the
loop propagators are contained in the set of loop propagators of
another diagram. The form factors of the most difficult diagram of
one topology can be reused for all pinched diagrams of the same topology
by using the internal cache of~\texttt{golem95}.

Besides the numerical code, the package is also
capable of producing a description of the process in \LaTeX{}
including all contributing Feynman diagrams drawn with
AxoDraw~\cite{Vermaseren:1994je} combined with an implementation
of the layout algorithm proposed in~\cite{Ohl:1995kr}
(See Figure~\ref{golem_main:fig:golem-feyn}).
We also plan to provide an interface to OPP like tensor reduction
algorithms~\cite{Ossola:2006us,Ossola:2007ax} where our program
provides the numerator function $N(q,\tilde{q}^2)$ of diagrams
or subamplitudes. Since our approach treats the numerator $n$-dimensional
it provides the full $\tilde{q}^2$ dependence which can be used for the
reconstruction of the rational term~$R_2$.

\subsection{The quark induced case of $pp\rightarrow b\bar{b}b\bar{b}$}
\label{golem_main:sec:results:qqbbbb}
For Higgs searches in many models beyond the Standard Model~(BSM),
processes with high multiplicities of $b$-quarks in the final state
become relevant.
For example, in a large part of
the parameter space of the Minimal Supersymmetric Standard Model (MSSM),
the light Higgs boson decays predominantly into $b\bar{b}$~pairs.
This opens up the possibility to measure the $hhH$ coupling at the
LHC through the process
$gg\rightarrow H\rightarrow hh\rightarrow b\bar{b}b\bar{b}$.
Experimental studies show, however, that such a measurement would be
extremely difficult, primarily due to the overwhelming QCD
background~\cite{Lafaye:2000ec}.
Another example where the $b\bar{b}b\bar{b}$ final state becomes important
are hidden valley models where the decay of exotic hadrons leads to
high multiplicities of $b$-quark.
The precise knowledge of the $b\bar{b}b\bar{b}$ final state
within the Standard Model is therefore an important factor
for the success of these measurements. The calculation of the NLO
corrections in $\alpha_s$ reduces the scale dependence of the prediction
and therefore greatly improves the precision of this prediction.
Here, we present the calculation of
$q\bar{q}\rightarrow b\bar{b}b\bar{b}$ with $q\in\{u,d,s,c\}$,
which is a subprocess of the reaction $pp\rightarrow b\bar{b}b\bar{b}$.

For the calculation of the virtual part of the amplitude we have applied
the setup as discussed in Section~\ref{golem_main:sec:golem-2.0}.
We have confirmed it by an independent implementation based on
FeynArts and FormCalc~\cite{Hahn:1998yk} and a symbolical reduction of the
tensor integrals to scalar integrals using the formalism described
in~\cite{Binoth:2005ff}.

The real corrections and the Born level amplitude as well as the
phase space integration of all parts have been computed
with MadGraph, MadEvent~\cite{Stelzer:1994ta,Maltoni:2002qb}
and MadDipole~\cite{Frederix:2008hu} and independently using
an adapted version of Whizard~\cite{Kilian:2007gr}.
In both cases the infrared singularities are treated by the subtraction
of Catani-Seymour dipoles~\cite{Catani:1996vz} with the improvements suggested
in~\cite{Nagy:2005gn}.

For all parts of the calculation we have used two independent implementations.
Additional checks such as the cancellation of the infrared
divergences, the symmetries of the amplitude and the independence on the
slicing parameter in the dipoles~\cite{Nagy:2005gn} have been performed
in order to ensure the correctness of our~results.

\begin{figure}[hbt]
\begin{center}
\includegraphics[width=0.45\textwidth]{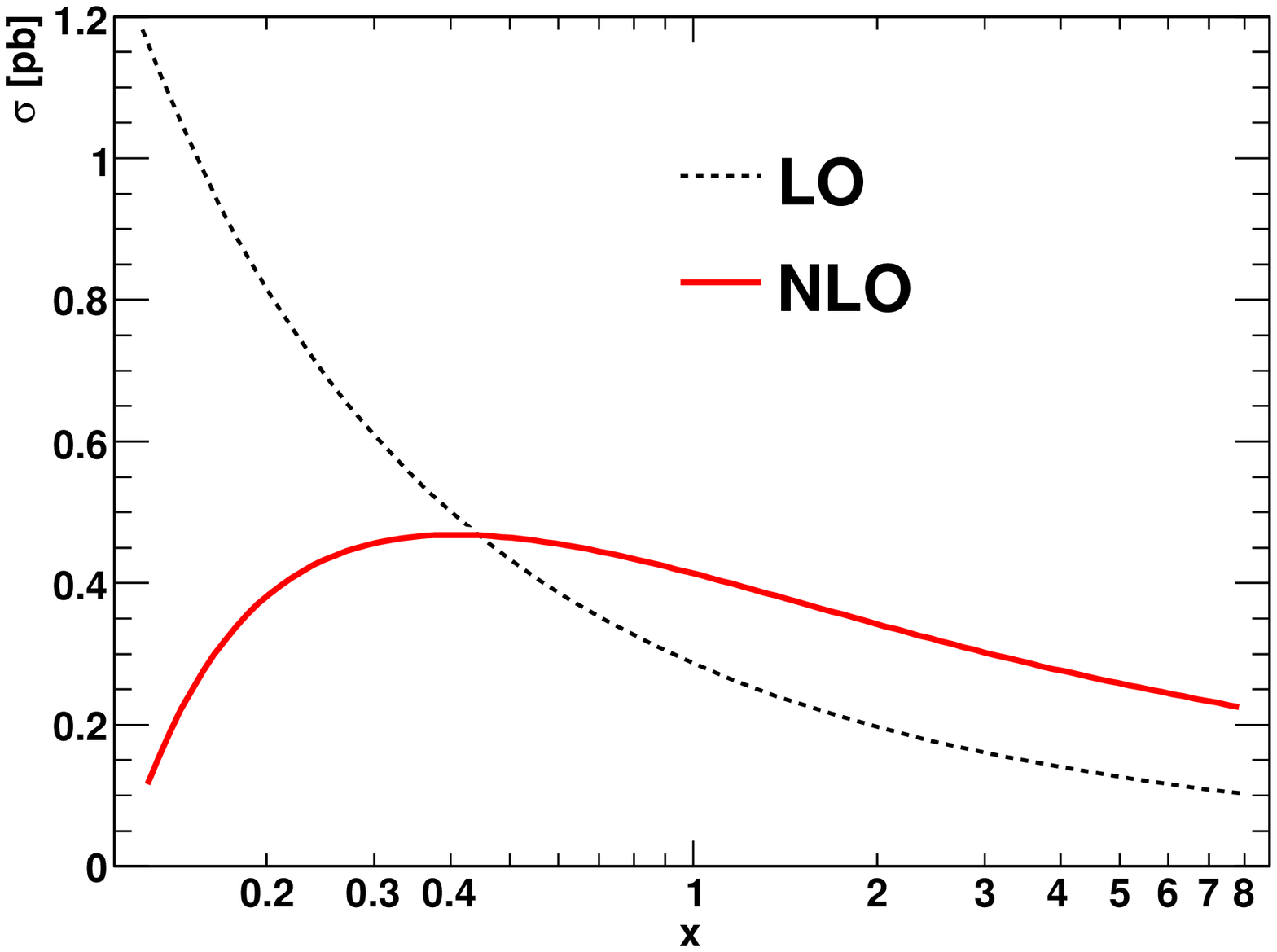}
\includegraphics[width=0.45\textwidth]{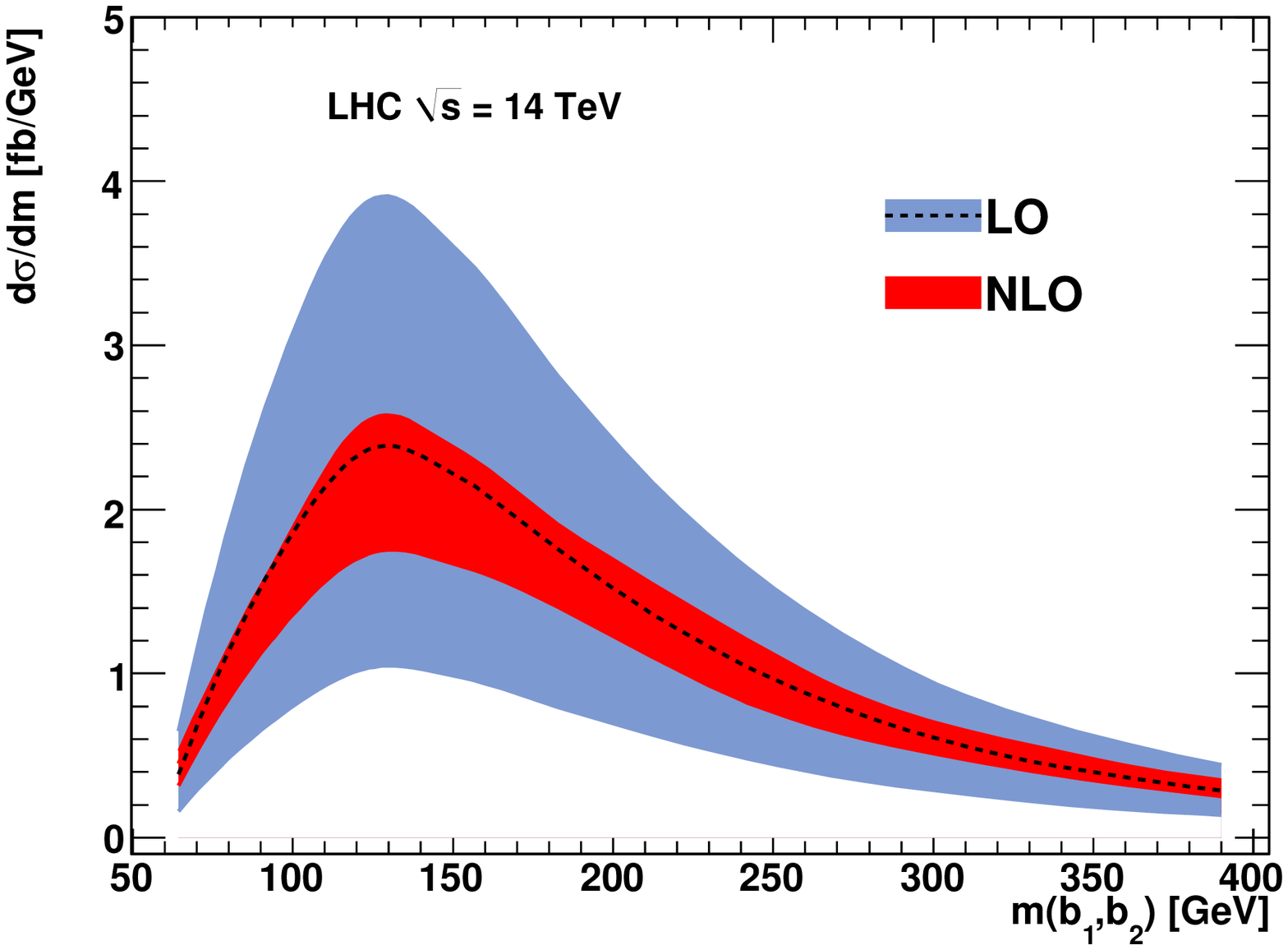}
\end{center}
\caption{
$pp(q\bar{q})\rightarrow b\bar{b}b\bar{b}+X$
at the LHC ($\sqrt{s}=14\;\mathrm{TeV}$).
(left)
The dependence of the cross-section on the renormalisation
scale $\mu_R=x\mu_0$ at fixed value of the factorisation
scale~$\mu_F=100\;\mathrm{GeV}$.
(right)
Invariant mass ($m_{bb}$) distribution of the two
leading $b$-jets (see text). The error bands are obtained from
a variation of the renormalisation scale $\mu_R$ between
$\mu_0/4$ and $2\mu_0$ with $\mu_0=\sqrt{\sum_j p_T^2(b_j)}$.
The dashed line shows the LO prediction for $\mu_R=\mu_0/2$.
} 
\label{golem_main:fig:qqbbbb-plots}
\end{figure}

Figure~\ref{golem_main:fig:qqbbbb-plots} shows some results obtained for
the LHC. We use a centre-of-mass energy of $\sqrt{s}=14\;\mathrm{TeV}$
and impose the following cuts on transverse momentum, rapidity and separation
$\Delta R(b_i,b_j)=\sqrt{(\Delta\Phi_{ij})^2+(\Delta\eta_{ij})^2}$:
\begin{equation}
\begin{array}{rcl}
p_T(b_i)&>&30\;\mathrm{GeV}\\
\vert\eta(b_i)\vert&<&2.5\\
\Delta R(b_i,b_j)&>&0.8
\end{array}
\end{equation}
Before cuts the $K_T$ algorithm~\cite{Blazey:2000qt} is applied to decide
if the extra gluon in the real emission part of the process can be resolved.
In the case of an unresolved gluon the momentum of the merged $b$-$g$-pair
$p_{b_i}+p_g$ is used as the momentum of the $b$-jet. For the initial state
we convolve with the $u$, $d$, $c$ and $s$-quark parton distribution
functions of the CTEQ6M set~\cite{Pumplin:2002vw}
with two-loop running of $\alpha_s$ for the
LO and the NLO part of the amplitude. We work in the limit $m_b=0$ and
$m_t\rightarrow\infty$.

The left plot of Figure~\ref{golem_main:fig:qqbbbb-plots} shows the
dependence of the total cross-section on the variation of the renormalisation
scale while the factorisation scale is fixed at~$\mu_F=100\;\mathrm{GeV}$.
The NLO curve clearly shows the expected improvement of the scale dependence
with a plateau region around a central value of $\mu_0/2$,
where $\mu_0$ has been defined as~$\mu_0=\sqrt{\sum_j p_T^2(b_j)}$.

The right plot shows the invariant mass distribution of the leading\footnote{
``leading'' is defined in terms of an ordering with respect to $p_T(b_j)$.}
two $b$-jets. The error bands obtained from a variation of the renormalisation
scale in the interval $\mu_0/4<\mu_R<2\mu_0$ confirm the expected
reduction of the renormalisation scale dependence.

\subsection*{CONCLUSIONS}
We have reported on the recent progress of the Golem collaboration.
We have started the development of a one-loop matrix element generator,
called \texttt{golem-2{.}0},
based on the Feynman diagrammatic expansion of helicity amplitudes.
This program has now been used to provide results for the order $\alpha_s$
virtual corrections of the amplitude $q\bar{q}\rightarrow b\bar{b}b\bar{b}$.
We have combined this program with MadGraph/MadEvent and Whizard to obtain
a complete next-to-leading order result for this~process. The calculation
of the remaining channels of $pp\rightarrow b\bar{b}b\bar{b}+X$ is
in~preparation.

The one-loop integral library \texttt{golem95} has been extended to the
case of massive loop propagators, which is necessary for many Standard
Model processes. Moreover, together with the interface
between \texttt{golem-2{.}0} and Feynman rule generators
such as FeynRules and LanHEP~\cite{Semenov:2008jy} we extended the
applicability of our programs to BSM~physics.

We focus on making the matrix element generator \texttt{golem-2{.}0}
publicly available
as a plug-in into existing Monte Carlo event generators, using the
Les Houches standard interface for one-loop programs~\cite{Binoth:2010xt},
which should be accomplished within the next six~months.

Our recent work demonstrates the potential of Feynman diagrams
and the ability of our approach to handle processes with up to six
external particles efficiently and in a highly automated manner
and gives confidence that even higher particle multiplicities are
within~reach.

\subsection*{ACKNOWLEDGEMENTS}
We would like to thank Thomas Gehrmann for interesting and useful discussions. 
T.B.~thanks the University of Freiburg and Nikhef in Amsterdam, 
where parts of this work have been done, for their hospitality. 
T.B.~is supported in parts by STFC, SUPA and the IPPP Durham. 
N.G.~thanks the Universities Edinburgh, Freiburg and the IPPP in Durham
for supporting various visits where work on the project was done.
N.G.~was supported by the Swiss National Science Foundation
(SNF) under contract 200020-126691.
G.H.~and G.C.~thank Nikhef for their hospitality.
G.H. and M.R. are supported by STFC.
N.K.~thanks the Higher Education Funding Council for England (HEFCE)
and the Science and Technology Facilities Council (STFC)
for financial support under the SEPnet Initiative.
IPPP Associates T.B.~and N.K.~thank the
Institute for Particle Physics Phenomenology (IPPP) Durham for support.
T.B.~and N.K.~thank the Galileo Galilei Institute for Theoretical
Physics for the hospitality and the INFN for partial support during the
completion of this work. 
T.R.~thanks the University of Z\"urich and
J.R.~wants to thank the Aspen and Les Houches Centers of Physics     
for their hospitality.
J.R.~was partially supported by the Baden-Wuerttemberg Ministry of Science 
and Culture. Part of the computations were done on the ECDF cluster at the 
University of Edinburgh.



%% file: schilling/schilling.tex










\subsection{INTRODUCTION}

Next-to-leading order (NLO) is the first order in perturbative QCD at
which the normalizations, and in some cases, the shapes, of cross
sections can be considered reliable.  As has been reported in the
proceedings of this workshop, there have been great advances in the
NLO calculations of processes with multi-parton final states. From an
experimental perspective, the ideal situation would be to have the NLO
matrix elements interfaced to a parton shower Monte Carlo. So far this
interface has been implemented only for a limited number of
processes. It is important to devise techniques
to allow this to be done for any NLO calculation, such as proposed 
in~\cite{Frixione:2007vw}.  

In the absence of a complete implementation of parton showering and
hadronization for all NLO processes, it is still useful to examine the
predictions from NLO calculations, at the parton level. A number of
NLO authors have made public code available for their programs; for
example, a large number of processes have been collected in
MCFM~\cite{mcfm}. There still remain, however, a number of important
calculations for which no public code is available. In lieu of a
public code, the authors can make parton 4-vector results from their
calculations available. Information
from any decay products (such as from W and Z bosons) can also be stored.
Even for processes for which public code is
available, it is still often useful to store the parton level event
information. A convenient approach for this type of storage is
ROOT~\cite{ROOT}. ROOT allows for data compression, indexed access to
specified events, immediate plotting of the variables, in addition to
having wide acceptance in the experimental community.  ROOT is one of
the output options in MCFM, through the FROOT subroutine provided by
Pavel Nadolsky. The format allows for the storage of all parton
4-vectors, the total event weight, the event weight by the initial
state partons, and the event weights for the parton distribution error
PDFs. The latter makes it easier to calculate the PDF uncertainty for
any observables, at the expense of additional storage space.

In this short contribution, we would like to generalize the FROOT
format, in order to provide a semi-official standard for NLO
output. This is generally compatible with LHEF2.0~\cite{LHEF2.0}, but
is much simplified, as less information is required.  We also provide
C++ classes to read and write the ntuple to disk, which shields the
user from the technical details of the ROOT interface.
At this workshop a standardized interface, the Binoth Les Houches
Accord, between Monte Carlo and one-loop programs was
developed~\cite{Binoth:2010xt}. However, this interface is not
directly applicable to the NLO event storage problem that we are
addressing here.

\subsection{NTUPLE STRUCTURE}

The ntuple structure in ROOT tree format is shown in
Table~\ref{tab:treevars}. Branches are available for the following
information:
\begin{itemize}
\item 4-vector information for the initial and final state
partons; 
\item the momentum fractions $x_1$ and $x_2$ and PDG identification codes $id1$ and $id2$ of the incoming partons;
\item factorization and renormalization scales;
\item total event weight; 
\item there is provision for additional
user-specified weights to be stored, for example for specific initial states;
\item the event weights for a set of error PDFs;
\item a unique event number, as well as event pointers are provided that 
allow relations between events to be stored.
\end{itemize}

Event relations (realized by pointers, see above) can be used, for
example, to associate each generated real emission event with its
counter-events resulting from Catani-Seymour dipole
subtractions~\cite{Catani:1996vz}.  This allows the possibility of
treating these events, which have potentially large cancellations
between them, together, e.g. for more easily calculating the
statistical error for the full sample, or any subset.  Such relations
could also prevent the inadvertent inclusion of an event without its
corresponding counter-events, for instance due to incomplete reading
of a ROOT tree.

All floating point variables are presently defined in double
precision, since in most NLO calculations double precision is used per
default. They could also be stored in single precision, which would
save a factor of roughly two in disk space for the produced trees.

\begin{table}
\caption{Variables stored in the proposed common ROOT ntuple output.
\label{tab:treevars}}
\centering
\begin{small}
\begin{tabular}{|l|l|}
\hline
ROOT Tree Branch & Description \\
\hline
\hline
\verb=Npart/I=  & number of partons (incoming and outgoing) \\
\verb=Px[Npart]/D= &  Px of partons \\
\verb=Py[Npart]/D= &  Py of partons \\
\verb=Pz[Npart]/D= &  Pz of partons \\
\verb=E[Npart]/D= &  E of partons \\
\verb=x1/D= & Bjorken-x of incoming parton 1 \\
\verb=x2/D= & Bjorken-x of incoming parton 2 \\
\verb=id1/I= & PDG particle ID of incoming parton 1 \\
\verb=id2/I= & PDF particle ID of incoming parton 2 \\
\verb=fac_scale/D= & factorization scale \\
\verb=ren_scale/D= & renormalization scale \\
\verb=weight/D= & global event weight \\
\verb=Nuwgt/I= & number of user weights \\
\verb=user_wgts[Nuwgt]/D= & user event weights \\
\verb=evt_no/L= & unique event number (identifier) \\
\verb=Nptr/I= & number of event pointers \\
\verb=evt_pointers[Nptr]/L= & event pointers (identifiers of related events) \\
\verb=Npdfs/I= & number of PDF weights \\
\verb=pdf_wgts[Npdfs]/D= & PDF weights \\
\hline
\end{tabular}
\end{small}
\end{table}

\subsection{C++ IMPLEMENTATION}

A set of C++ classes has been written for convenient input/output
of the above described ROOT trees. Class \verb=LhaNLOEvent=
provides a container for the event information to be
stored. The data members correspond to the Ntuple contents per
event. Member functions are provided which set or get the
event information. An example for storing the event information is
shown below
\begin{small}
\begin{verbatim}
LhaNLOEvent* evt = new LhaNLOEvent();
evt->addParticle(px1,py1,pz1,E1);
evt->setProcInfo(x1,id1,x2,id2);
evt->setRenScale(scale);
...
\end{verbatim}
\end{small}

Another class \verb=LhaNLOTreeIO= is responsible
for writing the events into the ROOT tree and outputting 
the tree to disk. In addition to the event-wise information
global data such as comments, cross sections etc can be written
as well. An example is shown below:

\begin{small}
\begin{verbatim}
LhaNLOTreeIO* writer = new LhaNLOTreeIO(); // create tree writer
writer->initWrite(''test.root'');
...
writer->writeComment(''W+4 jets at NNLO''); // write global comments
writer->writeComment(''total cross section: XYZ+/-IJK fb'');
...
writer->writeEvent(*evt); // write event to tree (in event loop)
...
writer->writeTree(); // write tree to disk
\end{verbatim}
\end{small}

Similarly, a tree can be read back from disk:

\begin{small}
\begin{verbatim}
LhaNLOTreeIO* reader = new LhaNLOTreeIO(); // init reader
ierr=reader->initRead("test.root");
if (!ierr) {
  for (int i=0; i< reader->getNumberofEvents();i++) {
    event->reset();
    ierr=reader->readEvent(i,*event);
    ...
  }
}
\end{verbatim}
\end{small}

It is important to note that the details of the
technical implementation of the tree and input/output
using ROOT are shielded from the user, who interacts
only with the \verb=LhaNLOEvent= and \verb=LhaNLOTreeIO=
classes. The only requirement is that the ROOT libraries
are provided when the program is compiled.

\subsection{EXAMPLE}

The aforementioned classes were interfaced with the C++ code for
calculating the NLO cross section for the production of a top-antitop
pair in association with one extra jet at the LHC
from~\cite{Dittmaier:2007wz,Dittmaier:2008uj}.  Ntuples were produced
for the leading-order (LO) contribution of the ttbar+jets process at a
centre-of-mass energy of 10 TeV.  The file size is of the order of 0.1
kilobytes/event (no PDF weights or event relations were used in this
test, which would lead to bigger event sizes).

\begin{figure}[ht]
\centering
\includegraphics[angle=270,width=0.8\linewidth]{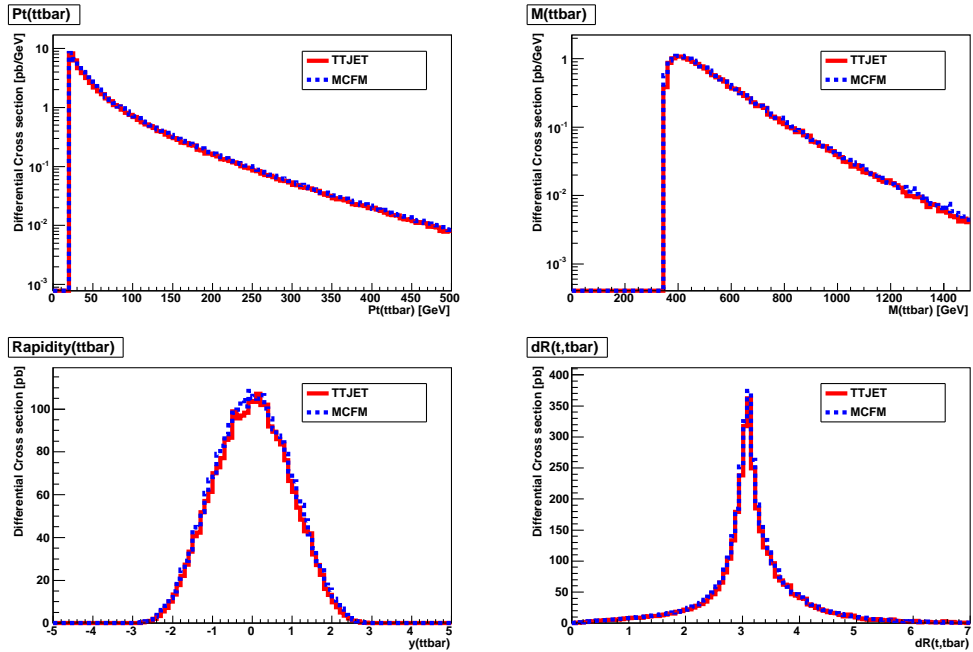}
\caption{Comparison between the LO contribution to the ttbar+jet process,
as calculated in~\cite{Dittmaier:2007wz,Dittmaier:2008uj} (TTJET),
and the NLO calculation of the total ttbar cross section
as calculated in MCFM. A cut on the existance of a third parton with $p_T>20 \rm\ GeV$ is applied in both cases, such that the predictions are comparable.
Shown are the $p_T$ (top left), invariant mass (top right) and 
rapidity of the ttbar system (bottom left),
as well as the dR between the top and the anti-top quark (bottom right).
\label{fig:fig1}}
\end{figure}

The results were compared in leading-order with MCFM, using ROOT trees
produced with the FROOT package. MCFM was set to calculate
the inclusive ttbar cross section at NLO. The comparison
was made for the region of phase space where 
a third parton is produced in addition to the top-antitop
pair (a cut $p_T>20 \rm\ GeV$ was applied on this third parton). In this
configuration the output of the two programs (TTJET LO vs MCFM NLO) 
should be identical, which is confirmed by Figure~\ref{fig:fig1}.

\subsection{CONCLUSIONS}

A common Ntuple output format for NLO cross section calculation codes
based on ROOT trees has been proposed.  It allows to make parton level
4-vector results of NLO calculations available even if no public code
exists, and thus constitutes a valuable tool for exchange between
theorists and experimentalists. The information provided in the
Ntuples is essentially a sub-set of the Les Houches Event Format
LHEF which is sufficient for applying cuts, plotting distributions etc.
The interface can be easily adopted by many of the existing
NLO codes.

In the future, a conversion tool between an LHE file and
the Ntuple format described here could also be provided.
Another possibility for an extension of the present proposal would be
the storage of event weights for different renormalization and
factorization scales, if the NLO programs were set up to provide this
information. The source code is available from~\cite{code}.



%% file: rodrigo/rodrigo.tex





\subsection{INTRODUCTION}
The duality method provides a method to numerically compute multi--leg
one--loop cross sections in perturbative field theories by defining a relation
between one--loop integrals and single phase--space integrals
\cite{meth,Gleisberg:2007zz,Catani:2008xa}. This is done by properly regularizing
propagators by a complex Lorentz--covariant prescription, which is different
from the customary $+i0$ prescription of the Feynman propagators. The duality
method is valid for massless as well as for real and virtual massive
propagators and can straightforwardly be applied not only for the evaluation
of basic one--loop integrals but also for complete one--loop quantities such
as Green's functions and scattering amplitudes \cite{Catani:2008xa}. An
extension to two--loop order is more involved and needs the treatment of
occurring dependences on one of the two integration momenta in the modified
$+i0$ description, which would lead to branch cuts in the complex energy
plane. This extension is currently under investigation.

One motivation for deriving the duality relation is given by the fact that the
computation of cross sections at next-to-leading order (NLO) requires the
separate evaluation of real and virtual radiative corrections.  Real (virtual)
radiative corrections are given by multi--leg tree--level (one--loop) matrix
elements to be integrated over the multi--particle phase space of the physical
process.  The loop--tree duality discussed here, as well as other methods that
relate one--loop and phase--space integrals, have the attractive feature that
they recast the virtual radiative corrections in a form that closely parallels
the contribution of the real radiative corrections
\cite{Soper:1998ye,Kramer:2002cd,meth,Kleinschmidt:2007zz,Moretti:2008jj}. This
close correspondence can help to directly combine real and virtual
contributions to NLO cross sections. In particular, using the duality
relation, one can apply mixed analytical and numerical techniques to the
evaluation of the one--loop virtual contributions \cite{meth}. The infrared or
ultraviolet divergent part of the corresponding dual integrals can be
analytically evaluated in dimensional regularization. The finite part of the
dual integrals can be computed numerically, together with the finite part of
the real emission contribution.  Partial results along these lines are
presented in Refs.~\cite{meth,Gleisberg:2007zz} and further work is in progress.

\subsection{THE DUALITY RELATION AT ONE--LOOP ORDER}

Consider a generic one--loop integral over Feynman propagators, where $q_i = q
+ \sum_{k=1}^i p_k$ are the momenta of the internal lines, $q$ being the loop
momentum, and $p_i$ ($\sum_{i=1}^N p_i = 0$) the external (outgoing and
clockwise ordered) momenta. The Feynman propagators have two poles in the
complex plane of the loop energy $q_0$, the pole with positive (negative)
energy being slightly displaced below (above) the real axis encoded by the
additional $+i0$ term in the propagator. Using the Cauchy residue theorem in
the complex $q_0$--plane, with the integration contour closed at $\infty$ in
the lower half--plane, we obtain a sum over terms given by the integral
evaluated at the poles with positive energy only. Hence a one--loop integral
with $N$ internal propagators leads to $N$ contributions, one for each
propagator for which the residue is taken. It can be shown that this residue
is equivalent to cutting that line by including the corresponding on--shell
propagator $\delta_+(q_i^2) = \theta(q_i^0) \delta(q_i^2)$. The remaining
propagators of the expression are shifted to
\begin{equation}
\left. \prod_{j\neq i} \,\frac{1}{q_j^2 + i0} \,  
\right|_{q_i^2=-i0}
= \prod_{j\neq i} \; \frac{1}{q_j^2 - i0 \,\eta (q_j-q_i)}~, 
\label{Duality_eta}
\end{equation}
where $\eta$ is a future-like vector, i.e.~a $d$-dimensional vector that can
be either light-like $(\eta^2=0)$ or time-like $(\eta^2 > 0)$ with positive
definite energy ($\eta_0\ge 0)$.  The calculation of the residue at the pole
of the $i^{\rm th}$ internal line modifies the $i0$ prescription of the
propagators of the other internal lines of the loop. This modified
regularization is named `dual' $i0$ prescription, and the corresponding
propagators are named `dual' propagators. The dual prescription arises,
because the original Feynman propagator $1/(q_j^2 +i0)$ is evaluated at the
{\em complex} value of the loop momentum $q$, which is determined by the
location of the pole at $q_i^2+i0 = 0$.  The presence of $\eta$ is a
consequence of the fact that the residue at each of the poles is not a
Lorentz--invariant quantity, since a given system of coordinates has to be
specified to apply the residue theorem. Different choices of the future-like
vector $\eta$ are equivalent to different choices of the coordinate system.
The Lorentz--invariance of the loop integral is, however, recovered after
summing over all the residues.  For a one--loop integral, the term $\eta
(q_j-q_i)$ is always solely proportional to external momenta and hence defines
a fixed pole in the $q_0$--plane.

Note that an extension to real and virtual massive propagators and full
scattering amplitudes is straightforward and described in detail in
Ref. \cite{Catani:2008xa}.

\subsection{FIRST STEPS TOWARDS TWO--LOOP ORDER}

The fact that the term $\eta (q_j-q_i)$ is proportional to external momenta
only, is not valid anymore once going to the next loop order and considering
a generic two--loop n--leg diagram. Taking the residues loop by loop for the
two integration momenta introduces in some cases a dependence on one of the
integration momenta in the difference of $\eta (q_j-q_i)$. Hence we encounter
not poles but rather branch cuts in the complex energy plane. To avoid this
and more generally to avoid any dependence on integration momenta in the $\eta
(q_j-q_i)$--terms demands a reformulation of the propagators into another
basis, which fulfills the required properties. First steps towards a
two--loop expression obtained by such a transformation have been undertaken,
while the full general two--loop expression is still under investigation.

\subsection*{ACKNOWLEDGEMENTS}

This work was supported by the Ministerio de Ciencia e Innovaci\'on under Grant
No. FPA2007-60323, CPAN (Grant No. CSD2007-00042), the Generalitat Valenciana
under Grant No. PROMETEO/2008/069, and by the European Commission MRTN
FLAVIAnet under Contract No. MRTN-CT-2006-035482 and by 
MRTN-CT-2006-035505 HEPTOOLS.



%% file: pittau/pittau.tex








\subsection{Introduction}

The associated production of a Higgs boson, with a $t\bar t$ pair,
is going to play an important role for precision measurements of the
Higgs boson Yukawa couplings at the LHC, especially in the range of masses 
$115~{\rm GeV} < M_H < 140~{\rm GeV}$, where the Higgs decays predominantly 
in $b \bar b$ pairs.
Whether or not it will also provide a discovery channel, very much depends on
the ratio between this signal and the main QCD $t \bar t b \bar b$ background.
A next-to-leading order (NLO) analysis of the inclusive $H t\bar t$ 
production performed by two
independent groups showed an increase of the leading order (LO) 
cross section by a factor of 1.2 at the central 
scale $\mu_0=m_t+m_H/2$ up to 1.4 at the threshold value $\mu=2\mu_0$, see 
\cite{Beenakker:2001rj,Beenakker:2002nc,Dawson:2002tg,Dawson:2003zu}. 
On the other hand, very recent 
calculations~\cite{Bredenstein:2008zb,Bredenstein:2009aj,Bevilacqua:2009zn}
showed a huge enhancement of the $t \bar t b \bar b$ 
background at NLO with a $K$ factor of the order of 1.77. 
Of course, much more detailed 
analyses are needed to establish the possibility
of detecting the Higgs in this channel, based on distributions, rather than
on a mere event counting, see e.g. \cite{Plehn:2009rk}. 
To this aim, an accurate description
of both signal and background is needed. A first step toward this, 
is the inclusion of the $H \to b \bar b$ decay directly
into the NLO calculation of the signal.
In this contribution, this is achieved by computing
the factorisable QCD corrections 
to the Higgs signal $p p \to t \bar t H \to t \bar t b \bar b$ process. 
We consider higher order corrections to both production and 
decay of the Higgs boson, where the latter is modeled by the propagator with a 
fixed width which we computed with \textsc{Hdecay}~\cite{Djouadi:1997yw}.

NLO QCD corrections have been calculated with the help of the  
\textsc{Helac-Nlo} system.  It consists of \textsc{Helac-Phegas}
\cite{Kanaki:2000ey,Papadopoulos:2000tt,Papadopoulos:2005ky,Cafarella:2007pc}, 
\textsc{Helac-Dipoles} \cite{Czakon:2009ss}, 
\textsc{OneLOop} \cite{vanHameren:2009dr}  and \textsc{Helac-1Loop} 
\cite{vanHameren:2009dr}, based on the OPP reduction technique 
\cite{Ossola:2006us} and the reduction 
code \textsc{CutTools}\cite{Ossola:2007ax}. 
The \textsc{Helac-Nlo} system has 
also been used to compute the $p p \to t \bar t b \bar b$
background \cite{Bevilacqua:2009zn}, 
allowing the comparisons presented in this work.

\subsection{Results}

For both signal and background, we consider the process 
$pp \rightarrow  t\bar{t} b\bar{b} + X$ at the LHC,
i.e. for  $\sqrt{s} = 14$ TeV. For the top-quark mass
we take $m_t = 172.6 $ GeV, whereas all other
QCD partons including b quarks are treated as massless. Higgs boson mass is
set to $m_H$ = 130 GeV.
Top quark mass  renormalization is performed in the on-shell scheme,
which  fixes the renormalization of the top quark Yukawa coupling. 
As far as the 
b-quark Yukawa coupling is concerned, we renormalize it in the 
$\overline{MS}$ scheme, which makes it proportional to the  
$\overline{MS}$ mass of the b-quark, $\overline{m}_b(\mu)$. Finally, we  
transform this parameter into $\overline{m}_b(m_H)$. While the difference is
of higher order, we are motivated by the fact that we work in the narrow
width approximation with $\Gamma_H$ calculated at $\mu=m_H$.   
 
We consistently use the CTEQ6  set of parton distribution functions 
(PDFs)~\cite{Pumplin:2002vw,Stump:2003yu} , i.e.
we take CTEQ6L1 PDFs with a 1-loop running $\alpha_s$ in LO and CTEQ6M PDFs
with a 2-loop running $\alpha_s$ in NLO, but the suppressed contribution from
b quarks in the initial state has been neglected. The number of active
flavors is $N_F = 5$, and the respective QCD parameters are $\Lambda^{LO}_5 =
165$ MeV and  $\Lambda^{MS}_5 = 226$ MeV. In the renormalization of the strong
coupling constant, the top-quark loop in the gluon self-energy is subtracted at
zero momentum. In this scheme the running of $\alpha_s$ is generated solely by
the contributions of the light-quark and gluon loops. By default, we set the 
renormalization and factorization scales, $\mu_{R}$  and $\mu_F$, to the 
common value $\mu_0 =m_t+m_H/2$ for the signal and $\mu_0 =m_t$ for the 
background.

All final-state b quarks and gluons with pseudorapidity $|\eta| <5$ are 
recombined into jets with separation $\sqrt{\Delta\phi^2 +\Delta y^2} > D 
= 0.8$ in the rapidity– azimuthal-angle plane via the IR-safe $k_T$-algorithm 
~\cite{Catani:1992zp,Catani:1993hr,Ellis:1993tq}. Moreover, we impose the 
following additional cuts on the transverse momenta and the rapidity of two 
recombined b-jets: $p_{T,b} > 20$ GeV, $|y_b|< 2.5$. The outgoing (anti)top
quarks are neither affected by the jet algorithm nor by phase-space cuts.  

\begin{figure}[!b]
\begin{center}
\includegraphics[width=0.40\textwidth]{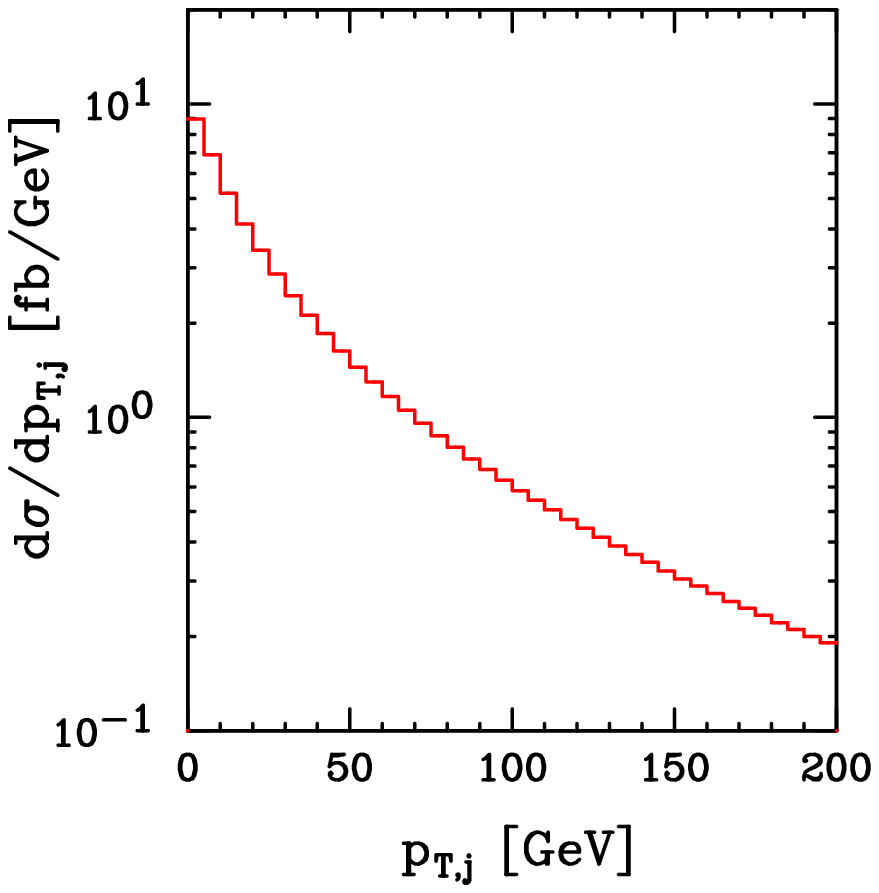}
\includegraphics[width=0.40\textwidth]{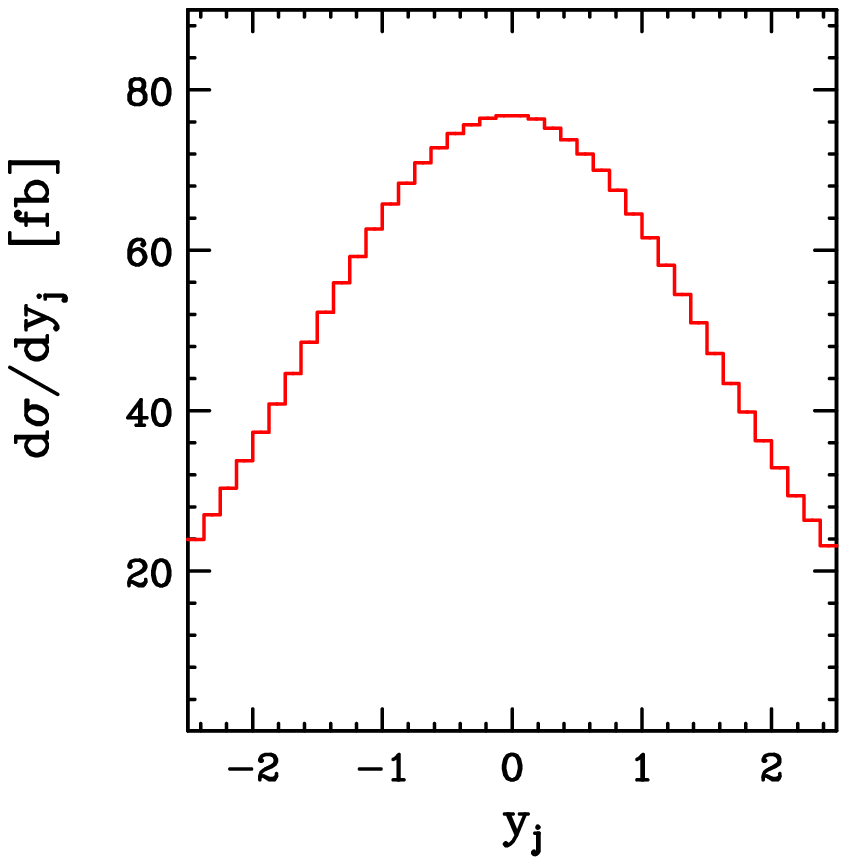}
\end{center}
\vspace{-0.2cm}
\caption{\it \label{fig1}  Distribution in  the transverse momentum 
(left panel) and in the rapidity (right panel) of the extra jet for 
$pp\to t\bar{t}H \rightarrow t\bar{t}b\bar{b}   + X$ at the LHC.}
\end{figure}
\begin{figure}[!h]
\begin{center}
\includegraphics[width=0.40\textwidth]{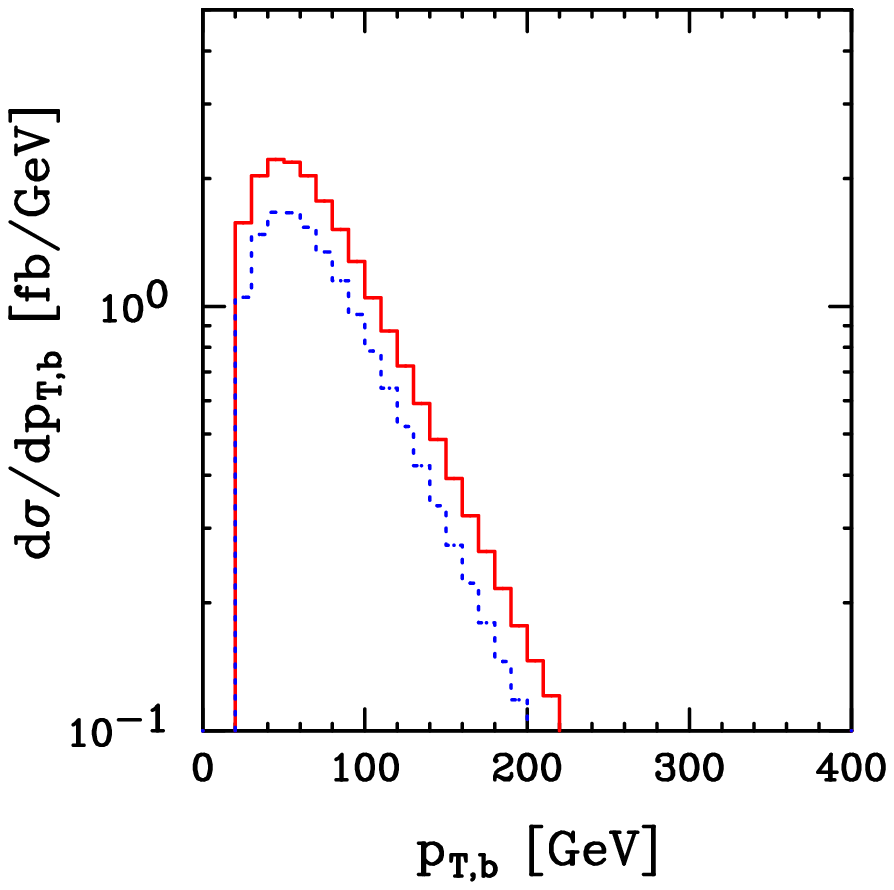}
\includegraphics[width=0.40\textwidth]{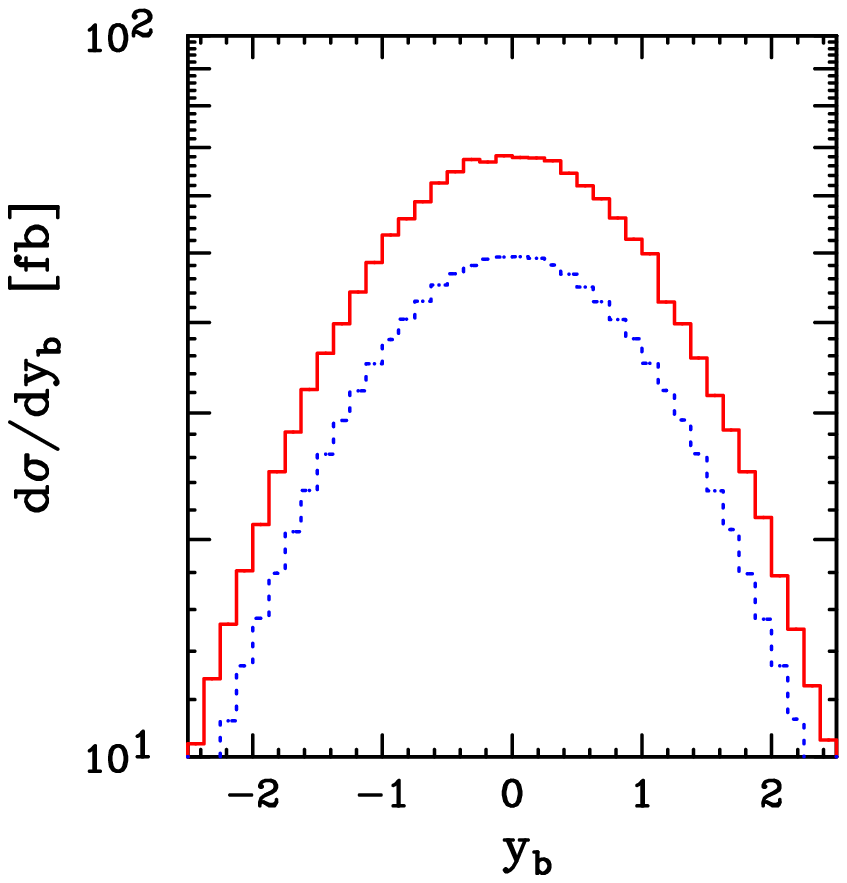}
\end{center}
\vspace{-0.2cm}
\caption{\it \label{fig2}  Averaged distributions of $p_T(b)$ and 
$p_T(\bar  b)$ (left panel) and averaged distributions of $y(b)$ and 
$y(\bar b)$ (right panel) for 
$pp\to t\bar{t}H \rightarrow t\bar{t}b\bar{b}   + X$ at the LHC. 
The red solid line refers to NLO and the blue dotted line to LO result.}
\end{figure}
\begin{figure}[!h]
\begin{center}
\includegraphics[width=0.40\textwidth]{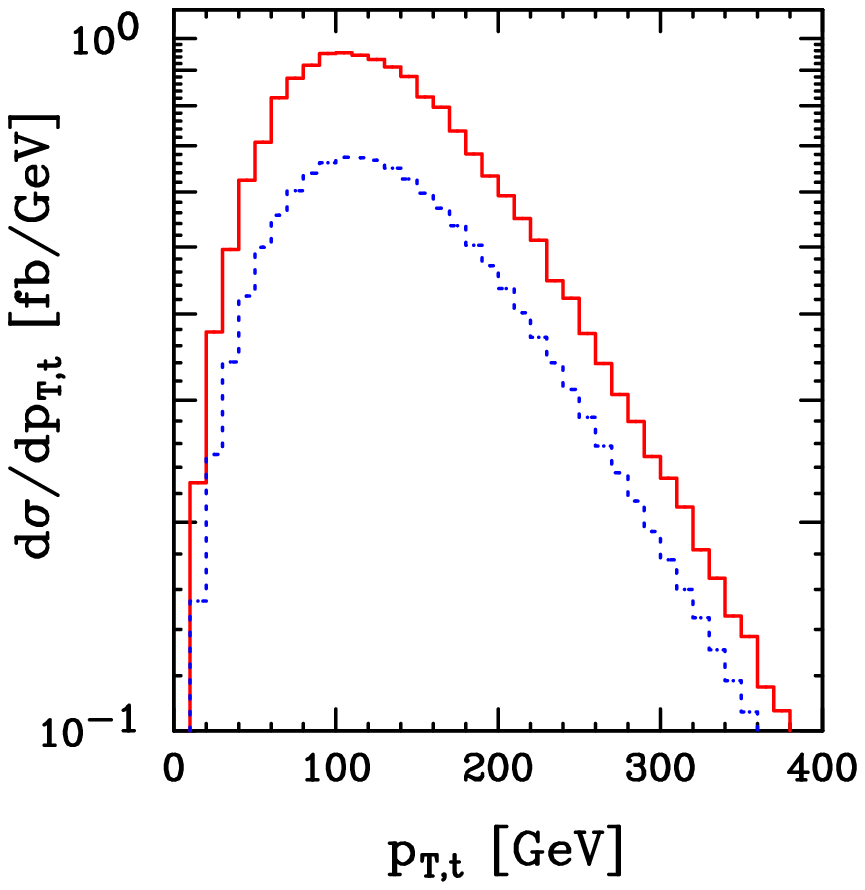}
\includegraphics[width=0.40\textwidth]{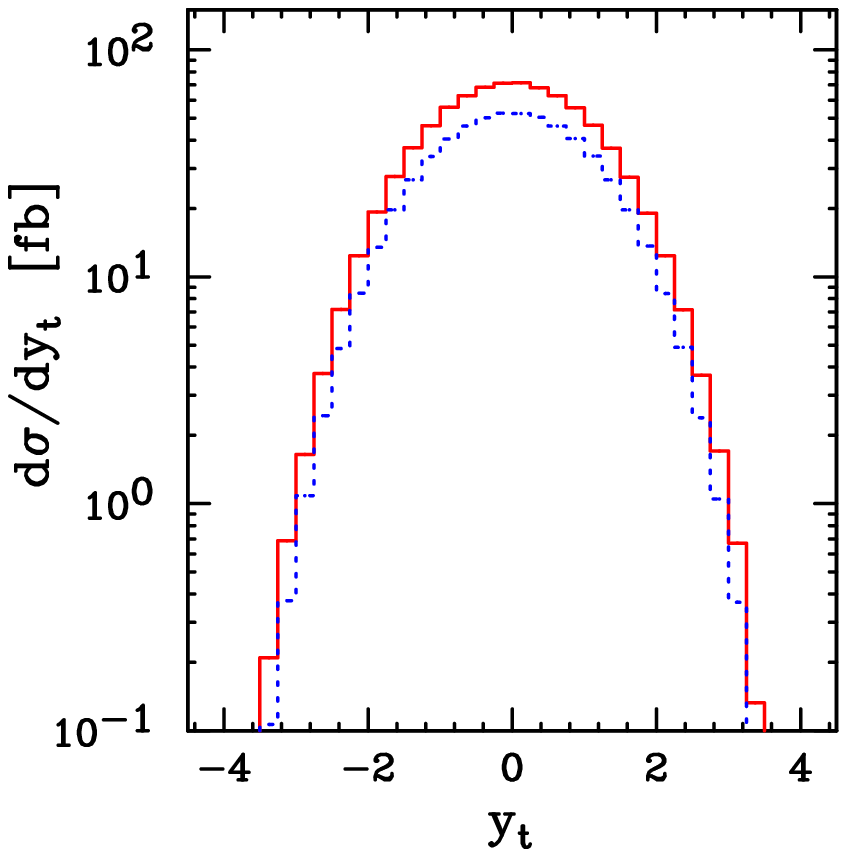}
\end{center}
\vspace{-0.2cm}
\caption{\it \label{fig3}  Averaged distributions of $p_T(t)$ and 
$p_T(\bar t)$ (left panel) and averaged distributions of 
$y(t)$ and $y(\bar t)$ (right panel) for 
$pp\to t\bar{t}H \rightarrow t\bar{t}b\bar{b}   + X$ at the LHC.
The red solid line refers to NLO and the blue dotted line to LO result.}
\end{figure}
\begin{figure}[!h]
\begin{center}
\includegraphics[width=0.40\textwidth]{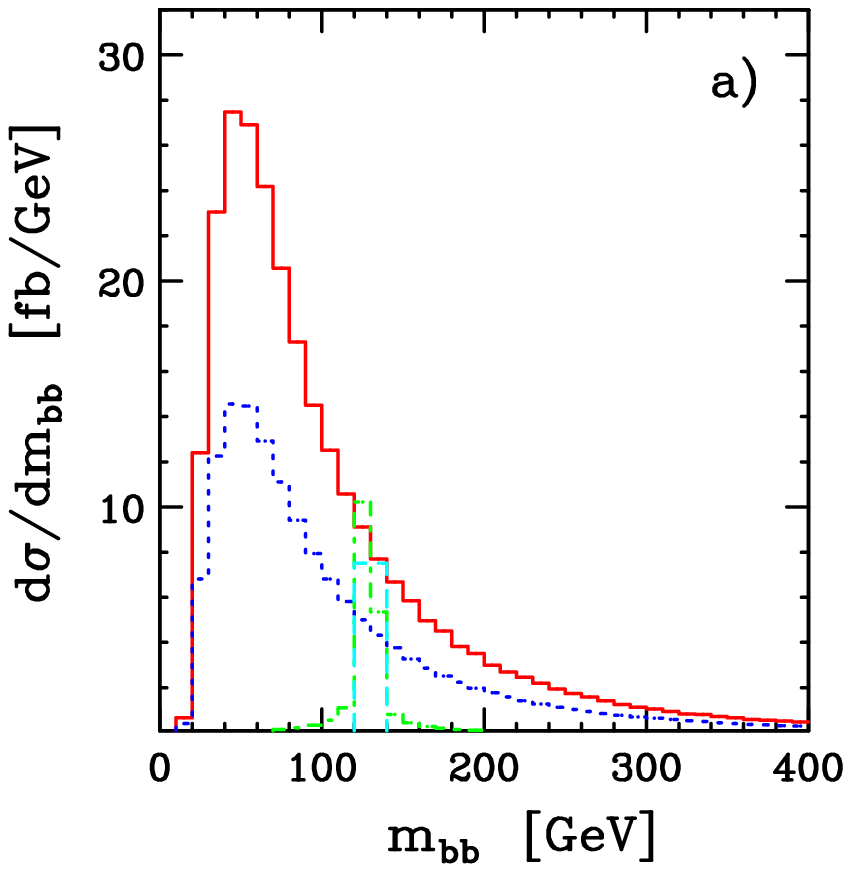}
\includegraphics[width=0.41\textwidth]{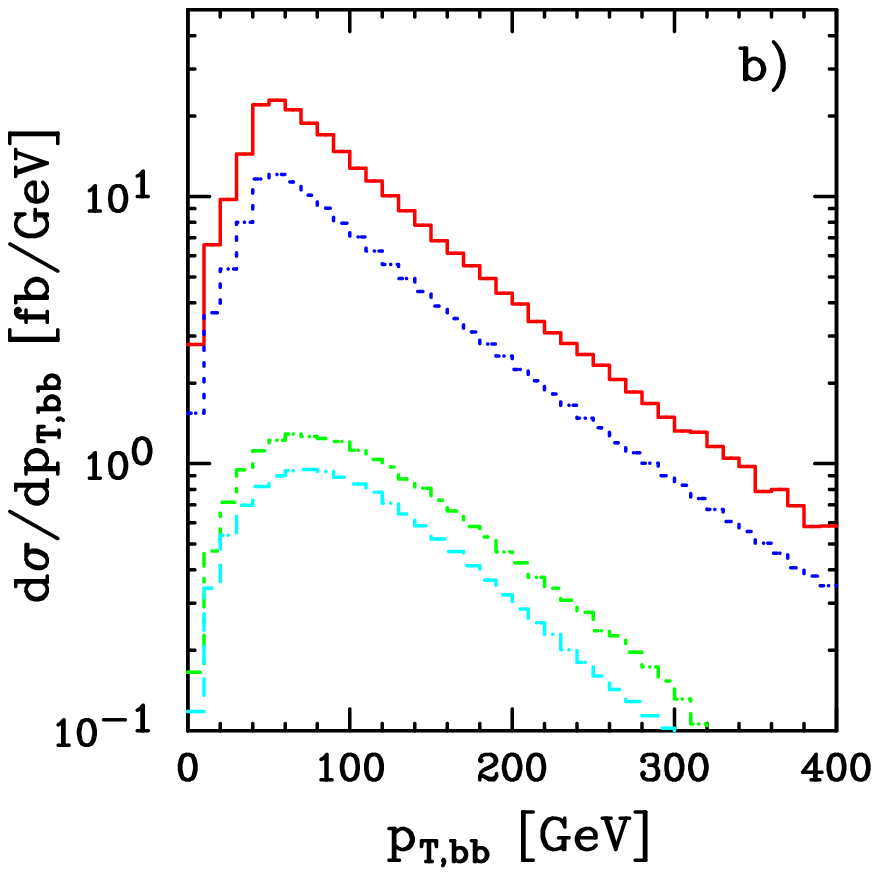}\\
\includegraphics[width=0.40\textwidth]{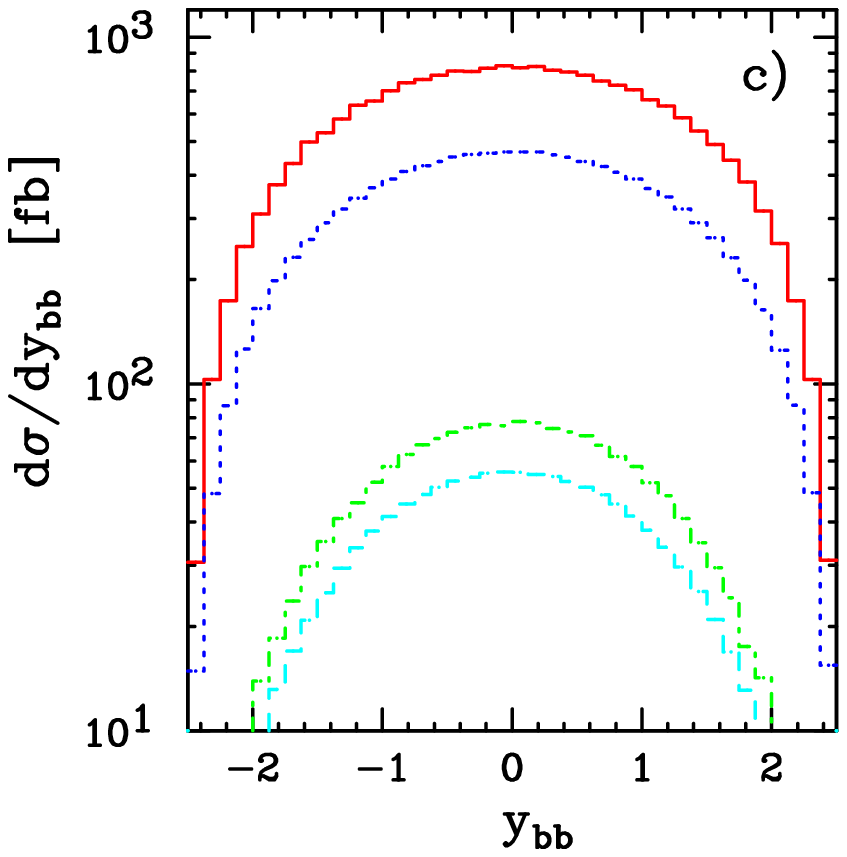}
\includegraphics[width=0.41\textwidth]{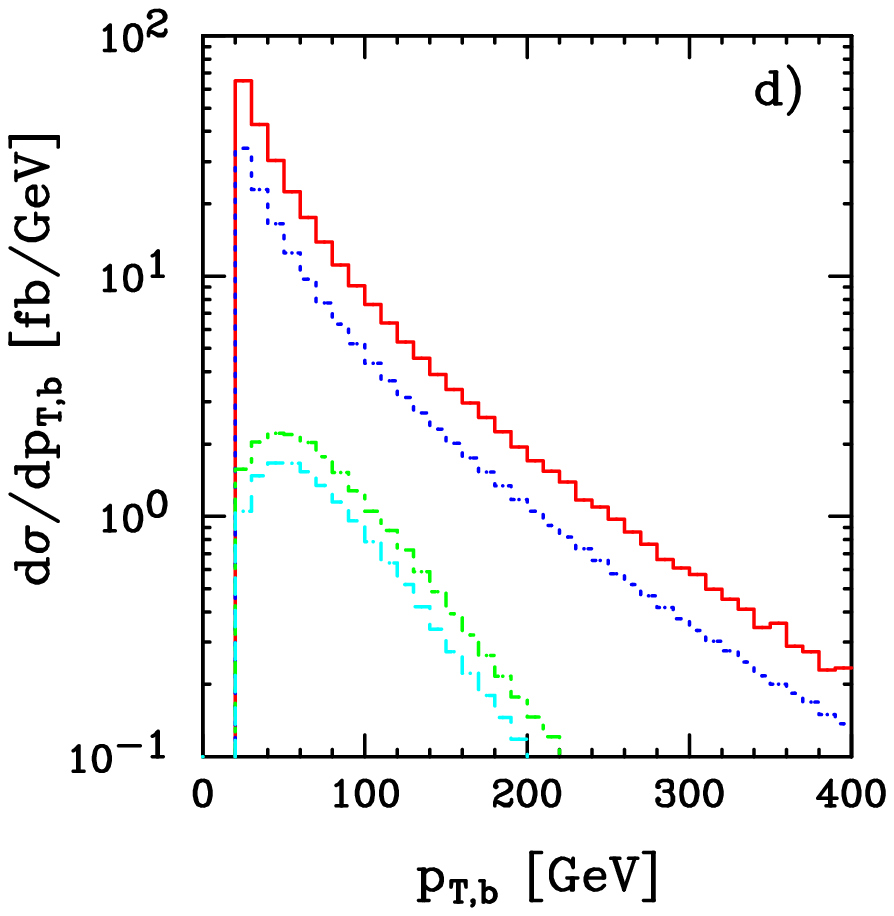}
\end{center}
\vspace{-0.2cm}
\caption{\it \label{fig4}  
  Distribution of the invariant mass 
  $m_{b\bar{b}}$ of the bottom-anti-bottom pair (a), 
  distribution in the transverse momentum  $p_{T_{b\bar{b}}}$  of 
  the bottom-anti-bottom 
  pair (b), distribution in the rapidity $y_{b\bar{b}}$ of the
  bottom-anti-bottom pair (c)  and  distribution in the transverse momentum
  $p_{T_{b}}$  of the  bottom quark (d) for 
  $pp\rightarrow t\bar{t}H \rightarrow t\bar{t}b\bar{b} +X$
  and $pp\rightarrow t\bar{t}b\bar{b}   + X$ at the LHC.
  The red solid line refers to the NLO QCD background, the blue dotted 
  line to the LO QCD background, while the green dash-dotted and cyan 
  dashed line to the NLO and LO signal,  respectively.}
\end{figure}
\begin{figure}[!h]
\begin{center}
\includegraphics[width=0.70\textwidth]{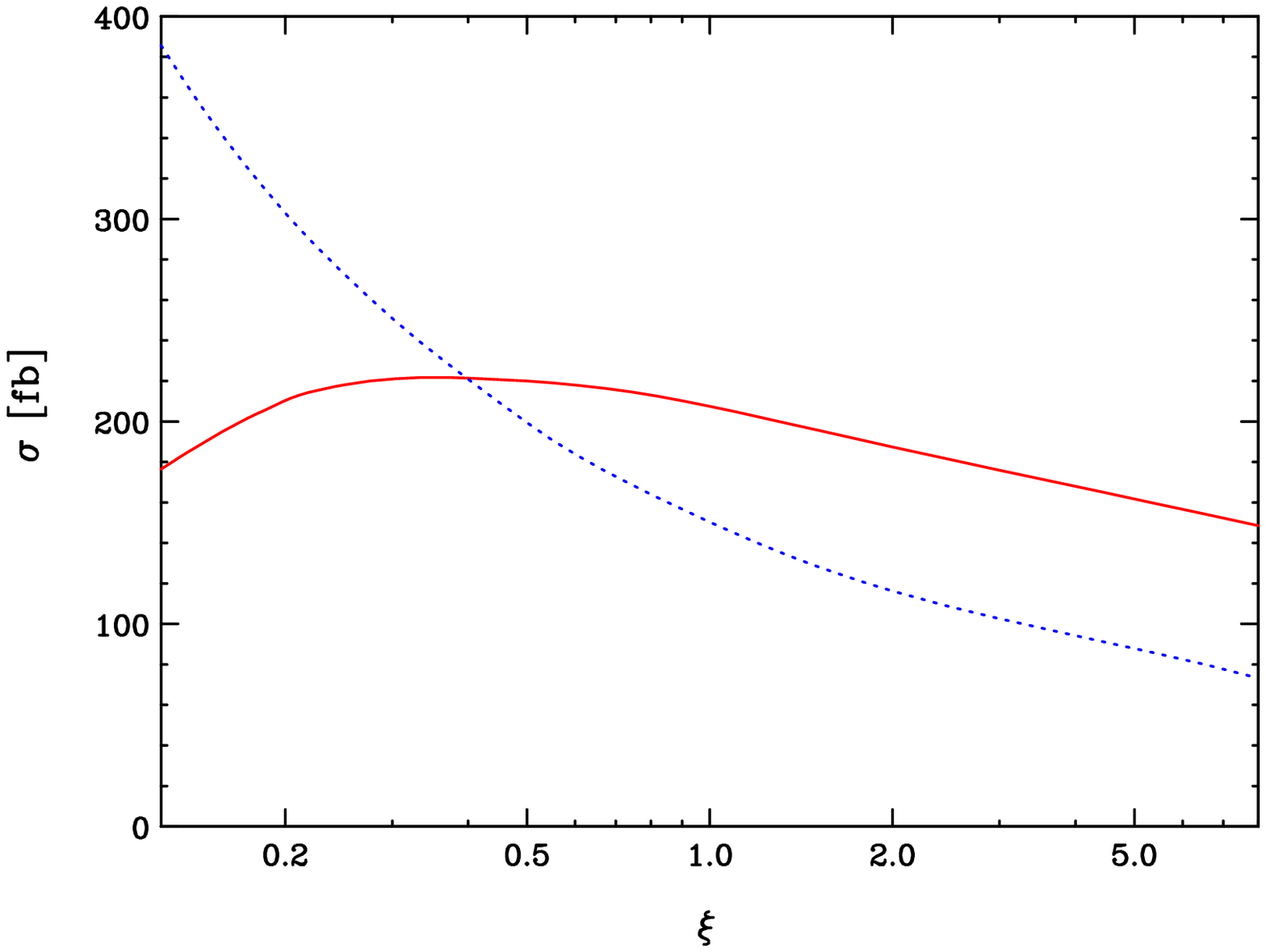} 
\end{center}
\vspace{-0.2cm}
\caption{\it \label{fig5} Scale dependence of the total cross
section for $pp\rightarrow t\bar{t}H \to  t\bar{t}b\bar{b} + X$ at the LHC  
with $\mu_R=\mu_F=\xi \cdot \mu_0$ where $\mu_0=m_t+m_H/2$. 
The blue dotted curve corresponds to 
the LO, whereas the red solid one to the NLO 
order result.}
\end{figure}
\begin{figure}[!h]
\begin{center}
\includegraphics[width=0.70\textwidth]{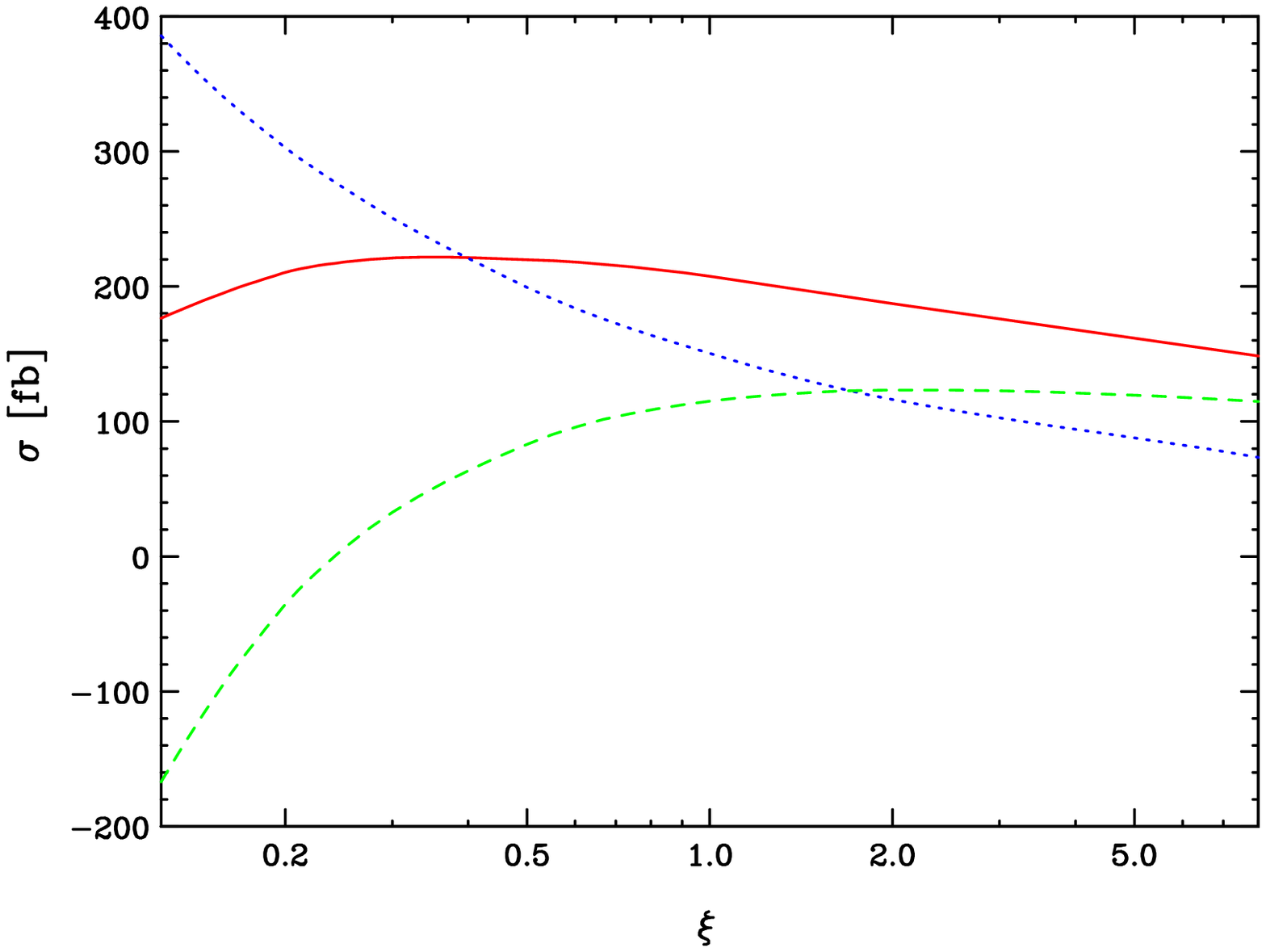}
\end{center}
\vspace{-0.2cm}
\caption{\it \label{fig6}  Scale dependence of the total cross
section for $pp\rightarrow t\bar{t}H \to  t\bar{t}b\bar{b} + X$ at the LHC  
with $\mu_R=\mu_F=\xi \cdot \mu_0$ where $\mu_0=m_t+m_H/2$. 
The blue dotted curve corresponds to 
the LO, the red solid one to the NLO 
order result whereas the green dashed one to the NLO
order result with a jet veto.  }
\end{figure}

We first study the impact of the NLO corrections on the 
$p p \to t \bar t H \to t \bar t b \bar b$ signal.
With the parameters and cuts specified above, the lowest order cross section 
is given by:
\begin{eqnarray} 
\sigma^{\rm S}_{\rm LO}= (150.375 \pm 0.077)  ~{\rm fb}\,.
\end{eqnarray} 
At the NLO we obtain 
\begin{equation} 
\sigma^{\rm S}_{\rm NLO}= (207.473  \pm 0.232) ~{\rm fb}\, 
~~~~{\rm for} ~~\alpha_{\rm{max}}= 0.01 \, ,
\end{equation}
\begin{equation}  
\sigma^{\rm S}_{\rm NLO}= (207.268 \pm 0.150) ~{\rm fb}\,
~~~~{\rm for}  ~~\alpha_{\rm{max}}= 1
\end{equation} 
which leave us with a  $K$ factor $K= 1.38$.  We run our code with two 
different values of $\alpha_{max}$, which is a common modification of
Catani-Seymour subtraction terms \cite{Catani:1996vz,Catani:2002hc}  
in the phase space region away from the singularity, 
see ~\cite{Bevilacqua:2009zn} for details, to check the independence of 
the final result on this value. 
This has to be compared with a LO and NLO $t \bar t b \bar b$ 
background cross sections given by~\cite{Bevilacqua:2009zn}: 
\begin{eqnarray} 
\sigma^{\rm B}_{\rm LO} &=&  (1489.2 \pm 0.9) ~{\rm fb}\, ,
\end{eqnarray} 
\begin{equation} 
\sigma^{\rm B}_{\rm NLO}=  (2636 \pm 3)~{\rm fb}\, 
~~~~{\rm for} ~~\alpha_{\rm{max}}= 0.01 \, ,
\end{equation}
\begin{equation}  
\sigma^{\rm B}_{\rm NLO}=  (2642 \pm 3)~{\rm fb}\,
~~~~{\rm for}  ~~\alpha_{\rm{max}}= 1
\end{equation} 
again for two different values of $\alpha_{max}$ parameter. At $\mu_0=m_t$ 
we obtained the $K$ factor $K=1.77$.

The transverse momentum and rapidity distributions of the extra jet for 
the $pp\to t\bar{t}H \to t\bar{t}b\bar{b}$ process 
are presented in Fig.~\ref{fig1}, from which
it is evident, that most of the extra radiation is at low $p_T$ and 
in the central region, as expected. It is therefore tempting to study 
the effect of a jet veto on the $K$ factor for the signal process.
With a jet veto of 50 GeV we obtain instead

\begin{equation} 
\sigma^{\rm S}_{\rm NLO-veto}=  (115.022 \pm 0.233) ~{\rm fb}\, 
~~~~{\rm for} ~~\alpha_{\rm{max}}= 0.01
\end{equation}
 \begin{equation} 
\sigma^{\rm S}_{\rm NLO-veto}= (114.880 \pm 0.152) ~{\rm fb}\, 
~~~~{\rm for} ~~\alpha_{\rm{max}}= 1
\end{equation} 
giving $K= 0.76$.
We therefore conclude that NLO QCD corrections are reduced from +38\%
down to -24\% when a jet veto of 50 GeV is applied on the additional jet.
For comparison, we also quote here the result presented by 
Bredenstein, Denner, Dittmaier and Pozzorini in~\cite{Bredenstein:2009aj}.
They find that a jet veto of 50 GeV reduces the 
NLO QCD corrections to the $t \bar t b \bar b$ background from +77\% 
down to +20\%, with respect to the tree level result.

The effect of the NLO corrections on the $p_T$ and rapidity distributions
of bottoms and tops is shown in Fig.~\ref{fig2} and Fig.~\ref{fig3}.
The distributions are similar for particles and anti-particles, therefore 
the average is taken in the figures.  The blue dotted 
curve corresponds to the LO, whereas the red solid one
to the NLO order result.

As for the comparisons between signal and background, we present, in
Fig.~\ref{fig4}, a few histograms, namely the invariant mass, 
transverse momentum and rapidity of the two-$b$-jet system, as well as 
the transverse momentum of the single $b$-jet.
In all figures the red solid line refers to
the NLO QCD background, the blue dotted line to the LO QCD background, while
the green dash-dotted and cyan dashed line to the NLO and LO signal, 
respectively. Apart from the invariant mass of the $b \bar b$ system 
and the $p_T$ spectrum of the $b$ quark, the shapes look very similar 
for signal and background. 

Finally the scale dependence of the total cross
section for $pp\rightarrow t\bar{t}H \to  t\bar{t}b\bar{b} + X$ at the LHC  
is presented graphically in Fig.~\ref{fig5}. 
The blue dotted curve corresponds to 
the LO, whereas the red solid one to the NLO 
order result. As expected, we observe a
reduction of the scale uncertainty while going from LO to NLO. Varying
the scale by a factor 2 changes the cross section by
+33\% and -23\% in the LO case, while in the NLO case we have
obtained a variation of the order +6\% and -10\%.

In Fig.~\ref{fig6} the impact of additional $p_T$ cut on the extra jet 
is shown. While for very small scales the scale dependence seems to have
deteriorated, for the large one within the usual range, 
the variation remains more or less
the same.  Varying the scale up and down by a factor 2 changes the cross 
section by -28\% and +7\% in this case.

\subsection{Conclusions}

A NLO study of the $t\bar{t}H \to t\bar{t}b\bar{b}$ signal 
and the QCD $t\bar{t}b\bar{b}$ background reveals that the signal over 
background ratio R passes from
R= 0.101 to R= 0.079, when inclusive NLO corrections are included. 
With a jet veto of 50 GeV, one obtains, instead,
R= 0.064. A more detailed analysis is needed to study the effects
of the jet veto procedure. However, our preliminary result shows that some
tuning is necessary to maximize R.
As for the distributions, the $p_T$ spectrum of the b quarks 
appears to be harder for the background
than for the signal. This fact, together with an accurate reconstruction of the
invariant $m_{b\bar{b}}$ mass, could be used as an extra handle to extract
the Higgs signal.

\subsection*{Acknowledgments}

This work was funded in part by the RTN European Programme 
MRTN-CT-2006-035505 HEPTOOLS - Tools and Precision Calculations 
for Physics Discoveries at Colliders. M.C. was  supported by the 
Heisenberg Programme of the Deutsche Forschungsgemeinschaft. M.W. was 
supported by the Initiative and Networking Fund of the Helmholtz 
Association, contract HA-101 ("Physics at the Terascale"). 
M.V.G. and R.P. thank the financial support of the MEC project FPA2008-02984.
M.V.G. was additionally supported by the INFN.



%% file: pozzorini/pozzorini.tex




%


{
\newcommand{\lsim}
{\mathrel{\raisebox{-.3em}{$\stackrel{\displaystyle <}{\sim}$}}}
\newcommand{\gsim}
{\mathrel{\raisebox{-.3em}{$\stackrel{\displaystyle >}{\sim}$}}}

\subsection{INTRODUCTION}

The discovery of the Higgs boson and the measurement of its interactions
with massive quarks and vector bosons represent a central goal of the 
Large Hadron Collider (LHC).
In the light-Higgs scenario, $M_{\mathrm{H}}\lsim 130\, \mathrm{GeV}$, associated 
$\mathrm{t\bar t H}$
production provides the opportunity to observe the
Higgs boson in the 
$\mathrm{H\to b\bar b}$ decay channel and to
measure the top-quark Yukawa coupling. 
However, the extraction of the 
$\mathrm{t\bar t H (H\to b\bar b)}$ signal 
from its large QCD backgrounds
represents a serious challenge.

The selection strategies elaborated by ATLAS and CMS
\cite{Kersevan:2002vu,Kersevan:2002dd,Cammin:2003,Aad:2009wy,Drollinger:2001ym,Cucciarelli:2006,Benedetti:2007sn,Ball:2007zza}
are based on the full reconstruction of the 
$\mathrm{t\bar t b\bar b}$
signature, starting from a final state with four $\mathrm{b}$ quarks and additional
light jets.  After imposing four  $\mathrm{b}$ -taggings, a reconstruction of the top
quarks is performed, which permits to identify two $\mathrm{b}$ quarks as top-decay
products.  The remaining two $\mathrm{b}$ quarks constitute a Higgs candidate, and
their invariant-mass distribution is the relevant observable to find the
Higgs signal.
However, the presence of multiple $\mathrm{b}$ quarks and light jets in the final state
represents a serious obstacle to the correct identification of the
$\mathrm{b\bar b}$ Higgs candidates.  Realistic simulations indicate that only
about 1/3 of the selected  $\mathrm{b}$ -quark pairs have correct combinatorics, while
the other Higgs candidates contain $\mathrm{b}$ jets from top decays or miss-tagged
light jets.  This so-called combinatorial background significantly dilutes
the Higgs signal and increases its background contamination.  The QCD
processes 
$\mathrm{pp\to t\bar t b\bar b}$ and 
$\mathrm{t\bar t jj}$ are the
main background components.
The latest ATLAS and CMS studies \cite{Aad:2009wy,Ball:2007zza}, for
$30\, \mathrm{fb}^{-1}$ and $60\, \mathrm{fb}^{-1}$, respectively, anticipate a statistical
significance around $2\sigma$ (ignoring systematic uncertainties) and a
fairly low signal-to-background ratio of order 1/10.  This calls for better
than 10\% precision in the background description, a very demanding
requirement both from the experimental and theoretical point of view.

More recently, an alternative strategy based on the selection of highly
boosted Higgs bosons, which decay into ``fat jets'' containing two
b~quarks, has opened new and very promising
perspectives~\cite{Plehn:2009rk}.  
This novel approach might enable a better background suppression and
increase the signal-to-background ratio beyond $1/3$. Moreover, three
 $\mathrm{b}$ -taggings would be sufficient to strongly suppress the 
$\mathrm{t\bar t jj}$ contamination. In this case the background would be completely dominated by
$\mathrm{t\bar t b\bar b}$ production.

The recent calculation of the NLO QCD corrections to the irreducible
$\mathrm{t\bar t b\bar b}$ background \cite{Bredenstein:2008zb,
Bredenstein:2009aj,Bredenstein:2010rs,Bevilacqua:2009zn} constitutes another
important step towards the observability of
$\mathrm{t\bar t H(H\to b\bar b)}$ at the LHC. These NLO predictions are
mandatory in order to reduce the huge scale uncertainty of the lowest-order
(LO) 
$\mathrm{t\bar t b\bar b}$ cross section, which can vary up to a factor
four if the QCD scales are identified with different kinematic
parameters~\cite{Kersevan:2002vu,Kersevan:2002dd}.
Previous results for five-particle processes that feature a signature
similar to 
$\mathrm{t\bar t b\bar b}$ indicate that setting the renormalization
and factorization scales equal to half the threshold energy,
$\mu_\mathrm{R,F}=E_{\mathrm{thr}}/2$, is a reasonable scale choice.
At this scale the NLO QCD corrections to 
$\mathrm{pp\to t\bar t H}$ ($K\simeq
1.2$)~\cite{Beenakker:2001rj,Beenakker:2002nc,Dawson:2002tg,Dawson:2003zu},
$\mathrm{pp\to t\bar t j}$
($K\simeq$1.1)~\cite{Dittmaier:2007wz,Dittmaier:2008uj}, and
$\mathrm{pp\to t\bar t Z}$ ($K\simeq 1.35$)~\cite{Lazopoulos:2008de}, are
fairly moderate. 
This motivated experimental groups to adopt the scale
$\mu_\mathrm{R,F}=E_{\mathrm{thr}}/2=m_{\mathrm{t}}+m_{\mathrm{b \bar b}}/2$ for the LO
simulation of the 
$\mathrm{t\bar t b\bar b}$
background~\cite{Kersevan:2002vu,Kersevan:2002dd,Cammin:2003,Aad:2009wy}.
However, at this scale the NLO corrections to 
$\mathrm{pp\to t\bar t b\bar b}$
turn out to be close to a factor of two $(K\simeq
1.8)$~\cite{Bredenstein:2009aj,Bevilacqua:2009zn}.%
\footnote{This NLO enhancement of the 
$\mathrm{t\bar t b\bar b}$ background has
already been taken into account in Ref.~\cite{Plehn:2009rk}.}
This sizable NLO correction suggests
the presence of large logarithms that tend to spoil the convergence of
the perturbative expansion. This is mainly due to the
fact that the scale $\mu_\mathrm{R,F}=E_{\mathrm{thr}}/2$ does not
provide an adequate description of the QCD dynamics of
$\mathrm{t\bar t b\bar b}$ production. 
To cure this problem we advocate the use of a new and more natural scale
choice~\cite{Bredenstein:2010rs}, which leads to a much smaller $K$ factor
and also reduces the residual scale dependence at NLO.
We then present a selection of the results of
Ref.~\cite{Bredenstein:2010rs}.
In particular we discuss the impact of a jet veto,
as well as NLO effects on distributions that are relevant for
the 
$\mathrm{t\bar t H}$
analysis, both within the traditional ATLAS/CMS approach
and in the boosted-Higgs framework.

\subsection{PREDICTIONS FOR THE  LHC}

We study the process 
$\mathrm{pp\to t\bar t b\bar b}+X$ at
$\sqrt{s}=14\, \mathrm{TeV}$.  For the top-quark mass, renormalized in the
on-shell scheme, we take the numerical value $m_{\mathrm{t}}=172.6\, \mathrm{GeV}$
\cite{Group:2008nq}. All other QCD partons, including b~quarks,
are treated as massless particles.  Collinear final-state
configurations are recombined into
infrared-safe jets using 
the \mbox{$k_\mathrm{T}$-algorithm} of
Ref.~\cite{Blazey:2000qt}. We recombine all final-state $\mathrm{b}$ quarks and
gluons with pseudorapidity $|\eta| < 5$ into jets with separation
$\sqrt{\Delta\phi^2+\Delta y^2}>D=0.4$. 
Requiring two $\mathrm{b}$ jets, this also
avoids collinear singularities resulting from massless
$\mathrm{g\to b\bar b}$  splittings.\footnote{ Note that, as compared to our
  previous analysis\cite{Bredenstein:2008zb,Bredenstein:2009aj}, we
  have reduced the jet-algorithm parameter from $D=0.8$ to $D=0.4$~\cite{Bredenstein:2010rs}.
  This is particularly important for highly boosted  $\mathrm{b}$ -quark pairs with
$m_{\mathrm{b \bar b}}\sim M_{\mathrm{H}}$, since $D=0.8$ would lead to their
  recombination into a single jet and, consequently, to their
  rejection.}
After recombination, we impose the following cuts on the 
transverse momenta and rapidities of the $\mathrm{b}$ jets:
\begin{equation}\label{tTbB_cuts}
p_{\mathrm{T,b}}> 
20\,\mathrm{GeV}, \qquad
|y_\mathrm{b}|< 
2.5.
\end{equation}
Since top decays are not included in our calculation,
we treat top quarks fully inclusively.
We consistently use the CTEQ6~\cite{Pumplin:2002vw,Stump:2003yu} set of PDFs,
i.e.~we take CTEQ6L1 PDFs with a one-loop running $\alpha_{\mathrm{s}}$ in
LO and CTEQ6M PDFs with a two-loop running $\alpha_{\mathrm{s}}$ in
NLO, but neglect the suppressed contributions from b~quarks in the
initial state.  The number of active flavours is $N_{\mathrm{F}}=5$,
and the respective QCD parameters are
$\Lambda_5^{\mathrm{LO}}=165\, \mathrm{MeV}$ and
$\Lambda_5^{\overline{\mathrm{MS}}}=226\, \mathrm{MeV}$.  In the renormalization
of $\alpha_{\mathrm{s}}$ the top-quark loop in the gluon
self-energy is subtracted at zero momentum. In this scheme, the
running of $\alpha_{\mathrm{s}}$ is generated solely by the
contributions of the light-quark and gluon loops.

In all recent ATLAS studies of 
$\mathrm{t\bar t H(H\to b\bar b)}$~\cite{Kersevan:2002vu,Kersevan:2002dd,Cammin:2003,Aad:2009wy}
the signal and its 
$\mathrm{t\bar t b\bar b}$ background are simulated by
setting the renormalization and factorization scales equal to half the
threshold energy, $E_{\mathrm{thr}}=2 m_{\mathrm{t}}+m_{\mathrm{b \bar b}}$.
Being proportional to $\alpha^4_{\mathrm{s}}$, these LO predictions are
extremely sensitive to the scale choice, and in Refs.~\cite{Bredenstein:2009aj,Bevilacqua:2009zn}
it was found that at $\mu_\mathrm{R,F}=E_{\mathrm{thr}}/2$ the NLO corrections
to 
$\mathrm{pp\to t\bar t b\bar b}$ are close to a factor of two.
This enhancement is due to the fact that 
$\mathrm{pp\to t\bar t b\bar b}$ is
a multi-scale process involving various scales well below
$E_{\mathrm{thr}}/2$.
While $m_{\mathrm{t}}$ sets a clear scale for the couplings to the top quarks, the
inspection of differential distributions reveals that the cross section is
saturated by $\mathrm{b}$ quarks with $p_{\mathrm{T,b}}\ll m_{\mathrm{t}}$ (see~Fig.~\ref{tTbB_mbb100}). 
In order to avoid large logarithms due to these different scales we advocate
the use of the dynamical scale
\begin{equation}\label{tTbB_scale}
\mu^2_{\mathrm{R}}=\mu^2_{\mathrm{F}}=m_{\mathrm{t}}\sqrt{p_{\mathrm{T,b}}p_{\mathrm{T,\bar
b}}}.
\end{equation}
As we will see from the reduction of the $K$ factor and the scale
uncertainties, the scale choice (\ref{tTbB_scale}) significantly improves
the convergence of the perturbative expansion as compared to
Refs.~\cite{Bredenstein:2009aj,Bevilacqua:2009zn}.
In Sections~\ref{tTbT_sec.a} and
\ref{tTbT_sec.b}
we present NLO distributions 
obtained with standard ATLAS/CMS cuts and in the regime 
of highly boosted 
$\mathrm{b\bar b}$ pairs, respectively.

\subsubsection{The regime of high 
$\mathrm{\mathbf{b\bar b}}$ invariant mass}
\label{tTbT_sec.a}

Here, after imposing the standard cuts (\ref{tTbB_cuts}), we select the
kinematic region $m_{\mathrm{b \bar b}}>100\,\mathrm{GeV}$, which is relevant for standard
ATLAS/CMS studies of 
$\mathrm{t\bar t H(H\to b\bar b)}$.
At the scale (\ref{tTbB_scale}) we obtain the integrated cross sections
$\sigma_{\mathrm{LO}}=786.3(2)\, \mathrm{fb}$ and
$\sigma_{\mathrm{NLO}}=978(3)\, \mathrm{fb}$,
corresponding to a $K$ factor of $1.24$. Factor-two scale variations shift
the LO and NLO predictions by 78\% and 21\%, respectively.  This is a
significant improvement with respect to the scale choice
of Ref.~\cite{Bredenstein:2009aj}, which yields $K=1.77$ and 33\% NLO scale
uncertainty.
The NLO cross section at the scale (\ref{tTbB_scale}) is a factor $2.18$
larger as compared to the LO prediction obtained with the ATLAS scale choice,
$\left.\sigma_{\mathrm{LO}}\right|_{\mu_{\mathrm{R,F}}=E_{\mathrm{thr}}/2}=448.7(1)\, \mathrm{fb}$.

\begin{figure}
\includegraphics[bb= 95 445 280 655, width=.40\textwidth]
{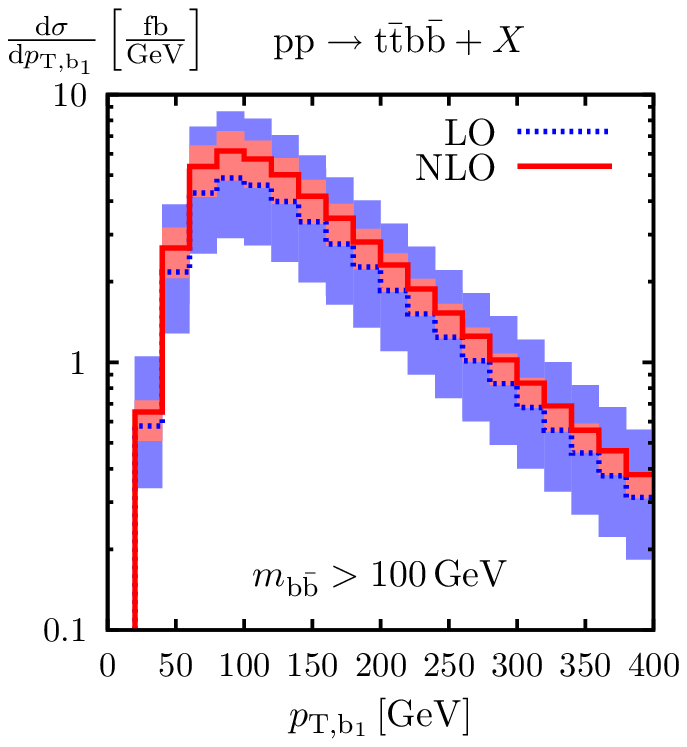}
\hfill
\includegraphics[bb= 95 445 280 655, width=.40\textwidth]
{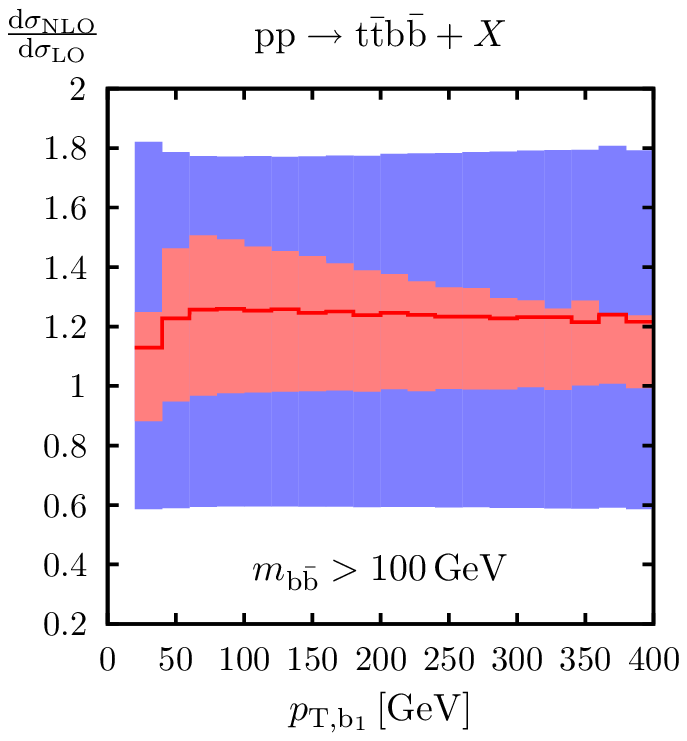}
\vspace{2mm}

\includegraphics[bb= 95 445 280 655, width=.40\textwidth]
{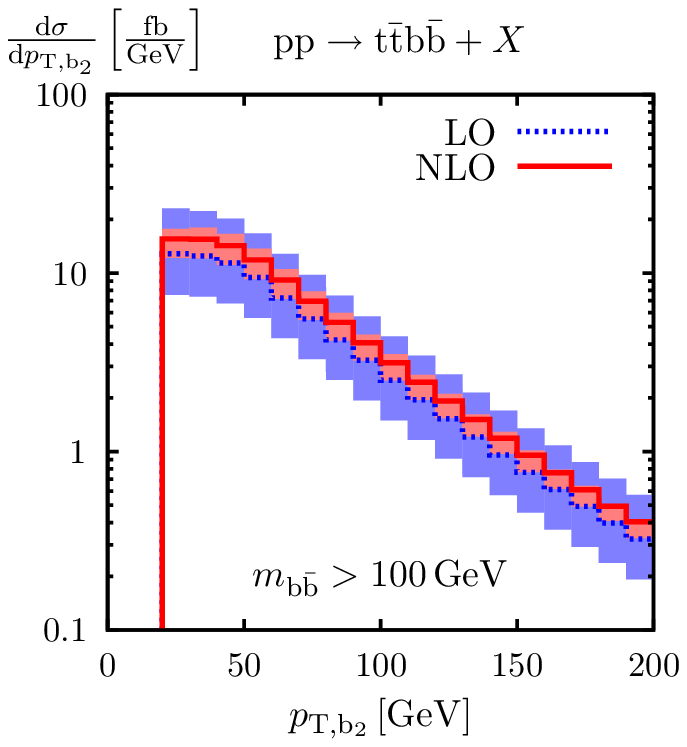}
\hfill
\includegraphics[bb= 95 445 280 655, width=.40\textwidth]
{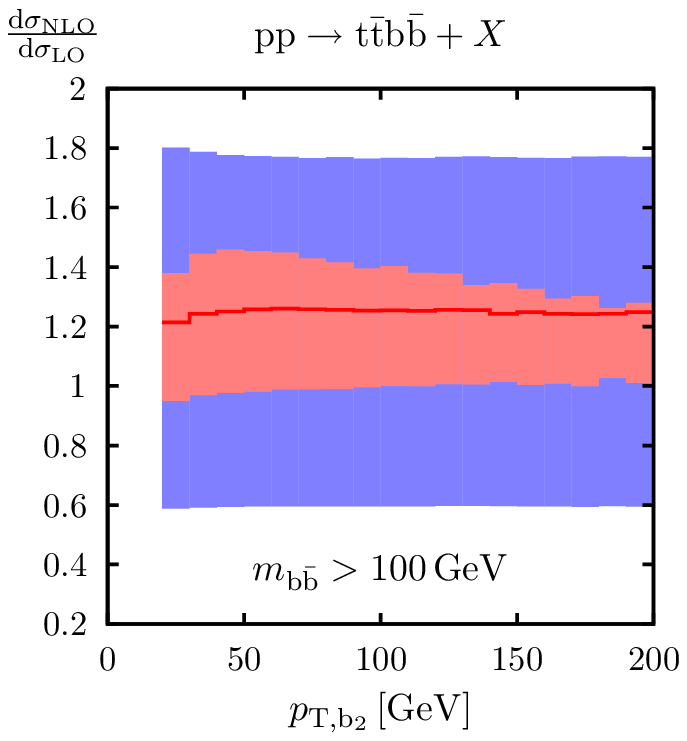}
\vspace{2mm}

\includegraphics[bb= 95 445 280 655, width=.40\textwidth]
{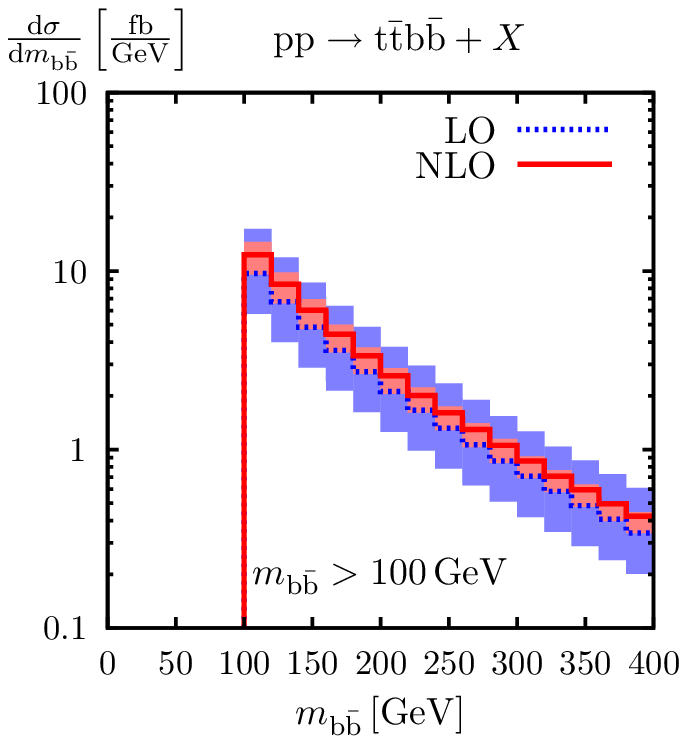}
\hfill
\includegraphics[bb= 95 445 280 655, width=.40\textwidth]
{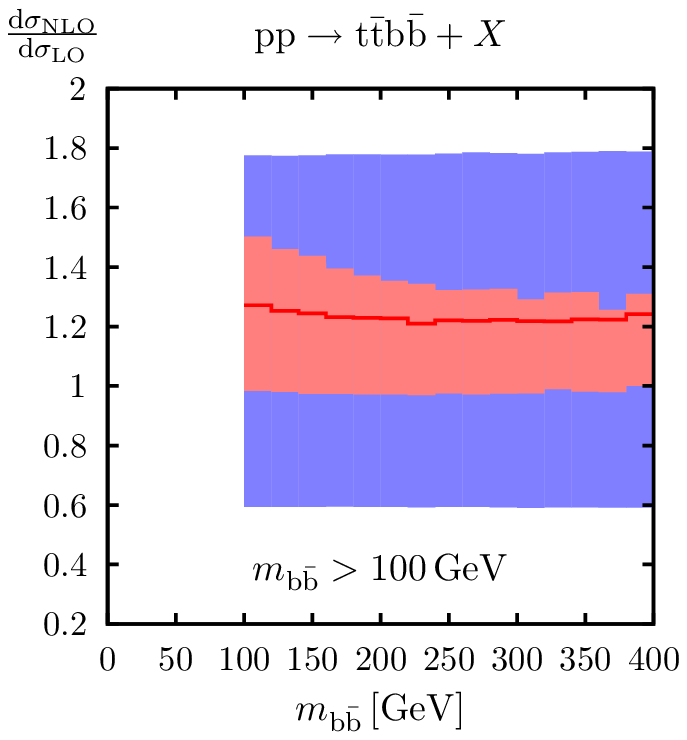}
\caption{Transverse-momentum 
of the harder ($\mathrm{b_1}$) and softer ($\mathrm{b_2}$) $\mathrm{b}$ jets 
and $m_{\mathrm{b \bar b}}$ distribution~\cite{Bredenstein:2010rs}:
absolute LO and NLO predictions (left) and NLO
$K$ factors (right) for $m_{\mathrm{b \bar b}}>100\, \mathrm{GeV}$.  
Uncertainty bands correspond to factor-two
  scale variations.}
\label{tTbB_mbb100}
\end{figure}
In Fig.~\ref{tTbB_mbb100} we present LO (blue) and NLO (red) results for various
distributions. Besides absolute predictions (left column) we show results
normalized to the LO distributions at the central scale (\ref{tTbB_scale})
(right column).  The bands correspond to factor-two variations of
$\mu_{\mathrm{R,F}}$.
The first two observables are the transverse momenta of the two $\mathrm{b}$ jets
ordered in $p_\mathrm{T}$. While the softer $\mathrm{b}$ jet ($p_{\mathrm{T,b_2}}$) tends to
saturate the lower bound of $20\, \mathrm{GeV}$, the harder
($p_{\mathrm{T,b_1}}$) behaves rather differently.  Its distribution has a maximum
around $100\, \mathrm{GeV}$ and a tail that extends up to fairly high transverse
momenta.
These shapes suggest that one of the two quarks is often emitted from
initial-state gluons, while the other one participates in the hard
scattering.  In contrast, none of the $\mathrm{b}$ quarks resulting from
$\mathrm{t\bar t H}$ originates from initial-state radiation.  This feature
 renders the cross section quite sensitive to $p_{\mathrm{T,b}}$ and might be
exploited to improve the separation of the 
$\mathrm{t\bar t H}$ signal.
The dynamical scale (\ref{tTbB_scale}) accounts for the different
kinematics of the two $\mathrm{b}$ jets and the extension of their transverse momenta
over a wide $p_\mathrm{T}$~range.  The goodness of this choice is confirmed by the
stability of the $K$ factor over the entire $p_\mathrm{T}$ spectrum.
A similarly stable behaviour is observed for the $m_{\mathrm{b \bar b}}$
distribution in Fig.~\ref{tTbB_mbb100}, where $1.21<K(m_{\mathrm{b \bar b}})<1.27$, as
well as for various other distributions~\cite{Bredenstein:2010rs}.

\begin{figure}
\includegraphics[bb= 95 445 280 655, width=.45\textwidth]
{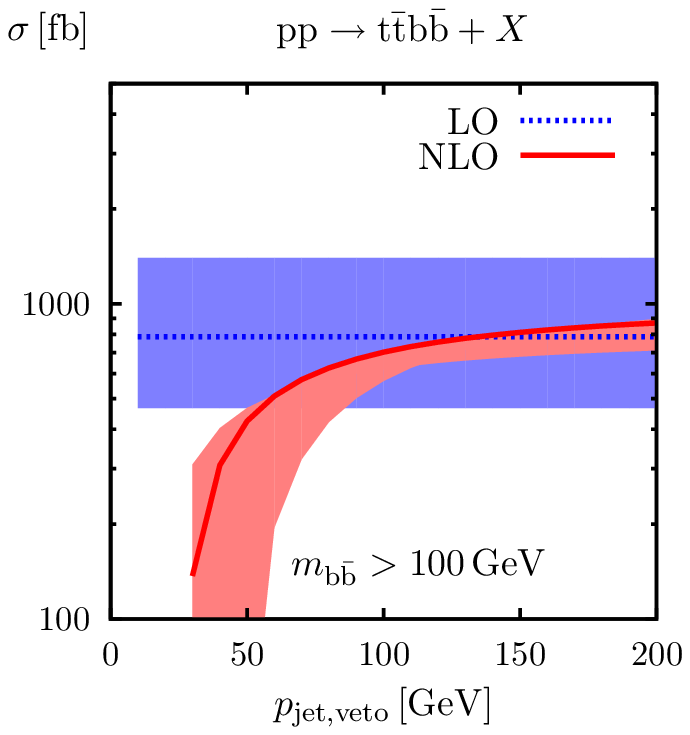}
\hfill
\includegraphics[bb= 95 445 280 655, width=.45\textwidth]
{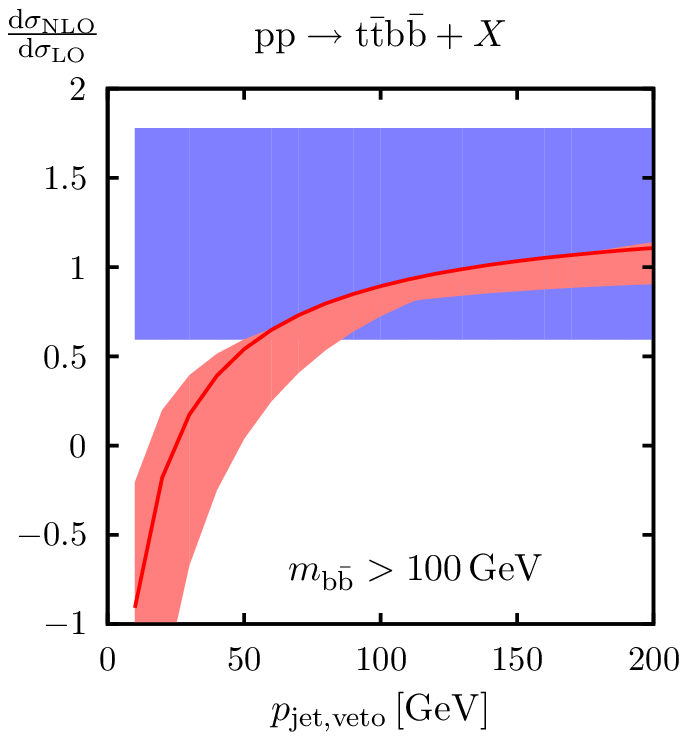}
\caption{Dependence of the
$\mathrm{pp\to t\bar t b\bar b}+X$ cross section 
on a jet veto  
($p_{\mathrm{T,jet}}<p_{\mathrm{jet,veto}}$)~\cite{Bredenstein:2010rs}:
absolute
LO and NLO predictions (left) and NLO $K$ factor (right)
for $m_{\mathrm{b \bar b}}>100\, \mathrm{GeV}$.
The bands correspond to factor-two scale variations.}
\label{tTbB_veto}
\end{figure}
In Fig.~\ref{tTbB_veto} we study the potential of a jet veto
($p_{\mathrm{T,jet}}<p_{\mathrm{jet,veto}}$) to suppress the large
$\mathrm{t\bar t b\bar b}$ background. The integrated cross section is plotted
versus $p_{\mathrm{jet,veto}}$, and since jet radiation takes place only at
NLO, the LO result is constant.
The NLO curve shows that a sizable reduction of the cross section requires a
jet veto well below $200\, \mathrm{GeV}$. For $p_{\mathrm{jet,veto}}=150,100$, and
$50\, \mathrm{GeV}$, the $K$ factor is reduced to $1.03$, $0.89$, and $0.54$,
respectively.
However, there is a trade-off between suppressing the NLO cross section and
increasing its perturbative uncertainty.  The jet veto tends to destroy the
cancellation between infrared logarithms of virtual and real origin and its
effect grows as
$-\alpha_{\mathrm{s}}^5\ln^2(E_{\mathrm{thr}}/p_{\mathrm{jet,veto}})$ when
$p_{\mathrm{jet,veto}}$ becomes small.
Since they are accompanied by an $\alpha_{\mathrm{s}}^5$ coefficient, these
logarithms can give rise to huge scale uncertainties already for moderate
values of $p_{\mathrm{jet,veto}}$. This is reflected by the dramatic
amplification of the uncertainty band in Fig.~\ref{tTbB_veto}.  Around
$p_{\mathrm{jet,veto}}=50\, \mathrm{GeV}$ the NLO band enters the pathologic regime of
negative cross sections, and perturbative predictions become completely
unreliable. Jet-veto values around $100\, \mathrm{GeV}$ provide a good compromise: the
reduction of the $K$ factor is already significant ($K\simeq 0.89$) and the
NLO scale uncertainty (19\%) is at the same level as for the total cross
section (21\%).

\subsubsection{The regime of highly boosted 
$\mathrm{\mathbf{b\bar b}}$ pairs}
\label{tTbT_sec.b}

Here, after imposing the standard cuts (\ref{tTbB_cuts}), we select
highly boosted 
$\mathrm{b\bar b}$ pairs with $p_{\mathrm{T,b\bar b}}>200\, \mathrm{GeV}$. This
permits to increase the separation between the Higgs signal and its
$\mathrm{t\bar t b\bar b}$ background~\cite{Plehn:2009rk}.
Although we do not impose any cut on the 
$\mathrm{b\bar b}$ invariant mass, the
cuts on $p_{\mathrm{T,b\bar b}}$, $p_{\mathrm{T,b}}$ and
$p_{\mathrm{T,\bar b}}$, and the
jet algorithm ($\Delta R_{\mathrm{b \bar b}}>0.4$), result into a lower bound
$m_{\mathrm{b \bar b}}\gsim 25\, \mathrm{GeV}$~\cite{Bredenstein:2010rs}.
At the scale (\ref{tTbB_scale}) we obtain the
integrated cross sections $\sigma_{\mathrm{LO}}=451.8(2)\, \mathrm{fb}$ and
$\sigma_{\mathrm{NLO}}=592(4)\, \mathrm{fb}$.  As compared to the previous setup
($m_{\mathrm{b \bar b}}>100\, \mathrm{GeV}$) the absolute cross section is reduced by about
$40\%$, the NLO correction is slightly increased ($K=1.31$), and the shifts
induced by factor-two scale variations are almost identical (79\% in LO and
22\% in NLO).  In presence of a jet veto of 100 GeV the $K$ factor and the
NLO uncertainty amount to 0.84 and 33\%, respectively.

\begin{figure}
\includegraphics[bb= 95 445 280 655, width=.40\textwidth]
{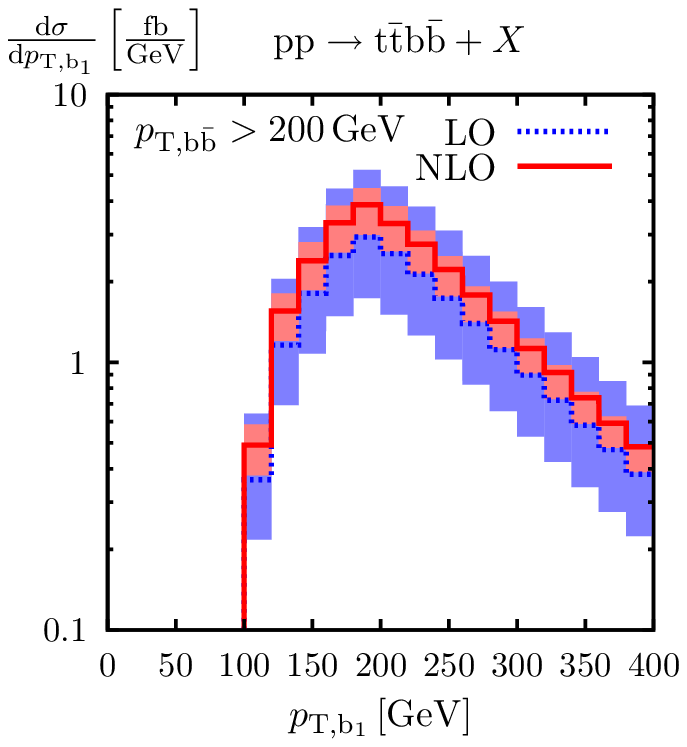}
\hfill
\includegraphics[bb= 95 445 280 655, width=.40\textwidth]
{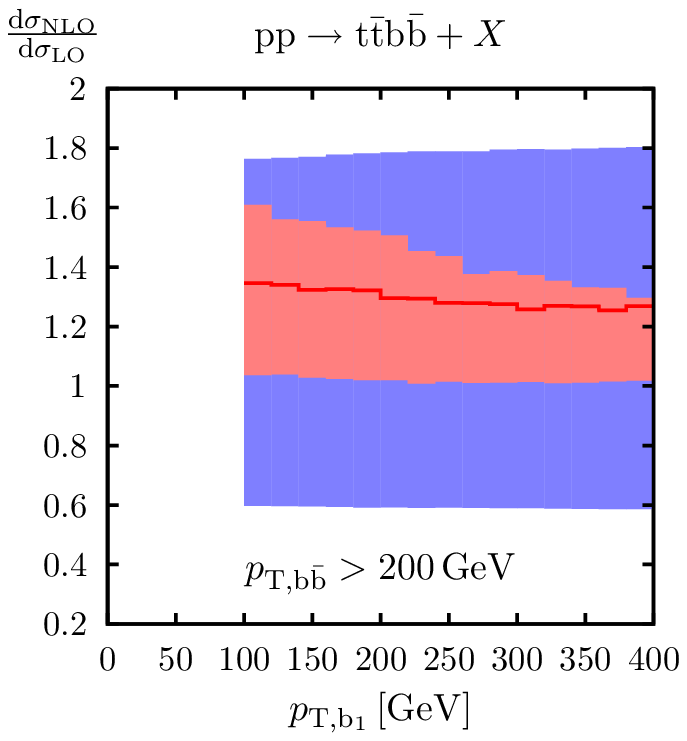}
\vspace{2mm}

\includegraphics[bb= 95 445 280 655, width=.40\textwidth]
{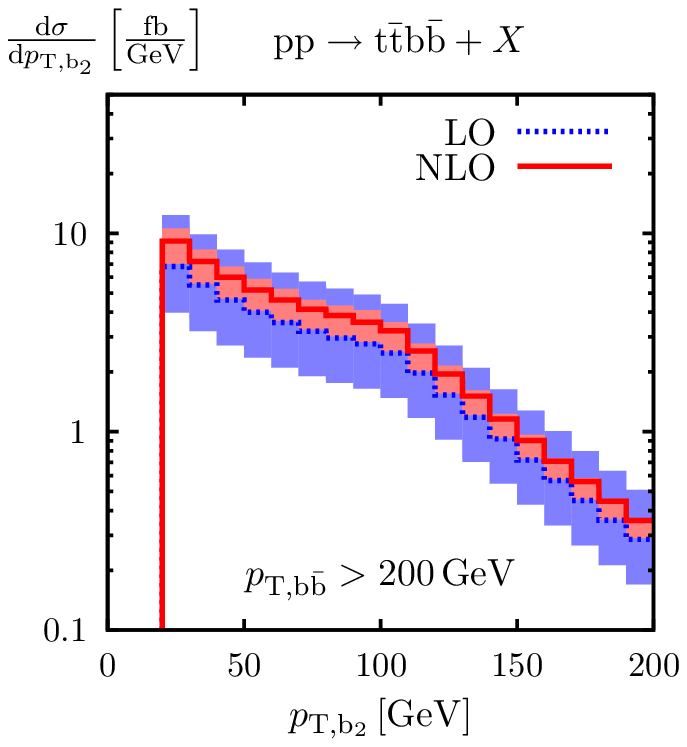}
\hfill
\includegraphics[bb= 95 445 280 655, width=.40\textwidth]
{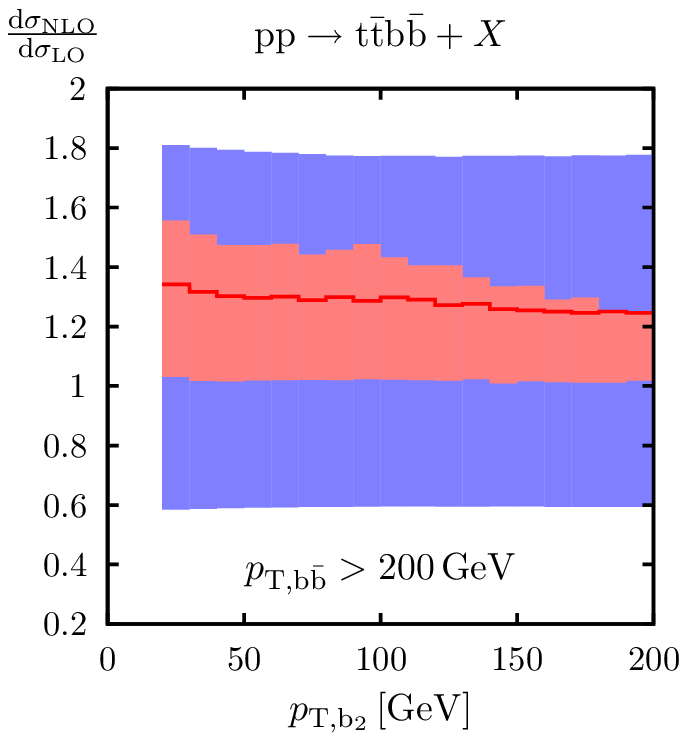}
\vspace{2mm}

\includegraphics[bb= 95 445 280 655, width=.40\textwidth]
{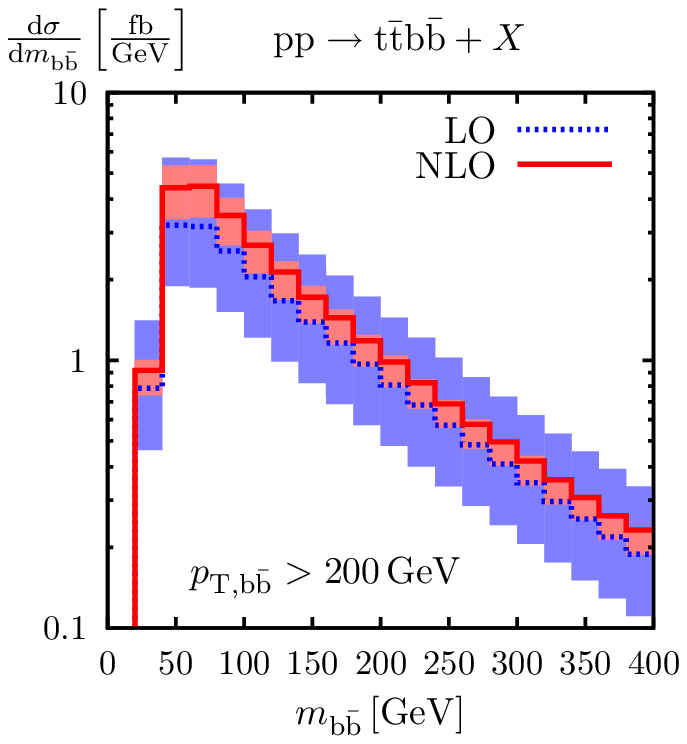}
\hfill
\includegraphics[bb= 95 445 280 655, width=.40\textwidth]
{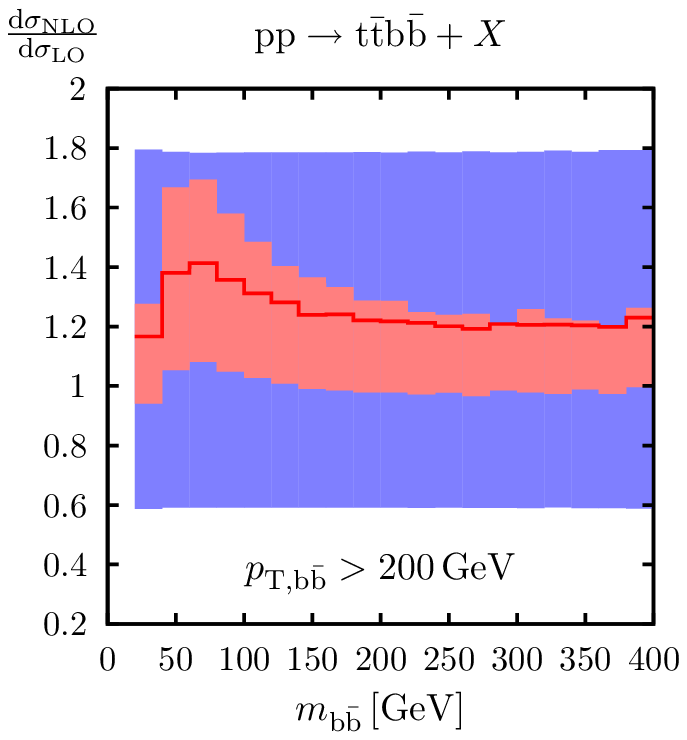}
\caption{Transverse-momentum 
of the harder ($\mathrm{b_1}$) and softer ($\mathrm{b_2}$) $\mathrm{b}$ jets 
and $m_{\mathrm{b \bar b}}$ distribution~\cite{Bredenstein:2010rs}:
absolute LO and NLO predictions (left) and NLO
$K$ factors (right) for $p_{\mathrm{T,b\bar b}}>200\, \mathrm{GeV}$.  Uncertainty bands correspond to factor-two
  scale variations.}
\label{tTbB_pTbb200}
\end{figure}
In Fig.~\ref{tTbB_pTbb200} we present the distributions of the transverse
momenta of the harder ($p_{\mathrm{T,b_1}}$) and softer ($p_{\mathrm{T,b_2}}$)
$\mathrm{b}$ jets
and the 
$\mathrm{b\bar b}$ invariant mass.  
As a consequence of the cut $p_{\mathrm{T,b\bar b}}>200\, \mathrm{GeV}$, the harder
$\mathrm{b}$ jet
is pushed to much higher $p_\mathrm{T}$ values as compared to Fig.~\ref{tTbB_mbb100}:
the maximum of its distribution is located around 200 GeV.  In contrast, the
softer $\mathrm{b}$ jet is much less sensitive to the $p_{\mathrm{T,b\bar b}}$ cut and is
predominantly produced in the region $20\, \mathrm{GeV}< p_{\mathrm{T,b_2}} <
100\, \mathrm{GeV}$.
This different kinematic behaviour of the two $\mathrm{b}$ jets might be
exploited to separate the 
$\mathrm{t\bar t b\bar b}$ background from the
$\mathrm{t\bar t H}$ signal, where both $\mathrm{b}$ jets are produced by the Higgs
boson and should thus have more similar $p_\mathrm{T}$-values.  
As a consequence of the dynamical scale choice, the NLO corrections feature
a mild dependence on the  $\mathrm{b}$-jet $p_\mathrm{T}$: the $K$ factor varies by about 10\%
within the plotted range.

The 
$\mathrm{b\bar b}$ invariant-mass distribution is displayed in the third row
of Fig.~\ref{tTbB_pTbb200}. Its behaviour in the region 
$m_{\mathrm{b \bar b}}\lsim
50\, \mathrm{GeV}$ reflects the abovementioned effective bound. Near the physically
interesting region of $m_{\mathrm{b \bar b}}\sim 100\, \mathrm{GeV}$ we observe that the NLO
corrections induce an appreciable shape distortion of about 20\%. Such
an effect tends to mimic a Higgs signal and should be carefully taken into
account in the 
$\mathrm{t\bar t H(H\to b\bar b)}$ analysis.

\subsection*{CONCLUSIONS}

The observation of the 
$\mathrm{t\bar t H(H\to b\bar b)}$ signal 
at the LHC requires a
very precise description of the 
$\mathrm{t\bar t b\bar b}$ irreducible
background. The NLO QCD corrections reveal that the scale choice adopted in
previous lowest-order simulations of 
$\mathrm{pp\to t\bar t b\bar b}$ does not account
for the multi-scale character of this process and underestimates its cross
section by a factor of two. We advocate the use of a new dynamical scale,
which significantly reduces both the $K$ factor and the residual NLO scale
uncertainty. 
In presence of standard ATLAS/CMS cuts NLO effects feature small 
kinematic dependence. But in the regime of highly boosted Higgs bosons 
we observe significant distortions in the shape of distributions.
Studying a jet veto as a possible strategy to suppress the
$\mathrm{t\bar t b\bar b}$ background, we find that jet-veto values well below
100 GeV lead to severe perturbative instabilities and should thus be
avoided. 
Further results are presented in Ref.~\cite{Bredenstein:2010rs}.
}



%% file: magnea/magnea.tex
\def \as {\relax\ifmmode\alpha_s\else{$\alpha_s${ }}\fi}
\def\e{\epsilon}
\def\b{\beta}
\def\M{{\cal M}}






\subsection{Introduction}
\label{magnea_intro}

The ultimate goal of our efforts as practitioners of quantum field theory
applied to high-energy physics is the calculation of the finite transition 
probabilities and cross sections that form the predictions to be tested
at particle colliders. It would be nice if we had methods to compute
directly these finite quantities, without having to deal with divergences 
in the intermediate stages of our calculations. Our tools are, however,
imperfect: perturbative methods in quantum field theory typically start 
out by mishandling both very short and very long wavelength excitations, 
and one finds divergent results along the way, which need to be regularized, properly interpreted, and finally cancelled in order to 
obtain the finite predictions that we need.
At the high-frequency end of the spectrum, this is achieved by renormalization of the parameters of the theory, a process that we understand well and that we can carry out explicitly at high
perturbative orders. At the low-frequency end of the spectrum, 
the solution is provided in principle by the construction of properly 
defined, sufficiently inclusive observables, as characterized by the 
Kinoshita-Lee-Nauenberg thorem. The practical implementation of
this theoretical insight has however proven challenging, especially 
in the context of confining non-abelian gauge theories like QCD.
Indeed, while we do have all-order perturbative proofs of factorization 
theorems and of the exponentiation of infrared and collinear 
divergences, not much was known until recently concerning the 
detailed structure of the anomalous dimensions that govern infrared exponentiation; on the other hand, at finite orders, the task of building 
a fully general and efficient subtraction algorithm to compute infrared-safe cross sections at NNLO in perturbation theory has been pursued
by several groups for many years, however complete results are 
available so far only for processes with no hadrons in the initial state.

There is thus much that we still need to understand concerning
infrared and collinear singularities, and it is perhaps worth recalling 
that we have strong motivations, both of theoretical and
phenomenological nature, to study the problem. At the simplest 
level, the all-order structure of singularities for scattering amplitudes 
provides `theoretical predictions' for a subset of terms arising in 
finite-order calculations, and these can be used to test the results
obtained order by order. Much more interestingly, in infrared-safe observables the cancellation of singularities leaves behind logarithms 
of ratios of kinematic invariants, and these finite contributions tend 
to dominate the cross sections in the vicinity of thresholds and in 
other situations where soft radiation is predominant. The way in 
which logarithms arise as finite remainders, after  summing 
contributions from real emission and from virtual corrections, is 
easily visualized in dimensional regularization. Setting $d = 4 - 2 \e$, 
with $\e <0$ for infrared regularization, the structure of the 
cancellation is typically of the form
\begin{equation}
\label{real_virt_canc}
 \qquad \underbrace{{\frac{1}{\epsilon}_{\,\,}}_{\,}}_{\rm virtual}\,\,\,\,
 + \, \underbrace{
 (Q^2)^{\epsilon} \int_0^{m^2} \frac{d k^2}{(k^2)^{1 + \epsilon}}
 }_{\rm real} \,\, \qquad \quad \Longrightarrow \,\,\,\,\,\, 
 \ln (m^2/Q^2) \, ,
\end{equation}
where $m$ is some scale associated with the chosen observable, 
say a jet mass. It is clear that the coefficients of the logarithms are 
dictated by the coefficients of the singular contributions, much as 
renormalization group logarithms are dictated by ultraviolet poles.

This tight relationship between singularities and logarithmic 
enhancements underlies a host of phenomenological applications:
since singularities exponentiate, so do logarithms, and, as a 
consequence, finite-order calculations can be used to perform all-order
resummations. These resummations actually take place in two rather 
different contexts, and with rather different tools. On the one hand,
for sufficiently inclusive observables, one can perform the real-virtual cancellation analytically, and provide all-order expressions for the
observable, to some definite logarithmic accuracy (the current 
standard being NNLL). On the other hand, data analysis at present colliders requires predictions for exclusive final states, in order to 
match the kinematic cuts dictated by experimental needs: to this 
end, one must supplement fixed order calculations with parton shower 
event generators, and subsequently hadronization. Parton showers
effectively resum the same collinear and infrared logarithms, but 
they do so without implementing real-virtual cancellations: rather, 
they impose cutoffs on real radiation, and they generate multiparticle 
final states by iteration, mimicking the pattern of exponentiation
of singularities. It should be emphasized that the coefficients of the relevant logarithms are the same: parton showers perform the 
resummation with suitable approximations (typically leading 
logarithmic, with the inclusion of some NLL terms, and typically 
taking the large-$N_c$ limit, as well as performing some sort
of angular averaging) which are necessary in order to turn the
quantum-mechanical time evolution into a Markov chain with
an iterative probabilistic interpretation. Improving our understanding 
of infrared and collinear singularities can be instrumental for both 
these applications: the analytic treatment of inclusive cross sections 
would be performed to a higher logarithmic accuracy, and extended 
to complex processes where more partons participate in the hard 
interaction. In these circumstances an accurate description requires 
taking full account of the interference between amplitudes of 
different colour flow. These colour correlations are encoded in 
the singularity structure of multi-leg amplitudes. At the same time, 
the approximations employed in the fully-exclusive parton shower
approach could be put under better theoretical control. Present 
day patron showers assume independent emission from individual
partons, and consequently fail to describe soft radiation at large
angles with respect to the jets. Upon considering multi-jet cross
sections, the independent-emission approximation becomes less 
reliable, since soft gluon radiation gets increasingly sensitive to the
underlying colour flow. In conclusion, in order to achieve the necessary 
precision for the complex QCD processes that will be under study at 
LHC, accurate predictions would be required for complex multi-jet 
cross section. These will be achieved both by extending the analytic 
approach to inclusive cross sections, and by improving the treatment 
of parton showers. Understanding the singularity structure is the first 
step towards achieving these goals.

One should remember, finally, that studying soft and collinear 
radiation means probing the long-distance behavior of the theory, 
and thus it is of great interest also from a purely theoretical 
standpoint. All massless gauge theories, at the perturbative level,
have a remarkably similar singularity structure, governed by a 
set of common anomalous dimensions: the most important 
differences arise from the different behavior of the running coupling
in different theories. In conformal gauge theories, such as ${\cal N} = 
4$ Super Yang-Mills (SYM), the pattern of exponentiation of 
divergences is greatly simplified by the simplicity of the running
coupling, and has been used as a guideline to study the exponentiation 
of finite contributions to scattering amplitudes~\cite{Bern:2005iz}.
In a confining theory like QCD, resummation displays the divergent
behavior of the perturbative series, which can be used to gauge 
the weight of non-perturbative effects in the kinematic 
regions in which their importance is enhanced. Good control on the 
size and shape of these power-suppressed effects has been achieved
for inclusive distributions in simple processes which are electroweak 
at tree level~\cite{Dasgupta:2003iq,Gardi:2006jc}: one may now 
hope that this level of understanding can be reached also for general 
multi-jet cross-sections.

With these motivations, we present below some recent striking
theoretical developments, implying that the structure of infrared
and collinear divergences in massless gauge theories, for amplitudes
with any number of colored partons, and to all orders in the $1/N_c$
expansion, is significantly simpler than previously expected. In 
Sect.~\ref{dipole} we present an expression organizing all divergences
as the exponentiation of a simple sum over contributions from color
dipoles, that was recently proposed in Refs.~\cite{Becher:2009cu,%
Gardi:2009qi,Becher:2009qa,Gardi:2009zv}, and that reproduces 
all known results of finite order perturbative calculations. In 
Sect.~\ref{corre} we explain how possible corrections to the dipole
formula are tightly constrained, and in fact forced to have a very 
specific functional dependence on kinematics, which correlates to 
their color structure. Whether these corrections do indeed arise, 
starting with three-loop, four-point amplitudes, is the subject of 
current studies.

\subsection{The dipole formula}
\label{dipole}

Our goal in this section is to illustrate the recent progress in our understanding of soft and collinear singularities. The main result 
is the establishment, to all orders in perturbation theory, and for 
any $N_c$, of a set of differential equations, which strongly 
constrain the soft anomalous dimensions for general multi-parton, 
fixed-angle amplitudes, in any massless gauge theory.
These equations  were derived from factorization, and by enforcing 
the invariance of soft gluon dynamics under the rescaling of hard 
parton momenta. The simplest solution to these constraints, which 
reproduces all known fixed-order perturbative results, is a compact
expression, encoding a simple correlation of color and kinematic 
degrees of freedom, and taking the form of a sum over color dipoles.
Below, we illustrate the structure of this `dipole formula': for 
the arguments suggesting its validity, and for detailed derivations
of the constraint equations, we refer the reader to 
Refs.~\cite{Gardi:2009qi,Becher:2009qa}.

Let ${\cal M} \left(p_i/\mu, \alpha_s (\mu^2), \epsilon \right)$ be
a renormalized scattering amplitude involving a fixed number $n$ of 
hard coloured partons carrying momenta $p_i$, $i = 1 \, \ldots \, n$, 
all lightlike, plus any number of additional non-coloured 
particles. The singularities of ${\cal M}$ depend on all the kinematic 
invariants that can be formed out of the hard parton momenta, $p_i 
\cdot p_j$, which are all assumed to be of the same parametric size, 
and all large compared to $\Lambda_{\rm QCD}^2$. Momentum 
conservation is not imposed between the coloured partons, allowing 
for any recoil momentum to be carried by non-coloured particles in 
both the initial and final states. As far as color is concerned, ${\cal M}$
should be thought of as a vector in the vector space spanned by the
color tensors available for the given set of hard partons. Soft and 
collinear factorization properties guarantee that all singularities can 
be absorbed into an overall multiplicative factor $Z$, acting as a 
matrix in color space. One writes formally
\begin{equation}
\label{introducing_Z}
 {\cal M} \left(p_i/\mu, \alpha_s (\mu^2), \epsilon \right)  \, = \,
 Z \left(p_i/\mu_f, \alpha_s (\mu_f^2), \epsilon \right) \,\,
 {\cal H} \left( p_i/\mu, \mu/\mu_f, \alpha_s (\mu^2), \e \right) \, ,
\end{equation}
where ${\cal H}$ is a vector in color space, which remains finite 
as $\e \to 0$, and we have introduced a factorization scale $\mu_f$
to isolate the infrared momentum region (below we will set $\mu_f = 
\mu$ for simplicity). Note that $\mu_f$ is introduced in the context 
of dimensional regularization and not as an explicit cutoff.

The matrix $Z$, accounting for all soft and collinear singularities, 
can be written in exponential form, as a consequence of appropriate 
evolution equations derived from factorization. The simplest 
expression for the logarithm of $Z$, satisfying all available constraints, 
and reproducing all known finite-order results, is a sum over color 
dipoles. $Z$ is thus conjectured to take the form
\begin{eqnarray}
  Z \left(p_i/\mu, \alpha_s (\mu^2), \epsilon \right)  
  \hspace{-1mm} & = & \hspace{-1mm}
  \exp\Bigg\{
  \int_0^{\mu^2}\frac{d\lambda^2}{\lambda^2} \Bigg[
  \frac18 \,
  \widehat{\gamma}_K\left(\alpha_s(\lambda^2,\epsilon)  \right)  
  \,\sum_{i \neq j} \,
  \ln\left(\frac{2 p_i\cdot p_j\,{\rm e}^{- {\rm i} \pi \phi_{ij} 
  }}{{\lambda^2}}\right) 
  \,  \mathrm{T}_i \cdot   \mathrm{T}_j \,  \nonumber \\ 
  &&  - \,\frac12\, \sum_{i=1}^n \, \gamma_{J_i} \left(\alpha_s 
  (\lambda^2, \epsilon) \right)  \Bigg]\Bigg\} \, .
\label{Z}  
\end{eqnarray}
Let us briefly explain the notations in Eq.~(\ref{Z}), and then pause 
to illustrate its physical content. The color structure of $Z$ is encoded 
in the color generators ${\rm T}_i$ associated with each hard parton, which
are linked in color dipoles by the products $\mathrm{T}_i \cdot 
\mathrm{T}_j \equiv \sum_a \mathrm{T}_i^a \cdot  \mathrm{T}_j^a$.
The matrices ${\rm T}^a$ depend on the identity of the hard parton: 
for a final-state quark or an initial-state antiquark they 
are just the generators of the fundamental representation, $T^a_{\alpha
\beta} = t^a_{\alpha \beta}$; for a final-state antiquark or an 
initial-state quark, $T^a_{\alpha \beta} = - t^a_{\beta \alpha}$; 
for a gluon, $T^a_{b c} = {\rm i} \, f_{cab}$. This ensures that color 
conservation is simply expressed by $\sum_{i = 1}^n \mathrm{T}_i^a 
= 0$. The phases $\phi_{ij}$ are introduced to display how the analytic 
properties of the amplitude change when the invariants $p_i \cdot  
p_j$ change from time-like to space-like: we define $p_i  \cdot p_j 
= - \left| p_i \cdot p_j \right| \, {\rm e}^{- {\rm i} \pi \phi_{ij}}$, 
where $\lambda_{ij} = 1$ if $i$ and $j$ are both initial-state partons 
or are both final--state partons, and $\lambda_{ij} = 0$ otherwise.

All singularities in Eq.~(\ref{Z}) are generated through the integration 
over the scale $\lambda^2$ of the $d$-dimensional running coupling
$\alpha_s (\lambda^2, \e)$~\cite{Magnea:1990zb,Magnea:2000ss}, 
and are governed by the finite anomalous dimensions 
$\widehat{\gamma}_K  (\as)$ and $\gamma_{J_i} (\as)$, which can 
be characterized as follows. $\gamma_{J_i} (\as)$ is simply the 
anomalous dimension of the field corresponding to hard parton $i$,
and is responsible for single collinear poles arising from radiation 
forming the virtual jet associated with that parton; it depends on 
parton spin as well as color, and it is known to three loops for quarks 
and gluons. $\widehat{\gamma}_K (\as)$, on the other hand, is 
simply related to the cusp anomalous dimension $\gamma_K^{(i)} 
(\as)$~\cite{Korchemsky:1985xj}, for Wilson lines in the 
color representation of parton $i$: in deriving Eq.~(\ref{Z}), 
we have assumed that the latter is proportional to all 
orders to the appropriate quadratic Casimir operator, according to 
$\gamma_K^{(i)} (\as) = C_i \, \widehat{\gamma}_K (\as)$, a fact 
which is established up to three loops; factoring out the quadratic 
Casimir operator $C_i$ is important in deriving Eq.~(\ref{Z}) since it 
can be expressed as $C_i  = {\rm T}_i \cdot {\rm T}_i$. 
The cusp factor in Eq.~(\ref{Z}) is responsible for all singularities 
associated with soft gluons, including double poles arising from the 
exchange of gluons that are both soft and collinear to some hard 
parton: these double poles arise from the integration over the scale 
$\lambda^2$, thanks to the explicit factor of $\log \lambda^2$.

\begin{figure}
\begin{center}
\includegraphics[width=0.3\textwidth]{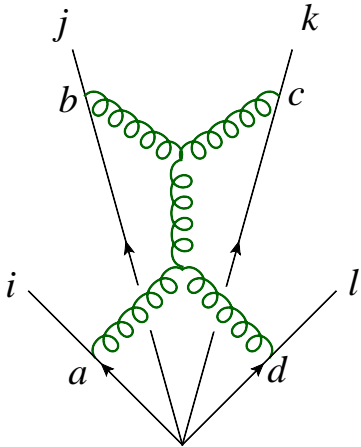}
  \caption{
  A three-loop gluon web connecting four hard partons. Such webs
  are the possible sources of the lowest order admissible violations
  of the dipole formula, since they may yield functions of the 
  conformal cross ratios $\rho_{ijkl}$.
  }
\label{magnea_figure}
\end{center}
\end{figure}

The striking feature of Eq.~(\ref{Z}) is that it correlates color and 
kinematic degrees of freedom, and it does so in an unexpectedly 
simple way. Indeed, correlations through dipoles only are what one expects, and finds, at one-loop order, where only one soft gluon can
be exchanged, correlating at most two hard partons. Beyond first 
order, the rules of eikonal exponentiation, as well as expectations 
from the analysis of ordinary Feynman diagrams, suggest that at 
$g$ loops one  should expect contributions from gluon webs linking
up to $g+1$ hard partons, and thus correlating $g+1$ momenta
$p_i$ and color generators $T_i$ (an example of such a web is 
shown in Fig.~\ref{magnea_figure}, for $g = 3$). Multiparton 
correlations in the exponent of Eq.~(\ref{Z}) were shown to vanish
at two loops in Ref.~\cite{Aybat:2006wq,Aybat:2006mz}, 
and at three loops for diagrams involving matter fields in 
Ref~\cite{Dixon:2009gx}. Although they cannot as yet be completely 
ruled out at higher orders, the derivation of Eq.~(\ref{Z}) shows that 
they are tightly constrained: they can only arise starting at three loops,
from webs correlating at least four partons, such as the one shown in 
Fig.~\ref{magnea_figure}, and they must have a very restricted 
dependence on color and kinematics, as briely discussed in 
Sect.~\ref{corre}

There at least three further features of Eq.~(\ref{Z}) that should 
be emphasized. The first point is to note that Eq.~(\ref{Z}) 
represents a very smooth generalization of the situation arising 
in the large $N_c$ limit: at large $N_c$, only planar diagrams 
contribute, which forces soft gluons to be exchanged only 
between adjacent hard partons in the specific color ordering 
being considered. This forces the color structure to be in the 
form of a sum over {\it adjacent} color dipoles, to all orders 
in perturbation theory; this color structure is in fact trivial 
and reduces to a product of color singlet form 
factors~\cite{Bern:2005iz,Sterman:2002qn,Dixon:2008gr}. 
Eq.~(\ref{Z}) generalizes this in a natural way, by simply extending 
the range of the sum to cover all possible dipoles, including 
non-adjacent ones: an extension which is sufficient to make the color 
exchange non-trivial, but simple enough to be determined at one loop.
The second observation stems from the fact that the correlated 
color and momentum structures appearing in the exponent 
of Eq.~(\ref{Z}) are fixed at one loop: the only effect of radiative 
corrections is to build up the anomalous dimensions 
$\widehat{\gamma}_K (\as)$ and $\gamma_{J_i} (\as)$. The consequence of this is that there is no need to introduce a path 
ordering in Eq.~(\ref{Z}), even though it is an expression arising
from the solution of a matrix evolution equation. When working
in a definite color basis in order to apply Eq.~(\ref{Z}) to some
specific problem, the diagonalization of the anomalous dimension 
matrix can be performed once and for all at the one loop level, after
which further radiative corrections affect just the anomalous 
dimensions. This leads to the third and final observation: 
Eq.~(\ref{Z}) appears to give greater weight and a more solid 
theoretical foundation to the idea of employing the cusp anomalous
dimension (or rather its representation-independent counterpart
$\widehat{\gamma}_K (\as)$, which is conjectured to be 
universal) as an effective charge  for soft and collinear 
gluon emission. This idea, brought forward in~\cite{Catani:1990rr} 
as a tool to improve the logarithmic accuracy of parton showers,
and subsequently developed in~\cite{Gardi:2002xm,Gardi:2007ma}, 
is generalized by Eq.~(\ref{Z}) beyond planar configurations, 
and appears to apply to the full complexity of QCD color exchange.

\subsection{Possible corrections to the dipole formula}
\label{corre}

Having described in detail the structure and implications of the 
dipole formula, we must also clearly indicate its limitations and
describe the corrections that may still arise at high perturbative 
orders, escaping the various constraints that have established it
as a credible all-order ansatz. 

First of all, let's emphasize again that the dipole formula arises 
as the simplest solution to a set of constraint equations. These
equations are derived by requiring the cancellation of an anomaly due 
to cusp singularities in the invariance of light-like Wilson lines under
rescalings of their velocity vectors. Since this rescaling invariance
is not anomalous  for Wilson lines off the light cone, the constraints 
do not apply to lines corresponding to massive partons. Indeed, 
an analysis of amplitudes involving massive partons has 
shown~\cite{Becher:2009kw} that the dipole formula breaks 
down for amplitudes involving at least two massive partons,
already at the two-loop level. At two-loops, in that case, 
three-parton correlations arise, parametrized by two functions
which have now been computed analytically at two 
loops~\cite{Ferroglia:2009ep,Ferroglia:2009ii}, and which 
vanish parametrically as the fourth power of the parton mass.

In the case of massless partons, the constraints arising from factorization and rescaling invariance are much more powerful.
While Eq.~(\ref{Z}) is established as the simplest solution to 
the constraint equations, the same constraints also indicate that
only two classes of corrections can arise at higher orders, going
beyond dipole correlations.
The first possible class of corrections arises if the cusp anomalous dimension $\gamma_K^{(i)} (\as)$ turns out not be proportional
to the quadratic Casimir $C_i$ to all orders. This can happen if
$\gamma_K^{(i)} (\as)$ receives contributions from higher-rank
Casimir operators of the gauge group, as is the case, for example,
for the four-loop QCD $\beta$ function. Contributions of this kind
can, in principle, arise starting at four loops. Arguments were given
in Ref.~\cite{Becher:2009qa}, based on symmetry and collinear 
consistency, suggesting that the contribution of the quartic Casimir operator is absent at four loops. On the other hand, studies of 
the cusp anomalous dimension at strong coupling, in the 
large-$N_c$ limit, in a special class of representations of the gauge
group~\cite{Armoni:2006ux}, suggest that higher-rank Casimir
operators should contribute at sufficiently high orders. Whether 
these rather special color correlations ultimately play a role in
the infrared structure of nonabelian gauge theories remains an
interesting open problem.

The second class of potential corrections to the dipole formula
arises from the fact that it is possible to respect the invariance
under rescalings of hard parton momenta, simply by considering 
functions of `conformally invariant cross ratios' of the form
\begin{equation}
\rho_{ijkl} \equiv \frac{p_i \cdot p_j \, p_k \cdot p_l}{p_i 
\cdot p_k \, p_j \cdot p_l}~.
\label{ccr}
\end{equation}
Since these cross-ratios are by themselves rescaling invariant,
they are not constrained by factorization. The rules of nonabelian
exponentiation, however, imply that contributions of this kind
must arise from gluon webs connecting directly at least four 
hard partons, which can only happen starting at three loops.

It should be emphasized, as noted in Ref.~\cite{Becher:2009qa},
that functions of conformal cross ratios are still strongly constrained
by requiring consistency with known collinear limits, and by imposing Bose symmetry: these further constraints, for example, rule out
functions linear in the logarithms of the cross ratios at three loops.
A more systematic analysis of the constraints on the available
functional forms was carried out in Ref.~\cite{Dixon:2009ur},
where the limits on the possible degree of transcendentality of the functions involved were also taken into account. It turns out
that functions satisfying all available constraints, and which might
plausibly arise in the calculation of high-order Feynman diagrams,
do exist, though they appear to form a very limited set. For example,
considering polynomials in the logarithms of the conformal cross 
ratios, precisely one function survives the constraints at three loops. 
If however one allows for polylogarithmic dependence, further examples 
can be  found. Whether indeed these `quadrupole' corrections to
the dipole formula do arise at three loops or  beyond remains at
the moment an open question.

Summarizing, a simple exponential all-order ansatz exists for the 
singularities of multi-parton amplitudes in massless gauge theories.
It has a solid theoretical foundation, arising from soft-collinear factorization, and it is exact at two loops in the exponent. 
Furthermore, possible corrections at higher perturbative 
orders can arise from only two sources, and must have very 
restricted structures in both color and kinematics. The resulting 
picture is much simpler than might have been expected from 
previous analyses, and the remaining potential corrections are 
actively studied. We are on track to gain full control of the soft 
and collinear dynamics of massless  gauge theories.

\subsection*{Acknowledgments}

We are indebted to Thomas Binoth --- who left us too soon --- for
drawing the attention of the Les Houches community to the work described in this contribution.
 
\noindent Work supported by by the European Community's Marie 
Curie Research Training Network `Tools and Precision Calculations 
for Physics Discoveries at Colliders'  (`HEPTOOLS'), under contract 
MRTN-CT-2006-035505.



%% file: weinzierl/weinzierl.tex




\subsection{INTRODUCTION}

The process $e^+ e^- \rightarrow 3 \; \mbox{jets}$ is of particular interest for the measurement of the
strong coupling $\alpha_s$. 
Three-jet events are well suited for this task because the leading term in a perturbative calculation 
of three-jet observables is already proportional to the strong coupling.
For a precise extraction of the strong coupling one needs in addition to a precise measurement of three-jet
observables in the experiment a precise prediction 
for this process from theory. This implies the calculation of higher order corrections.
The process $e^+ e^- \rightarrow 3 \; \mbox{jets}$ has been been calculated recently 
at next-to-next-to-leading order (NNLO) in 
QCD \cite{GehrmannDeRidder:2007hr,GehrmannDeRidder:2007bj,GehrmannDeRidder:2008ug,GehrmannDeRidder:2009dp,Weinzierl:2008iv,Weinzierl:2009ms,Weinzierl:2009yz}.
This was a very challenging calculation and I will report on some of the complications which occurred
during this computation.
The lessons we learned from this process have implications to other processes which will be calculated
at NNLO.
The two processes closest related to $e^+ e^- \rightarrow 3 \; \mbox{jets}$ are
$e^- p \rightarrow e^- + \;\mbox{2 jets}$ and $p p \rightarrow Z/W + \;\mbox{jet}$.
These are obtained from crossing final and initial state particles.
But also for processes like $p p \rightarrow \;\mbox{2 jets}$ and $p p \rightarrow t \bar{t}$
many techniques can be transferred.

\subsection{THE CALCULATION}

The master formula for the calculation of a three-jet observable at an 
electron-positron collider is
\begin{eqnarray}
\langle {\cal O} \rangle & = & \frac{1}{8 s}
             \sum\limits_{n \ge 3}
             \int d\phi_{n}
             {\cal O}_n\left(p_1,...,p_n,q_1,q_2\right)
             \sum\limits_{helicity} 
             \left| {\cal A}_{n} \right|^2,
\end{eqnarray}
where $q_1$ and $q_2$ are the momenta of the initial-state particles and $1/(8s)$ corresponds to the flux factor 
and the average over the spins of the initial state particles.
The observable has to be infrared safe, in particular this implies that in single and double
unresolved limits we must have
\begin{eqnarray}
{\cal O}_{4}(p_1,...,p_{4},q_1,q_2) & \rightarrow & {\cal O}_3(p_1',...,p_3',q_1,q_2)
 \;\;\;\;\;\;\mbox{for single unresolved limits},
 \nonumber \\
{\cal O}_{5}(p_1,...,p_{5},q_1,q_2) & \rightarrow & {\cal O}_3(p_1',...,p_3',q_1,q_2)
 \;\;\;\;\;\;\mbox{for double unresolved limits}.
\end{eqnarray}
${\cal A}_n$ is the amplitude with $n$ final-state partons.
At NNLO we need the following perturbative expansions of the amplitudes:
\begin{eqnarray}
& & 
 \left| {\cal A}_3 \right|^2
 =  
   \left. {\cal A}_3^{(0)} \right.^\ast {\cal A}_3^{(0)} 
 + 
   \left(
             \left. {\cal A}_3^{(0)} \right.^\ast {\cal A}_3^{(1)} 
           + \left. {\cal A}_3^{(1)} \right.^\ast {\cal A}_3^{(0)} 
       \right)
 + 
   \left(
             \left. {\cal A}_3^{(0)} \right.^\ast {\cal A}_3^{(2)} 
           + \left. {\cal A}_3^{(2)} \right.^\ast {\cal A}_3^{(0)}  
           + \left. {\cal A}_3^{(1)} \right.^\ast {\cal A}_3^{(1)} 
       \right),
 \nonumber \\
& &
  \left| {\cal A}_4 \right|^2
 =  
   \left. {\cal A}_4^{(0)} \right.^\ast {\cal A}_4^{(0)}
 + 
   \left(
          \left. {\cal A}_4^{(0)} \right.^\ast {\cal A}_4^{(1)} 
        + \left. {\cal A}_4^{(1)} \right.^\ast {\cal A}_4^{(0)}
 \right),
 \nonumber \\ 
& &
  \left| {\cal A}_5 \right|^2
 = 
   \left. {\cal A}_5^{(0)} \right.^\ast {\cal A}_5^{(0)}. 
\end{eqnarray}
Here ${\cal A}_n^{(l)}$ denotes an amplitude with $n$ final-state partons and $l$ loops.
We can rewrite symbolically
the LO, NLO and NNLO contribution as 
\begin{eqnarray}
\label{weinzierl_nnlo_def_LO_NLO_NNLO}
\langle {\cal O} \rangle^{LO} & = & \int {\cal O}_{3} \; d\sigma_{3}^{(0)},
 \nonumber \\
\langle {\cal O} \rangle^{NLO} & = & \int {\cal O}_{4} \; d\sigma_{4}^{(0)} + \int {\cal O}_{3} \; d\sigma_{3}^{(1)},
 \nonumber \\
\langle {\cal O} \rangle^{NNLO} & = & \int {\cal O}_{5} \; d\sigma_{5}^{(0)} 
                   + \int {\cal O}_{4} \; d\sigma_{4}^{(1)} 
                   + \int {\cal O}_{3} \; d\sigma_{3}^{(2)}.
\end{eqnarray}
The computation of the NNLO correction for the process $e^+ e^- \rightarrow \mbox{3 jets}$ 
requires the knowledge of the amplitudes
for the three-parton final state 
$e^+ e^- \rightarrow \bar{q} q g$ up to two-loops \cite{Garland:2002ak,Moch:2002hm},
the amplitudes of the four-parton final states
$e^+ e^- \rightarrow \bar{q} q g g$
and
$e^+ e^- \rightarrow \bar{q} q \bar{q}' q'$
up to one-loop \cite{Bern:1997ka,Bern:1997sc,Campbell:1997tv,Glover:1997eh}
and the five-parton final states
$e^+ e^- \rightarrow \bar{q} q g g g$
and
$e^+ e^- \rightarrow \bar{q} q \bar{q}' q' g$
at tree level \cite{Berends:1989yn,Hagiwara:1989pp,Nagy:1998bb}.
The most complicated amplitude is of course the two-loop amplitude. For the calculation of the two-loop
amplitude special integration techniques have been 
invented \cite{Gehrmann:1999as,Gehrmann:2000zt,Gehrmann:2001ck,Moch:2001zr,Weinzierl:2002hv,Moch:2005uc}. The analytic result can be expressed in terms of multiple polylogarithms, which in turn
requires routines for the numerical evaluation of these functions \cite{Gehrmann:2001pz,Gehrmann:2001jv,Vollinga:2004sn}.

\subsection{SUBTRACTION AND SLICING}

Is is well known that the individual pieces in the NLO and in the NNLO contribution of 
eq.~(\ref{weinzierl_nnlo_def_LO_NLO_NNLO}) are infrared divergent.
To render them finite, a mixture of subtraction and slicing is employed.
The NNLO contribution is written as
\begin{eqnarray}
\label{weinzierl_nnlo_nnlo_subtraction}
\langle {\cal O} \rangle^{NNLO} & = &
 \int \left( {\cal O}_{5} \; d\sigma_{5}^{(0)} 
             - {\cal O}_{4} \circ d\alpha^{single}_{4}
             - {\cal O}_{3} \circ d\alpha^{(0,2)}_{3} 
      \right) \nonumber \\
& &
 + \int \left( {\cal O}_{4} \; d\sigma_{4}^{(1)} 
               + {\cal O}_{4} \circ d\alpha^{single}_{4}
               - {\cal O}_{3} \circ d\alpha^{(1,1)}_{3}
        \right) \nonumber \\
& & 
 + \int \left( {\cal O}_{3} \; d\sigma_3^{(2)} 
               + {\cal O}_{3} \circ d\alpha^{(0,2)}_{3}
               + {\cal O}_{3} \circ d\alpha^{(1,1)}_{3}
        \right).
\end{eqnarray}
$d\alpha^{single}_{4}$ is the NLO subtraction term for $4$-parton configurations,
$d\alpha^{(0,2)}_{3}$ and $d\alpha^{(1,1)}_{3}$ are generic NNLO subtraction terms,
which can be further decomposed into
\begin{eqnarray}
 d\alpha^{(0,2)}_{3} & = & d\alpha^{double}_{3} + d\alpha^{almost}_{3} + d\alpha^{soft}_{3} 
                           - d\alpha^{iterated}_{3},
 \nonumber \\
 d\alpha^{(1,1)}_{3} & = & d\alpha^{loop}_{3} + d\alpha^{product}_{3}
                         - d\alpha^{almost}_{3} - d\alpha^{soft}_{3} + d\alpha^{iterated}_{3}.
\end{eqnarray}
In a hybrid scheme of subtraction and slicing the subtraction terms have to satisfy weaker
conditions as compared to a strict subtraction scheme.
It is just required that 
\begin{description}
\item{(a)} the explicit poles in the dimensional regularisation parameter $\varepsilon$
in the second line of eq.~(\ref{weinzierl_nnlo_nnlo_subtraction})
cancel after integration over unresolved phase spaces
for each point of the resolved phase space.
\item{(b)} the phase space singularities 
in the first and in the second line of eq.~(\ref{weinzierl_nnlo_nnlo_subtraction}) 
cancel after azimuthal averaging has been performed.
\end{description}
Point (b) allows the determination of the subtraction terms from spin-averaged matrix elements.
The subtraction terms can be found in \cite{Gehrmann-DeRidder:2005cm,GehrmannDeRidder:2007jk,Weinzierl:2006ij}.
The subtraction term $d\alpha^{(0,2)}_{3}$ without $d\alpha^{soft}_{3}$ would approximate 
all singularities except a soft single unresolved singularity. The subtraction term
$d\alpha^{soft}_{3}$ takes care of this last piece \cite{Weinzierl:2008iv,Weinzierl:2009nz}.
The azimuthal average is not performed in the Monte Carlo integration.
Instead a slicing parameter $\eta$ is introduced to regulate the phase space singularities
related to spin-dependent terms.
It is important to note that there are no numerically large contributions proportional
to a power of $\ln \eta$ which cancel between the 5-, 4- or 3-parton contributions.
Each contribution itself is independent of $\eta$ in the limit $\eta\rightarrow 0$.

\subsection{MONTE CARLO INTEGRATION}

The integration over the phase space is performed numerically with Monte Carlo techniques.
Efficiency of the Monte Carlo integration is an important issue, especially for the first moments
of the event shape observables. Some of these moments receive sizable contributions from the close-to-two-jet region.
In the 5-parton configuration this corresponds to (almost) three unresolved partons.
The generation of the phase space is done sequentially, starting from a 2-parton configuration.
In each step an additional particle is inserted.
In going from $n$ partons to $n+1$ partons, the $n+1$-parton phase space is partitioned into different channels.
Within one channel, the phase space is generated iteratively according to
\begin{eqnarray}
 d\phi_{n+1} & = & d\phi_n d\phi_{unresolved\;i,j,k}
\end{eqnarray}
The indices $i$, $j$ and $k$ indicate that the new particle $j$ is inserted between the hard radiators $i$ and $k$. 
For each channel we require that the product of invariants $s_{ij} s_{jk}$ is the smallest among all considered channels.
For the unresolved phase space measure we have
\begin{eqnarray}
 d\phi_{unresolved\;i,j,k}
 & = & \frac{s_{ijk}}{32 \pi^3} 
       \int\limits_0^1 dx_1 
       \int\limits_0^1 dx_2  
       \int\limits_0^{2\pi} d\varphi \; \Theta(1-x_1-x_2)
\end{eqnarray}
We are not interested in generating invariants smaller than $(\eta s)$, these configurations will be rejected by the slicing procedure.
Instead we are interested in generating invariants with values larger than $(\eta s)$ with a distribution which 
mimics the one of a typical matrix element.
We therefore generate the $(n+1)$-parton configuration from the $n$-parton configuration by using three random numbers
$u_1$, $u_2$, $u_3$ uniformly distributed in $[0,1]$ and by setting
\begin{eqnarray}
 x_1 = \eta_{PS}^{u_1}, \;\;\; x_2 = \eta_{PS}^{u_2} \;\;\; \varphi = 2 \pi u_3.
\end{eqnarray}
The phase space parameter $\eta_{PS}$ is an adjustable parameter of the order of the slicing parameter $\eta$.
The invariants are defined as
\begin{eqnarray}
 s_{ij} = x_1 s_{ijk},
 \;\;\;
 s_{jk} = x_2 s_{ijk},
 \;\;\;
 s_{ik} = (1-x_1-x_2) s_{ijk}. 
\end{eqnarray}
From these invariants and the value of $\varphi$ we can reconstruct the four-momenta of the $(n+1)$-parton configuration
\cite{Weinzierl:1999yf}.
The additional phase space weight due to the insertion of the $(n+1)$-th particle is
\begin{eqnarray}
 w & = & \frac{1}{16 \pi^2} \frac{s_{ij} s_{jk}}{s_{ijk}} \ln^2\eta_{PS}.
\end{eqnarray}
Note that the phase space weight compensates the typical eikonal factor $s_{ijk}/(s_{ij}s_{jk})$ of a single emission.
As mentioned above, the full phase space is constructed iteratively from these single emissions.



%% file: zanderighi/zanderighi.tex

\def\rc#1{{\red\it #1}}







\subsection{Introduction}\label{sec:intro}

Next-to-leading order (NLO) calculations are the first order at which
the normalization, and sometimes the shape, of perturbative
calculations can be considered reliable. On the other hand,
experimenters deal primarily with leading order (LO) calculations,
especially in the context of parton-shower Monte Carlo programs.
Given the better accuracy at fixed order in the coupling, the
predictions at NLO should be used (in appropriate regions of phase
space) as a benchmark for the various types of LO calculations. This
way the LO Monte Carlo programs can be validated, tuned and/or
improved before the actual comparison with data will be pursued.

Many of the interesting final states at the Tevatron and LHC involve
the production of a $W$\/ boson and multiple jets. Recently the NLO
calculation for the hadro-production of final states consisting of a
$W$\/ boson and 3 jets has been completed by two groups, one retaining
the full colour information \cite{Berger:2009zg,Berger:2009ep}, the
other using a leading-colour approximation \cite{Ellis:2009zw,%
  KeithEllis:2009bu,Melnikov:2009wh}.
This calculation provides, for the first time, a reliable prediction
for $W$\/ + 3 jet production both at the Tevatron and at the LHC.
Searches for new physics will of course benefit from these new
achievements, since they allow for a more detailed understanding of
one of the major backgrounds to beyond Standard Model signals.

At NLO, the dependence on the renormalization and factorization scales
is greatly reduced from that at LO. It has been noted that the use
of some scales in LO calculations for $W$\/ + 3 jets can result in
significant shape differences with NLO calculations. Conversely, the
use of other scales at LO can mimic the results obtained at NLO.
In Ref.~\cite{Berger:2009ep} it has been shown that the use of a scale
($H_T$) equal to the scalar sum of the missing transverse energy and
the transverse energies of the lepton and all jets in the event
reproduces the shape of the NLO calculation at LO for many relevant
kinematic distributions of a typical $W$\/ + 3 jet analysis, i.e.\
search cuts are applied in favour of $W$\/ production in association
with jets.\footnote{An optimized scale setting in the context of
  resumming logarithms in $PP\to V$\/ + jets has been also discussed
  in Ref.~\cite{Bauer:2009km} where the authors arrived at similar
  conclusions, but suggested as a scale
  $\sqrt{m^2_W+m^2_\mathrm{hadr}/4}$ where $m_\mathrm{hadr}$ is the
  hadronic mass of the final state.}
Dynamically generated scales, such as obtained with either the MLM or
CKKW procedures \cite{Alwall:2007fs}, are typically much smaller than $H_T$.
They usually are identified by backward clustering procedures that
locally analyze the relative transverse momenta (or similar
quantities) of pairs of hard matrix-element partons. As well known in
QCD, the scale of the coupling should be determined by the relative
$p_T$ of a branching that occurred in the perturbative evolution.
As for the case of optimal scale choices at LO, a similar improved
agreement with NLO predictions can be achieved by combining these
dynamical scale schemes with the necessary Sudakov rejection or
reweighting of events that are described by tree-level matrix elements
of higher order. Advantageously, these matching approaches are largely
independent of the kinematical cuts applied. Nevertheless, it is
interesting to try to understand why the two different procedures for
setting the scales lead to similar results. Of course, an NLO
prediction for $W$\/ + 3 jets provides a better description of the
cross section and kinematics, but experimenters are mostly confined to
the use of LO predictions, where the full event can be simulated.

\begin{figure}[t!]
  \centerline{
    \includegraphics[clip,width=0.71\columnwidth]{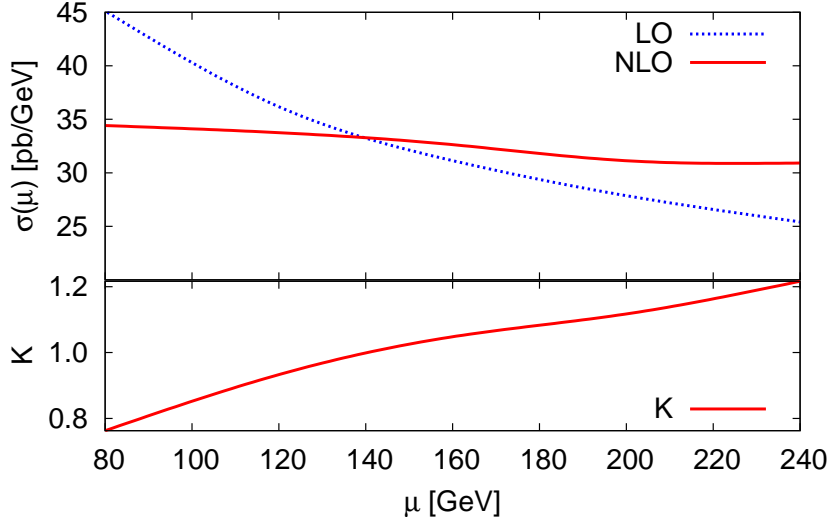}}
  \caption{Scale dependence of the $W^+$ + 3 jet inclusive production
    at the LHC (10~TeV) at LO and NLO. The lower panel displays the
    $K$-factor. The cuts used are given in Section~\ref{sec:calcs} At
    NLO the leading-colour adjustment procedure has been applied.}
  \label{fig:NLO}
\end{figure}

In Figure~\ref{fig:NLO}, we show the scale dependence for inclusive
$W^+$ + 3 jet production at the LHC (10~TeV) at LO and NLO. The cross
section at NLO has a smaller scale dependence than at LO as expected.
The maximum cross section occurs near a scale of $m_W$, but is
stable for a fairly wide range above that value. The LO cross
section, of course, decreases monotonically with increasing scale, but
the slope is less steep for larger scales.
As can be seen in Figure~\ref{fig:NLO} the choice of a
renormalization/factorization scale such as $2\,m_W$ places the $W$\/
+ 3 jet cross section in a region where the variation of the cross
section with scale is reduced. (An examination of the scale dependence
above and below the limits of this plot shows the usual parabolic
scale dependence.) For typical values of $H_T$, of the order of
$2\ldots4\,m_W$, and cuts as given below, the LO and NLO
inclusive cross sections will have a similar magnitude, i.e.\ the
total $K$-factor (NLO/LO) tends to be close to unity
($1.0\ldots1.25$)~\cite{huston:2010xp,Berger:2009ep}.

\subsection{Event generation parameters and description of the
  calculations}\label{sec:calcs}

In this study we are comparing predictions for the production of $W^+$
bosons in association with 3 or more jets at LHC energies of 10 TeV,
i.e.\ $PP\to e^+\nu_e$ + 3 jets at order
$\alpha^2_\mathrm{QED}\alpha^k_s$ where $k\ge3$.
We consider the vector boson to decay leptonically into a pair of
$e^+\nu_e$, hence all cross sections are given with the corresponding
branching ratio taken into account. For our comparison, we apply cuts
typical for signal studies. We require jets to have transverse
momenta $p_{T,j}>30$ GeV and pseudo-rapidities $|\eta_j|<3$. Jets are
defined according to the SISCone jet algorithm \cite{Salam:2007xv}
with $\Delta R=\sqrt{\Delta\eta^2+\Delta\phi^2}=0.5$ and merging parameter
$f=0.5$. For the leptonic sector, we cut on the lepton transverse
momentum, $p_{T,e}>20$ GeV, the lepton pseudo-rapidity, $|\eta_e|<2.4$,
the missing transverse energy, $\slashed{E}_T>15$ GeV, and the
vector-boson transverse mass, $m_{T,W}=
\sqrt{(|\vec p_{T,e}|+|\vec p_{T,\nu}|)^2-(\vec p_{T,e}+\vec p_{T,\nu})^2}>30$
GeV; we however do not impose lepton isolation cuts.

For all programs, we use the following parameters for the event
generation: the processes $PP\to e^+\nu_e+n$ partons are calculated
taking the quarks to be massless except for the top quark, which is
not considered. For the value of $\alpha_s(m_Z)$, we employ the fit
result as given by the respective PDF. At NLO (LO) the $\alpha_s(\mu)$
values are evaluated from two-loop (one-loop) running. The $W$\/ boson
mass is set to $m_W=80.419$ GeV, its couplings to fermions are
calculated from $\alpha_\mathrm{QED}(m_Z)=1/128.802$ and
$\sin^2\theta_\mathrm{W}=0.230$; the CKM matrix is taken to be the
identity matrix.

The different calculations used to obtain the $W^+$ + 3 jets
predictions of this study have been accomplished in the following way:
\begin{itemize}
\item {\sc BlackHat+Sherpa}~\cite{Berger:2008sj,Berger:2009zg,Berger:2009ep}:\\
  We have used two programs for this comparison: the virtual matrix
  elements are evaluated by {\sc BlackHat} \cite{Berger:2008sj}. The
  real part is computed within the {\sc Sherpa} framework
  \cite{Gleisberg:2008ta} using an automated Catani--Seymour dipole
  subtraction \cite{Catani:1996vz,Catani:1996erra,Gleisberg:2007md}.
  The phase-space integrations are entirely handled by {\sc Sherpa}.
  All subleading colour contributions have been included in the
  calculation. The renormalization and factorization scales are
  commonly set to
  $\hat{H}_T=\sum_\mathrm{partons}E_{T,p}+E_{T,e}+E_{T,\nu}$, which is
  determined dynamically on an event by event basis. The sum runs over
  all partons in the event, regardless of whether they will pass the
  jet cuts. This prevents a jet-algorithm and cut dependency on the
  scale choice. At LO, summing over the transverse energies of the
  jets is equivalent to summing those of the partons.
\item {\sc Rocket}~\cite{Giele:2008bc,Ellis:2009zw,KeithEllis:2009bu}:\\
  We closely follow Ref.~\cite{KeithEllis:2009bu} and perform
  calculations in the leading-colour approximation. The calculation
  relies heavily on the framework provided by {\sc MCFM}~\cite{mcfm}
  and uses one-loop amplitudes as calculated in
  Ref.~\cite{Ellis:2008qc}. We employ the Catani--Seymour dipole
  subtraction \cite{Catani:1996vz,Catani:1996erra} to compute the real
  emission corrections. The details of the implementation are given in
  \cite{Ellis:2009zw,Melnikov:2009wh}. We use the leading-colour
  adjustment (aLC) procedure described in the latter paper to correct
  for deficiencies of the leading-colour approximation, to the extent
  possible.\footnote{Briefly described, one multiplies the NLO results
    (both the virtual and the real part)
    by an overall adjustment parameter that is determined as the ratio
    of the full-colour over the leading-colour cross sections at LO.}
  The renormalization and factorization scales are chosen to be equal
  and given by the transverse energy of the $W$\/ boson, which has been
  defined as $E_{T,W}=\sqrt{m^2_W+p^2_{T,W}}$. The top quark is
  assumed to be infinitely heavy; the running of the strong coupling
  therefore evaluated in the five-flavour scheme.
\item {\sc Sherpa}~\cite{Gleisberg:2008ta}:\\
  For the {\sc Sherpa} event generation, we have used version 1.2.0
  \cite{sherpa}.\footnote{Effects induced by hadronization and the
    underlying event have not been taken into account. Their impact is
    tiny, furthermore we are not going to compare distributions at the
    particle level.}
  It incorporates a new strategy for merging tree-level higher-order
  matrix elements and parton showers, which we denote here as ME\&TS
  (matrix-element \& truncated-shower) merging
  \cite{Hoeche:2009rj,Hoeche:2009xc,Carli:2009cg}. This approach
  improves over the CKKW method owing to the incorporation of a
  consistent treatment of local scales that occur, on the one hand, in
  the matrix-element calculations and, on the other, in the parton
  showering. To ensure the strict ordering of the shower evolution,
  truncated shower algorithms are necessary for the ME\&TS approach to
  work properly. As a result the systematic uncertainties of the
  ME\&TS merging are greatly reduced with respect to CKKW.
  We have generated predictions from samples that merge matrix
  elements with up to $N^\mathrm{max}_\mathrm{ME}=2+3,2+4,2+5$
  particles, i.e.\ $PP\to e^+\nu_e+n$ partons where
  $n=0,\ldots,N^\mathrm{max}_\mathrm{ME}-2$. Notice that
  $N^\mathrm{max}_\mathrm{ME}$ denotes the maximum number of
  final-state particles of the matrix elements. For the evaluation of
  the PDF scales, the default scheme of ME\&TS has been employed. It
  is based on the identification of the most likely $2\to2$
  interaction that may lead to the actual $2\to2+n$ matrix-element final
  state; the factorization scale is then chosen according to the
  kinematics of that $2\to2$ core process
  \cite{Hoeche:2009rj}.\footnote{In most cases, the scale is set by
    the $\hat s^{1/2}$ of the identified $2\to2$ core process.}
  Scales of the strong couplings are entirely determined by the ME\&TS
  algorithm. The merging scale has been set to $Q_\mathrm{cut}=28$~GeV
  (to have it somewhat lower than the jet $p_T$ threshold). As for the
  comparison to the NLO results, the most relevant {\sc Sherpa}
  prediction relies on the $N^\mathrm{max}_\mathrm{ME}=2+4$ merged
  sample, since it contains the real-emission matrix elements for 3
  and 4 extra partons. For this case, we therefore have varied the
  default scales identified by the ME\&TS algorithm by factors of
  $1/2$ and $2$.\footnote{More exactly, in the case of reduced scales,
    we set $\mu_\mathrm{R}=0.5\,\mu_\mathrm{R}^\mathrm{ME\&TS}$, but
    used $\mu_\mathrm{F}=0.8\,\mu_\mathrm{F}^\mathrm{ME\&TS}$ for
    reasons of avoiding too low PDF scales in the shower evolution.}
  This leaves us with an estimate of the theoretical uncertainty of
  the ME\&TS results.
\end{itemize}
We would like to stress the major differences between the different
approaches: for the NLO case, the calculations only differ in their
treatment of colour (full colour for {\sc BlackHat+Sherpa} vs.\
leading colour for {\sc Rocket}) and choice of scales ($\hat{H}_T$ for
{\sc BlackHat+Sherpa} vs.\ $E_{T,W}$ for {\sc Rocket}). In the
{\sc Sherpa} case, it is evident that the virtual corrections to $W^+$
+ 3 jets are not completely taken into account, they only enter in
an approximate way through Sudakov form factor terms at
leading-logarithmic accuracy. The scales cannot be set globally as in
the NLO calculations, they have to be determined and set locally.

In addition to the predictions outlined above, we show LO $W^+$ + 3
jet parton-level results for two different choices of a common
factorization and renormalization scale,
$\mu=\mu_\mathrm{F}=\mu_\mathrm{R}$, defined by $\mu^2=\hat H^2_T$ and
$\mu^2=E^2_{T,W}=m^2_W+p^2_{T,W}$.\footnote{Note that the
  $W$\/ boson mass is taken as a parameter and not reconstructed from
  the momenta of the decay products.}
These results have been produced with the tree-level
matrix-element generator {\sc Comix}~\cite{Gleisberg:2008fv} by making
use of the {\sc Sherpa} event-generation framework.

In order to carry out a useful comparison, we tried to keep the
generation parameters as common as possible among the different
calculations. For example, in our main set of comparisons we use the
same PDF, CTEQ6M with $\alpha_s(m_Z)=0.118$, for both the NLO and
LO predictions in order to separate any differences induced by PDFs
from those resulting from the matrix elements. In all other cases the
LO computations employ the CTEQ6L and CTEQ6L1 PDF sets with
$\alpha_s(m_Z)=0.118$ and $\alpha_s(m_Z)=0.13$, respectively
\cite{Pumplin:2002vw,Nadolsky:2008zw}.

\subsection{Results of the comparison}\label{sec:results}

Before we discuss differential distributions, we list in
Table~\ref{tab:xsecs} the inclusive $e^+\nu_e$ + 3 jet cross sections
for LHC energies of 10 TeV that we have obtained from all calculations
outlined above. This gives us the possibility of rescaling the
different results to the {\sc Rocket} (aLC) NLO cross section, such
that we can comment on shape differences in the differential
distributions separately.
The LO cross sections generated with {\sc Comix} and given in
Table~\ref{tab:xsecs} vary by more than a factor of 2 for different
scale and PDF choices.
At NLO this reduces to a 20\% effect. This is still quite significant,
but can be understood as a consequence of the different scale choices
used in the NLO calculations: the $E_{T,W}$ choice of {\sc Rocket} is
found to yield average scales
$\langle\mu_\mathrm{F,R}\rangle\approx120$~GeV, whereas the $\hat H_T$
choice used by {\sc BlackHat+Sherpa} generates considerably larger
average values $\langle\mu_\mathrm{F,R}\rangle\approx390$~GeV. This
is more than a factor 3 higher. A change of 20\% over such a large
$\mu$\/ range seems reasonable, taking into account that the NLO cross
section shown in Figure~\ref{fig:NLO} already drops by about 10\%
between 120~GeV and 240~GeV.
The variation among {\sc Sherpa}'s ME\&TS cross sections (about 75\%
at most) turns out to be smaller compared to what we find at LO. One
should bear in mind that the two estimates are determined differently,
for ME\&TS, through varying the scales by constant factors and, for
the LO case, by choosing different dynamic though global scales. The
ME\&TS cross sections decrease by an overall factor of 35\% when
including matrix elements with larger numbers of partons. The
correction becomes weaker when adding in the 5-parton contributions
(15\% compared to 23\% in the first step) indicating, as expected, a
stabilization of the $W^+$ + $\ge3$ jet cross sections of the ME\&TS
approach. The scales chosen by the ME\&TS procedure reflect the local
$p_T$ at each vertex of the hard interaction and will almost always be
smaller than $H_T$. Nominally this results in a larger LO cross
section and thus a smaller $K$-factor, but the Sudakov rejection
applied with ME\&TS reduces the resultant cross section to something
smaller than that obtained at NLO (similar to that found at LO, cf.\
Table~\ref{tab:xsecs}, $18.6$~pb vs.\ $17.3$~pb).

For the LO and ME\&TS calculations, we also give results obtained from
Run II $k_T$ jet finding using $D=0.5$ \cite{Blazey:2000qt}. They are,
in all cases, larger than their respective SISCone counterparts.
Interestingly, the parton-shower corrections included by the ME\&TS
merging make the results from the two jet algorithms look very much
alike, including the shapes of the differential distributions
presented below.\footnote{It turns out that already at the ME\&TS
  parton level, before showering, the differences between the two jet
  algorithms start to wash out.}

\begin{table}[t!]\begin{center}
\begin{tabular}{|c||l|l|l|}
\hline\rule[-2mm]{0mm}{6mm} Order &
\multicolumn{3}{|c|}{and\qquad Specifics of calculation}\\\hline
\rule[-2mm]{0mm}{6mm} LO & \textbf{\textsc{Comix}} & \textbf{\textsc{Comix}}
& \textbf{\textsc{Comix}}\\[1mm]
&$\mu=E_{T,W}$, CTEQ6L1 & $\mu=E_{T,W}$, CTEQ6L & $\mu=H_T$, CTEQ6L\\[3mm]
\rule[-2mm]{0mm}{6mm}&\ \ 37.1 pb &\ \ 28.7 pb &\ \ 17.3 pb\\
\rule[-2mm]{0mm}{6mm}&\ \ 43.8 pb\ \ ($k_T$ jets) &\ \ 33.8 pb\ \
($k_T$ jets) &\ \ 20.6 pb\ \ ($k_T$ jets)\\[3mm]\hline\hline
\rule[-2mm]{0mm}{6mm} NLO & \textbf{\textsc{Rocket}} (aLC)
&& \textbf{\textsc{BlackHat+Sherpa}}\\[1mm]
&$\mu=E_{T,W}$, CTEQ6M & & $\mu=\hat H_T$, CTEQ6M\\[3mm]
\rule[-2mm]{0mm}{6mm}&\ \ 34.2 pb & &\ \ 28.6 pb\\[3mm]\hline\hline
\rule[-2mm]{0mm}{6mm} ME\&TS & \textbf{\textsc{Sherpa}}
& \textbf{\textsc{Sherpa}} & \textbf{\textsc{Sherpa}}\\[1mm]
&$N^\mathrm{max}_\mathrm{ME}=2+3$, CTEQ6L
&$N^\mathrm{max}_\mathrm{ME}=2+4$, CTEQ6L
&$N^\mathrm{max}_\mathrm{ME}=2+5$, CTEQ6L\\[3mm]
\rule[-2mm]{0mm}{6mm}&&\ \ 20.1 pb\ \ (CTEQ6M) &\\
\rule[-2mm]{0mm}{6mm}&&\ \ 14.3 pb\ \ ($\mu=\mu^\mathrm{ME\&TS}\cdot2$) &\\
\rule[-2mm]{0mm}{6mm}&\ \ 24.3 pb &\ \ 18.6 pb &\ \ 15.8 pb\\
\rule[-2mm]{0mm}{6mm}&&\ \ 24.7 pb\ \ ($\mu=\mu^\mathrm{ME\&TS}/2$) &\\
\rule[-2mm]{0mm}{6mm}&\ \ 24.4 pb\ \ ($k_T$ jets) &\ \ 18.8 pb\ \
($k_T$ jets) &\ \ 16.0 pb\ \ ($k_T$ jets)\\[3mm]\hline\hline
\end{tabular}
\caption{\label{tab:xsecs}
Inclusive $e^+\nu_e$ + 3 jet cross sections as obtained from the
different calculations used in this study. For cuts, parameter
settings and details of the calculations, cf.\ Section~\ref{sec:calcs}
If not stated otherwise, the SISCone jet algorithm \cite{Salam:2007xv}
has been used to identify the jets. The $k_T$ Run II jet finder
\cite{Blazey:2000qt} has been applied for evaluating the cross
sections labelled ``$k_T$ jets''. The ``aLC'' label expresses the fact
that we have used {\sc Rocket}'s adjusted leading-colour result
\cite{Melnikov:2009wh}.}
\end{center}
\end{table}

\begin{figure}[p!]
  \centerline{
    \includegraphics[clip,width=0.48\columnwidth]{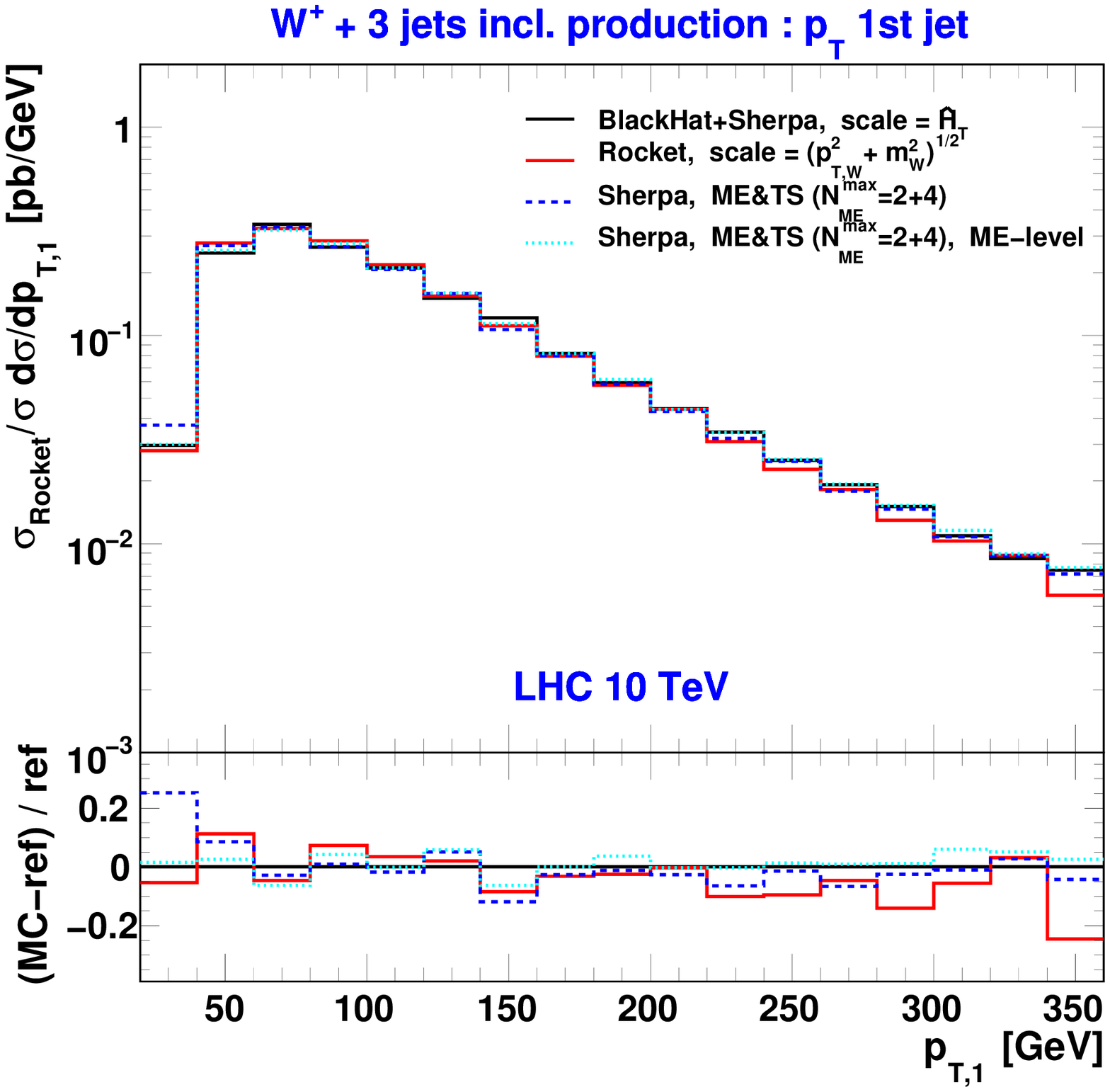}
    \includegraphics[clip,width=0.48\columnwidth]{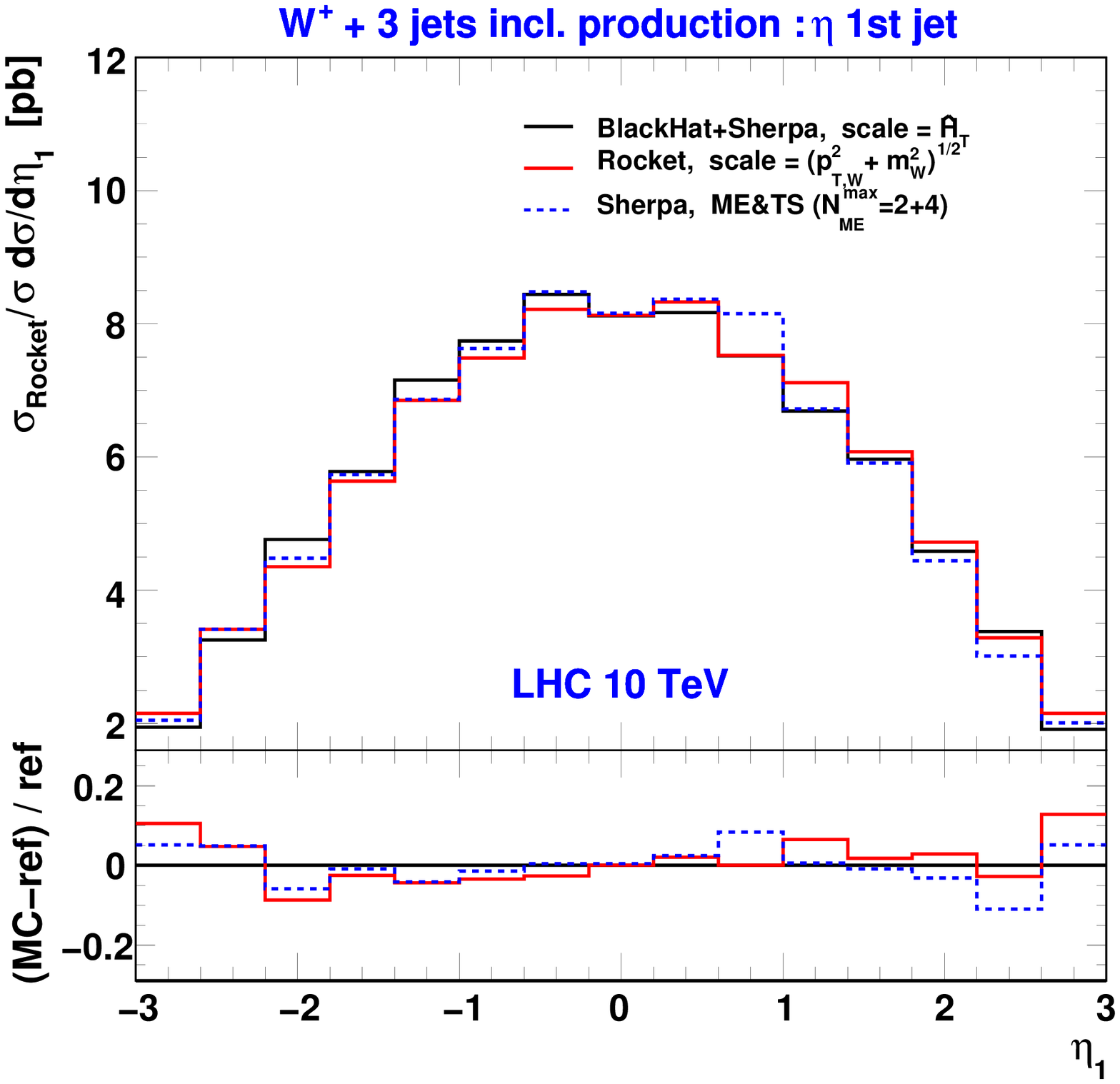}}
  \vspace{-3mm}
  \centerline{
    \includegraphics[clip,width=0.48\columnwidth]{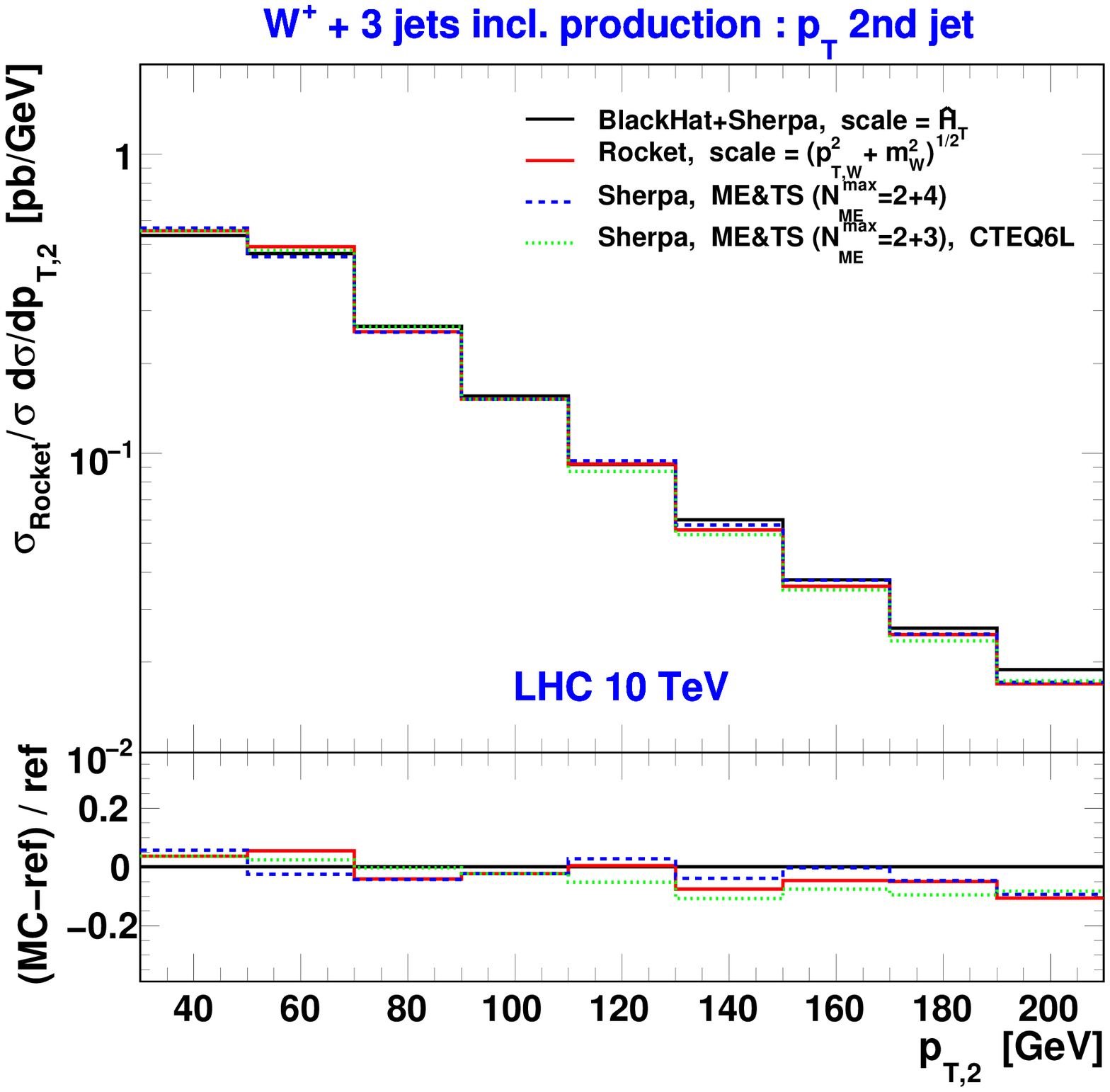}
    \includegraphics[clip,width=0.48\columnwidth]{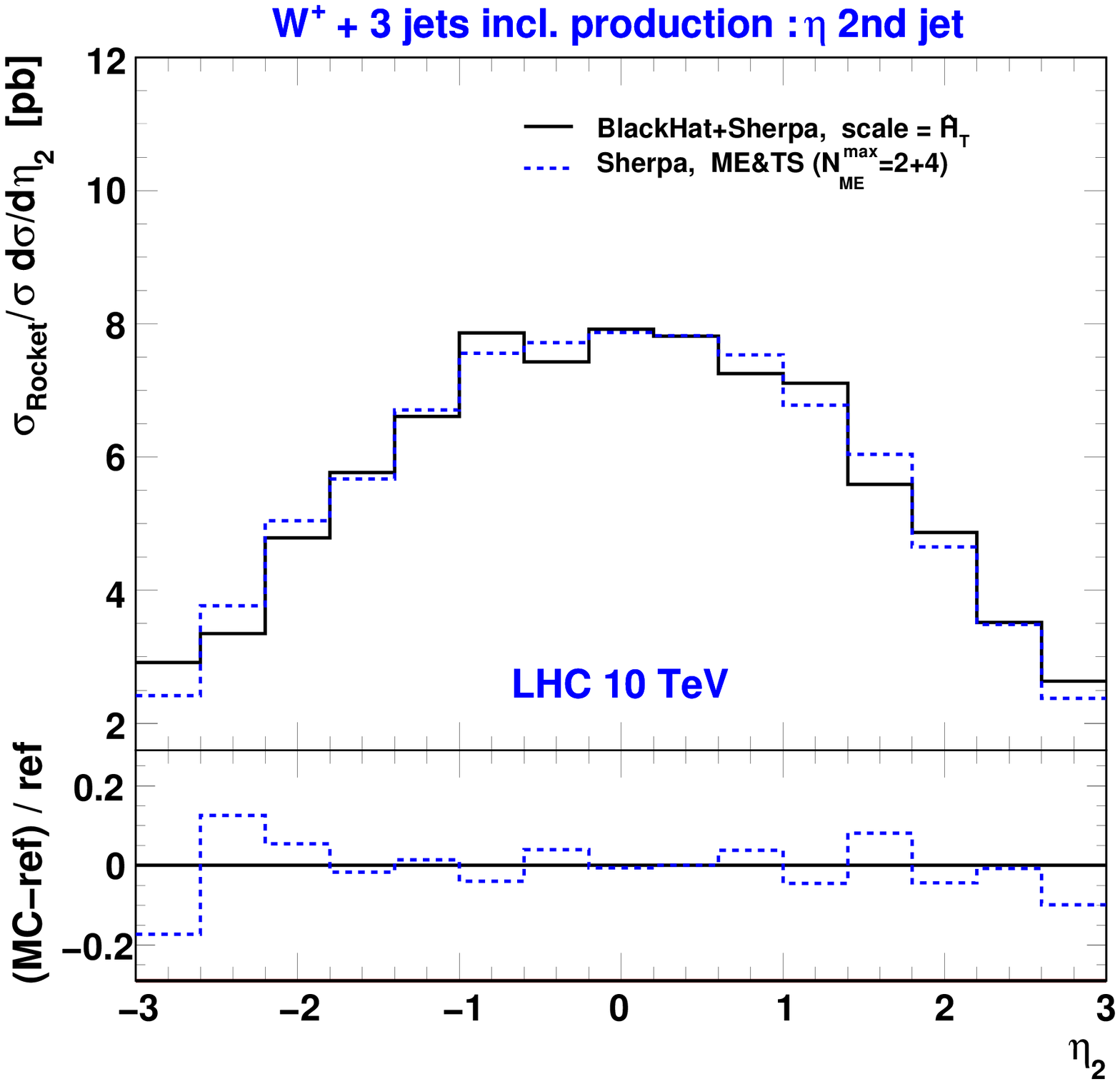}}
  \vspace{-3mm}
  \centerline{
    \includegraphics[clip,width=0.48\columnwidth]{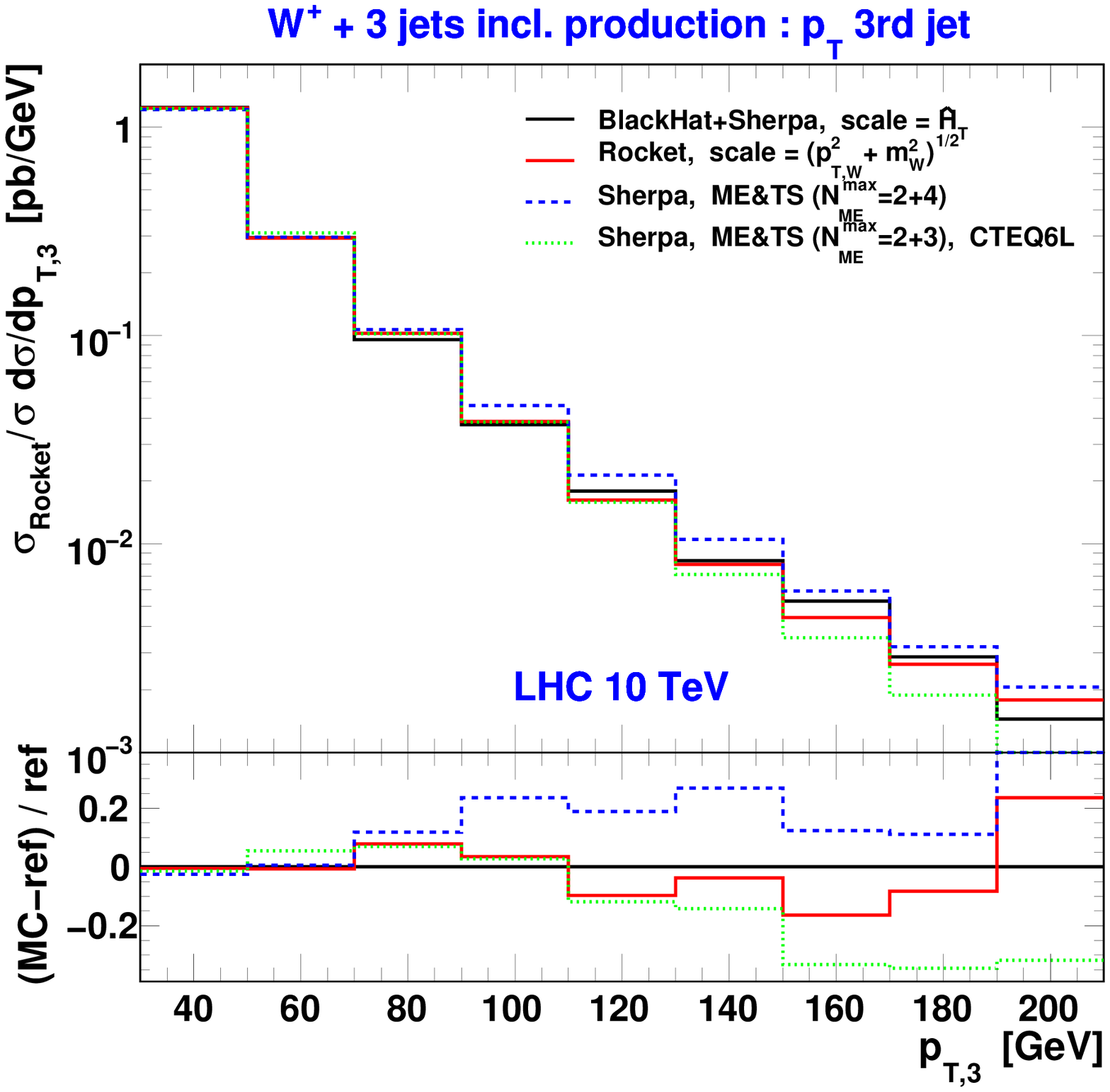}
    \includegraphics[clip,width=0.48\columnwidth]{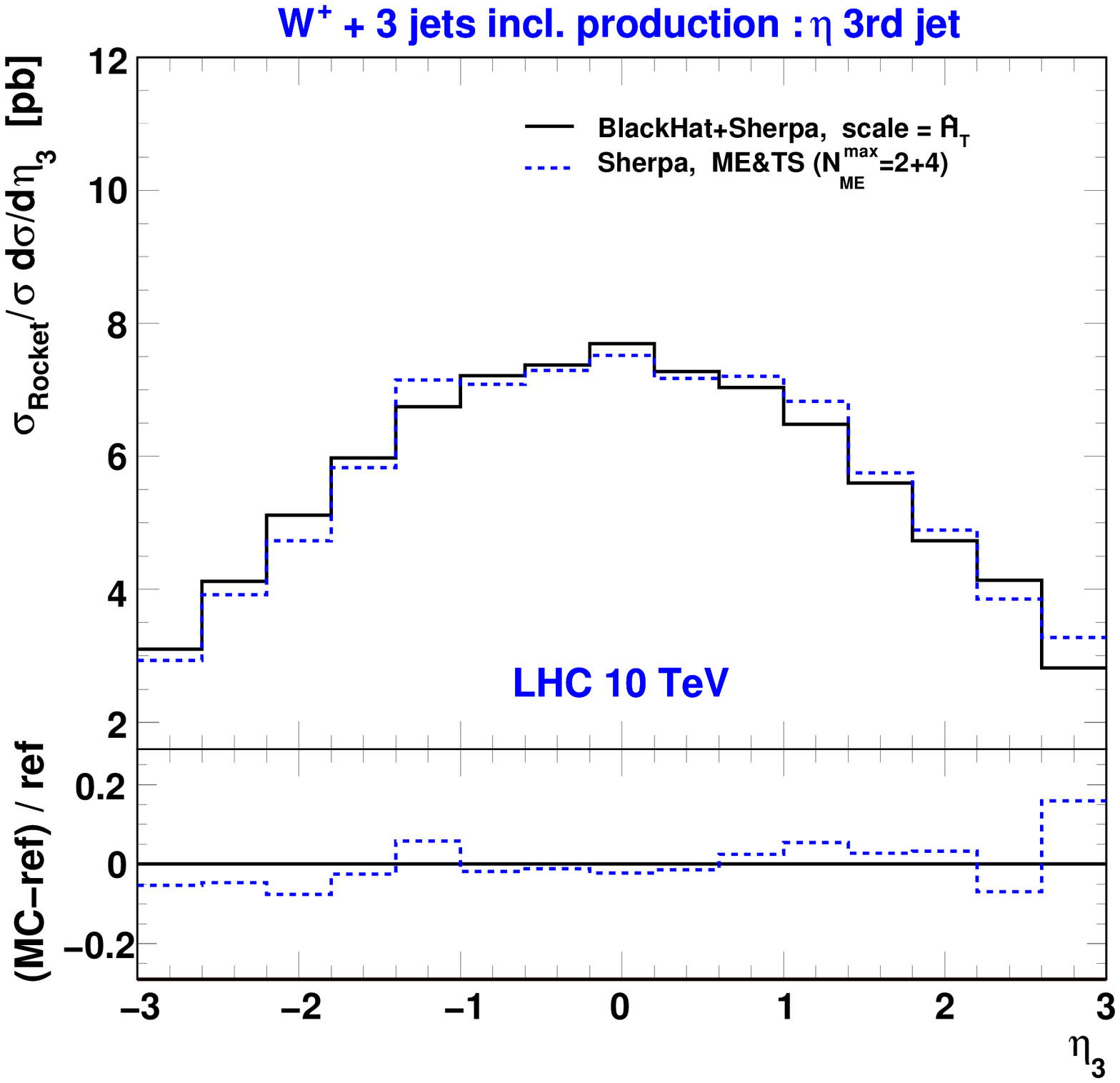}}
  \caption{The transverse momentum distributions (left) and
    pseudo-rapidity distributions (right) of the three hardest jets in
    $W^+$ + $\ge3$ jet production at the LHC. Predictions at NLO
    obtained from the {\sc BlackHat+Sherpa} (black line) and {\sc Rocket}
    (red line) codes are compared to LO results from {\sc Sherpa}
    using the ME\&TS merging. All curves have been rescaled to the
    {\sc Rocket} NLO cross section of Table~\ref{tab:xsecs};
    {\sc BlackHat+Sherpa} is used as the reference; cuts and
    parameters are detailed in Section~\ref{sec:calcs}}
  \label{fig:pt.eta.jets}
\end{figure}

\begin{figure}[t!]
  \centerline{
    \includegraphics[clip,width=0.48\columnwidth]{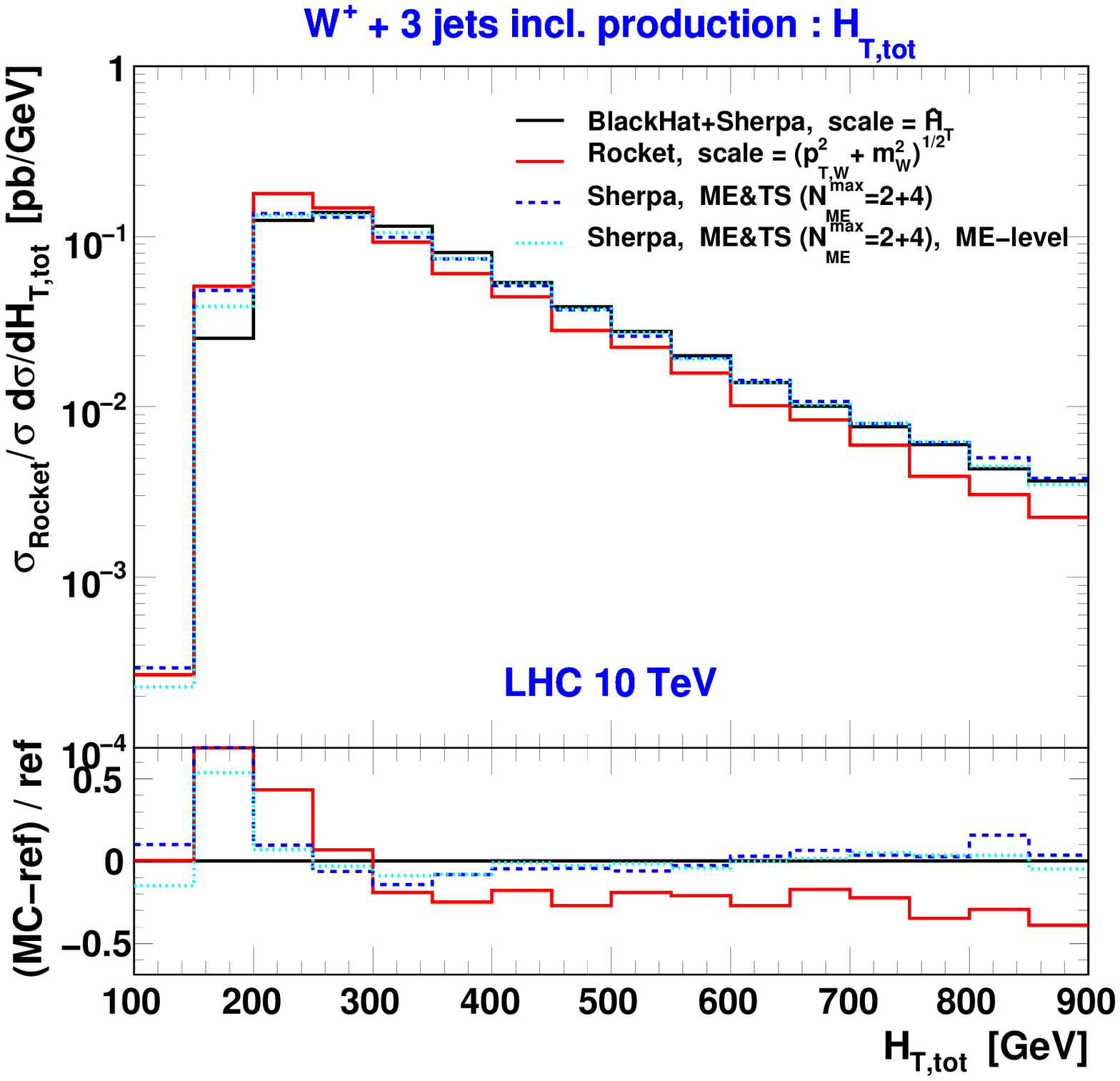}
    \includegraphics[clip,width=0.48\columnwidth]{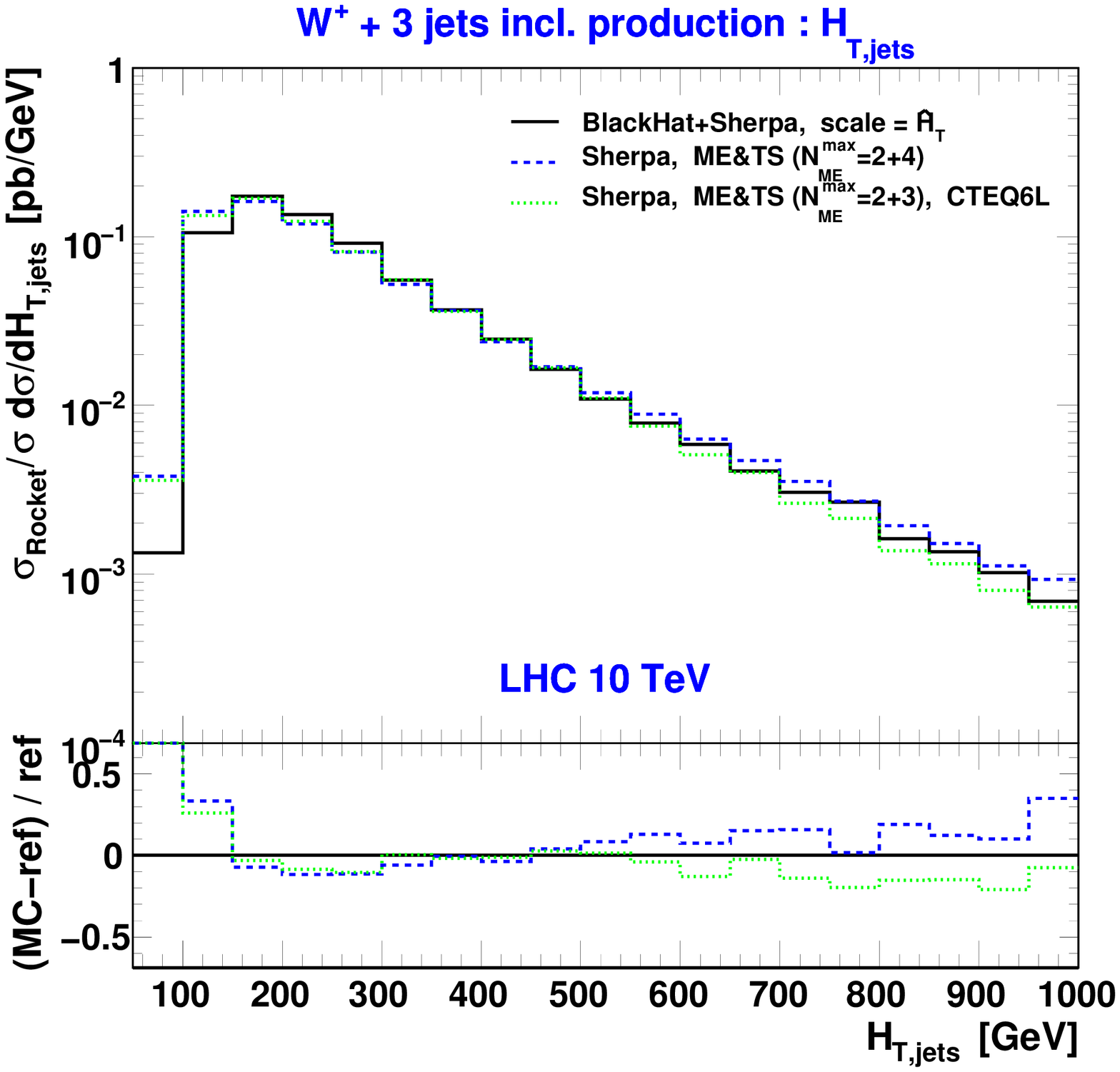}}
  \caption{$H_T$ and $H_{T,\mathrm{jets}}$ distributions in inclusive
    $W^+$ + 3 jet production at the LHC. NLO predictions obtained
    from {\sc BlackHat+Sherpa} (black line) and {\sc Rocket} (red
    line) are compared to LO results from {\sc Sherpa} using the
    ME\&TS merging. All curves have been rescaled to the {\sc Rocket}
    NLO cross section of Table~\ref{tab:xsecs}; the {\sc BlackHat+Sherpa}
    prediction is used as the reference; cuts and parameters are
    detailed in Section~\ref{sec:calcs}}
  \label{fig:ht}
\end{figure}

\begin{figure}[t!]
  \centerline{
    \includegraphics[clip,width=0.48\columnwidth]{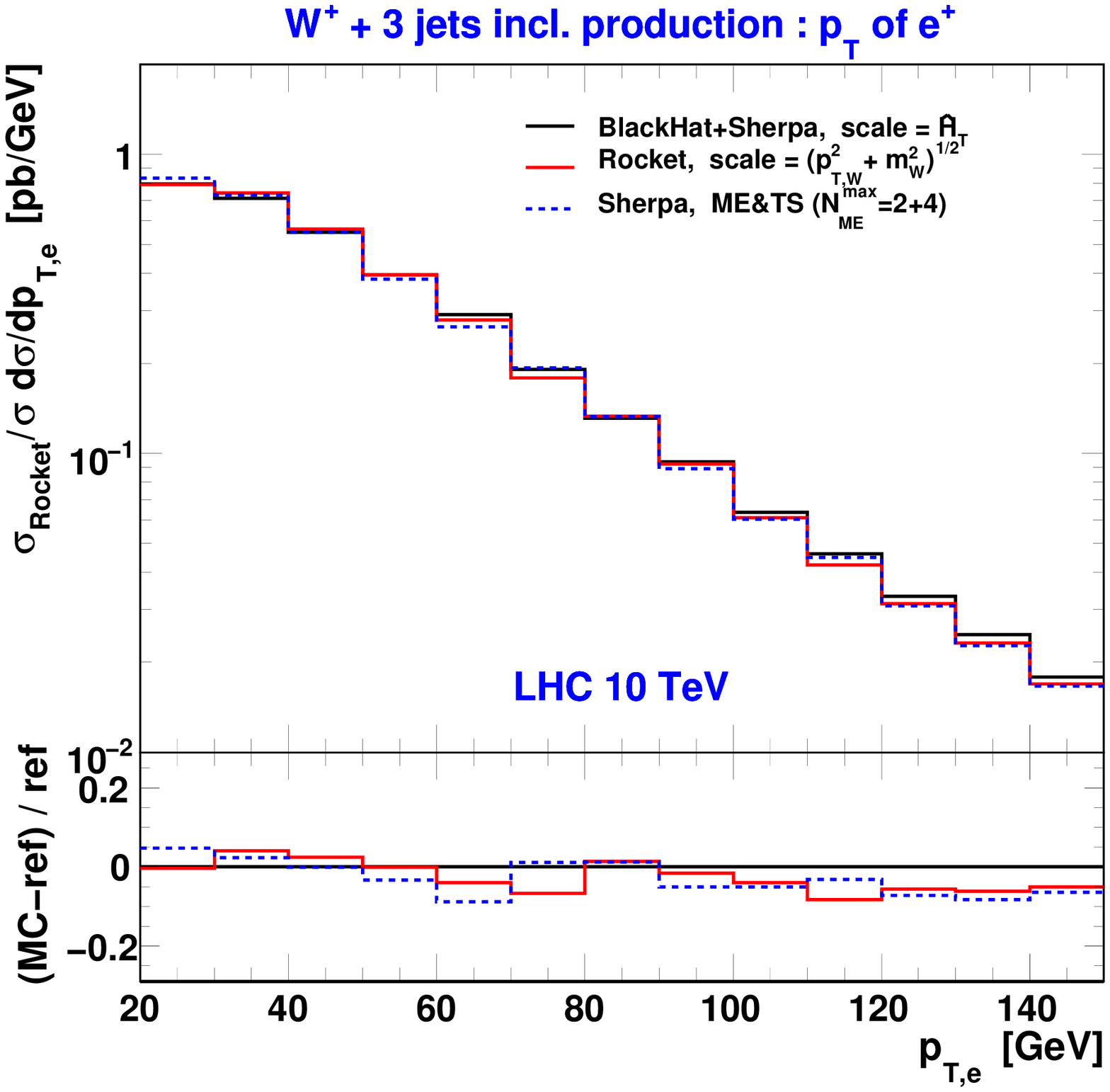}
    \includegraphics[clip,width=0.48\columnwidth]{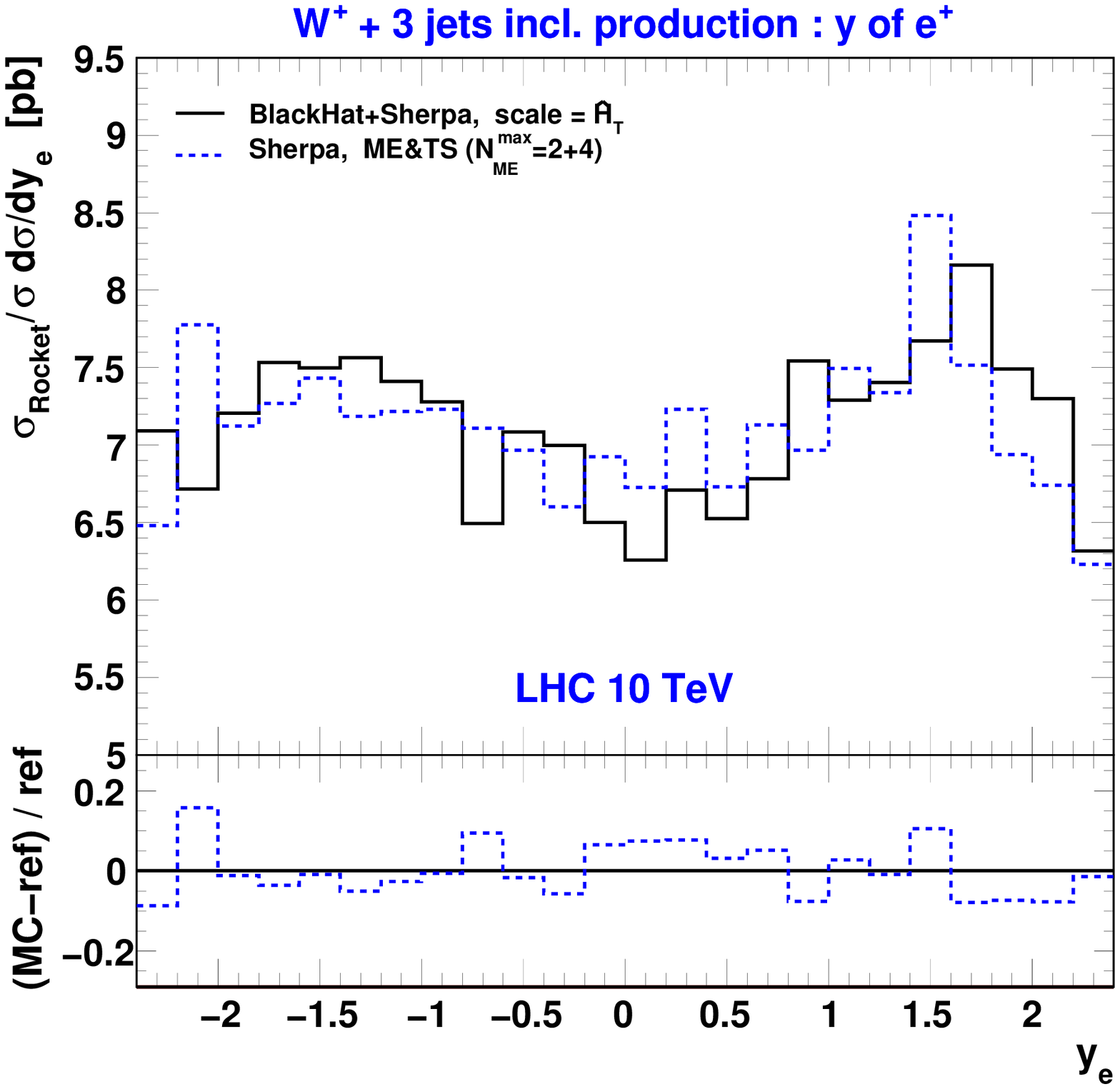}}
  \vspace{-3mm}
  \centerline{
    \includegraphics[clip,width=0.48\columnwidth]{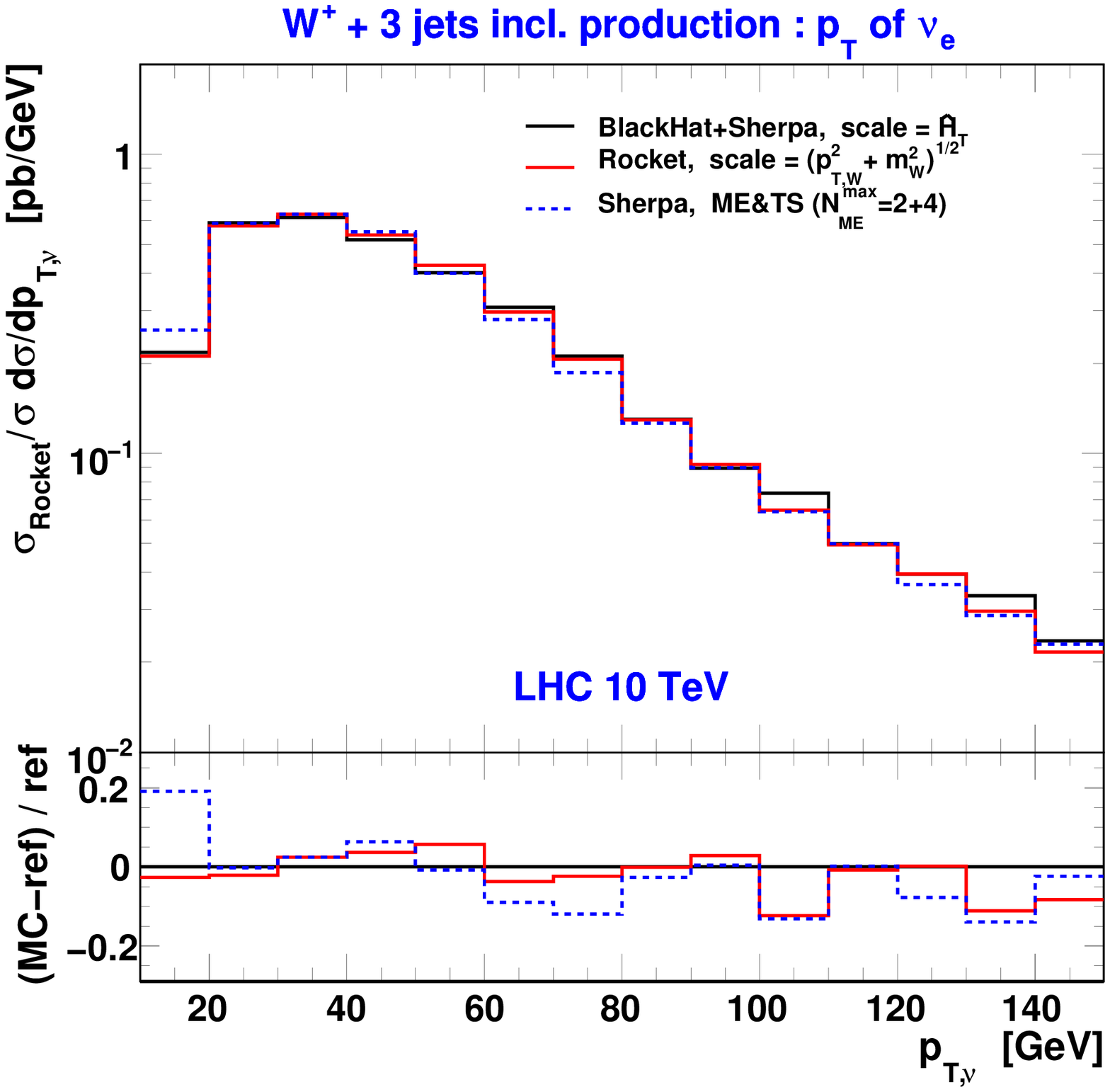}
    \includegraphics[clip,width=0.48\columnwidth]{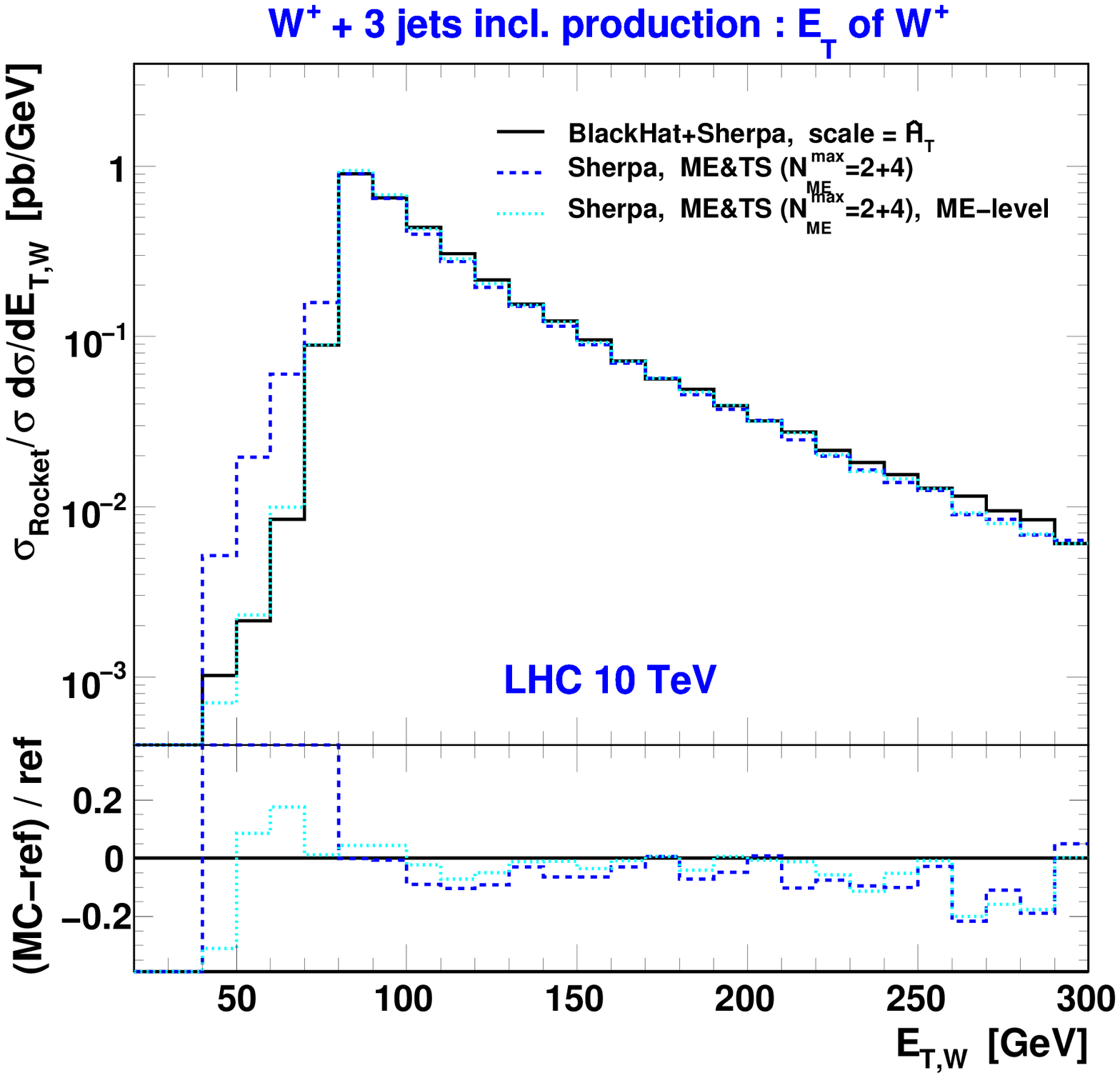}}
  \caption{The $p_T$ and rapidity spectra of the lepton (upper left
    and right), the neutrino's $p_T$ spectrum and the
    reconstructed $W^+$\/ transverse-energy distribution (lower left
    and right) in inclusive $e^+\nu_e$ + 3 jet production at the LHC.
    NLO predictions as given by {\sc BlackHat+Sherpa} (black line) and
    {\sc Rocket} (red line) are compared to LO results from
    {\sc Sherpa} using the ME\&TS merging. All curves have been
    rescaled to the {\sc Rocket} NLO cross section, cf.\
    Table~\ref{tab:xsecs}; the {\sc BlackHat+Sherpa} prediction is
    used as the reference; for cuts and parameters, see
    Section~\ref{sec:calcs}}
  \label{fig:pte.ye.ptnu.mtw}
\end{figure}

\begin{figure}[t!]
  \centerline{
    \includegraphics[clip,width=0.48\columnwidth]{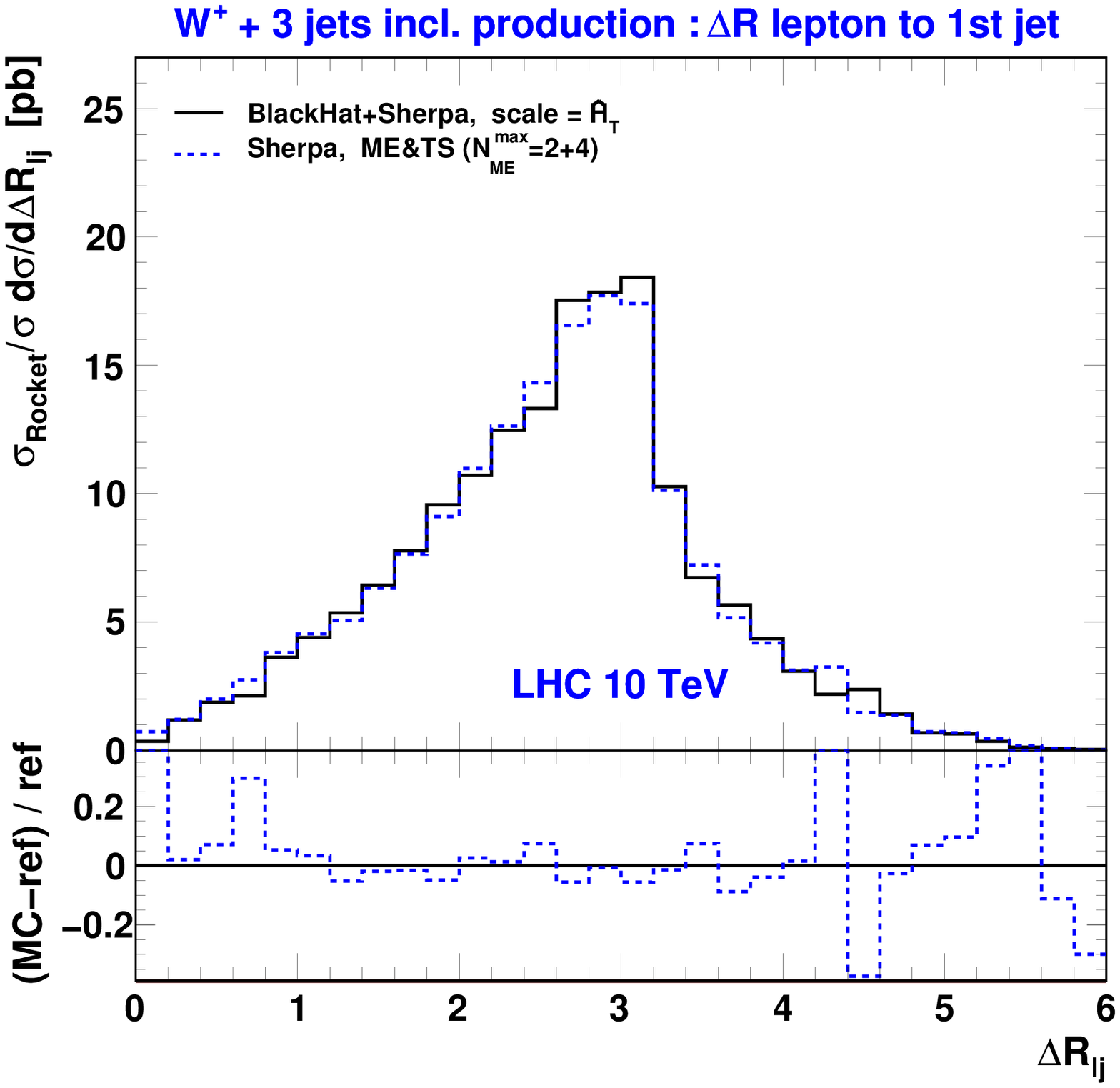}
    \includegraphics[clip,width=0.48\columnwidth]{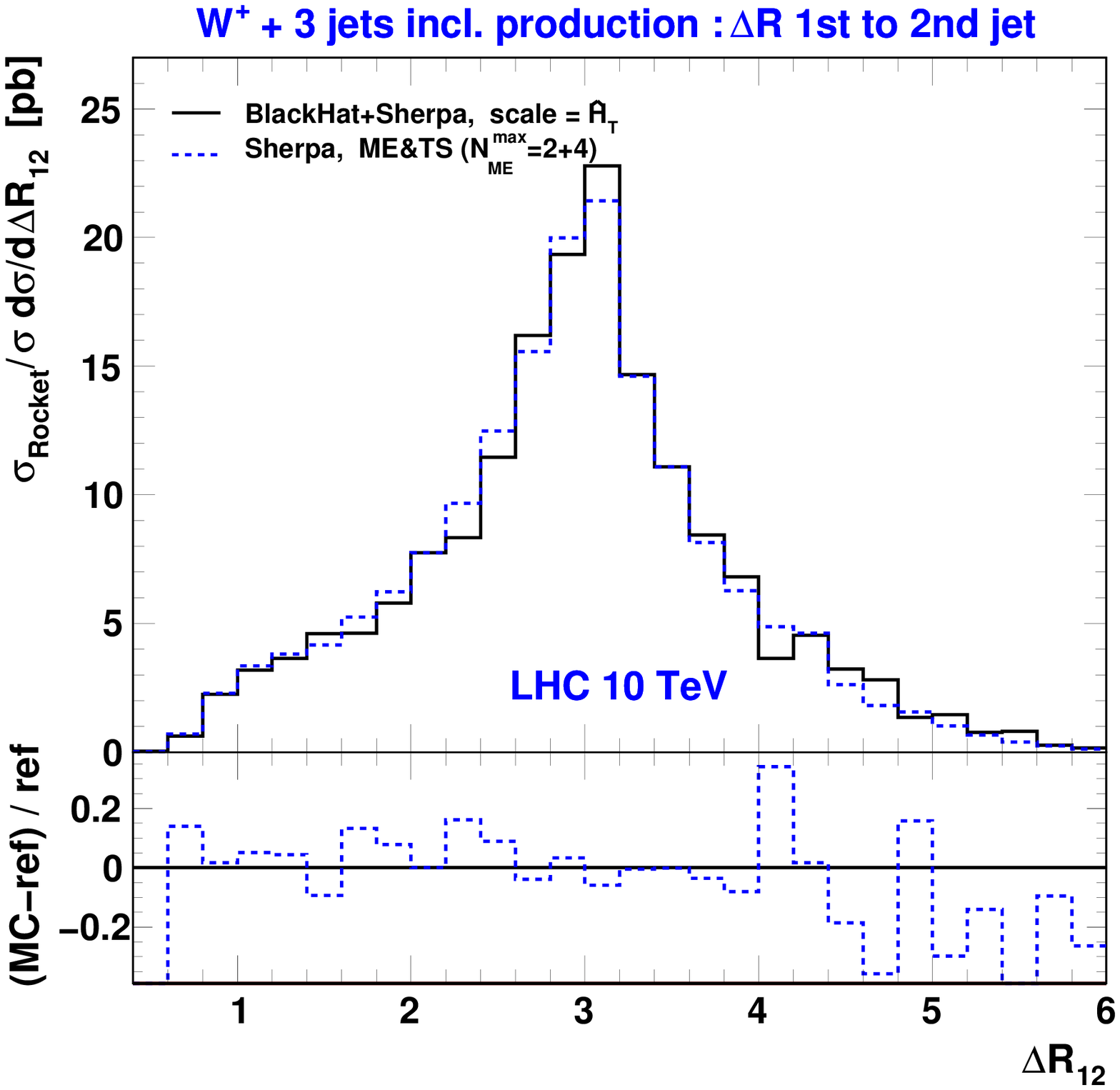}}
  \vspace{-3mm}
  \centerline{
    \includegraphics[clip,width=0.48\columnwidth]{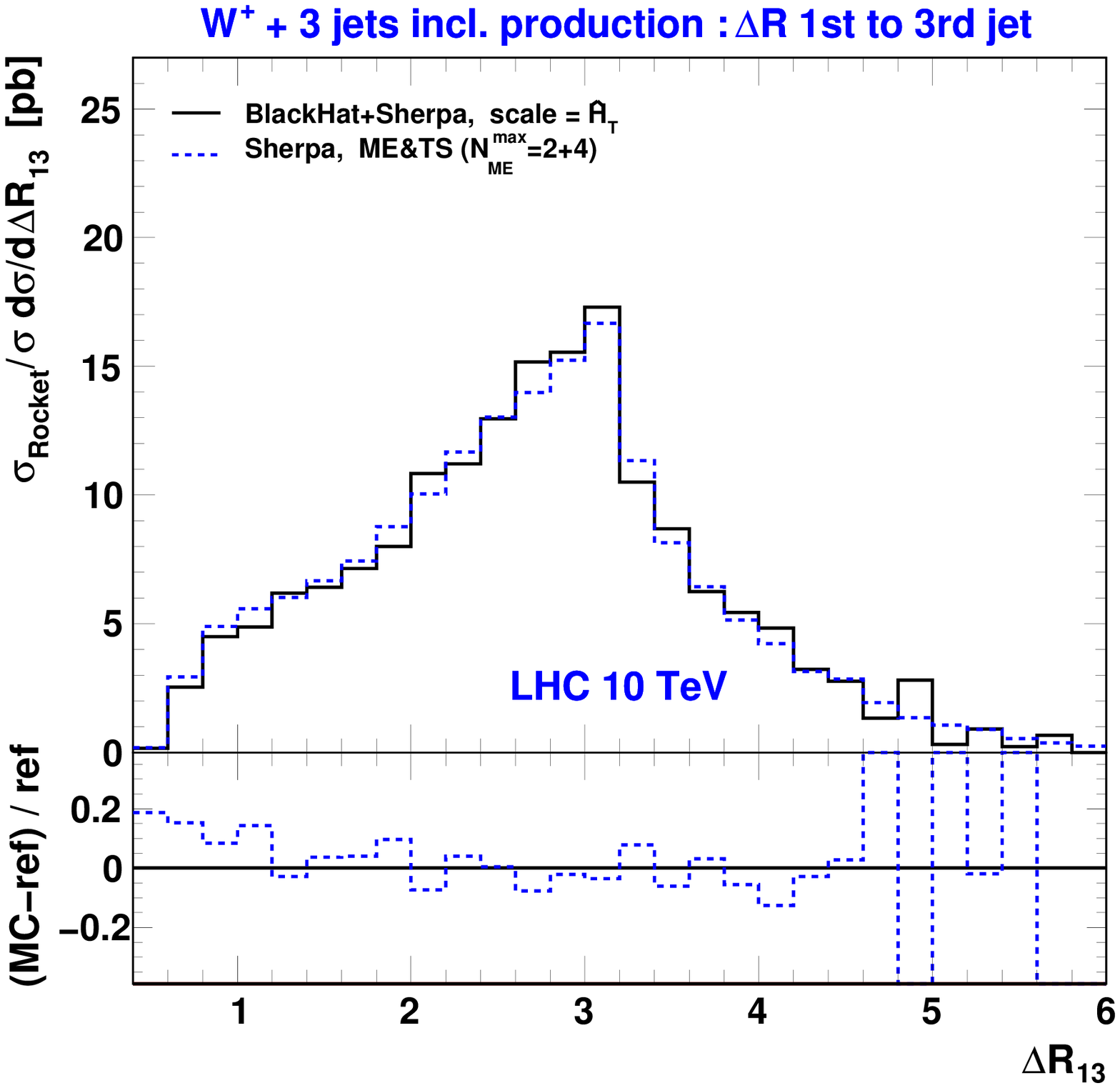}
    \includegraphics[clip,width=0.48\columnwidth]{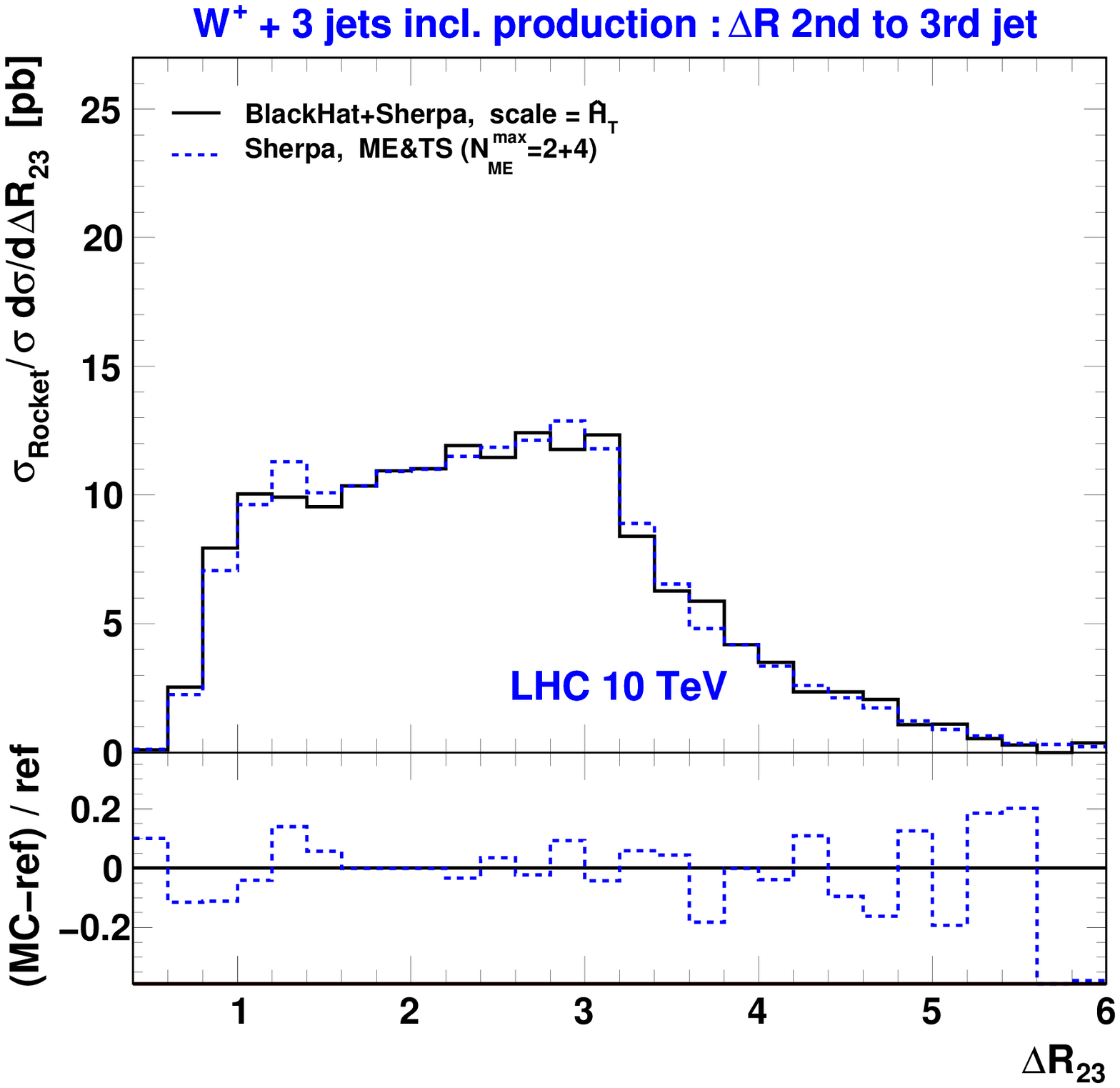}}
  \caption{Pairwise geometric separations between the lepton and
    hardest jet (upper left panel) and the three hardest jets
    in $W^+$ + $\ge3$ jet production at the LHC. {\sc BlackHat+Sherpa}
    results at NLO (black lines) are compared to those of {\sc Sherpa}'s
    ME\&TS merging. The normalization is still given by {\sc Rocket}'s
    NLO cross section of Table~\ref{tab:xsecs}; the {\sc BlackHat+Sherpa}
    prediction is used as the reference; for cuts and parameters, see
    Section~\ref{sec:calcs}}
  \label{fig:dr}
\end{figure}

The main set of differential distributions of our comparative study is
presented in Figures~\ref{fig:pt.eta.jets}--\ref{fig:dr}. For the
comparisons with {\sc Rocket}, we restrict ourselves to the
distributions available in \cite{Melnikov:2009wh}. We also include new
distributions only comparing {\sc BlackHat+Sherpa} with the ME\&TS
approach. For the latter, we always show, as the default, the
predictions obtained from the $N^\mathrm{max}_\mathrm{ME}=2+4$ merged
sample using the CTEQ6M PDF. Although the cross sections differ by
about 10\%, see Table~\ref{tab:xsecs}, we did not discover any
significant shape alterations induced by switching to the CTEQ6L PDF
set.

In Figure~\ref{fig:pt.eta.jets}, we show the transverse momentum
(left panels) and pseudo-rapidity (right panels) distributions for the
three leading jets in $W^+$ + $\ge3$ jet production. The two NLO
predictions agree well with each other for all three $p_T$ and the
leading-jet $\eta$\/ distributions, in spite of the different scales
used for each calculation; this is another manifestation of the
reduced scale dependence at NLO. {\sc Sherpa}'s ME\&TS $\eta$\/ curves
are in good agreement with the NLO prediction(s). This level of
agreement is also found for the first two hardest jets, for the third
jet it depends more on the details of the {\sc Sherpa} ME\&TS
generation. To this end we have added the predictions (dotted green
lines) from the $N^\mathrm{max}_\mathrm{ME}=2+3$ merged sample in the
second- and third-jet $p_T$ plots. As can be seen when omitting the
contributions of the real-emission matrix elements with four extra
partons, the corresponding transverse-momentum distributions fall
below that of the NLO calculations, most noticeably for the third-jet
$p_T$ spectrum. Once the matching is extended to
$N^\mathrm{max}_\mathrm{ME}=2+4$, the {\sc Sherpa} prediction for the
second jet improves with respect to the NLO results, while the one for
the third jet lies above those given at NLO.

To get an idea of the impact of parton showering, we can analyze the
matrix-element final states of the {\sc Sherpa} ME\&TS events (before
they undergo showering) and plot distributions at the hard-process
level, i.e.\ parton level, which we have labelled by ``ME-level'' in
the plots. In the top left panel of Figure~\ref{fig:pt.eta.jets},
we added the ME-level prediction (dotted turquoise line) to the first-jet
$p_T$ spectra. It is in remarkable agreement with the NLO prediction
of {\sc BlackHat+Sherpa} over the entire range of the spectrum. As
expected, the soft and collinear emissions added to the hard processes
slightly soften the distribution such that the lower $p_T$ bins lie
somewhat above the ME-level curve. There is almost no effect for bins
of large $p_T$ as expected from IR-safe observables describing hard
jets. Similar differences are found between {\sc Sherpa}'s parton-shower-
and ME-level curves for the $H_T$ observable presented in the left
panel of Figure~\ref{fig:ht} where we use the definition
$H_T=\sum_\mathrm{jets}p_{T,j}+p_{T,e}+\slashed{p}_T$.\footnote{Note
  the difference with the scale
  $\hat{H}_T=\sum_\mathrm{partons}E_{T,p}+E_{T,e}+E_{T,\nu}$ chosen
  for the {\sc BlackHat+Sherpa} results.}
With or without shower effects included, all three predictions
disagree for low $H_T$ values. Compared to the single-jet $p_T$s, the
$H_T$ observable takes the leptons as well as multi-jet multi-particle
correlations into account; it therefore contains more detailed
information about the structure of the events. Apparently, these
correlations and the generation of ($\ge$)~4-jet events are described
differently by the three calculations. To gain more insight, one may
investigate how the transverse momenta of the various jets are
correlated. For large $H_T$, the {\sc BlackHat+Sherpa} and ME\&TS
results agree quite well whereas the {\sc Rocket} curve lies lower. A
similar behaviour has been observed in \cite{Berger:2009ep} where
distributions for both scale choices $\hat{H}_T$ and $E_{T,W}$ have
been compared. The right panel of Figure~\ref{fig:ht} displays the
$H_{T,\mathrm{jets}}$ distribution, which does not include the lepton
and missing transverse momentum. As observed in the $p_T$ spectrum of
the third jet, here as well, the prediction from the
$N^\mathrm{max}_\mathrm{ME}=2+3$ ($N^\mathrm{max}_\mathrm{ME}=2+4$)
merged sample lies below (above) the {\sc BlackHat+Sherpa} curve.

Figure~\ref{fig:pte.ye.ptnu.mtw} shows in the top row the positron
transverse-momentum and rapidity distributions. The agreement between
the different curves is rather satisfactory. This is also true for the
missing transverse-momentum distribution shown in the lower left part
of Figure~\ref{fig:pte.ye.ptnu.mtw}. Here we do not anticipate larger
differences between the two NLO scale choices and the ME\&TS approach,
as the plotted variables are directly related to the $W$\/ boson decay.
In the lower right of Figure~\ref{fig:pte.ye.ptnu.mtw}, the transverse
energies of the reconstructed $W$\/ boson are compared.\footnote{Note
  that unlike in the scale choice, we here construct $E_{T,W}$ by using
  the invariant mass of the neutrino and positron pair instead of the
  fixed $m_W$ value. This is the reason for the non-vanishing
  distributions below 80 GeV.} The clear difference at low $E_{T,W}$
between the {\sc BlackHat+Sherpa} and {\sc Sherpa} ME\&TS curves is
explained by the fact that the showers in the latter approach broaden the
reconstructed mass peak of the $W$\/ boson. This is nicely confirmed
by the ME-level result (dotted turquoise line) extracted as before
from the ME\&TS matrix-element final states.

We complete our main comparison by presenting $\Delta R$\/ shapes as
given by the NLO computation of {\sc BlackHat+Sherpa} and the ME\&TS
approach implemented in {\sc Sherpa}. The results for the geometric
separations between the lepton and leading jet as well as between
pairs of the three hardest jets are shown in Figure~\ref{fig:dr}. The
predictions of both calculations are in remarkable
agreement.\footnote{In contrast to using SISCone jets, it turned out
  that the $k_T$ jet algorithm tends to identify more events of low
  $\Delta R$\/ separation.}

\begin{figure}[t!]
  \centerline{
    \includegraphics[clip,width=0.48\columnwidth]{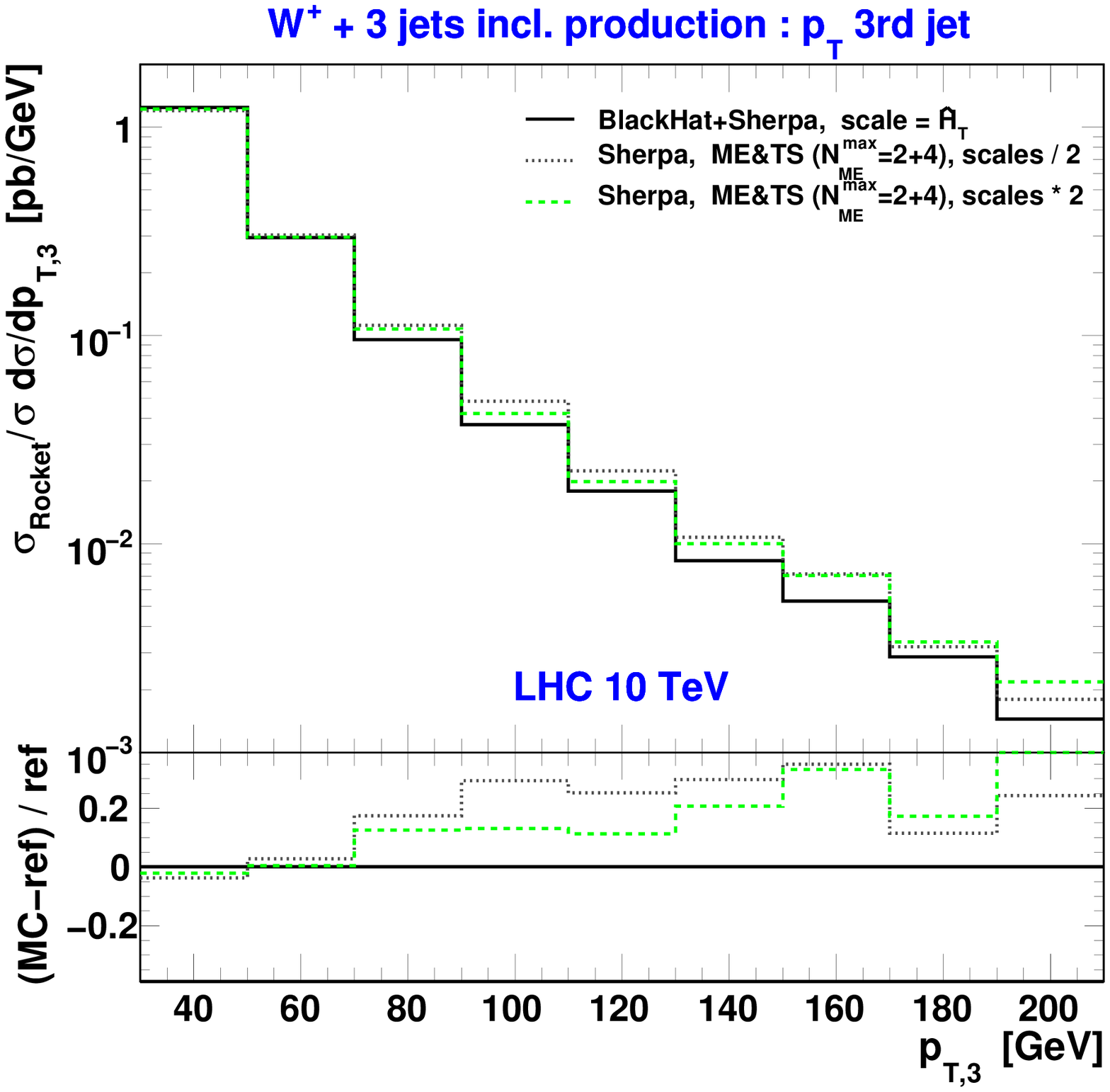}
    \includegraphics[clip,width=0.48\columnwidth]{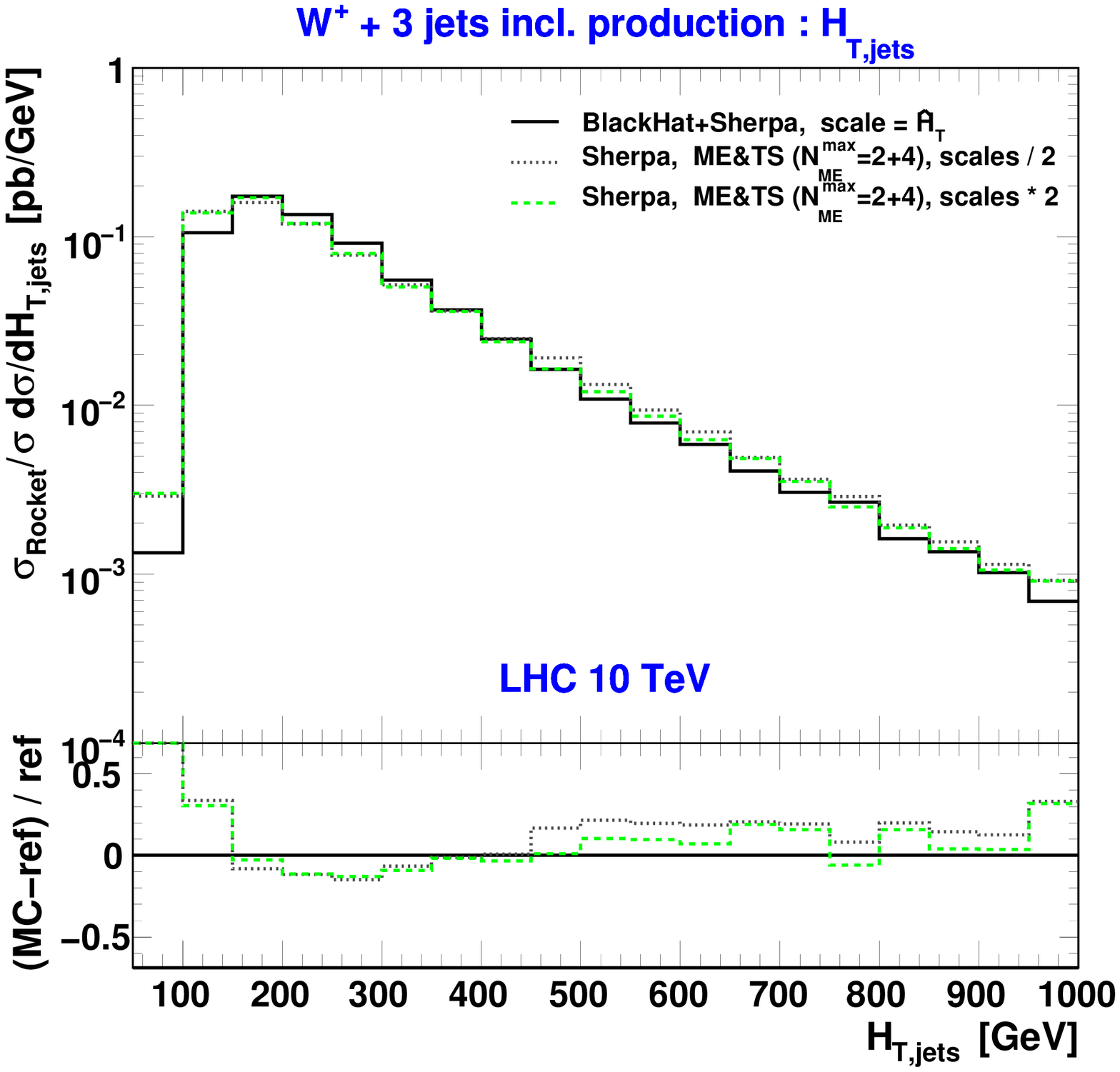}}
  \caption{Impact of the variation of PDF and strong-coupling scales
    on {\sc Sherpa}'s ME\&TS merging for up to 4 jets shown by means
    of the third-jet $p_T$ (left) and $H_{T,\mathrm{jets}}$
    distributions. The dotted dark and dashed green lines display the
    results for smaller and larger scales, respectively, also cf.\
    Section~\ref{sec:calcs} {\sc BlackHat+Sherpa} is used as the
    reference, with the absolute normalization of all curves again
    given by {\sc Rocket}'s NLO cross section even that no
    {\sc Rocket} curve is shown.}
  \label{fig:4v}
\end{figure}

As we have seen in Figure~\ref{fig:pt.eta.jets}, larger deviations
between the NLO and ME\&TS predictions appear in the third-jet
transverse-momentum and $H_{T,\mathrm{jets}}$ distributions. For these
observables, we present in Figure~\ref{fig:4v} scale variations of the
ME\&TS default scheme as described in the {\sc Sherpa} paragraph of
Section~\ref{sec:calcs} The {\sc Sherpa} shapes turn out to be rather
stable varying not more than 20\%. The reference curves given by
{\sc BlackHat+Sherpa} remain outside the uncertainty band. For a more
conclusive statement, one should however also investigate the
robustness of the NLO shapes under standard scale variations.

\begin{figure}
  \centerline{
    \includegraphics[clip,width=0.48\columnwidth]{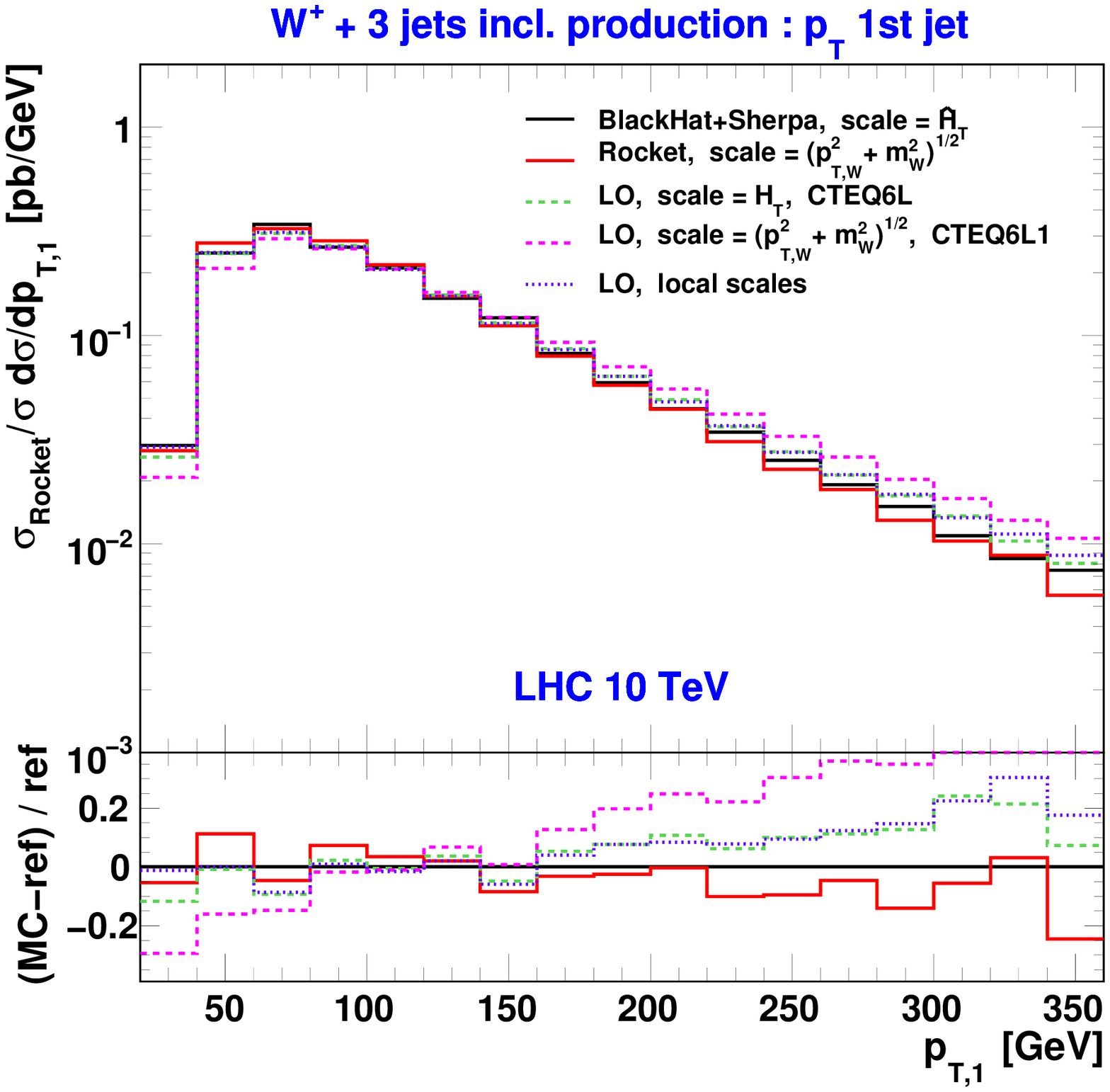}
    \includegraphics[clip,width=0.48\columnwidth]{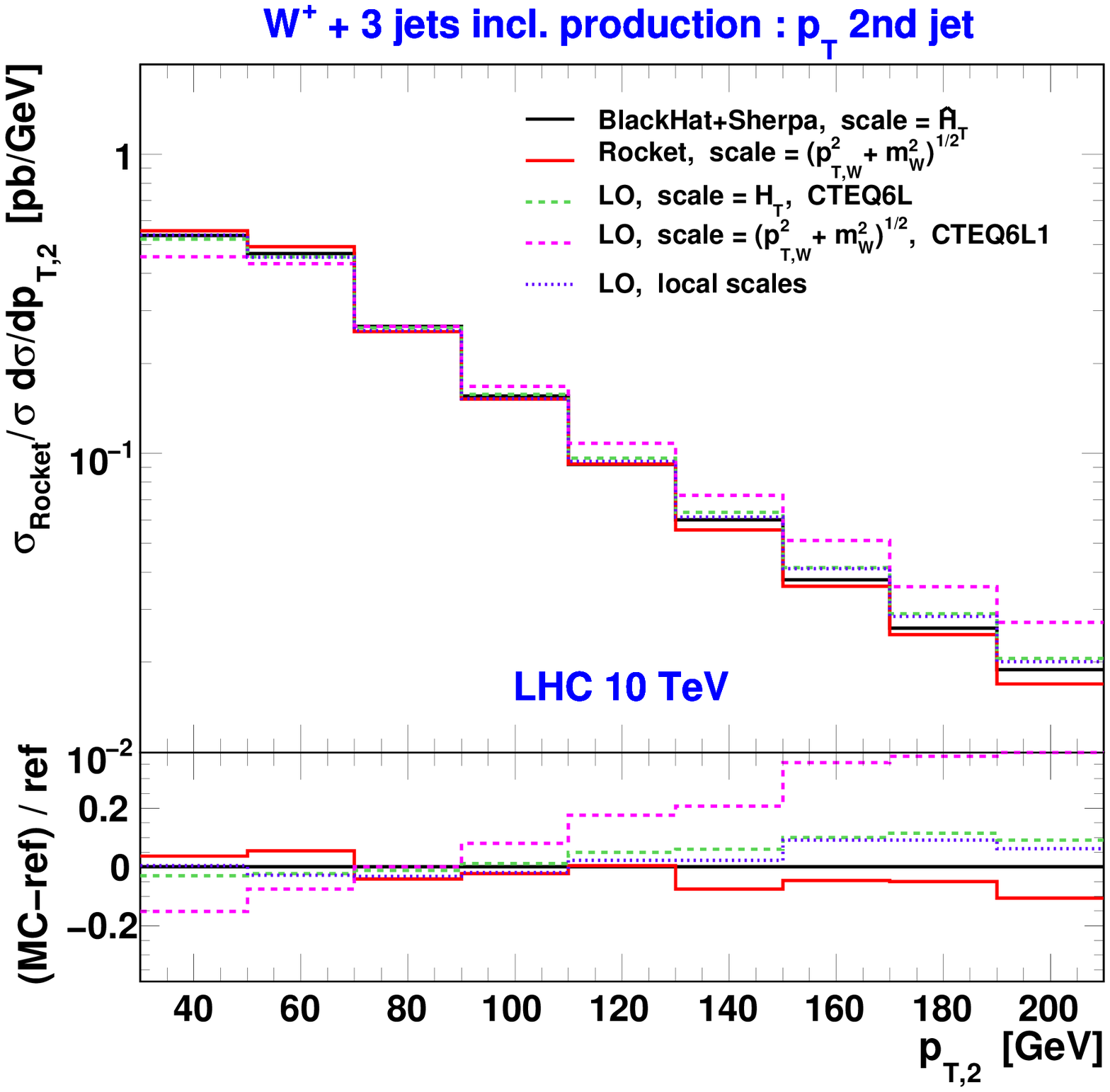}}
  \vspace{-3mm}
  \centerline{
    \includegraphics[clip,width=0.48\columnwidth]{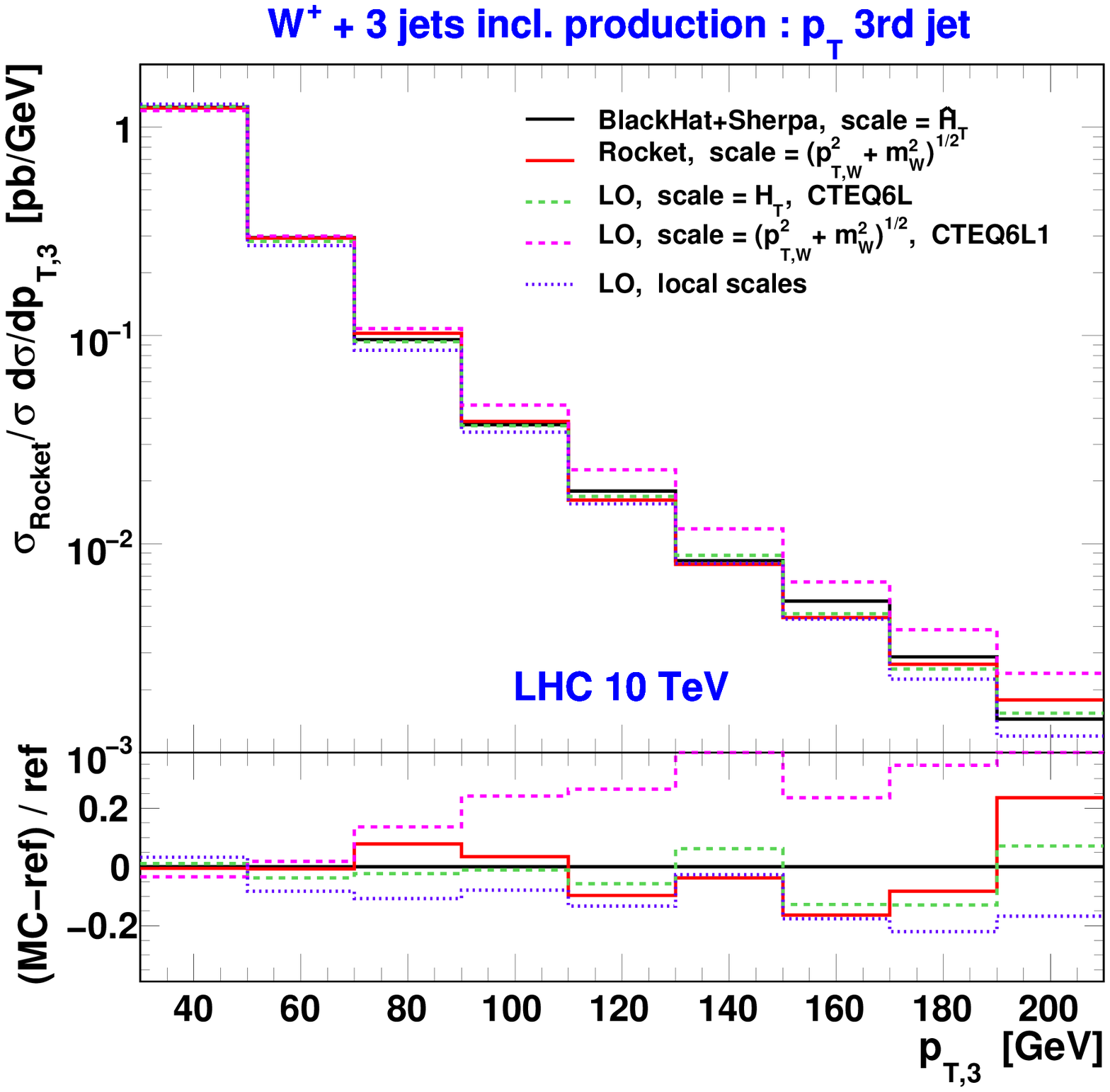}
    \includegraphics[clip,width=0.48\columnwidth]{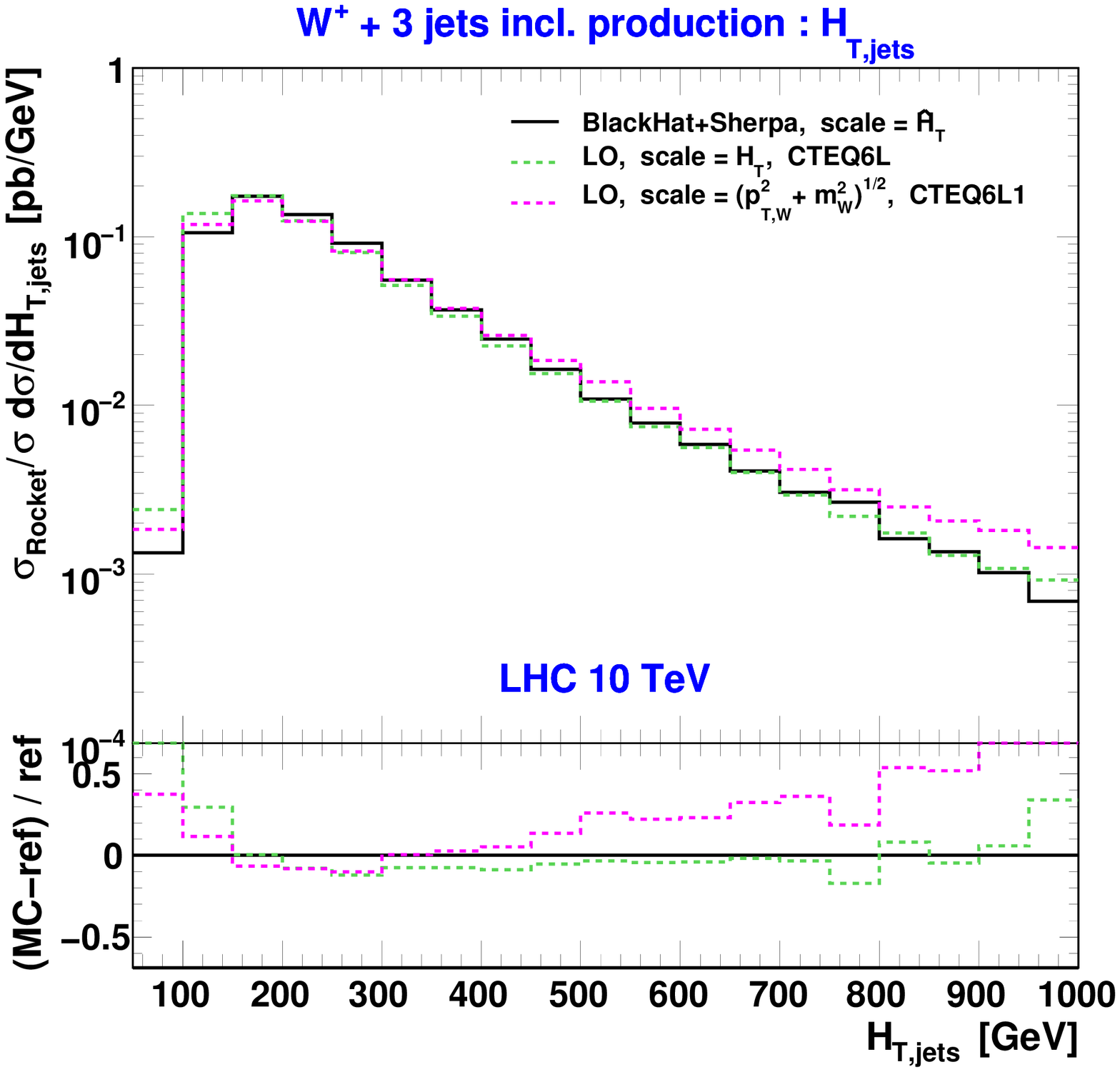}}
  \vspace{-3mm}
  \centerline{
    \includegraphics[clip,width=0.48\columnwidth]{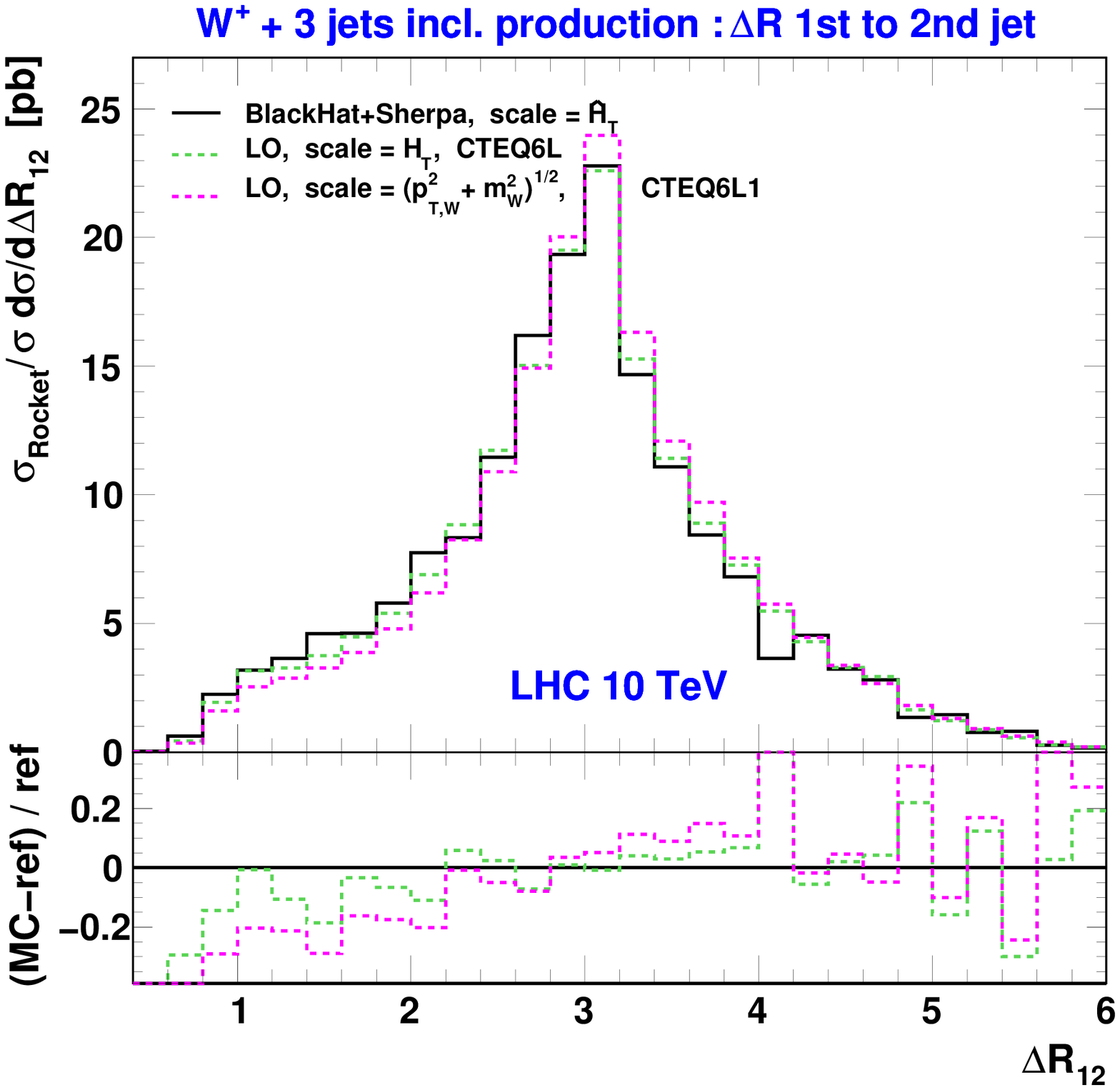}}
  \caption{Comparison between NLO predictions as given by
    {\sc BlackHat+Sherpa} (black lines) and {\sc Rocket} (red lines)
    and LO results generated by {\sc Comix} for the transverse momenta
    of the three hardest jets, $H_{T,\mathrm{jets}}$, the scalar sum
    of all jet $p_T$s, and the geometric separation $\Delta R_{12}$ of
    the two leading jets. The dashed magenta and green lines display
    LO results using $E_{T,W}$ and $H_T$ as a scale, respectively. The
    jet $p_T$ spectra also have coupling-reweighted LO curves taken
    from Ref.~\cite{Melnikov:2009wh}. The absolute normalization of
    all curves is again given by {\sc Rocket}'s NLO cross section; the
    reference is {\sc BlackHat+Sherpa}'s prediction.}
  \label{fig:lo}
\end{figure}

Finally, we turn to compare LO vs.\ NLO results for a subset of
observables of our main comparison. We select those exhibiting the
largest shape differences: the $p_T$ spectra of the hardest three
jets, the $H_{T,\mathrm{jets}}$ distribution and the $\Delta R$\/
separation between the leading and next-to-leading jets. All of which
are shown in Figure~\ref{fig:lo}. The LO predictions using the scale
choice $\mu=E_{T,W}$ lead in all cases to significant differences from
the corresponding NLO predictions; jet pairs being narrow in
$R$-space are predicted too low while the $p_T$ and $H_T$ spectra are
too hard. Conversely, the LO predictions using $\mu=H_T$ as a scale
are observed to produce relatively good agreement with NLO for the
third-jet $p_T$ and $H_{T,\mathrm{jets}}$ shapes. The spectra for the
leading and next-to-leading jets however overshoot the ones given at
NLO although they are softer with respect to those arising from the
$\mu=E_{T,W}$ scales. The $\Delta R_{12}$\/ curve has improved for low
separations, still remains below the NLO result. In addition to the
pure LO predictions, we have added to the $p_T$ spectra the
coupling-reweighted LO results (LO, local scales) as presented in
Ref.~\cite{Melnikov:2009wh}. They have been obtained at LO by purely
reweighting the initial strong couplings by those identified through
$k_T$ backward clustering. The implementation used in
\cite{Melnikov:2009wh} is in close spirit to the CKKW procedure
\cite{Catani:2001cc,Schalicke:2005nv}. The results look very similar
to the results at LO using $H_T$ as a scale, somewhat worse for the
$p_T$ spectrum of the third jet. Compared to
Figure~\ref{fig:pt.eta.jets} where we show the ME\&TS results, one can
conclude that the Sudakov rejections are the other important
ingredient of the merging approach to improve the shapes of
distributions and make them look similarly to those at NLO. The parton
showers of ME\&TS only correct in the soft/collinear phase-space
regions.


\subsection{Conclusions and outlook}\label{sec:conclus}

We have presented a comparison of predictions for $W^+$ + $\ge3$ jet
production at the LHC with $\sqrt s=10$~TeV between the NLO programs
{\sc BlackHat+Sherpa} and {\sc Rocket} and the ME\&TS method of
tree-level matrix-element plus parton-shower merging as implemented in
{\sc Sherpa}. This is the first time that results for this final state
have been compared to each other. Despite the different inputs to
these calculations, we have found an overall satisfactory agreement
among the various predictions for the $p_T$ and $\eta$\/ shapes of
jet and lepton distributions, and the jet--jet and jet-lepton $\Delta
R$\/ correlations. The largest shape differences, of the order of 20\%
and 40\%, are seen in the third-jet $p_T$ and $H_T$ distributions,
respectively. These uncertainties might still be worrisome in the
context of supersymmetry searches where $H_T$ is utilized as a major
discriminating variable. As a matter of fact, the scale dependence on
inclusive cross sections is considerably reduced at NLO; one still has
to be careful to what extent the higher-order correction improves the
predictions for more exclusive observables. The question for the
``right'' choice of scale(s) remains a tricky one to answer requiring
more detailed studies.

As shown in Ref.~\cite{Berger:2009ep}, choosing the factorization and
renormalization scales equal to the transverse energy of the $W$\/
boson can lead to unphysical (negative) results in the tails of some
distributions. The effects at Tevatron energies are far less dramatic,
as the range of the dynamical scales is much smaller there. It would
be suggestive to extend our comparison to include higher
transverse-energy bins in the plots to assess the potential danger of
the $E_{T,W}$ scale choice.

The inclusive $W^+$ + $\ge3$ jet cross sections given by the three
computations vary between 34.2~pb for {\sc Rocket}, 28.6~pb for
{\sc BlackHat+Sherpa} and 20.1~pb for {\sc Sherpa}'s ME\&TS
implementation (with merging up to 4 jets). The neglect of subleading
colour contributions in the {\sc Rocket} calculation has been
estimated to be less than 3\%.\footnote{The estimate has been taken
  from a comparison of full- and leading-colour NLO cross sections for
  $W$\/ + $1,2,3$ jet production at Tevatron energies.}
Therefore, at NLO the main reason for the deviations certainly lies in
the unequal scale choices, $E_{T,W}$ as used in {\sc Rocket} and
$\hat H_T$ as used in {\sc BlackHat+Sherpa}, generating rather
different average values for the $\mu_\mathrm{F}$ and $\mu_\mathrm{R}$
scales. The cross section given by the ME\&TS merging in {\sc Sherpa}
is of leading-order nature, however, compared to the pure LO
behaviour, it is more stable under the variation of scales and
inclusion of tree-level matrix elements of higher order.

The LO kinematic shape distributions resulting from $H_T$ scales,
rather than $E_{T,W}$ scales, resemble more closely those at NLO. In
particular, we have observed relatively uniform differential
$K$-factors for the third-jet $p_T$ and the $H_T$ variable. Hence,
$H_T$ seems to serve as a scale that more correctly describes the
overall hardness of the hard-scattering process.
In summary: there is satisfactory agreement among the NLO predictions,
even with the use of different scales, while there can be significant
disagreement between LO and NLO predictions unless care is taken with
the choice of the scales. Also, as a whole, the performance of the LO
predictions is worse than that of {\sc Sherpa}'s ME\&TS merging.
With sufficient tree-level matrix-element information, the ME\&TS
merging predictions given by {\sc Sherpa} agree (in shape) with the
NLO ones, indicating that the use of the correct local scale at each
vertex mimics, to some extent, the full NLO behaviour. That similar
results are obtained with the two very different scales is very
interesting, and deserves further investigation than possible in this
short write-up. It would be also very interesting to investigate the
agreement between the NLO and ME\&TS computations in a more detailed
study that could include scale variations, different PDF choices and
jet algorithms as well as a larger set of multi-particle correlations.

\subsection*{Acknowledgements}

The authors would like to thank Jennifer~Archibald, Carola~F.~Berger,
Zvi~Bern, John~M.~Campbell, Lance~J.~Dixon, R.~Keith~Ellis,
Fernando~Febres~Cordero, Darren~Forde, Walter~T.~Giele,
Tanju~Gleisberg, Harald~Ita, David~A.~Kosower, Frank~Krauss,
Zoltan~Kunszt, Joseph~D.~Lykken, Kirill~Melnikov, Gavin~P.~Salam,
Marek~Sch\"onherr, Steffen~Schumann and Frank~Siegert.




%% file: schwienhorst/schwienhorst.tex





\begin{figure}[!h!tbp]
\includegraphics[width=0.3\textwidth]{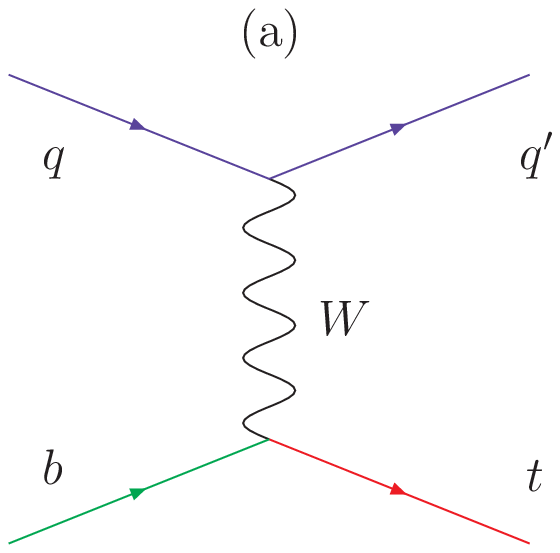}
\includegraphics[width=0.42\textwidth]{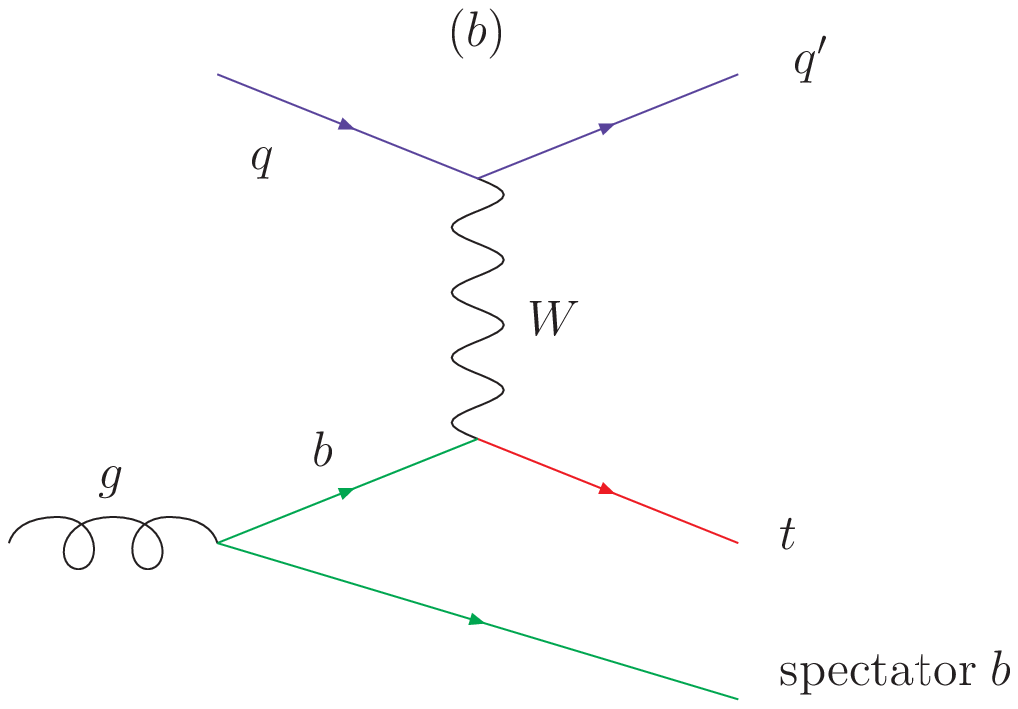}
\vspace{-0.1in}
\caption{Feynman diagram for $t$-channel single top quark production, in the $2\rightarrow 2$
scheme (a) and the $2\rightarrow 3$ scheme (b).}
\label{fig:singletopCompare_feynman}
\end{figure}

Single top quark production was recently observed for the first time by the D0~\cite{Abazov:2009ii}
and CDF~\cite{Aaltonen:2009jj} collaborations at the Tevatron proton-antiproton collider at Fermilab.
Two Feynman diagrams for single top quark production at a hadron collider are shown in 
Fig.~\ref{fig:singletopCompare_feynman}: (a) the leading order ($2\rightarrow 2$) exchange of a $W$~boson between
a light quark line and a heavy quark line, and (b) the ($2\rightarrow 3$) process where this
$b$~quark explicitly comes from gluon splitting. Diagram (a) is also referred to as single top 
production in the 5-flavor scheme because it utilizes the $b$~quark parton distribution 
function (PDF) in the proton. Diagram (b) is also referred to as $W$-gluon-fusion or single
top production in the 4-flavor scheme because the PDF of the gluon is required rather than the 
PDF of the $b$~quark. While the $2\rightarrow 3$ process (b) is one of the NLO corrections to
the LO process (a), it is an important contribution and the dominant contribution when
explicitly requiring three reconstructed jets.

A calculation of t-channel single top quark production at NLO in the $2\rightarrow 3$ 
scheme is now available, based on the MCFM NLO calculation~\cite{Campbell:2009ss}. 
This is the first calculation providing
$O(\alpha_s)$ corrections to the spectator $b$~quark from gluon splitting in the t-channel
process. Here we compare the distributions of this spectator $b$~quark with t-channel
single top samples generated with the CompHEP SingleTop generator~\cite{Boos:2006af,Boos:2004kh}. These
samples are used by the D0 experiment, for example in the observation of single top
quark production~\cite{Abazov:2009ii}.

In the CompHEP generation,
both $2 \rightarrow 2$ processes (eg $qb \rightarrow q't$, Fig.~\ref{fig:singletopCompare_feynman}(a)) and 
$2 \rightarrow 3$ processes (eg $qg \rightarrow q't\overline{b}$, Fig.~\ref{fig:singletopCompare_feynman}(b)) 
are included. 
The $2 \rightarrow 3$ process is relevant when the spectator $b$~quark is central and at high
$p_T$ and can be observed in the detector. The $2 \rightarrow 2$ processes start 
from a $b$-quark parton distribution function (PDF) and are
relevant when the spectator $b$~quark is soft and cannot be observed in the detector. These two 
contributions need to be combined
to provide one inclusive simulation sample. In the CompHEP matching approach~\cite{Boos:2006af,Boos:2004kh},
both samples are processed by {\sc Pythia}~\cite{Sjostrand:2006za}, and the $p_T$ distributions of the 
spectator $b$~quark produced by {\sc Pythia} are matched at a given spectator $b$~quark $p_T$ 
threshold. Below this threshold, the $2 \rightarrow 2$ sample is used, whereas above the threshold, 
the $2 \rightarrow 3$ sample is used. The threshold is chosen to produce a smooth $p_T$ distribution,
typically between 10GeV and 20GeV. In this particular example it is at 17~GeV.
The CompHEP sample was generated at a top quark mass of 172.5GeV, using the CTEQ~6.1 PDF 
set~\cite{Pumplin:2002vw}. 

The MCFM samples were generated at a top
quark mass of 170GeV and the CTEQ6M PDF set. There is a small difference in top quark mass 
between the two samples, but this has a negligible impact on the spectator $b$ quark kinematics.

\begin{figure}[!h!tbp]
\includegraphics[width=0.45\textwidth]{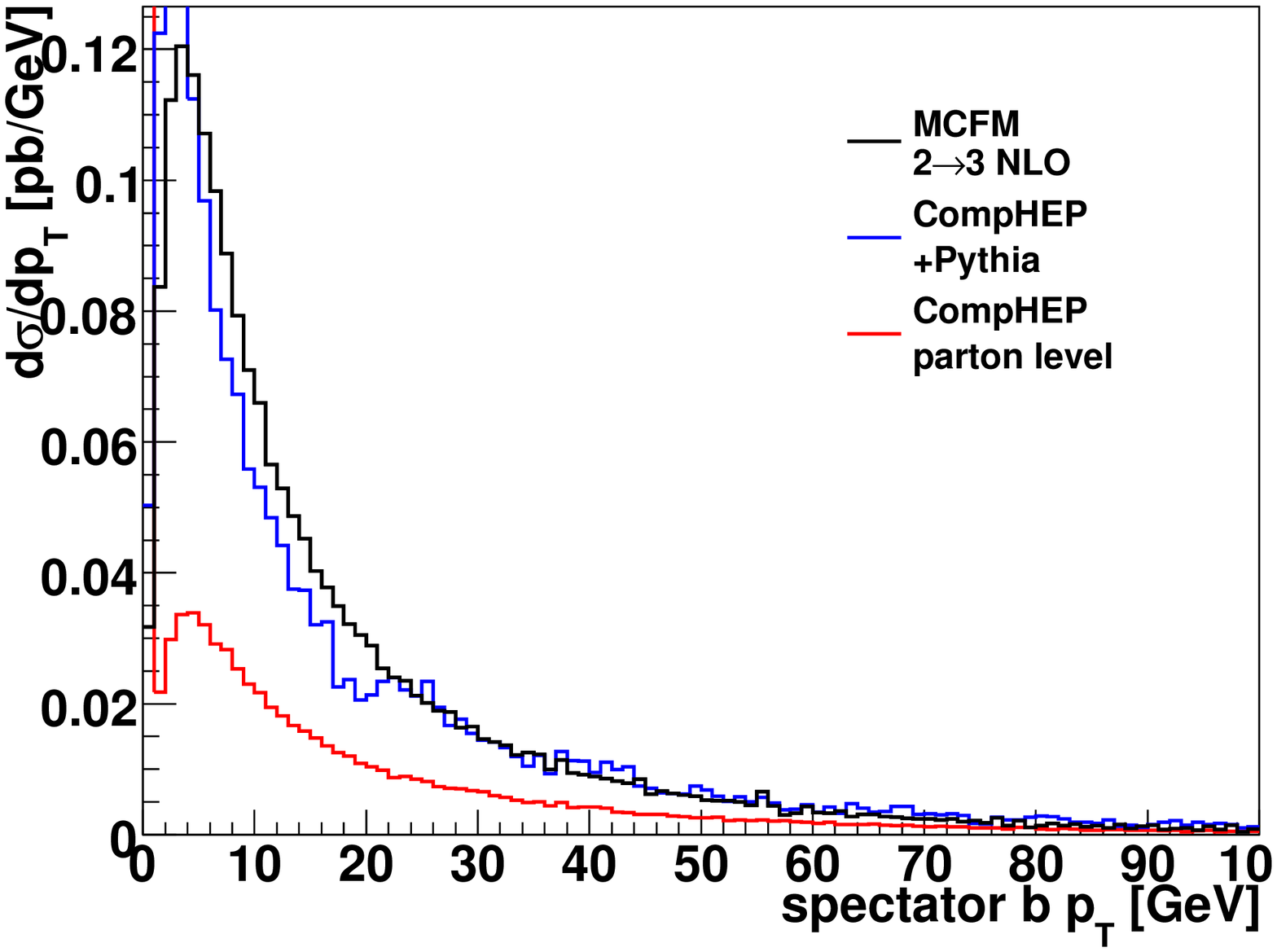}
\includegraphics[width=0.45\textwidth]{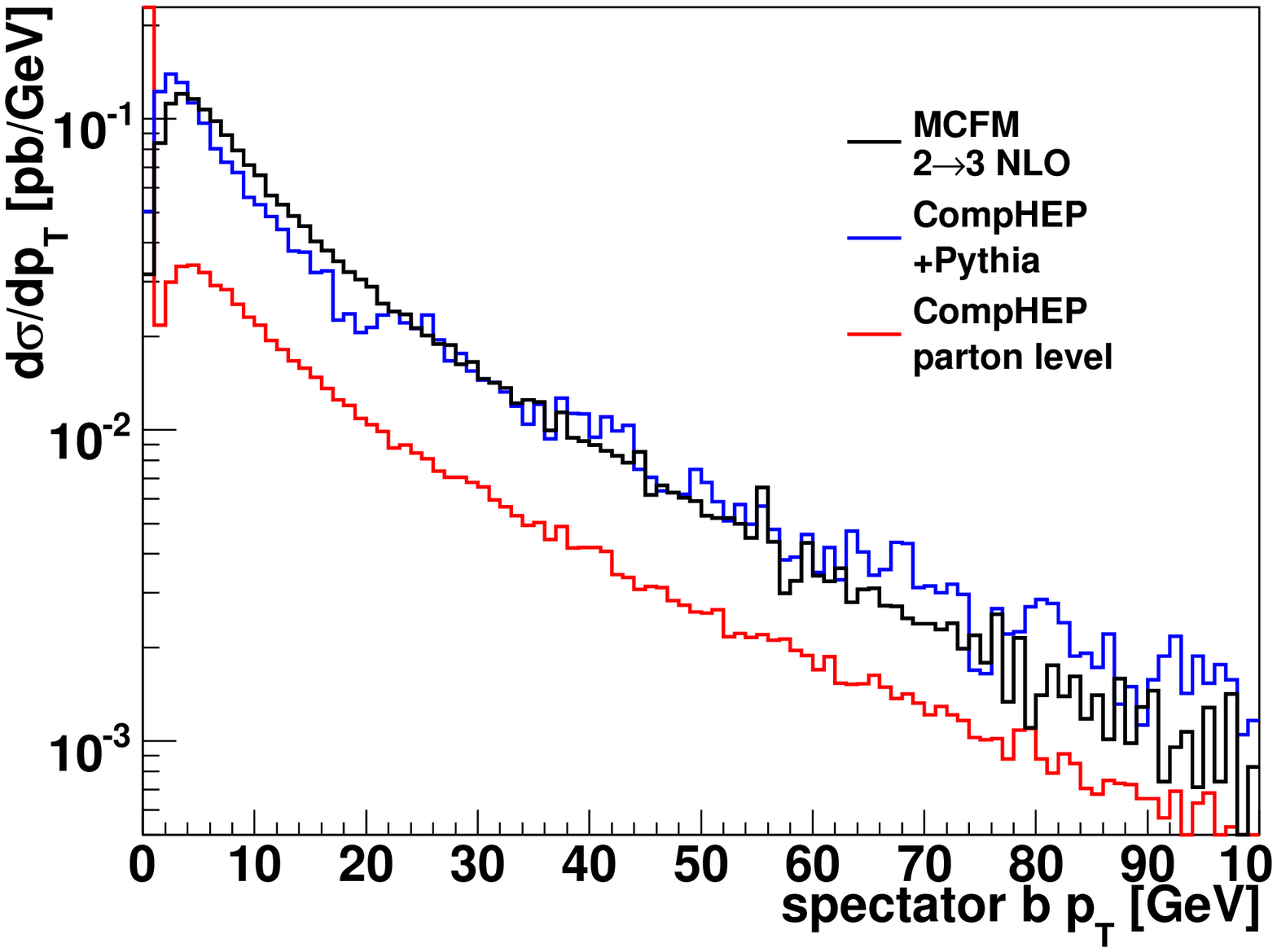}
\vspace{-0.1in}
\caption{Comparison of the $p_T$ of the spectator quark in $t$-channel single top quark
production between the MCFM $2\rightarrow 3$ NLO calculation and CompHEP interfaced to
{\sc Pythia}, in linear scale (left) and in log scale (right). Each distribution is normalized
to the NLO cross section.}
\label{fig:singletopCompare_pt}
\end{figure}
~
Figure~\ref{fig:singletopCompare_pt} shows a comparison of the transverse momentum ($p_T$) of the spectator quark.
No cuts have been applied.
The CompHEP parton level spectator $b$~quark $p_T$ distribution has a large spike at zero 
from $2\rightarrow 2$
events that have no spectator $b$~quark at parton level. The CompHEP $2\rightarrow 3$ contribution
at parton level is significantly below the MCFM calculation. However, once initial and final state 
gluon radiation is added by {\sc Pythia}, the spectator $b$~quark distribution agrees well
with the MCFM calculation. The main effect of {\sc Pythia} is to shift the $2\rightarrow 3$
contribution to the right and to fill in the low $p_T$ region with the $2\rightarrow 2$ calculation.

\begin{figure}[!h!tbp]
\includegraphics[width=0.45\textwidth]{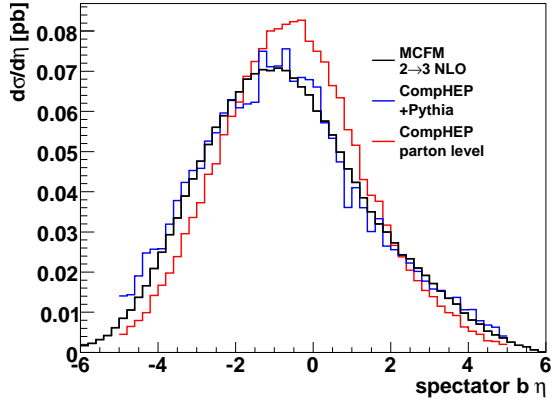}
\vspace{-0.1in}
\caption{Comparison of the pseudorapidity distribution of the spectator quark in $t$-channel 
single top quark production between the MCFM $2\rightarrow 3$ NLO calculation and 
CompHEP interfaced to {\sc Pythia}.}
\label{fig:singletopCompare_eta}
\end{figure}
~
Figure~\ref{fig:singletopCompare_eta} shows a comparison of the pseudorapidity of the spectator $b$~quark.
As explained above, the CompHEP parton level distribution consists only of $2\rightarrow 3$ events,
but even here the agreement with the NLO calculation is reasonable.
The {\sc Pythia} output again agrees well with the NLO calculation.

We have also calculated the acceptance for a spectator $b$~quark cut of $p_T>20$~GeV and a
pseudorapidity cut of $|\eta|<2.8$, following the approach of Ref.~\cite{Campbell:2009gj}. We obtain
an acceptance of 31.6\% for the {\sc Pythia} output samples, in good agreement with the MCFM
NLO calculation~\cite{Campbell:2009gj}.

In summary, we have presented a comparison of spectator $b$~quark in $t$-channel single top 
quark production
between the CompHEP parton-level event generator interfaced to {\sc Pythia} and the MCFM NLO 
calculation of the $t$-channel $2\rightarrow 3$ process. We find good agreement between the 
{\sc Pythia} output and MCFM for the spectator $b$~quark kinematics.



%% file: karg/karg.tex
\def\slsh{\rlap{$\;\!\!\not$}}     

\def\as{\alpha_s}
\def\eps{\epsilon}

\hfuzz 0.5pt




\subsection{Introduction}
The complicated hadron collider environment of the LHC requires not only 
sufficiently precise predictions for the expected signals, but also
reliable rates for complicated background reactions, especially
for those that cannot be entirely measured from data.
Among such background processes, several involve
three, four, or even more particles in the final state, rendering
the necessary next-to-leading-order (NLO) calculations in QCD 
technically challenging. At the previous Les Houches workshops
this problem led to the creation of a list of calculations that are 
a priority for LHC analyses, the so called 
''experimenters' wishlist for NLO calculations''
\cite{Buttar:2006zd,Campbell:2006wx,Bern:2008ef}.
The process class of 'electroweak gauge-boson pair with a hadronic jet' made 
it to the top of this list. Among other processes of that class, which have 
been addressed in earlier works~\cite{Dittmaier:2007th,Dittmaier:2009un,Campbell:2007ev,Campanario:2009um}, 
the process $pp\to ZZ{+}\text{jet}{+}X$ is of interest, for example as a background 
process to $H$+jet with the Higgs boson decaying into a pair of $Z$ bosons.

First results on the calculation of NLO QCD corrections to
$ZZ$+jet production have been presented in Ref.~\cite{Binoth:2009wk}.
A second calculation is in progress \cite{ppzzj_dittmaier} with 
some results already presented in Ref.~\cite{Kallweit:PhDthesis}.
In the following the key features of these two independent
calculations are described and results of an ongoing tuned comparison
are presented.

\subsection{Descriptions of the various calculations}

At leading order (LO), hadronic $ZZ$+jet production receives contributions
from the partonic processes $q\bar q\to ZZ g$, $qg\to ZZ q$,
and $\bar qg\to ZZ \bar q$, where $q$ stands for up- or down-type
quarks. All three channels are related by crossing symmetry.

The virtual corrections modify the partonic processes that are
already present at LO. At NLO these corrections
are induced by self-energy, vertex,
box (4-point), and pentagon (5-point) corrections, the latter being
the most complicated loop diagrams.
Apart from an efficient handling of the huge amount of algebra,
the most subtle point certainly is the numerically stable
evaluation of the numerous tensor loop integrals, in particular
in the vicinity of exceptional phase-space points.
The two calculations described below employ completely different
loop methods. Some of them are already briefly reviewed in
Ref.~\cite{Buttar:2006zd}, where
more details on problems in multi-leg loop calculations
and brief descriptions of proposed solutions can be found.

The real corrections are induced by the large variety of processes
that result from crossing any pair of QCD partons in
$0 \to ZZ   q \bar q g g$ and
$0 \to ZZ   q \bar q q' \bar q'$ into the initial state.
Here the main complication in the evaluation is connected to an
efficient phase-space integration with a proper separation
of soft and collinear singularities. For the separation of
singularities the two calculations
both employ the subtraction method\cite{Ellis:1980wv}
using the dipole subtraction formalism of Catani and Seymour\cite{Catani:1996vz}.
\subsubsection*{The calculation of DKU \cite{Kallweit:PhDthesis,ppzzj_dittmaier}}
This calculation is actually based on two completely independent evaluations of the
virtual and real corrections, referred to as DKU1 and DKU2 below. The $Z$~bosons are taken to be on~shell since the discussed 
results do not depend on the details of the $Z$~decays.
Both evaluations of loop diagrams are performed analogously 
to the calculations for the related process of $WW$+jet production, 
which are discussed in Refs.~\cite{Dittmaier:2007th,Dittmaier:2009un}.

The first calculation essentially follows the same strategy already
applied to the processes of $t\bar tH$
\cite{Beenakker:2002nc} and $t\bar t{+}\mathrm{jet}$ \cite{Dittmaier:2007wz}
production: The amplitudes are generated by {\sl Feyn\-Arts}~1.0 
\cite{Kublbeck:1990xc} and 
further processed with in-house {\sl Mathematica} routines,
which automatically create an output in {\sl Fortran}.
The IR (soft and collinear) singularities are treated in dimensional
regularization and analytically separated
from the finite remainder as described in
Refs.~\cite{Beenakker:2002nc,Dittmaier:2003bc}.
The pentagon tensor integrals are directly reduced to box
integrals following Refs.~\cite{Denner:2002ii,Denner:2005nn}.
Box and lower-point integrals are reduced
\`a la Passarino--Veltman \cite{Passarino:1978jh} to scalar integrals,
which are either calculated analytically or using the results of
Refs.~\cite{'tHooft:1978xw,Beenakker:1988jr,Denner:1991qq}.

The second loop calculation is based on {\sl Feyn\-Arts}~3.4 \cite{Hahn:2000kx} to generate the diagrams and 
{\sl FormCalc}~6.0 \cite{Hahn:1998yk} which
automatically produces {\sl Fortran} code.
The reduction of tensor to scalar integrals is done with the
help of the {\sl LoopTools} library \cite{Hahn:1998yk},
which also employs the method of Ref.~\cite{Denner:2002ii} for the
5-point tensor integrals, Passarino--Veltman \cite{Passarino:1978jh}
reduction for the lower-point tensors, and the {\sl FF} package
\cite{vanOldenborgh:1989wn,vanOldenborgh:1991yc} for the evaluation
of regular scalar integrals.
The dimensionally regularized soft or collinear singular 3- and 4-point
integrals had to be added to this library.

The first calculation of the real corrections employs analytical results
for helicity amplitudes obtained in a spinor formalism.
The phase-space integration is performed by the multi-channel Monte Carlo integrator~\cite{Berends:1994pv}
with weight optimization~\cite{Kleiss:1994qy} that has been written in {\sl C++} and tested in the calculation of $WW$+jet.
More details on this calculation and some numerical results can be found in Ref.~\cite{Kallweit:PhDthesis}.

The second evaluation of the real corrections
is based on scattering amplitudes calculated
with {\sl Madgraph} \cite{Stelzer:1994ta} generated code.
The code has been modified to allow for the extraction of the required colour and
spin structure. The latter enter the evaluation of the dipoles in the
Catani--Seymour subtraction method. The evaluation of the individual dipoles
was performed using a {\sl C++} library developed during the calculation of
the NLO corrections for $t\bar t{+}\mathrm{jet}$ \cite{Dittmaier:2007wz}.
For the phase-space integration a
simple mapping has been used where the phase space is generated from
a sequential splitting.

\subsubsection*{The calculation of BGKKS \cite{Binoth:2009wk}}

This calculation is based on two independent sets of amplitude expressions:~one 
generated manually starting from the Feynman graph representation, the other using 
QGRAF \cite{Nogueira:1991ex}.
Both representations employ the spinor helicity formalism of Ref.~\cite{Xu:1986xb}.  
Polarisation vectors have been represented via spinor traces, i.e.~kinematic
invariants up to global phases.  By obtaining an analytical representation for 
the full amplitude, we aim at promoting simplification via analytical cancellations.
Especially we employ that, apart from the rank one case, all pentagon tensor integrals 
are reducible, i.e.~can directly be written as simple combinations of box tensor integrals.   
For the remaining tensor integrals we employ the GOLEM-approach \cite{Binoth:1999sp,Binoth:2005ff,Binoth:2006hk}.
In this approach, the use of 6-dimensional IR finite box functions
allows to isolate IR divergences in 3-point functions. 
We use FORM \cite{Vermaseren:2000nd} and Maple to obtain tractable analytical expressions for the 
coefficients to the employed set of basis functions for each independent helicity amplitude,
and to further simplify them.  The basis functions are evaluated using the GOLEM95 
implementation \cite{Binoth:2008uq}.
We note that for the reduction of box topologies one obtains the same result
as with the Passarino-Veltman tensor reduction \cite{'tHooft:1978xw,Passarino:1978jh}. 
If one fully reduces all tensor integrals to a scalar integral representation, 
the difference between the two approaches results from the treatment of the pentagon 
integrals and the use of finite 6-dimensional box functions.

To treat $\gamma_5$ we employ the 't~Hooft-Veltman scheme 
\cite{'tHooft:1972fi,Breitenlohner:1975hg}, where the $\gamma^\mu$ are 
split into a 4-dimensional part that anti-commutes with $\gamma_5$ 
and a commuting remainder.
As is well known, to take into account differences between the QCD corrections 
to axial vector and vector currents, a finite renormalization has to be performed.
To enforce the correct chiral structure of the amplitudes, 
a finite counterterm for the axial part is included in the used gauge boson vertex 
(see e.g.~Refs.~\cite{Larin:1993tq,Trueman:1995ca,Harris:2002md}):
\[
V^\mu_{Vq\bar{q}} \sim g_v \, \gamma^\mu + Z_5 \, g_a \, \gamma^\mu \gamma_5 \quad\text{with}\quad
 Z_5 = 1 - C_F \,\frac{\alpha_s}{\pi}\,.
\]

We calculate with $N_{\mathrm{f}}=5$ and neglect quark mass effects.  
Our virtual amplitudes have been verified by comparing two independent internal
implementations, both generated using the GOLEM reduction.
We have verified that the relative 
contribution of graphs with quark loops to integrated results is typically well below 1\%.  
We therefore neglect this contribution.  
To calculate numerical results for the virtual contributions 
we employed the OmniComp-Dvegas package, which 
facilitates parallelised adaptive Monte Carlo integration
and was developed in the context of Ref.~\cite{Kauer:2001sp}.
We use the SHERPA implementation \cite{Gleisberg:2007md,Krauss:2001iv,Gleisberg:2008ta} to calculate
numerical results for the finite real corrections contribution.  All
amplitude and dipole contributions have been verified 
through comparison with results calculated with 
MadDipole/MadGraph \cite{Frederix:2008hu,Stelzer:1994ta}
and HELAC \cite{Czakon:2009ss,Kanaki:2000ey}.

\subsection{Tuned comparison of results}

The following results essentially employ the setup of
Ref.~\cite{Binoth:2009wk}.
The CTEQ6~\cite{Pumplin:2002vw,Stump:2003yu}
set of parton distribution functions (PDFs) is used throughout, i.e.\
CTEQ6L1 PDFs with a 1-loop running $\alpha_{\mathrm{s}}$ are taken in
LO and CTEQ6M PDFs with a 2-loop running $\alpha_{\mathrm{s}}$ in NLO.
In the strong coupling constant
the number of active flavours is $N_{\mathrm{f}}=5$, and we use the default  LHAPDF values
leading to
$\alpha_{\mathrm{s}}^{\mathrm{LO}}(91.188 \text{ GeV})=0.129783$ and
$\alpha_{\mathrm{s}}^{\mathrm{NLO}}(91.70 \text{ GeV})=0.1179$.
The top-quark loop in the gluon self-energy is
subtracted at zero momentum. The running of
$\alpha_{\mathrm{s}}$ is, thus, generated solely by the contributions of the
light quark and gluon loops. 
In all results shown in the following, the
renormalization and factorization scales are set to $M_Z$.
The top-quark mass is
$m_t=174.3$~GeV, the masses of all other quarks are neglected.
The weak boson masses are $M_Z=91.188$~GeV and $M_H=150$~GeV.
The weak mixing angle is set to its on-shell value, i.e.\
fixed by $s_w^2=0.222247$, and the electromagnetic
coupling constant is set to $\alpha=0.00755391226$. 

We apply the $k_\perp$ jet algorithm of Ref.~\cite{Catani:1993hr}
with covariant $E$-recombination scheme and $R=0.7$
for the definition of the tagged hard jet and
restrict the transverse momentum of the hardest jet by
$p_{\mathrm{T,jet}}>50$~GeV. 

\subsubsection{Results for a single phase-space point}

For the comparison the following set of four-momenta $(E, p_x, p_y, p_z)$ [GeV] is chosen,
\begin{equation}
\begin{aligned}
\label{eq:ppzzj_momenta}
p_1^\mu &= (250,0,0,250), \qquad p_2^\mu = (250,0,0,-250),\\
p_3^\mu &= (125.9335600344245,  -81.91900733932759,  -15.22986911133704, -24.52218428963296),\\
p_4^\mu &= (201.2131630027446,   37.57875773939030,  -105.1640094872687, 140.3561672919824),\\
p_5^\mu &= (172.8532769628309,   44.34024959993729,   120.3938785986057, -115.8339830023494),
\end{aligned}
\end{equation}
where incoming and outgoing particles are labelled as follows: $1, 2 \to 3, 4, 5$.

Table~\ref{tab:ppwwj_single_lo} shows some results for the
spin- and colour-summed squared LO matrix elements, where no factor
$1/2$ is included for the two identical $Z$~bosons in the final state.
The results of the two groups
\ agree within about 13 digits.
\begin{table}[bt]
\centering
\begin{tabular}{lc}\toprule
 & $|{\cal M}_{\mathrm{LO}}|^2/e^4/g_{\mathrm{s}}^2[\mathrm{GeV}^{-2}]$ \\
\midrule
$u \bar u \to ZZ g$ &  \\ 
\cmidrule(l){1-1}
BGKKS & $9.08160337631 1467 \cdot 10^{-4}$ \\ 

DKU1 & $9.081603376315696  \cdot 10^{-4}$ \\
DKU2 & $9.081603376315669  \cdot 10^{-4}$ \\
\midrule
$d \bar d \to ZZ g$ &  \\
\cmidrule(l){1-1}
BGKKS & $1.89258973073 5170 \cdot 10^{-3}$\\ 

DKU1 & $1.892589730736050  \cdot 10^{-3}$\\ 
DKU2 & $1.892589730736046  \cdot 10^{-3}$\\ 
\midrule
$u g \to ZZ g$ &  \\
\cmidrule(l){1-1}
BGKKS & $1.6876149896801 96\cdot 10^{-4}$\\

DKU1 & $ 1.687614989680182  \cdot 10^{-4}$ \\ 
DKU2 & $ 1.687614989680173  \cdot 10^{-4}$ \\ 
\midrule
$d g \to ZZ g$ &  \\
\cmidrule(l){1-1}
BGKKS & $ 3.5169591387734 90\cdot 10^{-4}$\\

DKU1 & $ 3.516959138773458  \cdot 10^{-4}$ \\ 
DKU2 & $ 3.516959138773441  \cdot 10^{-4}$ \\ 
\midrule
$g \bar u  \to ZZ g$ &  \\
\cmidrule(l){1-1}
BGKKS & $ 1.3192411141944 92\cdot 10^{-5}$ \\

DKU1 & $ 1.319241114194495  \cdot 10^{-5}$ \\ 
DKU2 & $ 1.319241114194489  \cdot 10^{-5}$ \\ 
\midrule 
$g \bar d \to ZZ g$ &  \\
\cmidrule(l){1-1}
BGKKS & $ 2.7492746397632 24\cdot 10^{-5}$ \\ 

DKU1 & $ 2.749274639763229 \cdot 10^{-5}$ \\ 
DKU2 & $ 2.749274639763217 \cdot 10^{-5}$ \\ 
\bottomrule
 \end{tabular}
\caption{Results for squared LO matrix elements at the phase-space point
(\ref{eq:ppzzj_momenta}).}
\label{tab:ppwwj_single_lo}
\end{table} 

\begin{table}[bt]
\centering
\setlength{\tabcolsep}{4pt}
\begin{tabular}{lrrr}\toprule
 &   $c^{\mathrm{bos}}_{0}[\mathrm{GeV}^{-2}]$  &  $c^{\mathrm{ferm1+2}}_{0}[\mathrm{GeV}^{-2}]$ &  $c^{\mathrm{ferm3}}_{0}[\mathrm{GeV}^{-2}]$\\
\midrule
\multicolumn{4}{l}{$u \bar u \to ZZ g$}\\ 

\cmidrule(l){1-1}
BGKKS &$ 2.57171837098 6939 \cdot 10^{-4}$&  $  2.77127 4006707126 \cdot 10^{-6}$ & \\

DKU1  &$ 2.571718370988091 \cdot 10^{-4}$&  $  2.771273991103833 \cdot 10^{-6}$ & $ 3.301195986341516 \cdot 10^{-6}$ \\
DKU2  &$ 2.571718370988072 \cdot 10^{-4}$&  $  2.771273991102529 \cdot 10^{-6}$ & $ 3.301195986341134 \cdot 10^{-6}$ \\
\midrule
\multicolumn{4}{l}{$d \bar d \to ZZ g$}\\ 

\cmidrule(l){1-1}
BGKKS &$ 5.33563785292 1577 \cdot 10^{-3}$&  $  3.5538049 47755081 \cdot 10^{-6}$ & \\

DKU1  &$ 5.335637852923933 \cdot 10^{-3}$&  $  3.553804924505993 \cdot 10^{-6}$ & $ -7.625169350877288 \cdot 10^{-6}$ \\
DKU2  &$ 5.335637852923915 \cdot 10^{-3}$&  $  3.553804924504350 \cdot 10^{-6}$ & $ -7.625169350877653 \cdot 10^{-6}$ \\
\midrule
\multicolumn{4}{l}{$u g \to ZZ g$}\\ 

\cmidrule(l){1-1}
BGKKS &$ 3.4553036909 23093 \cdot 10^{-4}$&  $  -1.5752777 09579237 \cdot 10^{-6}$ \\

DKU1  &$ 3.455303690940059 \cdot 10^{-4}$&  $  -1.575277712403393 \cdot 10^{-6}$ & $ -1.899597362881991 \cdot 10^{-6}$ \\
DKU2  &$ 3.455303690940080 \cdot 10^{-4}$&  $  -1.575277712403507 \cdot 10^{-6}$ & $ -1.899597362882020 \cdot 10^{-6}$ \\
\midrule
\multicolumn{4}{l}{$d g \to ZZ g$}\\ 

\cmidrule(l){1-1}
BGKKS &$ 7.1822187314 01221 \cdot 10^{-4}$&  $  -2.1348368 68278616 \cdot 10^{-6}$ & \\

DKU1  &$ 7.182218731436469 \cdot 10^{-4}$&  $  -2.134836871947412 \cdot 10^{-6}$ & $ 3.857433911012773 \cdot 10^{-6}$ \\
DKU2  &$ 7.182218731436517 \cdot 10^{-4}$&  $  -2.134836871947570 \cdot 10^{-6}$ & $ 3.857433911012694 \cdot 10^{-6}$ \\
\midrule
\multicolumn{4}{l}{$g \bar u \to ZZ g$}\\ 

\cmidrule(l){1-1}
BGKKS &$ 7.2840794 47744509 \cdot 10^{-5}$&  $  -3.8778568783 13408 \cdot 10^{-6}$ \\

DKU1  &$ 7.284079439746620 \cdot 10^{-5}$&  $  -3.877856878314387 \cdot 10^{-6}$ & $ -5.478348291183621 \cdot 10^{-7}$ \\
DKU2  &$ 7.284079439746720 \cdot 10^{-5}$&  $  -3.877856878314465 \cdot 10^{-6}$ & $ -5.478348291184200 \cdot 10^{-7}$ \\
\midrule 
\multicolumn{4}{l}{$g \bar d \to ZZ g$}\\ 

\cmidrule(l){1-1}
BGKKS &$ 1.50544875 6089957 \cdot 10^{-5}$&  $  -4.83914037543 5081 \cdot 10^{-6}$ & \\

DKU1  &$ 1.505448754415003 \cdot 10^{-5}$&  $  -4.839140375436319 \cdot 10^{-6}$ & $ 3.379222628266236 \cdot 10^{-7}$ \\
DKU2  &$ 1.505448754415026 \cdot 10^{-5}$&  $  -4.839140375436448 \cdot 10^{-6}$ & $ 3.379222628265571 \cdot 10^{-7}$ \\
\bottomrule
 \end{tabular}
\caption{Virtual corrections of the bosonic contributions,  the fermionic contributions of the two light  generations ($m_q=0$), and the fermionic contributions of the $3^{\rm{rd}}$ generation ($m_{\rm{b}}=0$, $m_{\rm{t}}=174.3$ GeV) at the phase-space point (\ref{eq:ppzzj_momenta}).}
\label{tab:virtual_contrib}
\end{table} 

In order to be independent of the subtraction scheme to cancel IR divergences, we found it useful to compare virtual results prior to any subtraction. The ${\cal O}(\alpha_s)$ contribution to the virtual, renormalized squared amplitude is given by the interference between tree-level and one-loop virtual amplitude, which we denote schematically as
\begin{equation}
\label{eq:virtonlydef}
2 \mathrm{Re}\{{\cal M}_{V}^*\cdot {\cal M}_{\mathrm{LO}}\} = e^4 g_s^2 
f(\mu_{\mathrm{ren}})
\left(c_{-2} \frac{1}{\eps^2} +c_{-1} \frac{1}{\eps} + c_0
\right)\,,
\end{equation}
with
$f(\mu_{\mathrm{ren}}) = \Gamma(1+\eps) (4\pi\mu_{\mathrm{ren}}^2/M_Z^2)^\eps$
and the number of space--time dimensions $D=4-2\eps$.
In the following we split the coefficients of
the double and single pole and for the constant part, $c_{-2}, c_{-1}$, and
$c_0$, into bosonic contributions (``bos'') without closed fermion loops
and the remaining fermionic parts. The fermionic corrections are further
split into contributions from the first two generations (``ferm1+2'')
and from the third generation (``ferm3'').

The results on $c_0$ obtained by the different groups typically agree 
within $8{-}12$ digits; the agreement between DKU1 and DKU2 results turns out to be within $12{-}14$ digits.\footnote{BGKKS show only one result in 
Table~\protect\ref{tab:virtual_contrib}, but our internal comparison of two 
independent implementations of the virtual amplitudes yielded agreement of 9-16 
significant digits for all contributions at two test phase space points.}
The values of $c_0$ for the different channels are collected in Table~\ref{tab:virtual_contrib} according to the splitting stated above. 
The coefficients of the poles have not been compared numerically since the cancellation of divergences can be checked analytically.

\subsubsection{Results for integrated cross sections}
Table~\ref{tab:ppzzj_cs} illustrates the agreement of the LO and NLO cross
sections for the LHC ($pp$, $\sqrt{s} = 14$ TeV) and Tevatron ($p\bar{p}$, $\sqrt{s} = 1.96$ TeV) calculated by both groups with the setup defined above. For the NLO observable labelled by 'excl', a veto on a $2^{\rm{nd}}$ hard jet ($p_{\mathrm{T,2^{nd}\,jet}}<50\,\rm{GeV}$) has been applied in the real-correction contribution.

Table~\ref{tab:ppzzj_ferm} provides individual contributions to the NLO cross section in Table~\ref{tab:ppzzj_cs}, as well as the contribution of the fermionic loops 
to the integrated NLO cross section---again subdivided into contributions of the 
two light generations and the third one---, which have not been taken into account 
in the cross sections of Table~\ref{tab:ppzzj_cs}. However, their size turns out to 
be well below the percent level, so that they may be neglected on the experimentally 
required level of accuracy.

We note that we also compared cross sections for different scale choices and distributions and found agreement.

\begin{table}[bt]
\centering
\begin{tabular}{llll}\toprule
$pp\to ZZ{+}\mathrm{jet}{+}X$ @ LHC & $\sigma_{\mathrm{LO}}$[fb] & $\sigma_{\mathrm{NLO}}$[fb] & $\sigma_{\mathrm{NLO,excl}}$[fb] \\
\midrule
BGKKS & $2697.82 \,[42]$  & $3644.5 \,[3.0]$  & $2627.5 \,[3.0]$   \\
DKU  & $2697.81 \,[18]$   & $3644.6 \,[1.0]$ & $2626.3 \,[1.1]$ \\
\midrule
$p\bar{p}\to ZZ{+}\mathrm{jet}{+}X$ @ Tevatron & $\sigma_{\mathrm{LO}}$[fb] & $\sigma_{\mathrm{NLO}}$[fb] & $\sigma_{\mathrm{NLO,excl}}$[fb]  \\ 
\midrule
BGKKS & $74.5589 \,[90]$ & $83.665 \,[62]$ & $78.824 \,[62]$ \\
DKU & $74.5664 \,[76]$ & $83.751 \,[47]$ & $78.915 \,[47]$ \\
\bottomrule
 \end{tabular}
\caption{Results for contributions to the integrated $ZZ$+jet cross sections at the LHC 
and Tevatron in LO and NLO. Only bosonic loop corrections are included here in the virtual
 part, i.e.~all fermion loops are neglected.}
\label{tab:ppzzj_cs}
\end{table} 

\begin{table}[bt]
\centering
\begin{tabular}{rrrrr}\toprule
&\multicolumn{2}{c}{$pp\to ZZ{+}\mathrm{jet}{+}X$ @ LHC} & \multicolumn{2}{c}{$p\bar{p}\to ZZ{+}\mathrm{jet}{+}X$ @ Tevatron}\\
\midrule
& \multicolumn{1}{c}{BGKKS} & \multicolumn{1}{c}{DKU} & \multicolumn{1}{c}{BGKKS} & \multicolumn{1}{c}{DKU} \\
\midrule
$\sigma_{\mathrm{born}}$[fb] & $2580.60\,\phantom{.}[39] $ & $ 2579.91\,[55]\phantom{0} $ & $ 70.0581\,[83] $ & $ 70.056\,[23]\phantom{00} $ \\
$\sigma_{\mathrm{coll}}$[fb] & $ 918.62\,\phantom{.}[54] $ & $ 917.59\,[39]\phantom{0} $ & $ 16.578\,[24]\phantom{0} $ & $ 16.592\,[17]\phantom{00} $ \\
$\sigma_{\mathrm{real}}$[fb] & $ -82.9\,[2.4]\phantom{0} $ & $ -82.79\,[72]\phantom{0} $ & $ -11.143 \,[26]\phantom{0} $ & $ -11.092 \,[36]\phantom{00} $ \\
$\sigma_{\mathrm{real,excl}}$[fb] & $ -1099.9 \,[2.4]\phantom{0} $ & $ -1101.09 \,[76]\phantom{0} $ & $ -15.983 \,[26]\phantom{0} $ & $ -15.928 \,[36]\phantom{00} $ \\
$\sigma_{\mathrm{virt,\, bose+I}}$[fb] & $ 228.1 \,[1.7]\phantom{0} $ & $ 229.92 \,[34]\phantom{0} $ & $ 8.171 \,[50]\phantom{0} $ & $ 8.1950 \,[88]\phantom{0} $ \\[1ex]
$\sigma_{\mathrm{virt,\, ferm\,1+2}}$[fb] & $  $ & $ -17.864 \,[28] $ & $  $ & $ -0.07527 \,[11] $ \\
$\sigma_{\mathrm{virt,\, ferm\,3}}$[fb] & $  $ & $ 6.750 \,[16] $ & $  $ & $ 0.18600 \,[14] $ \\
\bottomrule
\end{tabular}
\caption{Results for the born, sum of the $K$ and $P$ insertion operators, dipole subtracted real emissions, IR-finite sum of bosonic loops and the $I$ insertion operator and fermion loops contributions to the integrated $ZZ$+jet cross sections in NLO at the LHC and Tevatron.}
\label{tab:ppzzj_ferm}
\end{table} 

\subsection{Conclusions}

We have reported on a tuned comparison of calculations of the NLO QCD corrections to $ZZ$+jet production at the LHC and Tevatron. For a fixed phase-space point, the virtual corrections obtained by both groups using different calculational techniques agree at the level of $10^{-8}$ or better. The comparison of full NLO cross sections, which involve the non-trivial integration of virtual and real corrections over the phase space, shows agreement at the permille level.

\subsection*{Acknowledgements}

This work is supported in part by the EU's  Marie-Curie Research 
Training Network HEPTOOLS under contract MRTN-CT-2006-035505 and by Germany's 
DFG (SFB/TR9 and contract BI 1050/2) and BMBF (contract 05HT1WWA2), the
UK's HEFCE, STFC and SUPA and the US DOE (contract DE-AC02-76SF00515).



%% file: chachamis/chachamis.tex






\subsection{Introduction}

One of the primary goals for the
LHC is undoubtedly the discovery of the Higgs boson which
is responsible for the fermions and gauge bosons mass and also part
of the mechanism of dynamical breaking of
the Electroweak (EW) symmetry.
Another important aim for the LHC is the precise measurement of the
hadronic production of gauge boson pairs, 
$W W$, $W Z$, $Z Z$, $W \gamma$, $Z \gamma$, 
this in connection to the investigation
of the non-Abelian gauge structure of the SM.
W pair production,
\begin{equation}
q {\bar q} \rightarrow W^+ \, W^- \, ,
\end{equation}
plays an essential role as it serves as a signal process 
in the search for New Physics and also is the dominant irreducible
background to the promising Higgs discovery channel 
\begin{equation}
p p \rightarrow H \rightarrow W^* W^* 
\rightarrow l {\bar \nu} {\bar l}' \nu' \, 
\end{equation}
in the mass range $M_{\mathrm{Higgs}}$ 
between 140 and 180 GeV~\cite{Dittmar:1996ss}.

The process is currently known at next-to-leading order (NLO) 
accuracy~\cite{Brown:1978mq, Ohnemus:1991kk, Frixione:1993yp, Dixon:1998py, 
Dixon:1999di, Campbell:1999ah, Grazzini:2005vw }.
The NLO corrections were proven to be large
enhancing the tree-level by
almost 70\% which falls to a (still) large 30\% after 
imposing a jet veto. 
Therefore, if a theoretical estimate for  the
W pair production is to be compared against 
experimental measurements at the LHC, one is bound to 
go one order higher in the perturbative expansion, namely,
to the next-to-next-to-leading order (NNLO). This would 
allow, in principle, an accuracy of around 10\%.

High accuracy for the W pair production is also needed
when the process is studied as background to Higgs production
in order to match accuracies between signal and background.
The signal process for the Higgs discovery via
gluon fusion,
$g g \rightarrow H$, 
as well as the process
$H \rightarrow W W \rightarrow l {\bar \nu} {\bar l}' \nu'$
are known at 
NNLO~\cite{Spira:1995rr,Dawson:1990zj,Harlander:2002wh,Anastasiou:2002yz,
Ravindran:2003um,Catani:2001cr,Davatz:2004zg,Anastasiou:2004xq, 
Anastasiou:2007mz,Grazzini:2008tf},
whereas the EW corrections are known
beyond NLO~\cite{Bredenstein:2006rh}.
Another process that needs to be included in the background
is the W pair production
in the loop induced gluon fusion channel, 
\begin{equation}
g g \rightarrow W^+ W^- \, .
\end{equation}
The latter
contributes at $\mathcal{O}(\alpha_s^2)$ relative to the 
quark-anti-quark-annihilation channel but is 
nevertheless enhanced due to the large gluon flux
at the LHC~\cite{Binoth:2005ua, Binoth:2006mf}.

The first main difficulty in studying the NNLO QCD
corrections for W pair production is the calculation
of the two-loop virtual amplitude since it is
a $2 \rightarrow 2$ process with massive external particles.
We have already computed the virtual corrections
at the high energy 
limit~\cite{Chachamis:2007cy, Chachamis:2008yb, Chachamis:2008xu}.
However, this is not enough as it cannot
cover the kinematical region close to threshold. Therefore,
in order to cover all kinematical regions we proceed as follows.
We perform a deep expansion in the W mass around the
high energy limit which in combination with
the method
of numerical integration of differential
equations~\cite{Caffo:1998du, Boughezal:2007ny, Czakon:2007qi} 
allows us the numerical computation of the two-loop amplitude
with full mass dependence over the whole phase space.

\subsection{The high energy limit}

The methodology for obtaining the massive amplitude in the
high energy limit, namely the limit where all the invariants are
much larger than the W mass,
is similar to the one followed in Refs.~\cite{Czakon:2007ej, Czakon:2007wk}.
The amplitude is reduced to an expression
that only contains a small number of integrals (master integrals) 
with the help of the Laporta algorithm~\cite{Laporta:2001dd}. 
In the calculation for the two-loop amplitude there 
are 71 master integrals.  Next step is the construction, in a fully 
automatised way, of the Mellin-Barnes (MB) 
representations~\cite{Smirnov:1999gc, Tausk:1999vh} 
of all the master integrals by using the {\bf MBrepresentation} 
package~\cite{MBrepresentation}. 
The representations are then analytically continued in the number 
of space-time dimensions by means
of the {\bf MB} 
package~\cite{Czakon:2005rk}, thus revealing the full singularity 
structure. An asymptotic expansion in
the mass parameter (W mass) is performed by closing contours and the 
integrals are finally resummed, either
with the help of {\bf XSummer}~\cite{Moch:2005uc} or the 
{\bf PSLQ} algorithm~\cite{pslqAlg}.
The result is expressed in terms of harmonic polylogarithms.

\subsection{Power corrections and numerical evaluation}

The high energy limit by itself is not
enough, as was mentioned before. 
The next step, following the methods applied in Ref.~\cite{Czakon:2008zk},
is to compute power corrections in the W mass. 
Power corrections are good enough to
cover most of the
phase space, apart from the region near 
threshold as well as the regions corresponding
to small angle scattering.

We recapitulate here some of the notation of Ref.~\cite{Chachamis:2008xu}
for completeness.
The charged vector-boson 
production in the leading partonic scattering process
corresponds to
\begin{equation}
\label{chachamis_qqWW}
q(p_1) + {\overline q}(p_2) 
\:\:\rightarrow\:\: W^-(p_3,m) + W^+(p_4,m) \, ,
\end{equation}
where $p_i$ denote 
the quark and W momenta and $m$ is the mass of the W boson.

We have chosen to express the amplitude in terms
of the kinematic variables $x$ and $m_s$ which are defined to be
\begin{equation}
  x = -\frac{t}{s}, \;\; m_s = \frac{m^2}{s},
\end{equation}
where
\begin{equation}
s = (p_1+p_2)^2 \;\; {\rm and} \;\;t = (p_1-p_3)^2-m^2\,.
\end{equation}
The variation then of
$x$ within the range $ [ 1/2(1-\beta), 1/2(1+\beta) ] $, where
$\beta=\sqrt{1-4m^2/s}$ is the velocity, corresponds to angular variation
between the forward and backward scattering.

It should be evident that any master integral $M_i$ can be written then as
\begin{equation}
M_i = M_i \left( m_s, x, \epsilon \right) = \sum_{j=k}^l \epsilon^j {I_i}_j(m_s, x),
\label{chachamis_epReveal}
\end{equation}
where the lowest power of $\epsilon$ in the sum can be $-4$.

The crucial point now is that
the derivative of any Feynman integral with
respect to any kinematical variable is 
again a Feynman integral with possibly
higher powers of denominators or numerators which can also be
reduced to masters from the initial set of master integrals.
This means that one can construct a
partially triangular system of differential equations in the mass,
which can subsequently be solved in the form of a power series expansion,
with the expansion parameter in our case being $m_s$ following
the conventions above.

Let us differentiate with respect to $m_s$ and $x$, we will then have
respectively
\begin{equation}
m_s \frac{d}{dm_s} M_i(m_s,x,\epsilon) = 
\sum_j C_{i j}(m_s,x,\epsilon) ~M_j(m_s,x,\epsilon)
\label{chachamis_dms}
\end{equation}
and
\begin{equation}
x \frac{d}{dx} M_i(m_s,x,\epsilon) = 
\sum_j C_{i j}\sp{\prime}(m_s,x,\epsilon) ~M_j(m_s,x,\epsilon)\, .
\label{chachamis_dx}
\end{equation}
We use Eq.~(\ref{chachamis_dms}) to obtain the 
mass corrections for the masters 
calculating the power series expansion
up to order $m_s^{11}$ (see also Ref.~\cite{Czakon:2008zk} for more
details).
This deep expansion in $m_s$ should be sufficient for most
of the phase space but still not enough to
cover the whole allowed kinematical region. The way to
proceed from this point is to numerically integrate
the system of differential equations.

In particular, we choose to work with
the masters in the form of Eq.~(\ref{chachamis_epReveal}), where the
$\epsilon$ dependence is explicit. We can then work with the coefficients
of the $\epsilon$ terms and accordingly have
\begin{equation}
m_s \frac{d}{dm_s} I_i(m_s,x) = 
\sum_j J^M_{i j}(m_s,x) ~I_j(m_s,x)
\label{chachamis_I_dms}
\end{equation}
and
\begin{equation}
x \frac{d}{dx} I_i(m_s,x) = 
\sum_j J^X_{i j}(m_s,x) ~I_j(m_s,x),
\label{chachamis_I_dx}
\end{equation}
where the Jacobian matrices $J^M$ and $J^X$ have rational
function elements.

By using this last system of differential equations, one can obtain a full
numerical solution to the problem. 
What we are essentially dealing now with is an initial
value problem and the main requirement is 
to have the initial conditions to proper accuracy. 
The initial conditions, namely the values of the masters
at a proper kinematical point which we call initial
point, are provided by the power series expansion.
The initial point has to be chosen somewhere
in the high energy limit region, where
$m_s$ is small and therefore, the values obtained by the power series
are very accurate.
Starting from there, one can evolve to any other point
of the phase space by numerically integrating the system of differential
equations Eqs.~(\ref{chachamis_I_dms}) and~(\ref{chachamis_I_dx}).

We parametrise with a suitable grid of points the region close
to threshold and then we calculate the masters for all points of the
grid by evolving as described previously.
Given that the master integrals have to be
very smooth (we remain above all thresholds) one can use,
after having the values for the grid points,
interpolation to get the values at any point of the region.
We use 1600 points for the grid and take as initial conditions
the values of the master integrals at the
point $m_s = 5 \times 10^{-3}$, $x = 1/4$. The
relative errors at that point
were estimated not to exceed $10^{-18}$.

The numerical integration is performed by using
one of the most advanced
software packages implementing the variable coefficient multistep method
(ODEPACK)~\cite{ODEPACK}. We use quadruple precision to maximise
accuracy. 
The values at any single grid  point can be obtained
in about 15 minutes in average (with a typical 2GHz Intel Core 2 Duo system)
after compilation with the Intel Fortran compiler.
The accuracy is around 10 digits for most of the points
of the grid.
It is also worth noting that in order to perform the numerical
integration one needs to deform the contour in the complex plane
away from the real axis. This is due to the fact
that along the real axis there are spurious
singularities. 
We use an elliptic contour and 
we achieve a  better estimate of the final global error
by calculating more than once for each point of the grid, 
using each time different eccentricities.
Grids of solutions can actually be constructed, which
will be subsequently interpolated
when implemented as part of a Monte Carlo program.

We will not present here any results as this is only a report
on work in progress. The aim here was to describe the 
numerical method, 
the results of the study will be  presented in detail in a future 
publication~\cite{chachamis_czakon}.

\subsection{Conclusions}
W pair production via quark-anti-quark-annihilation
is an important signal process in the search for
New Physics as well as the dominant irreducible
background for one of the main Higgs discovery channels:
$H \rightarrow W W \rightarrow 4$ leptons. 
Therefore, the accurate knowledge
of this process is essential for the LHC. 
After having calculated
the two-loop and the one-loop-squared virtual QCD corrections
to the W boson pair production in the limit where all
kinematical invariants are large compared to the mass of the W boson
we proceed to the next step. Namely, we use
a combination of a deep expansion in the W mass around the
high energy limit and of numerical integration of differential
equations that
allows the computation of the two-loop amplitude
with full mass dependence over the whole phase space.



%% file: andersen/andersen.tex
\subsection{INTRODUCTION}
\label{sec:introduction}
The all-order QCD radiative corrections to processes involving QCD scattering
of two partons are known in the very exclusive limit of large invariant mass
between each hard (in transverse momentum), produced
parton\cite{Lipatov:1974qm,Fadin:1975cb,Kuraev:1976ge}. The simplification of
kinematic invariants in this limit restricts the dependence of partonic cross
sections to transverse components of the produced particles only. This
simplification permits the all-order inclusive corrections to be calculated
to logarithmic accuracy through the BFKL evolution
equation\cite{Balitsky:1979ap}. The evolution variable can be taken as the
rapidity length between the two scattered partons, and the evolution is
driven by additional emission, which has a flat density in rapidity. This
picture immediately leads to the expectation of a correlation between the
length of evolution (i.e.~rapidity difference between the most forward and
most backward jet) and the average number of hard partons. Such a correlation
was quantified for both the production of dijets and $W$+dijets in
Ref.~\cite{Andersen:2001ja,Andersen:2003gs} using an exclusive, recursive
solution to the BFKL equation. Within the simple BFKL picture, this
correlation is process-independent (up to effects from the parton density
functions) and applies to all processes which at lowest order allow a colour
octet exchange between two scattered partons.

The correlation between the rapidity span of the event and the average number
of hard jets was observed also in the framework for all-order perturbative
corrections developed in
Ref.\cite{Andersen:2008ue,Andersen:2008gc,Andersen:2009nu,Andersen:2009he}. This
framework is based on approximating the all-order perturbative corrections in
a simple formalism which reproduces the all-order, exact result in the limit of large
invariant mass between all produced particles, but crucially without
succumbing to the many kinematic approximations necessary within BFKL in
order to arrive at a formalism depending on transverse scales only.

The increase in the jet count with increasing rapidity span between the
forward and the backward jet has two simple origins: 1) the opening of phase
space for radiation in-between the jets, and 2) the ability of the process to
radiate in the rapidity interval. The first point will be identical for all
dijet processes, and indeed for all Monte Carlo descriptions
thereof. However, processes differ on the second point, giving rise to
different radiation patterns for e.g.~colour singlet and colour octet
exchanges\cite{Dokshitzer:1991he}. However, in the current contribution we
will concentrate on a single process, namely $W$-production in association
with at least two jets, and compare the description of several observables
as obtained in different approaches. The observables will be described in the
next section, followed by a brief discussion of the calculational models in
Section~\ref{sec:models}, before we present the results of the comparisons in
Section~\ref{sec:results}

\subsection{OBSERVABLES}
\label{sec:observables}
The relevant rapidity observable for exploring the correlation is the
rapidity difference between the most forward and most backward perturbative
jet, which we will denote by $\Delta y$. Note that this is not necessarily
the rapidity difference between the two hardest (in transverse momentum) jets
in the event.

In this contribution we studied the following observables, which all test the
description of the expected increase in hard radiation with increasing
$\Delta y$:
\begin{enumerate}
\item The average number of hard jets versus $\Delta{}y$.
\item $\frac{1}{\sigma_{W+n\mathrm{jets}}} \frac{d\sigma_{W+n\mathrm{jets}}}{d\Delta{}y}$ for
  the inclusive production of W plus two, three or four jets.
\item The exclusive rates in bins of increasing rapidity span.
\end{enumerate}

\subsection{Calculational Models}
\label{sec:models}
In this work we compared the modelling of $W+$dijets obtained in a variety of
generators, each based on different underlying perturbative models.  Four
main models were used: pure parton shower (PS) calculations, matched PS
calculations, NLO W+dijet (as implemented in MCFM\cite{Campbell:2003hd}), and
the scheme for resumming hard, perturbative corrections discussed in
Ref.\cite{Andersen:2008ue,Andersen:2008gc,Andersen:2009nu,Andersen:2009he},
which we will refer to as \emph{High Energy Jets} (HEJ).
While we refer the reader to the literature for the description of the two
latter approaches, we will here briefly describe the generation of events
used in the PS and PS+matched calculation.
\begin{itemize}
\item Shower Monte Carlos. This category is represented in this work by Pythia 6.421~\cite{Sjostrand:2006za}, Herwig 6.510~\cite{Corcella:2002jc} and Pythia 8.130~\cite{Sjostrand:2007gs}.
All these programs can produce W + 1 jet events at LO. They don't have a matrix element for W+dijet production, so the second jet is produced by the parton shower.
Even though these programs are not meant at describing multi jet final state we believe it is useful to check them as well in order to assess the differences with respect to more sophisticated multi jet calculations.
Also, we found some not negligible differences among them.
The shower formalism used in the three programs is different. Pythia 6.421 has a virtuality ordered shower, Herwig has an angular ordered shower, Pythia 8.130 has a transverse momentum shower.
\item Matched calculations. This category is represented by Alpgen~\cite{Mangano:2002ea}. Parton level events produced by Alpgen were showered and matched using Herwig 6.510.
We produced Alpgen samples for W plus 2, 3, 4, 5 partons, with a minimum $p_{T}$ for partons of 20~GeV.
Each sample was processed through Herwig shower, filtering events according to the MLM matching prescription.
All samples except for the highest multiplicity one were matched exclusively, while the highest multiplicity one was matched inclusively.
Plots resulting from the analysis of each sample were normalized to the cross section after the matching and then summed up.
\end{itemize}

\subsection{Results}

Events were produced for a 10~TeV $pp$-colllider and selected according to
the following cuts on W$^{+/-}$ decay products in the $(e,\nu_e)$-channel: charged lepton
$p_{T}$ higher than 20~GeV, charged lepton rapidity between -2.5 and 2.5, missing
transverse energy higher than 20~GeV. In the present study, jets were
reconstructed with the $k_{\perp}$ algorithm using a pseudo-radius parameter
of 0.7, minimum $p_{T}$ of 40~GeV and rapidity between -4.5 and 4.5.

\label{sec:results}
\begin{figure}
\begin{center}
\includegraphics[width=0.5\textwidth]{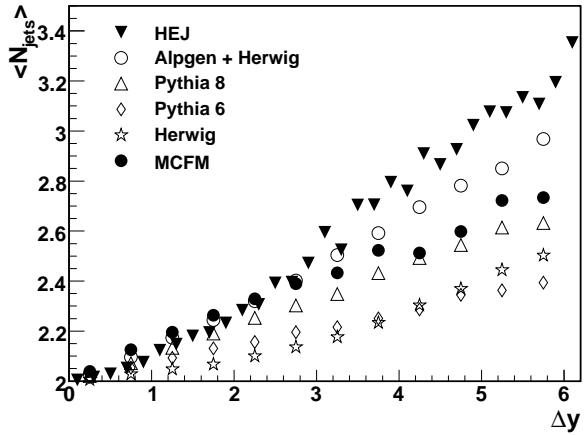}
 \caption{Average number of jets versus  the rapidity difference between the forward and the backward jets.
}
\label{fig:WJetComp_avnjets}
\end{center}
\end{figure}

The average number of jets as a function of $\Delta{}y$ is shown in
Fig.~\ref{fig:WJetComp_avnjets} for the 6 models considered.  All the models
show a strong correlation between the average number of hard jets and $\Delta
y$. We observe that the prediction for the level of hard radiation with
increasing $\Delta y$ is smallest for Pythia 6 and Herwig, and highest with
HEJ and the PS-matched calculation with Alpgen+HERWIG. The predictions
obtained using MCFM or Pythia 8 fall in-between, with MCFM agreeing well with
either HEJ or Alpgen+HERWIG out to around 3-4 units of
rapidity. Obviously, the maximum number of jets produced in the NLO
calculation of W+dijets implemented in MCFM is 3, and as we will see later,
the 4-jet rate peaks at around 3-4 units of rapidity. It is therefore perhaps
not surprising that the NLO calculation ``runs out of steam'' in increasing
the average jet count at 3-4 units of rapidity. This number coincides well
with the general observation of e.g.~Ref.\cite{Andersen:2008gc} that the High
Energy resummation produces one hard (40~GeV) jet every (roughly) two units of
rapidity span. Therefore one could expect the jet rate predictions obtained
in the resummation and the NLO calculation to agree up to 3-4 units of
rapidity, where-after the resummation will start producing more than the
maximum number of jets allowed in the NLO calculation.

The prediction for the average number of jets vs.~the rapidity span is
clearly sensitive to scale choices. For MCFM, we used $\mu_f\!=\!\mu_r\!=\!M_W$, in
HEJ we used $\mu_f\!=\!\mu_r\!=\!40$~GeV, while the parton shower predictions used
their inherent choices. A systematic study of the uncertainties is clearly
desirable.

The average number of jets obtained using MCFM is just two plus the ratio of
the (inclusive) 3-jet rate over the inclusive 2-jet rate. This ratio was
studied in Ref.\cite{Berger:2009dq}, but with both the 3-jet rate and the
2-jet rate calculated at both LO and NLO. Both show the same strong
correlation.

\begin{figure}
\begin{tabular}{ccc}
\includegraphics[width=0.31\textwidth]{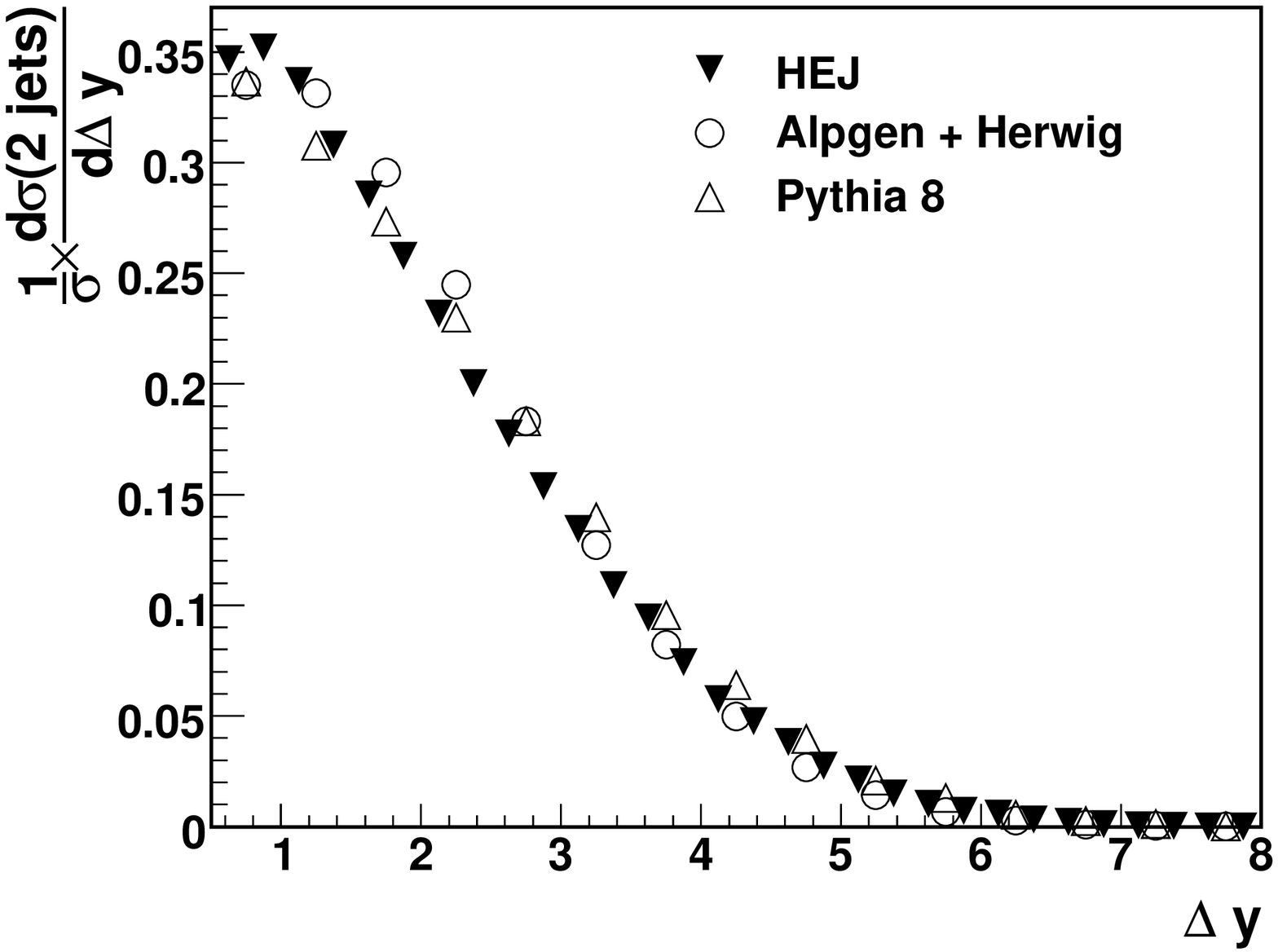} &
\includegraphics[width=0.31\textwidth]{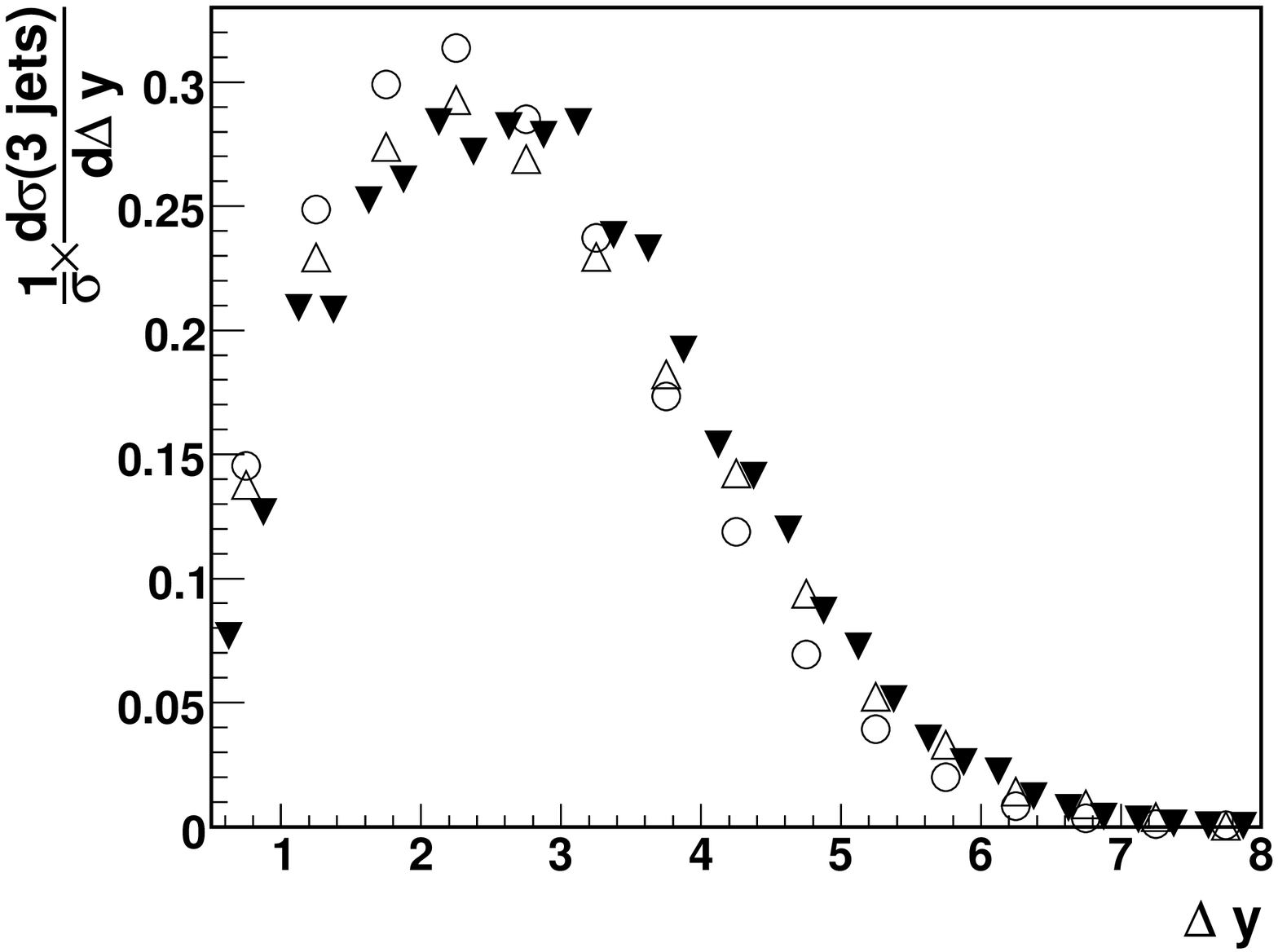} &
\includegraphics[width=0.31\textwidth]{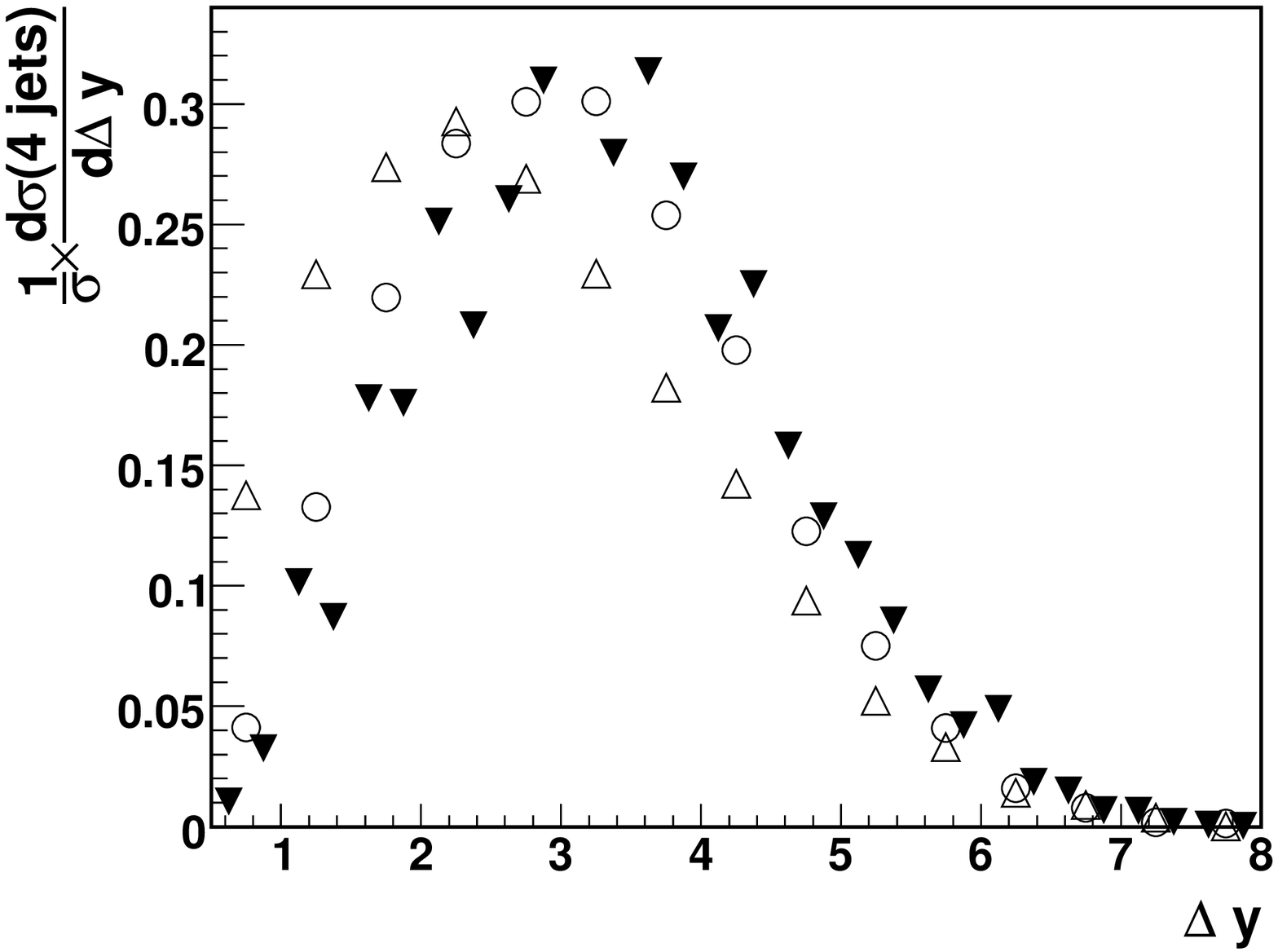} \\
(a) & (b) & (c)
\end{tabular} 
\caption{Spectrum of the rapidity difference between the forward and the backward jets for (a) at least two, (b) at least three, (c) at least four jets. Results for three generators are reported.}
\label{fig:WJetComp_xsec}
\end{figure}

The normalized differential cross-section $\frac{1}{\sigma_{W+n\mathrm{jets}}}
\frac{d\sigma_{W+n\mathrm{jets}}}{d\Delta{}y}$, for inclusive production of
two, three and four jets is shown in Fig.~\ref{fig:WJetComp_xsec} for Pythia
8, Alpgen+HERWIG and HEJ. We observe that for increasing jet count, the
cross-section peaks at an increasingly larger value of $\Delta{}y$ (about 1
unit of rapidity for each extra jet count). This is because of the opening of
phase space, and is observed also in the pure tree-level calculations.  It is
rather surprising how for two and three jets all models shows a similar
dependence on $\Delta{}y$. However, for four or more jets clear differences
appear. In fact, the spectrum for 4 jets produced with Pythia 8 peaks at the
same value of $\Delta y$ as for 3 jets produced with Pythia 8.
\begin{figure}
\begin{center}
\includegraphics[width=0.7\textwidth]{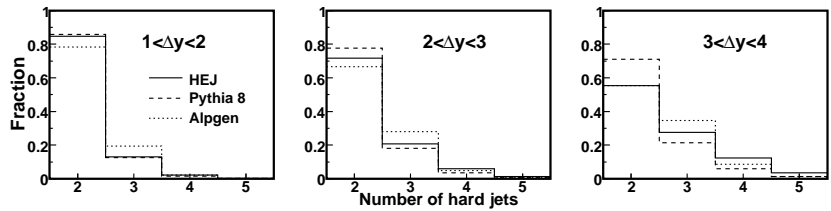}
 \caption{Relative exclusive jet rates for three rapidity intervals for different generators.
}
\label{fig:WJetComp_exclrates}
\end{center}
\end{figure}

To further investigate the radiation pattern as a funcion of $\Delta{}y$ we
concentrated on three bins of $\Delta{}y$ and compared the distribution of the
number of jets predicted by the different models, as shown in
Fig.~\ref{fig:WJetComp_exclrates}.  It appears that the matched parton shower
calculation predicts a larger number of three jets events compared to both
the parton shower and the high energy resummation for all rapidity spans, and
(unsurprisingly) the pure shower (represented by Pythia 8) has the largest
relative weight on exclusive two-jet events. For rapidity spans of more than
three, the high energy resummation predicts a significantly larger relative
weight on events with four or more jets than does Alpgen+HERWIG or Pythia 8.

\subsection{Conclusions}
\label{sec:conclusions}
In this contribution we have initiated a comparison of the description of the
multi-jet configurations in $W+$jets as obtained using various
approximations. Specifically, we have explained why there is a strong
correlation between the average number of jets and the rapidity difference
between the most forward and most backward jet, and why one can expect
differences in the description of this quantity within the generators
frequently used for studying LHC physics. We note that the observed
differences obtained in the predictions are stable against variations in both
the jet energy scale and the parameters used in defining the
jets. Specifically, we tried all four combinations with a transverse momentum
cut of 30~GeV or 40~GeV, and a parameter of D=0.4 or D=0.7 for the
$k_\perp$-jet algorithm. We estimate that the differences observed in the
predictions are sufficient that 1fb$^{-1}$ of $\sqrt{s}=7$~TeV data from the
first year of LHC running can discriminate between the models.

The universal behaviour observed in all the models of a strong correlation
between the rapidity span and the jet activity is (within the framework of
Ref.\cite{Lipatov:1974qm,Fadin:1975cb,Kuraev:1976ge,Balitsky:1979ap})
universal for all dijet processes. This means that information about jet
vetos in e.g.~Higgs boson production in association with dijets can be
extracted using e.g.~W+dijets as studied in this contribution.

%

%% file: dissertori/dissertori.tex





\subsection{INTRODUCTION}
Discovering the Higgs boson is one of the major goals of the hadron colliders Tevatron and LHC. It has been shown that if the Higgs mass lies in the region $m_\mathrm{H}\sim2\times m_\mathrm{W}$ the Higgs decay process $\mathrm{H}\to\mathrm{WW}$ serves as the most promising discovery channel. At both the colliders under consideration the main Higgs production process is gluon-fusion. There exists extensive literature about this process and its sensitivity to higher order QCD corrections~\cite{Harlander:2002wh,Anastasiou:2002yz,Ravindran:2003um}. In the following we investigate the impact of these higher order corrections in the specific case when the Higgs bosons decays into a pair of W bosons, which further decay into leptons.

\subsection{CROSS-SECTIONS AT THE LHC}
In this section we present the numbers for the Standard Model (SM) $\mathrm{H}\to\mathrm{WW}\to\ell\nu\ell\nu$ cross-section
via gluon-fusion in proton-proton collisions at a center of mass
energy of 14 TeV. As an example we choose a Higgs mass of
$m_\mathrm{H}=165\,\mathrm{GeV}$, where the decay into a pair of W-bosons
is dominating. It has been shown that for a  Higgs mass around
that value this is the most promising discovery channel for the SM
Higgs boson at the LHC experiments.

\subsubsection{INCLUSIVE CROSS-SECTION}
We compute the cross-sections for a center of mass energy of
$E_\mathrm{CM}=14\,\mathrm{TeV}$ and a Higgs mass of
$m_\mathrm{H}=165\,\mathrm{GeV}$ using the program {\tt FEHiP}~\cite{Anastasiou:2005qj}.
The renormalization and factorization scales are varied
within
$\mu_\mathrm{R}=\mu_\mathrm{F}\in[m_\mathrm{H}/2,2\,m_\mathrm{H}]$ to
estimate the level of convergence of the perturbative calculation. The
numbers, together with the $K$-factors\footnote{~The $K$-factors are
  defined as $K^\mathrm{(N)NLO}(\mu)=
  \sigma^\mathrm{(N)NLO}(\mu)/\sigma^\mathrm{LO}(\mu)$.} are shown in Tab.~\ref{tab:LHCinccross}.

\begin{table}[h]
  \begin{center}    
    \begin{tabular}{|l||c|c|c||c|c|}
      \hline
      $\sigma_\mathrm{inc}\;\;[\mathrm{fb}]$ & LO & NLO & NNLO & $K^\mathrm{NLO}$ & $K^\mathrm{NNLO}$\\\hline\hline
      $\mu=m_\mathrm{H}/2$ & $152.63\pm0.06$ & $270.61\pm0.25$ & $301.23\pm1.19$ & $1.773\pm0.001$ & $1.974\pm0.008$ \\
      $\mu=2\,m_\mathrm{H}$ & $103.89\pm0.04$ & $199.76\pm0.17$ & $255.06\pm0.81$ & $1.923\pm0.002$ & $2.455\pm0.008$\\
      \hline
    \end{tabular}
  \end{center}
  \caption{Inclusive cross sections for
    $m_\mathrm{H}=165\,\mathrm{GeV}$ and $E_\mathrm{CM}=14\,\mathrm{TeV}$, at various 
    orders in perturbation theory and for different
    scale choices.     \label{tab:LHCinccross}}
\end{table}

As expected, the impact of the higher order corrections is rather
large. Depending on the scale choice the effects are between $77\%$ and $92\%$ at NLO
and between $97\%$ and $145\%$ at NNLO. On the other hand the
uncertainty on the cross-sections under variation of the scales are
reduced, from $47\%$ at LO to $35\%$ at NLO and $18\%$ at NNLO. This
indicates that the perturbative series has not sufficiently converged
at NLO and the NNLO corrections have to be taken into account in order
to get a reliable cross-section prediction.

\subsubsection{IMPACT OF A JET-VETO}
\begin{figure}
\begin{center}
\includegraphics[width=0.48\textwidth]{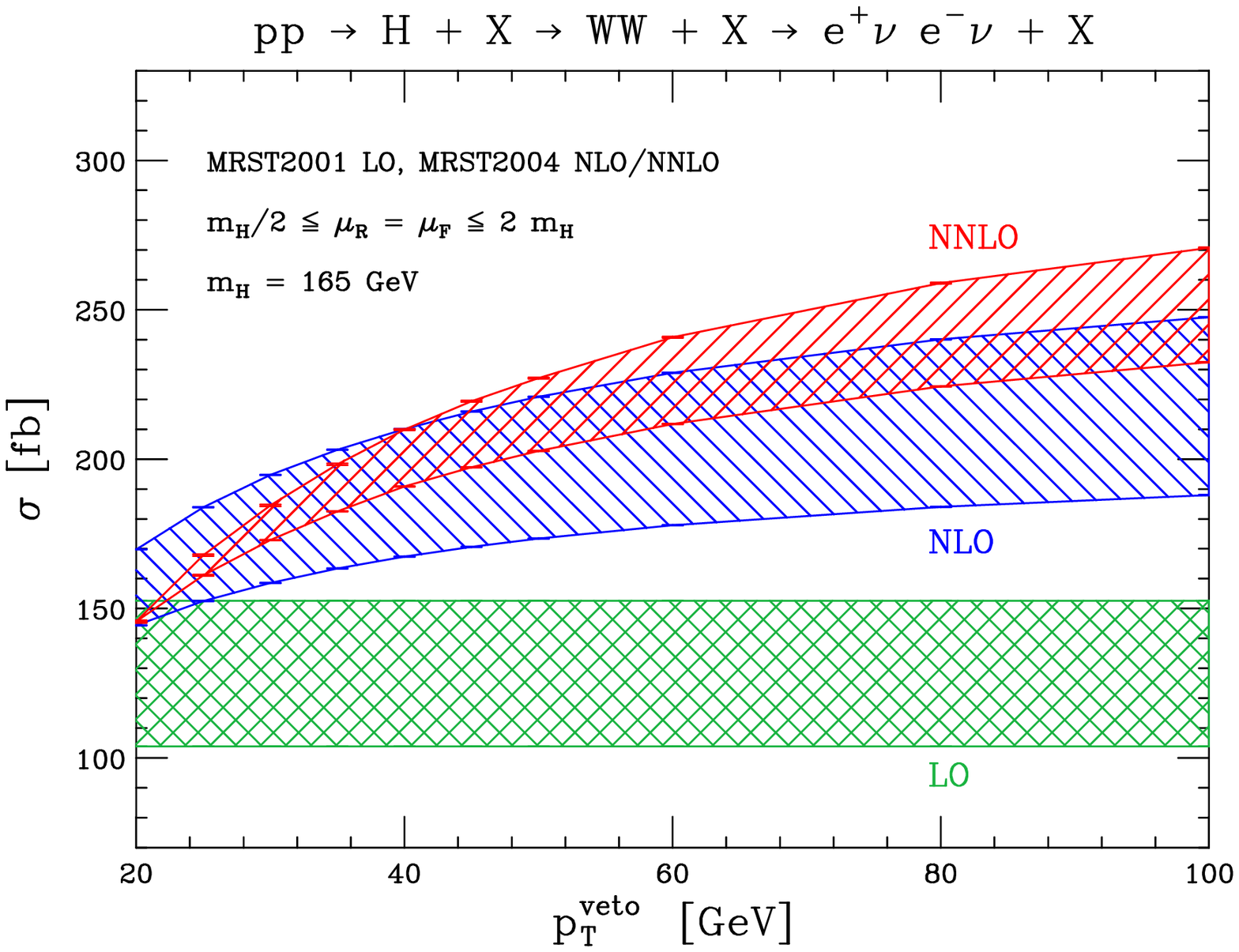}
\includegraphics[width=0.48\textwidth]{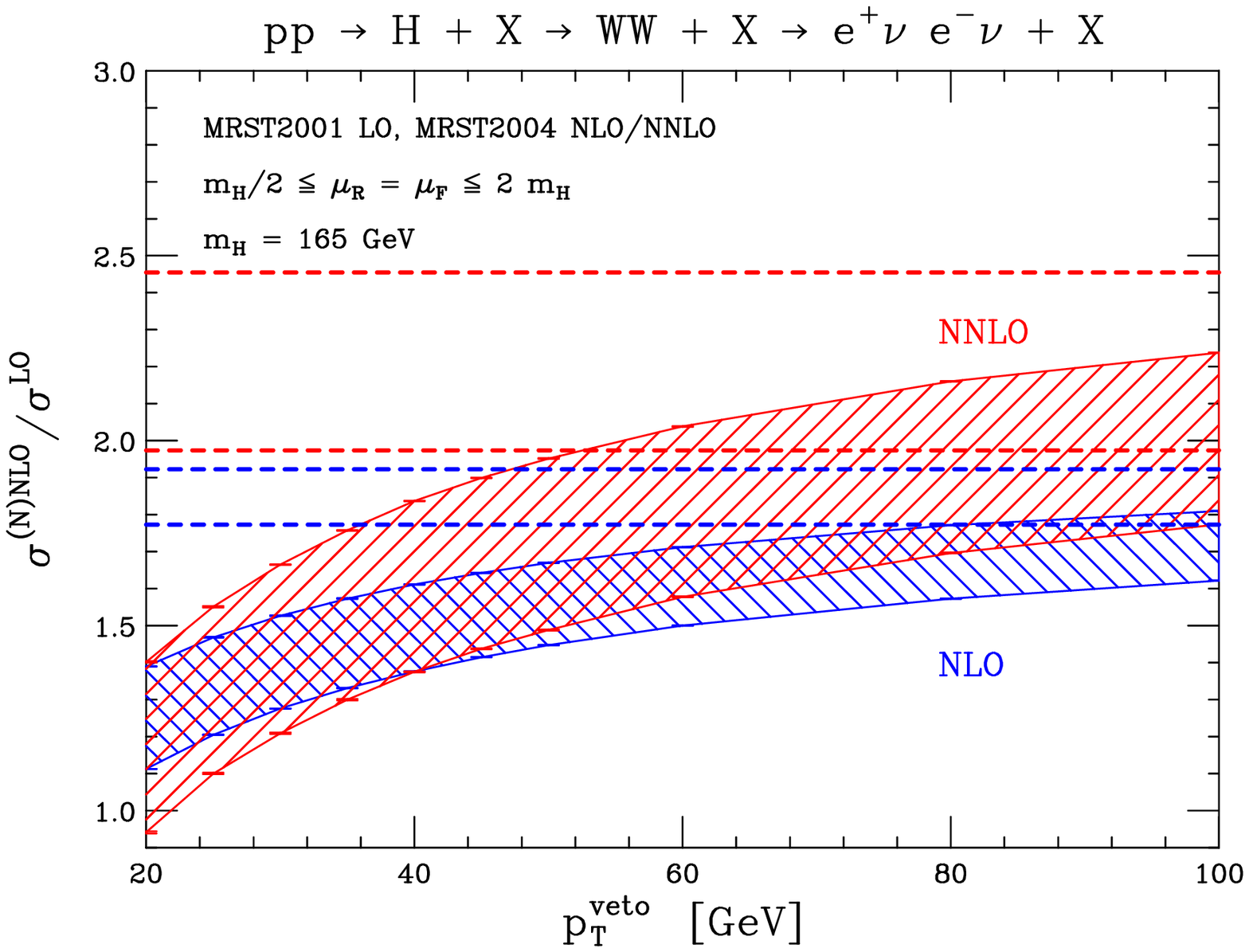}
\caption{\label{fig:jetveto}
  Cross-section (left) and $K-$factor (right) as a function of the
  jet-veto cut-off value $p_\mathrm{T}^\mathrm{veto}$ for a Higgs mass
  of $m_\mathrm{H}=165\,\mathrm{GeV}$.
}
\end{center}
\end{figure}

Here we demonstrate that applying experimental cuts can have a strong
impact on the $K-$factors in
Tab.~\ref{tab:LHCinccross}. As an example we apply a jet-veto. Such a jet-veto serves to reduce the background arising from top-pair production.
Jets are constructed by clustering partons into a jet when they
lie within a cone of radius $R=0.4$ and in the pseudo-rapidity range $|\eta|<2.5$.
The jet-veto procedure consists of vetoing any event that
contains at least one jet with transverse momentum
$p_\mathrm{T}^\mathrm{jet}$ larger than some cut-off value $p_\mathrm{T}^\mathrm{veto}$.
We show the cross-section, as well as the $K$-factors as a function
of this cut-off value in Fig.~\ref{fig:jetveto}.

It can be seen how that the (N)NLO cross-sections, and thus the
corresponding $K$-factors decrease when lowering the cut-off value
$p_\mathrm{T}^\mathrm{veto}$, while the LO cross-section stays
constant. Obviously this can be understood from the fact that at LO
there are no partons present in the final state, thus the jet-veto has
no impact on the cross-section. As a consequence, $K$-factors close to unity are found, 
if the jet-veto restricts the phase space strongly enough.
In addition, the scale uncertainty of the NNLO results decreases with lower cut-off $p_\mathrm{T}^\mathrm{veto}$. 
This can be understood from the fact that the jet-veto eliminates phase space regions were the one(two)-real-radiation diagrams are dominant. 
These diagrams however are especially sensitive to the variation of the scales.
A more detailed study can be found in
the dedicated papers~\cite{Anastasiou:2007mz,Anastasiou:2009bt}.

\subsection{CROSS-SECTIONS AT THE TEVATRON}
In this section we present numbers computed for proton-antiproton
collisions at a center of mass energy of $1.96\,\mathrm{TeV}$, as
currently produced at the Tevatron collider at Fermilab. Since the performance
of simple cut-based analyses is not sufficient to observe or exclude a
SM Higgs signal in those experiments, more involved, multi-variate
techniques have to be applied. Typical examples for such techniques
are Artificial Neural Networks (ANN).

Such techniques have been used in the recent past in order to exclude
a SM Higgs boson in the mass range
$m_\mathrm{H}\in[160\,\mathrm{GeV},170\,\mathrm{GeV}]$ at $95\%$
confidence level in a combined analysis of the two experiments CDF and D0~\cite{Bernardi:2008ee,TeVHiggs:2009pt}.
A substantial part of the exclusion power of these analyses comes 
from the decay mode discussed
here ($\mathrm{H}\to\mathrm{WW}\to\ell\nu\ell\nu$). Both experiments
present numbers for this mode, which allow the exclusion of a Higgs cross-section of about
$1.5-1.7$ times the SM cross-section at $95\%$ CL~\cite{Aaltonen:2008ec,CDFnote,D0note}.

In what follows we present the numbers for the inclusive
cross-section with $m_\mathrm{H}=160\,\mathrm{GeV}$, the cross-section
after a typical set of selection cuts, and investigate the impact
of the effects of the higher order corrections on the efficiencies of
such cuts and on typical input variables to ANN analyses, as used by
the Tevatron experiments. Further we discuss uncertainties on signal
yields when braking up the sample into jet multiplicity bins and
finally provide the output of an example ANN. For a more detailed
discussion we refer to the dedicated paper~\cite{Anastasiou:2009bt}.

\subsubsection{INCLUSIVE CROSS-SECTIONS AND PRE-SELECTION}
First we present the numbers for the inclusive
cross-section\footnote{~All numbers correspond to the cross-sections
  for one final state lepton
  combination,~e.g.~$\mu^+\nu\mu^-\bar{\nu}$.}, together
with the corresponding $K$-factors. The numbers are again computed using the
MRST2004 PDF sets at (N)NLO and MRST2001 PDF set at LO. The
uncertainty due to the choice of the renormalization and
factorization scales ($\mu:=\mu_\mathrm{R}=\mu_\mathrm{F}$) are
estimated by varying them simultaneously in the range
$\mu\in[m_\mathrm{H}/2,2\,m_\mathrm{H}]$ around the central value
$\mu=m_\mathrm{H}$. The quoted errors are the residual numerical
uncertainty from the MC integration.

\begin{table}[h]
  \begin{center}    
    \begin{tabular}{|l||c|c|c||c|c|}
      \hline
      $\sigma_\mathrm{inc}\;\;[\mathrm{fb}]$ & LO & NLO & NNLO & $K^\mathrm{NLO}$ & $K^\mathrm{NNLO}$\\\hline\hline
      $\mu=m_\mathrm{H}/2$ & $1.998\pm0.003$ & $4.288\pm0.004$ & $5.252\pm0.016$ & $2.149\pm0.008$ & $2.629\pm0.009$ \\
      $\mu=m_\mathrm{H}$ & $1.398\pm0.001$ & $3.366\pm0.003$ & $4.630\pm0.010$ & $2.412\pm0.002$ & $3.312\pm0.008$\\
      $\mu=2\,m_\mathrm{H}$ & $1.004\pm0.001$ & $2.661\pm0.002$ & $4.012\pm0.007$ & $2.651\pm0.008$ & $3.996\pm0.008$\\
      \hline
    \end{tabular}
  \end{center}
  \caption{Inclusive cross sections for
    $m_\mathrm{H}=160\,\mathrm{GeV}$ in $\mathrm{p}\bar{\mathrm{p}}$ collisions at $E_\mathrm{CM}=1.96\,\mathrm{TeV}$, at various 
    orders in perturbation theory and for different
    scale choices.     \label{tab:inccross}}
\end{table}

While the uncertainty from the scale variation is reduced when going
from LO $\left(^{+43\%}_{-28\%}\right)$ to NNLO $\left(^{+13\%}_{-13\%}\right)$, the effect
of the higher order corrections is rather large 
($\sigma^\mathrm{NNLO}\sim 1.4\times\sigma^\mathrm{NLO}\sim3.3\times\sigma^\mathrm{LO}$).

We now apply a typical set of pre-selection cuts. Such cuts are needed due to the detector geometry, like the limited coverage in
$\eta$, and in order to remove the first substantial part of background
events. The cuts we apply here are inspired by the ones applied in the
CDF analysis~\cite{Aaltonen:2008ec}, but are not identical to them.

\begin{enumerate}
\item 
Lepton selection: in the CDF experiment, the experimental acceptances for electrons and
muons are different. For this study we concentrate on the muon case only. 
First, one of the final-state leptons has to trigger the event read-out. 
This `trigger lepton' must have a transverse momentum $p_\mathrm{T}>20\,\mathrm{GeV}$.
The pseudo-rapidity coverage of the detector for measuring this trigger muon is $|\eta|<0.8$. 
In order to  pass a further lepton selection,  a second muon must be found with
$p_\mathrm{T}>10\,\mathrm{GeV}$ and $|\eta|<1.1$.

It is worth noting that the differences  in the muon and 
electron cuts are rather geometric, and should not alter  the convergence 
pattern of the  perturbative corrections.   \\[0.05cm]

\begin{enumerate}
\item Two opposite-sign leptons have to be found, fulfilling the requirements discussed above.
\item Both leptons have to be isolated, i.e. the additional transverse energy in a cone with radius $R=0.4$ 
         around the lepton has to be smaller than 10\,\% of the lepton transverse momentum.
\item In order to reduce the background from b~resonances, 
the invariant mass of the lepton pair has to be $m_{\ell\ell}>16\,\mathrm{GeV}$.\\[0.05cm]
\end{enumerate}

\item   The missing transverse energy (MET) is defined as the vectorial sum of the transverse momenta of the two neutrinos.
Then we can define the variable $\mathrm{MET}^*$ as 
\begin{equation}
\mathrm{MET}^*=\left\{\begin{array}{l} \mathrm{MET}\;\;\;\;\;\;\;\;\;\;\;\;,\;\phi\geq\pi/2  \\ \mathrm{MET}\times\sin{\phi}\;,\;\phi<\pi/2 \end{array}\right.,
\end{equation}
where $\phi$ is the angle in the transverse plane between MET and the nearest charged lepton or jet. 
We require $\mathrm{MET}^*>25$ GeV, which suppresses the background from Drell-Yan lepton pairs 
and removes contributions from mismeasured leptons or jets.
\item In order to suppress the $t{\bar t}$ background, we apply a veto on the number of jets in the event. 
Jets are found  using the  $k_\mathrm{T}$-algorithm with parameter $R=0.4$.  A jet must have
$p_\mathrm{T} >15\,\mathrm{GeV}$ and $|\eta|<3.0$.  Events are only accepted if there is no more than one such jet.
\end{enumerate}

\begin{table}[h]
  \begin{center}    
    \begin{tabular}{|l||c|c|c||c|c|}
      \hline
      $\sigma_\mathrm{acc}\;\;[\mathrm{fb}]$ & LO & NLO & NNLO & $K^\mathrm{NLO}$ & $K^\mathrm{NNLO}$\\\hline\hline
      $\mu=m_\mathrm{H}/2$ & $0.750\pm0.001$ & $1.410\pm0.003$ & $1.459\pm0.003$ & $1.880\pm0.005$ & $1.915\pm0.025$\\
      $\mu=m_\mathrm{H}$ & $0.525\pm0.001$ & $1.129\pm0.003$ & $1.383\pm0.004$ & $2.150\pm0.007$ & $2.594\pm0.052$\\
      $\mu=2\,m_\mathrm{H}$ & $0.379\pm0.001$ & $0.903\pm0.002$ & $1.242\pm0.001$& $2.383\pm0.008$ & $3.261\pm0.048$\\
      \hline
    \end{tabular}
  \end{center}
  \caption{
\label{tab:accxsec}
Accepted cross sections and $K$-factors after the application of all the 
selection cuts for $m_\mathrm{H}=160\,\mathrm{GeV}$ in $\mathrm{p}\bar{\mathrm{p}}$ collisions at $E_\mathrm{CM}=1.96\,\mathrm{TeV}$.}
\end{table}

The numbers and the $K$-factors after applying these cuts are shown in Tab.~\ref{tab:accxsec}.
The  impact of 
QCD radiative corrections is significantly reduced  when selection
cuts are applied.
For $\mu=m_\mathrm{H}$ the NLO and NNLO $K$-factors are reduced by $11\%$ and $22\%$, respectively. 
As a consequence, also the acceptance is reduced, since it is defined as the ratio
of the cross-section after cuts over the inclusive cross section.
At LO about $\sim 37.5\%$ of the  events are accepted.  At NLO, the efficiency drops to $33\%-34\%$ and at  NNLO to 
$28\% - 31\%$, depending on the  scale choice. 

\begin{figure}
\begin{center}
\includegraphics[width=0.48\textwidth]{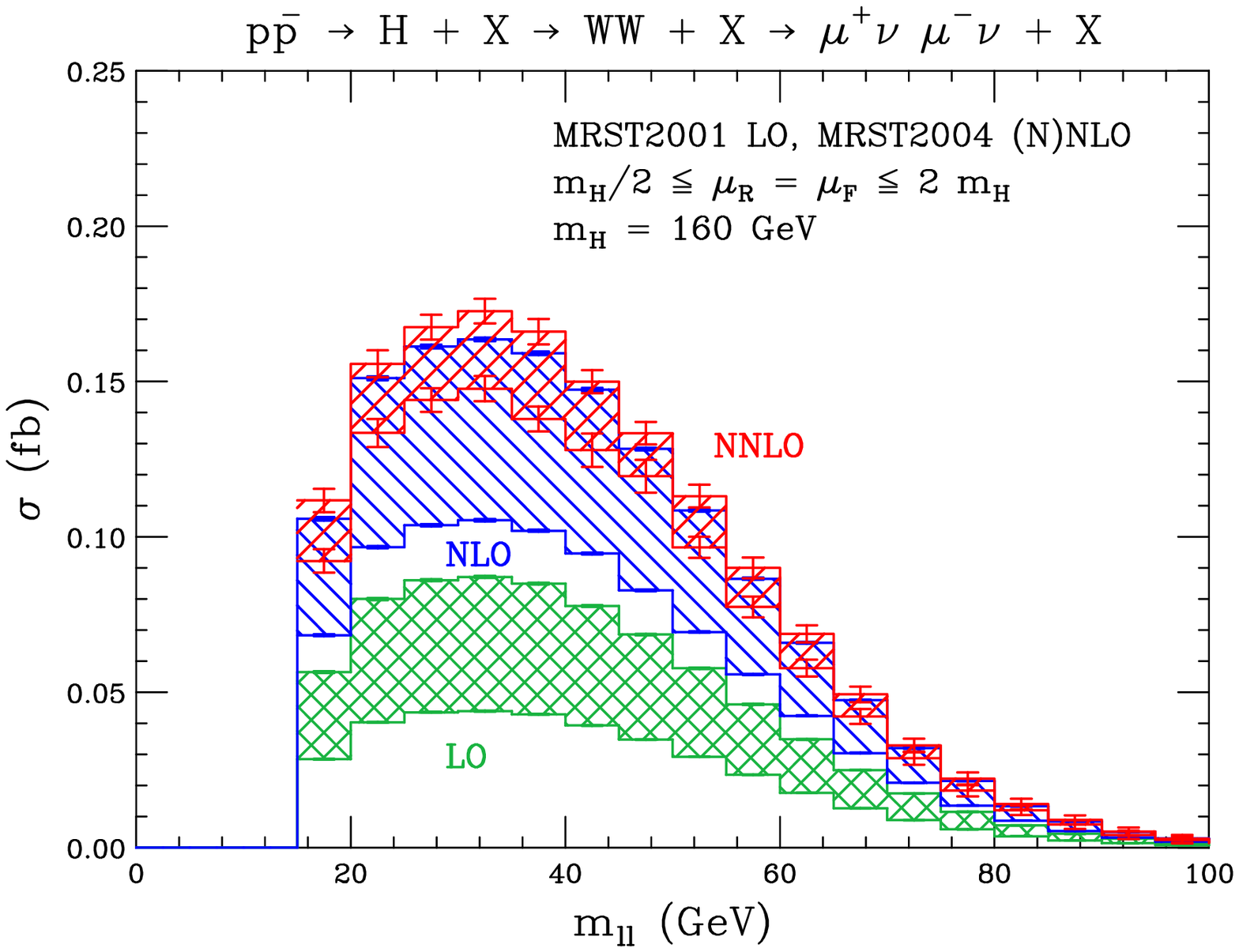}
\includegraphics[width=0.48\textwidth]{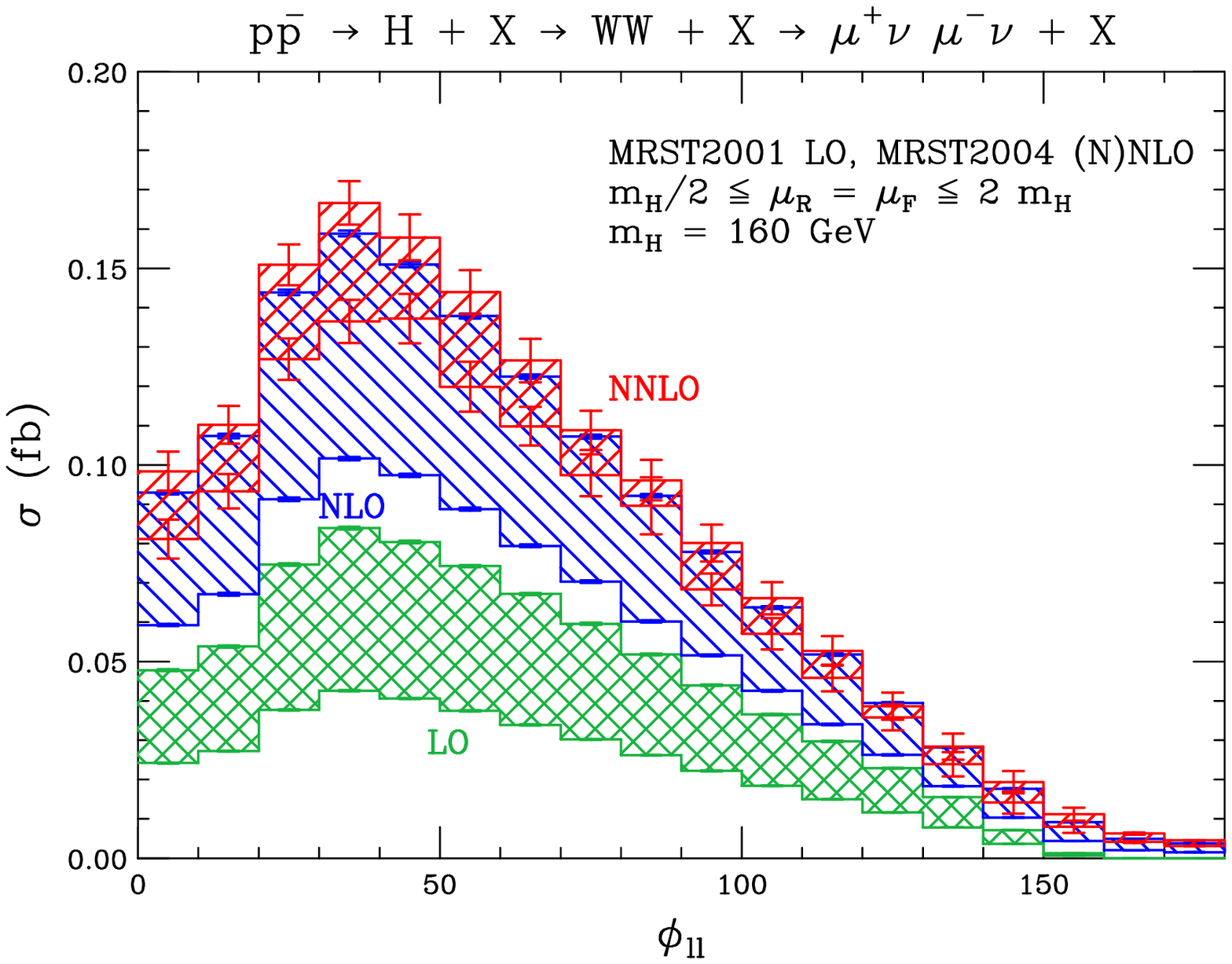}
\caption{\label{fig:fixord_distr}
  Kinematic distributions obtained at LO, NLO and NNLO in perturbative
  QCD. Shown are 
  the invariant mass of the two charged leptons ($m_{\ell\ell}$) 
  and the azimuthal separation of 
  the two charged leptons in the transverse plane ($\phi_{\ell\ell}$).
}
\end{center}
\end{figure}

In Fig.~\ref{fig:fixord_distr} we show kinematic distributions at different orders in perturbation theory after
applying the cuts described above. Distributions like this
are typical input variables for multi-variate analyses such as artificial neural networks (ANN).
On the left the invariant mass of the charged lepton pair is shown
($m_{\ell\ell}$), on the right the azimuthal angle between these
leptons ($\phi_{\ell\ell}$).
The uncertainty bands show again the variation of the distributions
under the variation of the ren.\ and fac.\ scales $\mu$. The
plots show a stable behavior with respect to both, the scale variation
and the addition of higher order corrections. This is no
surprise, since the leptonic final state variables are not expected to
be very sensitive to the higher order QCD corrections. This picture
changes when more involved, hadronic variables are under
consideration. When such variables (like the number of jets) are used,
e.g. as input to a ANN, care must be taken that these variables, and
especially the uncertainties on them are well understood and under
control. This is discussed in some detail in the next section.

\subsubsection{JET-MULTIPLICITY}

In an experimental analysis it might seem beneficial to divide the
experimental candidate data into different sub-samples,
distinguishable by the number of jets in the events. A possible
division is the one into the mutually exclusive '0-jet', '1-jet' and
'$\geq2$-jets' sub-samples. Further, for each sub-sample, a dedicated
analysis can be performed, benefiting from the different kinematic
behavior, not only of the signal, but especially for the background
events (as an example, the '0-jet' sub-sample suffers much less from
contamination by $\mathrm{t}\bar{\mathrm{t}}$ events). Such a strategy
is e.g. pursued by the CDF analysis~\cite{CDFnote}. It has to be
pointed out, that, if a jet is  required in all events, the ${\cal O}(\alpha_s^4)$ calculation includes  
matrix elements  through NLO only.  If two jets are required, only LO matrix-elements are taken into account.
More importantly, we find it inconsistent to use the theoretical uncertainty from the inclusive NNLO gluon fusion 
cross-section  as  the uncertainty of the samples with defined jet
multiplicities other than zero.

This is illustrated in the example below. We divide the signal
cross-section into the three jet-multiplicity bins described above,
where jets are defined using a $k_\mathrm{T}$-algorithm, and a jet is
identified as such when it has a minimal $p_\mathrm{T}$ of 15 GeV and
lies in the detector region $|\eta|<2$. We now compute the
cross-sections varying the ren.\ and fac.\ scales in the usual interval,
using either NNLO, NLO or LO parton density functions and $\alpha_s$
evolution from the MSTW2008 fit. The resulting numbers are shown in
Tab.~\ref{tab:jetmulti1}.

\begin{table}[h]
  \begin{center}    
    \begin{tabular}{|l||c|c|c|}
      \hline
      $\sigma\;\;[\mathrm{fb}]$ & LO (pdfs, $\alpha_s$) & NLO
      (pdfs, $\alpha_s$)& NNLO (pdfs, $\alpha_s$) \\ \hline\hline
      0-jet & $3.452^{+7\%}_{-10\%}$ & $2.883^{+4\%}_{-9\%}$ &
      $2.707^{+5\%}_{-9\%}$ \\ 
      1-jet & $1.752^{+30\%}_{-26\%}$ & $1.280^{+24\%}_{-23\%}$ &
      $1.165^{+24\%}_{-22\%}$ \\
      $\geq2$-jets & $0.336^{+91\%}_{-44\%}$ & $0.221^{+81\%}_{-42\%}$
      & $0.196^{+78\%}_{-41\%}$ \\
      \hline
    \end{tabular}
  \end{center}
  \caption{Inclusive cross-sections in the different jet-multiplicity
  bins. \label{tab:jetmulti1}}
\end{table}

From the numbers in Tab.~\ref{tab:jetmulti1} it can be seen that about
66.5\% of the events contain no jets, 28.6\% contain 1 jet and only
4.9\% contain at least 2 jets. From the scale uncertainties
listed in the table, the total scale uncertainty on the expected
signal yield can be reconstructed as
\begin{equation}\label{eq:eqone}
\frac{\Delta
  N_\mathrm{inc}}{N_\mathrm{inc}}=66.5\%\cdot\left(^{+5\%}_{-9\%}\right)+28.6\%\cdot\left(^{+24\%}_{-22\%}\right)+4.9\%\cdot\left(^{+78\%}_{-41\%}\right)=\left(^{+14.0\%}_{-14.3\%}\right),
\end{equation}
which agrees well with the $\sim\pm13\%$ uncertainty retrieved in the
previous section\footnote{~The residual difference can be explained by
  the different PDFs.}. The application of different selection cuts in
the three jet-multiplicity bins leads to a theoretical error estimate
for the total signal yield which is different from the value for the
inclusive NNLO cross-section. To illustrate this, we assume that after
applying further selection cuts, 60\% of the events belong to the
'0-jet' bin, 29\% to the '1-jet' bin and 11\% to the '$\geq2$-jet'
bin\footnote{~These numbers are taken from Tables 1-3
  in~\cite{CDFnote}.}. Recomputing the uncertainty on the total signal
yield along the lines of eq.~\ref{eq:eqone} results in
\begin{equation}\label{eq:eqtwo}
\frac{\Delta
  N_\mathrm{inc}}{N_\mathrm{inc}}=66\%\cdot\left(^{+5\%}_{-9\%}\right)+29\%\cdot\left(^{+24\%}_{-22\%}\right)+11\%\cdot\left(^{+78\%}_{-41\%}\right)=\left(^{+18.5\%}_{-16.3\%}\right),
\end{equation}
which is substantially larger then the errors on the inclusive
cross-sections. Moreover, a more consistent approach would be to
estimate the number of '1-jet' ('$\geq2$-jets') events using NLO (LO)
PDFs and $\alpha_s$ evolution correspondingly. This would lead to
\begin{equation}\label{eq:eqthree}
\frac{\Delta
  N_\mathrm{inc}}{N_\mathrm{inc}}=66\%\cdot\left(^{+5\%}_{-9\%}\right)+29\%\cdot\left(^{+24\%}_{-23\%}\right)+11\%\cdot\left(^{+91\%}_{-44\%}\right)=\left(^{+20.0\%}_{-16.9\%}\right).
\end{equation}

This demonstrates, that the theoretical uncertainty on the number of
signal events for different jet-multiplicities should not be estimated
collectively from the scale variation of the total cross-section. 

\subsubsection{ARTIFICIAL NEURAL NETWORK}
As a final result we present the fixed order prediction for an example
ANN. To train the ANN we have used events generated with
PYTHIA8~\cite{Sjostrand:2007gs}, for the signal process and the background
processes $\mathrm{pp}\to\mathrm{WW}$ and
$\mathrm{pp}\to\mathrm{t}\bar{\mathrm{t}}$. After this training, we
compute the neural-net output variable at different orders of
perturbation theory and under the variation of the scales. Obviously this ANN should not be understood as a full fletched analysis. In a real analysis, there a many more background processes that need to be considered, in addition we neglect any detector effects.

\begin{figure}
\begin{center}
\includegraphics[width=0.48\textwidth]{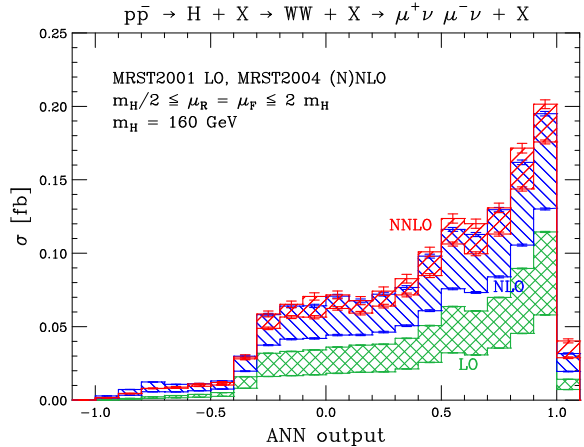}
\caption{\label{fig:fixord_ANN}
  ANN output obtained at LO, NLO and NNLO in perturbative
  QCD. The bands correspond to the variation of the ren.\ and
  fac.\ scales simultaneously in the range $\mu\in[m_\mathrm{H}/2,2\,m_\mathrm{H}]$.
}
\end{center}
\end{figure}

The resulting plots can be seen in Fig.~\ref{fig:fixord_ANN}. As
expected, the bulk of the Higgs cross-section falls into the
high-score bins of the ANN output. In addition, the higher order
corrections show a stable behavior within the full range of the output
variable. This is not surprising, since only 'well-behaved',
i.e.~leptonic final 
state variables were used as input to the ANN. However, this does not guarantee
that any ANN follows this stable
behavior. Especially when variables are used, that are very sensitive
to the inclusion of higher order corrections (e.g. hadronic variables)
care must be taken. The uncertainty under variation of the scales can
be increased in specific regions of the ANN output, depending on the
region of phase-space that is selected by the ANN variable.

\subsection*{CONCLUSIONS}
We have presented cross-sections for the process $\mathrm{H}\to\mathrm{WW}\to\ell\nu\ell\nu$ for proton-(anti)-proton collisions at the LHC ($E_\mathrm{CM}=14\,\mathrm{TeV}$) and the Tevatron ($E_\mathrm{CM}=1.96\,\mathrm{TeV}$) colliders. We have obtained predictions at different orders in perturbation theroy, from LO up to NNLO, and discussed the impact of these higher order corrections. 

As expected the corrections are large, reaching $\sim 300\%$ in certain cases. However, it turns out that the $K$-factors depend on the selection cuts that are applied. In the special case of applying a jet-veto the LO to NNLO $K$-factor can decrease to a value as low as $~1$. Further we discussed the uncertainties on the cross-section arising from varying the ren.\ and fac.\ scales. It was found that in general these decrease when including higher order diagrams in the calculations. We also point out that
 care has to be taken when dividing the sample into different jet-multiplicity bins. In that case, the uncertainty from the inclusive cross-section does not describe correctly the uncertainty on the cross-sections in the various jet-multiplicity bins.

Finally, for the first time we have computed a ANN output variable at NNLO. It turned out that the higher order corrections show a smooth behavior in our toy ANN, however, when creating a ANN care should be taken in the choice of the input variables to the ANN. We anticipate, without proof, that the ANN behavior will be smooth as long as all the input variables show a smooth behavior. This is typically the case for leptonic final state variables, but not guaranteed for hadronic variables.




%% file: heinrich/heinrich.tex
\def\d{{\rm d}}
\def\mom#1{\langle #1 \rangle}





\subsection{INTRODUCTION}

Event-shape observables describe topological properties of hadronic final states 
without the need to define jets, quantifying the  structure of an event by a single measure. 
This class of observables is also interesting because it shows  a rather 
strong sensitivity
to hadronisation effects, at least in phase-space regions characterised by 
soft and collinear gluon radiation, which correspond to certain limits 
for each event-shape variable.

Event-shape distributions in $\mathrm{e^+e^-}$ annihilation have
been measured with high accuracy by a number of 
experiments, most of them  at LEP at centre-of-mass energies
between 91 and 206\,GeV~\cite{Heister:2003aj,Buskulic:1996tt,Acton:1993zh,Alexander:1996kh,Ackerstaff:1997kk,Abbiendi:1999sx,Abbiendi:2004qz,Acciarri:1995ia,Acciarri:1997xr,Acciarri:1998gz,Achard:2002kv,Achard:2004sv,Abreu:1999rc,Abdallah:2003xz,Abdallah:2004xe}.
Mean values and higher moments also have been measured
by several experiments, most extensively by
{\small JADE}~\cite{MovillaFernandez:1997fr,Pahl:2008uc}
and {\small OPAL}~\cite{Abbiendi:2004qz}.

For a long time, the theoretical state-of-the-art description of
event-shape distributions over the full kinematic range was based on
the matching of the next-to-leading-logarithmic approximation
(NLLA)\newline\cite{Catani:1992ua} onto the fixed next-to-leading
order (NLO)  calculation~\cite{Ellis:1980wv,Kunszt:1980vt,eventKN}.
Recently,  NNLO results for event-shape distributions became 
available~\cite{GehrmannDeRidder:2007bj,GehrmannDeRidder:2007hr,Weinzierl:2009ms} 
and lead to the first determination of the strong coupling constant using 
NNLO predictions for hadronic event shapes 
in  $\mathrm{e}^+\mathrm{e}^-$ annihilations~\cite{Dissertori:2007xa}.
Soon after, the matching of the resummed result in the
next-to-leading-logarithmic approximation onto the NNLO calculation
has been performed~\cite{Gehrmann:2008kh} in the so-called $\ln\,R$-matching
scheme~\cite{Catani:1992ua}.
Based on these results, a determination of the strong coupling constant using
matched NNLO+NLLA predictions for hadronic event shapes has been carried 
out~\cite{Dissertori:2009ik}, 
together with a detailed investigation of hadronisation corrections.
Next-to-leading order electroweak corrections to event-shape distributions in 
$e^+e^-$ annihilation were also computed very recently~\cite{Denner:2009gx}.

A similar NNLO+NLLA study based on {\small JADE} data was done in
\cite{Bethke:2008hf}, while other NNLO determinations of $\alpha_s(M_Z)$
based on only the thrust distribution were presented in
\cite{Davison:2008vx,Becher:2008cf}.

Apart from distributions of event-shape observables,
one can also study mean values and higher moments, 
which are now available at NNLO accuracy~\cite{GehrmannDeRidder:2009dp,Weinzierl:2009yz}.
Moments are particularly attractive in view of studying
non-perturbative hadronisation corrections to event shapes.
In ref.~\cite{Gehrmann:2009eh}, NNLO perturbative QCD predictions have been combined with 
non-perturbative power corrections in a dispersive model~\cite{Dokshitzer:1995qm,Dokshitzer:1997ew,Dokshitzer:1998pt,Dokshitzer:1998qp}.
The resulting theoretical expressions  have been compared to experimental data from 
{\small JADE} and {\small OPAL}, and new values for both $\alpha_s(M_Z)$ 
and $\alpha_0$, the effective coupling  in the non-perturbative regime, have been determined.

The two approaches -- estimating the hadronisation corrections 
by general purpose Monte Carlo programs or modelling power corrections analytically
-- shed light on the subject of hadronisation corrections 
from two different sides and lead to some interesting insights 
which will be summarised in the following.

\subsection{THEORETICAL FRAMEWORK}\label{sec:theo}

We have studied
the six event-shape observables thrust $T$~\cite{Farhi:1977sg}
(respectively $\tau = 1-T$), heavy jet mass $M_H$~\cite{Clavelli:1981yh}, wide
and total jet broadening $B_W$ and $B_T$~\cite{Rakow:1981qn},
$C$-parameter~\cite{Parisi:1978eg,Donoghue:1979vi} and the 
two-to-three-jet transition parameter in the Durham algorithm, $Y_3$~\cite{Stirling:1991ds,Bethke:1991wk}. The definitions
of these variables, which we will denote collectively as $y$ in the
following, are summarised e.g.
in~\cite{GehrmannDeRidder:2007hr}.

\subsubsection{event-shape distributions}

The fixed-order QCD description of event-shape distributions
starts from the perturbative expansion
\begin{eqnarray}
\frac{1}{\sigma_{0}}\, \frac{\d\sigma}{\d y}
(y,Q,\mu) &=& \bar\alpha_s (\mu) \frac{\d {A}}{\d y}(y)
+ \bar\alpha_s^2 (\mu) \frac{\d {B}}{\d y} (y,x_\mu) +
\bar\alpha_s^3 (\mu) \frac{\d {C}}{\d y}(y,x_\mu) +
{\cal O}(\bar\alpha_s^4)\;, \label{eq:NNLO0mu}
\end{eqnarray}
where
$$
  \bar\alpha_s = \frac{\alpha_s}{2\pi}\;, \qquad
x_\mu = \frac{\mu}{Q}\;,
$$
and where $A$, $B$ and $C$ are the perturbatively calculated
coefficients~\cite{GehrmannDeRidder:2007hr} at LO, NLO and NNLO.

All coefficients are normalised to the tree-level cross section $\sigma_{0}$ for
$e^+e^-\rightarrow q \bar{q}$. For massless quarks, this
normalisation cancels all electroweak coupling factors, and the
dependence of (\ref{eq:NNLO0mu}) on the collision energy is only
through $\alpha_s$ and $x_\mu$. 
Predictions for the experimentally measured event-shape distributions
are then obtained by normalising to $\sigma_{{\rm had}}$ as
\begin{equation}
\frac{1}{\sigma_{{\rm had}}}\, \frac{\d\sigma}{\d y}(y,Q,\mu)
= \frac{\sigma_0}{\sigma_{{\rm had}}(Q,\mu)}\,
\frac{1}{\sigma_{0}}\, \frac{\d\sigma}{\d y}(y,Q,\mu)\;.
\label{eq:NNLOmu}
\end{equation}
In all expressions, the scale dependence of $\alpha_s$ is determined
according to the three-loop running:
\begin{equation}
\alpha_s(\mu^2) = \frac{2\pi}{\beta_0 L}\left( 1-
\frac{\beta_1}{\beta_0^2}\, \frac{\ln L}{L} + \frac{1}{\beta_0^2 L^2}\,
\left( \frac{\beta_1^2}{\beta_0^2}\left( \ln^2 L - \ln L - 1
\right) + \frac{\beta_2}{\beta_0}  \right) \right)\;,
\label{eq:runningas}
\end{equation}
where $L= 2\, \ln(\mu/\Lambda_{\overline{{\rm MS}}}^{(N_F)})$
and $\beta_i$ are the $\overline{{\rm MS}}$-scheme coefficients listed e.g. 
in~\cite{GehrmannDeRidder:2007hr}.

We take into account bottom mass effects by
retaining the massless $N_F=5$ expressions  and adding
 the difference between the massless and massive LO and NLO
coefficients $A$ and 
$B$~\cite{Brandenburg:1997pu,Bernreuther:1997jn,Rodrigo:1997gy,Nason:1997nw}, 
where a pole b-quark mass of $m_{\rm b}$ = 4.5\,GeV
was used.

In the limit $y\to 0$ one observes that the perturbative
contribution of order $\alpha_{s}^n$ to the cross section diverges like $\alpha_s^n L^{2n}$,
with $L=-\ln\, y$ ($L=-\ln\,(y/6)$ for $y=C$).
This leading logarithmic (LL) behaviour is due to multiple soft gluon
emission at higher orders, and the LL coefficients exponentiate, such that
they can be resummed to all orders.
For the event-shape observables considered here, and assuming
massless quarks, the next-to-leading logarithmic (NLL)
corrections can also be resummed to all orders in the coupling constant.

In order to obtain a reliable description of the event-shape
distributions over a wide range in $y$, it is mandatory to combine
fixed order and resummed predictions. However, in order to avoid the double
counting of terms common to both, the two predictions have to be
matched onto each other. A number of different matching procedures
have been proposed in the literature, see e.g. 
Ref.~\cite{Jones:2003yv} for a review. The most commonly used procedure
is the so-called $\ln\, R$-matching~\cite{Catani:1992ua}, 
which we used in two different variants  for our study on $\alpha_s$~\cite{Dissertori:2009ik}.
For more details about the  NLLA+NNLO
matching we refer the reader to Ref.~\cite{Gehrmann:2008kh}.

\subsubsection{Moments of event-shape observables}

The $n$th moment of an event-shape observable $y$ is
defined by
\begin{equation}
\mom{y^n}=\frac{1}{\sigma_{\rm{had}}}\,\int_0^{y_{\rm{max}}} y^n
 \frac{\d\sigma}{\d y} \d y \;,
\end{equation}
where $y_{\mathrm{max}}$ is the kinematically allowed upper limit of the
observable.
For moments of event shapes, one expects the hadronisation
corrections to be additive, such that they can be divided into a perturbative
and a non-perturbative contribution,
\begin{equation}
\mom{y^n}=\mom{y^n}_{\rm{pt}}+\mom{y^n}_{\rm{np}}\;,
\label{mom}
\end{equation}
where the non-perturbative contribution accounts for hadronisation effects.

In ref.~\cite{Gehrmann:2009eh}, 
the dispersive model derived in
Refs.~\cite{Dokshitzer:1995qm,Dokshitzer:1997ew,Dokshitzer:1998pt,Dokshitzer:1998qp}
has been used 
to estimate hadronisation corrections to event-shape moments 
by calculating analytical predictions for power corrections. It introduces
only a single new parameter $\alpha_0$, which can be interpreted
as the average strong coupling in the non-perturbative region: 
\begin{equation}
\frac{1}{\mu_I}\int_0^{\mu_I} dQ \,\alpha_{\rm{eff}}(Q^2)=\alpha_0(\mu_I)\;,
\label{alpha0}
\end{equation}
where below the  IR cutoff $\mu_{I}$ the strong
coupling is replaced by an effective coupling. 
This dispersive model for the strong coupling leads to a shift in the distributions
\begin{equation}
\frac{\d\sigma}{\d y}(y)=\frac{\d\sigma_{\rm{pt}}}{\d y}\,(y-a_y\,P)\;,
\label{eq:disp}
\end{equation}
where the numerical factor $a_y$ depends on the event shape, while
${P}$ is believed to be universal  and scales with the centre-of-mass energy like  $\mu_I/Q$.
Insertion of eq. (\ref{eq:disp}) into the definition of the moments leads to
\begin{eqnarray} 
\langle y^n \rangle &=& \int^{y_\mathrm{max}-a_yP}_{-a_yP}\mathrm{d}y\,(y+a_yP)^n\frac{1}{\sigma _\mathrm{tot}}\frac{\mathrm{d}\sigma_\mathrm{pt}}{\mathrm{d}y}(y)\nonumber \\
&\approx& \int^{y_\mathrm{max}}_0\mathrm{d}y\,(y+a_yP)^n\frac{1}{\sigma _\mathrm{tot}}\frac{\mathrm{d}\sigma_\mathrm{pt}}{\mathrm{d}y}(y)\;.
\end{eqnarray}
From this expression one can extract
the non-perturbative predictions for the moments of $y$.
To combine the dispersive model with the perturbative prediction at NNLO QCD,
the analytical expressions have been extended~\cite{Gehrmann:2009eh} 
to compensate for all scale-dependent terms at this order.

\subsection{DETERMINATION OF $\alpha_s$ AND $\alpha_0$}

\subsubsection{$\alpha_s$ from  distributions of hadronic event shapes}

We have used 
the six event-shape observables listed in section \ref{sec:theo} for our fits.
The measurements we use have been carried out by the {\small ALEPH} collaboration \cite{Heister:2003aj}
at eight different centre-of-mass energies between 91.2 and 
206\,GeV. The event-shape
distri\-butions were obtained from the reconstructed momenta and
energies of charged and neutral particles. The measurements have
been corrected for detector effects, i.e. the final distributions
correspond to the so-called particle (or hadron) level. 
In addition, at LEP2 energies
above the Z peak they were corrected for initial-state radiation
effects. At energies above 133 GeV,  backgrounds from 4-fermion
processes, mainly from W-pair production and also ZZ and
Z$\gamma^*$, were subtracted following the procedure given
in~\cite{Heister:2003aj}. The experimental uncertainties were
estimated by varying event and particle selection cuts. They are
below 1\% at LEP1 and slightly larger at LEP2. 

The perturbative QCD
prediction is corrected for hadronisation and resonance decays  by
means of a transition matrix, which is computed with the Monte Carlo
generators {\small PYTHIA}~\cite{Sjostrand:2000wi}, 
{\small HERWIG}~\cite{Corcella:2000bw}  and {\small ARIADNE}~\cite{Lonnblad:1992tz}, 
all tuned to global hadronic observables at $M_Z$~\cite{Barate:1996fi}. 
The parton level is defined by the quarks and
gluons present at the end of the parton shower in {\small PYTHIA} and 
{\small HERWIG}
and the partons resulting from the colour dipole radiation in
{\small ARIADNE}. Corrected measurements of event-shape distributions are
compared to the theoretical calculation at particle level.
For a detailed description of the determination and treatment of experimental
systematic uncertainties we refer to Refs.~\cite{Heister:2003aj,Dissertori:2007xa}.

We also made studies using 
the NLO+LL event generator {\small HERWIG++}~\cite{LatundeDada:2007jg}, 
which will be described in more detail below.

\begin{figure}
\begin{center}
\includegraphics[width=0.6\textwidth]{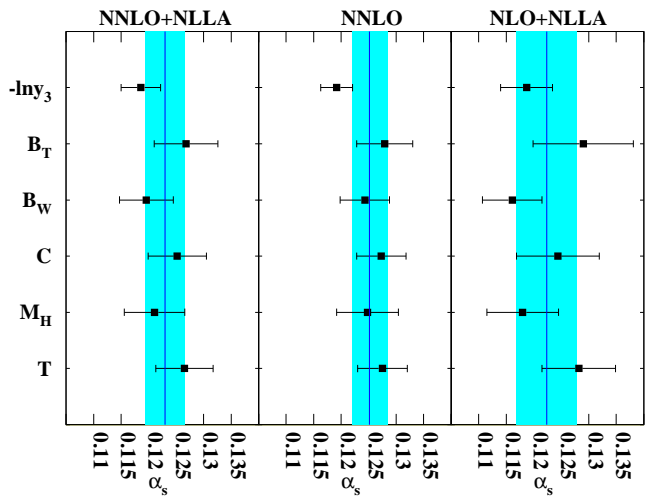}
\caption{\small The measurements of the strong coupling constant
$\alpha_s$ for the six event shapes, at $\sqrt{s}=M_Z$, when using
QCD predictions at different approximations in perturbation theory.
The shaded area corresponds to the total uncertainty.}
\label{fig:scatter}
\end{center}
\end{figure}

The value of $\alpha_s$ is determined at each energy using a binned
least-squares fit. The fit programs of  Ref.~\cite{Dissertori:2007xa}
have been extended to incorporate the NNLO+NLLA calculations.
Combining the results for six event-shape variables
and eight LEP1/LEP2 centre-of-mass energies, we obtain
\begin{center}
    $\alpha_s(M_Z) = 0.1224
    \;\pm\; 0.0009\,\mathrm{(stat)}
    \;\pm\; 0.0009\,\mathrm{(exp)}
    \;\pm\; 0.0012\,\mathrm{(had)}
    \;\pm\; 0.0035\,\mathrm{(theo)}\;.$
\end{center}
The fitted values of the coupling constant as found from event-shape variables calculated 
at various orders are shown in Fig.\ \ref{fig:scatter}.
Comparing our results to both the fit using purely fixed-order
NNLO predictions~\cite{Dissertori:2007xa}
and  the fits based on earlier NLLA+NLO calculations~\cite{Heister:2003aj},
we make the following observations:
\begin{itemize}
\item The central value 
is slightly lower than the central value of 0.1228 obtained from fixed-order NNLO only,
and slightly larger than the NLO+NLLA results. 
The fact that the central value is almost identical to
the purely fixed-order NNLO result could be anticipated from the
findings in Ref.~\cite{Gehrmann:2008kh}. There it is shown that
in the three-jet region, which provides the bulk of the fit range,
the matched NLLA+NNLO prediction is very close to the fixed-order NNLO calculation.
\item The dominant theoretical uncertainty on  $\alpha_s(M_Z)$, as estimated from
scale variations, is reduced by 20\% compared to NLO+NLLA.
However, compared to the fit based on purely fixed-order
NNLO predictions, the perturbative uncertainty is {\it increased} in the NNLO+NLLA fit.
The reason is that in the two-jet region the NLLA+NLO and NLLA+NNLO
predictions agree by construction, because the matching suppresses any
fixed order terms. Therefore, the renormalisation scale uncertainty
is dominated by the next-to-leading-logarithmic
approximation  in this region, which results in a larger overall
scale uncertainty in the $\alpha_s$ fit.
\item As already observed for the fixed-order NNLO results,
the scatter among the values of $\alpha_s(M_Z)$ extracted from the six
different event-shape variables is substantially reduced compared to the NLO+NLLA
case.
\item The matching of NLLA+NNLO introduces a mismatch in the cancellation
of renormalisation scale logarithms, since the NNLO expansion fully
compensates the renormalisation scale dependence up to two loops,
while NLLA only compensates it up to one loop. In order to assess
the impact of this mismatch, we have introduced the $\ln R(\mu)$ matching
scheme~\cite{Dissertori:2009ik}, which retains the two-loop renormalisation terms in the
resummed expressions and the matching coefficients. In this scheme,
a substantial reduction of the perturbative uncertainty from
$\pm0.0035$ (obtained in the default $\ln R$-scheme) to $\pm 0.0022$
is observed, which might indicate the size of the ultimately
reachable precision for a complete NNLO+NNLLA calculation. 
Although both schemes are in principle on the same
theoretical footing, it is the more conservative error
estimate obtained in the $\ln R$-scheme which should be taken as the
nominal value, since it measures the potential
impact of the yet uncalculated finite NNLLA-terms.
\item Bottom quark mass effects, which are numerically significant mainly at the LEP1 energy, 
were included through to NLO. Compared to a purely massless evaluation of the distributions, the inclusion of these mass effects enhances $\alpha_s(M_Z)$ by 0.8\%. 

\end{itemize}

\subsubsection*{Hadronisation corrections from LL+NLO event generators}

In recent years large efforts went into the development of 
modern Monte Carlo event generators which include in 
part NLO corrections matched to  parton showers 
at leading logarithmic accuracy (LL) for various processes. 
Here we use {\small HERWIG++}\,\cite{LatundeDada:2007jg,LatundeDada:2006gx} version 2.3 
for our investigations. 
Several  schemes for the implementation 
of NLO corrections are available~\cite{Frixione:2002ik,Nason:2004rx,Hamilton:2009ne}. 
We studied the {\small MCNLO} \cite{Frixione:2002ik} and {\small POWHEG} \cite{Nason:2004rx} schemes\footnote{We use 
the notation {\small MCNLO} for the {\it method}, while MC@NLO denotes the {\it program}.}. 


We  compared  the prediction for the 
event-shape distributions of {\small HERWIG++} to both the high precision data at LEP1 
from {\small ALEPH} and the predictions from the legacy generators 
{\small PYTHIA, HERWIG} and {\small ARIADNE}. We recall that the latter have 
all been tuned to the same global QCD observables measured by {\small ALEPH}~\cite{Barate:1996fi} 
at LEP1, which included event-shape variables similar to the ones analysed here.  
To investigate the origin of the observed differences between the 
generators, we decided to consider the parton-level predictions and 
the hadronisation corrections separately. 
Discussing the full details of our study is beyond the scope of this note; here we only 
mention some of our observations.
{\small HERWIG++} with {\small POWHEG} yields a similar shape as the legacy programs, but differs in the normalisation.
The other {\small HERWIG++} predictions differ most notably in shape from the former. 
The fit quality 
of {\small HERWIG++} with {\small POWHEG} is similar to the outcome of the legacy generators. 
Given the similar shape but different 
normalisation of {\small HERWIG++} with {\small POWHEG}, the resulting values of $\alpha_s$ are significantly lower, overall by 3\,$\%$.     For further details we refer to Ref.~\cite{Dissertori:2009ik}.

\vspace*{3mm}

From the study of hadronisation corrections we make the following important observation. 
It appears that there are two ``classes'' of variables.  
The first class contains thrust, C-parameter and total jet broadening, while the second
class consists of the heavy jet mass, wide jet broadening and the two-to-three-jet transition parameter  $Y_3$.
For the first class, using the standard hadronisation corrections from {\small PYTHIA}, 
we obtain $\alpha_s(M_Z)$ values around
$0.125 - 0.127$, some $5\%$ higher than those found from the second class of variables. In a study of 
higher moments of event shapes \cite{GehrmannDeRidder:2009dp}, 
indications were found that variables from the first class still suffer from sizable
missing higher order corrections, whereas the second class of observables have a better perturbative stability.
In Ref.~\cite{Dissertori:2009ik}, we observed that this first class
of variables gives a parton level prediction with {\small PYTHIA}, which is about $10\%$ higher than the NNLO+NLLA
prediction. 
The {\small PYTHIA} result is obtained with tuned parameters, where the tuning to data had been performed
at the hadron level. This tuning results in a rather large effective coupling in the parton shower, which might
partly explain the larger parton level prediction of {\small PYTHIA}. As the tuning has been performed
at hadron level, this implies that the hadronisation corrections come out to be smaller than what would have
been found by tuning a hypothetical Monte Carlo prediction with a parton level corresponding to the 
NNLO+NLLA prediction. This means that the {\small PYTHIA} hadronisation corrections, applied in the
$\alpha_s$ fit, might be too small, resulting in a larger $\alpha_s(M_Z)$ value. 
Since up to now the hadronisation uncertainties have been estimated
from the differences of parton shower based models, tuned to the data, it is likely that for these event shapes
the uncertainties were underestimated, missing a possible systematic shift.     
Such problems do not appear to exist for the second class of variables.

We would like to mention that a determination of $\alpha_s$ based on 
3-jet rates calculated at NNLO accuracy also 
has been performed recently~\cite{Dissertori:2009qa}, with the result
$\alpha_s(M_Z) = 0.1175 \pm 0.0020\, (\mathrm{exp}) \pm 0.0015\, (\mathrm{theo})$, 
which is also lower than the one obtained from fits to distributions of event shapes.
 
\subsubsection{$\alpha_s$ and $\alpha_0$ from moments of hadronic event shapes}

Now we turn to analytical models to estimate hadronisation corrections. 
The expressions derived in~\cite{Gehrmann:2009eh} 
match the dispersive model with the perturbative prediction at NNLO QCD.
Comparing these expressions 
with experimental data on event-shape moments, a combined determination
of the perturbative strong coupling constant $\alpha_s$ and the
non-perturbative parameter $\alpha_0$ has been performed~\cite{Gehrmann:2009eh}, 
based on data from the {\small JADE} and {\small OPAL} experiments~\cite{Pahl:2008uc}. 
The data consist of 18 points at centre-of-mass
 energies between 14.0 and 206.6 GeV for the first five moments of 
 $T$, $C$, $Y_3$, $M_H$, $B_W$ and $B_T$, and have been taken from \cite{Pahl:2007zz}.
For each moment the NLO as well as the NNLO prediction was
fitted with $\alpha_s(M_Z)$
 and $\alpha_0$ as fit parameters, except for the moments of $Y_3$, which
have no power correction and thus are independent of $\alpha_0$.

Compared to previous results at NLO,
inclusion of NNLO effects results in
a considerably improved consistency in the parameters determined from
different shape variables, and in a substantial reduction of the error on
$\alpha_s$.

We  further observe that the theoretical error on the extraction of $\alpha_S(M_Z)$ 
from $\rho$, $Y_3$ and $B_W$ 
is considerably smaller than from $\tau$, $C$ and $B_T$. As mentioned above and 
discussed in detail  in~\cite{GehrmannDeRidder:2009dp}, the moments of the former three shape 
variables receive moderate NNLO corrections for all $n$, while the NNLO 
corrections for the latter three are large already for $n=1$ and 
increase with $n$. Consequently, 
the theoretical description of the moments of $\rho$, $Y_3$ and $B_W$ 
displays a higher perturbative stability, which is reflected in the smaller 
theoretical uncertainty on  $\alpha_S(M_Z)$ derived from those variables. 

In a second step, we combine the $\alpha_s(M_Z)$ and $\alpha_0$ measurements 
obtained from different event-shape variables. 
Taking the weighted mean over all values except $B_W$ and $B_T$,  we obtain at NNLO:
\begin{eqnarray}
\alpha_s(M_Z)&=0.1153\pm0.0017(\mathrm{exp})\pm0.0023(\mathrm{th}),\nonumber \\
\alpha_0&=0.5132\pm0.0115(\mathrm{exp})\pm0.0381(\mathrm{th})\,,
\label{eq:final}
\end{eqnarray}
The moments of $B_W$ and $B_T$ have been excluded here 
since their theoretical description requires an additional contribution to the 
non-perturbative coefficient $P$ (see eq.~(\ref{eq:disp})) which is not 
available consistently to NNLO.

\begin{figure}[t]
\centering
\includegraphics[width=3.7cm]{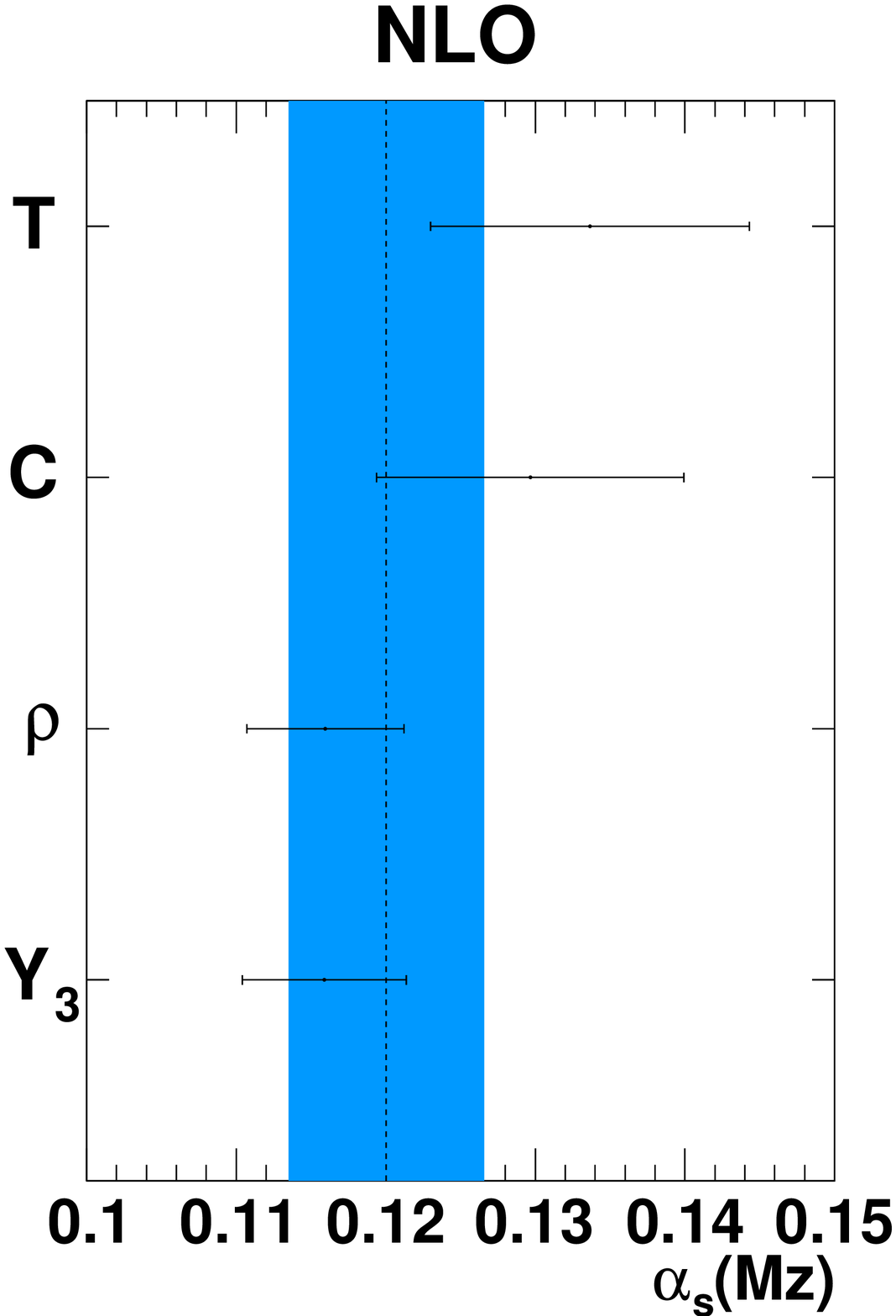}%
\qquad\qquad
\includegraphics[width=3.7cm]{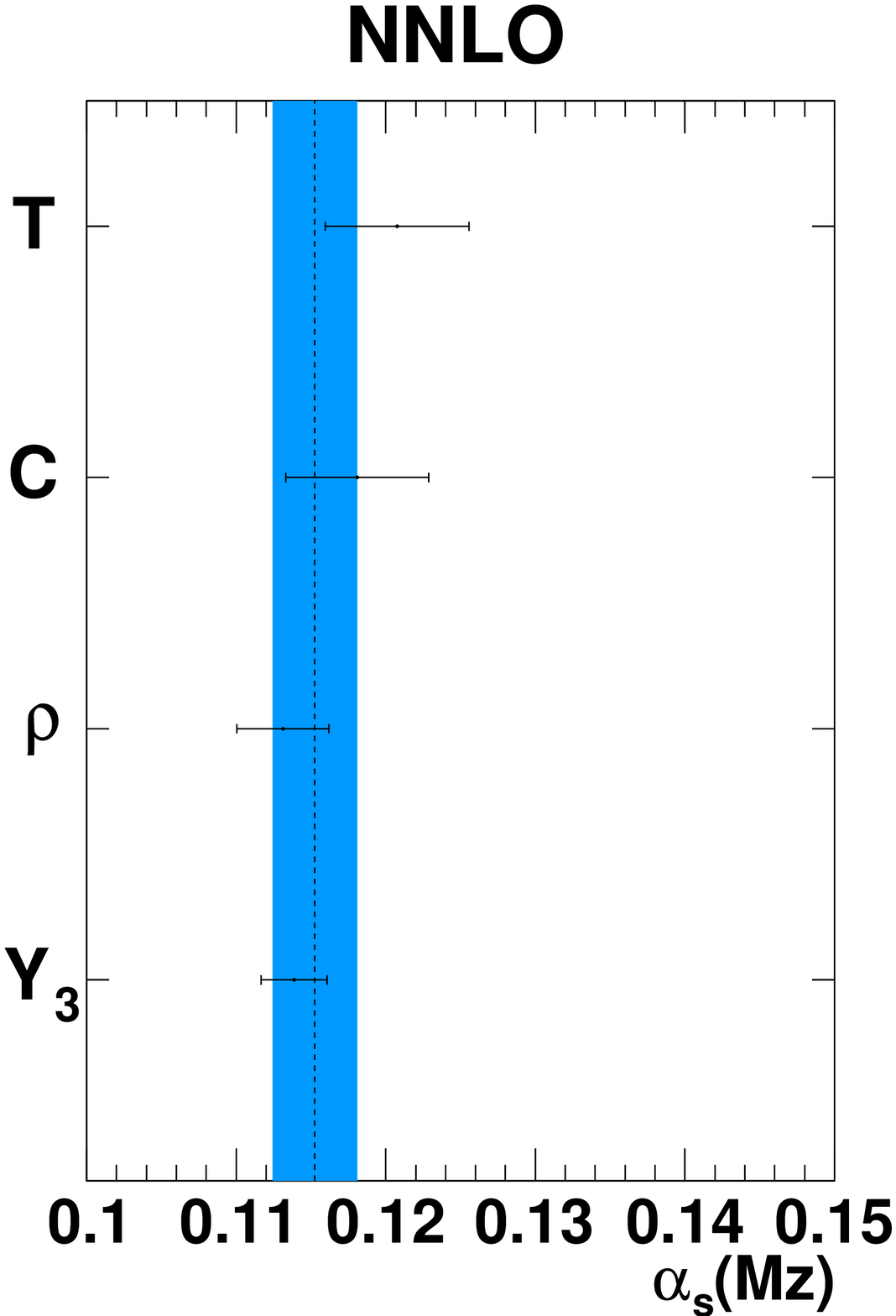}%
\\
\includegraphics[width=3.7cm]{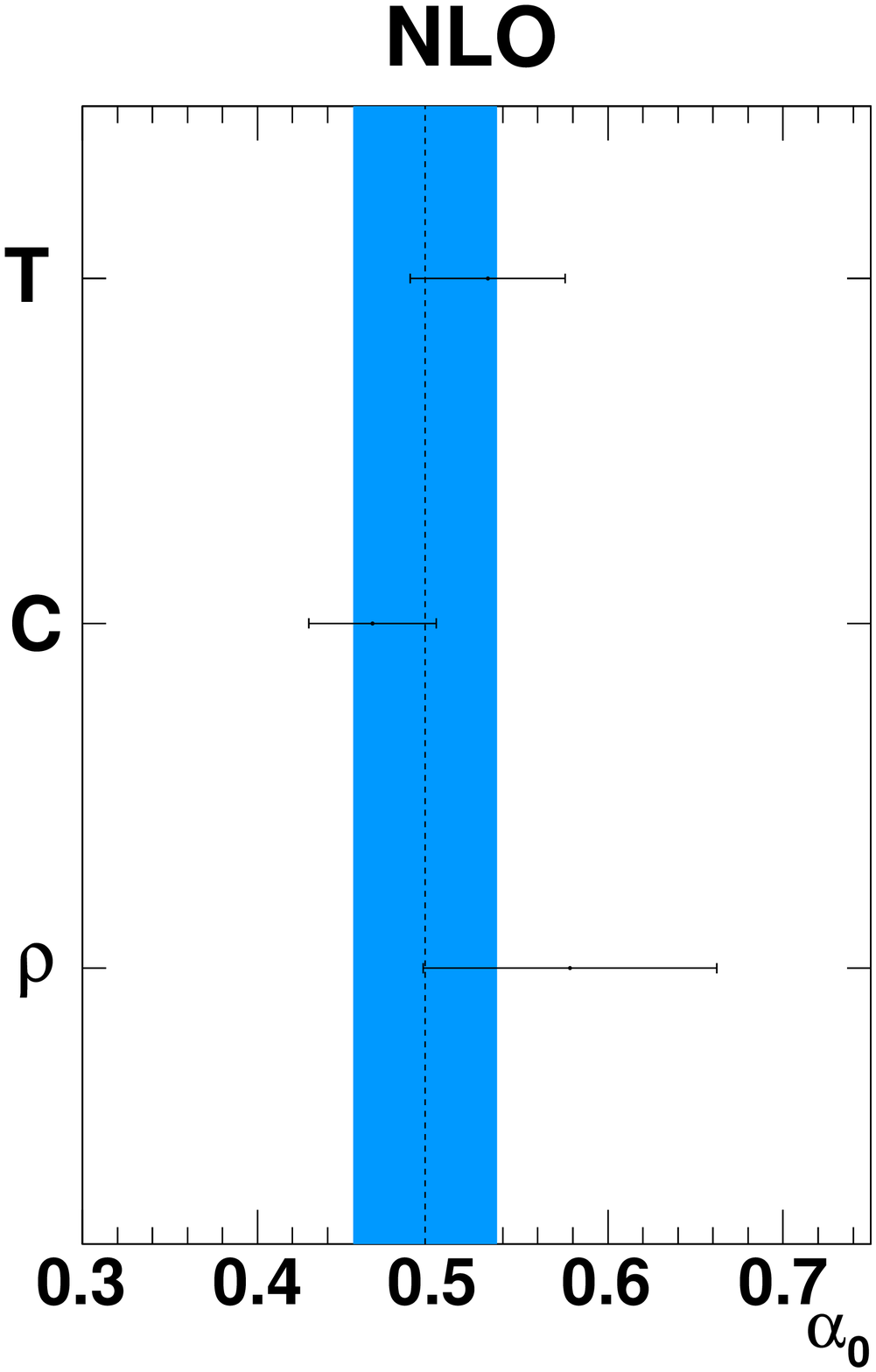}%
\qquad\qquad
\includegraphics[width=3.7cm]{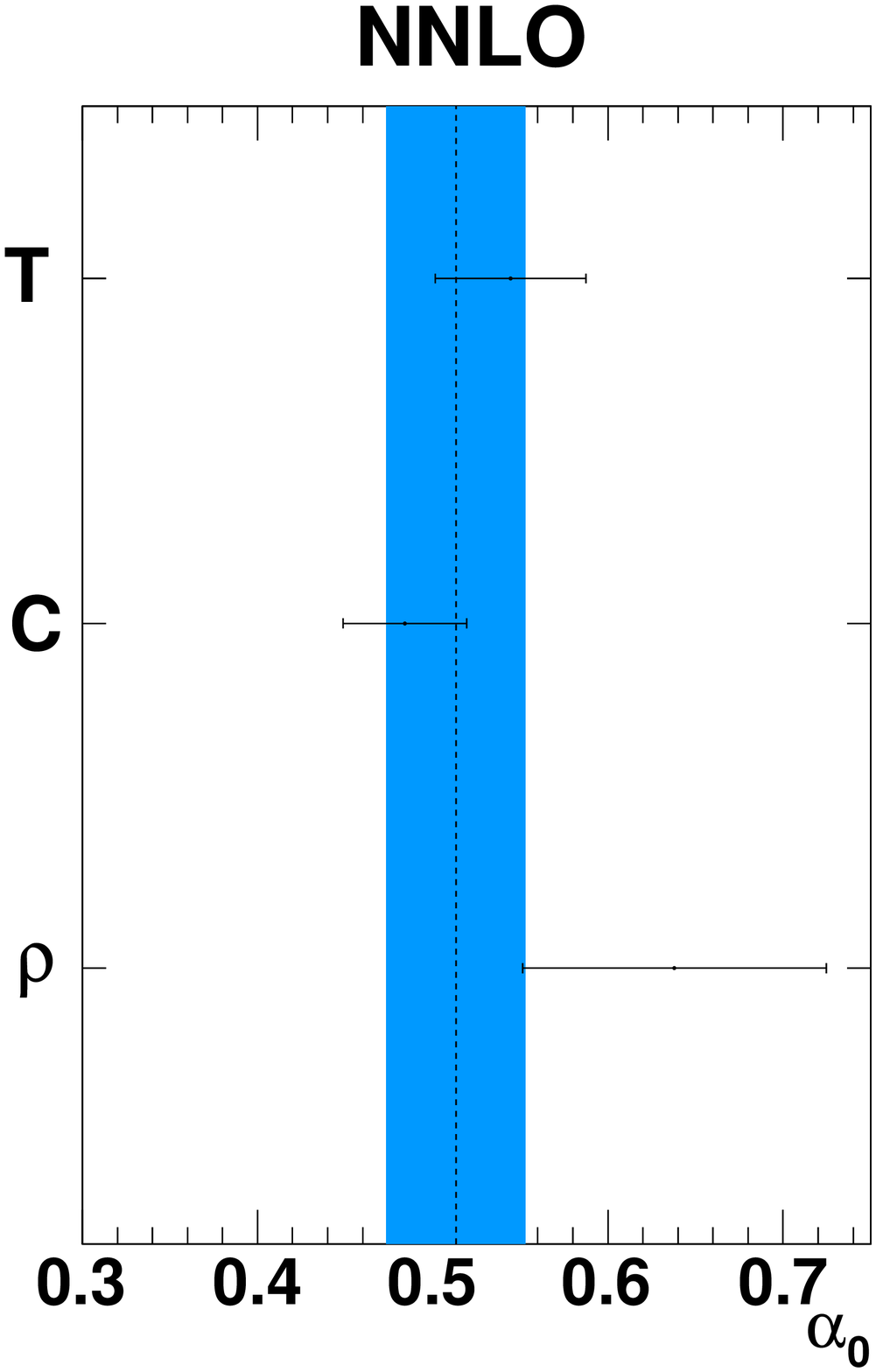}%
\caption{Error bands at NLO and NNLO for combinations of values for $\alpha_s$ 
and $\alpha_0$ obtained  from fits to moments of different event shapes.                                              The error on $\alpha_s$ is dominated by scale uncertainties,
while the largest contribution to the error on $\alpha_0$ comes from the uncertainty on the Milan factor.} 
\label{fig:final}
\end{figure}

To illustrate the improvement due to the inclusion of the NNLO 
corrections, we also quote the corresponding NLO results. Based on 
$\tau$, $C$, $\rho$ and $Y_3$, we obtain:
\begin{eqnarray}
\alpha^{{\rm NLO}}_s(M_Z)&=0.1200\pm0.0021(\mathrm{exp})\pm0.0062(\mathrm{th}),\nonumber \\
\alpha^{{\rm NLO}}_0&=0.4957\pm0.0118(\mathrm{exp})\pm0.0393(\mathrm{th})\,, \nonumber 
\end{eqnarray}
We compare the NLO and NNLO combinations in Figure~\ref{fig:final}. It can be 
seen very clearly that the measurements obtained from the different variables 
are consistent with each other within errors. The average of $\alpha_s(M_Z)$ is 
dominated by the measurements based on $\rho$ and $Y_3$, which have the 
smallest theoretical uncertainties. From NLO to NNLO, the error on 
 $\alpha_s(M_Z)$  is reduced by a factor of two.
 Analysing the different sources of the systematical errors, we observe that
the error on $\alpha_s(M_Z)$ is clearly dominated by the $x_{\mu}$ variation,
while the largest contribution to the error on $\alpha_0$ comes
from the uncertainty on the Milan factor ${\cal M}$~\cite{Dokshitzer:1998pt}. Since this
uncertainty has not been improved in the current study, it is understandable
that the systematic error on  $\alpha_0$ remains unchanged.

To quantify the difference of the dispersive model to hadronisation corrections from the 
legacy generators, we analysed the moments of (1-T)
with hadronisation corrections from {\small PYTHIA}. As a result, we obtained
fit results for $\alpha_s(M_{\mathrm Z})$ which are typically 4\% higher than by using the 
dispersive model, with a slightly worse quality of the fit.
Comparing perturbative and non-perturbative contributions at $\sqrt{s} = M_{{\mathrm Z}}$, we
observed that {\small PYTHIA} hadronisation corrections amount to less than half the power
corrections obtained in the dispersive model, thereby
explaining the tendency towards a larger value of $\alpha_s(M_{\mathrm Z})$,  
since the missing numerical magnitude of the power corrections must
be compensated by a larger perturbative contribution. 


\subsection*{CONCLUSIONS}

We have compared determinations of the strong coupling constant 
based on  hadronic event shapes measured at LEP using two different approaches:
\begin{enumerate}
\item a fit of perturbative QCD
results at  next-to-next-to-leading order (NNLO), matched to
resummation in the next-to-leading-logarithmic approximation
(NLLA), to {\small ALEPH} data where the hadronisation corrections 
have been estimated using Monte Carlo event generators
\item  a fit of perturbative QCD results at NNLO matched to non-perturbative power
corrections in the dispersive model, providing analytical parametrisations of hadronisation 
corrections, to {\small JADE}
and {\small OPAL} data. 
\end{enumerate} 
We find that the second approach results in a considerably lower value of $\alpha_s(M_Z)$ 
than the first one.

We conclude that apparently there are two ``classes'' of event-shape variables, the first class containing thrust, 
C-parameter and total jet broadening, the second
class containing  heavy jet mass, wide jet broadening and the two-to-three-jet transition parameter  $Y_3$.
Comparing parton level and hadron level predictions from {\small PYTHIA}, this first class
of variables gives a parton level prediction which is about $10\%$ higher than the NNLO+NLLA
prediction, where the {\small PYTHIA} curve has been obtained with tuned parameters, 
the tuning to data being performed
at the hadron level. This tuning results in a rather large effective coupling in the parton shower, 
such that the parton level prediction of {\small PYTHIA} turns out large. 
This may imply that the hadronisation corrections come out to be too small for these variables, 
resulting in a larger $\alpha_s(M_Z)$ value. 
This hypothesis is corroborated by the fact that 
the theoretical description of the moments of the variables thrust, 
C-parameter and total jet broadening displays a lower perturbative stability. 

For the moments of (1-T), we found that the legacy generators 
predict power corrections which are less than half of what is obtained in the dispersive model.
The large numerical
discrepancy between analytical power corrections and the estimate of hadronisation effects 
from the legacy generators
suggests to revisit the impact of hadronisation corrections on precision QCD observables.

\subsection*{ACKNOWLEDGEMENTS}
This research was supported in part by the Swiss National Science Foundation
(SNF) under  contracts PP0022-118864 and
200020-126691,
by the UK Science and Technology Facilities Council, 
by the European Commission's Marie-Curie Research Training Network 
MRTN-CT-2006-035505  and by the 
German Helmholtz Alliance ``Physics at the Terascale''.

%

%% file: warsinsky/warsinsky.tex





\subsection{INTRODUCTION}
\begin{figure}
\centerline{\includegraphics[width=0.9\textwidth]{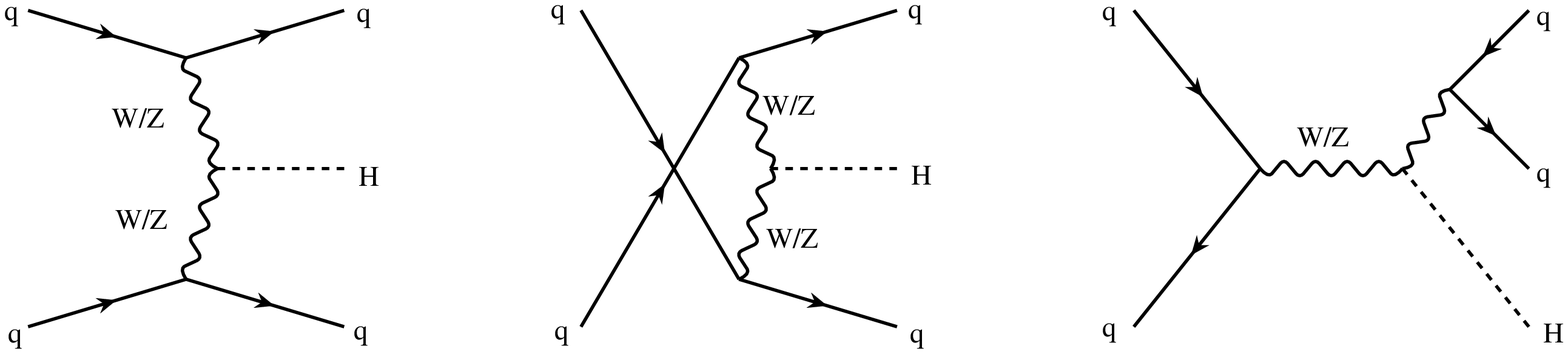}}
\caption{\label{vbf09_feynmangraphs}Feynman diagrams for the process $qq\to qqH$ at leading order. Left: t-channel, middle: u-channel, right: s-channel.}
\end{figure}

An important search channel for the Higgs boson at the LHC is Vector boson fusion (VBF), which is included in the process $qq\to qqH$, where the Higgs boson is produced via the coupling to the gauge bosons. Feynman diagrams for this process at leading order are depicted in Fig.~\ref{vbf09_feynmangraphs}. 

The experimental signature of the VBF process consists of two so-called tag-jets, which because the t- and u-channel diagrams are dominant, tend to be in the forward direction and in opposite detector hemispheres, the decay products of the Higgs boson in the central region and due to the absence of color flow only small additional hadronic activity in the central detector region. To accurately estimate cut efficiencies and acceptances higher order corrections are needed.

Higher order corrections for this process have been first evaluated in NLO QCD neglecting various interference terms and s-channel contributions~\cite{Han:1992hr,Spira:1997dg,Figy:2003nv,Figy:2004pt,Berger:2004pc}, and subsequently including NLO QCD
and NLO electroweak corrections in~\cite{Ciccolini:2007jr,Ciccolini:2007ec}, where all contributing diagrams, including s-channel diagrams have been taken into account. The electroweak corrections are about the same size as the strong corrections. The prediction of the total cross section has a scale uncertainty of only a few per cent.

However, for the simulation of the VBF process to evaluate experimental acceptances and cut efficiencies, a fixed order calculation is not sufficient, because it does not include a parton shower, hadronization, or an underlying event description. In experimental analyses rather parton-shower Monte Carlo generators (PS-MC) are used, which in general use a leading order matrix element\footnote{Until very recently there was no PS-MC available combining an NLO matrix element with a parton-shower for VBF. Within NLO QCD such a matching has been presented recently in the POWHEG scheme in \cite{Nason:2009ai}.}. It is thus difficult to make predictions of accepted cross sections after analysis cuts when one wants to make use of the precise higher order calculations. The accepted cross sections are necessary to evaluate the discovery potential at the LHC for the Higgs boson, and also, in case that no signal is observed, to place limits on the Higgs boson mass.

In addition to acceptance differences, the higher order calculations can also induce differences in the shapes of differential distributions that are used in experimental analyses. As long as no NLO PS-MC Monte Carlo generator is available, it is only possible to incorporate these by weighting the events of a PS-MC in a way that the kinematic distributions become as similar as possible to the NLO result.

In this note the acceptances of the PS-MC Herwig~\cite{Corcella:2002jc} and of the fixed order calculation in~\cite{Ciccolini:2007jr,Ciccolini:2007ec} are compared. In addition, comparisons of differential distributions of kinematic variables are made and a possible reweighting method to improve the modeling of the PS-MC is proposed.

\subsection{SETUP}
In the following comparisons are made for an assumed Higgs boson mass of $120\,\rm{GeV}$ and a LHC centre-of-mass energy of $14\,\rm{TeV}$.

The fixed order results shown in the following employ the program used in~\cite{Ciccolini:2007jr,Ciccolini:2007ec} with the imput parameters $M_W=80.425\,\rm{GeV}$, $\Gamma_W=2.124\,\rm{GeV}$, $M_Z=91.1876\,\rm{GeV}$, $\Gamma_Z=2.4952\,\rm{GeV}$, $G_\mu=1.16637\times 10^{-5}\rm{GeV}^{-2}$, 
$m_t=174.3\,\rm{GeV}$. All other input
  parameters are as in~\cite{Ciccolini:2007jr,Ciccolini:2007ec}. The strong coupling constant is chosen as the same as in the used parton density function, where for the leading order result the CTEQ6L1 set~\cite{Pumplin:2002vw} and for the NLO result the MRST2004qed set~\cite{Martin:2004dh} set is used. Processes with external b-quark contributions are excluded. A renormalization and factorization scale of $\mu_R=\mu_F=M_W$ is used.

As PS-MC generator Herwig 6.510~\cite{Corcella:2002jc} is used, using the same top quark mass as for the fixed order result. The Higgs boson is forced to decay into $ZZ$ and both $Z$ bosons are required to decay into neutrinos, in order not to introduce a sensitivity to the properties of the Higgs boson decay. The corresponding branching fractions are already removed from the cross sections quoted in the following for the Herwig PS-MC. The soft underlying event probability in Herwig was switched off.

Jets are reconstructed using a $k_T$-algorithm~\cite{Catani:1992zp}, as described in~\cite{Blazey:2000qt}, with a resolution parameter of $D=0.8$. For the fixed order result, all partons within $|y|<5$, where $y$ is the rapidity, are used as input for the jet algorithm. In the case of the PS-MC all stable particles after hadronization with $|y|<5$ are taken into account.

Typical experimental VBF cuts as in~\cite{Figy:2004pt} are used, requiring at least two jets with a transverse momentum of at least $20\,\rm{GeV}$ and $|y|<4.5$. The two jets with the highest transverse momentum passing these requirements are taken as tag-jets. These two tag-jets are required to be in opposite detector hemispheres ($y_1\cdot y_2<0$) and to have a separation in rapidity of at least 4 ($|y_1-y_2|>4$).

\subsection{COMPARISONS}

\subsubsection{Accepted Cross Sections}

Table~\ref{tab:vbf09_table1} shows a comparison of the cross section with and without VBF cuts along with the cut efficiency for the fixed order calculation from~\cite{Ciccolini:2007jr,Ciccolini:2007ec} and the Herwig parton shower generator. Compared to the full result from~\cite{Ciccolini:2007jr,Ciccolini:2007ec}, Herwig shows a too small cross section without cuts. When applying the VBF cuts, the cross section difference is much smaller. This is due to the fact that Herwig does not take s-channel contributions to the $qq\to qqH$ process into account. When comparing to the results from~\cite{Ciccolini:2007jr,Ciccolini:2007ec} with the s-channel contributions excluded, the difference becomes much smaller.

It should be expected that when the s-channel contributions are not taken into account, the cross section by Herwig should agree with the LO prediction from~\cite{Ciccolini:2007jr,Ciccolini:2007ec} using the CTEQ6L1 PDF set. However, this is not completely the case: Without cuts, Herwig predicts an about 4\% smaller cross section than the program of \cite{Ciccolini:2007jr,Ciccolini:2007ec}, and this difference increases to about 9\% when VBF cuts are applied. The overall normalization difference can for example arise from different scale choices in the Herwig parton shower compared to the fixed order calculation. The selection efficiency of the VBF cuts is also slightly smaller in Herwig than in the LO prediction of the fixed order calculation. However, the selection efficiency is in good agreement to the NLO result of the fixed order calculation. This is due to the fact that by the use of a parton shower already parts of the higher order corrections are taken into account.

Since the selection efficiency in Herwig is the same as the one in the fixed order calculation, the Herwig cross section can be scaled to the one from the fixed order calculation to obtain a prediction of the accepted cross section.


\begin{table}[htb]
\centerline{\begin{tabular}{cccccc}\hline\hline
program & order& PDF & $\sigma_{\mathrm{no~cuts}} [\mathrm{fb}]$  & $\sigma_{\mathrm{VBF~cuts}} [\mathrm{fb}]$  & $\epsilon_{\mathrm{cuts}}[\%]$\\\hline
fixed order& LO & CTEQ6L1 & 5406 & 1685 &  31.1\\ 
fixed order& NLO & MRST2004qed & 5872 & 1665 &  28.3\\
fixed order, no s-channel & LO & CTEQ6L1 & 4216 & 1685 &  40.0\\
fixed order, no s-channel & NLO & MRST2004qed & 4290 & 1656 &  38.5\\
Herwig & LO+PS & CTEQ6L1 & 4054 & 1547 & 38.2\\\hline\hline
\end{tabular}}
\caption{\label{tab:vbf09_table1}Cross sections with and without VBF cuts and cut efficiencies.}
\end{table}

\subsubsection{Differential Distributions}

In addition to the total cross section also the shape of differential distributions is in general changed by higher order corrections. In addition to the observables used in the VBF cuts, this is also important for experimental analysis for two reasons:

\begin{itemize}
\item The VBF cuts are rather soft, especially the cut of $20\,\mathrm{GeV}$ on the transverse momentum of the tagging jet. In general, additional kinematic variables like the invariant mass of the two tag-jets or the difference of their azimuthal angles is used in experimental analyses to extract the signal from the background. A good modeling of these variables is thus desirable.
\item Recently there has been some interest in analyses that use highly boosted Higgs bosons with a very high transverse momentum to discover the Higgs boson also in its decay into bottom quarks~\cite{Butterworth:2008iy,Plehn:2009rk}. While currently no such analysis exists for the VBF production channel, it might be of interest in the future. For such an analysis the accurate prediction of the transverse momentum of the produced Higgs boson is very important.
\end{itemize}

To assess the influence of higher order corrections on the shape of differential distributions and to compare the prediction of the Herwig PS-MC generator to the fixed order calculation, the following procedure has been applied: First, the VBF cuts as described in the previous section were applied. To compare Herwig to the LO prediction, it was decided to normalize the event sample to the LO cross section after cuts, thus removing the 9\% discrepancy that was observed.

\begin{figure}
\includegraphics[width=0.5\textwidth]{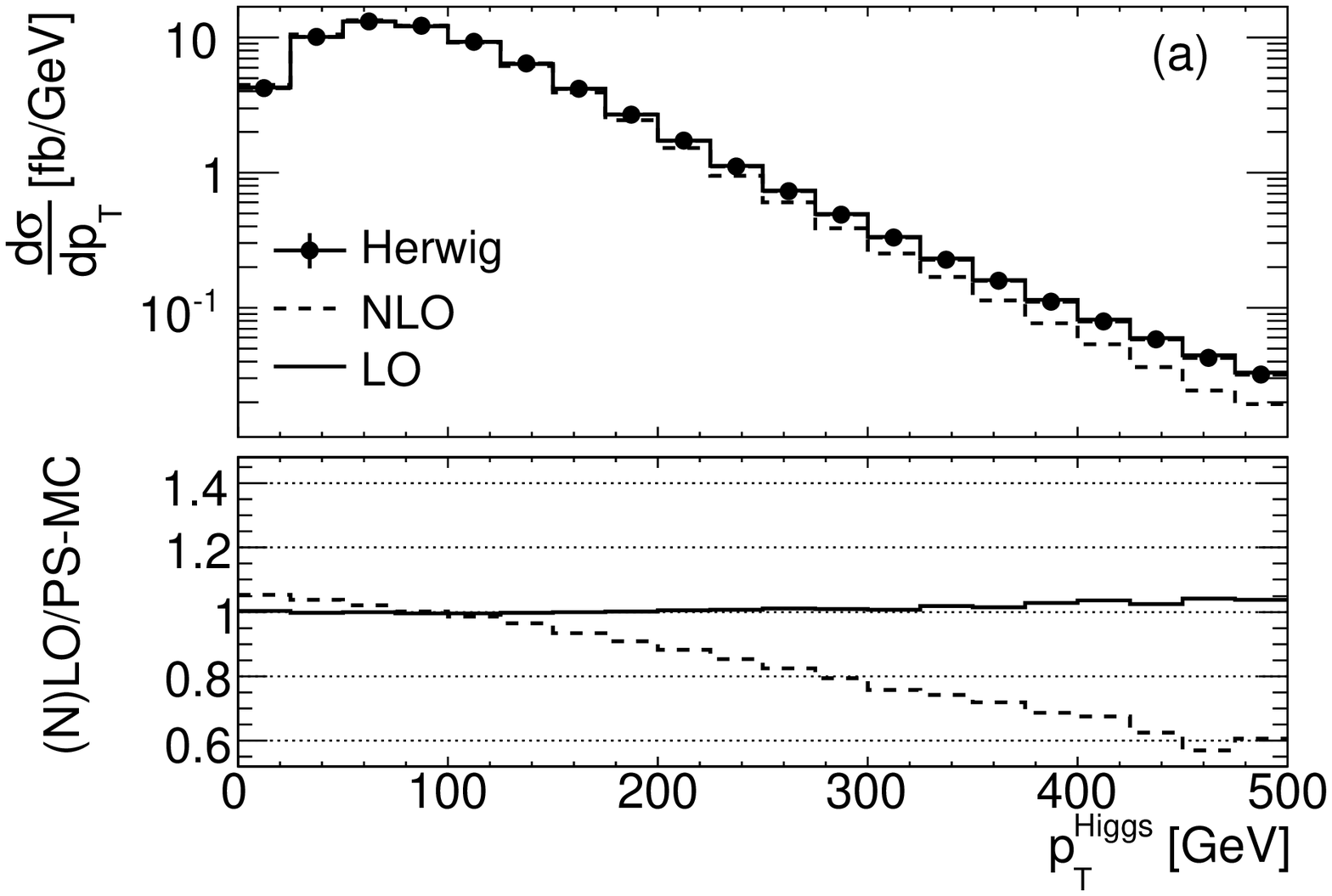}
\includegraphics[width=0.5\textwidth]{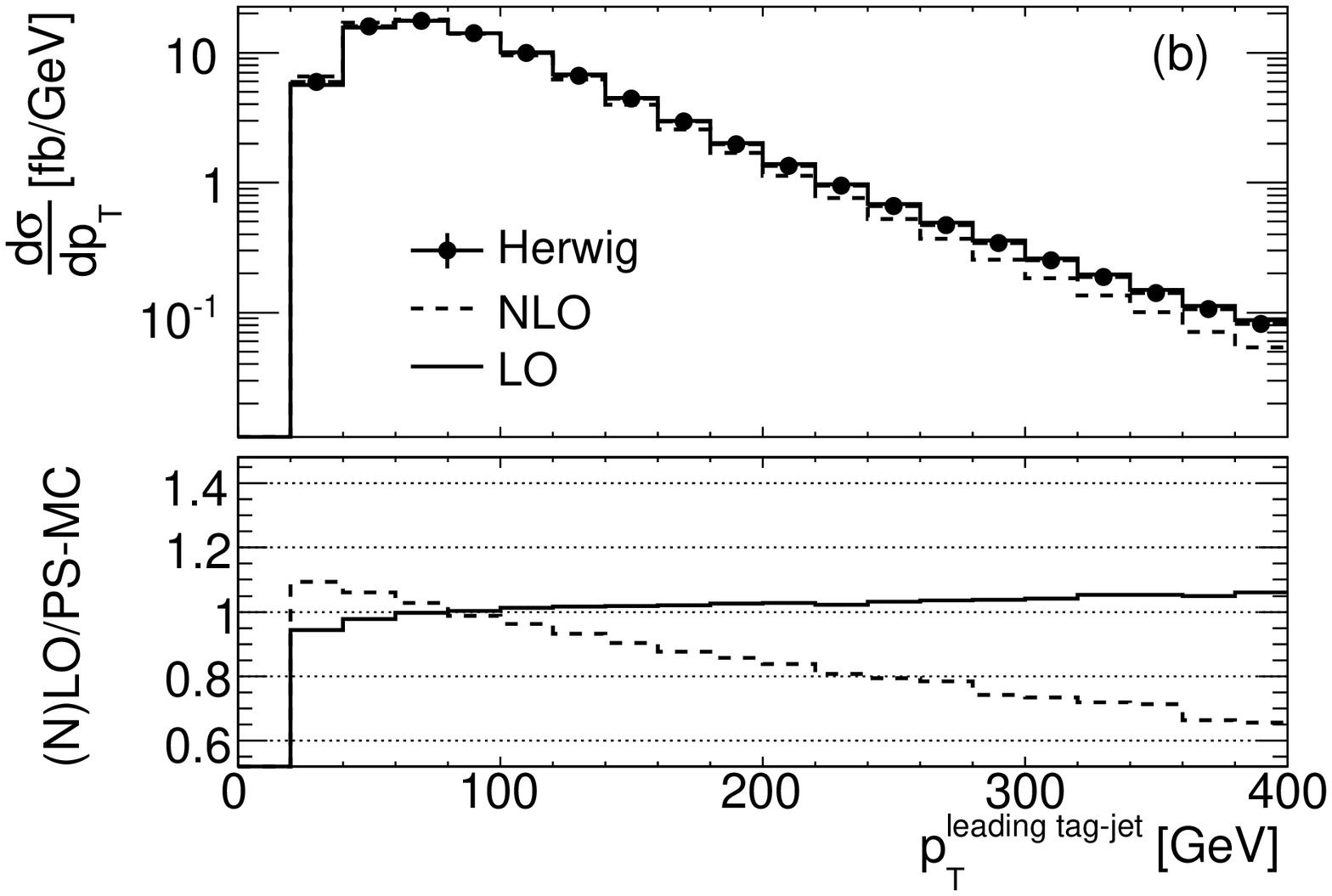}
\includegraphics[width=0.5\textwidth]{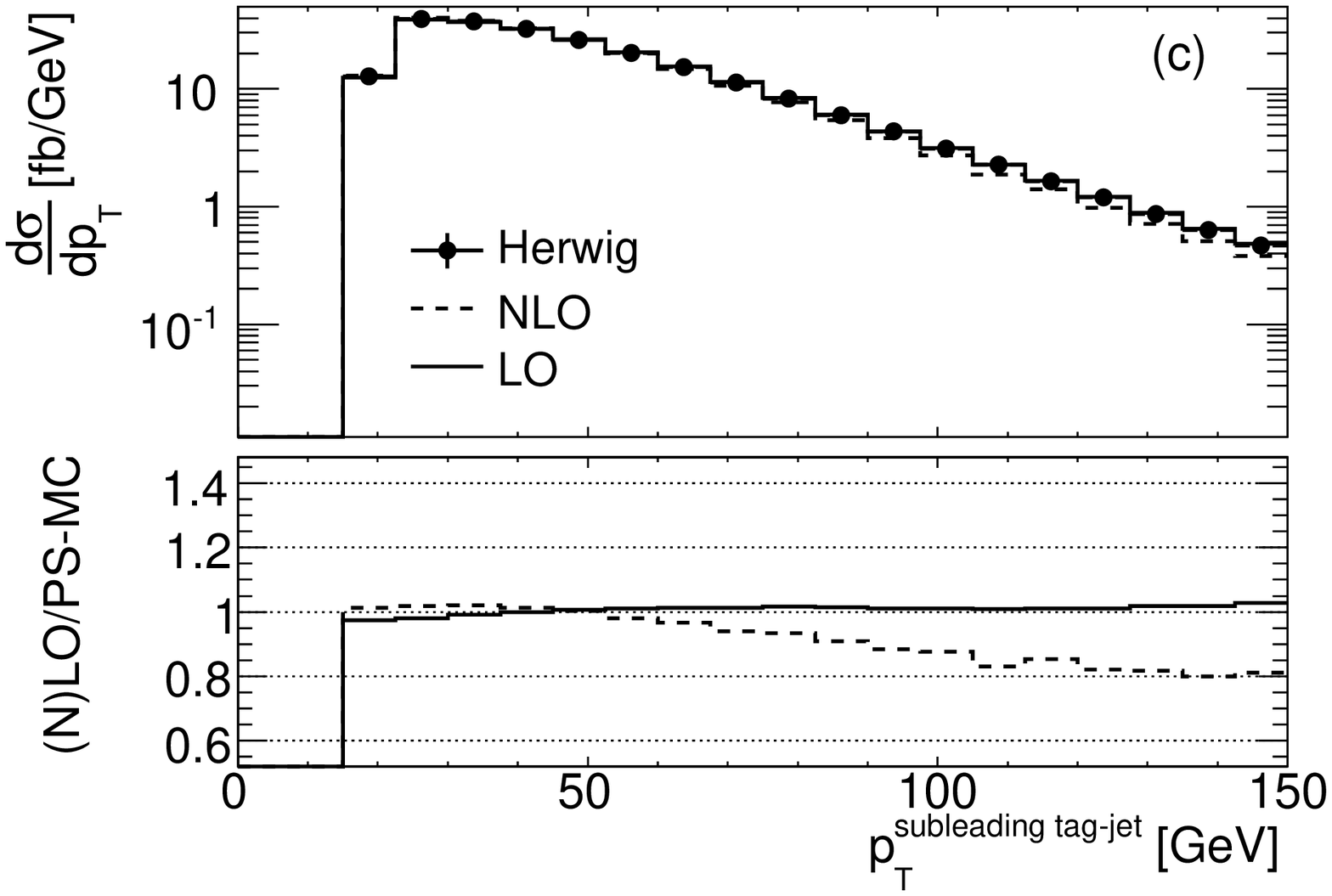}
\includegraphics[width=0.5\textwidth]{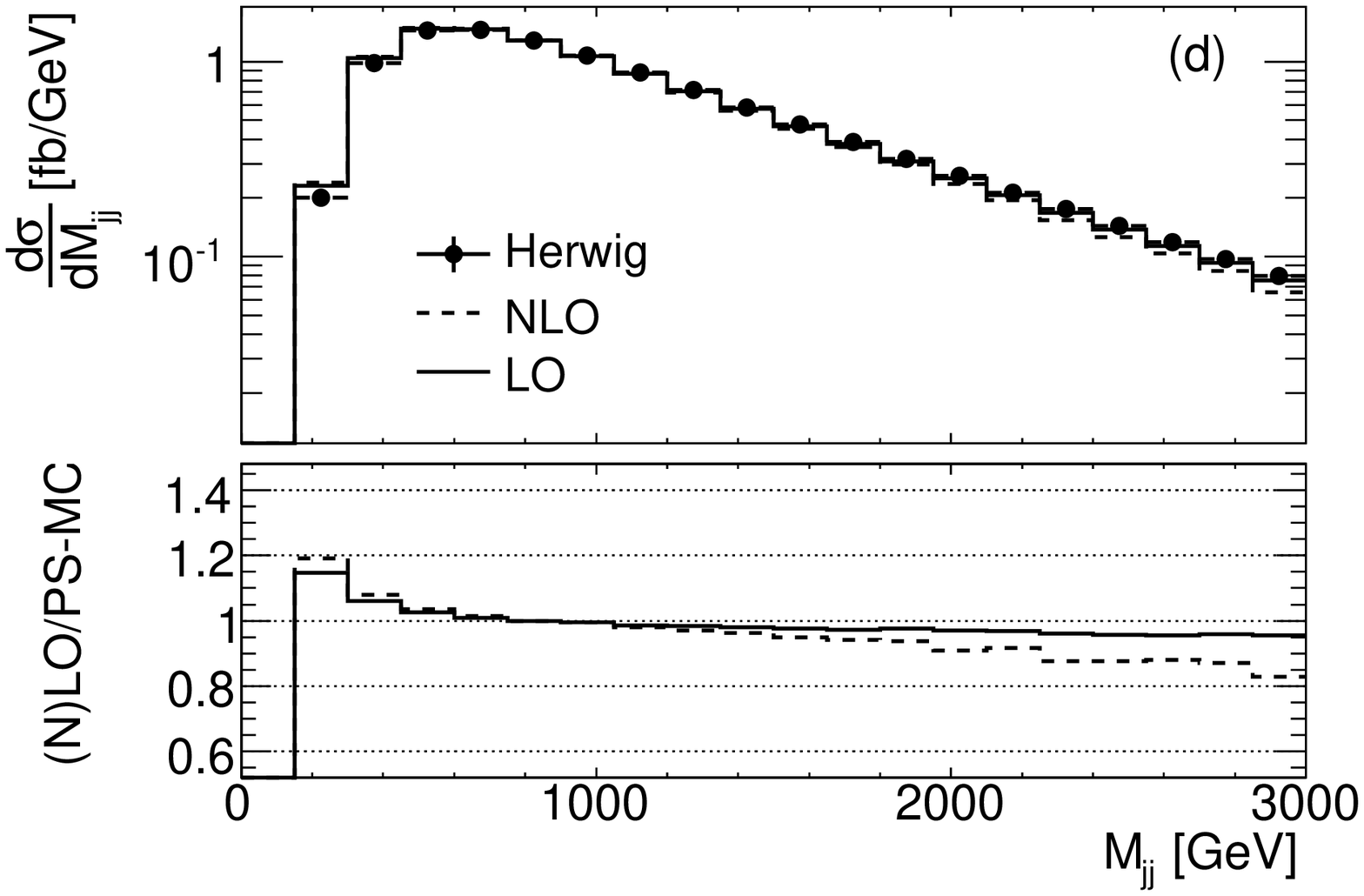}
\includegraphics[width=0.5\textwidth]{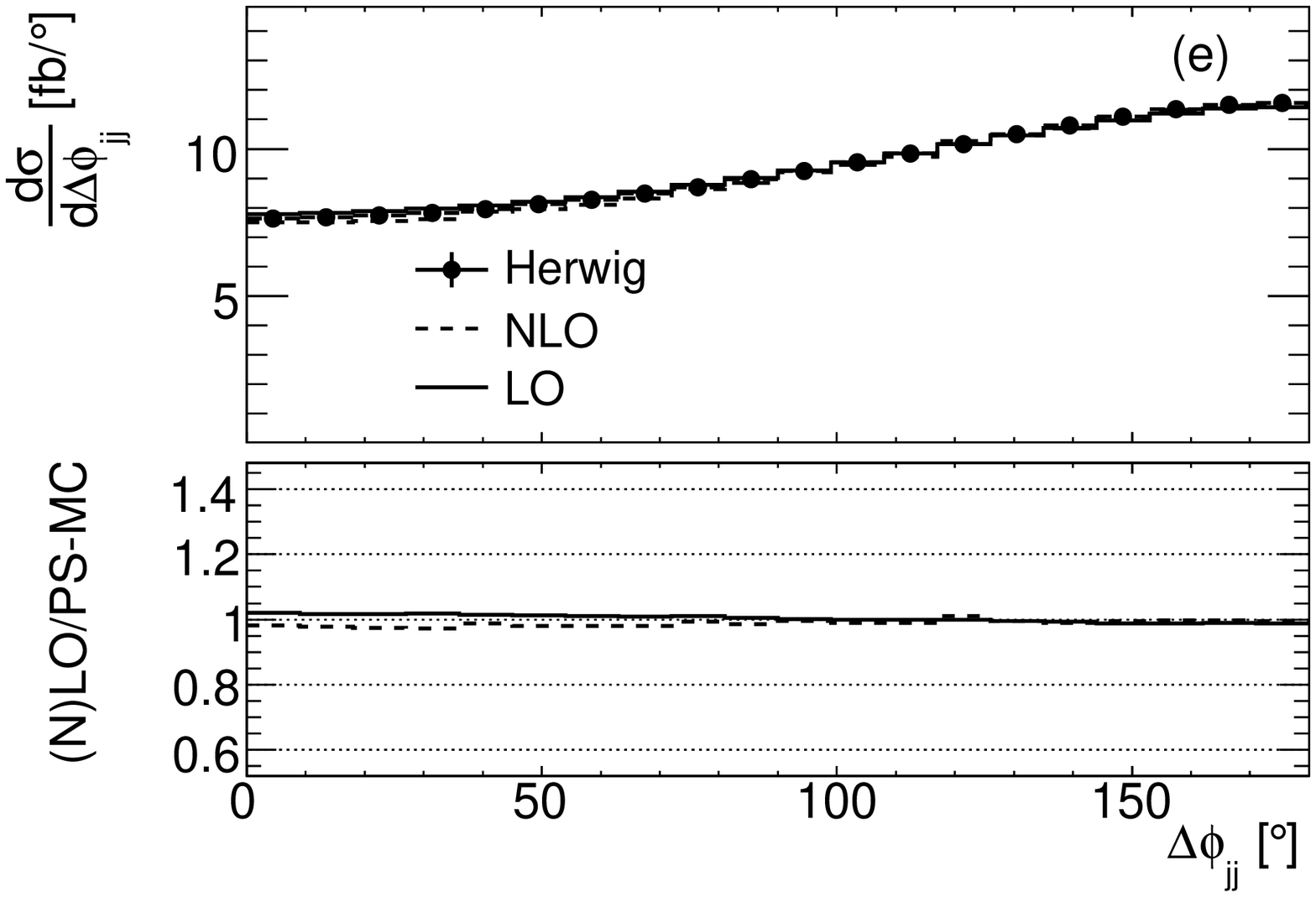}
\includegraphics[width=0.5\textwidth]{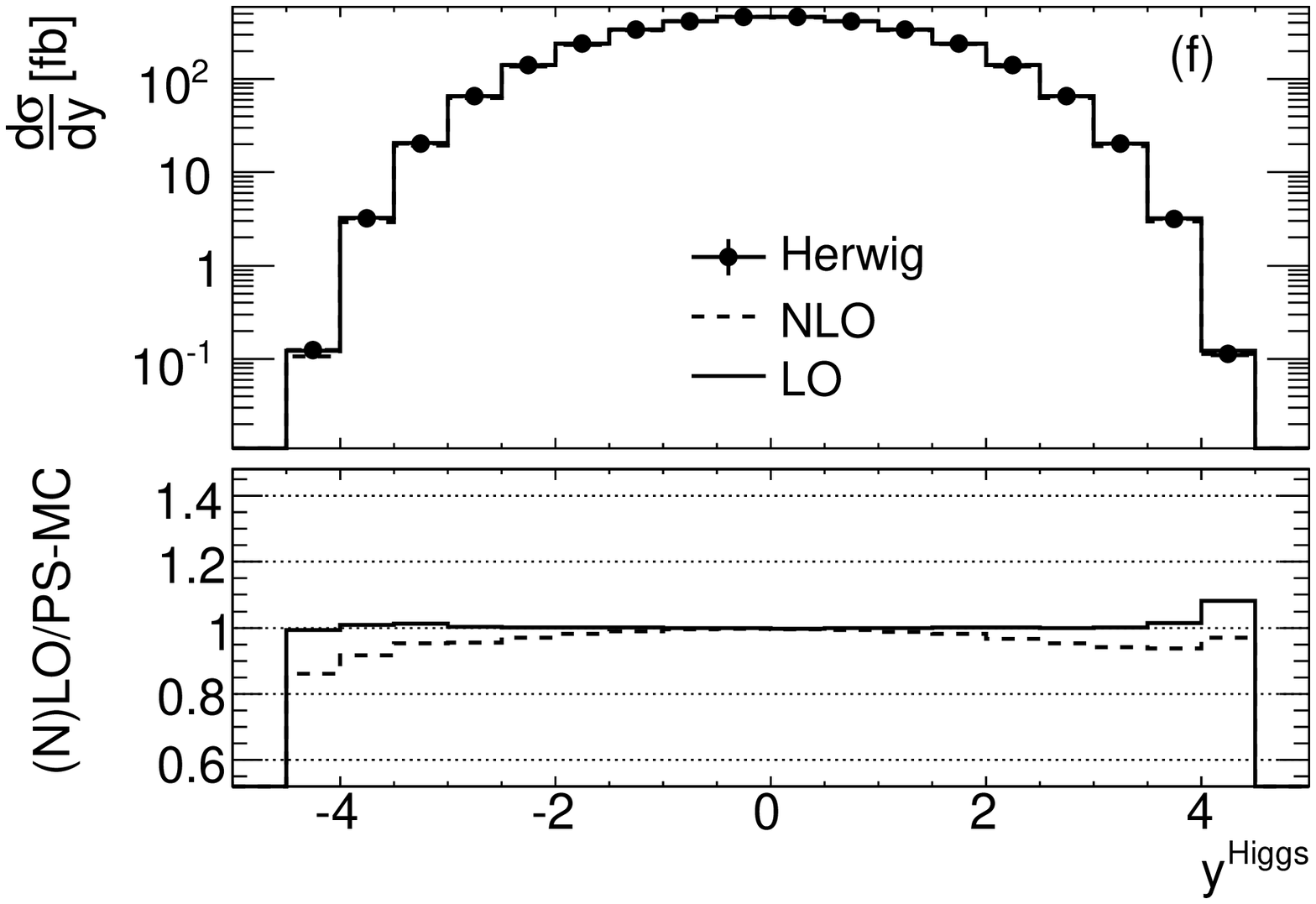}
\includegraphics[width=0.5\textwidth]{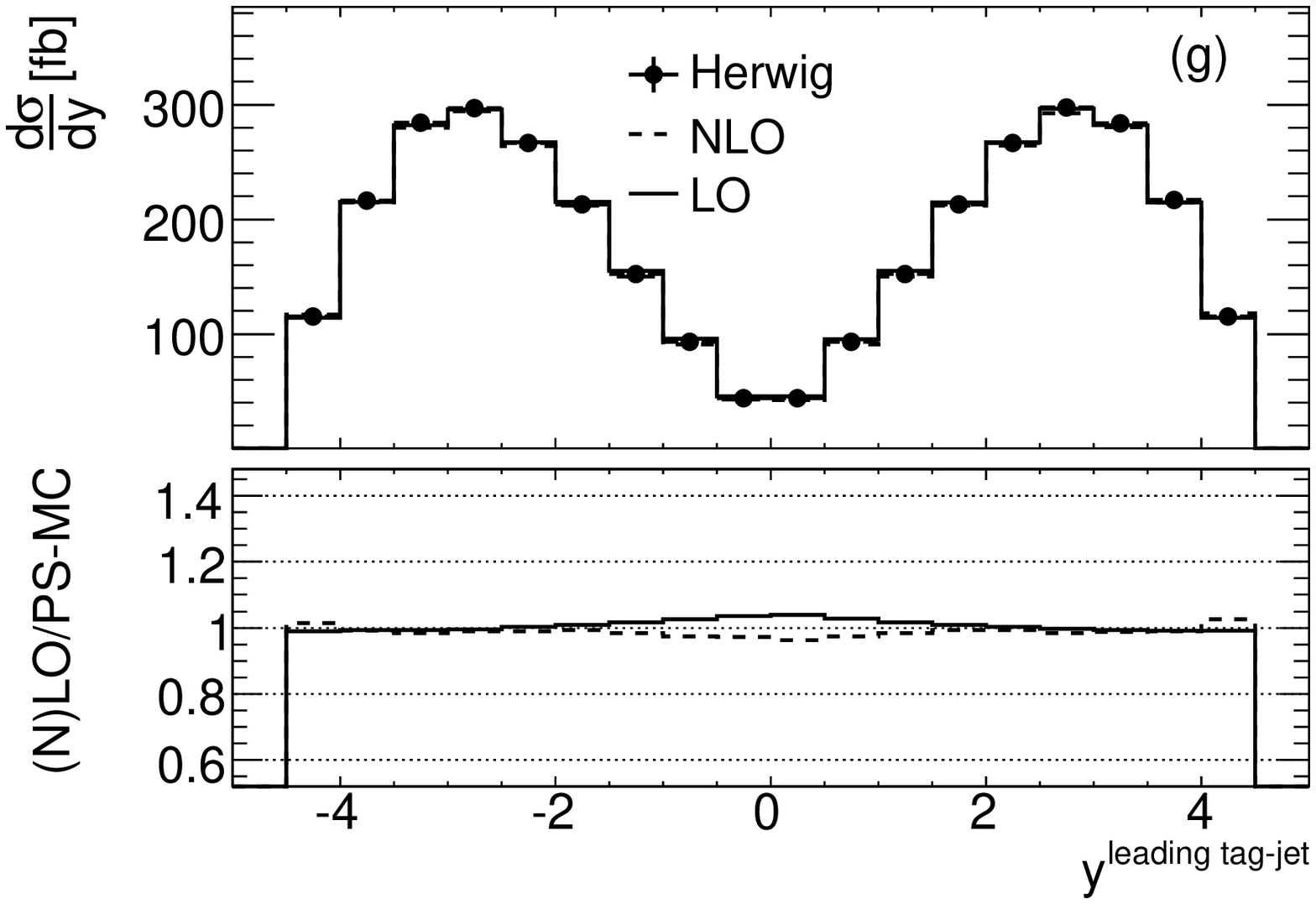}
\includegraphics[width=0.5\textwidth]{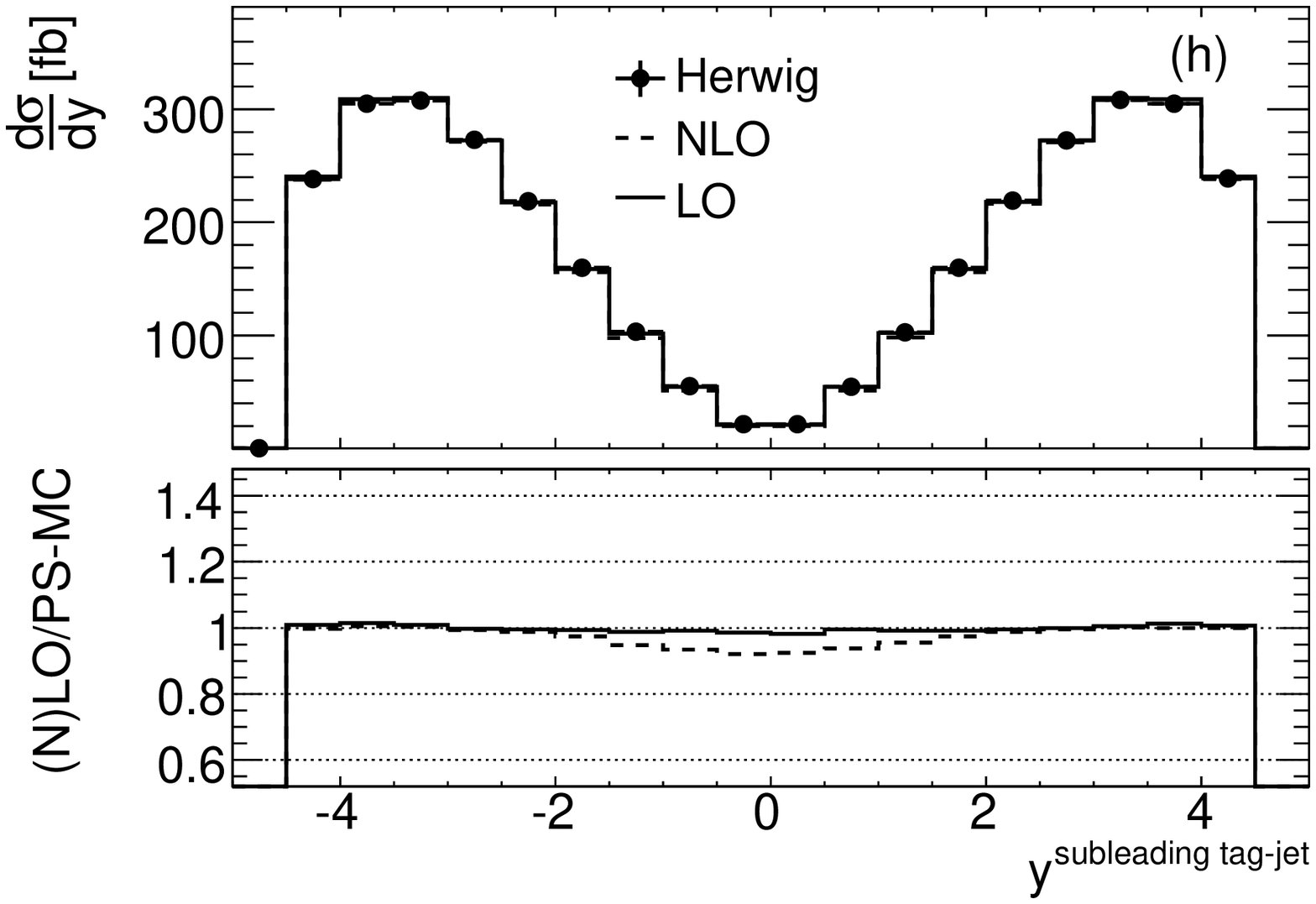}
\caption{\label{fig:vbf_09:beforereweighting}Differential distributions after VBF cuts, before the reweighting procedure described in the text.}
\end{figure}

The results of this comparison are shown in Figure~\ref{fig:vbf_09:beforereweighting}, where the transverse momenta and rapidities of the Higgs bosons and the tag-jets, the invariant mass of the tag-jets and the difference of their azimuthal angle is plotted. In the lower part of the sub-figures, the ratio of the fixed order calculation to Herwig is shown.

The shapes of the Herwig PS-MC prediction are very close to the LO prediction, as should be expected, as Herwig uses a LO matrix element. The parton shower does not seem to influence the shapes of the distributions significantly. The biggest difference can be seen in the invariant dijet mass of the tag-jets, but overall the agreement is within 10\%.

Compared to the NLO prediction, Herwig predicts a significantly harder transverse momentum spectrum both for the Higgs boson and for the tag-jets\footnote{Of course the transverse momenta of the Higgs boson and the tag-jets are correlated.}. Also the invariant dijet mass is preferred to be slightly larger in Herwig.

\subsubsection{Reweighting of PS-MC}

To partially account for the differences in the transverse momentum spectra, a simple reweighting method has been applied, where the Herwig events are weighted using the ratio to the NLO prediction in only one variable. This observable has been chosen to be the transverse momentum of the Higgs boson, since the differences are largest in this variable. The weights assigned to the Herwig events are chosen according to:

\begin{equation}
w=\frac{\frac{\mathrm{d}\sigma}{\mathrm{d}p_T^{\mathrm{H}}}(\mbox{NLO, MRST2004qed)}}
{\frac{\mathrm{d}\sigma}{\mathrm{d}p_T^{\mathrm{H}}}(\mbox{Herwig, CTEQ6L1)}~~~}
\end{equation}
The dashed Histogram in the lower part Figure~\ref{fig:vbf_09:beforereweighting} (a), which is the ratio between the NLO prediction from~\cite{Ciccolini:2007jr,Ciccolini:2007ec} and the Herwig prediction after VBF cuts, was fitted with a 3rd order polynomial in $p_T^H$ to be used as a reweighting function for the Herwig events. In principle also the LO prediction could be taken from~\cite{Ciccolini:2007jr,Ciccolini:2007ec}, but the shape of the transverse momentum distribution of the Higgs boson is identical to the one from Herwig in this case. 



\begin{figure}
\includegraphics[width=0.5\textwidth]{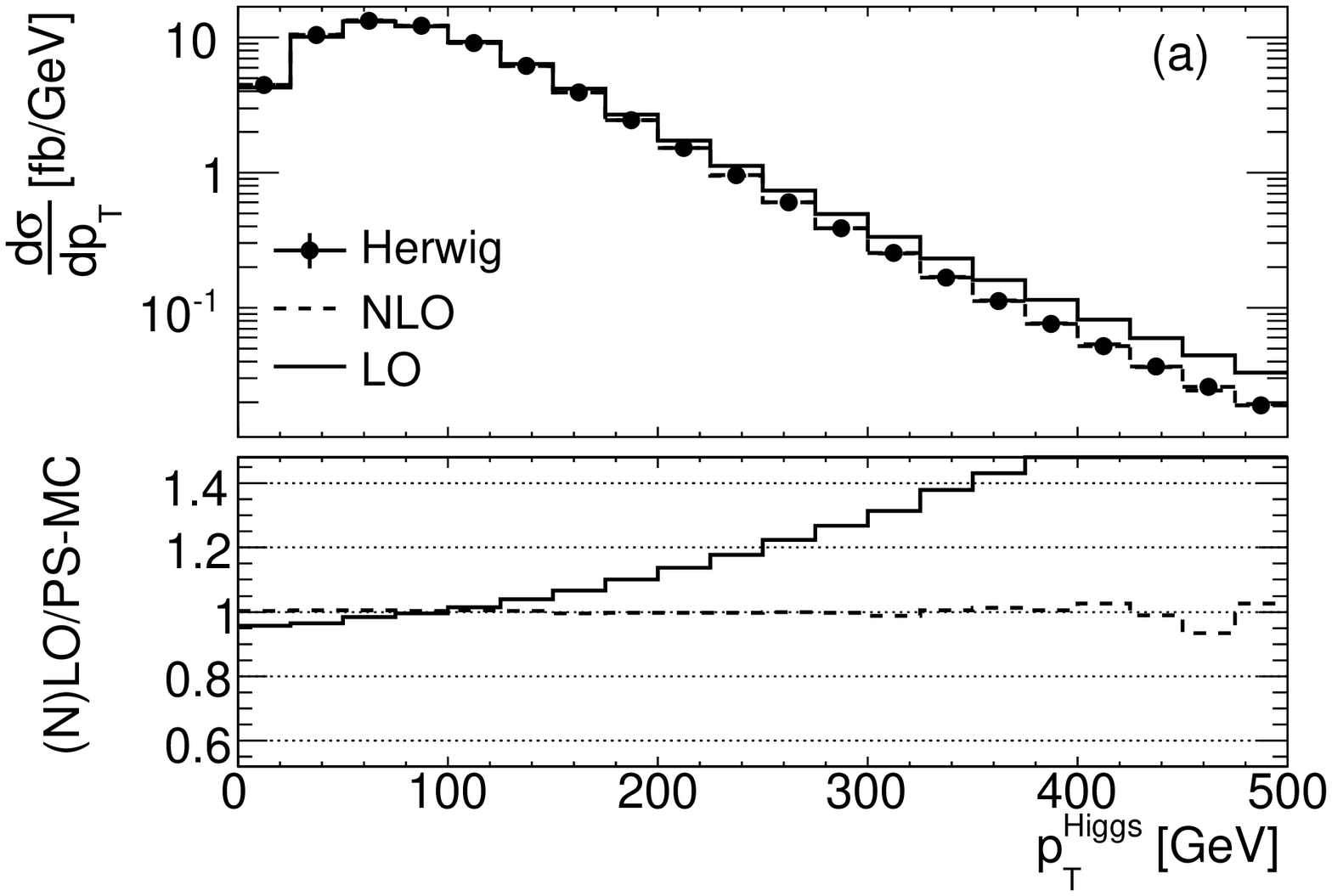}
\includegraphics[width=0.5\textwidth]{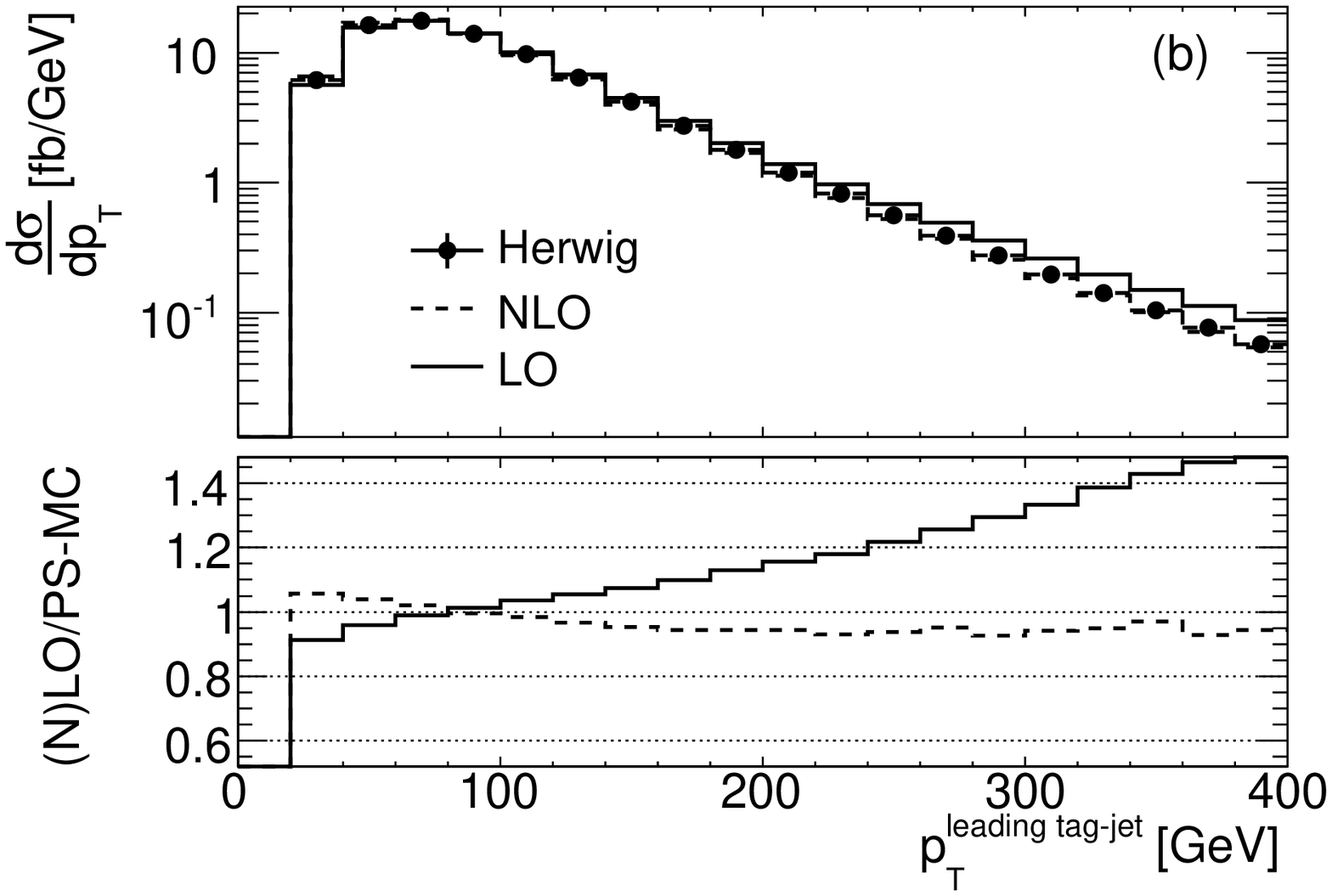}
\includegraphics[width=0.5\textwidth]{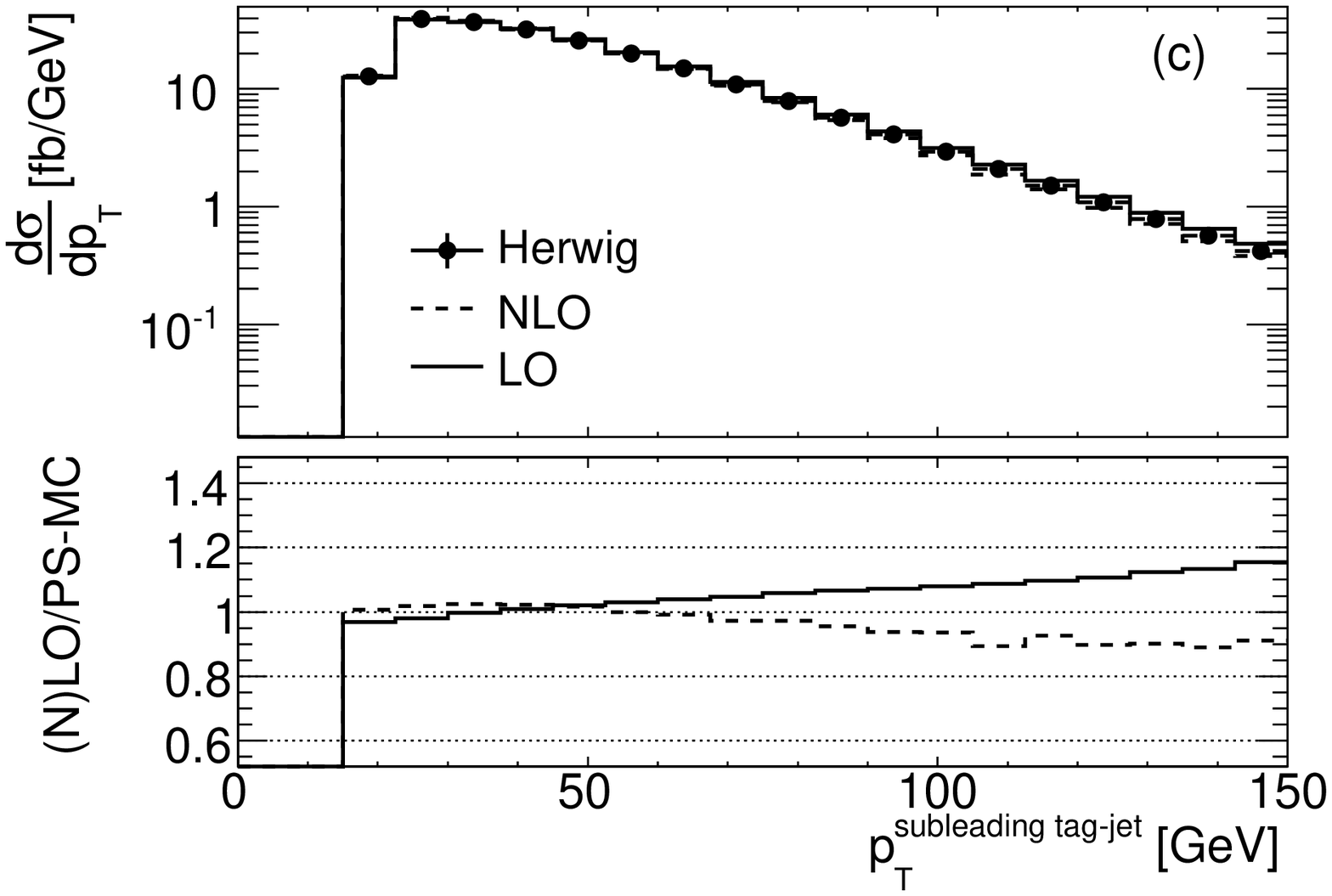}
\includegraphics[width=0.5\textwidth]{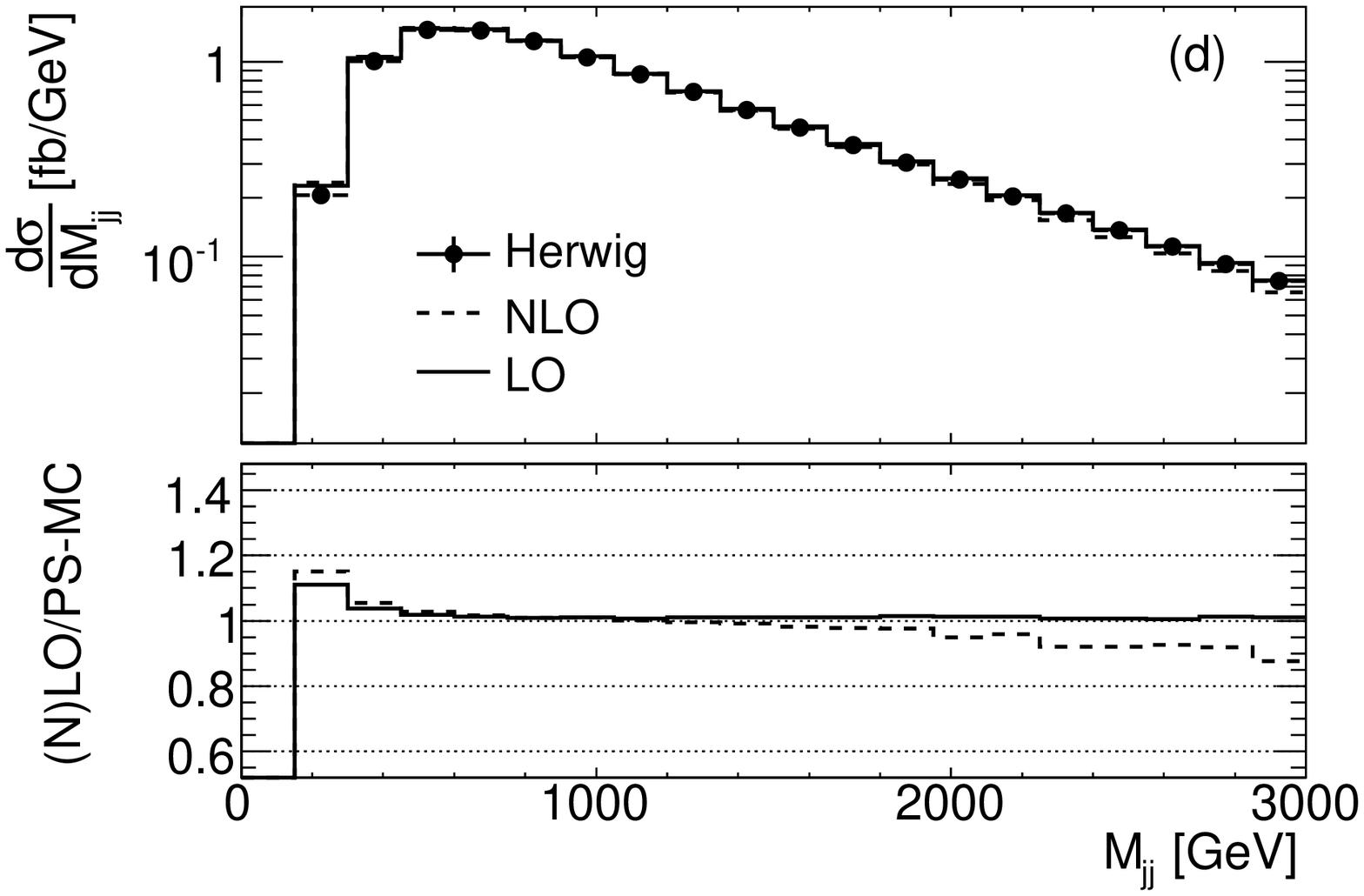}
\includegraphics[width=0.5\textwidth]{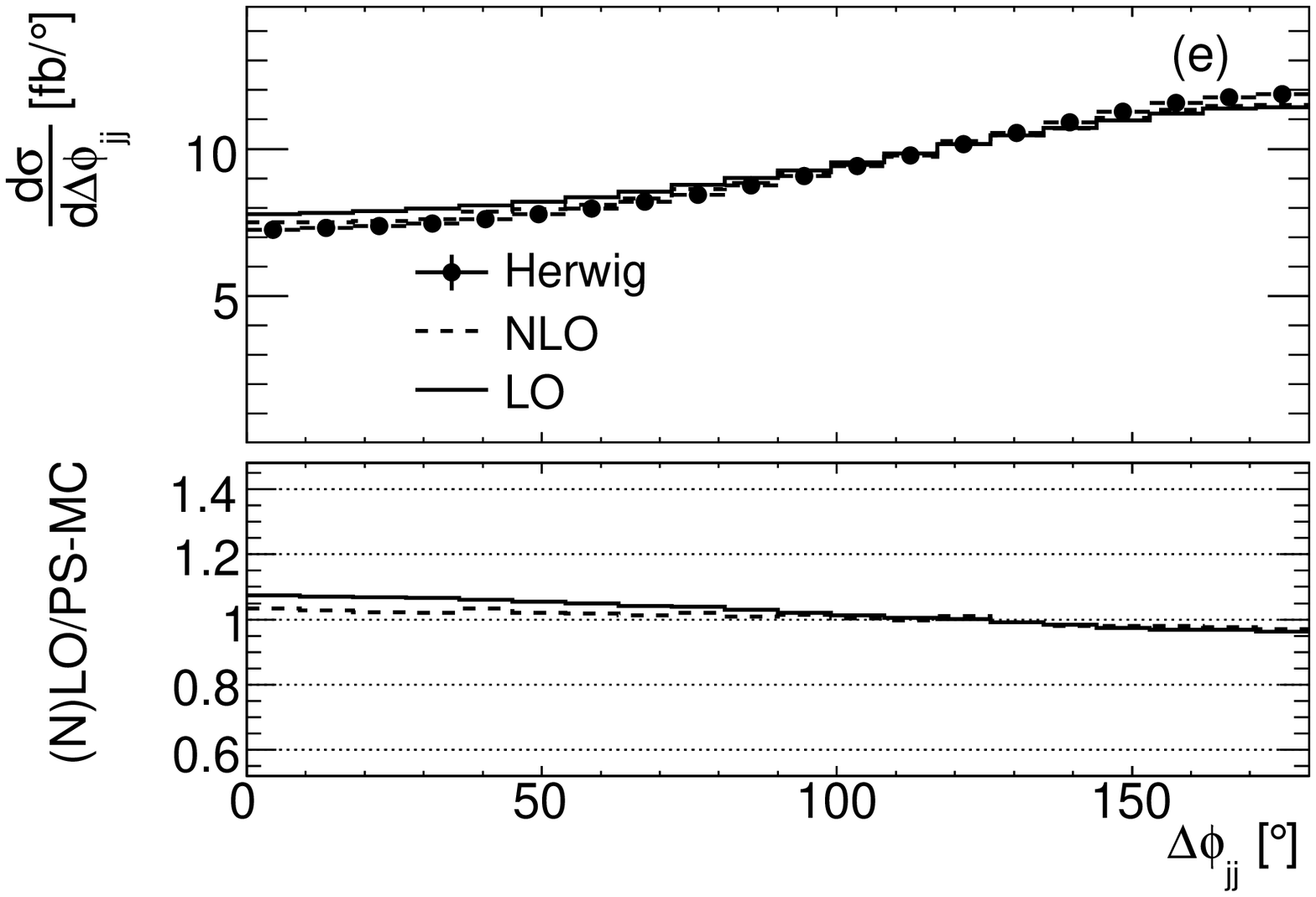}
\includegraphics[width=0.5\textwidth]{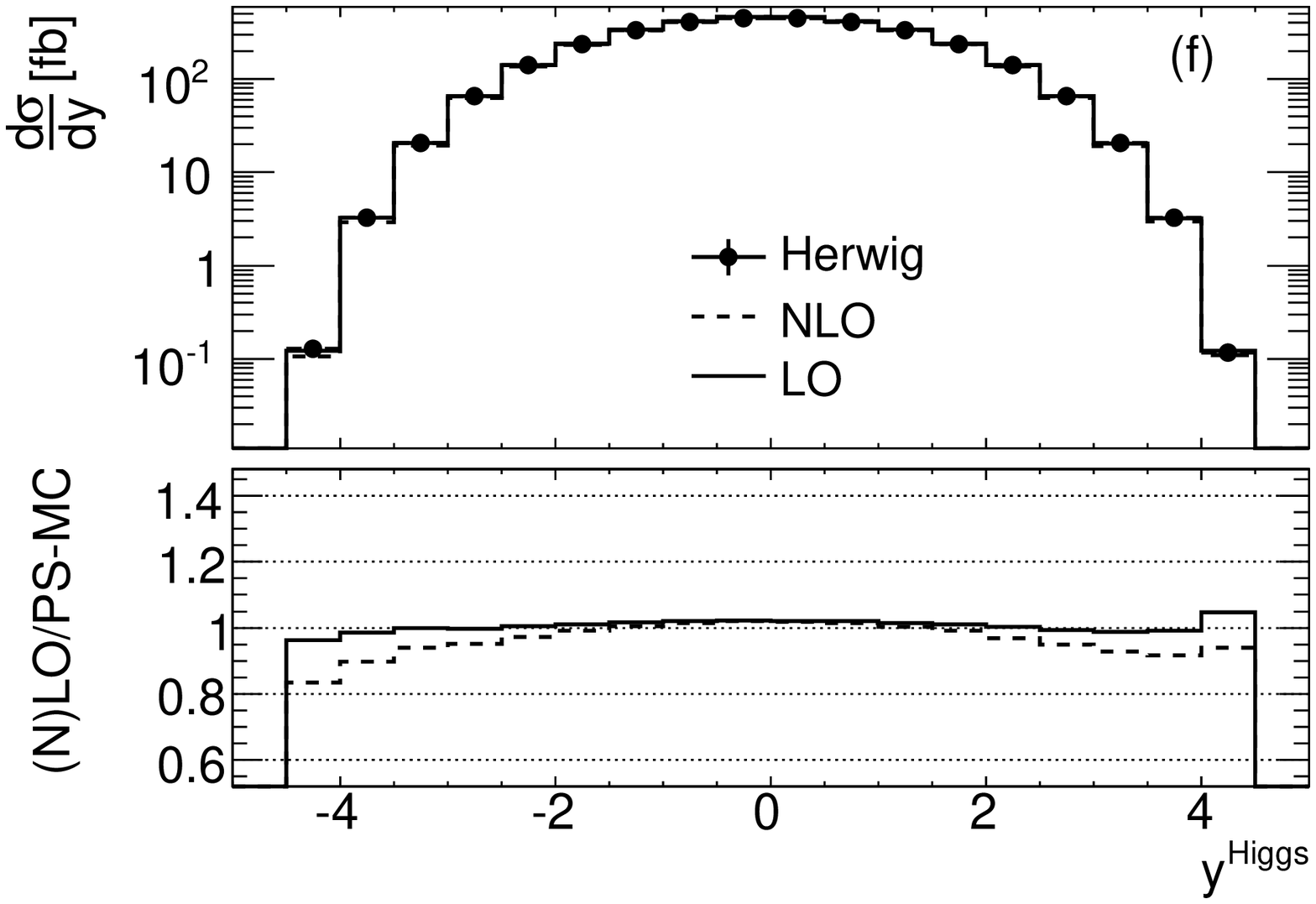}
\includegraphics[width=0.5\textwidth]{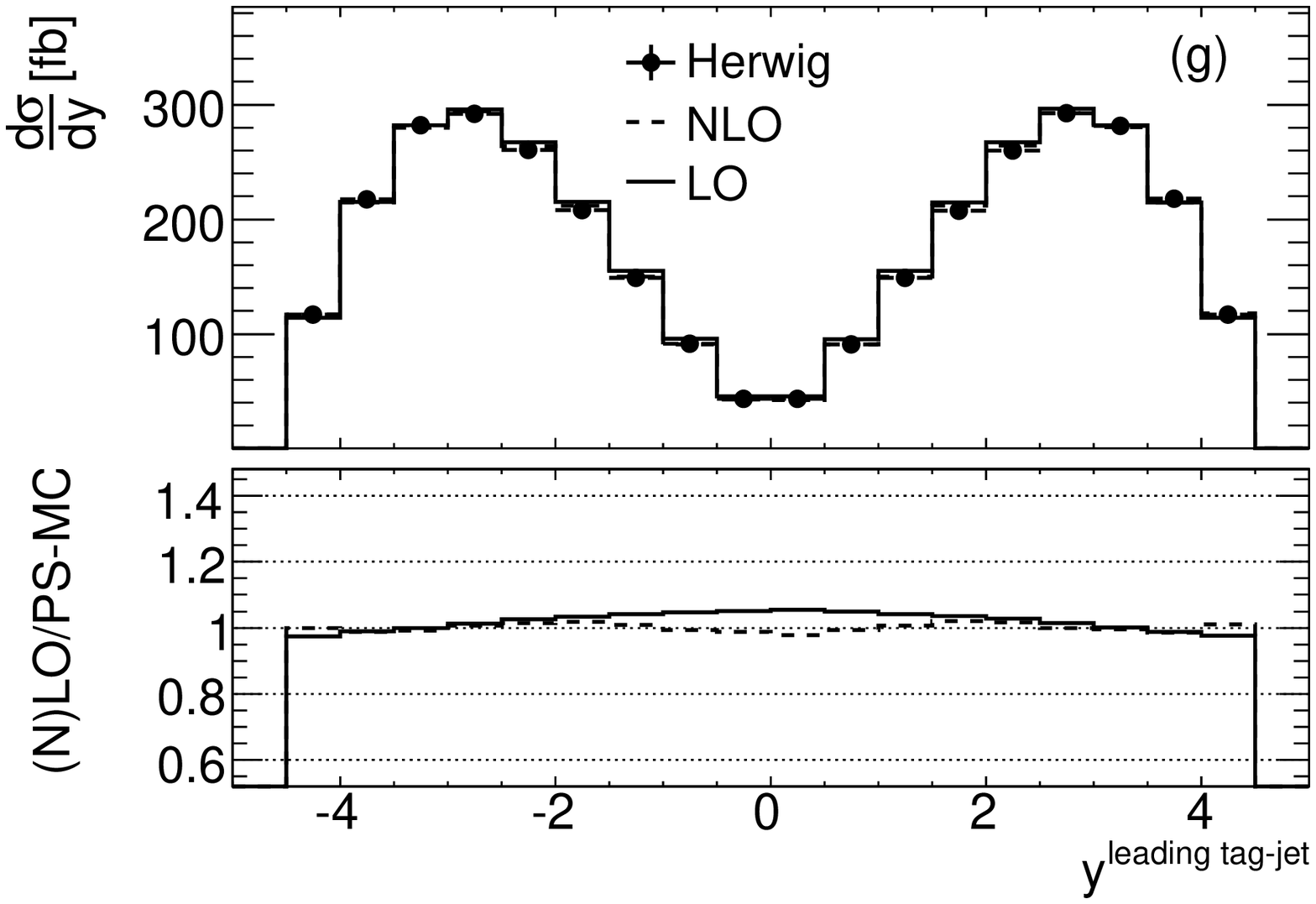}
\includegraphics[width=0.5\textwidth]{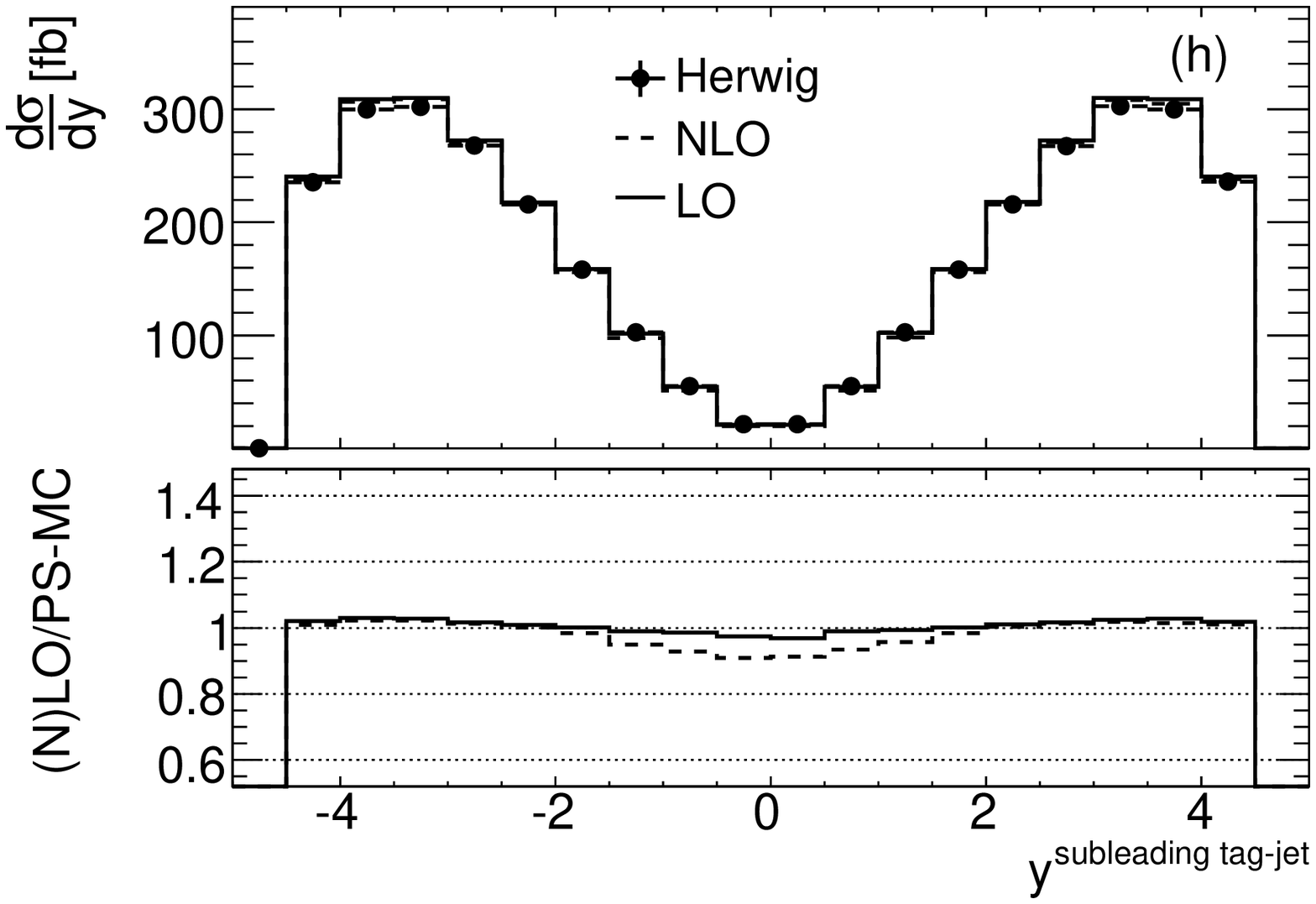}
\caption{\label{fig:vbf09:afterreweighting}Differential distributions in after VBF cuts, including the reweighting procedure described in the text.}
\end{figure}

Figure~\ref{fig:vbf09:afterreweighting} shows the  comparison of the differential distributions after the reweighting procedure. By construction, the Herwig prediction for the transverse momentum of the Higgs boson now fits exactly the one of the NLO prediction. But due to the kinematic correlations, also an improved description of the tag-jet transverse momenta and to a lesser extent the invariant dijet mass is obtained. The reweighted Herwig prediction is almost everywhere within 10\% of the NLO prediction.

\subsection*{CONCLUSIONS}
Acceptances and differential cross sections in the VBF process have been shown to agree between Herwig and the fixed order result from~\cite{Ciccolini:2007jr,Ciccolini:2007ec} in LO. The differences between the NLO predictions and the LO result in differential cross sections can be partially taken into account by a reweighting of Herwig events using a weight that depends on the transverse momentum of the generated Higgs boson. In this way an improved description can be obtained, though it would be better to have a fully merged NLO PS-MC available. It should be noted that in the VBF process, the electroweak NLO corrections have comparable influence to the QCD corrections on the cross section and on the shape of differential distributions~\cite{Ciccolini:2007jr,Ciccolini:2007ec}, thus a NLO PS-MC would have to take the electroweak corrections also into account to give the best available description of the VBF process.

\subsection*{ACKNOWLEDGMENTS}
M.~Warsinsky acknowledges the support of the Initiative and Networking Fund of the Helmholtz Association, contract HA-101 (``Physics at the Terascale''). We would like to acknowledge the use of the computing resources provided by the Black Forest Grid Initiative.



%% file: huston/huston.tex
\subsection{Introduction}
\label{sec:introduction}

Many interesting physics signatures at the LHC involve the presence of single or
multiple photons in the final state. These photons may either be produced
directly, through the fragmentation of a quark or gluon, or else through the
decay of a resonance - such as e.g. the Higgs boson. 
There are backgrounds to the measurement of photons, primarily through the 
fragmentation of jets to a leading $\pi^o$ or $\eta$, which carries most of the 
energy of the parent parton. Photon identification cuts, which examine the 
lateral and longitudinal development of the candidate photon shower, reject 
much of this background, with a typical efficiency for retaining real photons 
of the order of 80\% for photon transverse energy larger than 40 GeV/c.

\vspace{0.3cm}
\noindent
A large additional rejection factor can be obtained for this jet background by
the imposition of an isolation cut on the candidate photon~\cite{Aurenche:1989gv,Baer:1990ra}; 
in this isolation
cut, a restriction is placed on the amount of additional energy that can be
present in a cone about the photon direction. 
The tighter the isolation cut, the more background is removed 
from the photon candidate sample. The isolation cut also has the
effect of removing most of the photon contribution arising from the fragmentation
subprocesses, but should be structured so as to have a high efficiency 
for the retention of real, isolated photons. 
However, a tight isolation cut also has the
undesireable effect of making the theoretical prediction unstable, due to the
restriction of the available phase space for soft gluon emission. 
Typically, the
isolation cut may be formulated as requiring either the transverse energy in 
the isolation cone to be less than a fixed fraction $\epsilon_s$ of the 
candidate photon transverse energy, $E_{T}^{\gamma}$, or requiring there to be 
less than a fixed amount of additional energy present. The latter requirement 
is typically used at the Tevatron and is motivated by the fact that most of the 
energy in the isolation cone results from the underlying event (and pileup), 
and so is independent of the photon energy~\footnote{We note here that the description of underlying events at LHC, available in the event generators are yet to be tuned with LHC  data. Further the LHC is foreseen to be run at several energies and thus the underlying event will vary accordingly.}. 

\vspace{0.3cm}
\noindent
Another way to define direct photons is the so-called ``democratic approach"~\cite{Glover:1993xc,GehrmannDeRidder:1997wx}, 
where  photons   and QCD partons are treated on the same footing when being clustered 
into jets, and direct photons are then defined by jets containing a photon which carries 
a large fraction (typically more than 70\%) of the jet energy.
A detailed study of this approach  in the context of matrix element to parton shower merging 
has been performed  recently in~\cite{Hoeche:2009xc}.

\vspace{0.3cm}
\noindent
Another way of framing the isolation cut is due to Frixione~\cite{Frixione:1998jh}: a cone of fixed
radius $R_o$ (which typically has been of the order of 0.4) is drawn around the photon axis.
Then for all radii $R$ inside this cone, the amount of additional transverse 
energy, assumed to be due only to hadrons, inside the cone of radius R is required to satisfy the following condition
\begin{equation}\label{isol-e1}
E_T^{had} < f(R)    \;,                         
\end{equation}
where the energy profile $f(R)$ is some continuous function of $R$, growing with
$R$, and which falls to zero as $R \to 0$, typically like $R^{2n}$, 
for some $n > 0$. The following form~\footnote{It was namely the form used in 
\cite{carminati-study} in an earlier study, with $(n=0.2, \epsilon_s= 0.05)$.} for $f(R)$ has been used in this study:

\begin{equation}\label{isol-e2}
f(R) = 
\epsilon_s \,  E_{T}^{\gamma} \, 
\left[ \frac{1 - \cos R}{1- \cos R_o} \right] ^{n}
\end{equation}
In the formula above, $\epsilon_s$ and $n$ are positive numbers of order one.
This isolation criterion allows soft gluons to be arbitrarily close to the
photon direction, but requires the energy of partons emitted exactly collinear to the
photon direction, to vanish. This ideally prevents the appearance of 
any final state collinear divergence in the partonic calculation; as a result, it prevents the involvement of any fragmentation contribution, insofar as the 
latter is treated as a collinear process.
This greatly simplifies the theoretical calculation as the 
fragmentation  part is quite cumbersome to calculate at NLO; this is considered as one of the major advantages of the Frixione isolation criterion~\footnote{The fragmentation contribution also requires knowledge of the fragmentation functions at high $z$, a region where they are currently poorly known.}.  It is thus an important goal to be able to adapt both the theoretical and experimental
analysis machinery coherently at the LHC to be able to utilize this type of isolation.  This is the  major motivation for this contribution.

\subsection{Experimental Considerations}
\label{sec:considerations}

In order to adapt this criterion to the experimental situation, several 
considerations need to be taken into account. 
First, because of the finite size of the calorimeter cells used to measure the electromagnetic shower, the Frixione isolation 
cut must be applied only beyond a minimum distance of approximately 0.1 (in $\{ \Delta\eta, \Delta\phi \}$ space).
This allows a contribution from fragmentation in the innermost cone, and one 
has to check to which extent the fragmentation component is still suppressed.
In addition, the transverse energy in the experimental isolation cone is 
deposited in discrete cells of finite size and this granularity must be taken 
into account in the theoretical calculation.  
The continuity criterion, initially proposed by Frixione, has thus been 
replaced by a discretized version consisting of a finite number of nested cones, 
together with the collection of corresponding maximal values for the transverse 
energy allowed inside each of these cones.  

\vspace{0.3cm}
\noindent
As mentioned previously, the dominant contribution to the energy deposited in
the photon isolation cone is from the non-perturbative/semi-perturbative
underlying event (UE), and, at higher luminosities, from additional minimum bias
events occurring in the same beam crossing (pile-up) as foreseen in future LHC
running. These sources result in energy deposits of a fairly uniform density
over the area of the detector, which are uncorrelated with the collinear
fragmentation processes that the Frixione isolation cut is designed to remove.
Thus, it seems sensible to separate the analysis of the two sources of energy in
the isolation cone.  
Hence, a determination of the transverse energy
density may suffice for an estimation of the amount of underlying event/pileup
transverse energy inside the isolation cone. One convenient way of determining
this density was suggested by Cacciari, Salam and Soyez~\cite{Cacciari:2008gn}, in which the transverse
energy density is calculated, on an event-by-event basis, by measuring the
transverse energy in soft jets ($E_T <$ 20 GeV) using the $k_T$ algorithm with a
D-parameter of 0.5. As the harder jets are not included in this density
determination, the result is a measure of the amount of energy to be expected in
the isolation cone from sources independent of the production of the photon.
This energy can then be subtracted, as a flat background, from the amount of
energy found in the isolation cone of the photon candidate, and the Frixione
isolation criterion, modified for the experimental granularity, can then be run
on the remaining energy distribution~\footnote{It should be emphasized that this subtraction needs to take place independent of the type of isolation criterion that is applied.}. Otherwise, to avoid the occasional possibility of arriving at a negative
energy in the isolation cone, the UE/pileup energy determined by this
technique can be added, again as a flat background, to the amount determined
in the experimental analysis (along with the density allowed by the Frixione
profile). Again, it should be emphasized that the UE/pileup transverse energy
density determined in this manner is on an event-by-event basis, and thus
independent of the luminosity conditions or any fluctuations that may have
occurred in that particular crossing. Thus, to define an isolated photon for any theoretical calculation,  only
the Frixione isolation criterion needs to be applied, as the experimental/non-perturbative sources of transverse energy accumulation have already been  accounted for.  

\subsection{Implementation}
\label{sec:implementation}

It is not clear a priori what the best parameters for the Frixione isolation
criterion are. For this contribution to the Les Houches proceedings, we have
examined the impact of varying $\epsilon_s$ and $n$ in the isolation cut applied
to single photon production in the program Jetphox~\cite{jetphox,Aurenche:2006vj,Belghobsi:2009hx}. 
The parameter pairs examined
for ($n,\epsilon_s$) are: 
\begin{itemize}
\item (0.2,0.05)
\item (0.2,1.0) 
\item (1.0,1.0)
\item (1.0,0.5)
\item (1.0,0.05).
\end{itemize}
\noindent
We have calculated the direct and fragmentation components for single
photon production, after the imposition of the discretized version of the Frixione isolation criterion for the above parameter pairs. We considered $pp$ collisions at 10 TeV LHC operation\footnote{From the point-of-view of the photon background subtraction techniques, the comparisons presented here should be relatively independent of the center-of-mass energy.} and 
the photon transverse energy range of 60 GeV/c to 240 GeV/c, using 
CTEQ66 PDFs and a common factorization/renormalization/fragmentation scale 
of $p_T^\gamma/2$~\footnote{Up to small differences at NLO.}. 
The radius of the outermost isolation cone around the photon direction was set to $R_o = 0.4$. To simulate the detector granularity, 
we considered an isolation criterion made up of 6 nested cones of respective radii:
\begin{itemize}
\item $R_1 = 0.1$
\item $R_2= 0.16$
\item $R_3 = 0.22$ 
\item $R_4 = 0.28$ 
\item $R_5 = 0.34$ 
\end{itemize}
and $R_6 = R_o = 0.40$, with the corresponding maximal values of $E_T ^{had}$ 
allowed in each of these cones given by

\begin{equation}\label{isol-e3}
E_{T}^{ j} = 
\epsilon_s \,  p_{T \, \gamma} \, 
\left[ \frac{1 - \cos R_j}{1- \cos R_o} \right]^{n}
\end{equation}


\vspace{0.3cm}
\noindent
To carry out this study in practice, Jetphox has been modified in the following way. The discrete Frixione criterion has been parametrized in the form of a 2-dimensional array whose entries are
the radii of each of the nested cones and the corresponding maximal
transverse energy allowed inside each cone. The size of the array can be varied
up to a maximum of 10 and  is automatically
handled by a Perl script. These maximal energies are calculated as the values
 taken at the radii $R_j$ by a profile function which can be specified at will by
the user, and which was taken to be the  function presented above. The
criterion has been implemented at the level of the computation of the grid which is used for the partonic Monte Carlo event generation.
\vspace{0.3cm}
\noindent
\vspace{0.3cm}
\begin{figure}[htb]
\begin{minipage}[t]{.46\linewidth}
\includegraphics[scale=0.38]{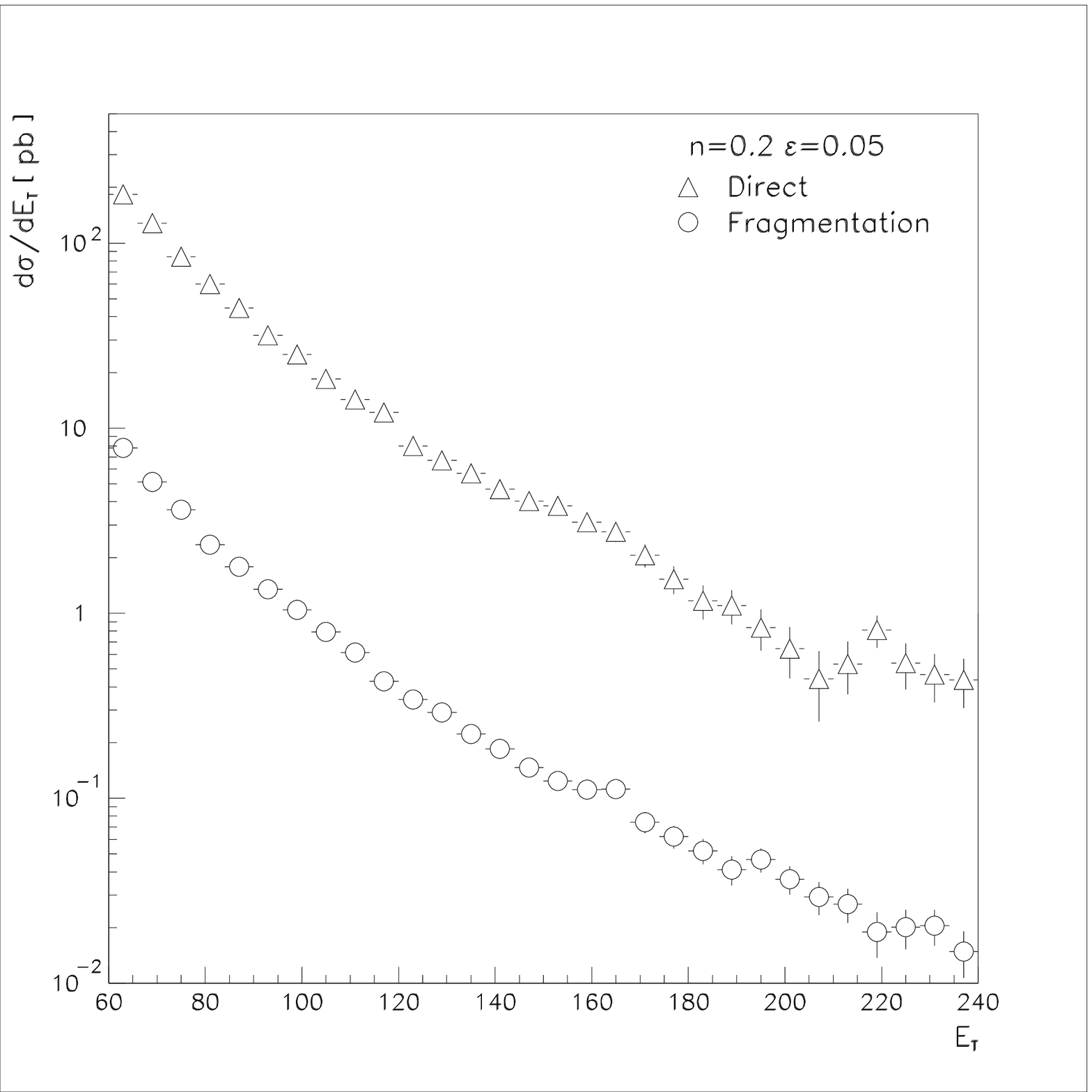}
\caption{\label{isol-fig1a} The Jetphox prediction for the photon $E_T$ 
distribution, for the parameter choice $n = 0.2, \epsilon_s = 0.05$ 
in the discrete form of the Frixione criterion.
The triangles denote the direct component, the circles the fragmentation
component.}
\end{minipage}
\hfill
\begin{minipage}[t]{.46\linewidth}
\includegraphics[scale=0.38]{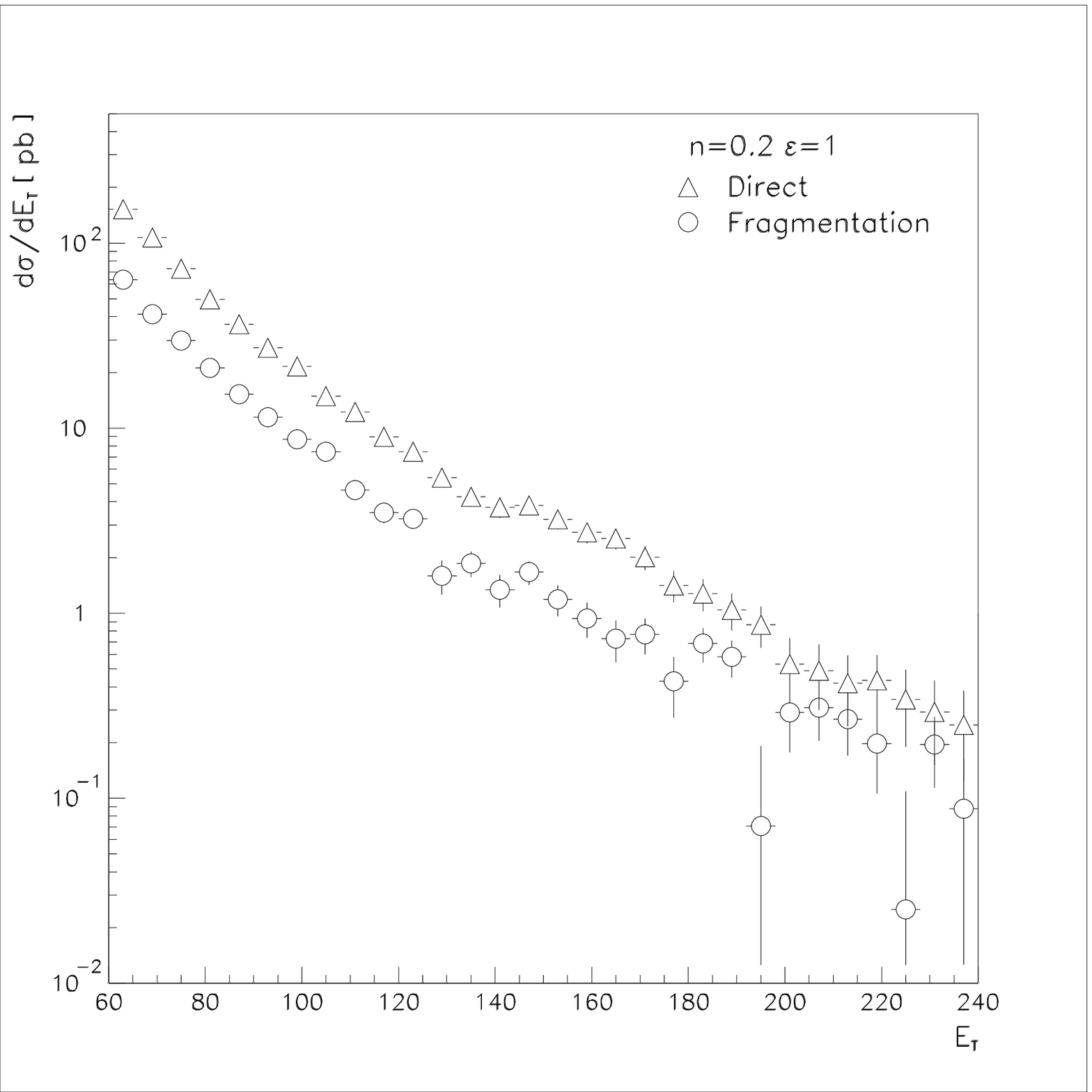}
\caption{\label{isol-fig1b}
The Jetphox prediction for the photon $E_T$ 
distribution, for the parameter choice $n = 0.2, \epsilon_s = 1.0$ 
in the discrete form of the Frixione criterion.
The triangles denote the direct component, the circles the fragmentation
component.}
\end{minipage}
\end{figure}

Jetphox acounts for the LO and NLO contributions for both the direct and
fragmentation contributions. For the direct contribution, isolation is not relevant at LO, since, due to transverse
momentum conservation, the recoiling parton lies
opposite the photon in $\phi$.  At NLO, at most one of the two final state partons can contribute to the energy in the  isolation region (the other parton recoils
 in the away-side region). For the fragmentation contribution, the collinear remnants of fragmentation lie completely inside the innermost cone of radius $R_1 = 0.1$,  and are accounted for in the calculation by the quantity $1-z$, where $z$ is defined as the fraction of the transverse energy of the
fragmenting parton carried away collinearly by the photon. 
At NLO, the extra parton, labelled ``5" in the figure below (the spectator w.r.t. the fragmentation process),  can be emitted at any angle with respect to the parent parton. Hence, this 5th parton  can fall either into the cone defined by 
$R<R_1$~\footnote{In this case, the parton will fall inside the electromagnetic shower created by the photon and will not be visible; depending on the energy of the parton, the manner in which it hadronizes and the specific identification cuts applied to the photon, the presence of this parton may cause the photon to be rejected.}, or into any of the annuli, $\{R_1 < R < R_2 \}$ 
to $\{ R_5 < R < R_6 \}$, or outside the cone defined by the maximal radius 
$R_6$. The implementation of the criterion on the fragmentation contribution amounts to the following possibilities:

\begin{itemize}
\item 
if the extra parton ``5" falls inside $R_1$, the criterion imposes
$$
\frac{1-z}{z}\, E_T^\gamma + p_T^5 < E_T^1
$$
\item
if the extra parton ``5" falls in the annulus $\{ R_{j} < R < R_{j+1} \}, j=1,\ldots, 5$, the
criterion imposes

\begin{eqnarray}
\frac{1-z}{z}\, E_T ^\gamma             & < &  E_T  ^1\nonumber\\
\frac{1-z}{z}\, E_T ^\gamma + p_T ^5& < & E_T ^{j+1} \nonumber
\end{eqnarray}
\item 
if the extra parton ``5" falls outside cone $R_6$, the criterion imposes
\[\frac{1-z}{z} E_T ^\gamma <  E_T ^1\]
\end{itemize}

\vspace{-0.4cm}

\begin{figure}[!h] 
\begin{center}
\includegraphics[width=0.32\textwidth]{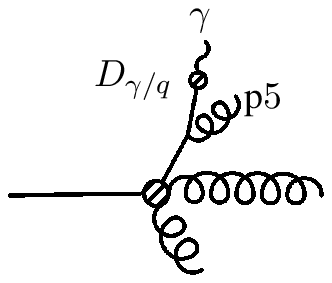}
\end{center}
\end{figure}

\vspace{-0.8cm}

\begin{figure}[htb]
\begin{minipage}[t]{.46\linewidth}
\includegraphics[scale=0.38]{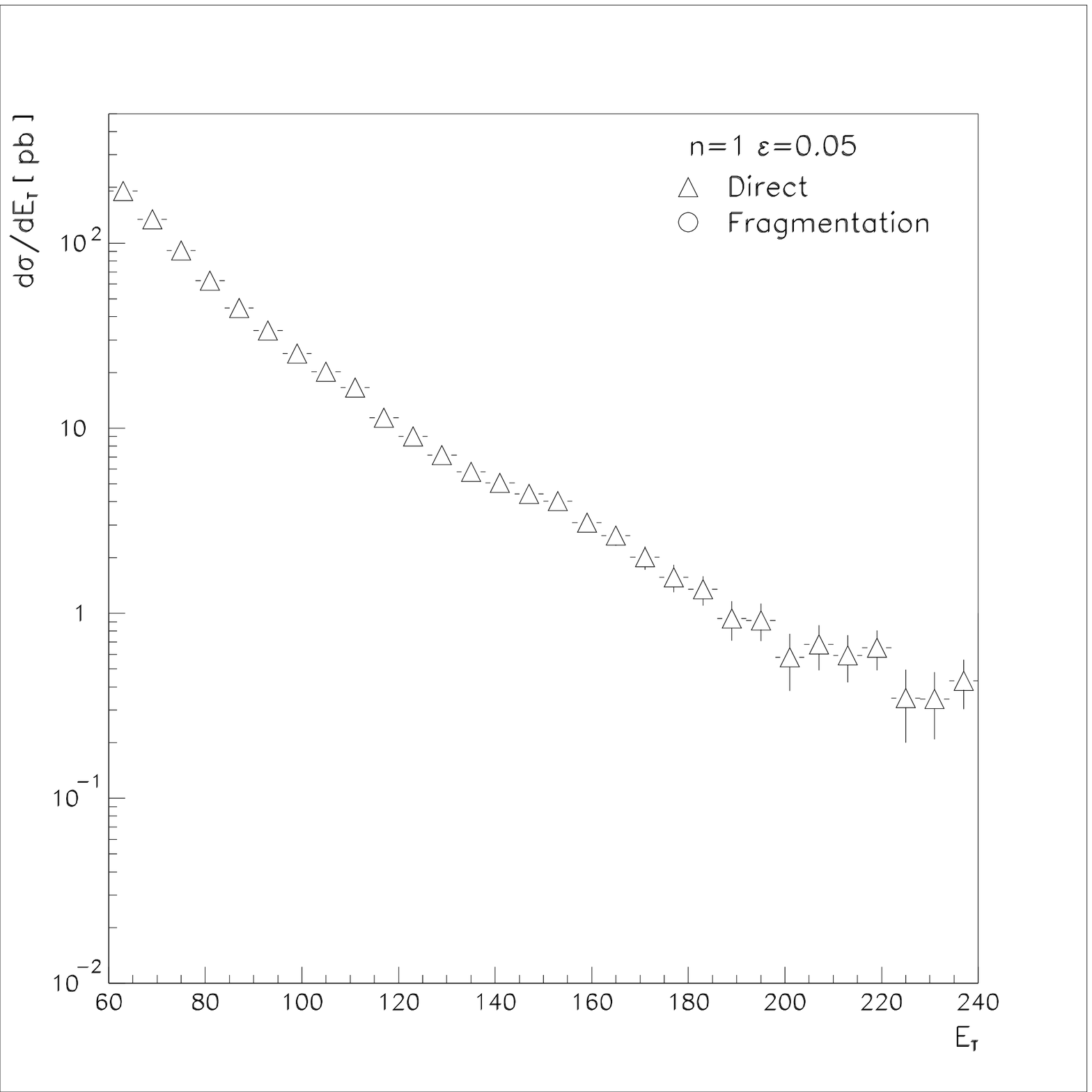}
\caption{\label{isol-fig1c} The Jetphox prediction for the photon $E_T$ 
distribution, for the parameter choice $n = 1, \epsilon_s = 0.05$.
Only the direct component is shown. The criterion was too stringent for the 
fragmentation component to be evaluated in this case.}
\end{minipage}
\hfill
\begin{minipage}[t]{.46\linewidth}
\includegraphics[scale=0.38]{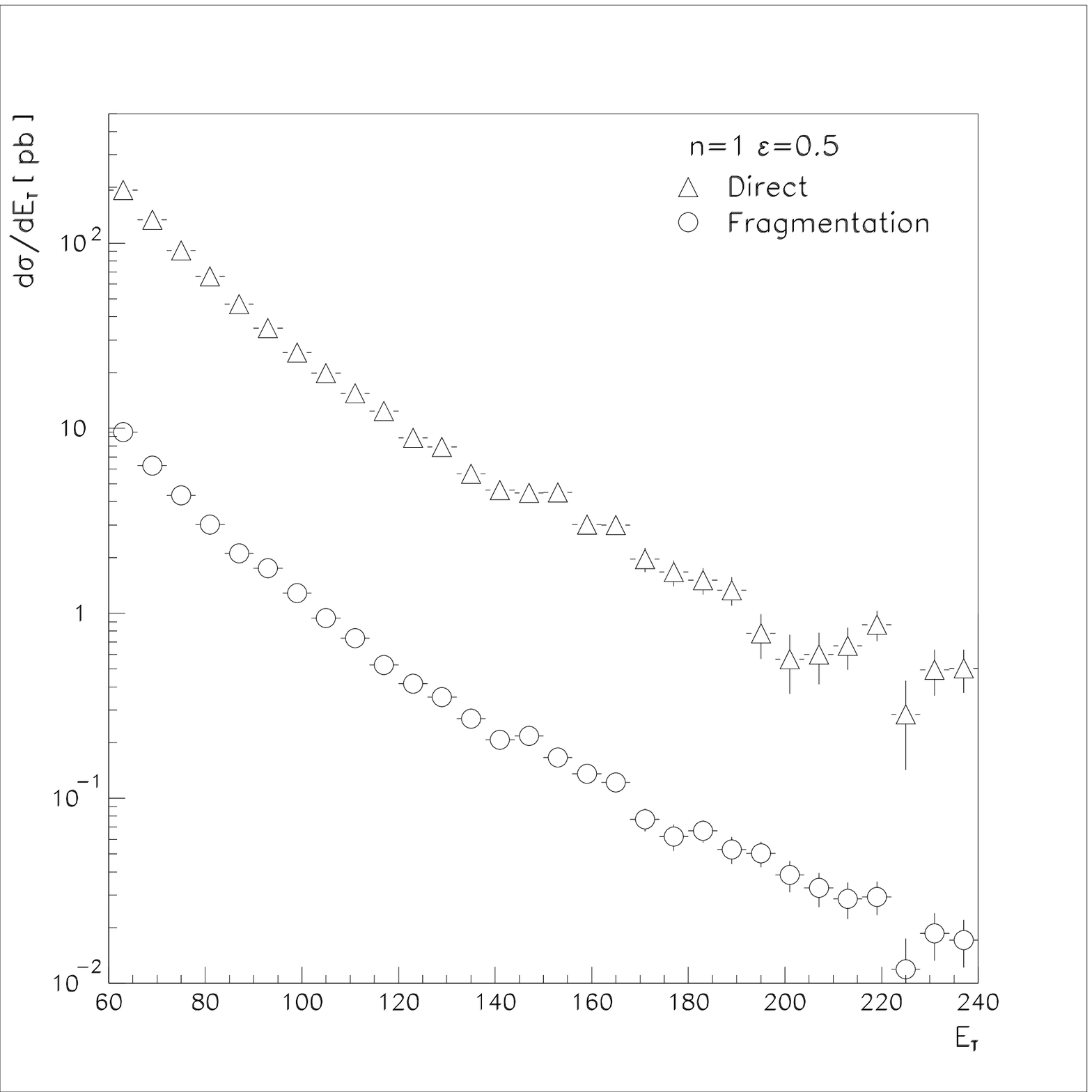}
\caption{\label{isol-fig1d} The Jetphox prediction for the photon $E_T$ 
distribution, for the parameter choice $n = 1, \epsilon_s = 0.5$. 
The triangles denote the direct component, the circles the fragmentation
component.}
\end{minipage}
\end{figure}

\subsection{Results}
\label{sec:results}

\begin{figure}[htb]
\begin{center}
\includegraphics[scale=0.38]{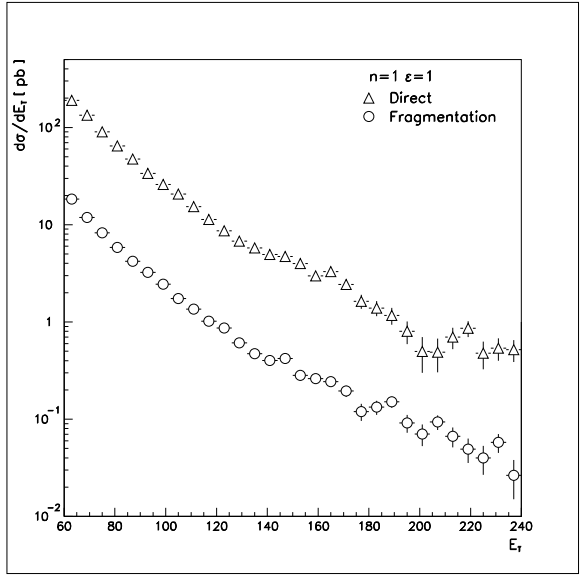}
\caption{\label{isol-fig1e} The Jetphox prediction for the photon $E_T$ 
distribution, for the parameter choice $n = 1, \epsilon_s = 1$. 
The triangles denote the direct component, the circles the fragmentation
component.}
\end{center}
\end{figure}

\noindent
The direct and fragmentation contributions are shown for the 5 parameter
combinations in Figs. \ref{isol-fig1a}-\ref{isol-fig1e}. 
As expected, changes in the isolation parameters
have little impact on the direct contributions (which are affected only at NLO,
by gluon radiation into the isolation cone), while most of the fragmentation
contribution can be eliminated by isolation - except with the choice 
$(n=0.2, \epsilon_s= 1)$ for which the photon isolation turns out to be very
loose.  In particular, in the innermost cone $R_1 = 0.1$, the photon can be
accompanied by as much as 58\% of the photon's transverse energy, i.e. the
accompanying hadronic $E_T$ in cone $R_1$ ranges from $\sim 35$ GeV for 
$E_{T}^{\gamma} = 60$ GeV to $\sim 138$ GeV for $E_{T}^{\gamma} = 240$ GeV.
The parameter choice $(n=1, \epsilon_s= 0.5)$ has a
similar isolation effect in the innermost cone, i.e. on fragmentation, as the 
choice $(n=0.2, \epsilon_s= 0.05)$ considered earler by \cite{carminati-study}, 
while the isolation energy profile of the former choice is much less stringent 
away from the photon's direction than the latter.

\begin{figure}[htb]
\begin{minipage}[t]{.46\linewidth}
\includegraphics[scale=0.38]{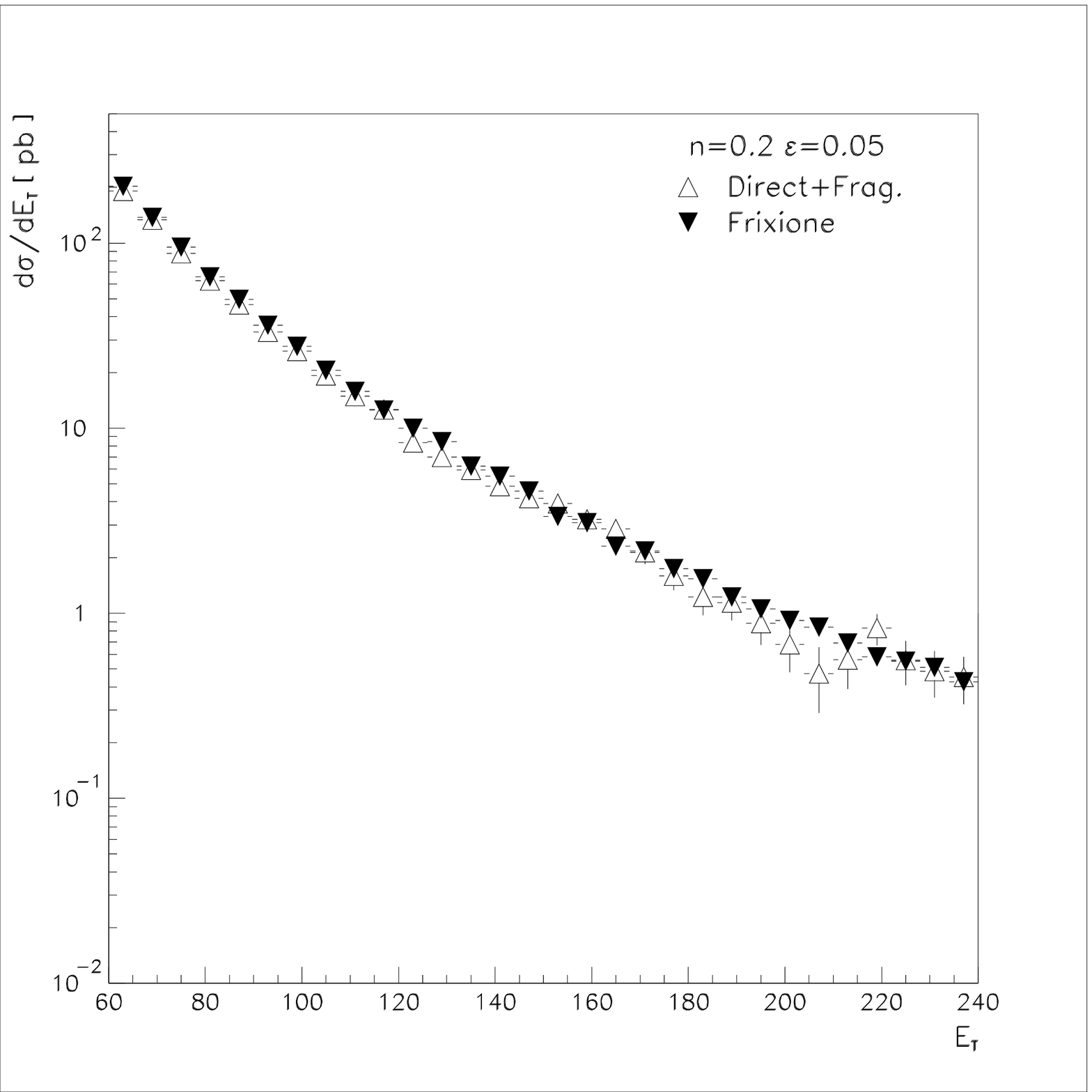}
\caption{\label{isol-fig2a} 
A comparison of the Jetphox $\{$Direct+Fragmentation$\}$ prediction for the photon $E_T$ distribution with the discrete
criterion (open triangles) vs.  Frixione with the continuous criterion (solid triangles),
for the parameter choice $n = 0.2, \epsilon_s = 0.05$.}
\end{minipage}
\hfill
\begin{minipage}[t]{.46\linewidth}
\includegraphics[scale=0.38]{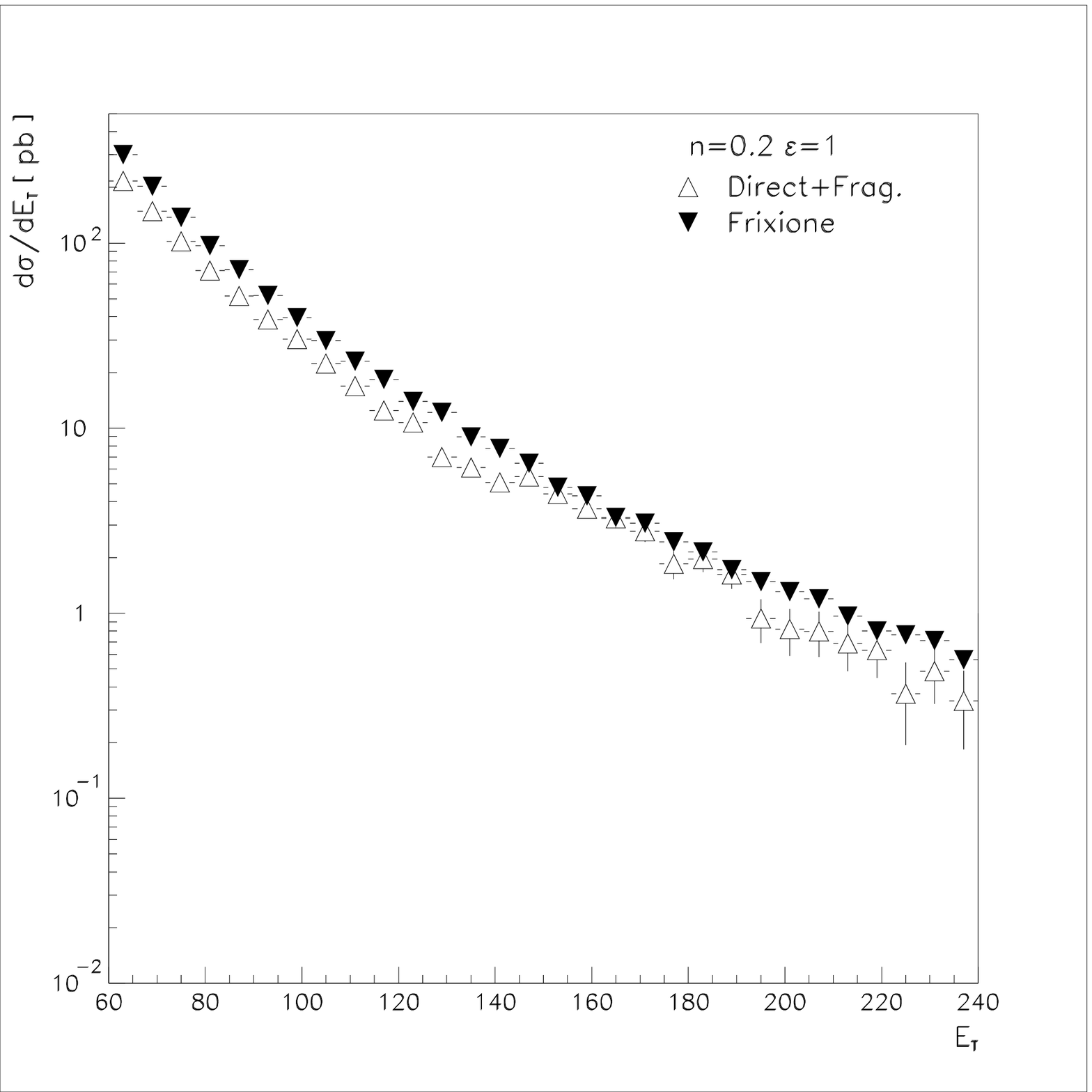}
\caption{\label{isol-fig2b} 
A comparison of the Jetphox $\{$ Direct + Fragmentation $\}$ prediction for the photon $E_T$ distribution with the discrete
criterion (open triangles) vs.  Frixione with the continuous criterion (solid triangles), for the parameter choice $n = 0.2, \epsilon_s = 1$.}
\end{minipage}
\end{figure}

\begin{figure}[htb]
\begin{minipage}[t]{.46\linewidth}
\includegraphics[scale=0.38]{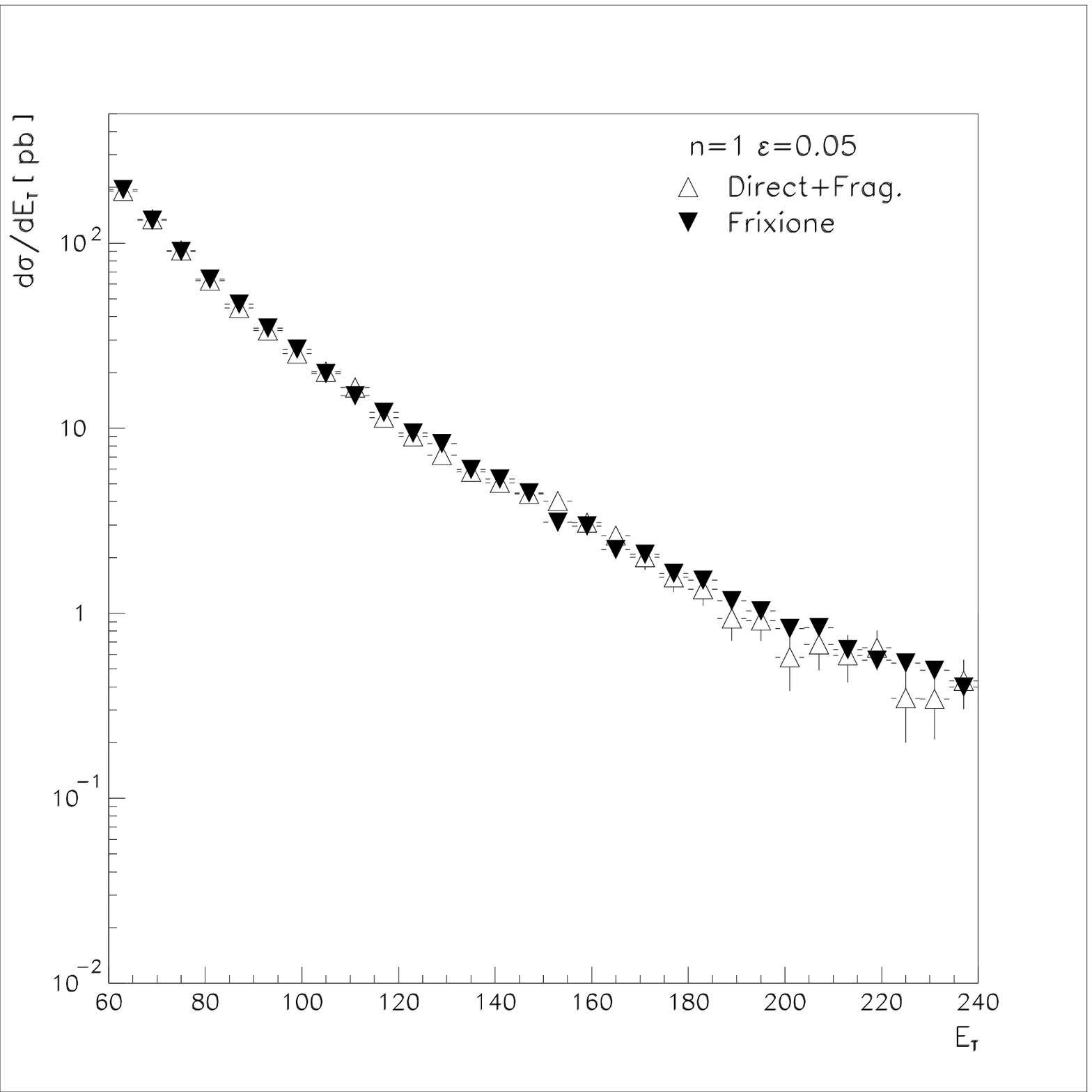}
\caption{\label{isol-fig2c} 
A comparison of the Jetphox $\{$Direct+Fragmentation$\}$ prediction for the photon $E_T$ distribution with the discrete
criterion (open triangles) vs.  Frixione with the continuous criterion (solid triangles),
for the parameter choice $n = 1, \epsilon_s = 0.05$.}
\end{minipage}
\hfill
\begin{minipage}[t]{.46\linewidth}
\includegraphics[scale=0.38]{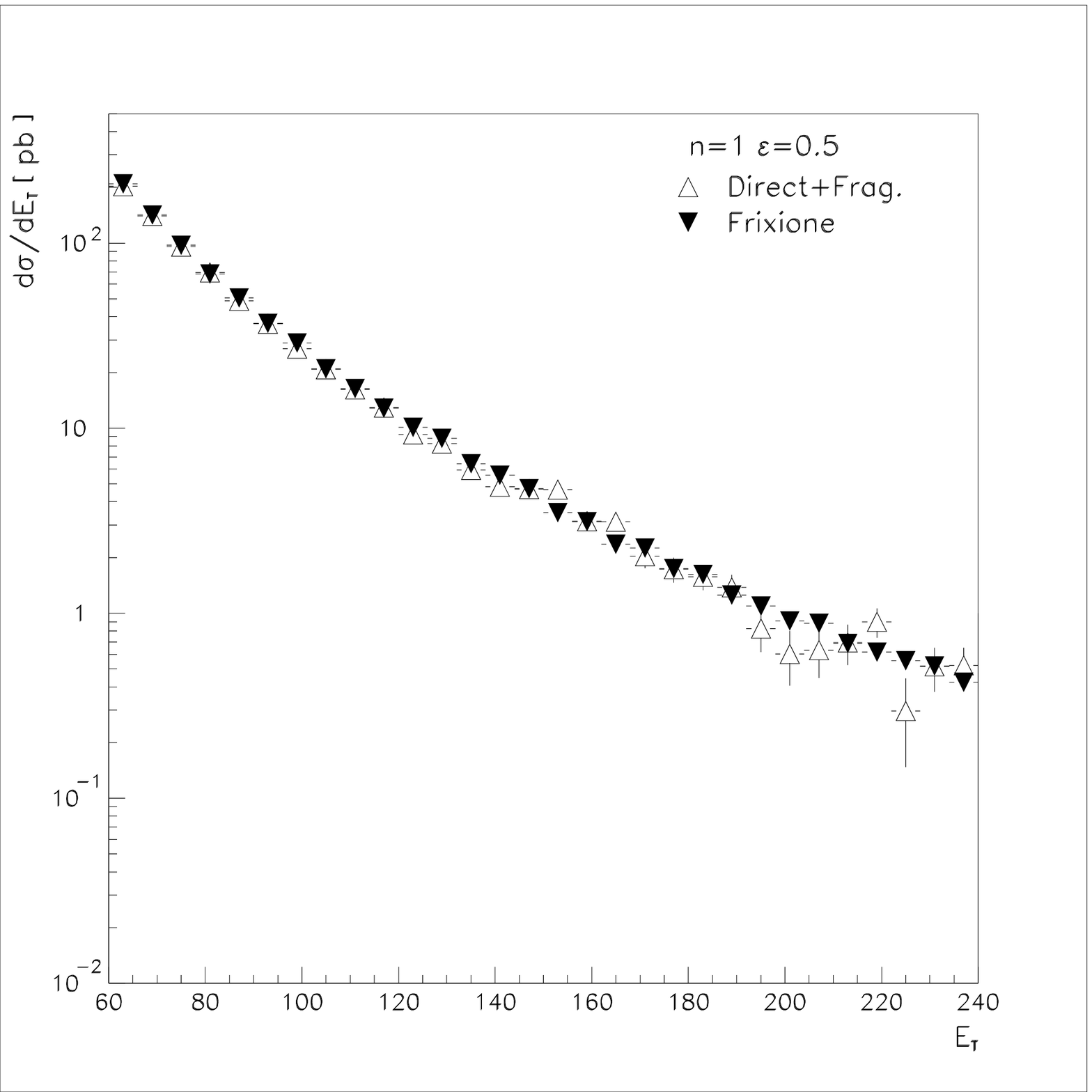}
\caption{\label{isol-fig2d}  
A comparison of the Jetphox $\{$Direct+Fragmentation$\}$ prediction for the photon $E_T$ distribution with the discrete
criterion (open triangles) vs.  Frixione with the continuous criterion (solid triangles),
for the parameter choice $n = 1, \epsilon_s = 0.5$.}
\end{minipage}
\end{figure}

\begin{figure}[b]
\begin{center}
\includegraphics[scale=0.38]{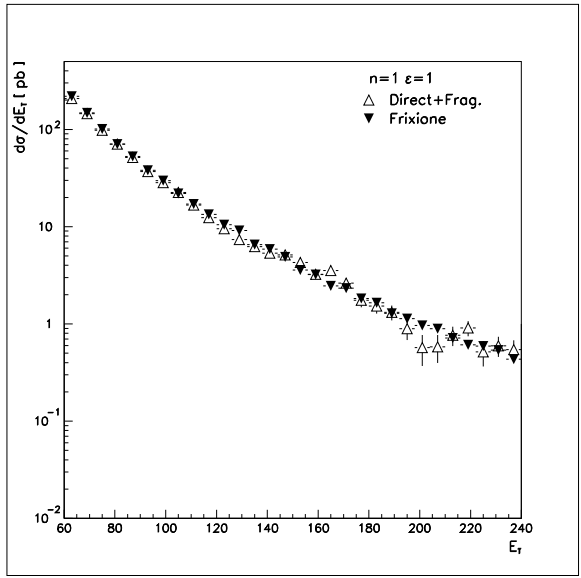}
\caption{\label{isol-fig2e} 
A comparison of the Jetphox $\{$Direct+Fragmentation$\}$ prediction for the photon $E_T$ distribution with the discrete
criterion (open triangles) vs.  Frixione with the continuous criterion (solid triangles),
for the parameter choice $n = 1, \epsilon_s = 1$.}
\end{center}
\end{figure}

\vspace{0.3cm}
\noindent
The comparison of the $\{$Direct+Fragmentation$\}$ Jetphox predictions with 
predictions using the continuous criterion as implemented in Frixione's 
code for the above five parameters choices $(n, \epsilon_s)$ are presented in 
Figs. \ref{isol-fig2a}-\ref{isol-fig2e}. The two codes used different scale choices, 
$\mu = M = M_F = E_{T}^{\gamma}/2$ for Jetphox vs. 
$(E_{T}^{\gamma} + E_{T}^{jet})/4$ for Frixione respectively. Notice 
however that these two scales coincide at the Born level. There may be
differences at NLO between the two, though hopefully not major ones.
In addition, Jetphox used a frozen $\alpha_{em}$ whereas Frixione used a running 
$\alpha_{em}$ (at the above scale choice). Frixione's choice for $\alpha_{em}$ 
systematically increases the prediction w.r.t. Jetphox, yet the net effect is
likely dominated by the QCD scale dependence.  The relative size of this effect is difficult to 
predict without performing a dedicated study. Nevertheless, the two
calculations yield similar results, illustrating that the 
discrete form of the Frixione criterion retains the features of the continuous 
criterion, at least at the partonic level, and as long as the discrete
criterion strongly suppresses the fragmentation component (i.e. all but
$(n=0.2,\epsilon=1)$ in the parameter choices considered for illustration).   

\clearpage

\vspace{0.3cm}
\noindent

\subsection{Some Theoretical Issues}
\label{sec:issues}

From a theoretical perspective,  one might be concerned by the use of cone radii as small as 0.1-0.2, for which earlier studies \cite{Catani:2002ny} revealed 
potential problems.
The earlier studies were carried out for the standard cone criterion, in the
limit of a narrow cone, $R \ll 1$, while allowing a given $R$-independent energy 
deposit inside the cone.  This lead to a collinear sensitivity in the form 
of a fairly large dependence on $\ln(1/R)$, which could make the prediction 
unreliable\footnote{This could even lead to an unphysical result such as an 
isolated cross section larger than the inclusive one, thereby violating 
unitarity.} unless these logarithms were resummed. With the Frixione criterion, at least in its continuous version, the amount of energy inside the 
cone is a function of $R$, decreasing to 0 as $R$ decreases; this correlates the energy and angular variables in such a way as to prevent the appearance of collinear divergences, and thus should avoid the concomitant appearance of troublesome logs. Therefore, the potential problem discussed in ~\cite{Catani:2002ny} regarding the appearance of large $\log R$ terms with the standard cone criterion, in the limit of narrow cone sizes ($R \ll 1$), is not 
expected in the present case, even though the discretization of the Frixione 
criterion mimics a standard cone criterion inside the innermost cone. 
In the discrete version studied 
here, the isolation in the innermost cone $R_1$ is effectively similar to the 
standard cone criterion with $(R_1, E_{T \, 1})$; however, unlike the situation
with the standard criterion, the region outside cone $R_1$ is still constrained 
by the isolation condition $(R_2, E_{T \, 2})$ and so on. This prevents any 
similar worrisome $\ln (1/R_j)$ dependence from developing.

\vspace{0.3cm}
\noindent
Another topic of concern might be the behaviour of the fragmentation functions 
into a photon (FFP) when $z \to 1$, a regime which is enhanced by the
requirement of a stringent isolation. It should be noted that the behaviour of 
the FFP in this regime is different from the corresponding one for hadrons.  The 
FFP are controlled by the so-called anomalous component induced by the 
inhomogenous terms in the DGLAP evolution equations, arising from the
point-like quark-photon coupling, and which are in principle fully calculable in 
perturbative QCD.  The non-perturbative, hadronic part is 
comparatively negligible in this regime. Unfortunately, the NLO calculation of the FFP is plagued by large logarithms of the form 
$\ln^{k}(1-z)$, $k = 1,2$ coming from both homogeneous and
inhomogeneous DGLAP kernels, and which make the predictions quantitatively 
unreliable. On the other hand, one expects that for the cross sections 
involving an integral over the fragmentation variable $z$, this sensitivity to 
these integrable logarithms $\ln^{k}(1-z)$ is smeared over a narrow domain in $z$, thus
yielding only a small contribution. We therefore expect that this issue is not 
too troublesome.


\subsection{Summary and outlook}
\label{sec:summary}

In this contribution, we have outlined an adaptation of the Frixione isolation criterion,
modified to take into account the experimental environment in which the photon measurements will be conducted at the LHC. The  resulting discrete version of the Frixione criterion provides isolated photon cross sections in good agreement with those obtained from the  continuous version. Much of the energy in the isolation cone results not from the hard process, but from the soft underlying event from the collision producing the photon, or from additional interactions taking place in the same crossing. A method was outlined to 
separate the energy from these soft processes with energies resulting from fragmentation
processes. With this separation, only the Frixione isolation criterion need be applied to any theoretical calculation. 

\vspace{0.3cm}
\noindent
In future studies, the techniques outlined here will be tested first against Monte Carlo data, and then against the early LHC data.



%% file: forte/forte.tex
\def\smallfrac#1#2{\hbox{${{#1}\over {#2}}$}}
\newcommand{\be}{\begin{equation}}
\newcommand{\ee}{\end{equation}}
\newcommand{\bea}{\begin{eqnarray}}
\newcommand{\eea}{\end{eqnarray}}
\newcommand{\bi}{\begin{itemize}}
\newcommand{\ei}{\end{itemize}}
\newcommand{\ben}{\begin{enumerate}}
\newcommand{\een}{\end{enumerate}}
\newcommand{\la}{\left\langle}
\newcommand{\ra}{\right\rangle}
\newcommand{\lc}{\left[}
\newcommand{\rc}{\right]}
\newcommand{\lp}{\left(}
\newcommand{\rp}{\right)}
\newcommand{\as}{\alpha_s}
\newcommand{\aq}{\alpha_s\left( Q^2 \right)}
\newcommand{\amz}{\alpha_s\left( M_Z^2 \right)}
\newcommand{\aqq}{\alpha_s \left( Q^2_0 \right)}
\newcommand{\aqz}{\alpha_s \left( Q^2_0 \right)}
\def\toinf#1{\mathrel{\mathop{\sim}\limits_{\scriptscriptstyle
{#1\rightarrow\infty }}}}
\def\tozero#1{\mathrel{\mathop{\sim}\limits_{\scriptscriptstyle
{#1\rightarrow0 }}}}
\def\toone#1{\mathrel{\mathop{\sim}\limits_{\scriptscriptstyle
{#1\rightarrow1 }}}}
\def\frac#1#2{{{#1}\over {#2}}}
\def\gsim{\mathrel{\rlap{\lower4pt\hbox{\hskip1pt$\sim$}}
    \raise1pt\hbox{$>$}}}         
\def\lsim{\mathrel{\rlap{\lower4pt\hbox{\hskip1pt$\sim$}}
    \raise1pt\hbox{$<$}}}         
\newcommand{\mrexp}{\mathrm{exp}}
\newcommand{\dat}{\mathrm{dat}}
\newcommand{\one}{\mathrm{(1)}}
\newcommand{\two}{\mathrm{(2)}}
\newcommand{\art}{\mathrm{art}} 
\newcommand{\rep}{\mathrm{rep}}
\newcommand{\net}{\mathrm{net}}
\newcommand{\stopp}{\mathrm{stop}}
\newcommand{\sys}{\mathrm{sys}}
\newcommand{\stat}{\mathrm{stat}}
\newcommand{\diag}{\mathrm{diag}}
\newcommand{\pdf}{\mathrm{pdf}}
\newcommand{\tot}{\mathrm{tot}}
\newcommand{\minn}{\mathrm{min}}
\newcommand{\mut}{\mathrm{mut}}
\newcommand{\partt}{\mathrm{part}}
\newcommand{\dof}{\mathrm{dof}}
\newcommand{\NS}{\mathrm{NS}}
\newcommand{\cov}{\mathrm{cov}}
\newcommand{\gen}{\mathrm{gen}}
\newcommand{\cut}{\mathrm{cut}}
\newcommand{\parr}{\mathrm{par}}
\newcommand{\val}{\mathrm{val}}
\newcommand{\tr}{\mathrm{tr}}
\newcommand{\checkk}{\mathrm{check}}
\newcommand{\reff}{\mathrm{ref}}
\newcommand{\extra}{\mathrm{extra}}
\newcommand{\draft}[1]{}
\newcommand{\comment}[1]{{\bf \it  #1}}






\subsection{Combined PDF and $\alpha_s$ uncertainties.}

The impact of the combined $\alpha_s-$PDF uncertainties
has been recently
investigated  by CTEQ~\cite{Pumplin:2005rh} and MSTW~\cite{Martin:2009bu}. 
In this contribution we discuss how the correlation between
the strong coupling and PDFs affects PDF determination and
uncertainties in physical observables within the
NNPDF 
approach~\cite{f2ns,f2p,DelDebbio:2007ee,Ball:2008by,Rojo:2008ke,Ball:2009mk,Guffanti:2009xk,t0,Ball:2010de} 
to
PDF determination. We show the impact that varying $\alpha_s$
has in the PDF determination, both for central values and for
uncertainties. We then quantify the correlation between $\alpha_s$
and the gluon. Finally, we discuss different procedures to combine
the uncertainties from PDFs and $\alpha_s$ in physical observables,
and compare these procedures for the important case of Higgs production
at the LHC.

\subsection{NNPDF1.2 with varying $\alpha_s(M_Z^2)$}
\label{LH_NNPDF_sec:nnpdf12alphas}

The strong coupling is determined from a global average from
a wide variety of different measurements. The current PDG value
gives~\cite{Amsler:2008zzb}
\be
\alpha_s\lp M_Z^2\rp=0.1176 \pm 0.002 \ ,
\ee
where the error is to be interpreted as
 a 1-$\sigma$ uncertainty. Another recent
world average~\cite{Bethke:2009jm} finds
\be
\alpha_s\lp M_Z^2\rp=0.1184 \pm 0.0007 \ .
\ee
In the rest of this contribution we will take
as reference value for $\alpha_s$ and its uncertainty the
following range:
\be
\label{LH_NNPDF_eq:alphasref}
\alpha_s\lp M_Z^2\rp= 0.119 \pm 0.0012 \ ,68\%\,{\rm C.L.} \qquad  (  \pm 0.0020 \ ,89\%\,{\rm C.L.})  \ ,
\ee
although the generalisation of the present study to any other value
of the strong coupling and its uncertainty is straightforward.

\begin{figure}[ht]
\begin{center}
\includegraphics[width=0.48\textwidth]{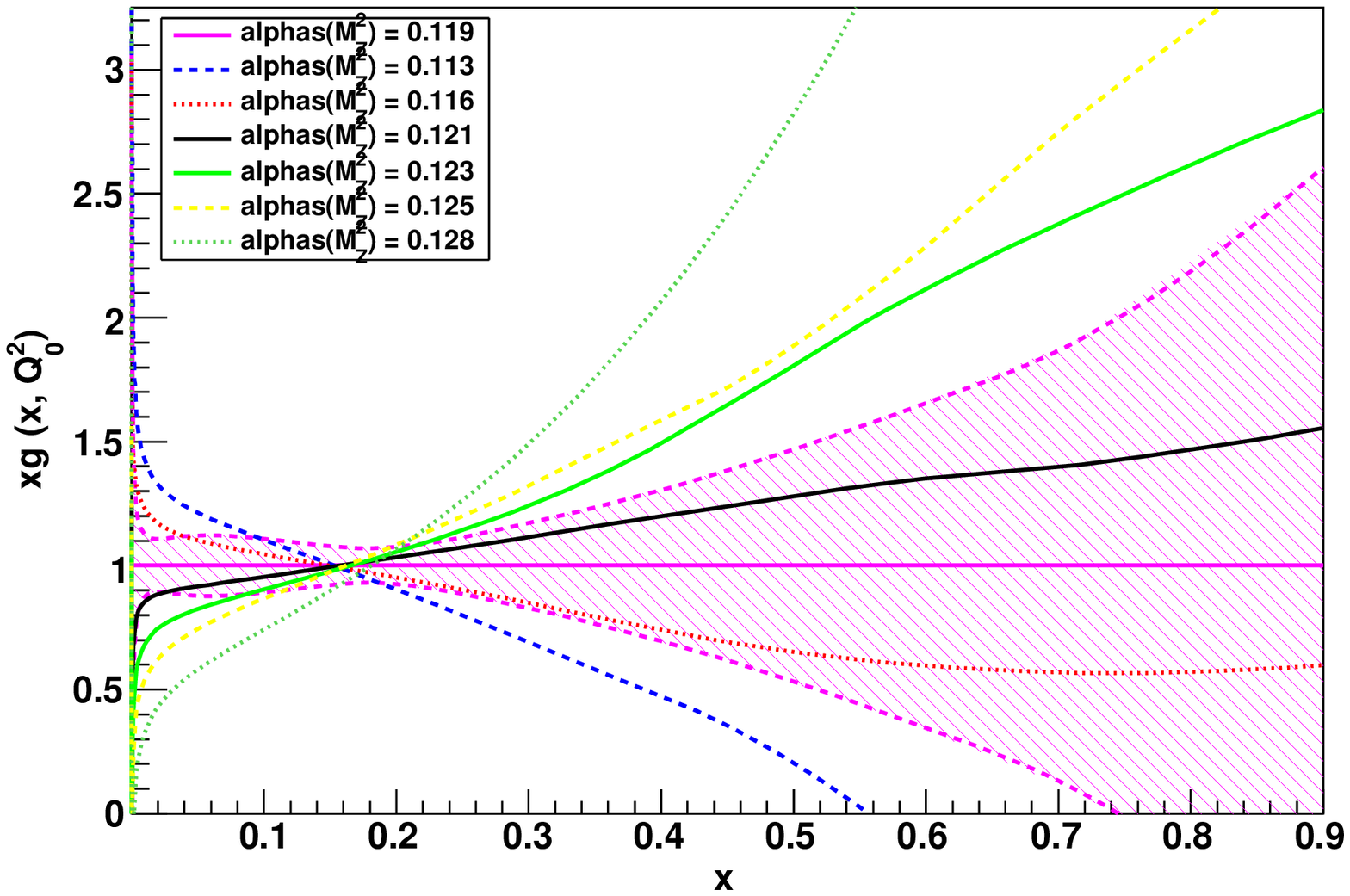}
\includegraphics[width=0.48\textwidth]{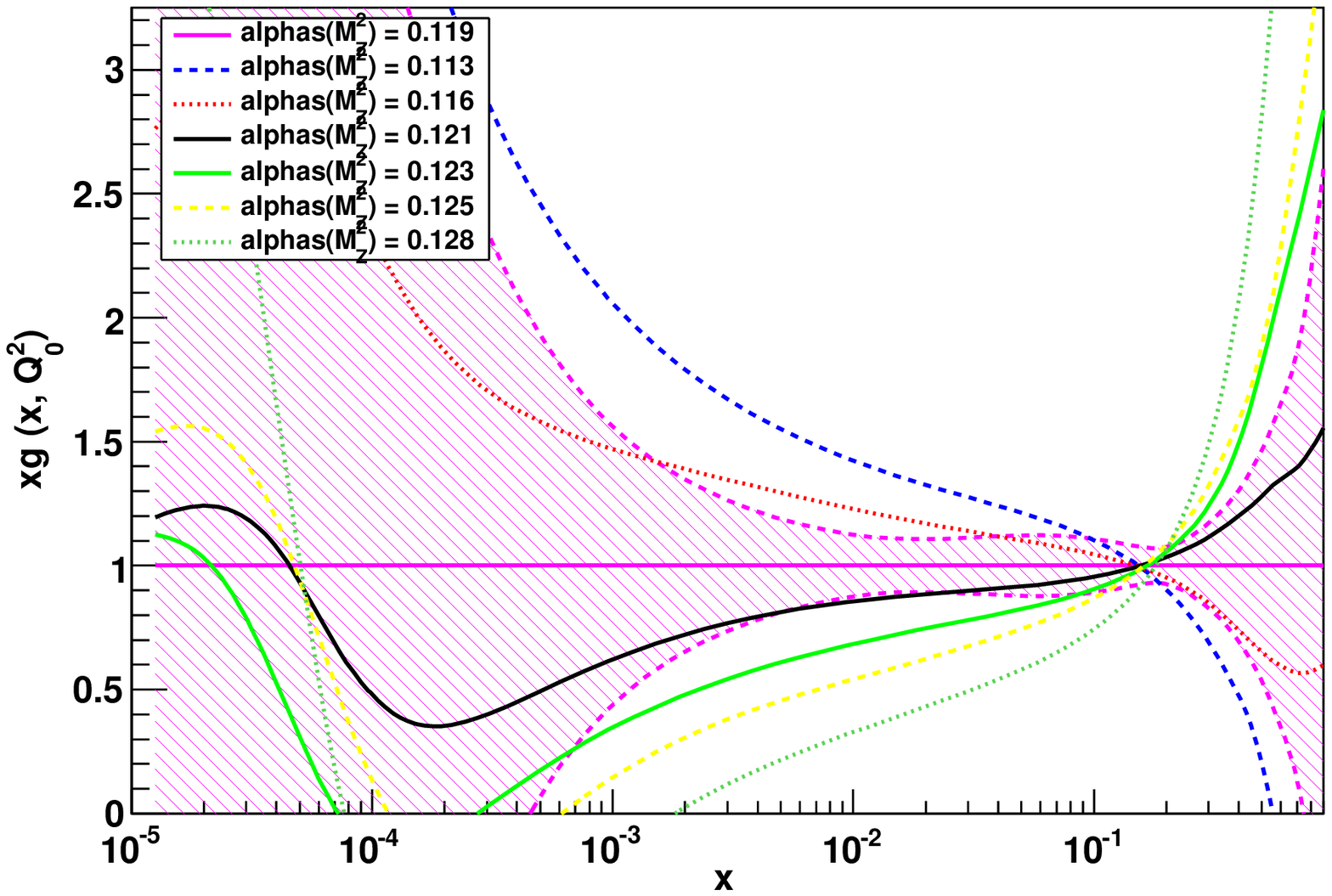}
 \caption{The ratios of the central gluons obtained in the
fits with varying $\alpha_s$ as compared to the reference
NNPDF1.2 gluon at the initial evolution scale
$Q^2_0=2$ GeV$^2$. The comparison
is shown both in a linear (left plot) and logarithmic (right plot)
scales. The dashed band corresponds to the NNPDF1.2 gluon relative
PDF uncertainty.
}
\label{LH_NNPDF_fig:xg_comp}
\end{center}
\end{figure}

The motivation of this contribution is to explore the impact
of the uncertainties in $\alpha_s$, Eq.~\ref{LH_NNPDF_eq:alphasref}, in PDF
determination and associated LHC observables. In order to do so,
taking as reference the NNPDF1.2 parton
determination~\cite{Ball:2009mk}, a set
of fits with different values of alphas were produced, together
with the associated PDF uncertainties in each case. 
In Fig.~\ref{LH_NNPDF_fig:xg_comp} we show
 the ratios of the central gluons obtained in these
fits with varying $\alpha_s$ as compared to the reference
NNPDF1.2 gluon with $\alpha_s=0.119$, 
together with the associated  PDF uncertainty
for this reference value. The sensitivity
with respect the chosen value of $\alpha_s$
is non-negligible, although for $\alpha_s$ variations 
within the assumed uncertainty range Eq.~\ref{LH_NNPDF_eq:alphasref} 
fall typically within the
PDF uncertainty band.

It is easy to understand the qualitative behaviour of the gluon
in Fig.~\ref{LH_NNPDF_fig:xg_comp}. In a DIS-only fit like NNPDF1.2~\cite{Ball:2009mk}, 
the gluon is essentially
determined at small-$x$ through the scaling violations of HERA structure
function data, and smaller values of $\alpha_s$ are compensated
with harder small-$x$ gluons. At large-$x$ there are no experimental
constrains on the gluon so it is essentially determined by the
momentum sum rule, and thus its behaviour is anti-correlated to that
of the small-$x$ region. In a global 
fit~\cite{Martin:2009bu,Pumplin:2005rh} the behaviour is essentially
 the same modulo 
some constrains from the Tevatron inclusive
jet data on the large-$x$ gluon.

\begin{figure}[ht]
\begin{center}
\includegraphics[width=0.48\textwidth]{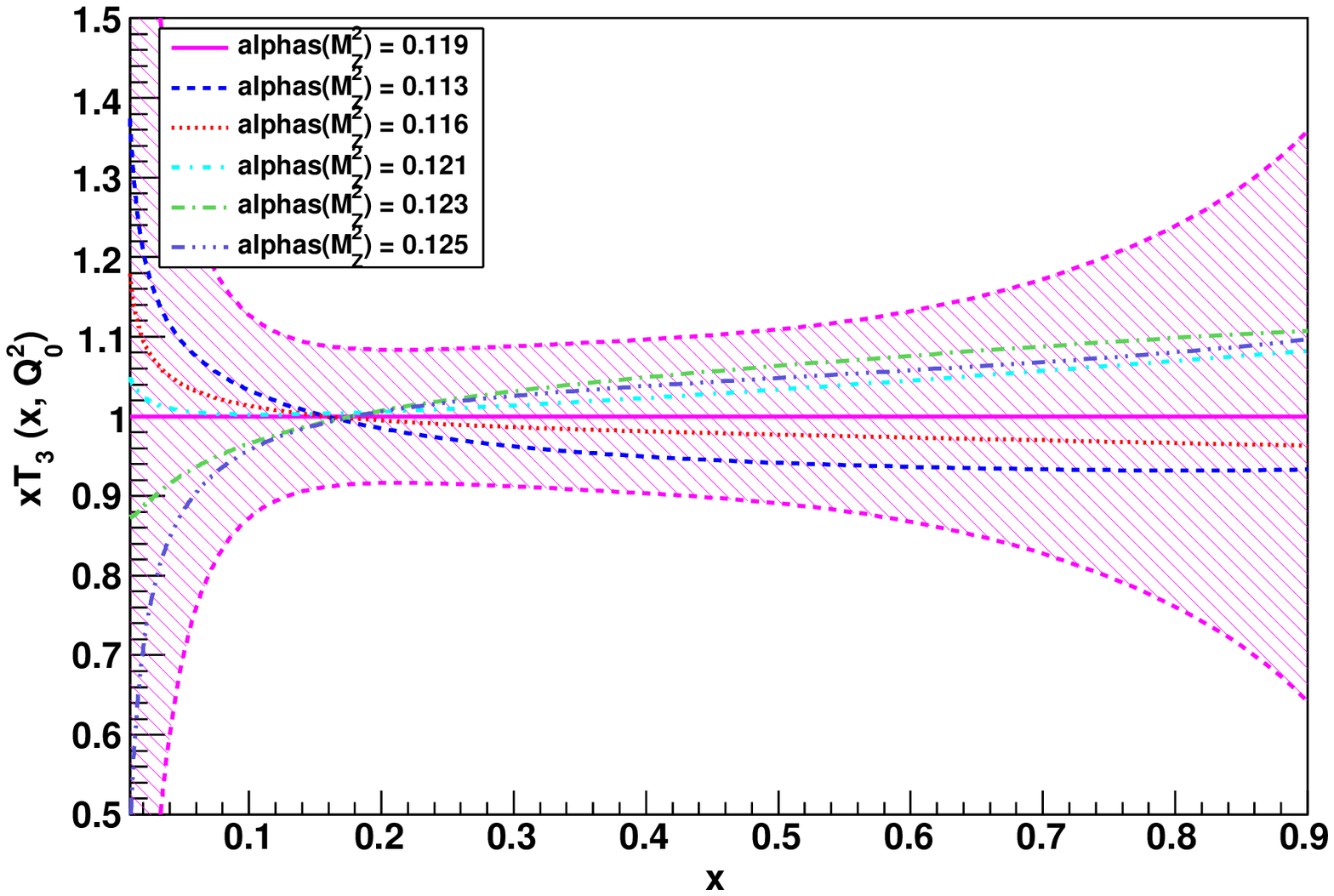}
\includegraphics[width=0.48\textwidth]{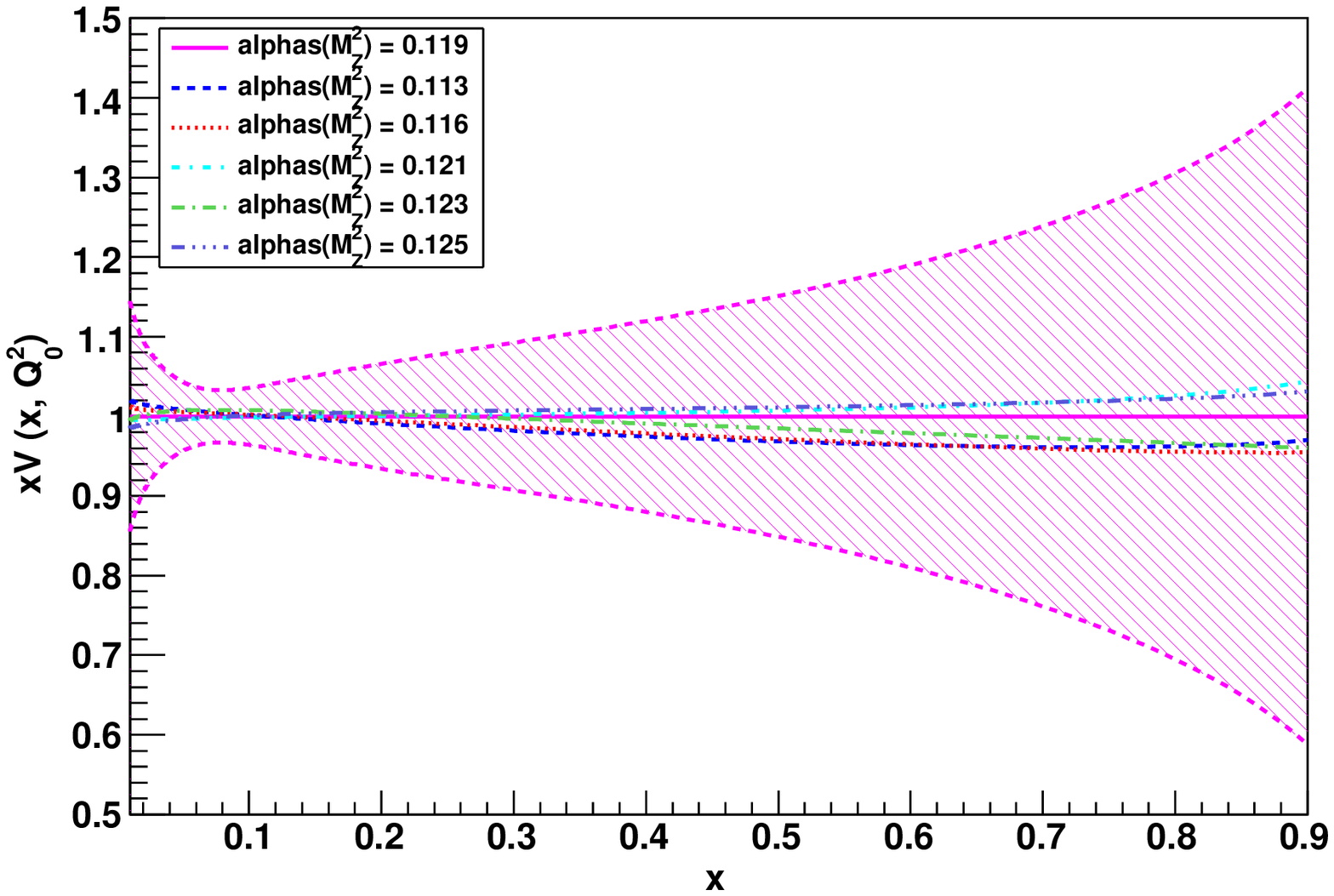}
\includegraphics[width=0.48\textwidth]{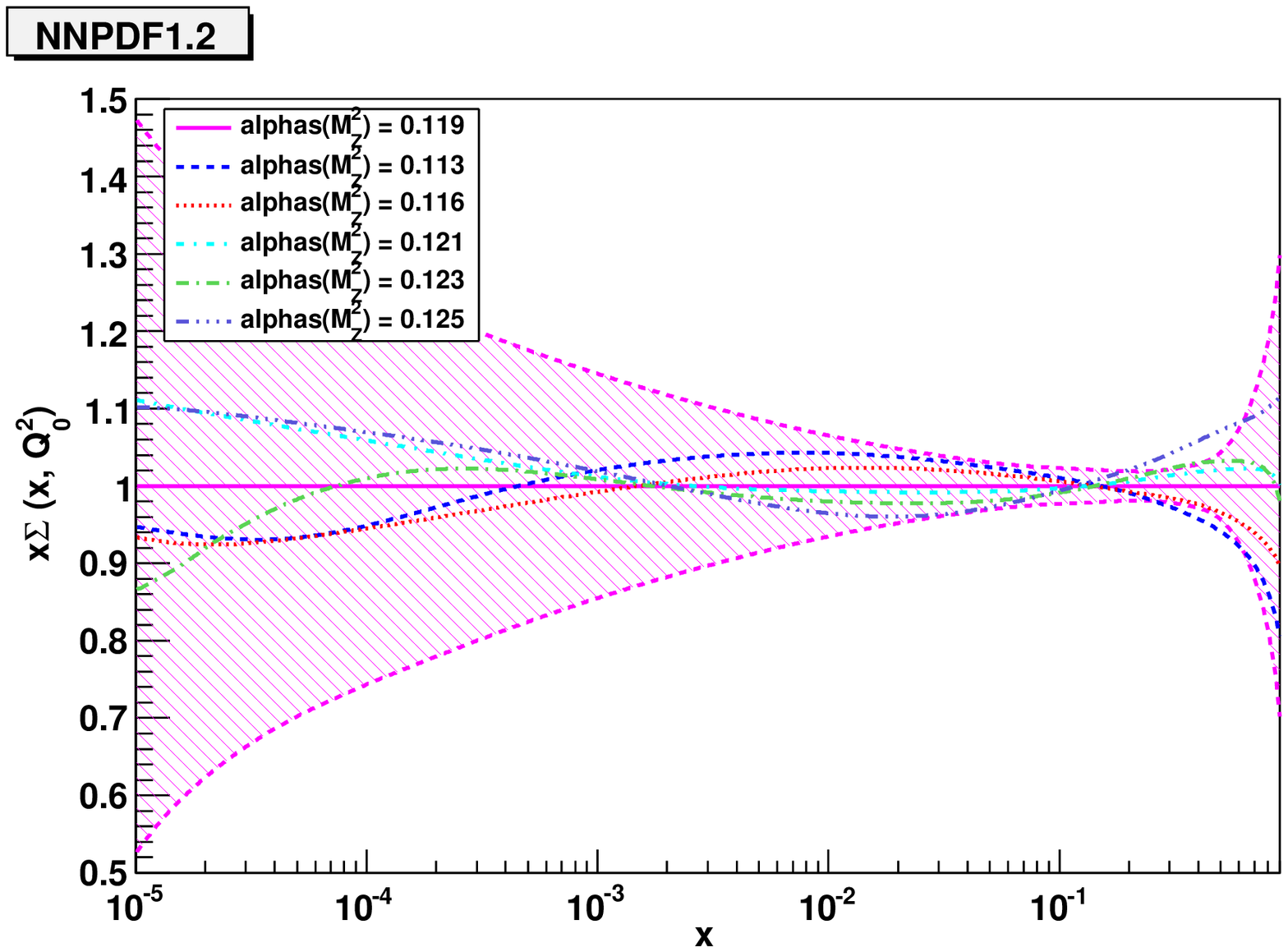}
\includegraphics[width=0.48\textwidth]{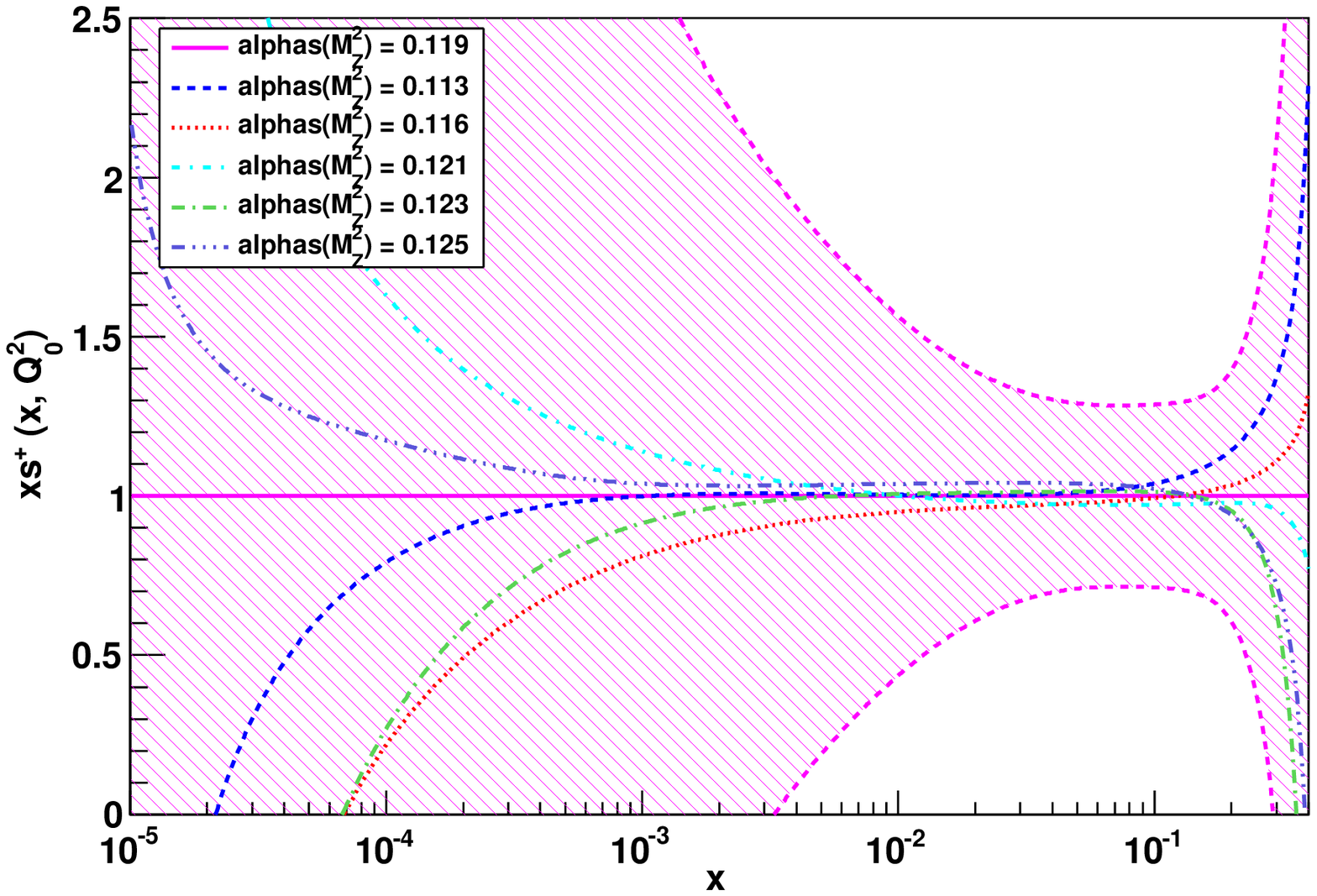}
 \caption{The ratios of the central PDFs obtained in the
fits with varying $\alpha_s$ as compared to the reference
NNPDF1.2 PDF, including PDF uncertainties, at the initial evolution scale
$Q^2_0=2$ GeV$^2$. The PDFs shown here
are from top to bottom and from left to right: the triplet $T_3$ in a linear
scale, the total valence $V$ in a linear scale, the singlet $\Sigma$
in a log scale and the strange sea $s^+$ in a log scale.}
\label{LH_NNPDF_fig:xpdf_comp}
\end{center}
\end{figure}

Other PDFs
are affected to a much lesser extent, as shown in 
Fig.~\ref{LH_NNPDF_fig:xpdf_comp}. For example, the $\alpha_s$ dependence
of the triplet or the total valence is clearly negligible
when compared with the respective PDF
uncertainties.  The only possible exception is the singlet $\Sigma(x,Q^2)$, 
determined with precision from HERA data and which
is coupled to variations in the gluon through the momentum sum rule. However,
even in this case variations are rather smaller than PDF
uncertainties.

\begin{figure}[ht]
\begin{center}
\includegraphics[width=0.48\textwidth]{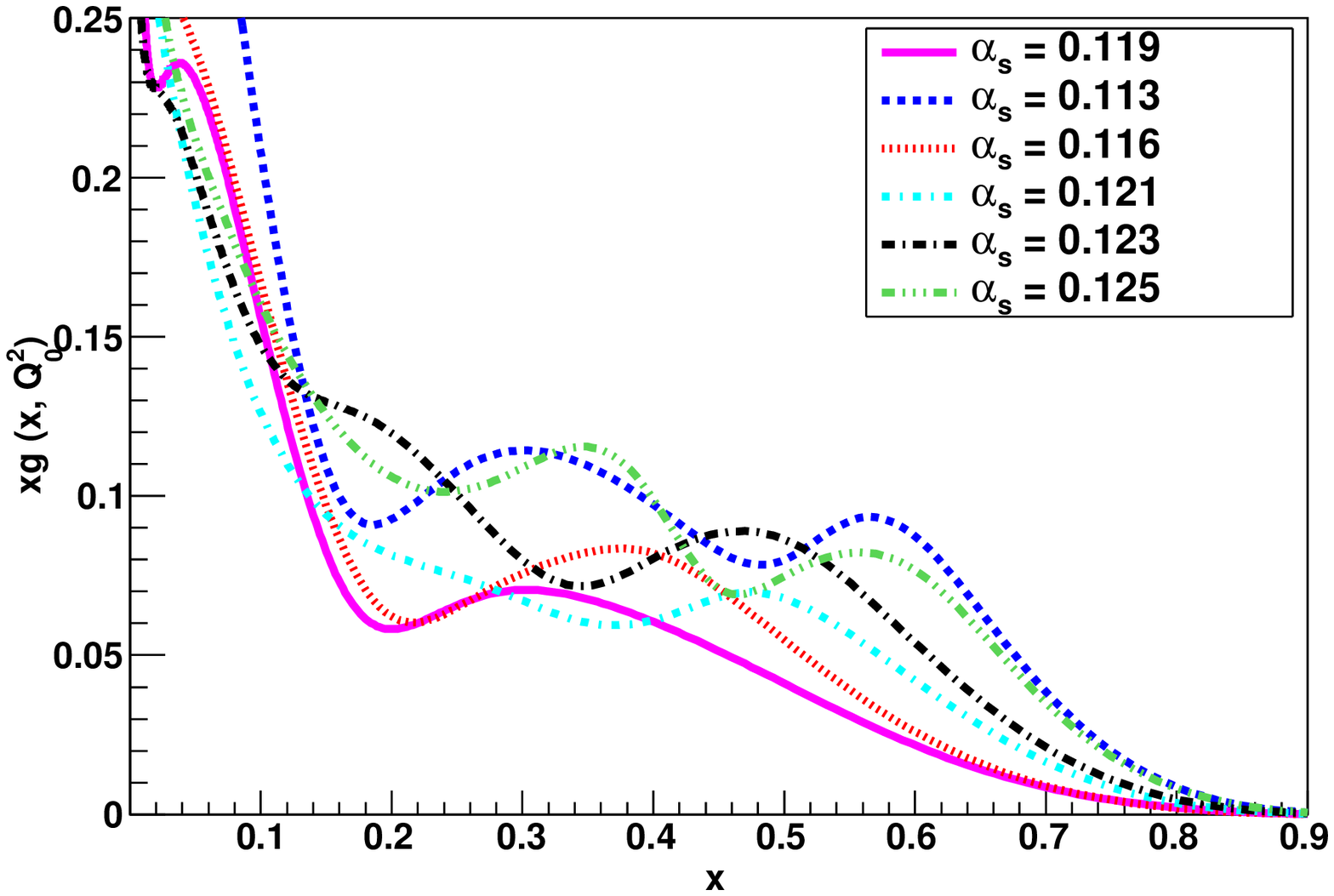}
\includegraphics[width=0.48\textwidth]{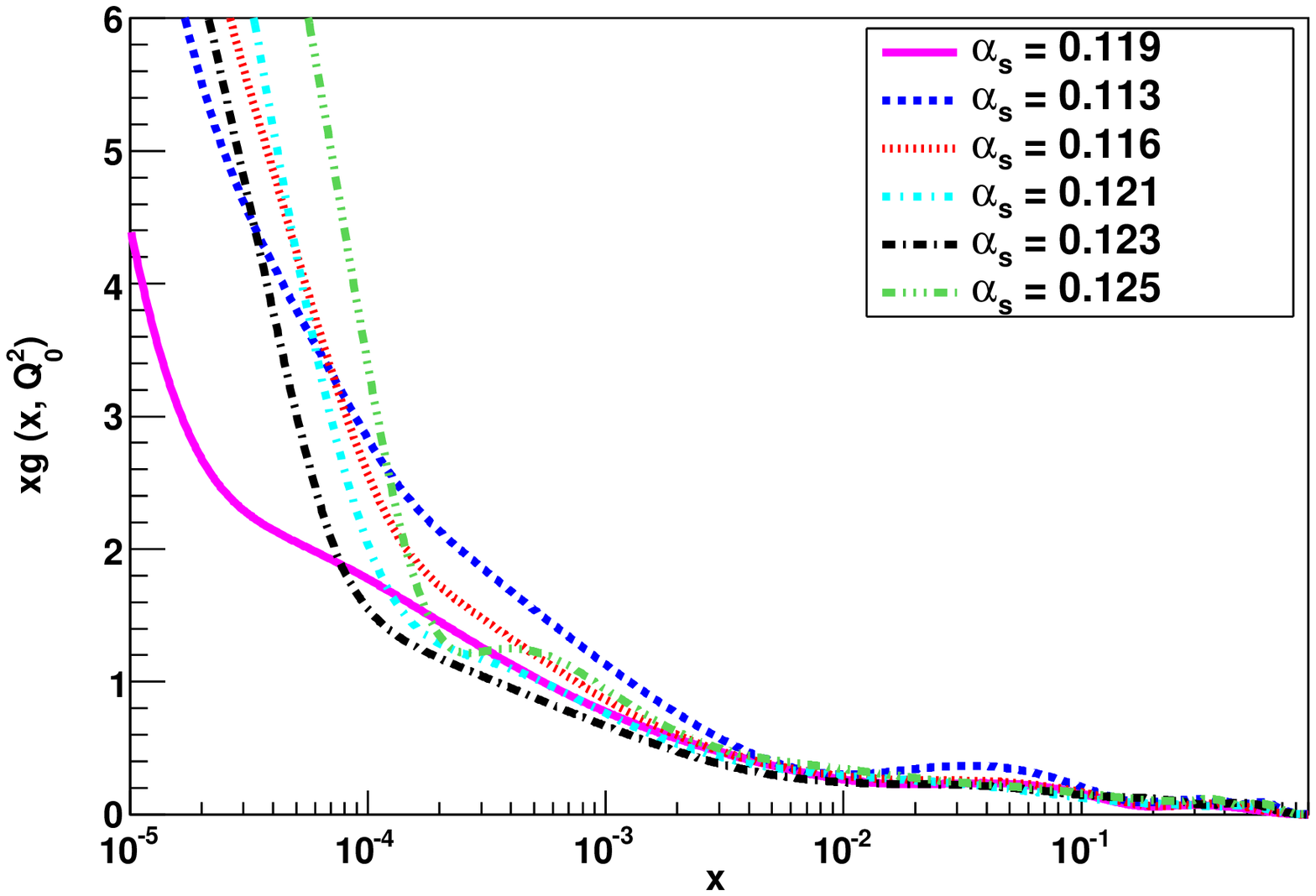}
 \caption{\small Comparison of the absolute gluon PDF uncertainty obtained in the
fits with varying $\alpha_s$ as compared to the reference
NNPDF1.2 gluon. The comparison
is shown both in a linear (left plot) and logarithmic (right plot)
scales. Note that what is shown in the uncertainty on the PDF
and not the PDF itself.
}
\label{LH_NNPDF_fig:xg_comp_errors}
\end{center}
\end{figure}

\begin{figure}[ht]
\begin{center}
\includegraphics[width=0.48\textwidth]{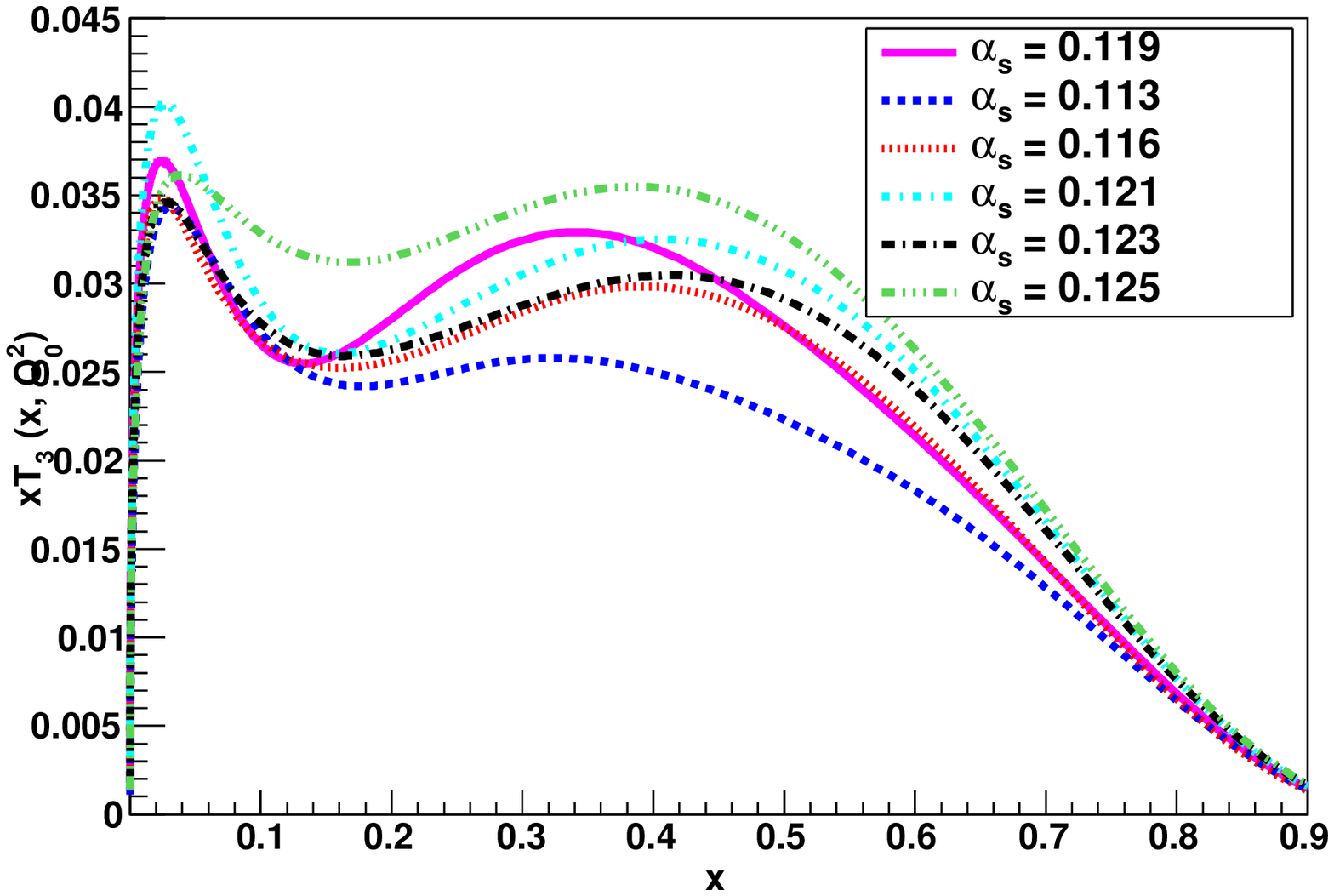}
\includegraphics[width=0.48\textwidth]{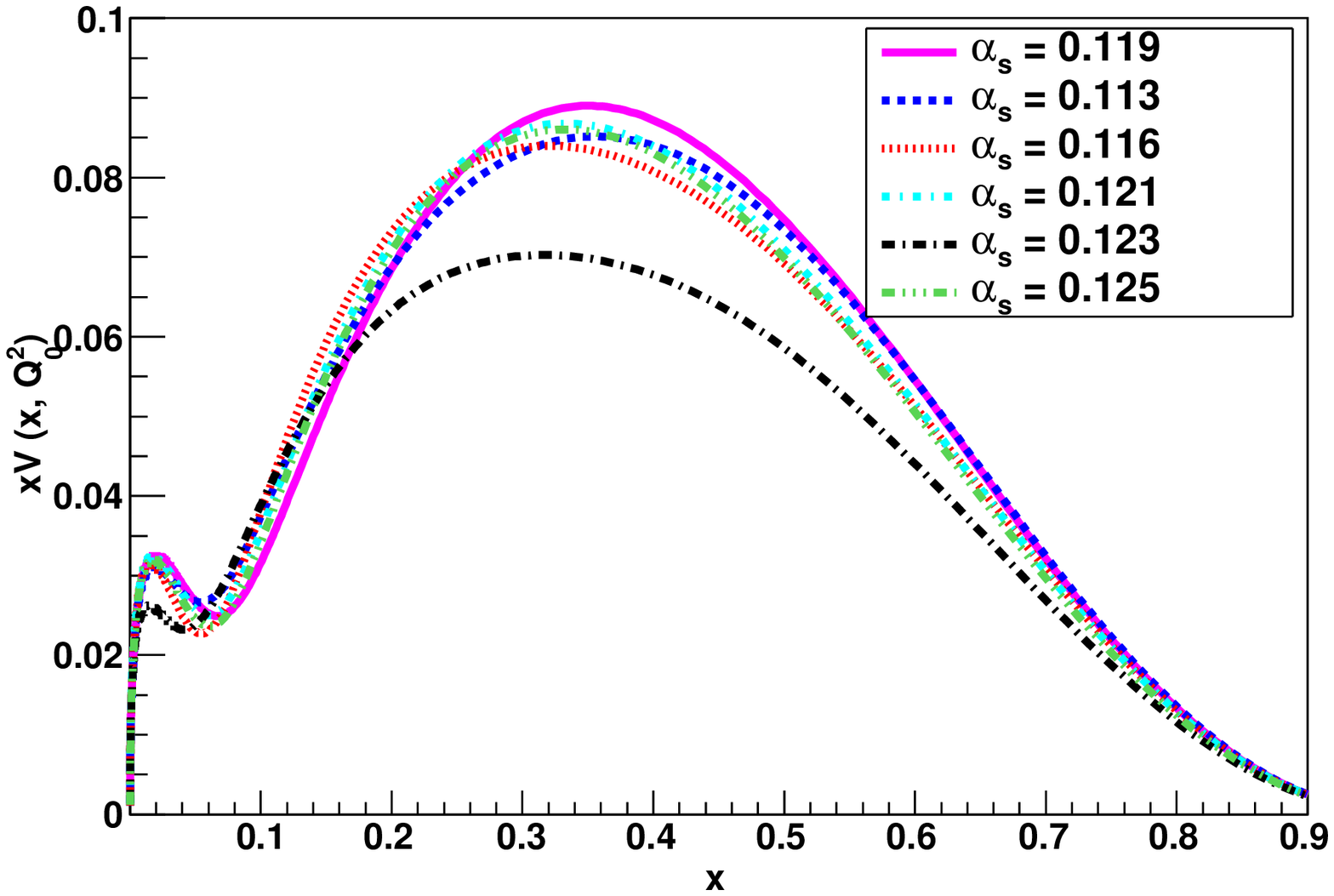}
\includegraphics[width=0.48\textwidth]{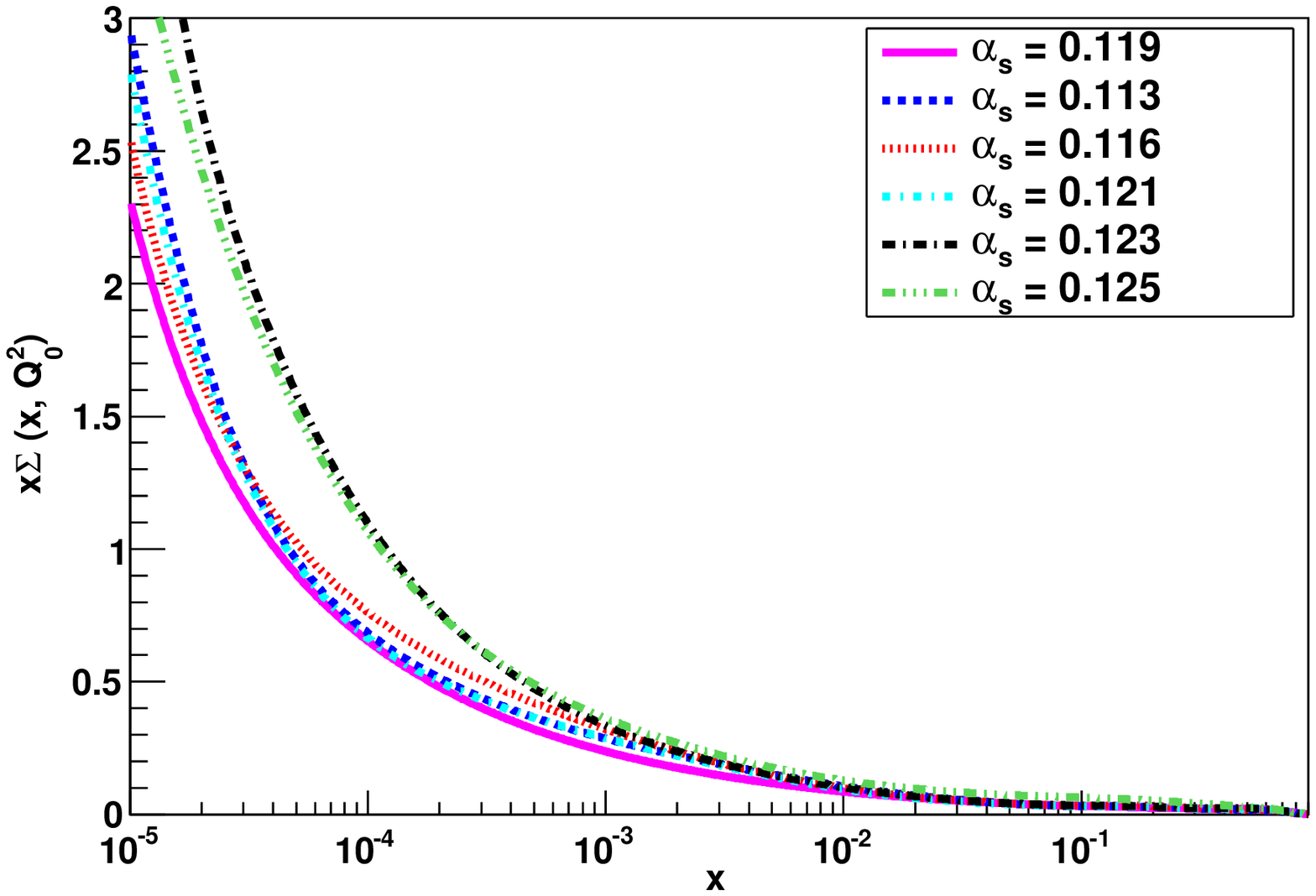}
\includegraphics[width=0.48\textwidth]{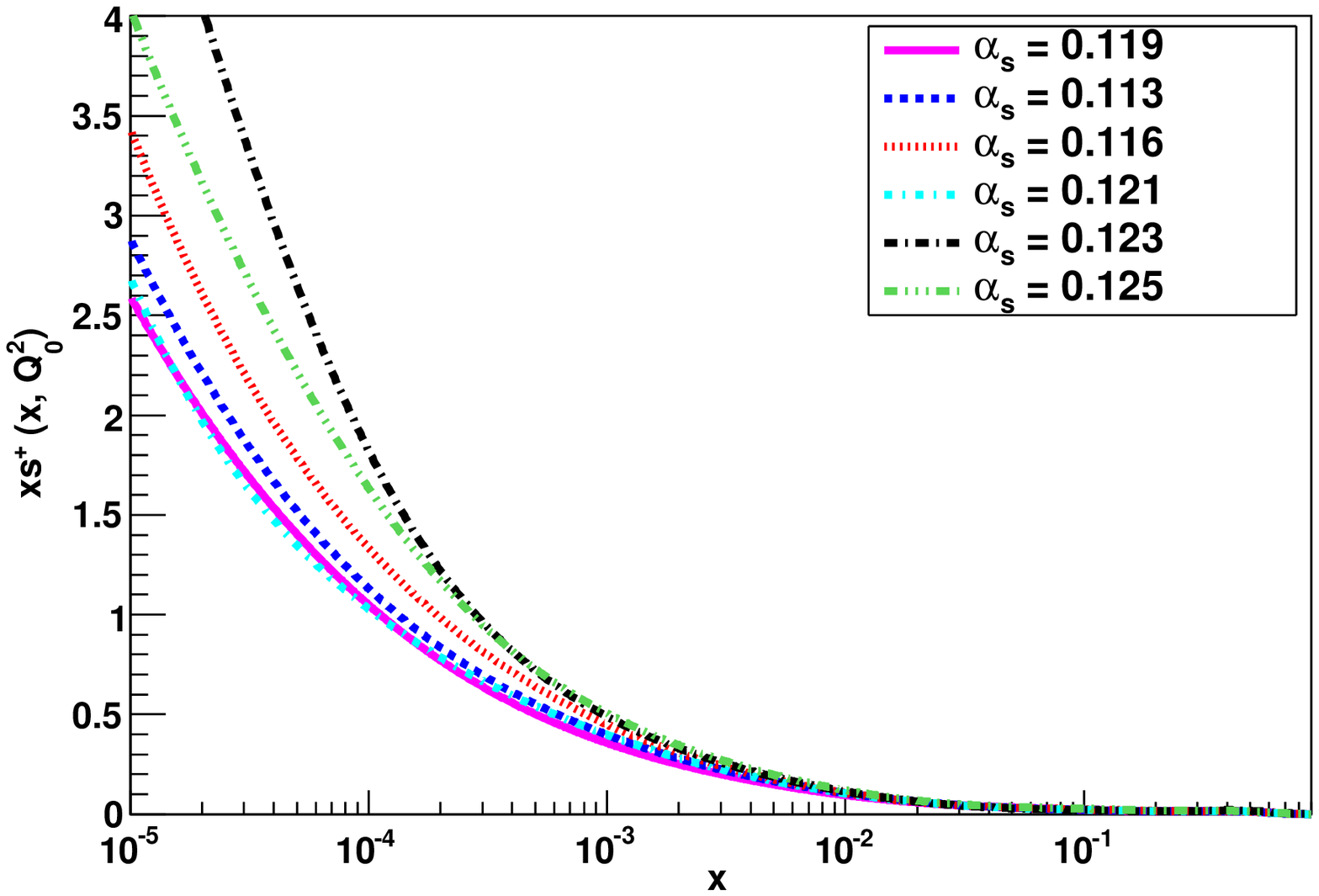}
 \caption{\small Comparison of PDF uncertainties obtained in the
fits with varying $\alpha_s$ as compared to the reference
NNPDF1.2 PDF. The absolute PDF uncertainties shown here
are from top to bottom and from left to right: the triplet $T_3$ in a linear
scale, the total valence $V$ in a linear scale, the singlet $\Sigma$
in a log scale and the strange sea $s^+$ in a log scale. Note that what 
is shown are the uncertainty on the PDFs
and not the PDF themselves.}
\label{LH_NNPDF_fig:xpdf_comp_errors}
\end{center}
\end{figure}

On top of the
impact of variations in $\alpha_s$ in the
PDF central values, also the PDF uncertainties
are in principle modified by these variations. 
Within the  Hessian approach (see for example Ref.~\cite{Martin:2009bu}),
which implies a simultaneous determination of $\alpha_s$
and the PDFs,
$\alpha_s$ variations from the best fit value result
in PDFs with reduced uncertainties by construction. 
This is however not necessarily the case if one does not assume a
quadratic approximation of the $\chi^2$ as both PDF
parameters and $\alpha_s$ are varied.

To assess quantitatively 
how PDF uncertainties are affected by $\alpha_s$ variations
within the NNPDF approach, 
we show in Fig.~\ref{LH_NNPDF_fig:xg_comp_errors} the absolute PDF
uncertainties for the gluon for the different values of
$\alpha_s$ obtained using the NNPDF1.2-like fits, 
and in Fig.~\ref{LH_NNPDF_fig:xpdf_comp_errors} the same
for other PDFs which are much less affected by $\alpha_s$
variations (see Figs.~\ref{LH_NNPDF_fig:xg_comp}-\ref{LH_NNPDF_fig:xpdf_comp}
respectively). 
In the case of the gluon, it seems that the reference
value $\alpha_s\lp M_Z^2\rp=0.119$ tends to have the smaller PDF uncertainties,
although as will be shown below essentially all values
of $\alpha_s$ result in similar PDF uncertainties once
fluctuations in the PDF uncertainties themselves are
taken into account. For the other PDFs, Fig.~\ref{LH_NNPDF_fig:xpdf_comp_errors},
no such pattern can be identified and in any case the
dependence of
PDF uncertainties on $\alpha_s$ is much milder.

\begin{figure}[ht]
\begin{center}
\includegraphics[width=0.99\textwidth]{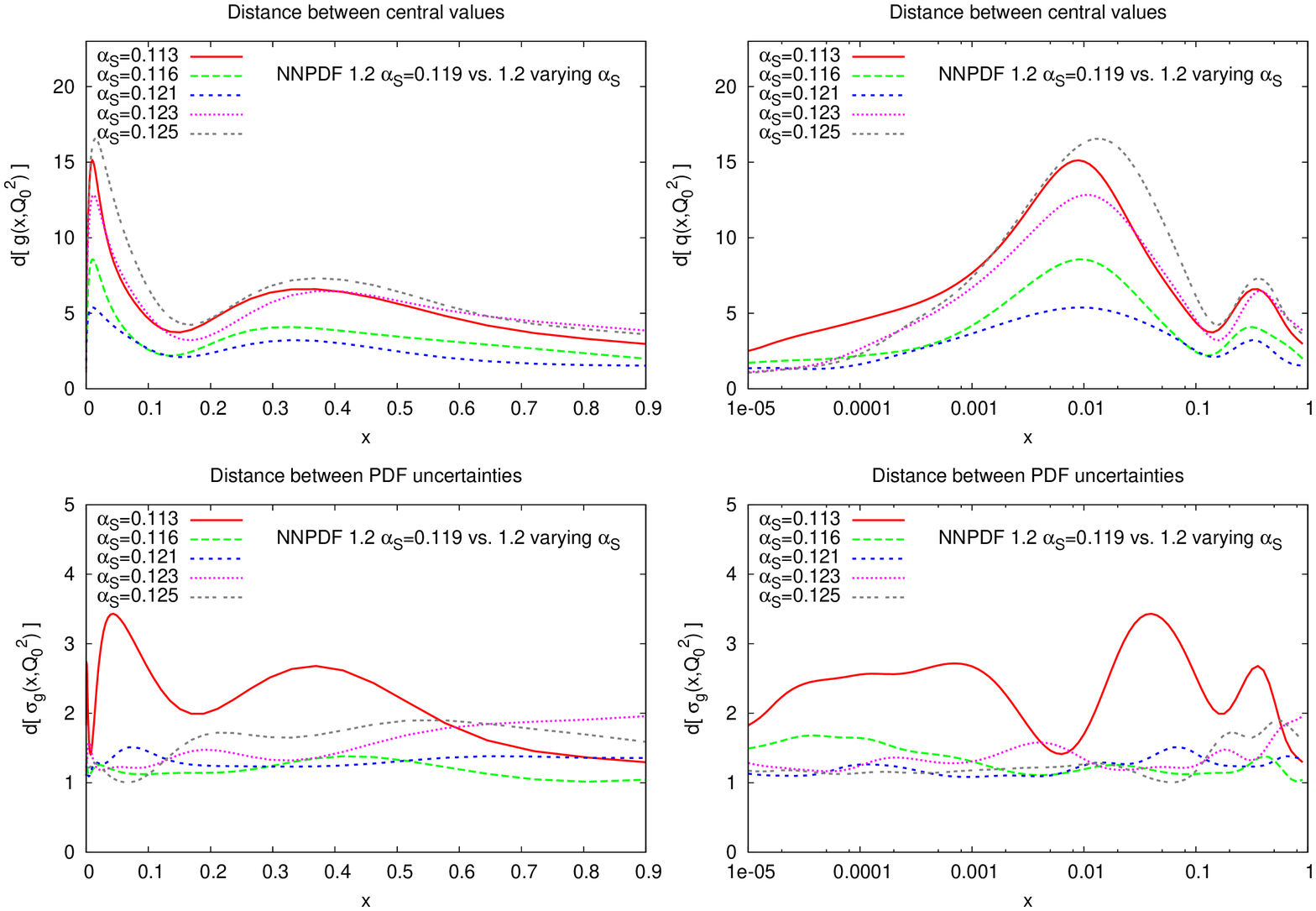}
 \caption{The distance between
the NNPDF1.2 gluon and the NNPDF1.2-like
gluons with different values of $\alpha_s\lp M_Z^2\rp$, both
for central values (upper plots) and standard deviations
(lower plots),
as a function of $x$. All distances are
computed from sets of  
$N_{\rm rep}=100$ replicas. A distance $d\sim 1$ corresponds
to two fits which are statistically equivalent, while 
a distance $d\sim 10$ correspond to fits which differ
by approximately 1-$\sigma$~\cite{DelDebbio:2007ee,t0}.
}
\label{LH_NNPDF_fig:distances}
\end{center}
\end{figure}

In order  to determine whether such variations of the
PDF uncertainty when $\alpha_s$ is varied in the fit are
statistically significant, we need to compute the error on the
PDF error itself. This is done automatically using the distance
estimator, as for example done in Ref.~\cite{t0}. Therefore
we show in Fig.~\ref{LH_NNPDF_fig:distances} the distances for central values and
uncertainties for the gluons with different $\alpha_s$
as compared with the reference NNPDF1.2 gluon. We observe that 
for all the values of $\alpha_s$ the uncertainties in the
gluon PDF are statistically equivalent, with the possible
exception of the rather extreme value $\alpha_s\lp M_Z^2\rp=0.113$.

\subsection{The correlation between $g(x,Q^2)$ and $\alpha_s\lp M_Z^2\rp$}

In order to make more quantitative the qualitative statements
about the correlation between PDFs and $\alpha_s$, we can
compute their correlation coefficient for any given
values of $x$ and $Q^2$.
The correlation between the strong coupling and the
gluon (or in general any other PDF) is defined as the
usual correlation between two probability distributions, namely
\be
\label{LH_NNPDF_eq:gcorr}
\rho \lc  \alpha_s\lp M_Z^2\rp,g\lp x,Q^2\rp\rc=
\frac{\la \alpha_s\lp M_Z^2\rp g\lp x,Q^2\rp \ra_{\rep}-
\la \alpha_s\lp M_Z^2\rp\ra_{\rep}\la g\lp x,Q^2\rp \ra_{\rep}
}{\sigma_{\alpha_s\lp M_Z^2\rp}\sigma_{g\lp x,Q^2\rp}} \ .
\ee
Note that the computation of this
correlation takes into account not only
the central gluons of the fits with different $\alpha_s$ but
also the corresponding uncertainties in each case.

Whereas the distribution of gluon distributions in
Eq.~(\ref{LH_NNPDF_eq:gcorr}) is given by the Monte Carlo sample, the 
distribution of $\alpha_s$ values is given by the procedure with which
$\alpha_s$ is determined. Because we take $\alpha_s$ as determined
from a global fit~\cite{Amsler:2008zzb,Bethke:2009jm} we assume its
value to be gaussianly distributed, with the mean and standard
deviation given by Eq.~(\ref{LH_NNPDF_eq:alphasref}).
We then fix the total number of PDF replicas to be used as
\be
N_{\rm rep} = \sum_{j=1}^{N_{\alpha_s}}N^{\alpha_s^{(j)}}_{\rm rep} \ ,
\ee
where $N^{\alpha_s^{(j)}}_{\rm rep}$  is the number of PDF replicas,
 randomly
selected
from the fit obtained with the corresponding value of
$\alpha_s$, $\alpha_s^{(j)}$, and $N_{\alpha_s}$ is the number
of PDF determinations with different values of
$\alpha_s$ which have been performed. The number of replicas
for each different value of $\alpha_s$ to be used
is thus, for a gaussian distribution,
\be
N^{\alpha_s^{(j)}}_{\rm rep}\propto \exp\lp 
-\frac{\lp \alpha_s^{(j)}- \alpha_s^{(0)}\rp^2}{
2 \delta_{\alpha_s}^2}\rp \ .
\ee
with $\alpha_s^{(0)}$ and $\delta_{\alpha_s}$ given in 
Eq.~\ref{LH_NNPDF_eq:alphasref}. 

The average over Monte Carlo replicas 
of a general quantity which depends on both
$\alpha_s$ and the PDFs, $\mathcal{F}\lp  {\rm PDF},\alpha_s\rp$,
for example that of Eq.~\ref{LH_NNPDF_eq:gcorr},
has to be understood schematically as follows
\be
\label{LH_NNPDF_eq:avrep}
\la \mathcal{F}\ra_{\rep} =\frac{1}{N_{\rep}}\sum_{j=1}^{N_{\alpha}}
\sum_{k_j=1}^{N_{\rm rep}^{\alpha_s^{(j)}}} 
\mathcal{F}\lp  {\rm PDF}^{(k_j,j)},\alpha_s^{(j)}\rp \ ,
\ee
where ${\rm PDF}^{(k_j,j)}$ stands for the replica $k_j$ of the
PDF fit obtained using $\alpha_s^{(j)}$ as the value of the
strong coupling.

\begin{figure}[ht]
\begin{center}
\includegraphics[width=0.49\textwidth]{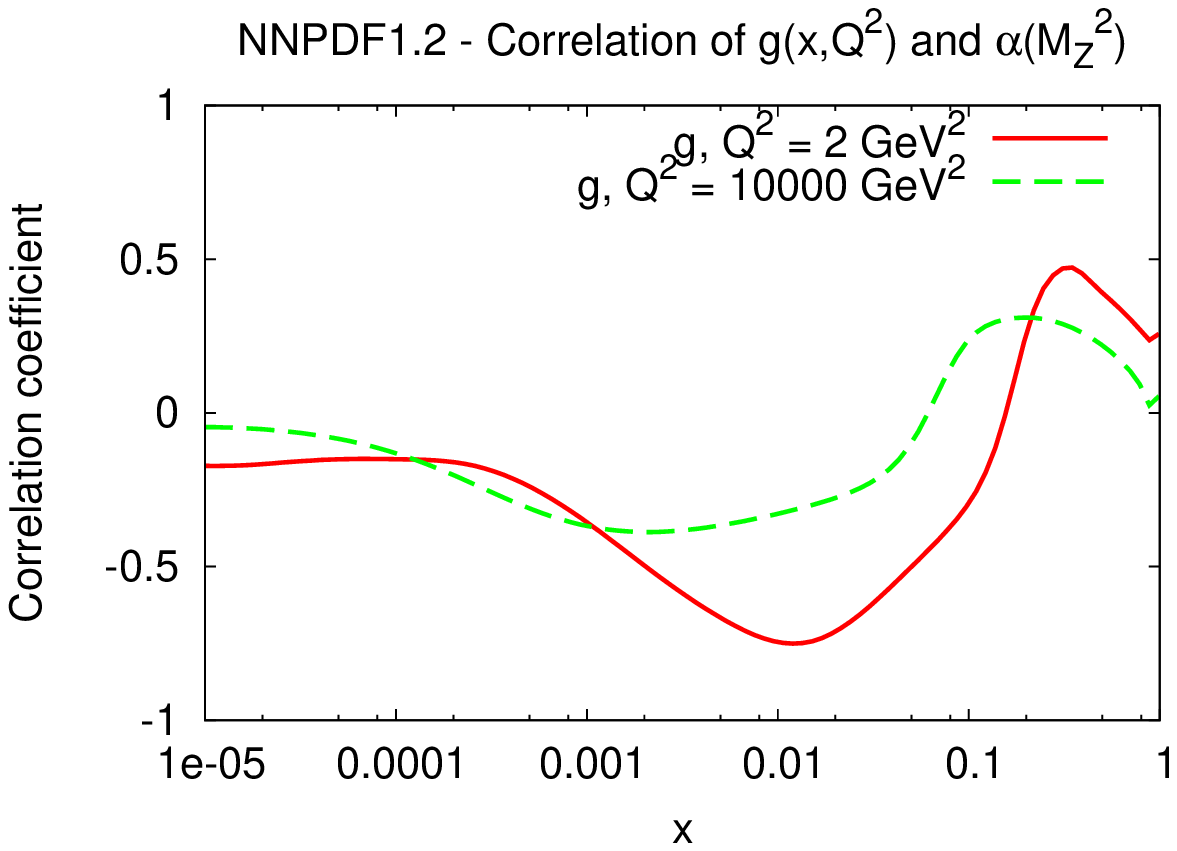}
\includegraphics[width=0.49\textwidth]{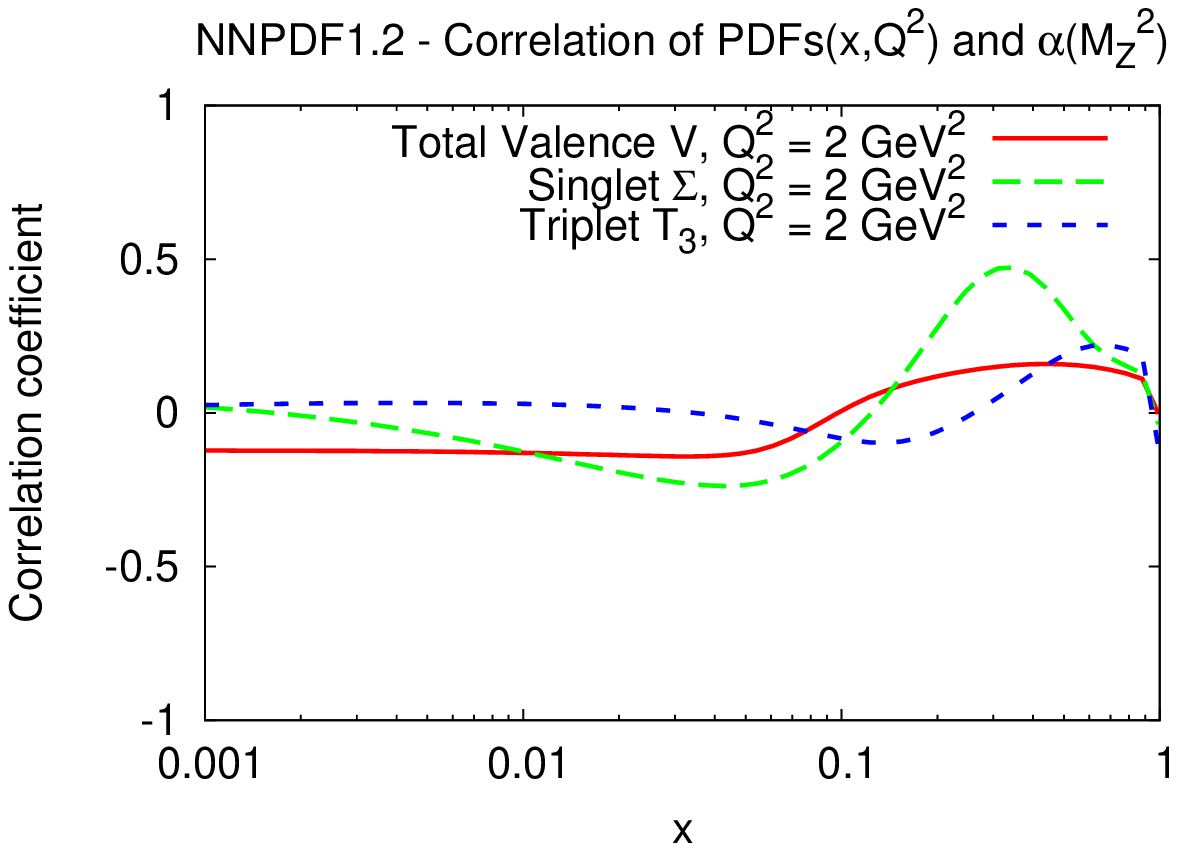}
 \caption{Left plot: The correlation coefficient,
defined in Eq.~\ref{LH_NNPDF_eq:gcorr}, between the gluon and $\alpha_s(M_Z^2)$
as a function of $x$, both at the input evolution scale
$Q_0^2=2$ GeV$^2$ and at a typical LHC scale $Q^2=10^4$ GeV$^2$.
Right plot: the analogous correlation coefficient for the
singlet, triplet and total valence PDFs. The distribution
of $\alpha_s\lp M_Z^2\rp$ has been assumed to be a gaussian
with mean and standard deviation given by Eq.~\ref{LH_NNPDF_eq:alphasref}.
}
\label{LH_NNPDF_fig:gluon-alphas-corr}
\end{center}
\end{figure}

Our results for the correlation coefficient
 between the gluon and $\alpha_s(M_Z^2)$
as a function of $x$, computed using Eq.~\ref{LH_NNPDF_eq:gcorr} 
both at the input evolution scale
$Q_0^2=2$ GeV$^2$ and at a typical LHC scale $Q^2=10^4$ GeV$^2$
are shown in 
Fig.~\ref{LH_NNPDF_fig:gluon-alphas-corr}. It is interesting to note 
how evolution
decorrelates the gluon from the strong coupling. 
We also show in Fig.~\ref{LH_NNPDF_fig:gluon-alphas-corr} the correlation
coefficient for other PDFs: as expected for the triplet and
valence PDFs it is essentially zero, that is, in NNPDF1.2 these
PDFs show no sensitivity to $\alpha_s$, as was clear from
Fig.~\ref{LH_NNPDF_fig:xpdf_comp}. 

The 
correlation coefficient Fig.~\ref{LH_NNPDF_fig:gluon-alphas-corr}
quantifies the qualitative observations of
Figs.~\ref{LH_NNPDF_fig:xg_comp}-\ref{LH_NNPDF_fig:xpdf_comp}. This correlation
coefficient could be used to correct the sum in quadrature of PDf and
$\alpha_s$ uncertainties, though in practice it is simpler to just use
the exact formula Eq.~(\ref{LH_NNPDF_eq:avrep}).

\subsection{Strong coupling uncertainty in Higgs production}

Let us consider a physical cross section which depends
both on the PDFs and $\alpha_s$, and which will
be denoted schematically by $\sigma\lp {\rm PDF},\alpha_s\rp$.
This cross-section has associated a PDF uncertainty 
$\delta\sigma_{\rm PDF}$, obtained from a fixed value $\alpha_s^{(0)}$.
Different PDF groups provide different recipes to estimate
this uncertainty.\footnote{
See for example appendix B of Ref.~\cite{Ball:2008by}.}
On the other hand, this cross section
 also has associated an uncertainty due to our limited
knowledge of $\alpha_s$, 
$\delta\sigma_{\alpha_s}$. The simplest way of estimating this uncertainty
is keeping the PDFs fixed to their central value, PDF$^{(0)}$, which
gives the following relative uncertainty
\be
\label{LH_NNPDF_eq:deltaalphas}
\frac{\lp \delta \sigma \rp_{\alpha_s}^{\pm} }{\sigma}
=
\frac{  \sigma\lp {\rm PDF}^{(0)},\alpha_s^{(0)}\pm \delta_{\alpha_s} \rp}{\sigma \lp {\rm PDF}^{(0)},\alpha_s^{(0)} \rp }  \ ,
\ee
where $\delta_{\alpha_s}$ is the assumed 68\% confidence level range for
$\alpha_s$, in our case given by Eq.~\ref{LH_NNPDF_eq:alphasref}.

Taking into account the presence of these two sources of
uncertainties, PDFs and $\alpha_s$, there are at least 
three different recipees to determine the combined
uncertainty in
the cross-section $\sigma$, denoted by
 $\lp \delta \sigma \rp_{{\rm PDF}+\alpha_s}^{\pm}$.
They can be ordered in  order of
formal accuracy

\begin{itemize}
\item The simplest approach consist in adding in quadrature the PDF and
$\alpha_s$ uncertainties, where the latter is defined by 
Eq.~\ref{LH_NNPDF_eq:deltaalphas}. In this case the combined uncertainty
will be given by
\be
\label{LH_NNPDF_eq:deltasigma}
\lp \delta \sigma \rp_{{\rm PDF}+\alpha_s}^{\pm} =\sqrt{
\lc \lp \delta \sigma \rp_{\alpha_s}^{\pm}\rc^2 +
\lc \lp \delta \sigma \rp_{\rm PDF}^{\pm}\rc^2 } \ .
\ee
The main drawback of this approach is that it neglects the correlation
between the PDFs and $\alpha_s$, which as we have seen
in Sect.~\ref{LH_NNPDF_sec:nnpdf12alphas} is not negligible in principle.
\item A more refined approach requires using PDFs
obtained from different values of  $\alpha_s$: this way it is
possible to take properly into account the correlations between
$\alpha_s$ and the PDFs. In this case, instead of using the
approximation Eq.~\ref{LH_NNPDF_eq:deltaalphas}, the $\alpha_s$ uncertainty
is evaluated with PDF sets obtained with the corresponding value
of $\alpha_s$, namely,
\be
\label{LH_NNPDF_eq:deltaalphas2}
\frac{\lp \delta \sigma \rp_{\alpha_s}^{\pm}}{\sigma} 
=\frac{\sigma\lp {\rm PDF}^{(\pm)},\alpha_s^{(0)}\pm \delta_{\alpha_s} \rp}{\sigma\lp {\rm PDF}^{(0)},\alpha_s^{(0)} \rp}  \ ,
\ee
where {\rm PDF}$^{(\pm)}$ stands schematically for the PDFs obtained
when $\alpha_s$ is varied within its 1-$\sigma$ range, $\alpha_s^{(0)}
\pm \delta_{\alpha_s}$.
Then the overall combined uncertainty will be given again by 
Eq.~\ref{LH_NNPDF_eq:deltasigma}, but with 
Eq.~\ref{LH_NNPDF_eq:deltaalphas2} for the
$\alpha_s$ uncertainties.
This approach, while being a clear improvement with respect
to the former, still misses some information on the correlations
between $\alpha_s$ and the PDFs: it assumes that PDFs obtained
with any value of $\alpha_s$ have  the same
uncertainties.
\item The third and more accurate option is given
by full correlated propagation of the PDF and $\alpha_s$ uncertainties
into the cross section $\sigma$. The details of this approach
will be different depending on the method used to determine the
PDF uncertainties. Within the NNPDF approach (or more in general
for any approach which uses the Monte Carlo method to estimate
PDF uncertainties), this combined uncertainty is simply
given by
\be
\label{LH_NNPDF_eq:exacterror}
\lp \delta \sigma \rp_{{\rm PDF}+\alpha_s}^{\pm} = \sqrt{\la  \sigma^2\ra_{\rm rep}
-\la  \sigma\ra^2_{\rm rep}} \ ,
\ee
where the average over replicas (which include PDFs with different
$\alpha_s$) is defined in Eq.~\ref{LH_NNPDF_eq:avrep} (note that here $\sigma$
denotes a cross--section, and $\delta \sigma$ the uncertainty on it).
\end{itemize}

As an illustration of the different procedures for the
combined treatment of PDFs and
$\alpha_s$ uncertainties within NNPDF, we have studied the specific case of
Higgs production through gluon-gluon fusion, computing the cross-section
uncertainties with the three different methods described above.
 As
in the rest of the contribution, the range of $\alpha_s$ is taken
to be that of Eq.~\ref{LH_NNPDF_eq:alphasref}, namely $\delta_{\alpha_s}=0.0012$
as a 68\% confidence level.

As described below, in the simplest approach
of sum in quadrature of the two uncertainties
one needs to compute first the PDF uncertainty
at fixed $\alpha_s$. In the particular case of the 
Higgs boson production cross section
 the PDF 
uncertainty  can be estimated
by computing the gluon-gluon luminosity,
\be
\label{LH_NNPDF_eq:higgs_lumi}
\Phi\lp m_H^2\rp \equiv \frac{1}{S}\int_{\tau}^1
\frac{dx_1}{x_1} g\lp x_1,M_H^2\rp g\lp x_2=\tau/x_1,m_H^2\rp \ ,
 \ee
with $\tau = m_H^2/S$ and $\sqrt{S}$ the center of
mass energy. At leading order, the Higgs cross section 
is simply proportional to Eq.~\ref{LH_NNPDF_eq:higgs_lumi}.

\begin{figure}[ht]
\begin{center}
\includegraphics[width=0.49\textwidth]{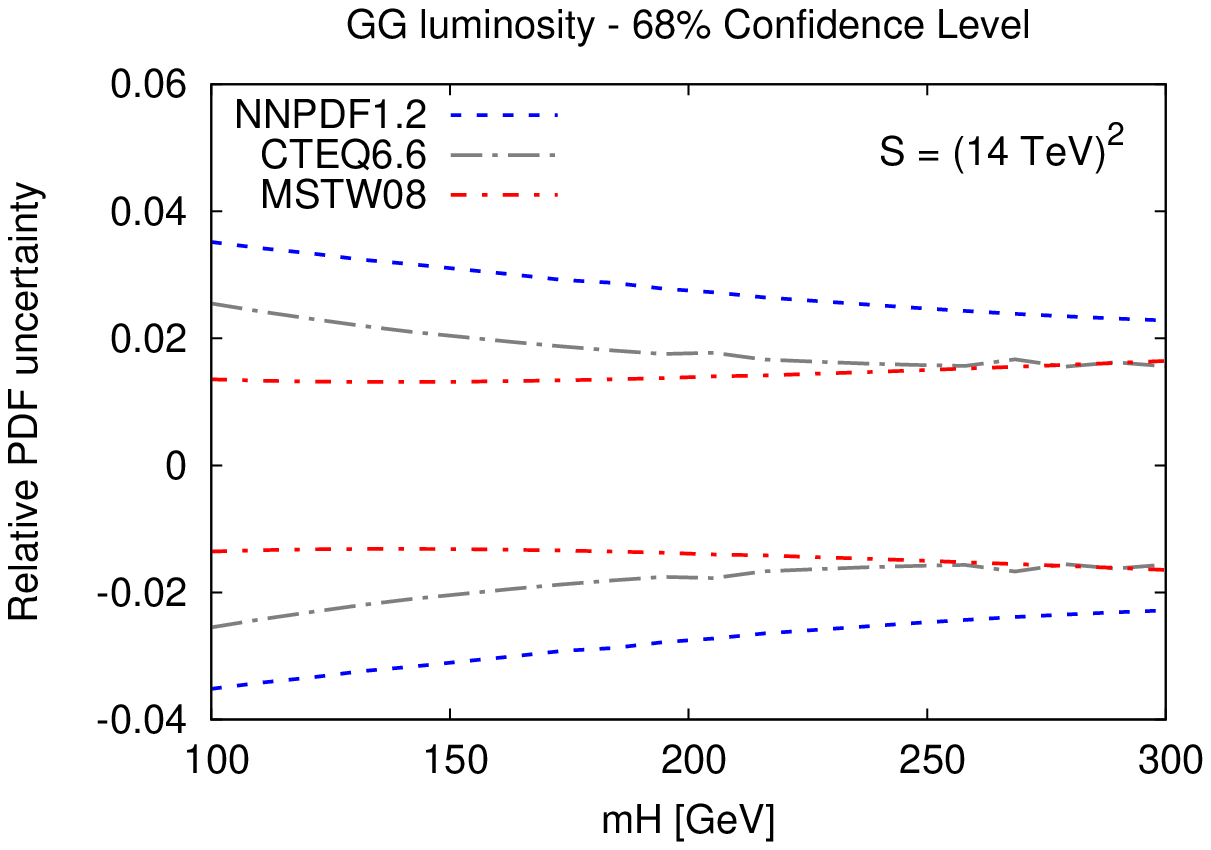}
\includegraphics[width=0.49\textwidth]{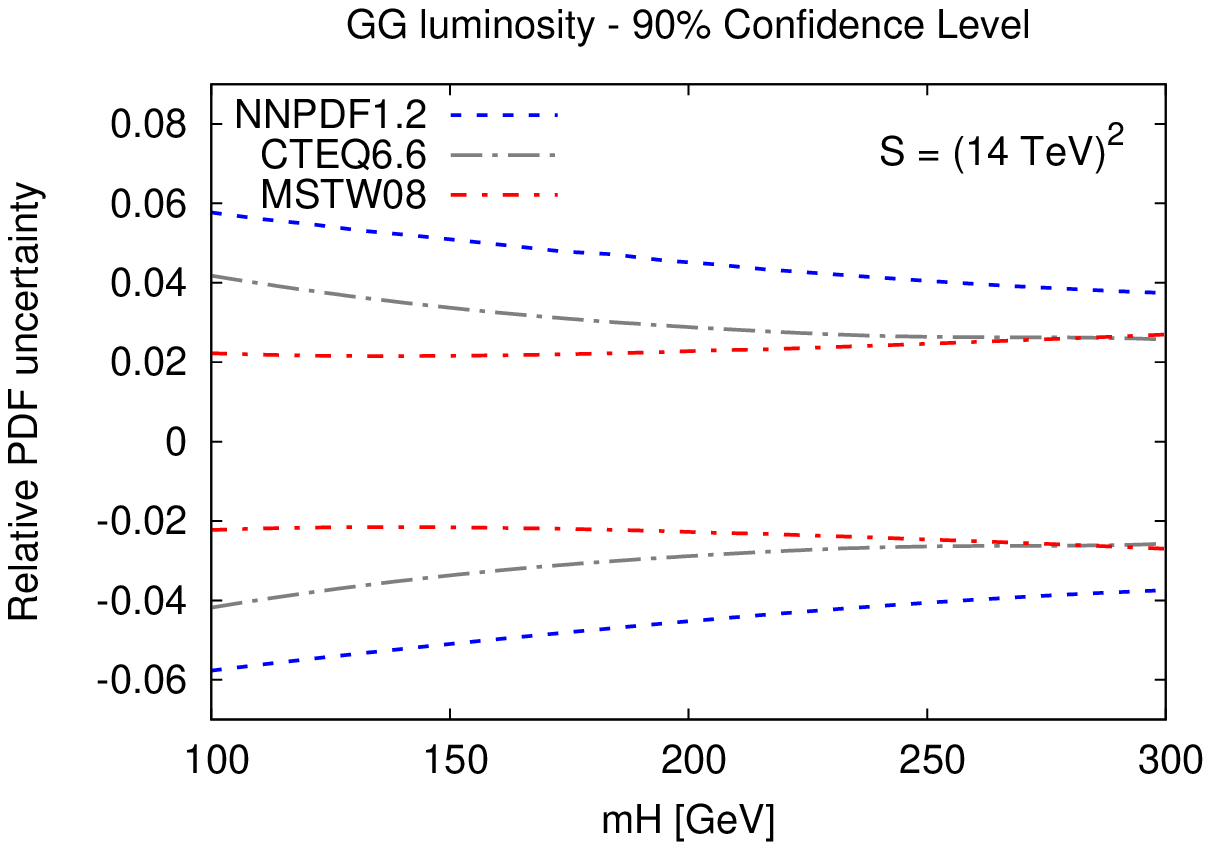}
\includegraphics[width=0.49\textwidth]{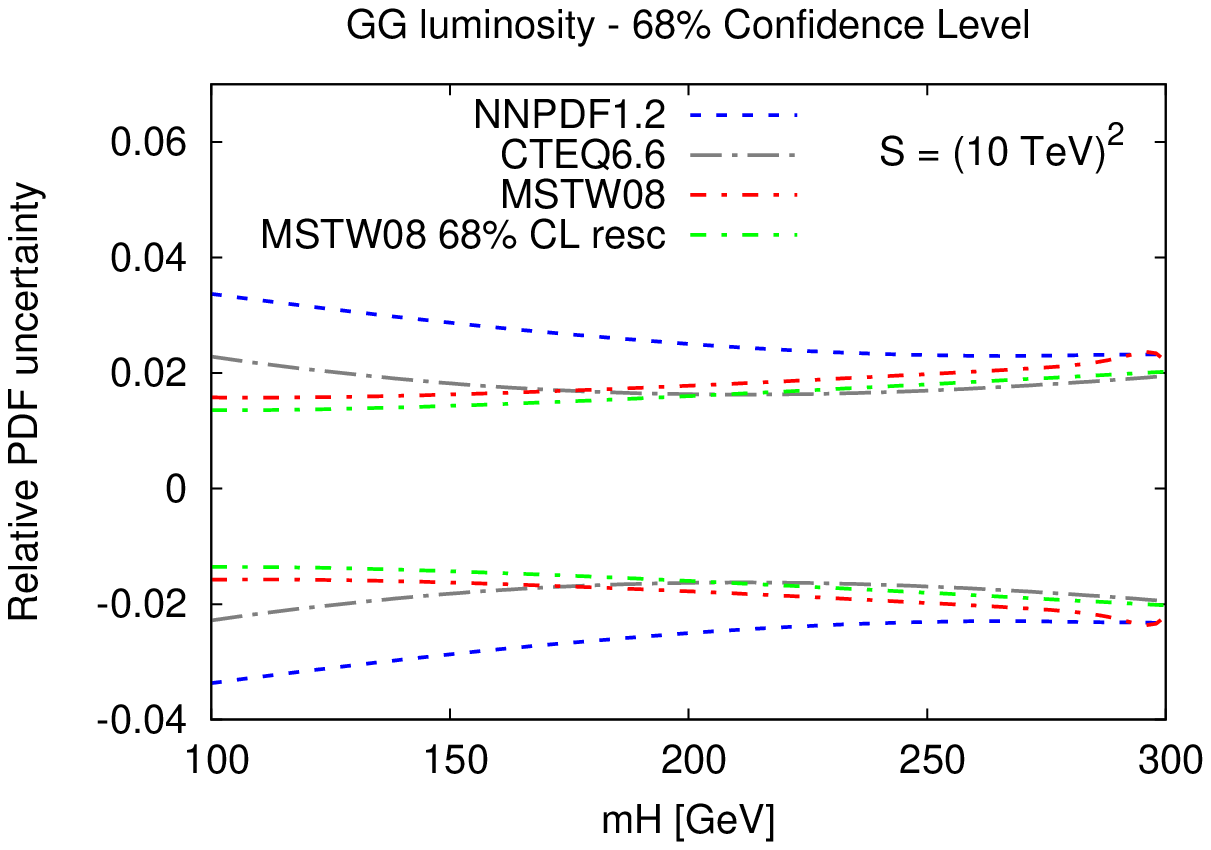}
\includegraphics[width=0.49\textwidth]{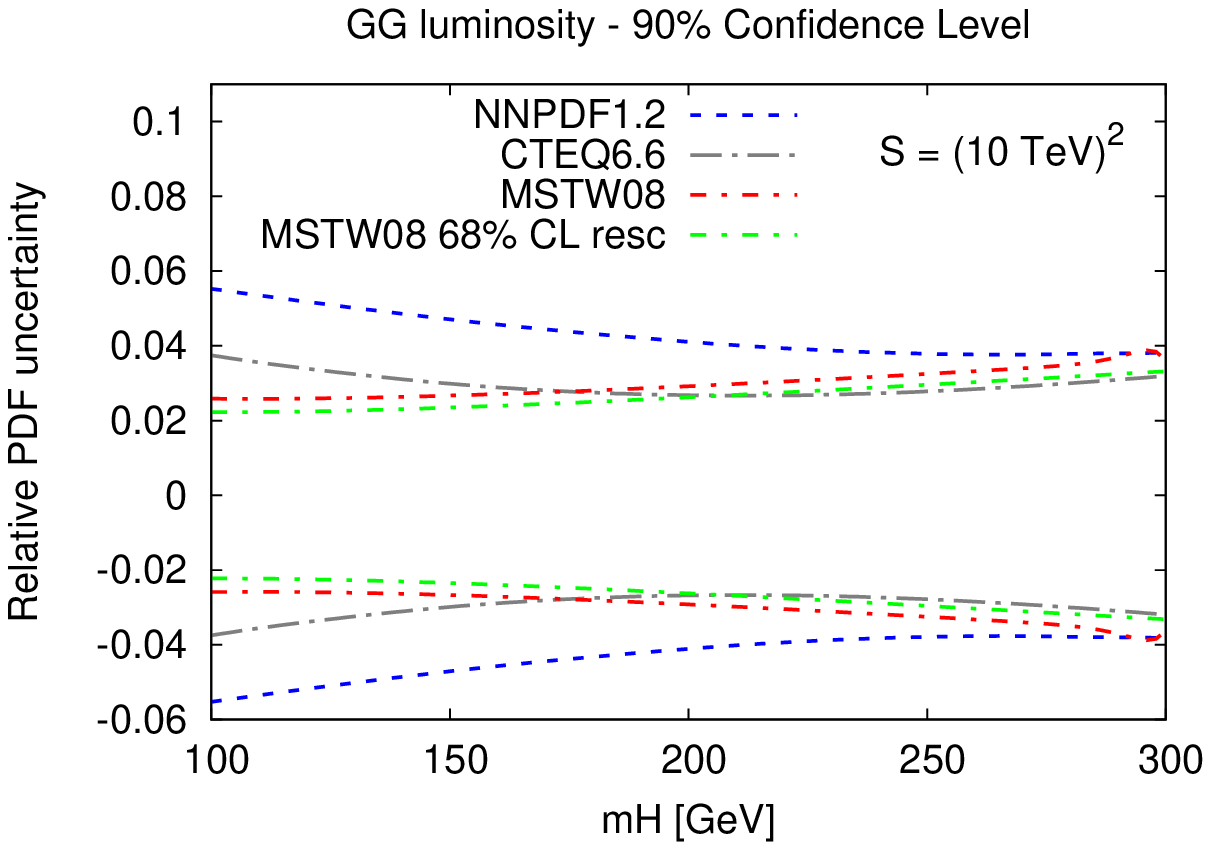}
 \caption{\small The gluon-gluon luminosity, Eq.~\ref{LH_NNPDF_eq:higgs_lumi},
computed at the LHC for $\sqrt{S}=14$ TeV (upper plots)
and $\sqrt{S}=10$ TeV (lower plots)  as a function of the
Higgs boson mass for NNPDF1.2, as well as for MSTW08 and CTEQ6.6.
In both cases we show both the 68\% (left plots) and the 90\% confidence
levels (right plots). As expected, the impact of the different CMS
energy in the PDF uncertainties is very reduced.}
\label{LH_NNPDF_fig:gglumi}
\end{center}
\end{figure}

 This effective
gluon-gluon luminosity as a function of the Higgs boson mass
at the LHC with $\sqrt{S}=14$ and  $\sqrt{S}=10$ TeV is shown in 
Fig.~\ref{LH_NNPDF_fig:gglumi}, where it
is also compared to the same quantity from two other
global PDF determinations: 
CTEQ6.6~\cite{Nadolsky:2008zw} and 
MSTW08~\cite{Martin:2009iq}. We can see that at large $m_H$ the CTEQ6.6
and MSTW08 uncertainties are identical, while CTEQ6.6 is
larger at small $m_H$. The NNPDF1.2 analysis results in the
largest uncertainties, partially at least
because the constrains from the hadronic
data included in the other global analyses is not
included.

\begin{figure}[ht]
\begin{center}
\includegraphics[width=0.49\textwidth]{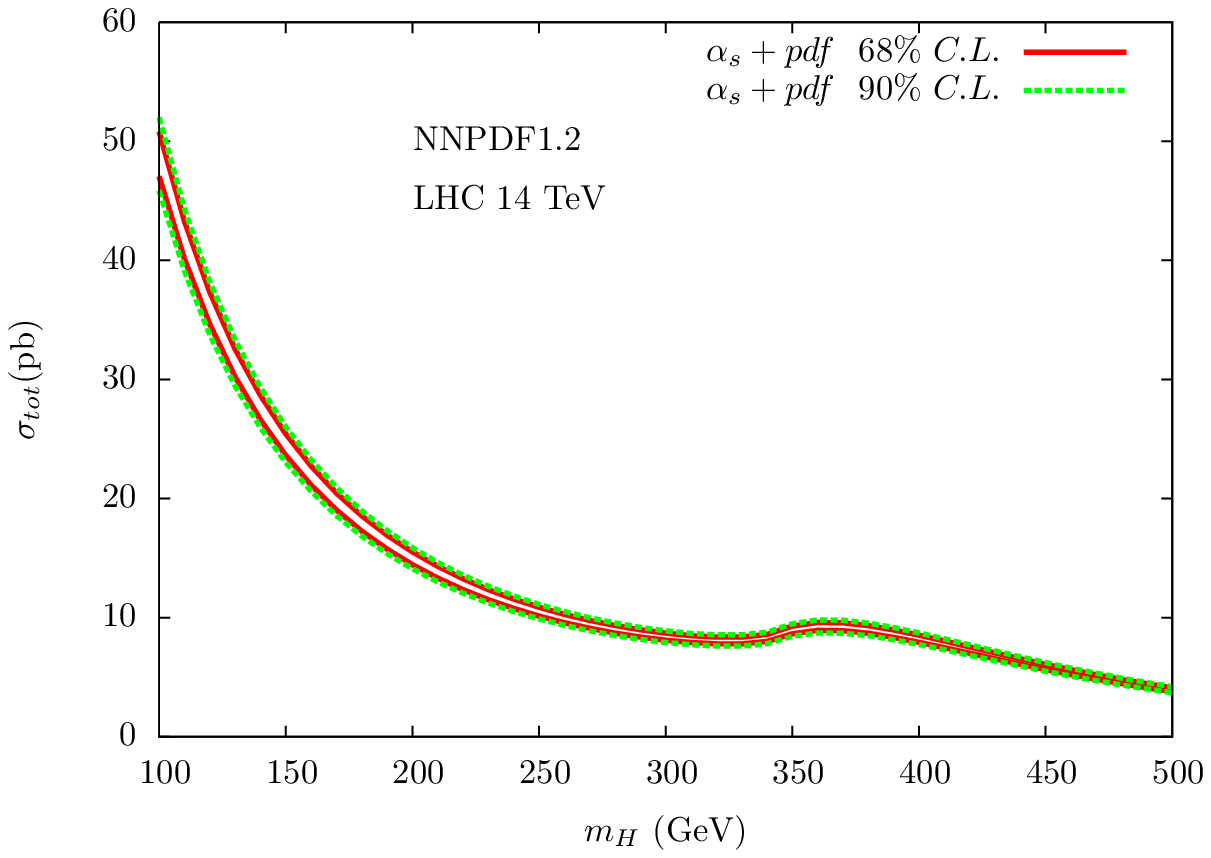}
\includegraphics[width=0.49\textwidth]{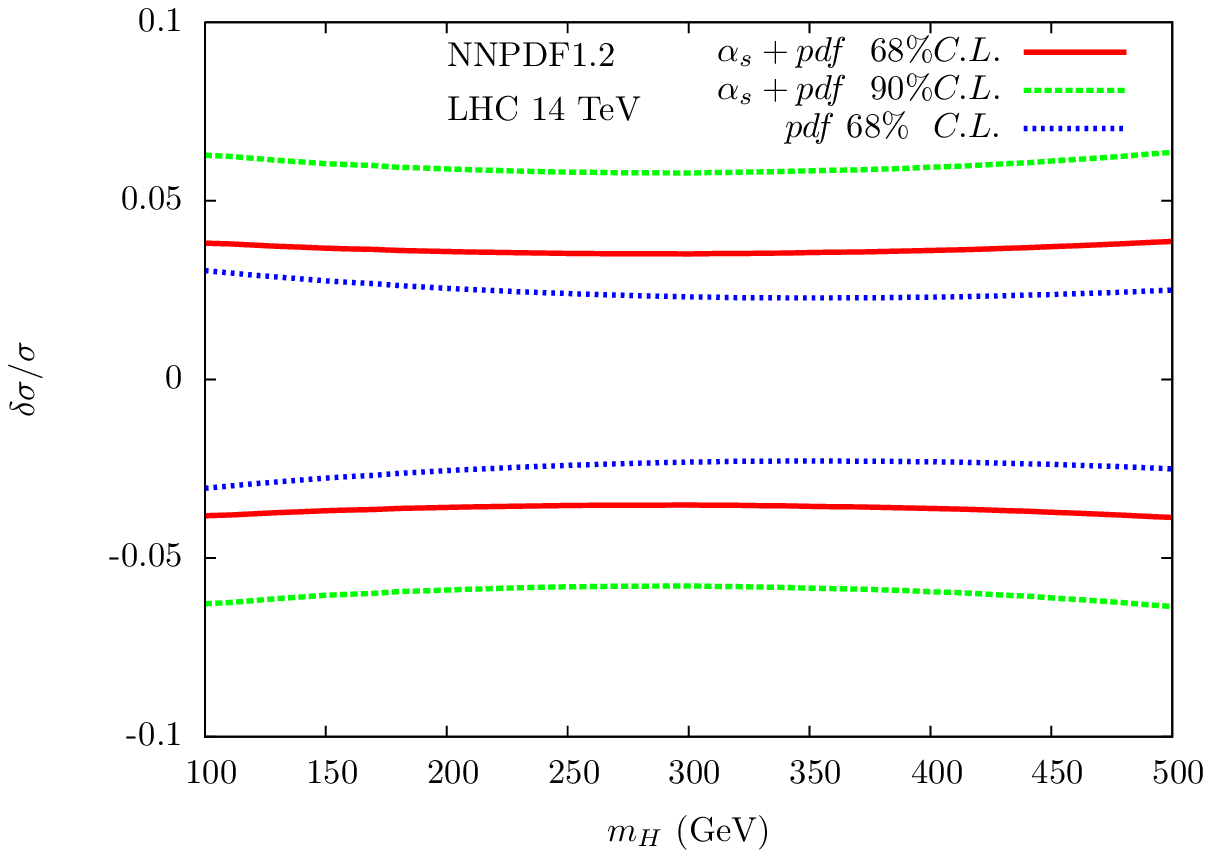}
 \caption{\small Left figure: The total cross section for
Higgs boson production at the LHC as a function of $m_H$, computed
with the NNPDF1.2 set. The red band is the combined PDF and $\alpha_s$
uncertainty obtained exact error propagation (see text for details) at
68\% confidence level, while the green band is the corresponding
90\% confidence level. Right figure: the relative uncertainties
in the total cross section as computed from PDFs only 
(blue line) and combined PDFs and $\alpha_s$ uncertainties
(red line) at 68\% C.L., always with exact error
propagation Eq.~\ref{LH_NNPDF_eq:exacterror}.}
\label{LH_NNPDF_fig:xsec_tot}
\end{center}
\end{figure}

Now we turn to a discussion of the effect of the
combination of PDF and $\alpha_s$ uncertainties in Higgs
boson production. All numerical results discussed below have
been obtained at NLO 
using the code of Refs.~\cite{Aglietti:2006tp,Bonciani:2007ex}.
In Fig.~\ref{LH_NNPDF_fig:xsec_tot} we show the total cross section for
Higgs boson production at the LHC as a function of $m_H$, computed
with the NNPDF1.2 set, with the uncertainty band obtained
by exact combination of the $\alpha_s$ and PDF uncertainties, both at
68\% and 90\% C.L..
The same figure also shows the relative uncertainties at
68\%
in the total cross section as computed from PDFs only.

\begin{figure}[ht]
\begin{center}
\includegraphics[width=0.75\textwidth]{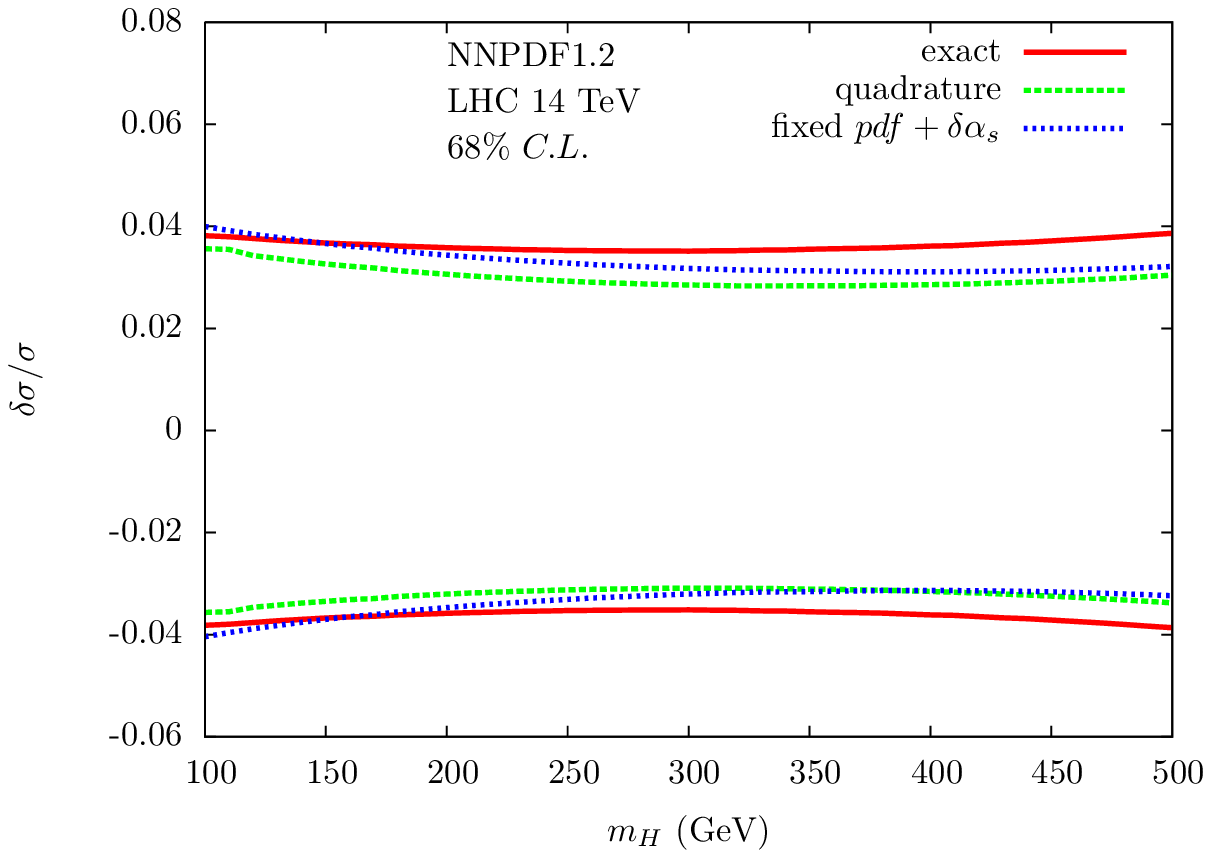}
 \caption{\small A comparison of the 68\% C.L. uncertainties in the Higgs
boson production cross section as a function of $m_H$ for
NNPDF1.2 with the
combined PDFs+$\alpha_s$ uncertainties, using the various
methods for the combination discussed in the text.  The red line corresponds
to exact error propagation Eq.~\ref{LH_NNPDF_eq:exacterror},
the green line to the sum
in quadrature of the two uncertainties in the case in which
$\alpha_s$ and also the PDFs are varied simultaneously, and
finally
the blue line is the sum in quadrature where the
PDFs are kept fixed to their value obtained in the fit
with central $\alpha_s$, Eq.~\ref{LH_NNPDF_eq:deltasigma}, see  the text for a more
detailed discussion.
It is clear that the sum of uncertainties in quadrature even
for fixed PDFs already provides a reasonable approximation to the
full result.}
\label{LH_NNPDF_fig:nnpdf_pdf+as_exact_quadr_nf6}
\end{center}
\end{figure}

In Fig.~\ref{LH_NNPDF_fig:nnpdf_pdf+as_exact_quadr_nf6} we show
a comparison of the 68\% C.L. in the Higgs
boson production cross section as a function of $m_H$ with the
combined PDFs+$\alpha_s$ uncertainties, were exact error propagation
is compared to the sum in quadrature of the two uncertainties. The sum
in quadrature is done either by keeping the PDF fixed when $\alpha_s$
varied, or eles by taking the central best fit PDF set for each value
of $\alpha_s$. Clearly, even  the simplest sum in quadrature provides a very
reasonable
approximation to the exact result obtained with full
error propagation. Therefore, one can conclude that,
at least for the range of variation of $\alpha_s$ assumed,
Eq.~\ref{LH_NNPDF_eq:alphasref}, and with the NNPDF1.2 parton fit,
the naive sum in quadrature of the PDF and
$\alpha_s$ uncertainties seems to be a good enough approximation
to the full result for most practical purposes.

A more detailed study of the interplay between $\alpha_s$ and
PDF uncertainties in Higgs production for various PDF
sets will be presented elsewhere~\cite{higgspdfs}

\subsection*{CONCLUSIONS}
In this contribution we have studied the interplay between the strong
coupling and PDF determination, and assessed the impact of the
combined uncertainty one of the most sensitive processes to $\alpha_s$
variations, Higgs production through gluon-gluon fusion. The main
result is that, at least within the
NNPDF approach, even in a worst--case scenario like Higgs production
at the LHC, the sum in quadrature of PDF and $\alpha_s$
uncertainties provides an excellent approximation to the full
result obtained from exact error propagation.

%

%% file: rojo/rojo.tex
%
%




\newcommand{\tmop}[1]{\ensuremath{\operatorname{#1}}}
\newcommand{\tmtextit}[1]{{\itshape{#1}}}
\newcommand{\tmtextrm}[1]{{\rmfamily{#1}}}
\newcommand{\tmtexttt}[1]{{\ttfamily{#1}}}
\def\smallfrac#1#2{\hbox{${{#1}\over {#2}}$}}
\newcommand{\be}{\begin{equation}}
\newcommand{\ee}{\end{equation}}
\newcommand{\bea}{\begin{eqnarray}}
\newcommand{\eea}{\end{eqnarray}}
\newcommand{\bi}{\begin{itemize}}
\newcommand{\ei}{\end{itemize}}
\newcommand{\ben}{\begin{enumerate}}
\newcommand{\een}{\end{enumerate}}
\newcommand{\la}{\left\langle}
\newcommand{\ra}{\right\rangle}
\newcommand{\lc}{\left[}
\newcommand{\rc}{\right]}
\newcommand{\lp}{\left(}
\newcommand{\rp}{\right)}
\newcommand{\as}{\alpha_s}
\newcommand{\aq}{\alpha_s\left( Q^2 \right)}
\newcommand{\amz}{\alpha_s\left( M_Z^2 \right)}
\newcommand{\aqq}{\alpha_s \left( Q^2_0 \right)}
\newcommand{\aqz}{\alpha_s \left( Q^2_0 \right)}
\def\toinf#1{\mathrel{\mathop{\sim}\limits_{\scriptscriptstyle
{#1\rightarrow\infty }}}}
\def\tozero#1{\mathrel{\mathop{\sim}\limits_{\scriptscriptstyle
{#1\rightarrow0 }}}}
\def\toone#1{\mathrel{\mathop{\sim}\limits_{\scriptscriptstyle
{#1\rightarrow1 }}}}
\def\frac#1#2{{{#1}\over {#2}}}
\def\gsim{\mathrel{\rlap{\lower4pt\hbox{\hskip1pt$\sim$}}
    \raise1pt\hbox{$>$}}}         
\def\lsim{\mathrel{\rlap{\lower4pt\hbox{\hskip1pt$\sim$}}
    \raise1pt\hbox{$<$}}}         
\newcommand{\MS}{$\overline{\text{\tmtextrm{MS}}}$}
\newcommand{\mrexp}{\mathrm{exp}}
\newcommand{\dat}{\mathrm{dat}}
\newcommand{\one}{\mathrm{(1)}}
\newcommand{\two}{\mathrm{(2)}}
\newcommand{\art}{\mathrm{art}} 
\newcommand{\rep}{\mathrm{rep}}
\newcommand{\net}{\mathrm{net}}
\newcommand{\stopp}{\mathrm{stop}}
\newcommand{\sys}{\mathrm{sys}}
\newcommand{\stat}{\mathrm{stat}}
\newcommand{\diag}{\mathrm{diag}}
\newcommand{\pdf}{\mathrm{pdf}}
\newcommand{\tot}{\mathrm{tot}}
\newcommand{\minn}{\mathrm{min}}
\newcommand{\mut}{\mathrm{mut}}
\newcommand{\partt}{\mathrm{part}}
\newcommand{\dof}{\mathrm{dof}}
\newcommand{\NS}{\mathrm{NS}}
\newcommand{\cov}{\mathrm{cov}}
\newcommand{\gen}{\mathrm{gen}}
\newcommand{\cut}{\mathrm{cut}}
\newcommand{\parr}{\mathrm{par}}
\newcommand{\checkk}{\mathrm{check}}
\newcommand{\reff}{\mathrm{ref}}
\newcommand{\extra}{\mathrm{extra}}
\newcommand{\draft}[1]{}
\newcommand{\comment}[1]{{\bf \it  #1}}
\definecolor{grey}{rgb}{0.5,0.5,0.5}

\newcommand{\grey}{\color{grey}}

\subsection{Introduction}

Interest in the inclusion of heavy flavour contributions to deep--inelastic
electroproduction structure functions was recently revived by the
discovery~\cite{Tung:2006tb} that mass-suppressed
terms in global parton fits can affect predictions for the total $W$
and $Z$ production at the LHC by almost 10~\%. 
A technique for the inclusion of these mass-suppressed
contributions
to structure functions was developed long 
ago~\cite{Collins:1998rz,Aivazis:1993pi},
based upon a 
renormalization scheme with
explicit heavy quark decoupling~\cite{Collins:1978wz}.
Several variants of this  method  (usually called ACOT) 
were subsequently proposed, such as 
S-ACOT~\cite{Kramer:2000hn} and
ACOT-$\chi$~\cite{Tung:2001mv,Kretzer:2003it}.
However,  the ACOT method was 
first used for an actual general-purpose global parton fit only recently, in
Refs.~\cite{Kretzer:2003it,Tung:2006tb}.\footnote{
It had been however used in specific studies
in the CTEQ HQ series of fits, HQ4~\cite{Lai:1997vu}, 
HQ5~\cite{Lai:1999wy} and HQ6~\cite{Kretzer:2003it}.}

An alternative method (sometimes called TR) has also been
advocated~\cite{Thorne:1997ga,Thorne:1997uu},
and used for all MRST parton fits until
2004~\cite{Martin:1998sq,Martin:2001es,Martin:2002aw,Martin:2004ir}.
Recently, however, the methods used by
the CTEQ~\cite{Tung:2006tb} and MRST/MSTW~\cite{Martin:2007bv,Martin:2009iq}
groups for their current parton fits, based respectively on the
ACOT~\cite{Collins:1998rz,Aivazis:1993pi} and
TR$^\prime$~\cite{Thorne:2006qt} procedures, have adopted at
least in part a common framework: they have been compared recently in
Refs.~\cite{Thorne:2008xf,Olness:2008px}, thereby elucidating
differences and common aspects.

A somewhat different technique for the inclusion of heavy quark
effects, the so-called FONLL method, 
was introduced in Ref.~\cite{Cacciari:1998it} in the context of
hadroproduction of  heavy
quarks. The FONLL method only relies on
standard QCD factorization and calculations with massive quarks in the
decoupling scheme of Ref.~\cite{Collins:1978wz} and with massless quarks in the
 \MS\ scheme. The name FONLL is motivated by the fact
that the method was originally 
used to combine a fixed (second) order calculation
with a next-to-leading log one; however, the method is entirely
general, and it can be used to combine consistently a fixed order with
a resummed
calculations to any order of either. The application of the
FONLL scheme to deep--inelastic structure functions was
recently presented in Ref.~\cite{Forte:2010ta}. Thanks to
its simplicity, the FONLL method  provides a
framework for understanding differences between other existing
approaches, and for a study of the effect of different choices in the
inclusion of subleading terms.

It is the aim of this contribution to update previous
comparisons of GM-VFN schemes in DIS~\cite{Thorne:2008xf,Olness:2008px} 
from a rather more
quantitative point of view. Therefore, after unique settings have
been adopted for all parties involved, the heavy quark structure
functions $F_{2c}$ and $F_{Lc}$, as implemented in the various
available approaches, have been computed and
compared in detail. This comparison is of extreme importance
in order to understand how parton distribution sets obtained
from different schemes might differ, and what are the associated
implications for LHC observables.

The outline of this contribution is the following. First of
all, we present the benchmark settings for the computation
of charm structure functions. Then we present the results for
the comparison between the different schemes considered: ACOT, TR$^\prime$
and FONLL, and discuss their similarities and differences.
Finally we summarize and provide benchmark tables which should
be used for other GM-VFN schemes not considered here, either
existing, updated or completely new.

\subsection{Benchmark settings}
\label{LH_HQ_sec:intro}

Let us discuss now the settings for the benchmark comparisons
between different GM-VFN schemes. These settings have been
designed to isolate only the potential similarities and differences
between GM-VFN schemes, while other choices that are generally varied
between PDF fitting groups (like, for example, the value of $\alpha_s$)
are shared among all the parties.

The goal of the benchmark
comparison is to produce and
compare results for the  charm structure functions $F_{2c}$ and $F_{Lc}$
(for which we adopt the notation of Ref.~\cite{Forte:2010ta}),
computed at different values of $x$ and $Q^2$ from a variety of
GM-VFN schemes. 

 These settings which we adopt for the
benchmark comparison are the following:

\begin{itemize}
\item As input PDF set,
the Les Houches initial conditions~\cite{Dittmar:2005ed,Giele:2002hx} 
are used, with the initial scale for the PDF and $\alpha_s$
evolution taken to be $Q_0^2=2$~GeV$^2$.  The initial $\alpha_s(Q_0^2)=0.35$.
\item The charm mass is taken to be $m_c=Q_0=\sqrt{2}$~GeV at NLO.  At NNLO, both PDFs and $\alpha_s$ are discontinuous at $Q^2=m_c^2$ in a VFN scheme.  We take the input PDFs and $\alpha_s$ at $Q_0=\sqrt{2}$~GeV to correspond to $N_f=3$, i.e.~the charm mass is taken to be $m_c=(\sqrt{2}+\epsilon)$~GeV, for infinitesimal $\epsilon$, so that the appropriate NNLO discontinuities present at $Q^2=m_c^2$ are added to the input values before evolving to higher $Q^2$.
\item The PDFs have been evolved with HOPPET~\cite{Salam:2008qg},
an $x-$space PDF evolution code,  and interpolated in grids
for easier interfacing with the various programs. Any other
evolution code whose accuracy has been benchmarked with HOPPET
like PEGASUS~\cite{pegasus} would be equally valid.
\item The charm quark is the only heavy quark present
in the theory; the bottom and top
quark masses are taken to infinity. This way complications arising from
the presence of multiple heavy quarks are not considered.
\item The $Q^2$
range of these benchmarks 
is from $Q^2=4$ GeV$^2$ (near the heavy quark threshold)
to $Q^2=100$ GeV$^2$ (which is close to the asymptotic
limit for  practical purposes).
 Appropriate intermediate values are $Q^2=10$ and 24 GeV$^2$.
\item The strong coupling constant $\alpha_s(Q^2)$ is computed by means
of exact numerical integration of the evolution equations
(as usually done in $x-$space codes like HOPPET)
instead
of one of the various possible expanded solutions.
The initial $\alpha_s(Q_0^2)=0.35$ at both NLO and NNLO.  Again, we take
$m_c=(\sqrt{2}+\epsilon)$~GeV at NNLO, so that the input scale
$Q_0=\sqrt{2}$~GeV corresponds to $N_f=3$.
With this choice, the values of $\alpha_s$ to be used
in the benchmark computations will be given by:
\bea
\alpha_s\lp Q^2 = 4\,{\rm GeV}^2\rp&=&0.295\,\, (0.295) \ , \nonumber \\
\alpha_s\lp Q^2 = 10\,{\rm GeV}^2\rp&=&0.245\,\, (0.244) \ , \nonumber \\
\alpha_s\lp Q^2 = 24\,{\rm GeV}^2\rp&=&0.212\,\, (0.211) \ , \\
\alpha_s\lp Q^2 = 100\,{\rm GeV}^2\rp&=&0.174\,\, (0.173) \ ,\nonumber
\eea
at NLO (and NNLO) respectively for the values of $Q^2$ used
for the benchmarks.
\item As discussed in Refs.~\cite{Chuvakin:1999nx,Forte:2010ta}, 
from $\mathcal{O}\lp \alpha_s^2\rp$ 
there is an ambiguity in the definition of the heavy
quark structure functions from terms
in which a light quark
couples to the virtual photon. For these benchmarks, $F_{2c}$ and $F_{Lc}$, 
the heavy quark structure functions, are
always defined as the sum of the contributions in which a charm 
quark is struck by the virtual photon, as opposed to the
widely used experimental definition, which is the sum of all
the contributions in which a charm 
quark is present in the final state. This definition avoids
the presence of infrared unsafe terms from the non-cancellation
of mass singularities.
\end{itemize}

\subsection{General--Mass heavy quark schemes}

As discussed in the introduction, the aim of the benchmark
comparison is to identify similarities and differences between
the GM-VFN schemes which are, have  been or will be used in
global PDF determinations. Without the purpose
of being comprehensive, we present now a brief introduction of
the three approaches which are compared in this contribution:
ACOT (used in the CTEQ family of PDF fits), TR/TR$^{\prime}$ 
(used in the MRST/MSTW
family) and FONLL (currently being implemented in the
NNPDF family). The interested reader can find all relevant
technical details in the quoted bibliography.

For simplicity,  all the discussion in this
section assumes a single heavy quark with mass $m_c$, since the
case of the charm quark is the one with the most phenomenological
importance.

\subsubsection{ACOT}
\label{LH_HQ_sec:acot}

The ACOT renormalization scheme~\cite{Aivazis:1993kh,Aivazis:1993pi}
provides a mechanism to incorporate the heavy quark mass into the
theoretical calculation of heavy quark production both kinematically
and dynamically. This is built upon the Collins-Wilczek-Zee
(CWZ)\cite{Collins:1978wz} renormalization procedure which provides a
formal foundation for the ACOT scheme which is valid to all
orders. The CWZ renormalization ensures there are no large logarithms
of the form $\ln(m_{c}/Q)$, and yields manifest decoupling of the
heavy quarks in the $m_{c}\gg Q$ limit. In 1998
Collins~\cite{Collins:1998rz} extended the factorization theorem to
address the case of heavy quarks; this work ensures we can compute
heavy quark processes  to all orders.
Thus, the ACOT scheme yields the complete quark mass dependence from
the low to high energy regime; for $m_{c}\gg Q$ it ensures manifest
decoupling, and in the limit $m_{c}\ll Q$ it reduces precisely to the
$\overline{{\rm MS}}$ scheme \emph{without} any finite renormalization
terms.\footnote{This has been demonstrated both analytically and
numerically; {\it e.g.}  the Fortran code used in the current comparison has been
numerically verified with the $\overline{{\rm MS}}$ results for QCDNUM
version 16.12.}

As a result of the Collins~\cite{Collins:1998rz} proof, it was
observed that the heavy quark mass could be set to zero in certain
pieces of the hard scattering terms without any loss of accuracy.
This modification of the ACOT scheme goes by the name Simplified-ACOT
(S-ACOT) and can be summarized as follows.

\begin{description}
\item [{{{S-ACOT:}}}] For hard-scattering processes with incoming
heavy quarks or with internal on-shell cuts on a heavy quark line, the
heavy quark mass can be set to zero ($m_{c}=0$) for these
pieces.~\cite{Kramer:2000hn}
\end{description}
If we consider the case of NLO DIS heavy quark production, this means
we can set $m_{c}=0$ for both the LO terms ($QV\to Q$) and the NLO
quark-initiated terms (both the real $QV\to Qg$ and the virtual $QV\to
Q$) as this involves an incoming heavy quark. We can also set
$m_{c}=0$ for the subtraction terms as this has an on-shell cut on an
internal heavy quark line; this includes both the gluon-initiated
process: $(g\to Q\bar{Q})\otimes(QV\to Q)$ as well as the quark
initiated process: $(Q\to gQ)\otimes(QV\to Q)$. Hence, the only
contribution which requires calculation with $m_{c}$ retained is the
NLO $gV\to Q\bar{Q}$ process.

In the conventional implementation of the heavy quark PDFs, we must
{}``rescale'' the Bjorken $x$ variable as we have a massive parton in
the final state. The original rescaling procedure is to make the
substitution $x\to x(1+m_{c}^{2}/Q^{2})$ which provides a kinematic
penalty for producing the heavy charm quark in the final
state~\cite{Barnett:1976ak}.  As the charm is pair-produced by the
$g\to c\bar{c}$ process, there are actually two charm quarks in the
final state---one which is observed in the semi-leptonic decay, and
one which goes down the beam pipe with the proton remnants. Thus, the
appropriate rescaling is not $x\to x(1+m_{c}^{2}/Q^{2})$ but instead
$x\to\chi=x(1+(2m_{c})^{2}/Q^{2})$; this rescaling is implemented in
the ACOT--$\chi$ scheme, for
example~\cite{Amundson:1998zk,Amundson:2000vg,Tung:2001mv}.  The
factor $(1+(2m_{c})^{2}/Q^{2})$ represents a kinematic suppression
factor which will suppress the charm process relative to the lighter
quarks.

\subsubsection{Thorne-Roberts (TR/TR$^\prime$)}
\label{LH_HQ_sec:mstw}

The TR scheme was introduced in Refs.~\cite{Thorne:1997ga,Thorne:1997uu} as 
an alternative to ACOT~\cite{Aivazis:1993pi} with more emphasis on 
correct threshold behaviour. Like the ACOT scheme it is 
based on there being two different regions separated by a transition point, 
by default $Q^2=m_c^2$.
Below this point a heavy quark is not an active parton 
but is generated in the final state using fixed-flavour (FF) 
coefficient functions,
while above this point the heavy quark becomes a new parton, evolving 
according to the massless evolution equations, and structure functions 
are obtained using variable-flavour (VF) coefficient functions which must 
tend to the 
correct massless $\overline{\rm MS}$-scheme limits as $Q^2/m_c^2 \to \infty$,
up to possibly higher-order corrections. 
The relationships between the partons below and above the transition point are
obtained from the transition matrix elements $A_{ij}(z,m_c^2/\mu^2)$ 
calculated to 
${\cal O}(\alpha_s^2)$ in Ref.~\cite{Buza:1996wv}, which fortuitously result in
continuity up to NLO in the $\overline{\rm MS}$-scheme. 

The definition of the scheme is therefore equivalent to the definition of 
the VF coefficient functions. These are found by imposing the 
exact all-orders equivalence of the structure functions described using the 
FF scheme and the VF scheme. This provides a 
relationship between the coefficient functions in the two cases via the 
equation
\begin{equation}
C^{\rm FF}_{j} = \sum_i C^{\rm VF}_i \otimes A_{ij},
\label{LH_HQ_eq:vfdef}
\end{equation}
where the sum is over all the different partons in the VF 
description. This equivalence effectively defines the subtraction of 
the large logarithms in $Q^2/m_c^2$ in 
the FF coefficient functions in the correct manner. 
It was applied to obtain relationships in the asymptotic limit in
Refs.~\cite{Buza:1995ie,Buza:1996wv}, and used to define the BMSN scheme 
in Ref.~\cite{Buza:1996wv},
but in Refs.~\cite{Thorne:1997ga,Thorne:1997uu} it was used to define the 
VF coefficient functions 
for all $Q^2 > m_c^2$.  The definition is not unique because 
there are more coefficient functions on the right than on the left of
Eq.~\ref{LH_HQ_eq:vfdef}, because of
the extra heavy quark coefficient function on the right-hand side.
As $Q^2/m_c^2 \to \infty$ all 
VF coefficient functions must tend to the 
massless $\overline{\rm MS}$-scheme limit, but at
finite $Q^2$ there is a freedom in the heavy quark coefficient functions, 
beginning with the 
zeroth-order ($C^{{\rm VF},(0)}_{c}$). Via Eq.~\ref{LH_HQ_eq:vfdef} this
affects other coefficient functions, e.g.
\begin{equation}
C^{{\rm FF},(1)}_{2c,g} \equiv C^{{\rm VF},(0)}_{2c,c} \otimes A^{(1)}_{cg}
+ C^{{\rm VF},(1)}_{2c,g},
\end{equation} 
so the choice of  $C^{{\rm VF},(0)}_{2c,c}$ also defines
$C^{{\rm VF},(1)}_{2c,g}$.

In the TR scheme~\cite{Thorne:1997ga,Thorne:1997uu} the approach is
to make a choice where all coefficient 
functions obey the correct threshold $W^2\geq 4m_c^2$ for heavy quark pair 
production. This was first imposed by defining the heavy quark coefficient
functions such that the evolution $\partial F_{2c}/\partial\ln Q^2$ is
continuous
order-by-order at the transition point (possible only in the gluon sector 
beyond LO). This was used in subsequent MRST global analyses up to 
MRST 2004~\cite{Martin:2004ir}.  However, it 
results in expressions which become increasingly complicated at higher order.  

In Ref.~\cite{Tung:2001mv} the correct threshold behaviour was achieved
by using the simple approach of replacing the limit of $x$ 
for convolution 
integrals with $\chi=x(1+4m_c^2/Q^2)$. In the case that the heavy flavour 
coefficient functions are just the massless ones with this restriction one 
obtains the S-ACOT($\chi$) approach. A very similar definition for
heavy flavour coefficients was adopted in Ref.~\cite{Thorne:2006qt}, resulting
in the TR$^{\prime}$ scheme, and extended explicitly to NNLO. 
This TR$^{\prime}$ scheme 
was first used in the MRST 2006 analysis~\cite{Martin:2007bv}, 
and has been used in all subsequent MSTW analyses
(see Sect.~4 of Ref.~\cite{Martin:2009iq}).

%
%

There is one other aspect to the TR/TR$^\prime$ scheme definition. 
For $F_{2c}$ 
the relative order 
of the FF and VF coefficient functions is different, 
i.e.~the former begin at first order in $\alpha_s$ and the latter at zeroth 
order. One cannot simply adopt the correct ordering above and below the 
transition point since then there would be a discontinuity in the structure 
function at $Q^2=m_c^2$, and because higher order effects are large at 
small $x$ and $Q^2$, this would be phenomenologically significant. The 
procedure adopted is to freeze the highest order part of the FF
expression (i.e.~${\cal O}(\alpha_s)$ at LO, ${\cal O}(\alpha_s^2)$ at NLO, 
etc.), and keep this in the expression above $Q^2=m_c^2$. Hence, there is an 
additional, strictly higher order contribution in this region which becomes 
less important as $Q^2$ increases, but never vanishes even at asymptotic 
$Q^2$. At NNLO this requires producing a model 
for the ${\cal O}(\alpha_s^3)$ FF coefficient functions from 
the known small-$x$ and threshold limits~\cite{Thorne:2006qt}.


\subsubsection{FONLL}
\label{LH_HQ_sec:fonll}

The FONLL scheme was first introduced in the context of heavy flavour
hadroproduction in Ref.~\cite{Cacciari:1998it}.
It is based upon the idea of looking at
both the massless and massive scheme calculations
as power expansions in the strong coupling constant, and replacing
the coefficient of the expansion in the former with their exact massive
counterpart in the latter, when available.
A detailed description of the FONLL method
for DIS has been given in Ref.~\cite{Forte:2010ta}.

In Ref.~\cite{Forte:2010ta} 
three FONLL scheme implementations have been proposed:
scheme A, where one uses the NLO massless scheme calculation, matched
with the LO (i.e.~${\cal O}(\alpha_s)$) massive scheme calculation;
scheme B, where one uses the NLO massless scheme calculation, matched
with the NLO (i.e.~${\cal O}(\alpha_s^2)$) massive scheme calculation;
and scheme C, where one uses the NNLO massless scheme calculation, matched
with the NLO massive scheme calculation.

Among the three schemes, scheme B has a peculiarity in the way the matching
is performed. In fact, the massless scheme calculation of $F_{2c}$
at NLO expanded up to order $\alpha_s^2$ has the form
$\alpha_s+\alpha_s L +\alpha_s^2 L^2+ \alpha_s^2 L$, with
$L\equiv \ln Q^2/m_c^2$, i.e.~terms of order
$\alpha_s^2$ with no logarithms are missing. On the other hand, in the
massive coefficients the full $\alpha_s^2$ term is present.
In this case, what one subtracts from the massless result is not simply
the massless limit of the massive result, but only a part of it,
not including the constant (i.e.~without factors of $L$) term of
order $\alpha_s^2$. As a consequence, these order $\alpha_s^2$ terms 
are not subtracted from the massive coefficient functions and persist
as strictly higher-order contributions at high $Q^2$. 

It is easily seen that the scheme A in the FONLL calculation should be
equivalent to the S-ACOT scheme. If a $\chi$-scaling prescription is
applied to all terms computed in the massless approximation, scheme A
should become equivalent to the S-ACOT-$\chi$ prescription.
Scheme B does not correspond to any S-ACOT calculation. It is more
reminiscent of the TR method, where at the NLO level
the full NLO massive result is also used. However, as we will shown
below, this is only true at $Q^2=m_c^2$, since in the TR method
the higher order term in the massive calculation is frozen
at threshold.
We conjecture that scheme C should again be equivalent to 
a NNLO generalization of the S-ACOT scheme.

Finally, let us mention that the 
FONLL GM-VFN scheme is currently being implemented
in the NNPDF family of 
fits~\cite{DelDebbio:2007ee,Ball:2008by,Rojo:2008ke,Ball:2009mk},
which up to now have been obtained in the zero--mass scheme for heavy quarks.
Both schemes (A and B) will be implemented in the NLO NNPDF fits,
and the theoretical uncertainties arising from the inherent
ambiguities in the matching procedure will be thoroughly studied.

\subsection{Results and discussion}

We turn now to discuss the results of the quantitative 
comparison between the GM-VFN approaches described above, using the
benchmark settings introduced in Sect.~\ref{LH_HQ_sec:intro}
First of all, we will discuss the comparison between FONLL and
S-ACOT: we will show that FONLL-A is identical to S-ACOT, with
and without threshold prescriptions. Having settled this
point, we will turn to studying the similarities and differences between
the FONLL and TR$^\prime$ schemes, this time both at NLO and at NNLO.
This last set of comparisons are also equivalent to comparing
S-ACOT with TR$^\prime$.

First of all, however, let us mention that there exist at least
two different implementations of the $\chi$-scaling threshold
prescription used in the literature.
The first form of $\chi-$scaling is given by
\be
\label{LH_HQ_eq:chi-sc-1}
  F^{({ \chi})}_{2c} (x, Q^2) \equiv 
x\int_{\chi(x, Q^2)}^1\frac{dy}{y} C\left(\frac{\chi(x,
  Q^2)}{y},\alpha_s(Q^2)\right) f(y,Q^2) \ ,
\ee
which is adopted by default in FONLL and also in the 
CTEQ6.5/6.6~\cite{Tung:2006tb,Nadolsky:2008zw}
PDF fits. On the other hand,
one can use an alternative form of $\chi-$scaling,
\be
\label{LH_HQ_eq:chi-sc-2}
  F^{({ \chi,{\rm v2}})}_{2c} (x, Q^2) \equiv
\chi\int_{\chi(x, Q^2)}^1\frac{dy}{y} C\left(\frac{\chi(x,
  Q^2)}{y},\alpha_s(Q^2)\right) f(y,Q^2) \ , 
\ee
which is used, for example, in the TR$^\prime$ definition, and is equivalent
to the unambiguous result in charged-current charm production from
strange quarks where $F_{2c}(x,Q^2) = \xi s(\xi, Q^2)$ at leading order
with $\xi = x(1+m_c^2/Q^2)$.  For neutral-current scattering, in both cases,
Eq.~\ref{LH_HQ_eq:chi-sc-1} and Eq.~\ref{LH_HQ_eq:chi-sc-2}, the scaling variable
is given by
\be
\label{LH_HQ_eq:chidef}
\quad \chi(x,Q^2) = x \left( 1 + \frac{4 m^2_c}{Q^2} \right).
\ee
It is clear that the two forms of the prescription, Eqs.~\ref{LH_HQ_eq:chi-sc-1} 
and~\ref{LH_HQ_eq:chi-sc-2}, differ only
by a mass suppressed term $\lp 1+4m_c^2/Q^2\rp$, and therefore are formally
equivalent, although can be numerically quite different
depending on the matching scheme adopted. These differences
represent an inherent ambiguity of the matching procedure. 
Let us finally note that
even the choice of scaling variable 
Eq.~\ref{LH_HQ_eq:chidef} is arbitrary: indeed, in 
Ref.~\cite{Nadolsky:2009ge} a one-parameter family of such scaling
variables was explored.

\subsubsection{Comparison of FONLL and S-ACOT}

\begin{figure}[ht]
\begin{center}
\epsfig{file=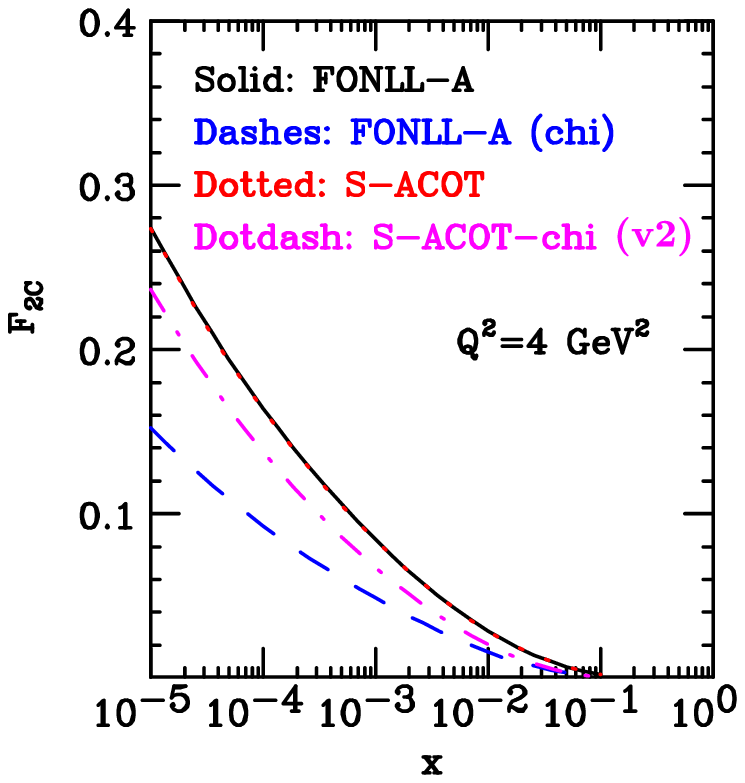,width=0.40\textwidth,angle=0}
\epsfig{file=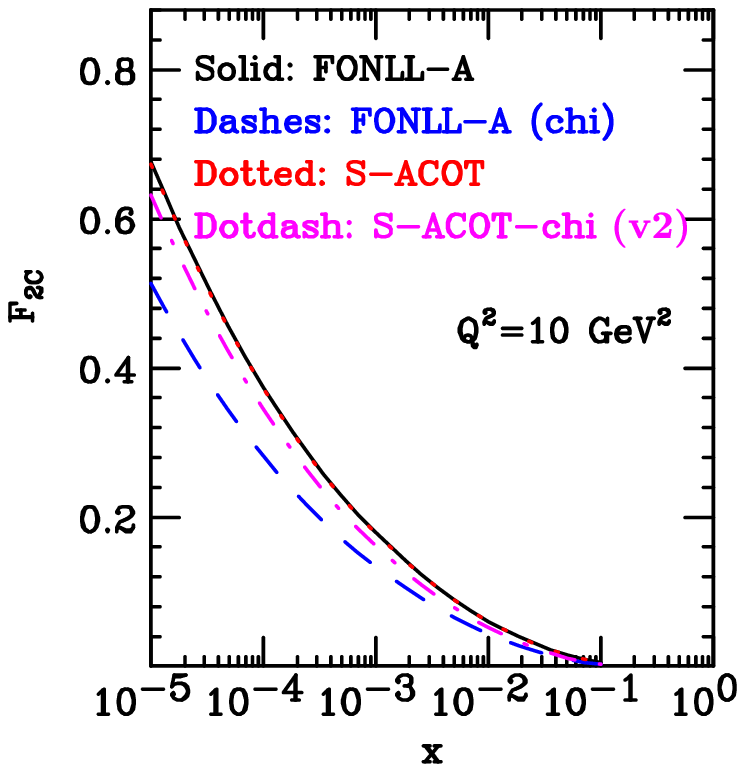,width=0.40\textwidth,angle=0}\\
\epsfig{file=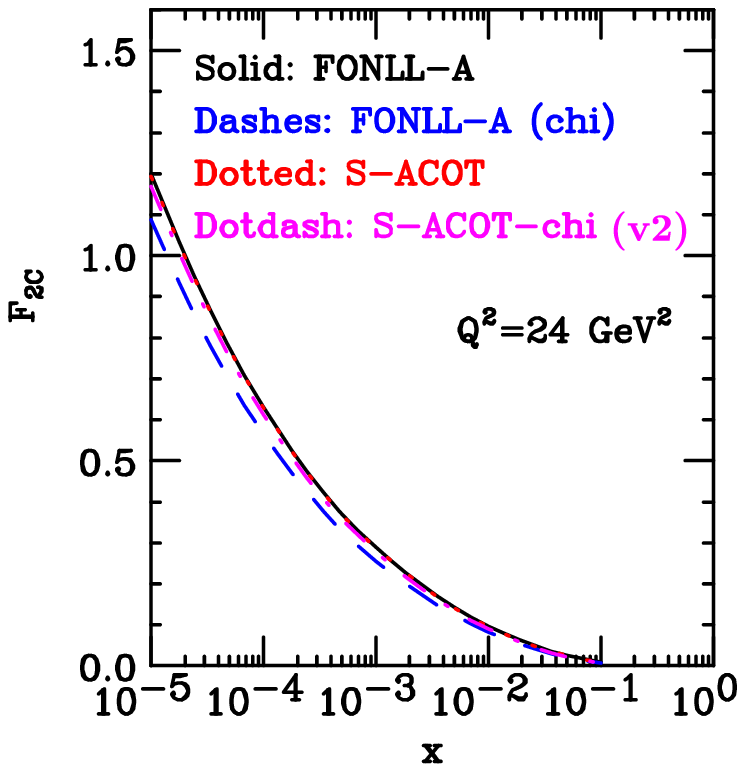,width=0.40\textwidth,angle=0}
\epsfig{file=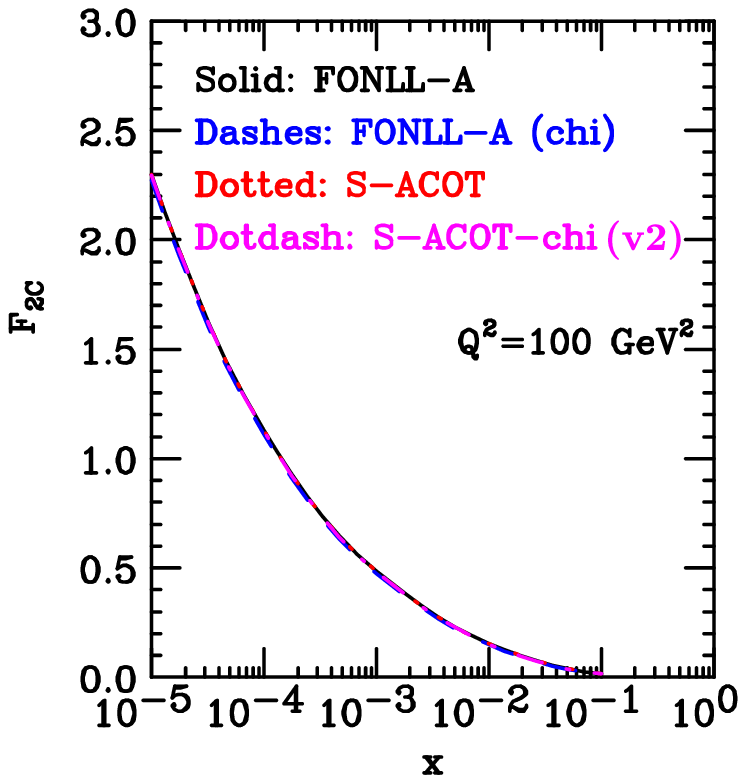,width=0.40\textwidth,angle=0}
\caption{\small 
\label{LH_HQ_fig:F2c-comp-acot} The $F_{2c}$ structure function for $Q^2=4$,
10, 24 and 100 GeV$^2$ 
in the FONLL scheme A (plain and with $\chi-$scaling) 
compared to the Simplified ACOT (S-ACOT)
and S-ACOT-$\chi$ schemes. Note that the FONLL-A-$\chi$
scheme implements the threshold prescription 
as in Eq.~\ref{LH_HQ_eq:chi-sc-1}, and 
 the S-ACOT-$\chi$ (v2) curve is computed with the
definition of Eq.~\ref{LH_HQ_eq:chi-sc-2}.}
\end{center}
\end{figure}

\begin{figure}[ht]
\begin{center}
\epsfig{file=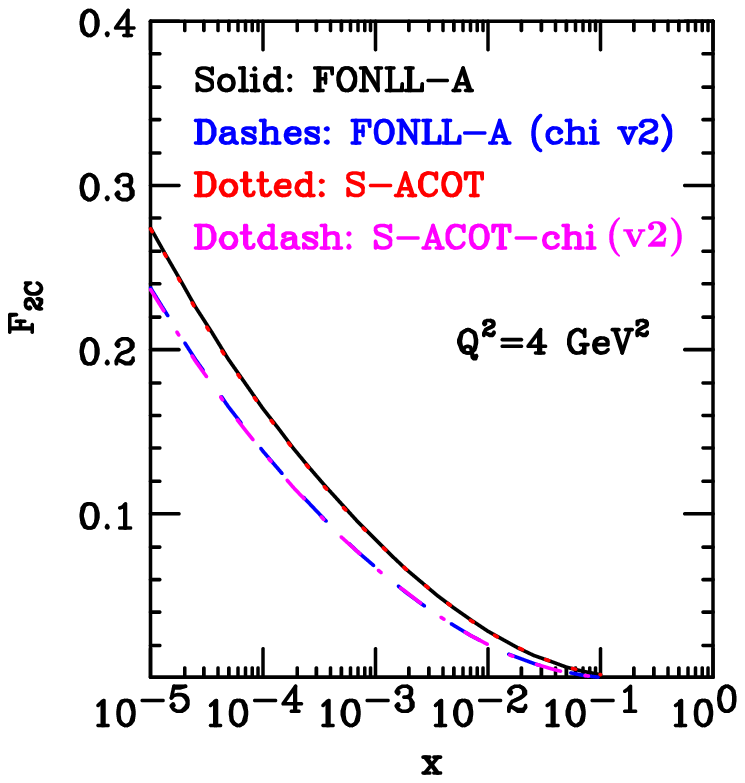,width=0.40\textwidth,angle=0}
\epsfig{file=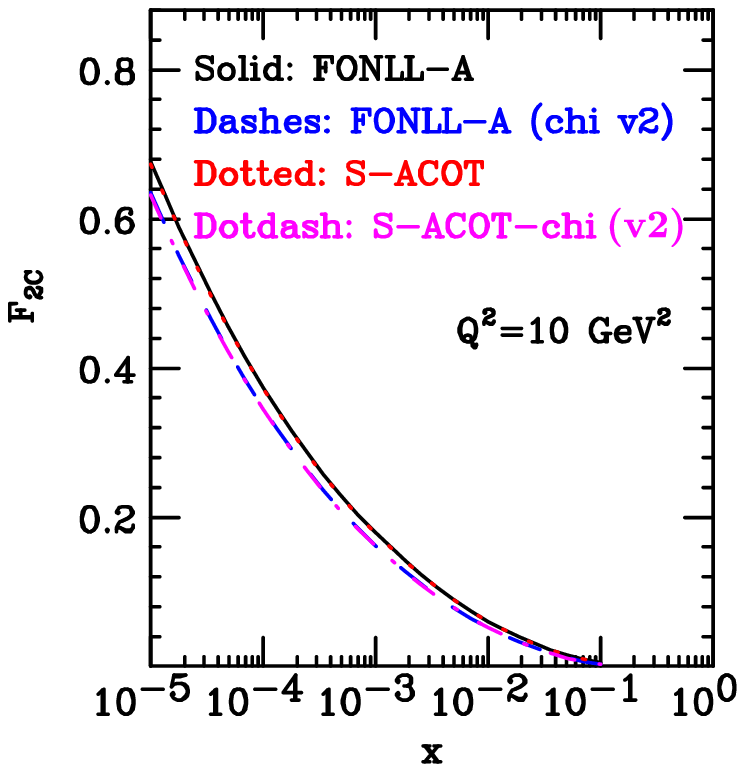,width=0.40\textwidth,angle=0}\\
\epsfig{file=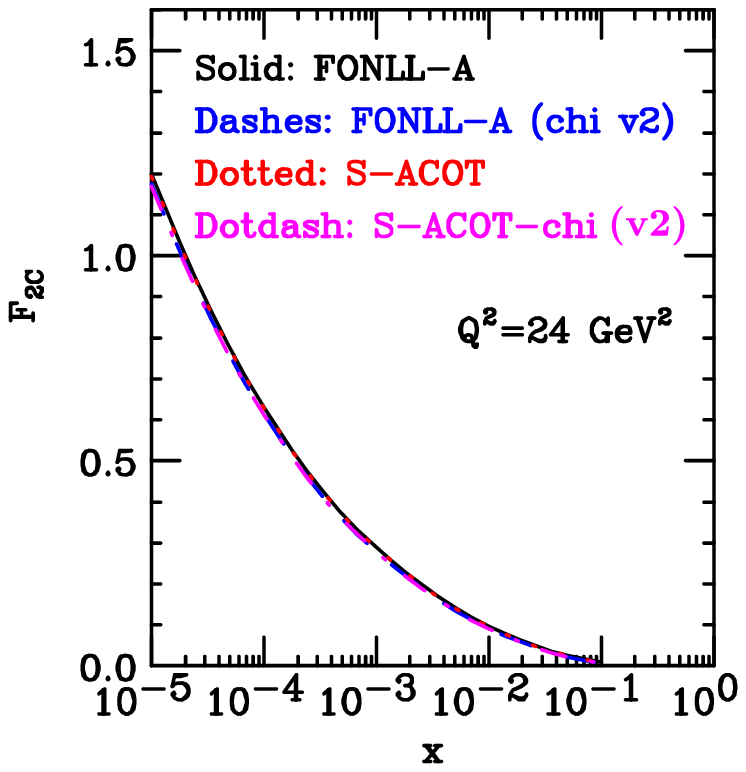,width=0.40\textwidth,angle=0}
\epsfig{file=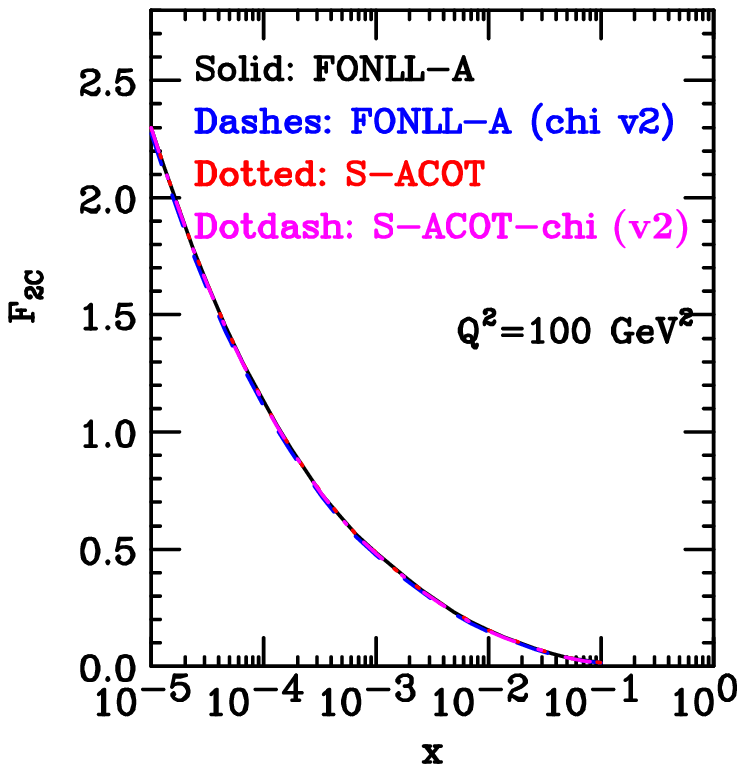,width=0.40\textwidth,angle=0}
\caption{\small 
\label{LH_HQ_fig:F2c-comp-acot2} Same as
Fig.~\ref{LH_HQ_fig:F2c-comp-acot}, but now the FONLL-A-$\chi$
scheme implements the threshold prescription 
as in Eq.~\ref{LH_HQ_eq:chi-sc-2}.}
\end{center}
\end{figure}

Let us begin with the comparison between
FONLL and S-ACOT for $F_{2c}(x,Q^2)$. Since the ACOT scheme has only
been implemented at NLO, we restrict the comparison
to the FONLL-A scheme. The ACOT scheme is extensible
to higher orders, but the NNLO is only in progress.
First of all, the Simplified ACOT (S-ACOT) scheme,
introduced in Sect.~\ref{LH_HQ_sec:acot}, is compared to FONLL scheme A
(see Sect.~\ref{LH_HQ_sec:fonll})
in Figs.~\ref{LH_HQ_fig:F2c-comp-acot} and~\ref{LH_HQ_fig:F2c-comp-acot2}
 at the benchmark kinematical points.
The only difference 
between Figs.~\ref{LH_HQ_fig:F2c-comp-acot} and~\ref{LH_HQ_fig:F2c-comp-acot2}
is the choice of $\chi-$scaling threshold prescription adopted in FONLL
scheme-A: while in Fig.~\ref{LH_HQ_fig:F2c-comp-acot} the default
FONLL choice Eq.~\ref{LH_HQ_eq:chi-sc-1} is adopted, in
Fig.~\ref{LH_HQ_fig:F2c-comp-acot2} the 
alternative form Eq.~\ref{LH_HQ_eq:chi-sc-2} is used instead. 
In both figures, the S-ACOT-$\chi$ (v2) curve is computed with the
definition of Eq.~\ref{LH_HQ_eq:chi-sc-2}.

From this comparison it is clear that, without any
threshold prescriptions, $F_{2c}$ in S-ACOT is identical to FONLL-A,
and moreover that S-ACOT-$\chi$ is identical to FONLL-A-$\chi$, once
$\chi$-scaling is understood as in Eq.~\ref{LH_HQ_eq:chi-sc-2}. 
The comparison between the FONLL-A-$\chi$ curves in 
Figs.~\ref{LH_HQ_fig:F2c-comp-acot} and~\ref{LH_HQ_fig:F2c-comp-acot2} shows
that in this scheme (as in S-ACOT) the impact of the choice
of arbitrary threshold prescription at low and 
moderate $Q^2$ can be as large as the resummation itself.
Again, let us emphasize that when the same
threshold prescriptions are applied  the S-ACOT and
FONLL-A schemes give always the same results 
(within minor numerical differences like integration errors).

\begin{figure}[ht]
\begin{center}
\epsfig{file=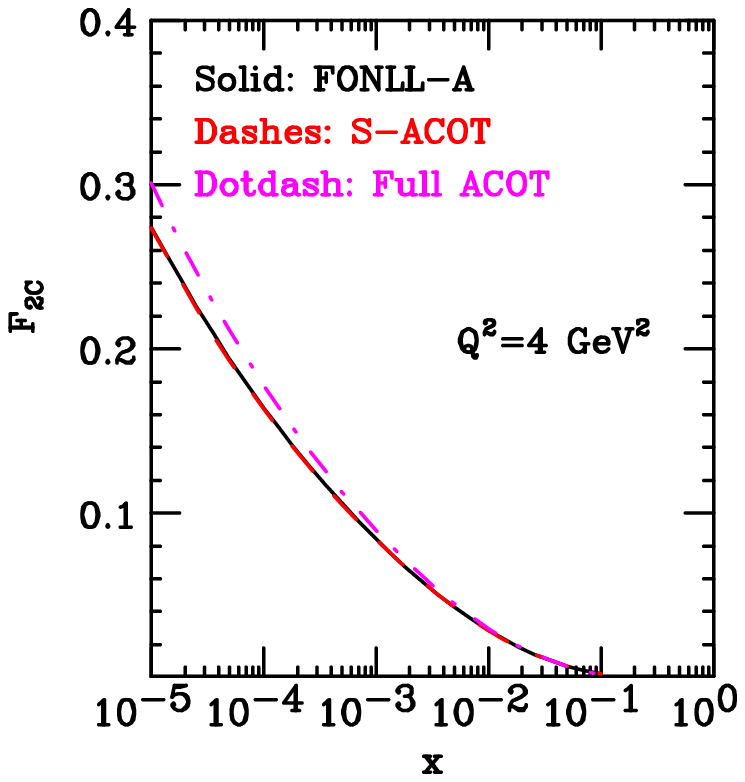,width=0.40\textwidth,angle=90}
\epsfig{file=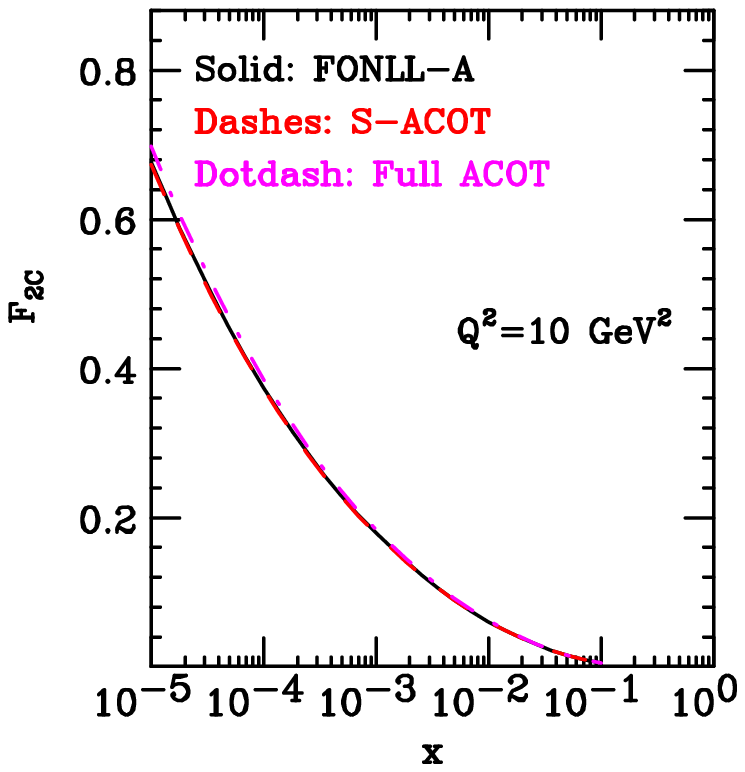,width=0.40\textwidth,angle=90}\\
\epsfig{file=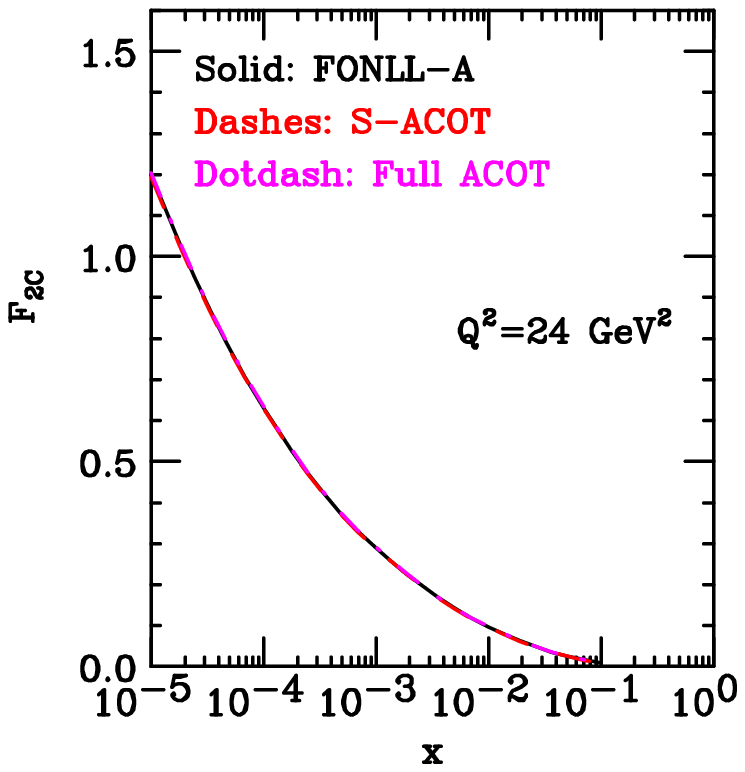,width=0.40\textwidth,angle=90}
\epsfig{file=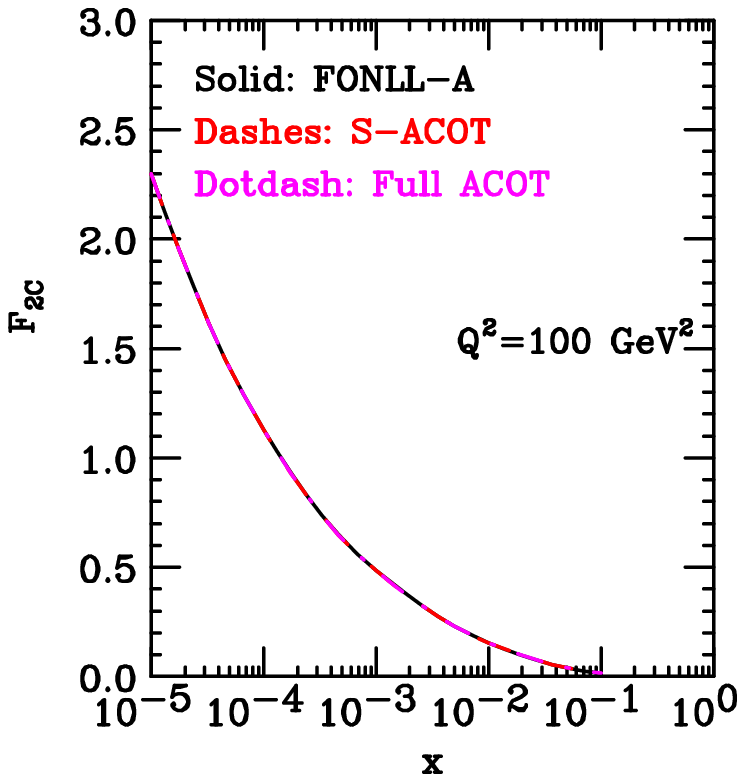,width=0.40\textwidth,angle=90}
\caption{\small 
\label{LH_HQ_fig:F2c-comp-fullacot} The $F_{2c}$ structure function for $Q^2=4$,
10, 24 and 100 GeV$^2$ 
in the FONLL scheme A (plain) 
compared to the Simplified ACOT (S-ACOT)
and full ACOT schemes.}
\end{center}
\end{figure}

It is also interesting to compare the S-ACOT scheme with the full ACOT
scheme~\cite{Aivazis:1993kh,Aivazis:1993pi}. 
In Fig.~\ref{LH_HQ_fig:F2c-comp-fullacot}
we show the results of this comparison. Since ACOT and
S-ACOT differ only by mass-suppressed terms, their
difference turns out to be as
expected very small, and essentially vanishes
already for $Q^2\sim 10$ GeV$^2$. Therefore, for 
$Q^2\gsim 10$ GeV$^2$ the full ACOT calculation
is identical to the FONLL-A scheme.

We would like to note that  the full ACOT scheme
was used only in the CTEQ HQ series
 (HQ4~\cite{Lai:1997vu}, HQ5~\cite{Lai:1999wy} and HQ6~\cite{Kretzer:2003it}),
while instead the zero-mass  approximation was adopted in the
general purpose CTEQ4M~\cite{Lai:1996mg}, CTEQ5M~\cite{Lai:1999wy} and 
CTEQ6M~\cite{Pumplin:2002vw} PDF sets. When CTEQ
adopted a GM-VFN as default in their fits (starting from
CTEQ6.5~\cite{Tung:2006tb}),  the GM-VFN scheme adopted
was instead S-ACOT-$\chi$. As it is clear from the
comparison between Figs.~\ref{LH_HQ_fig:F2c-comp-acot} and~\ref{LH_HQ_fig:F2c-comp-acot2},
the main difference is the choice of threshold prescription,
formally subleading but
which can be numerically as large as the whole effect of the resummation
itself, and which could have important phenomenological
implications.

\begin{figure}[ht]
\begin{center}
\epsfig{file=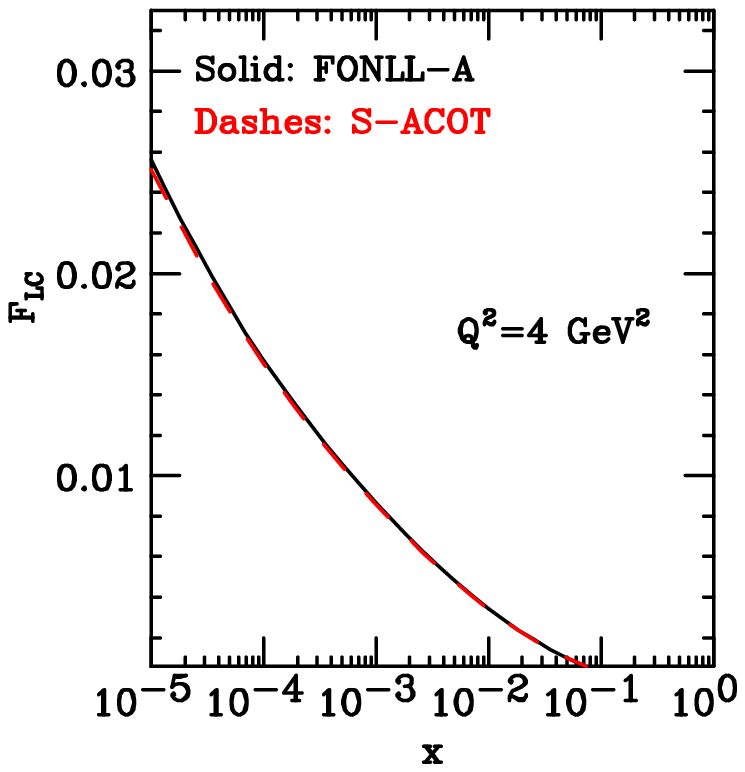,width=0.40\textwidth,angle=90}
\epsfig{file=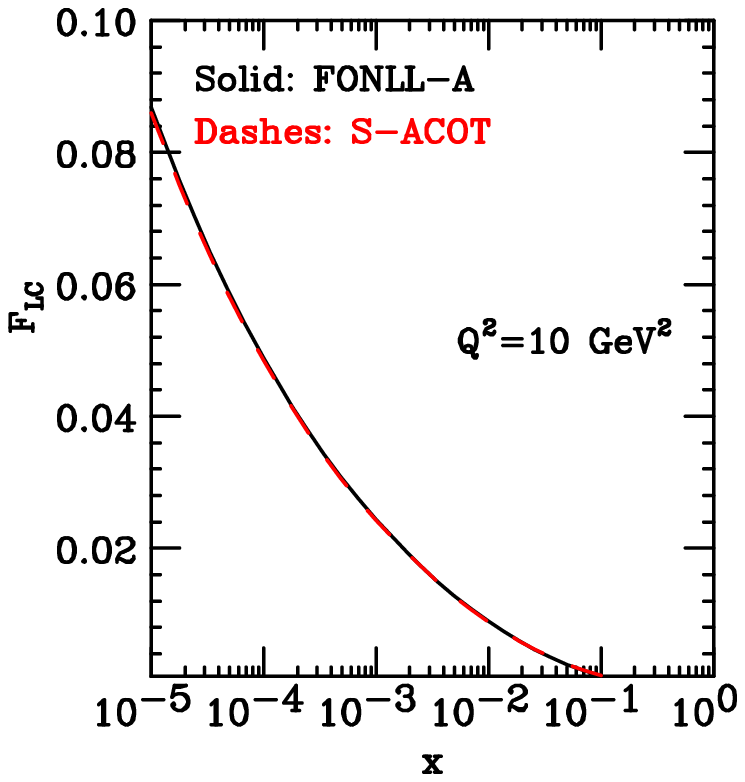,width=0.40\textwidth,angle=90}\\
\epsfig{file=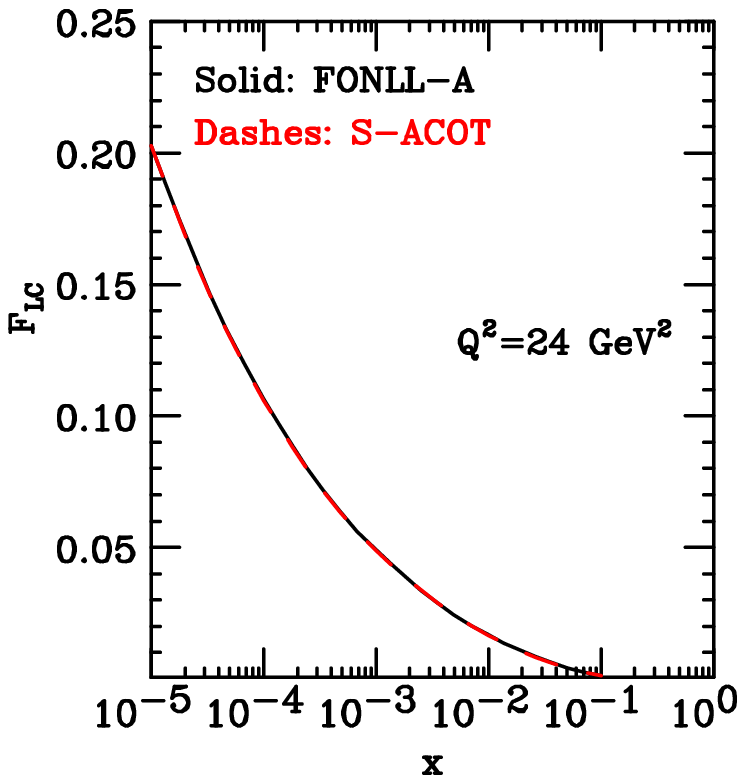,width=0.40\textwidth,angle=90}
\epsfig{file=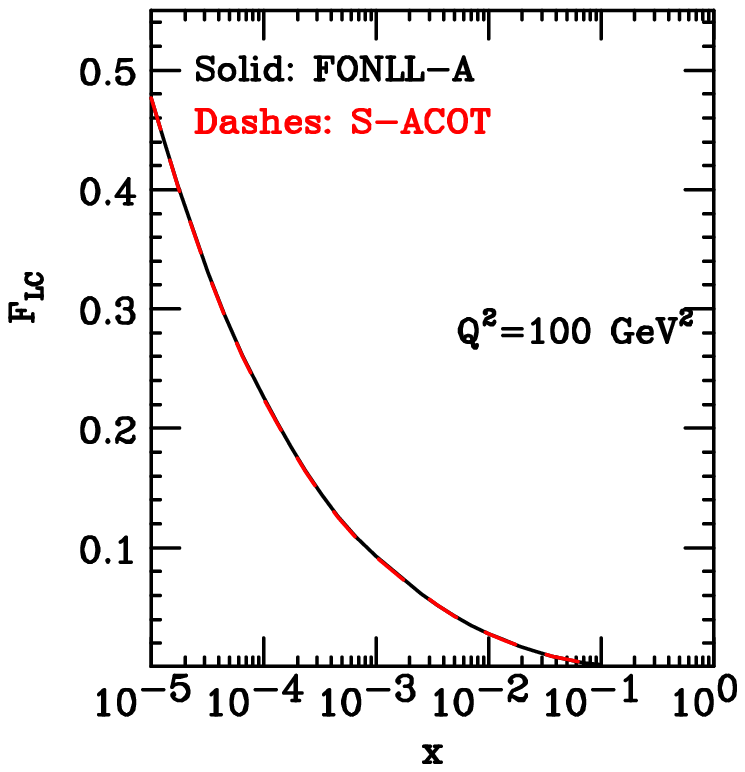,width=0.40\textwidth,angle=90}
\caption{\small 
\label{LH_HQ_fig:FLc-comp-acot} The $F_{Lc}$ structure function for $Q^2=4$,
10, 24 and 100 GeV$^2$ 
in the FONLL scheme A (plain) and Simplified ACOT (S-ACOT)
schemes.}
\end{center}
\end{figure}

Finally, let us compare the results for the
charm component of the longitudinal structure
function $F_{Lc}(x,Q^2)$. The results of these
comparisons are shown in Fig.~\ref{LH_HQ_fig:FLc-comp-acot} for
the case of FONLL-A and S-ACOT without threshold prescriptions.
We can see that both schemes  coincide also in this case,
as was the case for $F_{2c}$.

In summary, we have shown that when threshold prescriptions are
switched off, FONLL-A and S-ACOT are completely identical, both
for $F_{2c}$ and $F_{Lc}$. This is also the case when $\chi-$scaling
is adopted as a threshold prescription, but only when the same
of the two possible implementations
 Eqs.~\ref{LH_HQ_eq:chi-sc-1}-\ref{LH_HQ_eq:chi-sc-2} is consistently used 
in both cases. Finally, we have shown that the full ACOT result
is numerically very close to S-ACOT (and thus FONLL-A), being
numerically equivalent for $Q^2\ge 10$ GeV$^2$.


\subsubsection{Comparison of FONLL and TR$^\prime$}

Let us now discuss the results of the comparison between
FONLL and the TR$^\prime$ scheme
which has been used in the MSTW 2008 NLO and NNLO parton 
fits~\cite{Martin:2009iq}, as is introduced in 
Sect.~\ref{LH_HQ_sec:mstw}. As shown in the previous section,
all the results of this comparison apply both to
FONLL-A and S-ACOT, which are numerically identical.

\begin{figure}[ht]
\begin{center}
\epsfig{file=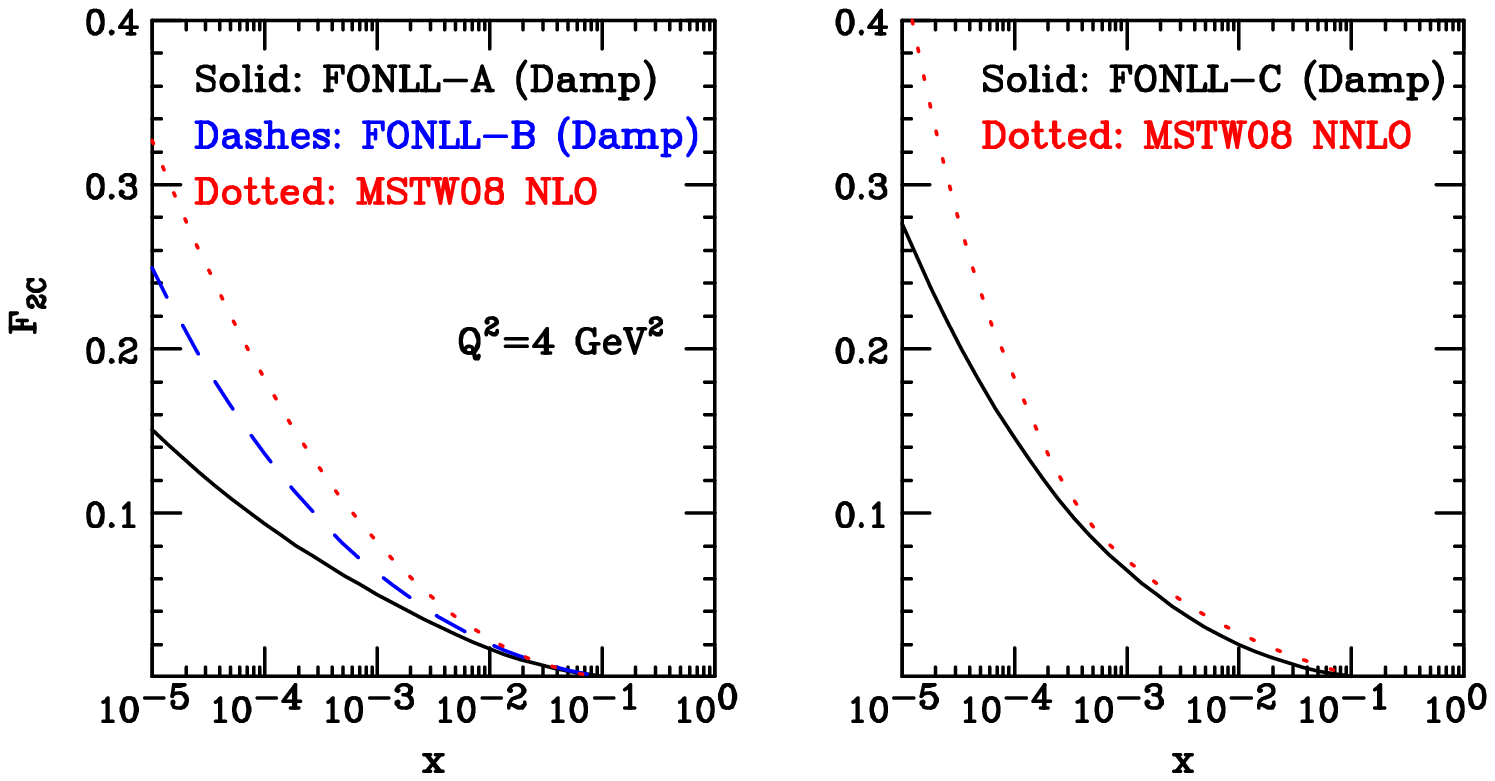,width=0.40\textwidth,angle=90}
\epsfig{file=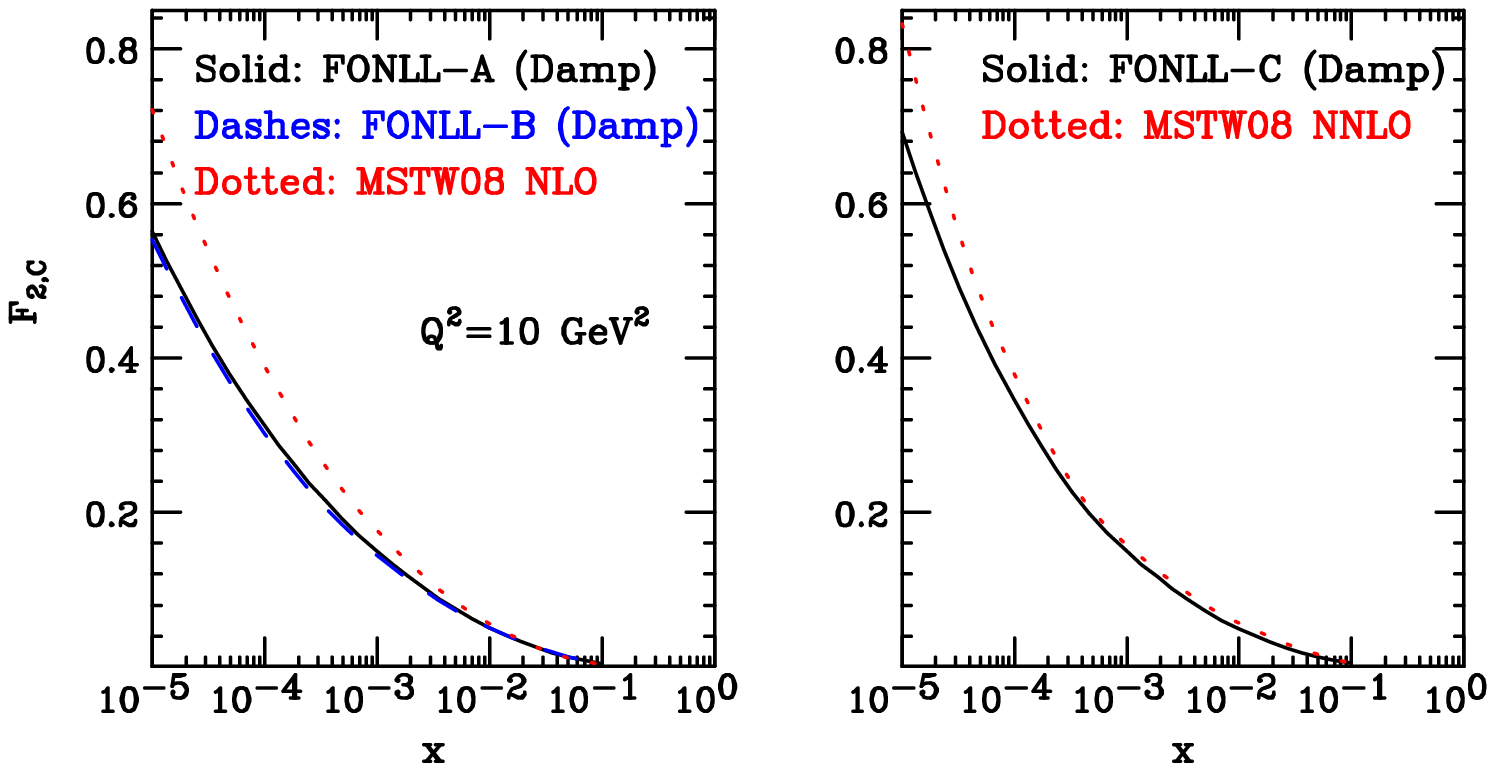,width=0.40\textwidth,angle=90}
\caption{\small 
\label{LH_HQ_fig:F2c-comp} The $F_{2c}$ structure function for $Q^2=4$ and
10 GeV$^2$ 
in FONLL and in TR$^\prime$, both for the NLO schemes (left plots) and
for the NNLO schemes (right plots). In both cases the default
threshold prescriptions are used: $\chi$-scaling using Eq.~\ref{LH_HQ_eq:chi-sc-2}
for TR$^\prime$ and a damping factor for FONLL.}
\end{center}
\end{figure}

\begin{figure}[ht]
\begin{center}
\epsfig{file=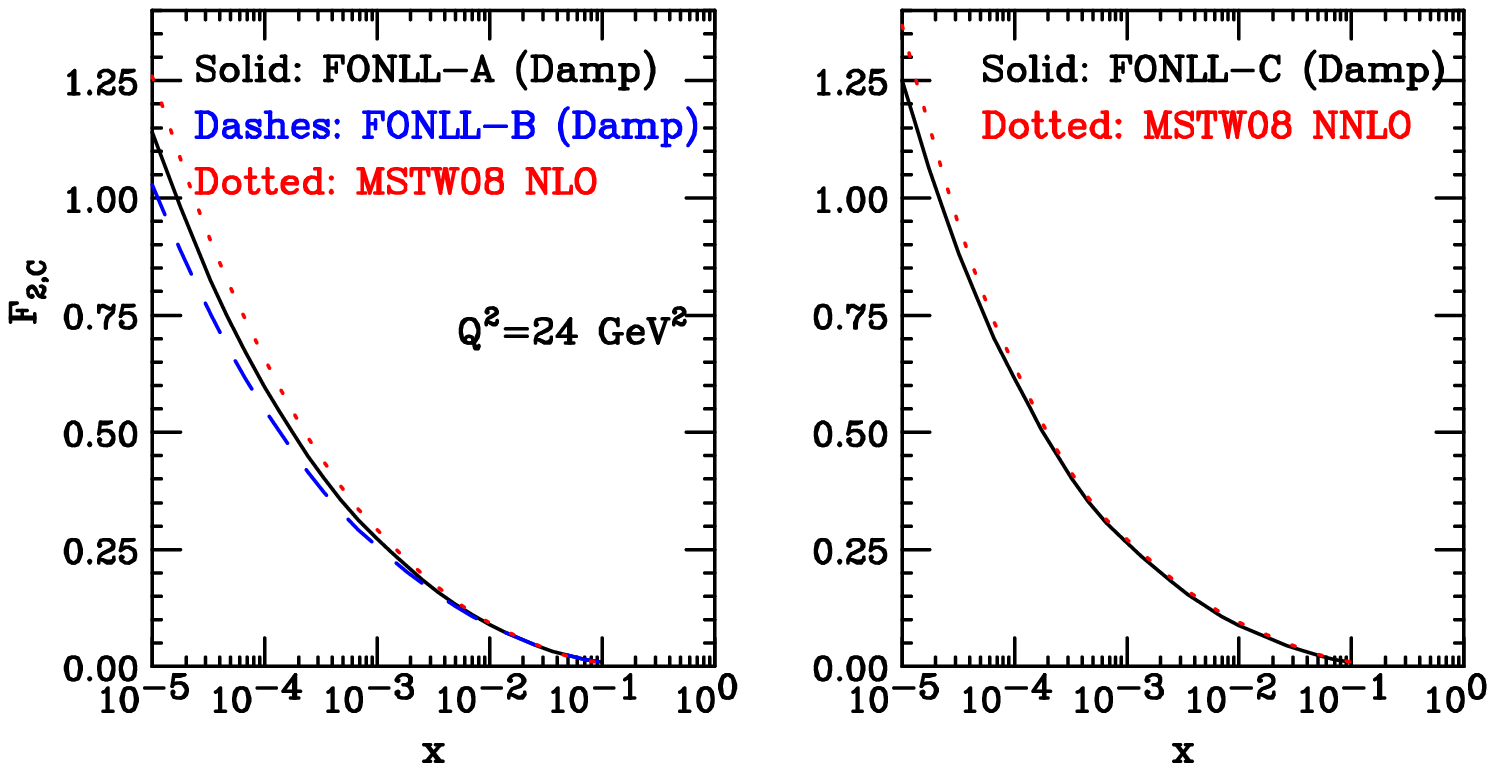,width=0.40\textwidth,angle=90}
\epsfig{file=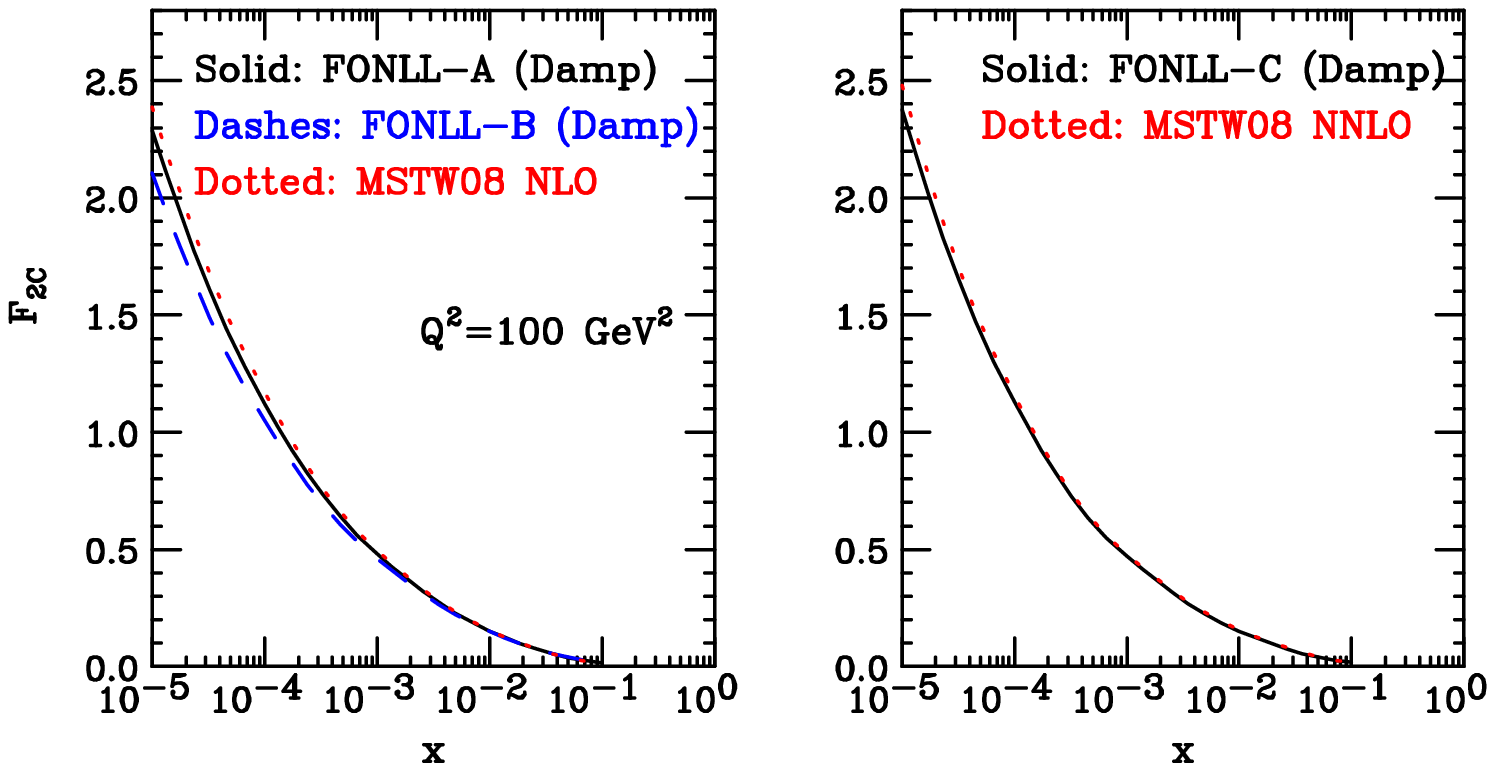,width=0.40\textwidth,angle=90}
\caption{\small 
\label{LH_HQ_fig:F2c-comp2} Same as Fig.~\ref{LH_HQ_fig:F2c-comp} for $Q^2=24$ and
100 GeV$^2$.}
\end{center}
\end{figure}

First 
of all, we show in Figs.~\ref{LH_HQ_fig:F2c-comp}-\ref{LH_HQ_fig:F2c-comp2} a comparison 
of the FONLL results with the results from the TR$^\prime$
NLO and NNLO GM-VFN schemes for the $F_{2c}$ structure function
at the benchmark kinematical points. Unlike the ACOT case, since
TR$^\prime$ has been formulated also up to NNLO, now we compare
both the FONLL NLO schemes (A and B) with TR$^\prime$ NLO and
separately the FONLL NNLO scheme (denoted by C) with TR$^\prime$
NNLO. From Figs.~\ref{LH_HQ_fig:F2c-comp}-\ref{LH_HQ_fig:F2c-comp2} no
obvious similarities can be identified between the two
families of schemes, neither at NLO nor at NNLO, apart from the
obvious remark that differences between schemes decrease when
$Q^2$ is increased.

\begin{figure}[ht]
\begin{center}
\epsfig{file=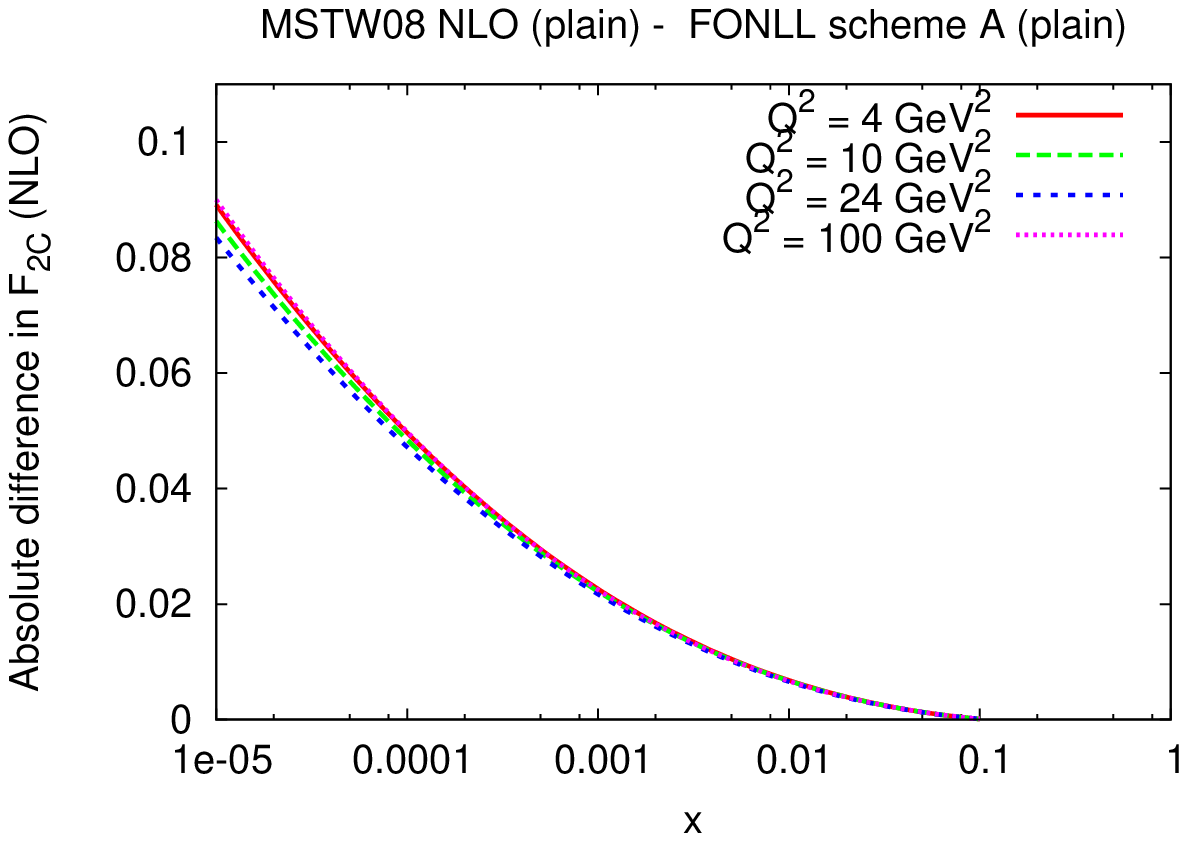,width=0.49\textwidth}
\epsfig{file=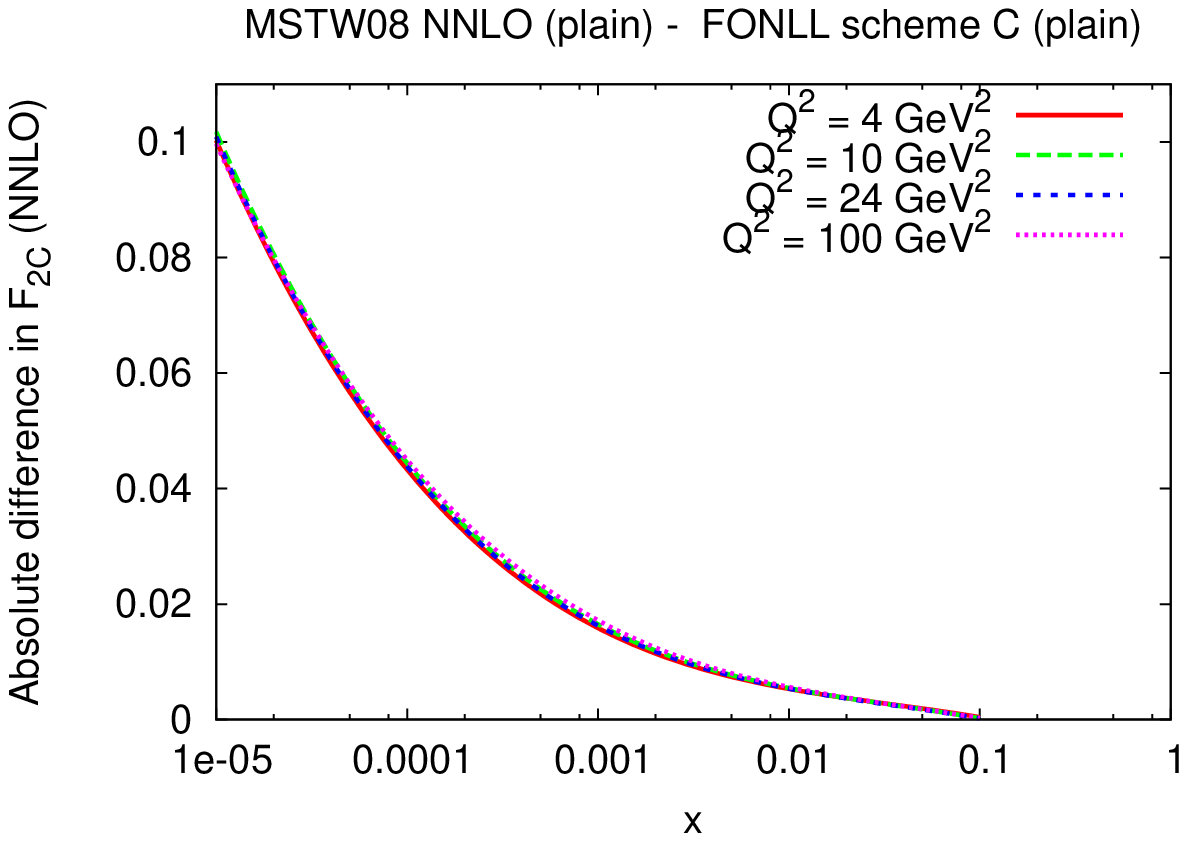,width=0.49\textwidth}
\caption{\small 
Left plot: Absolute difference between $F_{2c}$ in the TR$^\prime$
NLO and FONLL-A schemes, in both cases without any 
threshold damping factor. Right plot: same between
 the TR$^\prime$
NNLO and FONLL C-schemes. \label{LH_HQ_fig:fonll-mstw-diff} }
\end{center}
\end{figure}

\begin{figure}[ht]
\begin{center}
\epsfig{file=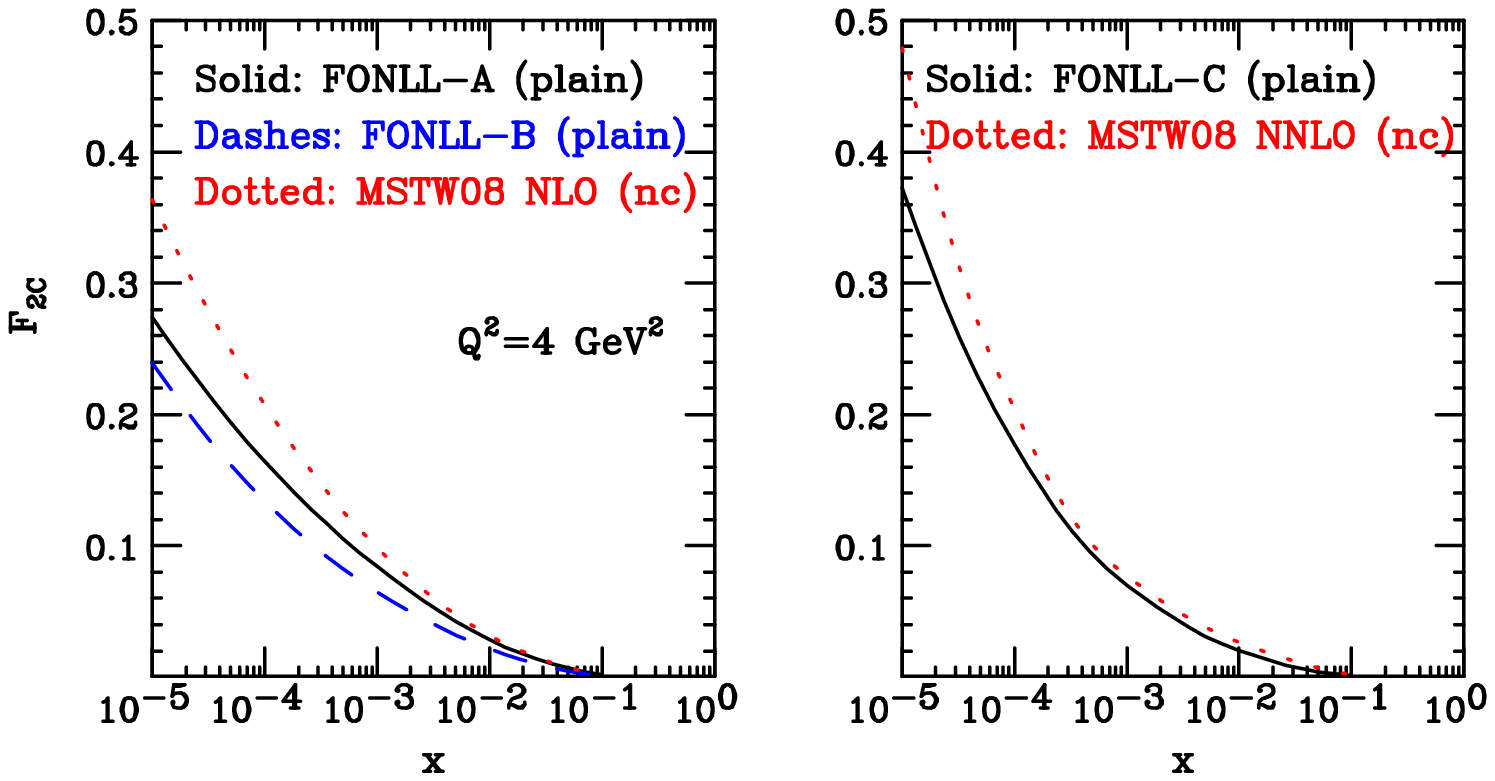,width=0.40\textwidth,angle=90}
\epsfig{file=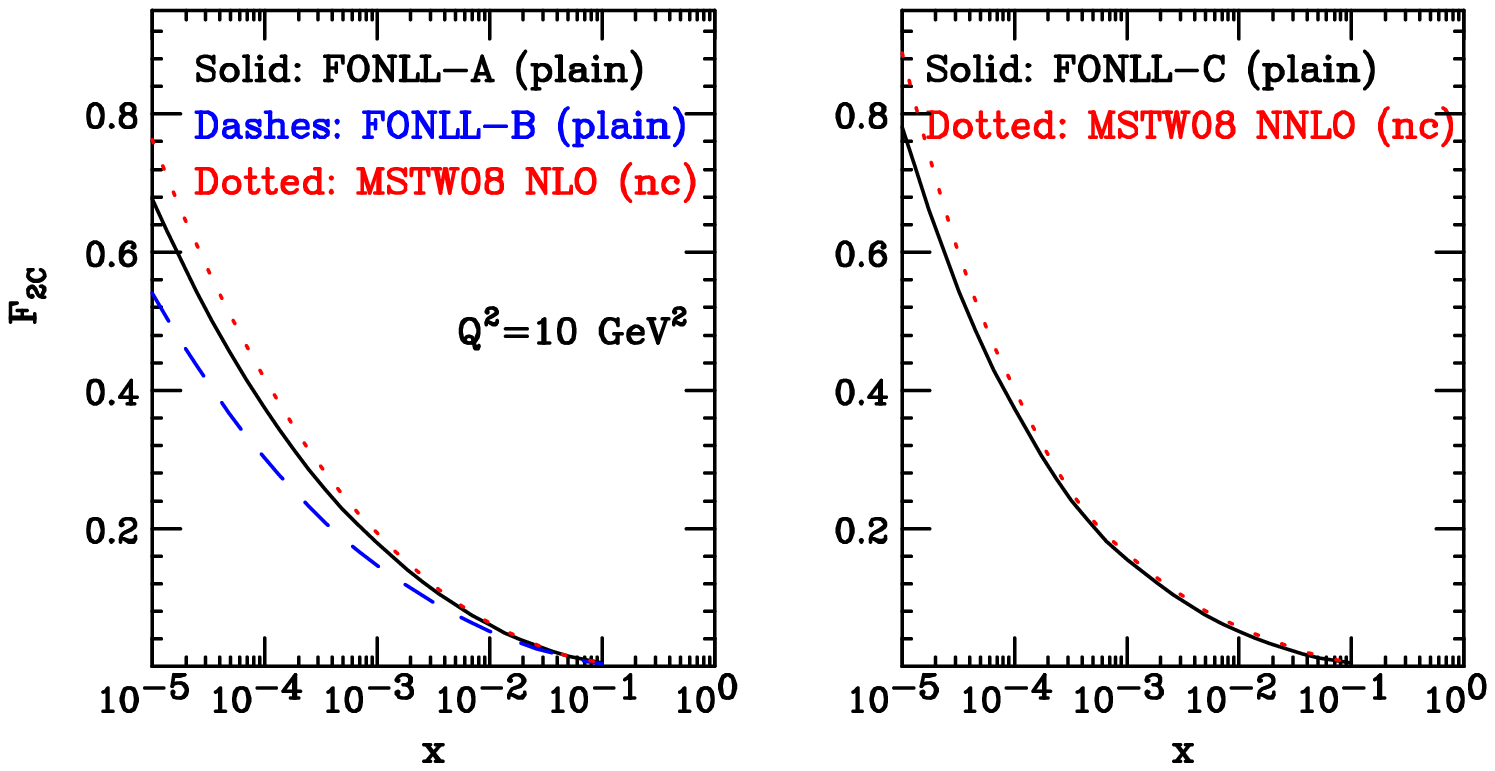,width=0.40\textwidth,angle=90}
\caption{\small 
Same as Fig.~\ref{LH_HQ_fig:F2c-comp} \label{LH_HQ_fig:F2c-comp-mstw}
 with all threshold prescriptions switched off
in the FONLL and TR$^\prime$ schemes.}
\end{center}
\end{figure}

\begin{figure}[ht]
\begin{center}
\epsfig{file=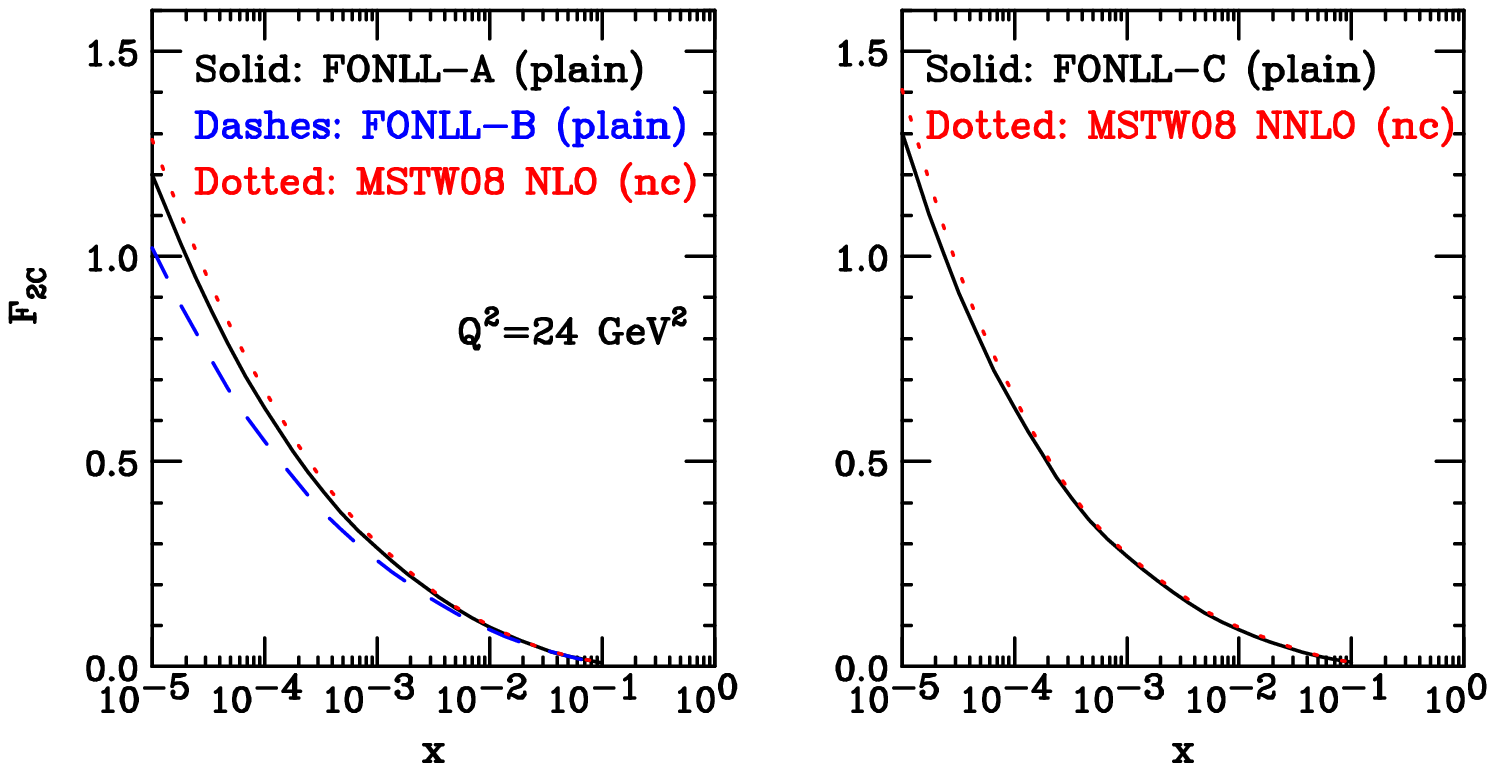,width=0.40\textwidth,angle=90}
\epsfig{file=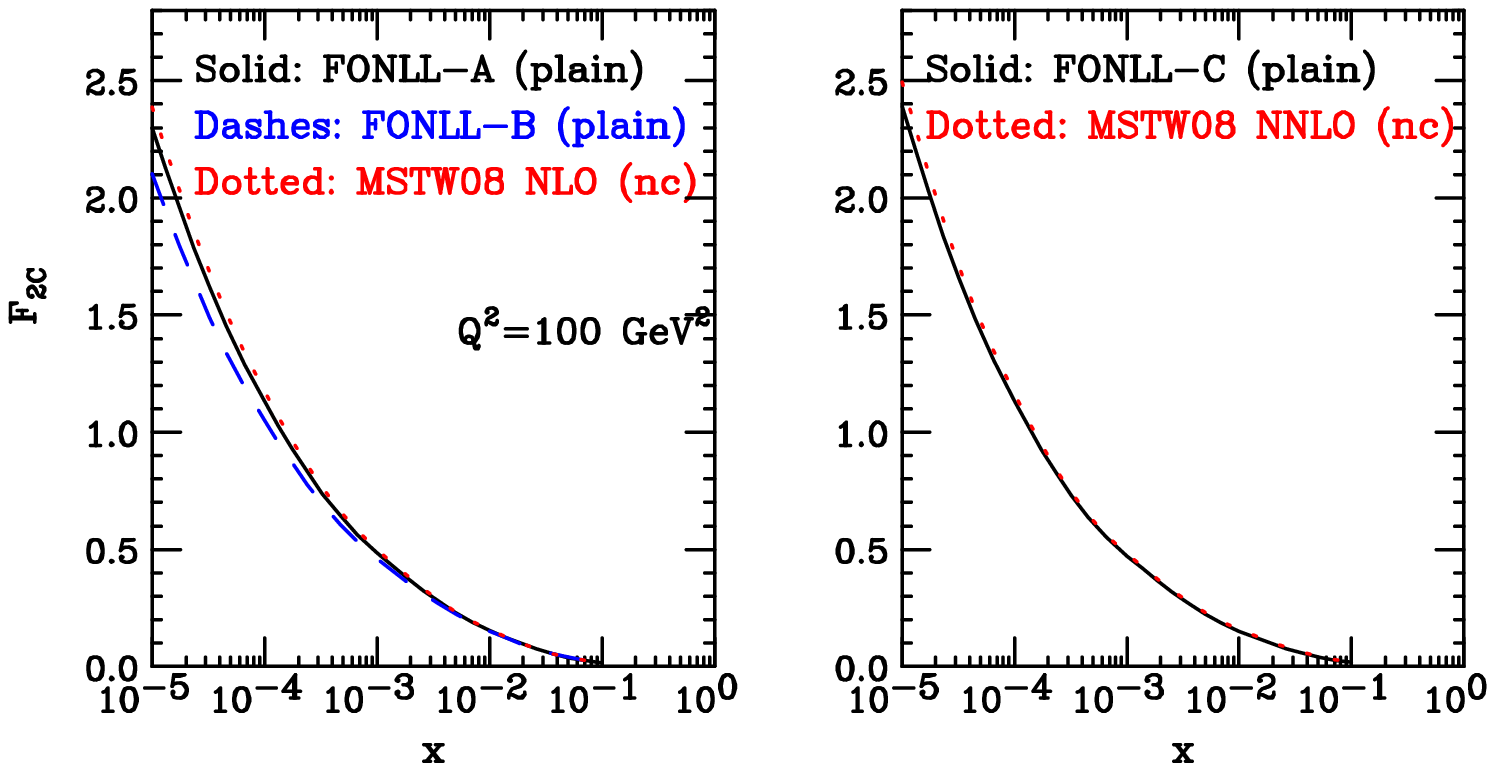,width=0.40\textwidth,angle=90}
\caption{\small 
Same as Fig.~\ref{LH_HQ_fig:F2c-comp2} \label{LH_HQ_fig:F2c-comp2-mstw}
 with all threshold prescriptions switched off
in the FONLL and TR$^\prime$ schemes.}
\end{center}
\end{figure}

In order
to render the comparison more meaningful, the threshold
prescriptions are switched off in both cases in 
Figs.~\ref{LH_HQ_fig:F2c-comp-mstw}-\ref{LH_HQ_fig:F2c-comp2-mstw}.
Having done this, it is clear that FONLL-A is rather close
to TR$^\prime$ NLO, while in turn FONLL-C is rather close
to TR$^\prime$ NNLO. Indeed, it can be shown that the two
schemes differ only by a constant ($Q^2$-independent) term
which is formally higher order and that is included in the
TR$^\prime$ schemes in order to ensure continuity of physical
observables at the
heavy quark threshold. This is verified explicitly in
Fig.~\ref{LH_HQ_fig:fonll-mstw-diff}.

Let us be more precise about this latter point.
The TR$^\prime$ scheme, discussed in Sect.~\ref{LH_HQ_sec:mstw}, 
defines the default prescription from Ref.~\cite{Thorne:2006qt}. 
By construction, at NLO
it should be similar for $F_{2c}$ to the S-ACOT-$\chi$ scheme and 
correspondingly also to
FONLL-A (when the same
form of $\chi$-scaling threshold prescription is used
consistently in all cases). 
The only difference is the additional (subleading) 
${\cal O}(\alpha_s^2)$ contribution which is a constant for 
$Q^2/m_c^2\to \infty$, and
which is numerically significant for low $Q^2$. 
This leads to a slightly larger $F_{2c}$, though the relative
difference disappears as $Q^2/m_c^2\to\infty$.

On the other hand, the NNLO definition in TR$^\prime$ should be the 
same for $F_{2c}$ as in FONLL-C up to the additional  
${\cal O}(\alpha_s^3)$ term from the massive
coefficient function, which stems
from the matching between the massive and
GM schemes at $Q^2=m_c^2$,  provided that as usual
the threshold prescription is the same is both cases.
These expectations are explicitly verified both at NLO and at NNLO in 
Figs.~\ref{LH_HQ_fig:F2c-comp-mstw}-\ref{LH_HQ_fig:F2c-comp2-mstw}.

Finally, the NLO TR$^{\prime}$ will
be the same as the FONLL-B scheme only at
$Q^2=m_c^2$  and will differ from it for any other value
$Q^2 > m_c^2$, since the ${\cal O}(\alpha_s^2)$ term
included in both cases is frozen at $Q^2=m_c^2$ in
TR$^{\prime}$ but runs as usual with $Q^2$ in
FONLL-B. As noted earlier, both TR$^\prime$ and FONLL-B schemes
at high $Q^2$ contain strictly higher order 
(beyond ${\cal O}(\alpha_s)$) terms which are of the
form $\alpha^2_s(m_c^2)g(m_c^2)$ in the former case and
$\alpha^2_s(Q^2)g(Q^2)$ in the latter case (with analogous singlet
quark terms). These turn out to be of opposite sign in the two cases
although they both originate from the order $\alpha_s^2$ massive
coefficient functions.
  
Note that this implies that NLO TR$^{\prime}$ is, as discussed
before, identical to S-ACOT and FONLL-A up to a constant
subleading term. In particular, the same non-negligible dependence
on the choice of threshold prescription which is
present in  S-ACOT and FONLL-A will be present in NLO TR$^{\prime}$.
This is opposite to what happens for FONLL-B, since as
discussed extensively in Ref.~\cite{Forte:2010ta}, in this case the
matched results turns out to be essentially independent
of the choice of arbitrary threshold prescription, and the
FONLL result coincides with the massive result for low
and moderate $Q^2$.

 A variety of modifications of
the default TR$^{\prime}$ scheme have been 
explored in Ref.~\cite{diffVFNS}, along with their consequences, 
and a new ``optimal choice'' suggested. 
In particular, the higher order FF part was 
frozen in the original TR definition in order 
to obtain the exact continuity 
of the evolution of $F_2$, but this is not required
in the TR$^\prime$ definition. Allowing this term  
instead to fall like a power of $Q^2$ results in exactly the same results as 
S-ACOT or FONLL-A (or FONLL-C at NNLO) in the limit that
$Q^2/m_c^2 \to \infty$, where
those terms causing the differences in the default scheme now vanish.

\begin{figure}[ht]
\begin{center}
\epsfig{file=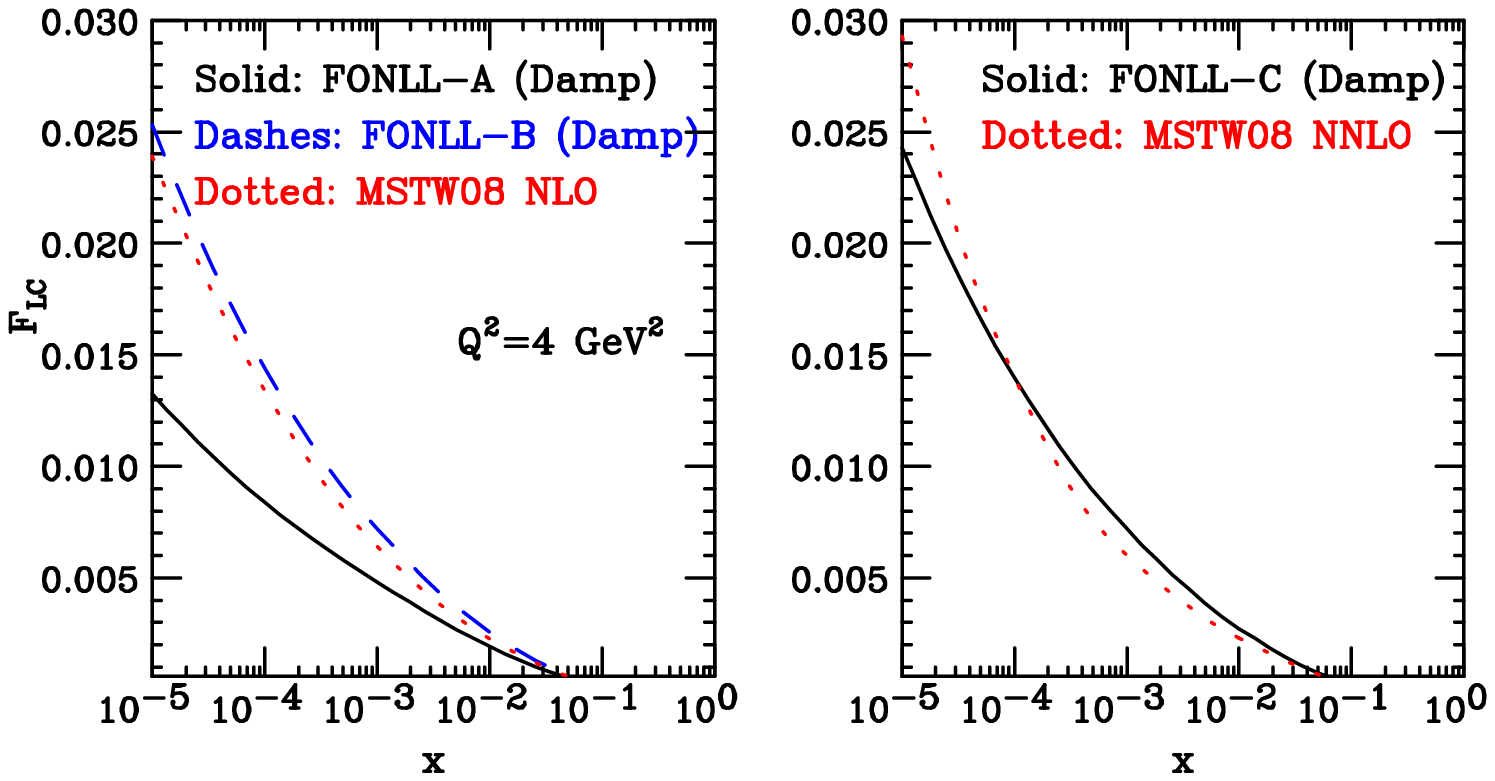,width=0.40\textwidth,angle=90}
\epsfig{file=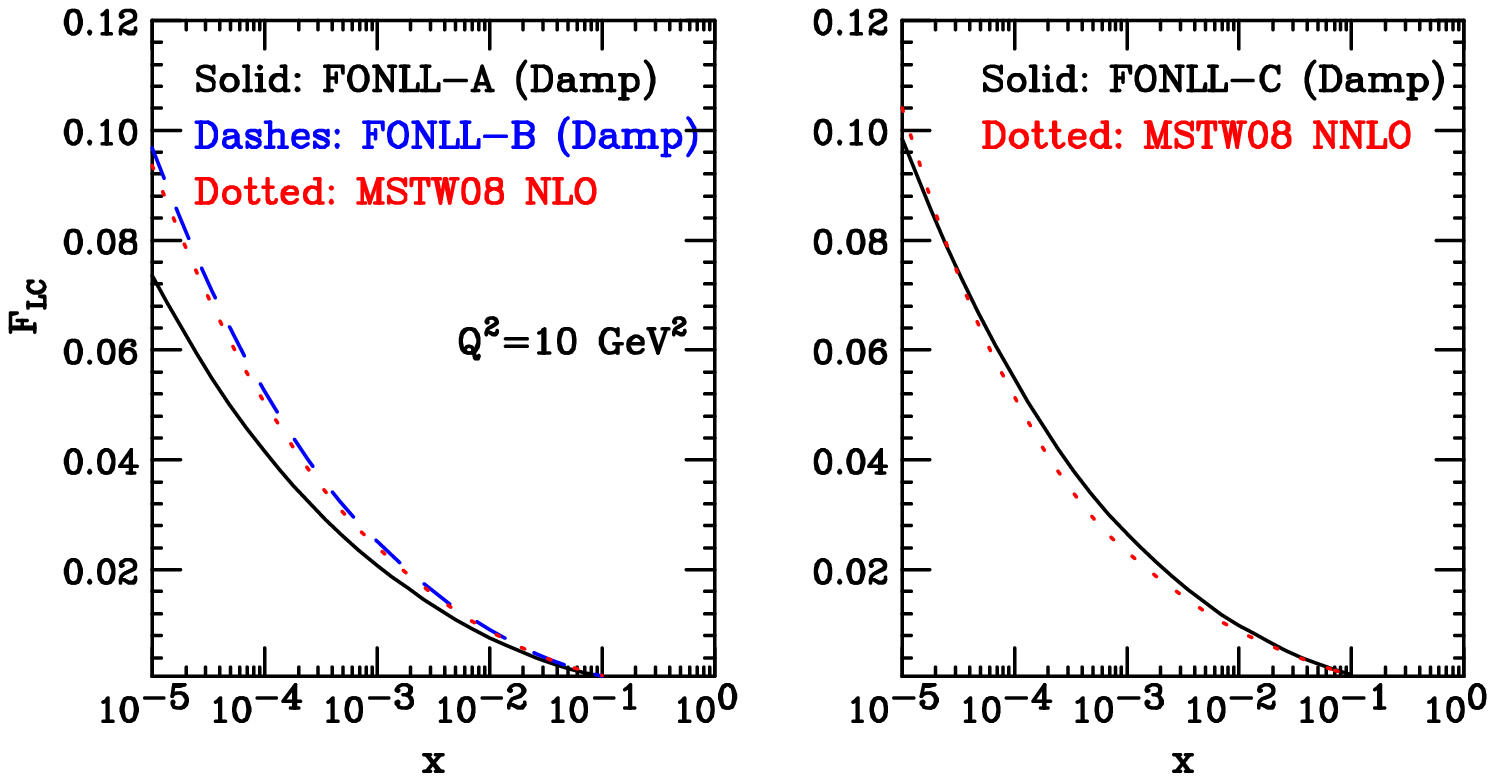,width=0.40\textwidth,angle=90}
\caption{\small 
\label{LH_HQ_fig:FLc-comp} The $F_{Lc}$ structure function for $Q^2=4$ and
10 GeV$^2$ 
from FONLL and TR$^\prime$, both for the NLO schemes (left plots) and
for the NNLO schemes (right plots).}
\end{center}
\end{figure}

\begin{figure}[ht]
\begin{center}
\epsfig{file=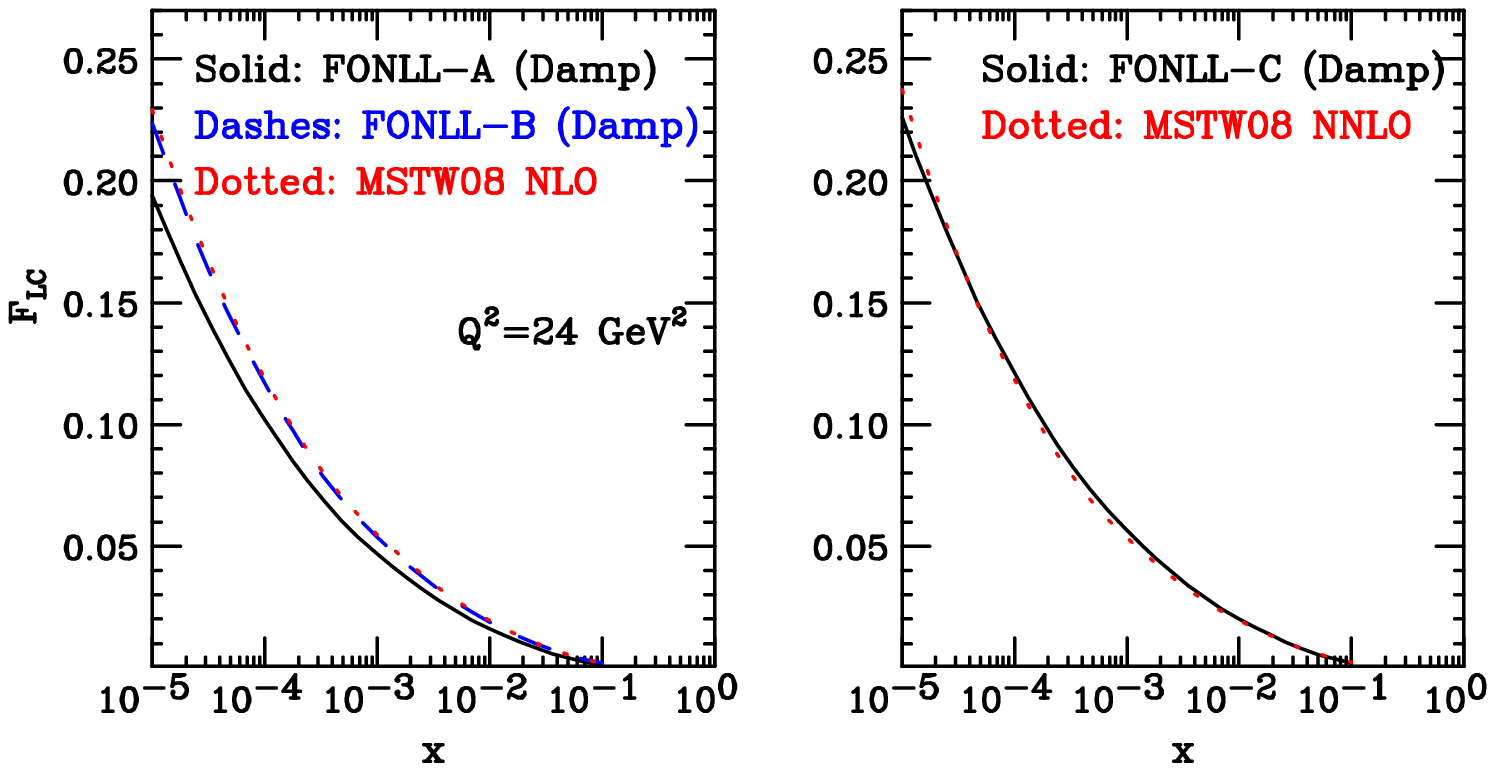,width=0.40\textwidth,angle=90}
\epsfig{file=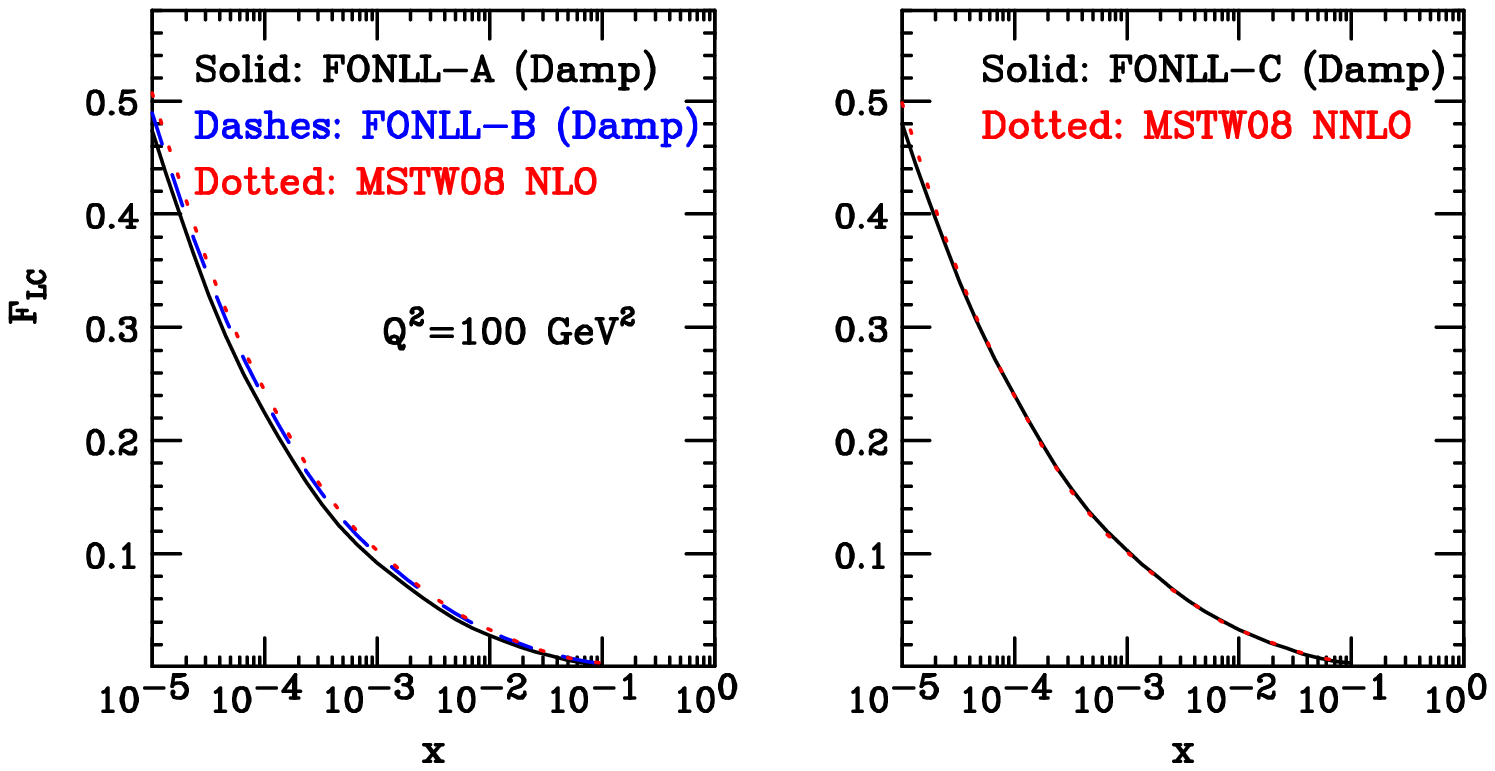,width=0.40\textwidth,angle=90}
\caption{\small 
\label{LH_HQ_fig:FLc-comp2} Same as Fig.~\ref{LH_HQ_fig:FLc-comp} for $Q^2=24$ and
100 GeV$^2$.}
\end{center}
\end{figure}

After this detailed discussion about the
 comparison for $F_{2c}$ between the FONLL and TR$^\prime$
schemes, we now turn to discuss the results of the comparison
for the longitudinal structure function 
  $F_{Lc}$. These results are shown
in Figs.~\ref{LH_HQ_fig:FLc-comp}-\ref{LH_HQ_fig:FLc-comp2}. In this case,
due to the perturbative ordering of the TR/TR$^\prime$ schemes, FONLL-B
turns out to be very similar to TR$^\prime$ NLO (which now
includes also the full running $\alpha_s^2$ massive term) for
any $Q^2$. In the case of the NNLO schemes, 
 FONLL-C
is somewhat different to TR$^\prime$ NNLO. This is likely due to the additional
$\alpha_s^3$ term included in the latter. Differences 
tend to wash out with $Q^2$, but an order $\alpha_s^3$ discrepancy, 
depending on 
the massless coefficient functions at this order, persists even at high $Q^2$.



\subsection{Conclusions}

In this contribution we have performed a detailed quantitative
comparison between GM-VFN heavy quark schemes in deep-inelastic 
scattering. We have compared the heavy quark schemes
adopted by the three main global PDF fitting groups.
The main results of this benchmark
comparison have been the following:

\begin{enumerate}
\item FONLL-A (plain) is identical to S-ACOT, and FONLL-A-$\chi$
is identical to  S-ACOT-$\chi$, 
both for $F_{2c}$ and $F_{Lc}$, when the same $\chi-$scaling
threshold prescription is used in both cases.
\item The only difference between FONLL-A (plain) (and S-ACOT)
and TR$^\prime$ NLO for $F_{2c}$ is a subleading $Q^2$-independent
matching term present in the TR$^\prime$ scheme, whose relative
impact decreases with $Q^2$.
\item  Similarly, the only difference between FONLL-C (plain) 
and TR$^\prime$ NNLO for $F_{2c}$ 
is a subleading $Q^2$-independent 
matching term present in TR$^\prime$.
\item FONLL scheme B is only identical to
 TR$^\prime$ NLO for $ F_{2c}$ for $Q^2=m_c^2$, being
different for any other $Q^2 > m_c^2$, since the
higher order term included in TR$^\prime$ is frozen
at the heavy quark threshold.
\item FONLL-B, as
discussed in Ref.~\cite{Forte:2010ta},  is to a very
good approximation  (to the order $\alpha_S^2$ massive result as 
$Q^2$ increases above $m_c^2$) independent of the choice of
arbitrary threshold prescription. The other NLO schemes,
S-ACOT, TR$^\prime$ NLO and FONLL-A are, on the other hand, much
more sensitive to this choice, with differences that can be
as large as the effect of the resummation itself.
\item Only the ACOT type (ACOT, S-ACOT, S-ACOT-$\chi$) and 
the FONLL-A schemes reduce to exactly the order
$\alpha_S$ NLO massless limit at high $Q^2$, without some strictly 
higher-order $\alpha_S^2$ contributions.
\item Due to the TR ordering, FONLL scheme B is very close
to TR$^\prime$ NLO for the case of the longitudinal
structure function $F_{Lc}$.
\end{enumerate}

In order to provide easy access to the present benchmark comparisons
for future developments,  we have summarized the values of
the DIS charm structure function $F_{2c}$ at the benchmark kinematical
points obtained from the various possible approaches discussed
in the text. The benchmark tables for
the FONLL schemes are given in 
Tables~\ref{LH_HQ_tab:tablebench-fonll-1} 
and~\ref{LH_HQ_tab:tablebench-fonll-2}, those
of the ACOT scheme in Table~\ref{LH_HQ_tab:tablebench-acot-1}, and finally those of
the TR$^\prime$ schemes in Table~\ref{LH_HQ_tab:tablebench-mstw-1}.

Whenever new or improved GM-VFN schemes are proposed,
they should be compared with the results of these benchmark
tables, in order to quantitatively compare with previous approaches and
avoid confusion.  The corresponding benchmark numbers
for the $F_{Lc}$ charm structure function are available
from the authors upon request.

\begin{table}[ht]
\begin{center}
\scriptsize
 \begin{tabular}{|c|c|c|c|c|c|c|}
 \hline
 $x$ & FONLL-A plain & FONLL-A-damp & FONLL-A-$\chi$
& FONLL-B plain & FONLL-B-damp & FONLL-B-$\chi$\\
\hline
 \hline
\multicolumn{7}{|c|}{$Q^2=4$ GeV$^2$} \\
 \hline
$10^{-5}$ &     0.27426 & 0.15066 & 0.15220  & 0.23919 & 0.24905 & 0.25011 \\
 $10^{-4}$  &   0.16424 & 0.09356 & 0.09273  & 0.13421 & 0.13606 & 0.13672 \\
 $10^{-3}$  &   0.08424 & 0.05055 & 0.04875  & 0.06483 & 0.06375 & 0.06391 \\
  $10^{-2}$  &  0.02859 & 0.01738 & 0.01591  & 0.02161 & 0.02063 & 0.02058 \\
  $10^{-1}$  &  0.00207 & 0.00072 & 0.00031  & 0.00093 & 0.00068 & 0.00059 \\
\hline
\hline
\multicolumn{7}{|c|}{$Q^2=10$ GeV$^2$} \\
 \hline
$10^{-5}$ &     0.67714 & 0.56354 & 0.51346 & 0.54126 & 0.55361 & 0.56340 \\
 $10^{-4}$  &   0.37430 & 0.31206 & 0.28187 & 0.30220 & 0.30225 & 0.30449 \\
 $10^{-3}$  &  0.17900 & 0.14995 & 0.13374  & 0.14685 & 0.14405 & 0.14330 \\
  $10^{-2}$  & 0.06051 & 0.05056 & 0.04357 & 0.05196 & 0.05016 & 0.04913 \\
  $10^{-1}$  & 0.00562 & 0.00423 & 0.00252 & 0.00430 & 0.00392 & 0.00346 \\
\hline
\hline
\multicolumn{7}{|c|}{$Q^2=24$ GeV$^2$} \\
 \hline
$10^{-5}$ &     1.19978 & 1.13985 & 1.08785 & 1.02189 & 1.02970 & 1.04182 \\
 $10^{-4}$  &  0.63020 & 0.59690 & 0.56568 & 0.55006 & 0.54938 & 0.55112 \\
 $10^{-3}$  &  0.28826 & 0.27221 & 0.25568 & 0.25873 & 0.25637 & 0.25476 \\
  $10^{-2}$  & 0.09642 & 0.09051 & 0.08307  & 0.09010 & 0.08869 & 0.08716 \\ 
  $10^{-1}$  &  0.00997 & 0.00908 & 0.00708 & 0.00924 & 0.00895 & 0.00831 \\
\hline
\hline
\multicolumn{7}{|c|}{$Q^2=100$ GeV$^2$} \\
 \hline
$10^{-5}$ &    2.29879 & 2.28636 & 2.27201 & 2.10444 & 2.10708 & 2.11399 \\
 $10^{-4}$  &  1.13024 & 1.12186 & 1.11165  & 1.04894 & 1.04875 & 1.05111 \\
 $10^{-3}$  &  0.48483 & 0.48008 & 0.47343 & 0.46063 & 0.45974 & 0.45920 \\
  $10^{-2}$  & 0.15406 & 0.15207 & 0.14862 & 0.15111 & 0.15051 & 0.14962 \\
  $10^{-1}$  & 0.01646 & 0.01615 & 0.01510 &  0.01639 & 0.01627 & 0.01588 \\
 \hline
 \end{tabular}
\end{center}
\caption{\small Results of the benchmark comparison for the
$F_{2c}(x,Q^2)$ structure function in the
two NLO FONLL schemes, denoted by scheme A and scheme B.
In the two cases we provide the results without threshold
prescription and with two different threshold prescriptions,
$\chi-$scaling and a damping factor. Results are provided at the
benchmark kinematical points in $x,Q^2$.
\label{LH_HQ_tab:tablebench-fonll-1}}
\end{table}

\begin{table}[ht]
\begin{center}
\scriptsize
 \begin{tabular}{|c|c|c|c|}
 \hline
 $x$ & FONLL-C plain & FONLL-C-damp & FONLL-C-$\chi$\\
 \hline
\hline
\multicolumn{4}{|c|}{$Q^2=4$ GeV$^2$} \\
 \hline
$10^{-5}$  &    0.37255 & 0.27609 & 0.27096 \\
 $10^{-4}$ &    0.17702 & 0.14585 & 0.14235 \\
 $10^{-3}$  &  0.07001 & 0.06492 & 0.06381 \\
  $10^{-2}$ &  0.02027 & 0.02004 & 0.02019 \\
  $10^{-1}$ &  0.00149 & 0.00078 & 0.00069 \\
 \hline
\hline
\multicolumn{4}{|c|}{$Q^2=10$ GeV$^2$} \\
 \hline
$10^{-5}$ &   0.78141 & 0.69206 & 0.64037 \\
 $10^{-4}$  &  0.37400 & 0.34507 & 0.32514 \\
 $10^{-3}$  &  0.15564 & 0.14932 & 0.14446 \\
  $10^{-2}$  & 0.05106 & 0.04938 & 0.04869 \\
  $10^{-1}$  & 0.00547 & 0.00462 & 0.00387 \\
 \hline
\hline
\multicolumn{4}{|c|}{$Q^2=24$ GeV$^2$} \\
 \hline
$10^{-5}$ &   1.30170 & 1.25019 & 1.19206 \\
 $10^{-4}$  & 0.63138 & 0.61466 & 0.59361 \\
 $10^{-3}$  &   0.26822 & 0.26401 & 0.25864 \\
  $10^{-2}$  & 0.09009 & 0.08856 & 0.08717 \\
  $10^{-1}$   & 0.01067 & 0.01012 & 0.00915 \\
 \hline
\hline
\multicolumn{4}{|c|}{$Q^2=100$ GeV$^2$} \\
 \hline
$10^{-5}$  &    2.39198 & 2.37634 & 2.35015 \\
 $10^{-4}$ &    1.13357 & 1.12863 & 1.11943 \\
 $10^{-3}$  &   0.47207 & 0.47058 & 0.46798 \\
  $10^{-2}$ &  0.15213 & 0.15146 & 0.15054 \\
  $10^{-1}$ &  0.01804 & 0.01784 & 0.01729 \\
 \hline
 \end{tabular}
\end{center}
\caption{\small Results of the benchmark comparison for the
$F_{2c}(x,Q^2)$ structure function in the
 NNLO FONLL scheme, denoted by scheme C.
As before, we provide the results without threshold
prescription and with two different threshold prescriptions,
$\chi-$scaling and a damping factor.
\label{LH_HQ_tab:tablebench-fonll-2}}
\end{table}

\begin{table}[ht]

\begin{centering}
\scriptsize 
\begin{tabular}{|c|c|c|c|c|}
\hline 
{\scriptsize $x$ } & {\scriptsize S-ACOT plain } & {\scriptsize S-ACOT-$\chi$ } & {\scriptsize S-ACOT-$\chi$ (v2) } & {\scriptsize Full ACOT plain }\tabularnewline
\hline
\hline 
{\scriptsize $Q^{2}=4$ GeV$^{2}$ } &  &  &  & \tabularnewline
\hline 
{\scriptsize $10^{-5}$ } & {\scriptsize 0.27339 } & {\scriptsize 0.15780} & {\scriptsize 0.23670 } & {\scriptsize 0.30121 }\tabularnewline
{\scriptsize $10^{-4}$ } & {\scriptsize 0.16327 } & {\scriptsize 0.09171} & {\scriptsize 0.13749 } & {\scriptsize 0.17774}\tabularnewline
{\scriptsize $10^{-3}$ } & {\scriptsize 0.08386 } & {\scriptsize 0.04497} & {\scriptsize 0.06734 } & {\scriptsize 0.08956 }\tabularnewline
{\scriptsize $10^{-2}$ } & {\scriptsize 0.02839 } & {\scriptsize 0.01360} & {\scriptsize 0.02030 } & {\scriptsize 0.02950 }\tabularnewline
{\scriptsize $10^{-5}$ } & {\scriptsize 0.00203 } & {\scriptsize 0.00026} & {\scriptsize 0.00038 } & {\scriptsize 0.00169}\tabularnewline
\hline
\hline 
{\scriptsize $Q^{2}=10$ GeV$^{2}$ } &  &  &  & \tabularnewline
\hline 
{\scriptsize $10^{-5}$ } & {\scriptsize 0.67349 } & {\scriptsize 0.52648} & {\scriptsize 0.63177 } & {\scriptsize 0.69789 }\tabularnewline
{\scriptsize $10^{-4}$ } & {\scriptsize 0.37254 } & {\scriptsize 0.28768} & {\scriptsize 0.34503 } & {\scriptsize 0.38545 }\tabularnewline
{\scriptsize $10^{-3}$ } & {\scriptsize 0.17826 } & {\scriptsize 0.13482} & {\scriptsize 0.16153 } & {\scriptsize 0.18395 }\tabularnewline
{\scriptsize $10^{-2}$ } & {\scriptsize 0.06024 } & {\scriptsize 0.04338} & {\scriptsize 0.05185 } & {\scriptsize 0.06211 }\tabularnewline
{\scriptsize $10^{-5}$ } & {\scriptsize 0.00554 } & {\scriptsize 0.00259} & {\scriptsize 0.00305 } & {\scriptsize 0.00547 }\tabularnewline
\hline
\hline 
{\scriptsize $Q^{2}=24$ GeV$^{2}$ } &  &  &  & \tabularnewline
\hline 
{\scriptsize $10^{-5}$ } & {\scriptsize 1.19413 } & {\scriptsize 1.07983} & {\scriptsize 1.16981 } & {\scriptsize 1.20362 }\tabularnewline
{\scriptsize $10^{-4}$ } & {\scriptsize 0.62805 } & {\scriptsize 0.56516} & {\scriptsize 0.61188 } & {\scriptsize 0.63403 }\tabularnewline
{\scriptsize $10^{-3}$ } & {\scriptsize 0.28739 } & {\scriptsize 0.25642} & {\scriptsize 0.27736 } & {\scriptsize 0.29070 }\tabularnewline
{\scriptsize $10^{-2}$ } & {\scriptsize 0.09588 } & {\scriptsize 0.08382} & {\scriptsize 0.09046 } & {\scriptsize 0.09742 }\tabularnewline
{\scriptsize $10^{-5}$ } & {\scriptsize 0.00986 } & {\scriptsize 0.00736} & {\scriptsize 0.00785 } & {\scriptsize 0.00997 }\tabularnewline
\hline
\hline 
{\scriptsize $Q^{2}=100$ GeV$^{2}$ } &  &  &  & \tabularnewline
\hline 
{\scriptsize $10^{-5}$ } & {\scriptsize 2.29983 } & {\scriptsize 2.25162} & {\scriptsize 2.29665 } & {\scriptsize 2.29853}\tabularnewline
{\scriptsize $10^{-4}$ } & {\scriptsize 1.12920 } & {\scriptsize 1.10453} & {\scriptsize 1.12588 } & {\scriptsize 1.12988 }\tabularnewline
{\scriptsize $10^{-3}$ } & {\scriptsize 0.48339 } & {\scriptsize 0.47203} & {\scriptsize 0.48072 } & {\scriptsize 0.48442 }\tabularnewline
{\scriptsize $10^{-2}$ } & {\scriptsize 0.15346 } & {\scriptsize 0.14918} & {\scriptsize 0.15161 } & {\scriptsize 0.15415 }\tabularnewline
{\scriptsize $10^{-5}$ } & {\scriptsize 0.01629 } & {\scriptsize 0.01531} & {\scriptsize 0.01540 } & {\scriptsize 0.01640 }\tabularnewline
\hline
\end{tabular}
\par\end{centering}{\scriptsize \par}

\caption{{\small Results of the benchmark comparison for the $F_{2c}(x,Q^{2})$
structure function for the ACOT family of NLO GM-VFN schemes. We provide
the results for both full ACOT and simplified ACOT (S-ACOT) without
any threshold prescriptions, as well as those for S-ACOT-$\chi$ with
the $\chi-$scaling threshold prescription of Eq.~\ref{LH_HQ_eq:chi-sc-1},
and S-ACOT-$\chi$ (v2) with the $\chi-$scaling threshold prescription
of Eq.~\ref{LH_HQ_eq:chi-sc-2}, Results are provided at the benchmark
kinematical points in $x,Q^{2}$. \label{LH_HQ_tab:tablebench-acot-1}}}

\end{table}

\begin{table}[ht]
\begin{center}
\scriptsize
 \begin{tabular}{|c|c|c|c|c|}
 \hline
 $x$ & MSTW08 NLO plain &  MSTW08 NLO $\chi$  & 
 MSTW08 NNLO plain &  MSTW08 NNLO $\chi$ \\
 \hline
 \hline
 $Q^2=4$ GeV$^2$ & & & & \\
 \hline
 $10^{-5}$ & 0.36337 & 0.32667 & 0.47899 & 0.42824 \\
 $10^{-4}$ & 0.20751 & 0.18173 & 0.20366 & 0.18164 \\
 $10^{-3}$ & 0.09873 & 0.08220 & 0.07654 & 0.07139 \\
 $10^{-2}$ & 0.03174 & 0.02364 & 0.02649 & 0.02690 \\
 $10^{-1}$ & 0.00215 & 0.00046 & 0.00192 & 0.00128 \\
 \hline
 \hline
 $Q^2=10$ GeV$^2$ & & & & \\
 \hline
 $10^{-5}$ & 0.76342 & 0.72170 & 0.88803 & 0.83214 \\
 $10^{-4}$ & 0.41689 & 0.38936 & 0.40160 & 0.37813 \\
 $10^{-3}$ & 0.19326 & 0.17652 & 0.16315 & 0.15704 \\
 $10^{-2}$ & 0.06370 & 0.05530 & 0.05738 & 0.05679 \\
 $10^{-1}$ & 0.00571 & 0.00319 & 0.00575 & 0.00468 \\
 \hline
 \hline
 $Q^2=24$ GeV$^2$ & & & & \\
 \hline
 $10^{-5}$ & 1.28397 & 1.25966 & 1.40641 & 1.36811 \\
 $10^{-4}$ & 0.67251 & 0.65634 & 0.65687 & 0.64135 \\
 $10^{-3}$ & 0.30255 & 0.29252 & 0.27505 & 0.27089 \\
 $10^{-2}$ & 0.09946 & 0.09403 & 0.09593 & 0.09520 \\
 $10^{-1}$ & 0.01008 & 0.00806 & 0.01090 & 0.00998 \\
 \hline
 \hline
 $Q^2=100$ GeV$^2$ & & & & \\
 \hline
 $10^{-5}$ & 2.38877 & 2.38560 & 2.49312 & 2.47939 \\
 $10^{-4}$ & 1.17358 & 1.17027 & 1.16278 & 1.15769 \\
 $10^{-3}$ & 0.49872 & 0.49601 & 0.47953 & 0.47822 \\
 $10^{-2}$ & 0.15719 & 0.15534 & 0.15837 & 0.15802 \\
 $10^{-1}$ & 0.01658 & 0.01568 & 0.01836 & 0.01789 \\
 \hline
 \end{tabular}
\end{center}
\caption{\small Results of the benchmark comparison for the
$F_{2c}(x,Q^2)$ structure function for the 
 NLO and NNLO TR$^\prime$ schemes.
In the two cases we provide the results without threshold
prescription and with the TR$^\prime$ 
$\chi-$scaling threshold prescription
which is implemented in the MSTW08 parton fits. Results are provided at the
benchmark kinematical points in $x,Q^2$.
\label{LH_HQ_tab:tablebench-mstw-1}}
\end{table}



%% file: campbell/campbell.tex
\newcommand\Sherpa{S\scalebox{1.0}{HERPA}\xspace}

\newcommand{\mc}[1]{\mathcal{#1}}
\newcommand{\sbr}[1]{\left[ #1\right]}








\subsection{Introduction}
\label{sec:introduction}
The description of Higgs production in association with jets, through gluon fusion,
is important for several reasons. Firstly, a clean extraction of the coupling of the
Higgs boson to weak bosons in the weak-boson-fusion channel requires that the gluon fusion
contribution be suppressed as much as possible. Secondly, by focusing directly on the
gluon fusion contribution, it is possible to extract not just the absolute
size, but also the $CP$-structure of the Higgs boson coupling to gluons
induced through a top-loop\cite{Klamke:2007cu}. In both cases the
description of further hard radiation from the lowest-order process is
important. The different radiation pattern observed in
weak-boson-fusion and gluon-fusion can be used to reject the latter. Moreover, the
tree-level observations on the extraction of $CP$-properties could be spoilt
by decorrelations due to further hard emissions, unless observables which are
robust against these higher order corrections are
employed in the analysis\cite{Andersen:2010zx}.

In this contribution we will compare the description of the final
state obtained using a tree-level matched parton shower~\cite{Hoeche:2009rj,Hoeche:2009xc,Carli:2009cg} (represented by
\Sherpa\cite{Gleisberg:2003xi,Gleisberg:2008ta}), a next-to-leading order (NLO) calculation of $hjj$ production
through gluon fusion (as implemented in
MCFM\cite{Campbell:2006xx,Campbell:2010cz}) and finally the HEJ (\emph{High
  Energy Jets}) resummation
scheme based on the factorisation of scattering amplitudes in the high energy
limit\cite{Andersen:2008ue,Andersen:2008gc,Andersen:2009nu,Andersen:2009he}.


\subsection{Inclusive predictions}
\label{sec:higgs-boson-prod}
We start by comparing predictions for distributions in a sample obtained by
requiring at least two jets of at least 40~GeV transverse momentum and a
rapidity less then 4.5 obtained with the $k_t$-jet algorithm of
Ref.\cite{Cacciari:2005hq} with $D=0.7$. The rapidity of the Higgs boson is
also required to be less then 4.5. For this study we have chosen a putative
Higgs boson mass of 120~GeV and set the centre-of-mass energy, $\sqrt s =10$~TeV.

\begin{figure}
  \begin{minipage}{0.49\textwidth}
    \includegraphics[width=\linewidth]{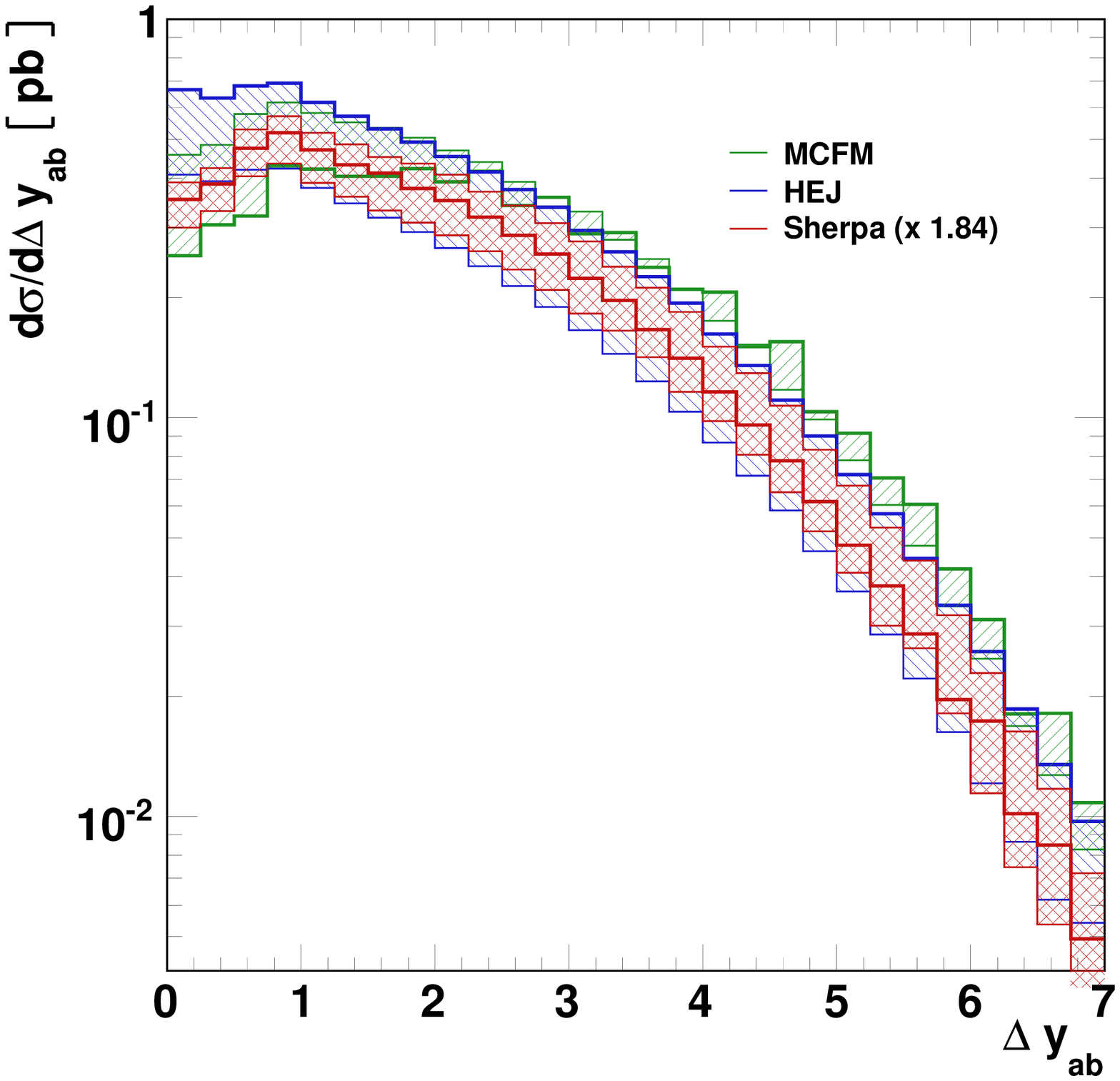}
    \caption{The distribution with respect to the rapidity difference
      $\Delta y_{ab}$ between the most forward and most backward jet of more
      than 40~GeV. \label{HiggsJets_Dy}}
  \end{minipage}\hfill
  \begin{minipage}{0.49\textwidth}
    \includegraphics[width=\linewidth]{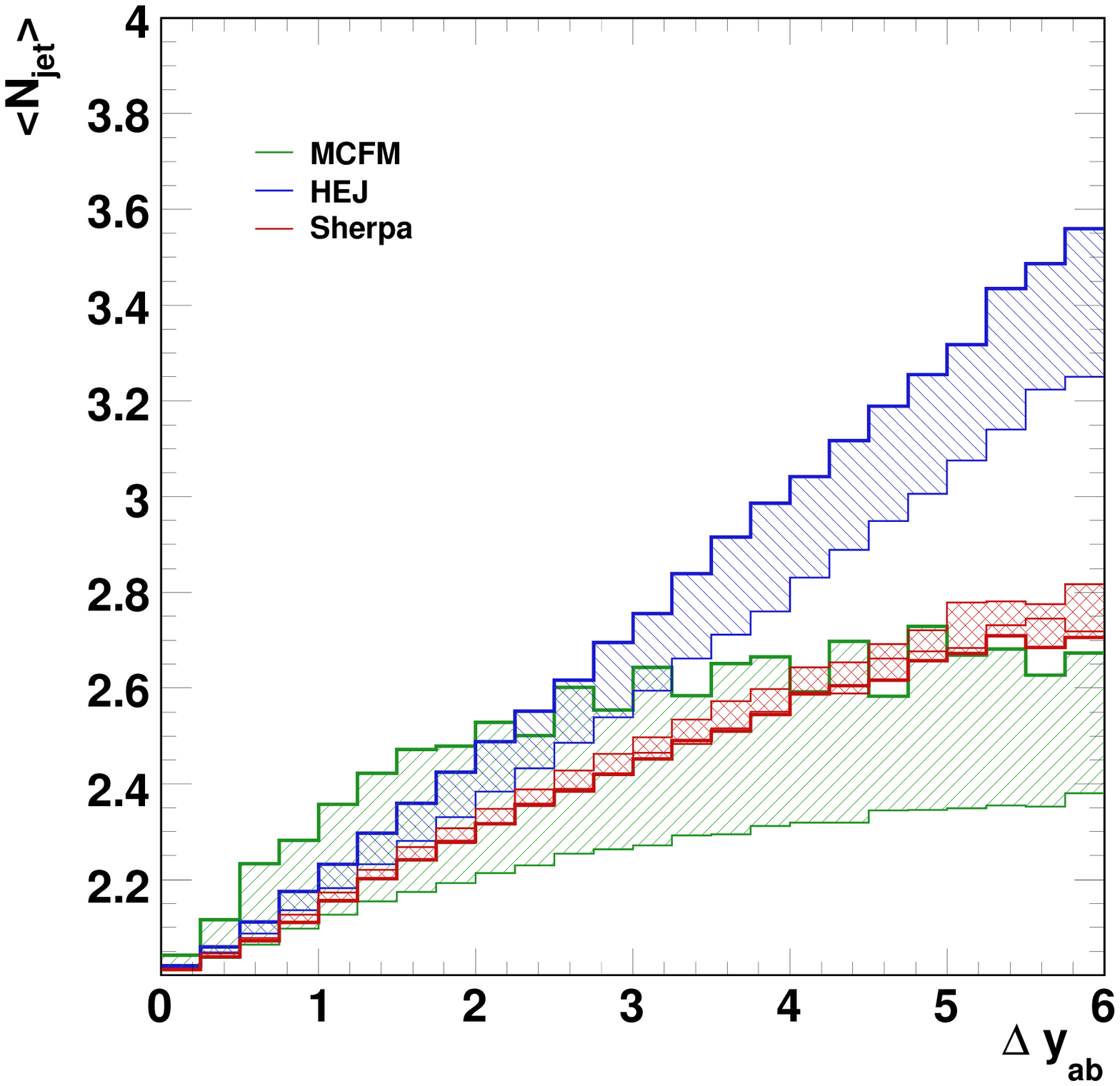}
    \caption{The average number of hard jets (above 40~GeV transverse momentum) obtained in the $k_t$ jet algorithm,
             as a function of the largest rapidity span $\Delta y_{ab}$ between two jets.\label{HiggsJets_NJet}}
  \end{minipage}
\end{figure}
In Fig.~\ref{HiggsJets_Dy} we compare the differential distribution of the
rapidity span between the most forward and most backward jet, $\Delta y_{ab}$, obtained within
the three descriptions. For MCFM and HEJ the bands indicate the variation
obtained by changing the renormalisation and factorisation scale between
40~GeV and 120~GeV (but with two powers of $\alpha_s$ always evaluated at the
scale of the Higgs boson mass). The uncertainty of the predictions made
with \Sherpa is estimated by varying the prefactor of the nodal scales
within the parton shower (and the higher-order tree-level matrix elements)
between 1/2 and 2. We see that the
shapes are all in good agreement with each other, and that the scale
uncertainty is smallest for the full NLO calculation of MCFM.

In Fig.~\ref{HiggsJets_NJet} we analyse directly the average number of hard
jets (above 40~GeV transverse momentum) obtained in the $k_t$ jet algorithm
as a function of the rapidity span $\Delta y_{ab}$. The same observable was
analysed for $W+$dijets in
Ref.\cite{andersen10:_simpl_radiat_patter_hard_multi}. As discussed there,
the framework of Ref.\cite{Lipatov:1974qm,Fadin:1975cb,Kuraev:1976ge} implies,
for a certain class of processes (to which both $pp\to hjj$ through gluon fusion
and $pp\to Wjj$ belong), an increase in jet count with increasing rapidity
span. In turn, such an increase in the amount of hard radiation necessitates the inclusion of increasingly higher order
corrections in order to obtain stable predictions for increasing $\Delta y_{ab}$. The modelling of this increase is
important for the application of a jet veto to suppress the gluon fusion
contribution to $h+$dijets. While all three calculations show an increase of
the jet count with increasing $\Delta y_{ab}$, the amounts of increase
differ. Using HEJ, the average number of jets found in events with a rapidity
span of 5 is just slightly more than 3 (compared to 2.8 in the case of
$W+$dijets studied in
Ref.\cite{andersen10:_simpl_radiat_patter_hard_multi}). \Sherpa produces less
of the hard radiation, with an average of 2.8 hard jets when the rapidity
span is $\Delta y_{ab}=5$. The rise in the average number of jets with
increasing rapidity span obtained with \Sherpa is slightly slower than linear
-- and slower than the predictions obtained with
ALPGEN\cite{Mangano:2002ea}+HERWIG\cite{Corcella:2002jc} for
$W+$dijets. Finally, the prediction of NLO QCD (obtained from MCFM) indicates 
an even smaller average number of jets at large rapidity spans.
For a scale choice of 40~GeV, the
average number of hard jets in events with a span of 5 units of rapidity is
comparable to that obtained in NLO QCD for
$W$+dijets\cite{andersen10:_simpl_radiat_patter_hard_multi}.
There is a large variation induced by a variation of factorisation and renormalisation scale,
which is to be expected since this observable only becomes non-trivial at NLO.
This variation is particularly noticeable at small rapidity spans, where the extrema
of the band represent rises that are either stronger or weaker than those obtained in the
other two predictions.

\subsection{Predictions at large rapidity span}
We now impose an additional cut, to focus on the region where two of the jets are separated by a large
rapidity span, $\Delta y_{ab}>4$. This situation arises naturally for the
weak-boson-fusion process, where the gluon-fusion channel discussed here will
act as a background in extracting the coupling between the weak and the Higgs boson.

\begin{figure}
  \begin{minipage}{0.49\textwidth}
    \includegraphics[width=\linewidth]{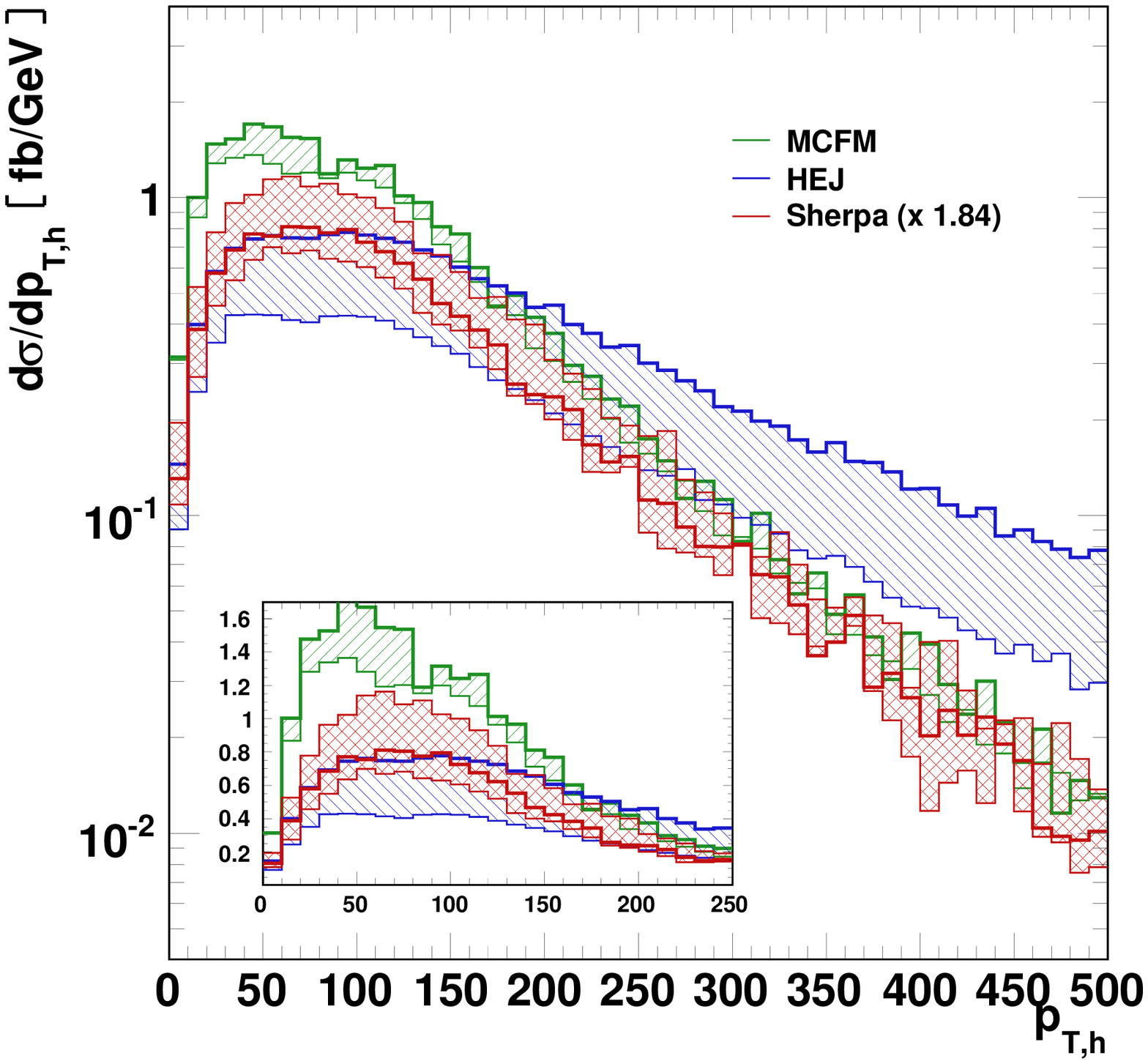}
    \caption{The transverse momentum of the Higgs boson, in the region of large jet rapidity span
    between two jets defined by $\Delta y_{ab}>4$. The inset shows the same quantity on a linear scale,
    over a smaller range.\label{HiggsJets_PTh}}
  \end{minipage}
  \hspace{0.3cm}
  \begin{minipage}{0.49\textwidth}
    \includegraphics[width=\linewidth]{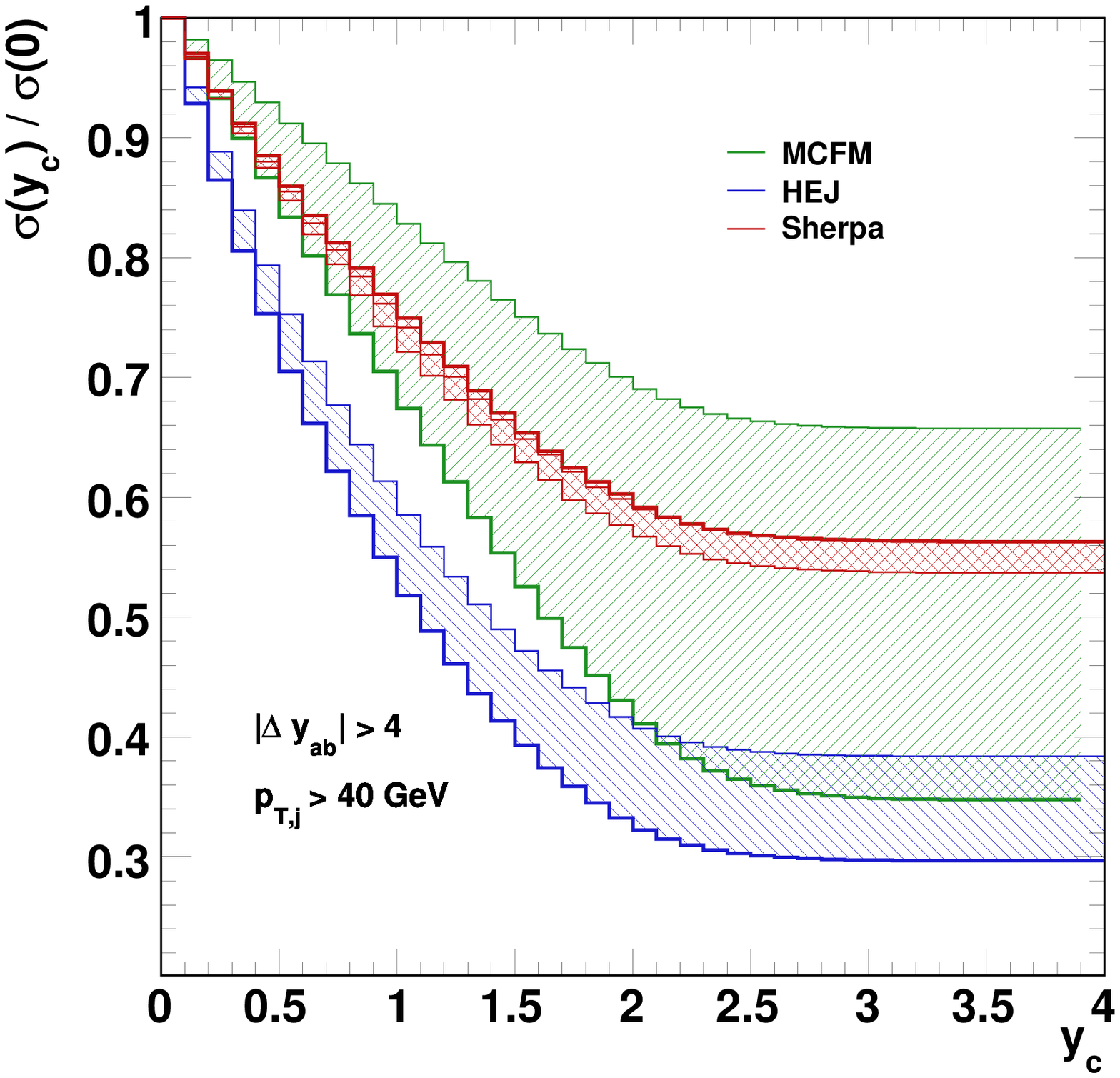}
    \caption{Dependence of the normalized cross section on the jet veto parameter $y_c$, defined in
    the text. For large $y_c$ this normalized cross section gives the prediction for the
    ratio of exclusive dijet and inclusive dijet+X cross sections.\label{HiggsJets_Yc}}
  \end{minipage}
\end{figure}
In Fig.~\ref{HiggsJets_PTh} we show the predictions for the transverse momentum spectrum of the Higgs
boson. The HEJ approach predicts a much harder spectrum than that obtained in \Sherpa and MCFM.
As expected, the latter two predictions agree with each other quite well in this region, with the 
NLO curve consistently narrower.

Finally, in Fig.~\ref{HiggsJets_Yc} we investigate the effect of vetoing additional jet activity beyond
the two that define the rapidity span. This is interesting for suppressing
the gluon-fusion contribution. We parametrize the veto as in Ref.~\cite{Andersen:2008gc} by using the
variable $y_c$ as follows. Given the rapidity span $\Delta y_{ab}= |y_a - y_b|$, we compute the midpoint
of rapidities of the two jets furthest apart in rapidity, $y_0 = (y_a+y_b)/2$. An event is only removed by the veto if it contains a
further jet (with transverse momentum at least 40~GeV) at rapidity $y^\prime$ with $|y^\prime - y_0| < y_c$.
With this definition we see that the
cross section $\sigma(y_c = 0)$ corresponds to applying no veto at all, whilst the limit of large $y_c$
corresponds to vetoing all additional jets. The $y$-axis on
Fig.~\ref{HiggsJets_Yc} indicates the fraction of the cross section with no
jet-veto which survives as a function of $y_c$. Once more there is a clear difference between the HEJ
prediction and that of \Sherpa and MCFM, both in the dependence on $y_c$ as it increases from zero, and
in the asymptotic value for large $y_c$. This is simply a reflection of the
differences observed in the predictions for the amount of hard jets for
larger rapidity spans, as indicated in Fig.~\ref{HiggsJets_NJet}. As in that
figure, the MCFM band indicates a large scale uncertainty due to the essentially leading order nature of the
prediction for this observable.

\subsection{Conclusions}
\label{sec:conclusions}
In this contribution we have performed an introductory study of the additional radiation expected in Higgs
+ dijet events at the LHC. Understanding the characteristics of this radiation would be crucial to
determining the nature of the Higgs boson, such as its couplings to matter and its $CP$-properties.
We have compared the predictions of three very different theoretical approaches:
a matrix-element improved parton-shower (\Sherpa), NLO QCD (MCFM) and a
calculation based on behaviour of the matrix elements to all orders in the
high energy limit (the HEJ resummation).

The predictions obtained for the rapidity difference between
the most forward and most backward hard jet are in good
agreement between all three calculations. Furthermore, the
predictions for the number of additional jets are in
reasonable agreement between the models for smaller rapidity
spans (up to 2). For larger rapidity spans, the HEJ
resummation scheme leads to more hard jets than in either
NLO QCD or \Sherpa. This induces a difference in the effect
of a veto on further jet activity. Using HEJ, the Higgs
boson tends to be accompanied by more additional central
radiation that would be subject to a jet veto.

The work presented here should be contrasted with the
similar study of $W+$dijet events in
Ref.\cite{andersen10:_simpl_radiat_patter_hard_multi}. A
more detailed analysis of predictions for additional
radiation in $W$, $Z$ and $h$+dijet events is required in
order to assess the potential for applying results
from the study of $W$,$Z$+dijets to that of $h$+dijets. One
might expect that a relatively early study of radiation in
$W$ and $Z$+dijet events could pave the way for later
studies of the Higgs boson with more integrated luminosity.



%% file: passarino/passarino.tex



%
\newcommand{\nl}{\nonumber\\}
\newcommand{\nn}{\nonumber}
\newcommand{\ds}{\displaystyle}
\newcommand{\mpar}[1]{{\marginpar{\hbadness10000%
                      \sloppy\hfuzz10pt\boldmath\bf#1}}%
                      \typeout{marginpar: #1}\ignorespaces}
\def\mnew{\mpar{\hfil NEW \hfil}\ignorespaces}
\newcommand{\lpar}{\left(}                            
\newcommand{\rpar}{\right)} 
\newcommand{\lrbr}{\left[}
\newcommand{\rrbr}{\right]}
\newcommand{\lcbr}{\left\{}
\newcommand{\rcbr}{\right\}} 
\newcommand{\rbrak}[1]{\lrbr#1\rrbr}
\newcommand{\bq}{\begin{equation}}                    
\newcommand{\eq}{\end{equation}}
\newcommand{\bqa}{\arraycolsep 0.14em\begin{eqnarray}}
\newcommand{\eqa}{\end{eqnarray}}
\newcommand{\ba}[1]{\begin{array}{#1}}
\newcommand{\ea}{\end{array}}
\newcommand{\ben}{\begin{enumerate}}
\newcommand{\een}{\end{enumerate}}
\newcommand{\bei}{\begin{itemize}}
\newcommand{\eei}{\end{itemize}}
\newcommand{\eqn}[1]{Eq.(\ref{#1})}
\newcommand{\eqns}[2]{Eqs.(\ref{#1})--(\ref{#2})}
\newcommand{\eqnss}[1]{Eqs.(\ref{#1})}
\newcommand{\eqnsc}[2]{Eqs.(\ref{#1}) and (\ref{#2})}
\newcommand{\eqnst}[3]{Eqs.(\ref{#1}), (\ref{#2}) and (\ref{#3})}
\newcommand{\eqnsf}[4]{Eqs.(\ref{#1}), (\ref{#2}), (\ref{#3}) and (\ref{#4})}
\newcommand{\eqnsv}[5]{Eqs.(\ref{#1}), (\ref{#2}), (\ref{#3}), (\ref{#4}) and (\ref{#5})}
\newcommand{\tbn}[1]{Tab.~\ref{#1}}
\newcommand{\tabn}[1]{Tab.~\ref{#1}}
\newcommand{\tbns}[2]{Tabs.~\ref{#1}--\ref{#2}}
\newcommand{\tabns}[2]{Tabs.~\ref{#1}--\ref{#2}}
\newcommand{\tbnsc}[2]{Tabs.~\ref{#1} and \ref{#2}}
\newcommand{\fig}[1]{Fig.~\ref{#1}}
\newcommand{\figs}[2]{Figs.~\ref{#1}--\ref{#2}}
\newcommand{\sect}[1]{Section~\ref{#1}}
\newcommand{\sects}[2]{Section~\ref{#1} and \ref{#2}}
\newcommand{\sectm}[2]{Section~\ref{#1} -- \ref{#2}}
\newcommand{\subsect}[1]{Subsection~\ref{#1}}
\newcommand{\subsectm}[2]{Subsection~\ref{#1} -- \ref{#2}}
\newcommand{\appendx}[1]{Appendix~\ref{#1}}
\newcommand{\hsp}{\hspace{.5mm}}
\def\negs{\hspace{-0.26in}}
\def\negsh{\hspace{-0.13in}}
%
%
\newcommand{\TeV}{\mathrm{TeV}}                     
\newcommand{\GeV}{\mathrm{GeV}}
\newcommand{\MeV}{\mathrm{MeV}}
\newcommand{\eV}{\mathrm{eV}}
\newcommand{\nb}{\mathrm{nb}}
\newcommand{\pb}{\mathrm{pb}}
\newcommand{\fb}{\mathrm{fb}}
\def\Re{\mathop{\operator@font Re}\nolimits}
\def\Im{\mathop{\operator@font Im}\nolimits}
\newcommand{\ord}[1]{{\cal O}\lpar#1\rpar}
\newcommand{\group}{SU(2)\otimes U(1)}
\newcommand{\ib}{i}
\newcommand{\asums}[1]{\sum_{#1}}
\newcommand{\asumt}[2]{\sum_{#1}^{#2}}
\newcommand{\asum}[3]{\sum_{#1=#2}^{#3}}
%
%
\newcommand{\tmi}{\times 10^{-1}}
\newcommand{\tmii}{\times 10^{-2}}
\newcommand{\tmiii}{\times 10^{-3}}
\newcommand{\tmiv}{\times 10^{-4}}
\newcommand{\tmfv}{\times 10^{-5}}
\newcommand{\tmfvi}{\times 10^{-6}}
\newcommand{\tmfvii}{\times 10^{-7}}
\newcommand{\tmfviii}{\times 10^{-8}}
\newcommand{\tmfix}{\times 10^{-9}}
\newcommand{\tmfx}{\times 10^{-10}}
%
%
\newcommand{\fer}{{\rm{fer}}}
\newcommand{\bos}{{\rm{bos}}}
\newcommand{\lep}{{l}}
\newcommand{\had}{{h}}
\newcommand{\gen}{\rm{g}}
\newcommand{\dbl}{\rm{d}}
\newcommand{\philone}{\phi}
\newcommand{\philoneb}{\phi_{0}}
\newcommand{\phiind}[1]{\phi_{#1}}
\newcommand{\gBi}[2]{B_{#1}^{#2}}
\newcommand{\gBn}[1]{B_{#1}}
%
%
\newcommand{\ph}{\gamma}
\newcommand{\ab}{A}
\newcommand{\abr}{A^r}
\newcommand{\abb}{A^{0}}
\newcommand{\abi}[1]{A_{#1}}
\newcommand{\abri}[1]{A^r_{#1}}
\newcommand{\abbi}[1]{A^{0}_{#1}}
\newcommand{\wb}{W}            
\newcommand{\wbi}[1]{W_{#1}}           
\newcommand{\wbp}{W^{+}}
\newcommand{\wbm}{W^{-}}
\newcommand{\wbpm}{W^{\pm}}
\newcommand{\wbpi}[1]{W^{+}_{#1}}
\newcommand{\wbmi}[1]{W^{-}_{#1}}
\newcommand{\wbpmi}[1]{W^{\pm}_{#1}}
\newcommand{\wbli}[1]{W^{[+}_{#1}}
\newcommand{\wbri}[1]{W^{-]}_{#1}}
\newcommand{\zb}{Z}
\newcommand{\zbi}[1]{Z_{#1}}
\newcommand{\vb}{V}
\newcommand{\vbi}[1]{V_{#1}}      
\newcommand{\vbiv}[1]{V^{*}_{#1}}      
\newcommand{\Pb}{P}
\newcommand{\Sb}{S}
\newcommand{\Bb}{B}
%
%
\newcommand{\hk}{K}
\newcommand{\hKi}[1]{K_{#1}}
\newcommand{\hkg}{\phi}
\newcommand{\hkn}{\phi^{0}}                 
\newcommand{\hkp}{\phi^{+}}
\newcommand{\hkm}{\phi^{-}}
\newcommand{\hkpm}{\phi^{\pm}}
\newcommand{\hkmp}{\phi^{\mp}}
\newcommand{\hki}[1]{\phi^{#1}}
\newcommand{\hb}{H}
\newcommand{\hbi}[1]{H_{#1}}
\newcommand{\hkl}{\phi^{[+\cgfi\cgfi}}
\newcommand{\hkr}{\phi^{-]}}
%
%
\newcommand{\fpx}{X}
\newcommand{\fpy}{Y}
\newcommand{\fpxp}{X^+}
\newcommand{\fpxm}{X^-}
\newcommand{\fpxpm}{X^{\pm}}
\newcommand{\fpxi}[1]{X^{#1}}
\newcommand{\fpyZ}{Y^{\ssZ}}
\newcommand{\fpyA}{Y^{\ssA}}
\newcommand{\fpyZA}{Y_{\ssZ,\ssA}}
\newcommand{\fpbxi}[1]{{\overline{X}}^{#1}}
\newcommand{\fpbyZ}{{\overline{Y}}^{\ssZ}}
\newcommand{\fpbyA}{{\overline{Y}}^{\ssA}}
\newcommand{\fpbyZA}{{\overline{Y}}^{\ssZ,\ssA}}
%
%
\newcommand{\Flone}{F}
\newcommand{\fpsi}{\psi}
\newcommand{\fpsii}[1]{\psi^{#1}}
\newcommand{\fpsib}{\psi^{0}}
\newcommand{\fpsir}{\psi^r}
\newcommand{\fpsiL}{\psi_{_L}}
\newcommand{\fpsiR}{\psi_{_R}}
\newcommand{\fpsiLi}[1]{\psi_{_L}^{#1}}
\newcommand{\fpsiRi}[1]{\psi_{_R}^{#1}}
\newcommand{\fpsiLbi}[1]{\psi_{_{0L}}^{#1}}
\newcommand{\fpsiRbi}[1]{\psi_{_{0R}}^{#1}}
\newcommand{\fpsiLR}{\psi_{_{L,R}}}
\newcommand{\fbpsi}{{\overline{\psi}}}
\newcommand{\fbpsii}[1]{{\overline{\psi}}^{#1}}
\newcommand{\fbpsir}{{\overline{\psi}}^r}
\newcommand{\fbpsiL}{{\overline{\psi}}_{_L}}
\newcommand{\fbpsiR}{{\overline{\psi}}_{_R}}
\newcommand{\fbpsiLi}[1]{\overline{\psi_{_L}}^{#1}}
\newcommand{\fbpsiRi}[1]{\overline{\psi_{_R}}^{#1}}
\newcommand{\fe}{e}
\newcommand{\ff}{f}
\newcommand{\fep}{e^{+}}
\newcommand{\fem}{e^{-}}
\newcommand{\fepm}{e^{\pm}}
\newcommand{\fp}{f^{+}}
\newcommand{\fm}{f^{-}}
\newcommand{\fhp}{h^{+}}
\newcommand{\fhm}{h^{-}}
\newcommand{\fh}{h}
\newcommand{\flm}{\mu}
\newcommand{\flmp}{\mu^{+}}
\newcommand{\flmm}{\mu^{-}}
\newcommand{\fll}{l}
\newcommand{\fllp}{l^{+}}
\newcommand{\fllm}{l^{-}}
\newcommand{\flt}{\tau}
\newcommand{\fltp}{\tau^{+}}
\newcommand{\fltm}{\tau^{-}}
\newcommand{\fq}{q}
\newcommand{\fqi}[1]{\fq_{#1}}
\newcommand{\bfqi}[1]{\barq_{#1}}
\newcommand{\ffQ}{Q}
\newcommand{\fu}{u}
\newcommand{\fd}{d}
\newcommand{\fc}{c}
\newcommand{\fs}{s}
\newcommand{\fqp}{q'}
\newcommand{\fup}{u'}
\newcommand{\fdp}{d'}
\newcommand{\fcp}{c'}
\newcommand{\fsp}{s'}
\newcommand{\fdpp}{d''}
\newcommand{\ffi}[1]{f_{#1}}
\newcommand{\bffi}[1]{{\overline{f}}_{#1}}
\newcommand{\ffpi}[1]{f'_{#1}}
\newcommand{\bffpi}[1]{{\overline{f}}'_{#1}}
\newcommand{\ft}{t}
\newcommand{\ffb}{b}
\newcommand{\ffp}{f'}
\newcommand{\fft}{{\tilde{f}}}
\newcommand{\fl}{l}
\newcommand{\fli}[1]{\fl_{#1}}
\newcommand{\fnu}{\nu}
\newcommand{\fU}{U}
\newcommand{\fD}{D}
\newcommand{\fUc}{\overline{U}}
\newcommand{\fDc}{\overline{D}}
\newcommand{\fnul}{\nu_l}
\newcommand{\fnue}{\nu_e}
\newcommand{\fnum}{\nu_{\mu}}
\newcommand{\fnut}{\nu_{\tau}}
\newcommand{\fbe}{{\overline{e}}}
\newcommand{\fbu}{{\overline{u}}}
\newcommand{\fbv}{{\overline{v}}}
\newcommand{\fbd}{{\overline{d}}}
\newcommand{\fbf}{{\overline{f}}}
\newcommand{\fbfp}{{\overline{f}}'}
\newcommand{\fbl}{{\overline{l}}}
\newcommand{\fbnu}{{\overline{\nu}}}
\newcommand{\fbnul}{{\overline{\nu}}_{\fl}}
\newcommand{\fbnue}{{\overline{\nu}}_{\fe}}
\newcommand{\fbnum}{{\overline{\nu}}_{\flm}}
\newcommand{\fbnut}{{\overline{\nu}}_{\flt}}
\newcommand{\fuL}{u_{_L}}
\newcommand{\fdL}{d_{_L}}
\newcommand{\ffL}{f_{_L}}
\newcommand{\fbuL}{{\overline{u}}_{_L}}
\newcommand{\fbdL}{{\overline{d}}_{_L}}
\newcommand{\fbfL}{{\overline{f}}_{_L}}
\newcommand{\fuR}{u_{_R}}
\newcommand{\fdR}{d_{_R}}
\newcommand{\ffR}{f_{_R}}
\newcommand{\fbuR}{{\overline{u}}_{_R}}
\newcommand{\fbdR}{{\overline{d}}_{_R}}
\newcommand{\fbfR}{{\overline{f}}_{_R}}
%
%
\newcommand{\barf}{\overline f}                
\newcommand{\barh}{\overline h}                
\newcommand{\barl}{\overline l}
\newcommand{\barq}{\overline q}
\newcommand{\barqp}{\overline{q}'}
\newcommand{\barb}{\overline b}
\newcommand{\bart}{\overline t}
\newcommand{\barc}{\overline c}
\newcommand{\baru}{\overline u}
\newcommand{\bard}{\overline d}
\newcommand{\bars}{\overline s}
\newcommand{\barv}{\overline v}
\newcommand{\barnu}{\overline{\nu}}
\newcommand{\barne}{\overline{\nu}_{\fe}}
\newcommand{\barnm}{\overline{\nu}_{\flm}}
\newcommand{\barnt}{\overline{\nu}_{\flt}}
%
%
\newcommand{\glu}{g}
%
%
\newcommand{\prot}{p}
\newcommand{\aprot}{{\bar{p}}}
\newcommand{\Nucln}{N}
%
%
\newcommand{\tM}{{\tilde M}}
\newcommand{\tMs}{{\tilde M}^2}
\newcommand{\tW}{{\tilde \Gamma}}
\newcommand{\tWs}{{\tilde\Gamma}^2}
\newcommand{\fphi}{\phi}
\newcommand{\fJpsi}{J/\psi}
\newcommand{\fgpsi}{\psi}
\newcommand{\Glone}{\Gamma}
\newcommand{\Gloni}[1]{\Gamma_{#1}}
\newcommand{\Glones}{\Gamma^2}
\newcommand{\Glonec}{\Gamma^3}
\newcommand{\glone}{\gamma}
\newcommand{\glones}{\gamma^2}
\newcommand{\gloneq}{\gamma^4}
\newcommand{\gloni}[1]{\gamma_{#1}}
\newcommand{\glonis}[1]{\gamma^2_{#1}}
\newcommand{\Grest}[2]{\Gamma_{#1}^{#2}}
\newcommand{\grest}[2]{\gamma_{#1}^{#2}}
\newcommand{\resampl}{A_{_{\rm{R}}}}
\newcommand{\resasyi}[1]{{\cal{A}}_{#1}}
\newcommand{\sSrest}[1]{\sigma_{#1}}
\newcommand{\Srest}[2]{\sigma_{#1}\lpar{#2}\rpar}
\newcommand{\Gdist}[1]{{\cal{G}}\lpar{#1}\rpar}
\newcommand{\sGdist}{{\cal{G}}}
\newcommand{\Aarea}{A_{0}}
\newcommand{\Aareai}[1]{{\cal{A}}\lpar{#1}\rpar}
\newcommand{\sAarea}{{\cal{A}}}
\newcommand{\resolw}{\sigma_{\ssE}}
\newcommand{\resolws}{\sigma^2_{\ssE}}
\newcommand{\chizer}{\chi_{0}}
\newcommand{\ini}{\rm{in}}
\newcommand{\fin}{\rm{fin}}
\newcommand{\ifi}{\rm{if}}
\newcommand{\ipf}{\rm{i+f}}
\newcommand{\tot}{\rm{T}}
\newcommand{\FB}{\rm{FB}}
\newcommand{\Nn}{\rm{N}}
\newcommand{\Bac}{\rm{Q}}
\newcommand{\Res}{\rm{R}}
\newcommand{\Int}{\rm{I}}
\newcommand{\NRe}{\rm{NR}}
\newcommand{\ratoe}{\delta}
\newcommand{\ratoes}{\delta^2}
%
%
\newcommand{\Fbox}[2]{f^{\rm{box}}_{#1}\lpar{#2}\rpar}
\newcommand{\Dbox}[2]{\delta^{\rm{box}}_{#1}\lpar{#2}\rpar}
\newcommand{\Bbox}[3]{{\cal{B}}_{#1}^{#2}\lpar{#3}\rpar}
%
%
\newcommand{\phm}{\lambda}
\newcommand{\phms}{\lambda^2}
\newcommand{\mV}{M_{\ssV}}
\newcommand{\mw}{M_{_W}}
\newcommand{\mX}{M_{_X}}
\newcommand{\mY}{M_{_Y}}
\newcommand{\LM}{M}
\newcommand{\mz}{M_{_Z}}
\newcommand{\bzm}{M_{_0}}
\newcommand{\mh}{M_{_H}}
\newcommand{\bhm}{M_{_{0H}}}
\newcommand{\mf}{m_f}
\newcommand{\mfp}{m_{f'}}
\newcommand{\mfh}{m_{h}}
\newcommand{\mt}{m_t}
\newcommand{\me}{m_e}
\newcommand{\mm}{m_{\mu}}
\newcommand{\mtau}{m_{\tau}}
\newcommand{\muq}{m_u}
\newcommand{\md}{m_d}
\newcommand{\muqp}{m'_u}
\newcommand{\mdqp}{m'_d}
\newcommand{\mc}{m_c}
\newcommand{\ms}{m_s}
\newcommand{\mb}{m_b}
\newcommand{\mup}{M_u}                              
\newcommand{\mdp}{M_d}
\newcommand{\mcp}{M_c}
\newcommand{\msp}{M_s}
\newcommand{\mbp}{M_b}
%
%
\newcommand{\mls}{m^2_l}
\newcommand{\mVs}{M^2_{_V}}
\newcommand{\mws}{M^2_{_W}}
\newcommand{\mwc}{M^3_{_W}}
\newcommand{\LMs}{M^2}
\newcommand{\LMc}{M^3}
\newcommand{\mzs}{M^2_{_Z}}
\newcommand{\mzc}{M^3_{_Z}}
\newcommand{\bzms}{M^2_{_0}}
\newcommand{\bzmc}{M^3_{_0}}
\newcommand{\bhms}{M^2_{_{0H}}}
\newcommand{\mhs}{M^2_{_H}}
\newcommand{\mhc}{M^3_{_H}}
\newcommand{\mhz}{M^6_{_H}}
\newcommand{\mfs}{m^2_f}
\newcommand{\mfc}{m^3_f}
\newcommand{\mfps}{m^2_{f'}}
\newcommand{\mfhs}{m^2_{h}}
\newcommand{\mfpc}{m^3_{f'}}
\newcommand{\mts}{m^2_t}
\newcommand{\mes}{m^2_e}
\newcommand{\mms}{m^2_{\mu}}
\newcommand{\mmc}{m^3_{\mu}}
\newcommand{\mmfour}{m^4_{\mu}}
\newcommand{\mmf}{m^5_{\mu}}
\newcommand{\mmfive}{m^5_{\mu}}
\newcommand{\mmsix}{m^6_{\mu}}
\newcommand{\mtsix}{m^6_{\ft}}
\newcommand{\mminv}{\frac{1}{m_{\mu}}}
\newcommand{\mtaus}{m^2_{\tau}}
\newcommand{\mus}{m^2_u}
\newcommand{\mds}{m^2_d}
\newcommand{\muqps}{m'^2_u}
\newcommand{\mdqps}{m'^2_d}
\newcommand{\mcs}{m^2_c}
\newcommand{\mss}{m^2_s}
\newcommand{\mbs}{m^2_b}
\newcommand{\mups}{M^2_u}
\newcommand{\mdps}{M^2_d}
\newcommand{\mcps}{M^2_c}
\newcommand{\msps}{M^2_s}
\newcommand{\mbps}{M^2_b}
%
%
\newcommand{\muf}{\mu_{\ff}}
\newcommand{\mufs}{\mu^2_{\ff}}
\newcommand{\mufq}{\mu^4_{\ff}}
\newcommand{\mufx}{\mu^6_{\ff}}
\newcommand{\muz}{\mu_{_{\zb}}}
\newcommand{\muw}{\mu_{_{\wb}}}
\newcommand{\mut}{\mu_{\ft}}
\newcommand{\muzs}{\mu^2_{_{\zb}}}
\newcommand{\muws}{\mu^2_{_{\wb}}}
\newcommand{\muts}{\mu^2_{\ft}}
\newcommand{\mubs}{\mu^2_{\ffb}}
\newcommand{\muSW}{\mu^2_{_{\wb}}}
\newcommand{\muwq}{\mu^4_{_{\wb}}}
\newcommand{\muwsx}{\mu^6_{_{\wb}}}
\newcommand{\muwms}{\mu^{-2}_{_{\wb}}}
\newcommand{\muhs}{\mu^2_{_{\hb}}}
\newcommand{\muhq}{\mu^4_{_{\hb}}}
\newcommand{\muhsx}{\mu^6_{_{\hb}}}
\newcommand{\mutq}{\mu^4_{_{\hb}}}   
\newcommand{\mutsx}{\mu^6_{_{\hb}}}  
\newcommand{\muL}{\mu}
\newcommand{\muS}{\mu^2}
\newcommand{\muQ}{\mu^4}
\newcommand{\muizs}{\mu^2_{0}}
\newcommand{\muizq}{\mu^4_{0}}
\newcommand{\muis}{\mu^2_{1}}
\newcommand{\muiis}{\mu^2_{2}}
\newcommand{\muiiis}{\mu^2_{3}}
\newcommand{\muii}[1]{\mu_{#1}}
\newcommand{\muisi}[1]{\mu^2_{#1}}
\newcommand{\muiqi}[1]{\mu^4_{#1}}
\newcommand{\muixi}[1]{\mu^6_{#1}}
\newcommand{\zm}{z_m}
\newcommand{\mwzs}{M^2_{\ssW,\ssZ}}
\newcommand{\ri}[1]{r_{#1}}
\newcommand{\xw}{x_w}
\newcommand{\xws}{x^2_w}
\newcommand{\xwc}{x^3_w}
\newcommand{\xth}{x_t}
\newcommand{\xths}{x^2_t}
\newcommand{\xthc}{x^3_t}
\newcommand{\xthf}{x^4_t}
\newcommand{\xthv}{x^5_t}
\newcommand{\xthx}{x^6_t}
\newcommand{\xh}{x_h}
\newcommand{\xhs}{x^2_h}
\newcommand{\xhc}{x^3_h}
\newcommand{\Rl}{R_{\fl}}
\newcommand{\Rb}{R_{\ffb}}
\newcommand{\Rc}{R_{\fc}}
%
%
\newcommand{\mwq}{M^4_{_\wb}}
\newcommand{\mwf}{M^4_{_\wb}}
\newcommand{\LMq}{M^4}
\newcommand{\mzq}{M^4_{_Z}}
\newcommand{\bzmq}{M^4_{_0}}
\newcommand{\mhq}{M^4_{_H}}
\newcommand{\mfq}{m^4_f}
\newcommand{\mfpq}{m^4_{f'}}
\newcommand{\mtq}{m^4_t}
\newcommand{\meq}{m^4_e}
\newcommand{\mmq}{m^4_{\mu}}
\newcommand{\mtauq}{m^4_{\tau}}
\newcommand{\muqq}{m^4_u}
\newcommand{\mdq}{m^4_d}
\newcommand{\mcq}{m^4_c}
\newcommand{\msq}{m^4_s}
\newcommand{\mbq}{m^4_b}
\newcommand{\mupq}{M^4_u}
\newcommand{\mdpq}{M^4_d}
\newcommand{\mcpq}{M^4_c}
\newcommand{\mspq}{M^4_s}
\newcommand{\mbpq}{M^4_b}
%
%
\newcommand{\mwx}{M^6_{_W}}
\newcommand{\mzx}{M^6_{_Z}}
\newcommand{\mfx}{m^6_f}
\newcommand{\mfpx}{m^6_{f'}}
\newcommand{\LMx}{M^6}
%
%
\newcommand{\mer}{m_{er}}
\newcommand{\mlep}{m_l}
\newcommand{\mleps}{m^2_l}
\newcommand{\mzero}{m_0}
\newcommand{\mone}{m_1}
\newcommand{\mtwo}{m_2}
\newcommand{\mtre}{m_3}
\newcommand{\mfor}{m_4}
\newcommand{\mfiv}{m_5}
\newcommand{\mlone}{m}
\newcommand{\mloneb}{\bar{m}}
\newcommand{\mind}[1]{m_{#1}}
\newcommand{\mones}{m^2_1}
\newcommand{\mtwos}{m^2_2}
\newcommand{\mtres}{m^2_3}
\newcommand{\mfors}{m^2_4}
\newcommand{\mlones}{m^2}
\newcommand{\minds}[1]{m^2_{#1}}
\newcommand{\moneq}{m^4_1}
\newcommand{\mtwoq}{m^4_2}
\newcommand{\mtreq}{m^4_3}
\newcommand{\mforq}{m^4_4}
\newcommand{\mloneq}{m^4}
\newcommand{\mindq}[1]{m^4_{#1}}
\newcommand{\mlonev}{m^5}
\newcommand{\mindv}[1]{m^5_{#1}}
\newcommand{\monex}{m^6_1}
\newcommand{\mtwox}{m^6_2}
\newcommand{\mtrex}{m^6_3}
\newcommand{\mforx}{m^6_4}
\newcommand{\mlonex}{m^6}
\newcommand{\mindx}[1]{m^6_{#1}}
\newcommand{\Mone}{M_1}
\newcommand{\Mtwo}{M_2}
\newcommand{\Mtre}{M_3}
\newcommand{\Mfor}{M_4}
\newcommand{\Mlone}{M}
\newcommand{\Mlonep}{M'}
\newcommand{\Miind}{M_i}
\newcommand{\Mind}[1]{M_{#1}}
\newcommand{\Minds}[1]{M^2_{#1}}
\newcommand{\Mindc}[1]{M^3_{#1}}
\newcommand{\Mindf}[1]{M^4_{#1}}
\newcommand{\Mones}{M^2_1}
\newcommand{\Mtwos}{M^2_2}
\newcommand{\Mtres}{M^2_3}
\newcommand{\Mfors}{M^2_4}
\newcommand{\Mlones}{M^2}
\newcommand{\Mloneps}{M'^2}
\newcommand{\Miinds}{M^2_i}
\newcommand{\Mlonec}{M^3}
\newcommand{\Monec}{M^3_1}
\newcommand{\Mtwoc}{M^3_2}
\newcommand{\Moneq}{M^4_1}
\newcommand{\Mtwoq}{M^4_2}
\newcommand{\Mtreq}{M^4_3}
\newcommand{\Mforq}{M^4_4}
\newcommand{\Mloneq}{M^4}
\newcommand{\Miindq}{M^4_i}
\newcommand{\Monex}{M^6_1}
\newcommand{\Mtwox}{M^6_2}
\newcommand{\Mtrex}{M^6_3}
\newcommand{\Mforx}{M^6_4}
\newcommand{\Mlonex}{M^6}
\newcommand{\Miindx}{M^6_i}
\newcommand{\meb}{m_0}
\newcommand{\mebs}{m^2_0}
%
%
\newcommand{\Mq }{M_q  }
\newcommand{\MqS}{M^2_q}
\newcommand{\Ms }{M_s  }
\newcommand{\MsS}{M^2_s}
\newcommand{\Mc }{M_c  }
\newcommand{\McS}{M^2_c}
\newcommand{\Mb }{M_b  }
\newcommand{\MbS}{M^2_b}
\newcommand{\Mt }{M_t  }
\newcommand{\MtS}{M^2_t}
%
%
\newcommand{\mq}{m_q}
\newcommand{\mqs}{m^2_q}
\newcommand{\mqS}{m^2_q}
\newcommand{\mqQ}{m^4_q}
\newcommand{\mqX}{m^6_q}
\newcommand{\mqp}{m'_q }
\newcommand{\mqpS}{m'^2_q}
\newcommand{\mqpQ}{m'^4_q}
%
%
\newcommand{\lL}{l}
\newcommand{\lLi}[1]{l_{#1}}
\newcommand{\ls}{l^2}
\newcommand{\LL}{L}
\newcommand{\LcalL}{\cal{L}}
\newcommand{\LS}{L^2}
\newcommand{\LC}{L^3}
\newcommand{\LQ}{L^4}
\newcommand{\lw}{l_w}
\newcommand{\Lw}{L_w}
\newcommand{\Lws}{L^2_w}
\newcommand{\Lz}{L_z}
\newcommand{\Lzs}{L^2_z}
\newcommand{\Li}[1]{L_{#1}}
\newcommand{\Lis}[1]{L^2_{#1}}
\newcommand{\Lic}[1]{L^3_{#1}}
%
%
\newcommand{\sman}{s}
\newcommand{\tman}{t}
\newcommand{\uman}{u}
\newcommand{\smani}[1]{s_{#1}}
\newcommand{\bsmani}[1]{{\bar{s}}_{#1}}
\newcommand{\smans}{s^2}
\newcommand{\tmans}{t^2}
\newcommand{\umans}{u^2}
\newcommand{\shat}{{\hat s}}
\newcommand{\that}{{\hat t}}
\newcommand{\uhat}{{\hat u}}
%
%
\newcommand{\smanp}{s'}
\newcommand{\smanpi}[1]{s'_{#1}}
\newcommand{\tmanp}{t'}
\newcommand{\umanp}{u'}
\newcommand{\kappi}[1]{\kappa_{#1}}
\newcommand{\zetai}[1]{\zeta_{#1}}
%
%
%
\newcommand{\Phaspi}[1]{\Gamma_{#1}}
\newcommand{\rbetai}[1]{\beta_{#1}}
\newcommand{\ralphai}[1]{\alpha_{#1}}
\newcommand{\rbetais}[1]{\beta^2_{#1}}
\newcommand{\Lambdi}[1]{\Lambda_{#1}}
\newcommand{\Nomini}[1]{N_{#1}}
\newcommand{\smlone}{\frac{-\sman-\ib\ep}{\mlones}}
%
%
\newcommand{\theti}[1]{\theta_{#1}}
\newcommand{\delti}[1]{\delta_{#1}}
\newcommand{\phigi}[1]{\phi_{#1}}
\newcommand{\acoli}[1]{\xi_{#1}}
\newcommand{\scats}{s}
\newcommand{\scatss}{s^2}
\newcommand{\scatsi}[1]{s_{#1}}
\newcommand{\scatsis}[1]{s^2_{#1}}
\newcommand{\scatst}[2]{s_{#1}^{#2}}
\newcommand{\scatc}{c}
\newcommand{\scatcs}{c^2}
\newcommand{\scatci}[1]{c_{#1}}
\newcommand{\scatcis}[1]{c^2_{#1}}
\newcommand{\scatct}[2]{c_{#1}^{#2}}
\newcommand{\angamt}[2]{\gamma_{#1}^{#2}}
\newcommand{\scatsin}{\sin\theta}
\newcommand{\scatsins}{\sin^2\theta}
\newcommand{\scatcos}{\cos\theta}
\newcommand{\scatcoss}{\cos^2\theta}
%
%
\newcommand{\Regia}{{\cal{R}}}
\newcommand{\Iconi}[2]{{\cal{I}}_{#1}\lpar{#2}\rpar}
\newcommand{\sIcon}[1]{{\cal{I}}_{#1}}
\newcommand{\betaf}{\beta_{\ff}}
\newcommand{\betafs}{\beta^2_{\ff}}
\newcommand{\Kfact}[2]{{\cal{K}}_{#1}\lpar{#2}\rpar}
%
%
\newcommand{\Struf}[4]{{\cal D}^{#1}_{#2}\lpar{#3;#4}\rpar}
\newcommand{\sStruf}[2]{{\cal D}^{#1}_{#2}}
\newcommand{\Fluxf}[2]{H\lpar{#1;#2}\rpar}
\newcommand{\Fluxfi}[4]{H_{#1}^{#2}\lpar{#3;#4}\rpar}
\newcommand{\sFluxf}{H}
\newcommand{\Bflux}[2]{{\cal{B}}_{#1}\lpar{#2}\rpar}
\newcommand{\bflux}[2]{{\cal{B}}_{#1}\lpar{#2}\rpar}
\newcommand{\Fluxd}[2]{D_{#1}\lpar{#2}\rpar}
\newcommand{\fluxd}[2]{C_{#1}\lpar{#2}\rpar}
\newcommand{\Fluxh}[4]{{\cal{H}}_{#1}^{#2}\lpar{#3;#4}\rpar}
\newcommand{\Sluxh}[4]{{\cal{S}}_{#1}^{#2}\lpar{#3;#4}\rpar}
\newcommand{\Fluxhb}[4]{{\overline{{\cal{H}}}}_{#1}^{#2}\lpar{#3;#4}\rpar}
\newcommand{\sFluxhb}{{\overline{{\cal{H}}}}}
\newcommand{\Sluxhb}[4]{{\overline{{\cal{S}}}}_{#1}^{#2}\lpar{#3;#4}\rpar}
\newcommand{\sSluxhb}[2]{{\overline{{\cal{S}}}}_{#1}^{#2}}
\newcommand{\fluxh}[4]{h_{#1}^{#2}\lpar{#3;#4}\rpar}
\newcommand{\fluxhs}[3]{h_{#1}^{#2}\lpar{#3}\rpar}
\newcommand{\sfluxhs}[2]{h_{#1}^{#2}}
\newcommand{\fluxhb}[4]{{\overline{h}}_{#1}^{#2}\lpar{#3;#4}\rpar}
\newcommand{\Strufd}[2]{D\lpar{#1;#2}\rpar}
%
%
\newcommand{\rMQ}[1]{r^2_{#1}}
\newcommand{\rMQs}[1]{r^4_{#1}}
\newcommand{\hf}{h_{\ff}}
\newcommand{\rf}{w_{\ff}}
\newcommand{\zf}{z_{\ff}}
\newcommand{\rfs}{w^2_{\ff}}
\newcommand{\zfs}{z^2_{\ff}}
\newcommand{\rfc}{w^3_{\ff}}
\newcommand{\zfc}{z^3_{\ff}}
\newcommand{\df}{d_{\ff}}
\newcommand{\rfp}{w_{\ffp}}
\newcommand{\rfps}{w^2_{\ffp}}
\newcommand{\rfpc}{w^3_{\ffp}}
\newcommand{\Lrfp}{L_{w}}
\newcommand{\rt}{w_{\ft}}
\newcommand{\rts}{w^2_{\ft}}
\newcommand{\rb}{w_{\ffb}}
\newcommand{\rbs}{w^2_{\ffb}}
\newcommand{\dt}{d_{\ft}}
\newcommand{\dts}{d^2_{\ft}}
\newcommand{\rh}{r_{h}}
\newcommand{\Lnrt}{\ln{\rt}}
\newcommand{\Rw}{R_{_{\wb}}}
\newcommand{\Rws}{R^2_{_{\wb}}}
\newcommand{\Rz}{R_{_{\zb}}}
\newcommand{\Rzp}{R^{+}_{_{\zb}}}
\newcommand{\Rzm}{R^{-}_{_{\zb}}}
\newcommand{\Rzs}{R^2_{_{\zb}}}
\newcommand{\Rzc}{R^3_{_{\zb}}}
\newcommand{\Rv}{R_{_{\vb}}}
\newcommand{\rhw}{w_h}
\newcommand{\rhz}{z_h}
\newcommand{\rhws}{w^2_h}
\newcommand{\rhzs}{z^2_h}
%
%
\newcommand{\vqrato}{z}
\newcommand{\vqrats}{w}
\newcommand{\vqratq}{w^2}
\newcommand{\seyrat}{z}
\newcommand{\sexrat}{w}
\newcommand{\sexrats}{w^2}
\newcommand{\sehrat}{h}
\newcommand{\sewrat}{w}
\newcommand{\sezrat}{z}
\newcommand{\zetav}{\zeta}
\newcommand{\zetavi}[1]{\zeta_{#1}}
\newcommand{\bpo}{\beta^2}
\newcommand{\bpos}{\beta^4}
\newcommand{\bpt}{{\tilde\beta}^2}
\newcommand{\lap}{\kappa}
\newcommand{\hw}{w_h}
\newcommand{\hz}{z_h}
%
%
\newcommand{\ec}{e}
\newcommand{\ecs}{e^2}
\newcommand{\ect}{e^3}
\newcommand{\ecq}{e^4}
\newcommand{\ecb}{e_{_0}}
\newcommand{\ecbs}{e^2_{_0}}
\newcommand{\ecbq}{e^4_{_0}}
\newcommand{\eci}[1]{e_{#1}}
\newcommand{\ecis}[1]{e^2_{#1}}
\newcommand{\hate}{{\hat e}}
\newcommand{\gss}{g_{_S}}
\newcommand{\gsss}{g^2_{_S}}
\newcommand{\gssb}{g^2_{_{S_0}}}
\newcommand{\als}{\alpha_{_S}}
\newcommand{\as}{a_{_S}}
\newcommand{\ass}{a^2_{_S}}
\newcommand{\gf}{G_{\ssF}}
\newcommand{\gfs}{G^2_{\ssF}}
\newcommand{\gb}{g} 
\newcommand{\gbi}[1]{g_{#1}}
\newcommand{\gbb}{g_{0}}
\newcommand{\gbs}{g^2}
\newcommand{\gbc}{g^3}
\newcommand{\gbf}{g^4}
\newcommand{\gpb}{g'}
\newcommand{\gpbs}{g'^2}
\newcommand{\vc}[1]{v_{#1}}
\newcommand{\ac}[1]{a_{#1}}
\newcommand{\vcc}[1]{v^*_{#1}}
\newcommand{\acc}[1]{a^*_{#1}}
\newcommand{\hatv}[1]{{\hat v}_{#1}}
\newcommand{\vcs}[1]{v^2_{#1}}
\newcommand{\acs}[1]{a^2_{#1}}
\newcommand{\gcv}[1]{g^{#1}_{\ssV}}
\newcommand{\gca}[1]{g^{#1}_{\ssA}}
\newcommand{\gcp}[1]{g^{+}_{#1}}
\newcommand{\gcm}[1]{g^{-}_{#1}}
\newcommand{\gcpm}[1]{g^{\pm}_{#1}}
\newcommand{\vci}[2]{v^{#2}_{#1}}
\newcommand{\aci}[2]{a^{#2}_{#1}}
\newcommand{\vceff}[1]{v^{#1}_{\rm{eff}}}
\newcommand{\hvc}[1]{\hat{v}_{#1}}
\newcommand{\hvcs}[1]{\hat{v}^2_{#1}}
\newcommand{\Vc}[1]{V_{#1}}
\newcommand{\Ac}[1]{A_{#1}}
\newcommand{\Vcs}[1]{V^2_{#1}}
\newcommand{\Acs}[1]{A^2_{#1}}
\newcommand{\vpa}[2]{\sigma_{#1}^{#2}}
\newcommand{\vma}[2]{\delta_{#1}^{#2}}
\newcommand{\vfw}{\sigma^{a}_{\ff}}
\newcommand{\vfpw}{\sigma^{a}_{\ffp}}
\newcommand{\vfwi}[1]{\sigma^{a}_{#1}}
\newcommand{\vfwsi}[1]{\lpar\sigma^{a}_{#1}\rpar^2}
\newcommand{\vvfw}{v^{a}_{\ff}}
\newcommand{\vvew}{v^{a}_{\fe}}
\newcommand{\vzm}{v^{-}_{\ssZ}}
\newcommand{\vzp}{v^{+}_{\ssZ}}
\newcommand{\vzpm}{v^{\pm}_{\ssZ}}
\newcommand{\vzmp}{v^{\mp}_{\ssZ}}
\newcommand{\vam}{v^{-}_{\ssA}}
\newcommand{\vap}{v^{+}_{\ssA}}
\newcommand{\vapm}{v^{\pm}_{\ssA}}
\newcommand{\gv}{g_{_V}}
\newcommand{\ga}{g_{_A}}
\newcommand{\gve}{g^{\fe}_{_{V}}}
\newcommand{\gae}{g^{\fe}_{_{A}}}
\newcommand{\gvf}{g^{\ff}_{_{V}}}
\newcommand{\gaf}{g^{\ff}_{_{A}}}
\newcommand{\gva}{g_{_{V,A}}}
\newcommand{\gvae}{g^{\fe}_{_{V,A}}}
\newcommand{\gvaf}{g^{\ff}_{_{V,A}}}
\newcommand{\sGv}{{\cal{G}}_{_V}}
\newcommand{\cGa}{{\cal{G}}^{*}_{_A}}
\newcommand{\cGv}{{\cal{G}}^{*}_{_V}}
\newcommand{\sGa}{{\cal{G}}_{_A}}
\newcommand{\Gvf}{{\cal{G}}^{\ff}_{_{V}}}
\newcommand{\Gaf}{{\cal{G}}^{\ff}_{_{A}}}
\newcommand{\Gvaf}{{\cal{G}}^{\ff}_{_{V,A}}}
\newcommand{\Gve}{{\cal{G}}^{\fe}_{_{V}}}
\newcommand{\Gae}{{\cal{G}}^{\fe}_{_{A}}}
\newcommand{\Gvae}{{\cal{G}}^{\fe}_{_{V,A}}}
\newcommand{\gvl}{g^{\fl}_{_{V}}}
\newcommand{\gal}{g^{\fl}_{_{A}}}
\newcommand{\gval}{g^{\fl}_{_{V,A}}}
\newcommand{\gvb}{g^{\ffb}_{_{V}}}
\newcommand{\gab}{g^{\ffb}_{_{A}}}
\newcommand{\fvf}{F_{_V}^{\ff}}
\newcommand{\faf}{F_{_A}^{\ff}}
\newcommand{\fvl}{F_{_V}^{\fl}}
\newcommand{\fal}{F_{_A}^{\fl}}
\newcommand{\corat}{\kappa}
\newcommand{\corats}{\kappa^2}
%
%
\newcommand{\dr}{\Delta r}
\newcommand{\drl}{\Delta r_{_L}}
\newcommand{\drh}{\Delta{\hat r}}
\newcommand{\drhw}{\Delta{\hat r}_{_W}}
\newcommand{\rhou}{\rho_{_U}}
\newcommand{\rhoz}{\rho_{_\zb}}
\newcommand{\rZ}{\rho_{_\zb}}
\newcommand{\rhob}{\rho_{_0}}
\newcommand{\rZf}{\rho^{\ff}_{_\zb}}
\newcommand{\rhoe}{\rho_{\fe}}
\newcommand{\rhof}{\rho_{\ff}}
\newcommand{\rhoi}[1]{\rho_{#1}}
\newcommand{\kZf}{\kappa^{\ff}_{_\zb}}
\newcommand{\rWf}{\rho^{\ff}_{_\wb}}
\newcommand{\brWf}{{\bar{\rho}}^{\ff}_{_\wb}}
\newcommand{\rHf}{\rho^{\ff}_{_\hb}}
\newcommand{\brHf}{{\bar{\rho}}^{\ff}_{_\hb}}
\newcommand{\rhoR}{\rho^{\ssR}_{_{\zb}}}
\newcommand{\hatrh}{{\hat\rho}}
\newcommand{\ku}{\kappa_{\ssU}}
\newcommand{\rZdf}[1]{\rho^{#1}_{_\zb}}
\newcommand{\kZdf}[1]{\kappa^{#1}_{_\zb}}
\newcommand{\rdfL}[1]{\rho^{#1}_{_L}}
\newcommand{\kdfL}[1]{\kappa^{#1}_{_L}}
\newcommand{\rdfR}[1]{\rho^{#1}_{\rm{rem}}}
\newcommand{\kdfR}[1]{\kappa^{#1}_{\rm{rem}}}
\newcommand{\bark}{\overline\kappa}
%
%
\newcommand{\stw}{s_{\theta}}             
\newcommand{\ctw}{c_{\theta}}
\newcommand{\stws}{s_{\theta}^2}
\newcommand{\stwc}{s_{\theta}^3}
\newcommand{\stwf}{s_{\theta}^4}
\newcommand{\stwq}{s_{\theta}^4}
\newcommand{\stwx}{s_{\theta}^6}
\newcommand{\ctws}{c_{\theta}^2}
\newcommand{\ctwc}{c_{\theta}^3}
\newcommand{\ctwf}{c_{\theta}^4}
\newcommand{\ctwq}{c_{\theta}^4}
\newcommand{\ctwx}{c_{\theta}^6}
\newcommand{\ctwvi}{c_{\theta}^6}
\newcommand{\stwfiv}{s_{\theta}^5}
\newcommand{\ctwfiv}{c_{\theta}^5}
\newcommand{\stwsix}{s_{\theta}^6}
\newcommand{\ctwsix}{c_{\theta}^6}
%
%
\newcommand{\siw}{s_{_W}}           
\newcommand{\cow}{c_{_W}}
\newcommand{\siws}{s^2_{_W}}
\newcommand{\cows}{c^2_{_W}}
\newcommand{\siwc}{s^3_{_W}}
\newcommand{\cowc}{c^3_{_W}}
\newcommand{\siwf}{s^4_{_W}}
\newcommand{\cowf}{c^4_{_W}}
\newcommand{\siwx}{s^6_{_W}}
\newcommand{\cowx}{c^6_{_W}}
\newcommand{\sons}{s_{_W}}
\newcommand{\sonss}{s^2_{_W}}
\newcommand{\cons}{c_{_W}}
\newcommand{\cooss}{c^2_{_W}}
%
%
\newcommand{\szs}{{\overline s}^2}
\newcommand{\szq}{{\overline s}^4}
\newcommand{\czs}{{\overline c}^2}
\newcommand{\sbs}{s_{_0}^2}
\newcommand{\cbs}{c_{_0}^2}
\newcommand{\dss}{\Delta s^2}
\newcommand{\snes}{s_{\nu e}^2}
\newcommand{\cnes}{c_{\nu e}^2}
\newcommand{\shs}{{\hat s}^2}
\newcommand{\chs}{{\hat c}^2}
\newcommand{\chl}{{\hat c}}
\newcommand{\seffs}{s^2_{\rm{eff}}}
\newcommand{\seffsf}[1]{\sin^2\theta^{#1}_{\rm{eff}}}
\newcommand{\sress}{s^2_{\rm res}}                
\newcommand{\sR}{s_{_R}}
\newcommand{\sRs}{s^2_{_R}}
\newcommand{\ctwe}{c_{\theta}^6}
\newcommand{\sany}{s}
\newcommand{\cany}{c}
\newcommand{\sanys}{s^2}
\newcommand{\canys}{c^2}
%
%
\newcommand{\sip}{u}                             
\newcommand{\siap}{{\bar{v}}}                    
\newcommand{\sop}{{\bar{u}}}                     
\newcommand{\soap}{v}                            
\newcommand{\ip}[1]{u\lpar{#1}\rpar}             
\newcommand{\iap}[1]{{\bar{v}}\lpar{#1}\rpar}    
\newcommand{\op}[1]{{\bar{u}}\lpar{#1}\rpar}     
\newcommand{\oap}[1]{v\lpar{#1}\rpar}            
%
%
\newcommand{\ipp}[2]{u\lpar{#1,#2}\rpar}         
\newcommand{\ipap}[2]{{\bar v}\lpar{#1,#2}\rpar} 
\newcommand{\opp}[2]{{\bar u}\lpar{#1,#2}\rpar}  
\newcommand{\opap}[2]{v\lpar{#1,#2}\rpar}        
\newcommand{\upspi}[1]{u\lpar{#1}\rpar}
\newcommand{\vpspi}[1]{v\lpar{#1}\rpar}
\newcommand{\wpspi}[1]{w\lpar{#1}\rpar}
\newcommand{\ubpspi}[1]{{\bar{u}}\lpar{#1}\rpar}
\newcommand{\vbpspi}[1]{{\bar{v}}\lpar{#1}\rpar}
\newcommand{\wbpspi}[1]{{\bar{w}}\lpar{#1}\rpar}
\newcommand{\udpspi}[1]{u^{\dagger}\lpar{#1}\rpar}
\newcommand{\vdpspi}[1]{v^{\dagger}\lpar{#1}\rpar}
\newcommand{\wdpspi}[1]{w^{\dagger}\lpar{#1}\rpar}
\newcommand{\Ubilin}[1]{U\lpar{#1}\rpar}
\newcommand{\Vbilin}[1]{V\lpar{#1}\rpar}
\newcommand{\Xbilin}[1]{X\lpar{#1}\rpar}
\newcommand{\Ybilin}[1]{Y\lpar{#1}\rpar}
\newcommand{\up}[2]{u_{#1}\lpar #2\rpar}
\newcommand{\vp}[2]{v_{#1}\lpar #2\rpar}
\newcommand{\ubp}[2]{{\overline u}_{#1}\lpar #2\rpar}
\newcommand{\vbp}[2]{{\overline v}_{#1}\lpar #2\rpar}
\newcommand{\Pje}[1]{\frac{1}{2}\lpar 1 + #1\,\gfd\rpar}
\newcommand{\Pj}[1]{\Pi_{#1}}
\newcommand{\trace}{\mbox{Tr}}
%
%
\newcommand{\Poper}[2]{P_{#1}\lpar{#2}\rpar}
\newcommand{\Loper}[2]{\Lambda_{#1}\lpar{#2}\rpar}
\newcommand{\proj}[3]{P_{#1}\lpar{#2,#3}\rpar}
\newcommand{\sproj}[1]{P_{#1}}
\newcommand{\Nden}[3]{N_{#1}^{#2}\lpar{#3}\rpar}
\newcommand{\sNden}[1]{N_{#1}}
\newcommand{\nden}[2]{n_{#1}^{#2}}
%
%
\newcommand{\vwf}[2]{e_{#1}\lpar#2\rpar}             
\newcommand{\vwfb}[2]{{\overline e}_{#1}\lpar#2\rpar}
\newcommand{\pwf}[2]{\epsilon_{#1}\lpar#2\rpar}      
\newcommand{\sla}[1]{/\!\!\!#1}
\newcommand{\slac}[1]{/\!\!\!\!#1}
%
%
\newcommand{\iemom}{p_{_-}}                    
\newcommand{\ipmom}{p_{_+}}
\newcommand{\oemom}{q_{_-}}                    
\newcommand{\opmom}{q_{_+}}
%
%
\newcommand{\spro}[2]{{#1}\cdot{#2}}
%
%
\newcommand{\gfour}{\gamma_4}                    
\newcommand{\gfd}{\gamma_5}                    
\newcommand{\gap}{\lpar 1+\gamma_5\rpar}
\newcommand{\gam}{\lpar 1-\gamma_5\rpar}
\newcommand{\gdp}{\gamma_+}
\newcommand{\gdm}{\gamma_-}
\newcommand{\gdpm}{\gamma_{\pm}}
\newcommand{\gad}{\gamma}
\newcommand{\gapm}{\lpar 1\pm\gamma_5\rpar}
\newcommand{\gadi}[1]{\gamma_{#1}}
\newcommand{\gadu}[1]{\gamma_{#1}}
\newcommand{\gapu}[1]{\gamma^{#1}}
\newcommand{\sigd}[2]{\sigma_{#1#2}}
\newcommand{\sumsp}{\overline{\sum_{\mbox{\tiny{spins}}}}}
%
%
\newcommand{\li}[2]{\mathrm{Li}_{#1}\lpar\displaystyle{#2}\rpar} 
\newcommand{\sli}[1]{\mathrm{Li}_{#1}} 
\newcommand{\etaf}[2]{\eta\lpar#1,#2\rpar}
\newcommand{\lkall}[3]{\lambda\lpar#1,#2,#3\rpar}       
\newcommand{\slkall}[3]{\lambda^{1/2}\lpar#1,#2,#3\rpar}
\newcommand{\segam}{\Gamma}                             
\newcommand{\egam}[1]{\Gamma\lpar#1\rpar}               
\newcommand{\ebe}[2]{B\lpar#1,#2\rpar}                  
\newcommand{\ddel}[1]{\delta\lpar#1\rpar}               
\newcommand{\drii}[2]{\delta_{#1#2}}                    
\newcommand{\driv}[4]{\delta_{#1#2#3#4}}                
\newcommand{\intmomi}[2]{\int\,d^{#1}#2}
\newcommand{\intmomii}[3]{\int\,d^{#1}#2\,\int\,d^{#1}#3}
\newcommand{\intfx}[1]{\int_{\scriptstyle 0}^{\scriptstyle 1}\,d#1}
\newcommand{\intfxy}[2]{\int_{\scriptstyle 0}^{\scriptstyle 1}\,d#1\,
                        \int_{\scriptstyle 0}^{\scriptstyle #1}\,d#2}
\newcommand{\intfxyz}[3]{\int_{\scriptstyle 0}^{\scriptstyle 1}\,d#1\,
                         \int_{\scriptstyle 0}^{\scriptstyle #1}\,d#2\,
                         \int_{\scriptstyle 0}^{\scriptstyle #2}\,d#3}
\newcommand{\intfxyzz}[4]{\int_{\scriptstyle 0}^{\scriptstyle 1}\,d#1\,
                         \int_{\scriptstyle 0}^{\scriptstyle #1}\,d#2\,
                         \int_{\scriptstyle 0}^{\scriptstyle #2}\,d#3\,
                         \int_{\scriptstyle 0}^{\scriptstyle #3}\,d#4}
\newcommand{\Beta}[2]{{\rm{B}}\lpar #1,#2\rpar}
\newcommand{\sBeta}{\rm{B}}
\newcommand{\sign}[1]{{\rm{sign}}\lpar{#1}\rpar}
\newcommand{\ointfxy}[2]{\int_{\scriptstyle 0}^{\scriptstyle 1}\,d#1\,d#2}
%
%
\newcommand{\gn}{\Gamma_{\nu}}
\newcommand{\gel}{\Gamma_{\fe}}
\newcommand{\gmu}{\Gamma_{\mu}}
\newcommand{\gff}{\Gamma_{\ff}}
\newcommand{\gt}{\Gamma_{\tau}}
\newcommand{\gl}{\Gamma_{\fl}}
\newcommand{\gq}{\Gamma_{\fq}}
\newcommand{\gu}{\Gamma_{\fu}}
\newcommand{\gd}{\Gamma_{\fd}}
\newcommand{\gc}{\Gamma_{\fc}}
\newcommand{\gs}{\Gamma_{\fs}}
\newcommand{\gbq}{\Gamma_{\ffb}}
\newcommand{\gz}{\Gamma_{_{\zb}}}
\newcommand{\gw}{\Gamma_{_{\wb}}}
\newcommand{\gh}{\Gamma_{_{h}}}
\newcommand{\ghb}{\Gamma_{_{\hb}}}
\newcommand{\gi}{\Gamma_{\rm{inv}}}
\newcommand{\gzs}{\Gamma^2_{_{\zb}}}
%
%
\newcommand{\tcie}{I^{(3)}_{\fe}}
\newcommand{\tcim}{I^{(3)}_{\flm}}
\newcommand{\tcif}{I^{(3)}_{\ff}}
\newcommand{\tciq}{I^{(3)}_{\fq}}
\newcommand{\tcib}{I^{(3)}_{\ffb}}
\newcommand{\tcih}{I^{(3)}_h}
\newcommand{\tcii}{I^{(3)}_i}
\newcommand{\tcift}{I^{(3)}_{\tilde f}}
\newcommand{\tcifp}{I^{(3)}_{f'}}
\newcommand{\wispt}[1]{I^{(3)}_{#1}}
\newcommand{\ql}{Q_l}
\newcommand{\qe}{Q_e}
\newcommand{\qu}{Q_u}
\newcommand{\qd}{Q_d}
\newcommand{\qb}{Q_b}
\newcommand{\qt}{Q_t}
\newcommand{\qup}{Q'_u}
\newcommand{\qdp}{Q'_d}
\newcommand{\qmu}{Q_{\mu}}
\newcommand{\qes}{Q^2_e}
\newcommand{\qec}{Q^3_e}
\newcommand{\qus}{Q^2_u}
\newcommand{\qds}{Q^2_d}
\newcommand{\qbs}{Q^2_b}
\newcommand{\qts}{Q^2_t}
\newcommand{\qbc}{Q^3_b}
\newcommand{\qf}{Q_f}
\newcommand{\qfs}{Q^2_f}
\newcommand{\qfc}{Q^3_f}
\newcommand{\qff}{Q^4_f}
\newcommand{\qep}{Q_{e'}}
\newcommand{\qfp}{Q_{f'}}
\newcommand{\qfps}{Q^2_{f'}}
\newcommand{\qfpc}{Q^3_{f'}}
\newcommand{\qq}{Q_q}
\newcommand{\qqs}{Q^2_q}
\newcommand{\qi}{Q_i}
\newcommand{\qis}{Q^2_i}
\newcommand{\qj}{Q_j}
\newcommand{\qjs}{Q^2_j}
\newcommand{\QW}{Q_{_\wb}}
\newcommand{\QWs}{Q^2_{_\wb}}
\newcommand{\Qd}{Q_d}
\newcommand{\Qds}{Q^2_d}
\newcommand{\Qu}{Q_u}
\newcommand{\Qus}{Q^2_u}
\newcommand{\vi}{v_i}
\newcommand{\vis}{v^2_i}
\newcommand{\ai}{a_i}
\newcommand{\ais}{a^2_i}
%
%
\newcommand{\piv}{\Pi_{_V}}
\newcommand{\pia}{\Pi_{_A}}
\newcommand{\piva}{\Pi_{_{V,A}}}
\newcommand{\pivi}[1]{\Pi^{({#1})}_{_V}}
\newcommand{\piai}[1]{\Pi^{({#1})}_{_A}}
\newcommand{\pivai}[1]{\Pi^{({#1})}_{_{V,A}}}
\newcommand{\pih}{{\hat\Pi}}
\newcommand{\sgh}{{\hat\Sigma}}
\newcommand{\Pgg}{\Pi_{\ph\ph}}
\newcommand{\Ptg}{\Pi_{_{3Q}}}
\newcommand{\Ptt}{\Pi_{_{33}}}
\newcommand{\Pzg}{\Pi_{_{\zb\ab}}}
\newcommand{\Pzga}[2]{\Pi^{#1}_{_{\zb\ab}}\lpar#2\rpar}
\newcommand{\Pf}{\Pi_{_F}}
\newcommand{\Sgg}{\Sigma_{_{\ab\ab}}}
\newcommand{\Szg}{\Sigma_{_{\zb\ab}}}
\newcommand{\SVV}{\Sigma_{_{\vb\vb}}}
\newcommand{\USvv}{{\hat\Sigma}_{_{\vb\vb}}}
\newcommand{\Sww}{\Sigma_{_{\wb\wb}}}
\newcommand{\Swwg}{\Sigma^{_G}_{_{\wb\wb}}}
\newcommand{\Szz}{\Sigma_{_{\zb\zb}}}
\newcommand{\Shh}{\Sigma_{_{\hb\hb}}}
\newcommand{\Spzz}{\Sigma'_{_{\zb\zb}}}
\newcommand{\Stg}{\Sigma_{_{3Q}}}
\newcommand{\Stt}{\Sigma_{_{33}}}
\newcommand{\bSww}{{\overline\Sigma}_{_{WW}}}
\newcommand{\bStg}{{\overline\Sigma}_{_{3Q}}}
\newcommand{\bStt}{{\overline\Sigma}_{_{33}}}
\newcommand{\Sssn}{\Sigma_{_{\hkn\hkn}}}
\newcommand{\Sssc}{\Sigma_{_{\phi\phi}}}
\newcommand{\Szn}{\Sigma_{_{\zb\hkn}}}
\newcommand{\Swc}{\Sigma_{_{\wb\hkg}}}
\newcommand{\mix}[2]{{\cal{M}}^{#1}\lpar{#2}\rpar}
\newcommand{\bmix}[2]{\Pi^{{#1},\ssF}_{_{\zb\ab}}\lpar{#2}\rpar}
\newcommand{\hPgg}[2]{{\hat{\Pi}^{{#1},\ssF}}_{\ph\ph}\lpar{#2}\rpar}
\newcommand{\hmix}[2]{{\hat{\Pi}^{{#1},\ssF}}_{_{\zb\ab}}\lpar{#2}\rpar}
\newcommand{\Dz}[2]{{\cal{D}}_{_{\zb}}^{#1}\lpar{#2}\rpar}
\newcommand{\bDz}[2]{{\cal{D}}^{{#1},\ssF}_{_{\zb}}\lpar{#2}\rpar}
\newcommand{\hDz}[2]{{\hat{\cal{D}}}^{{#1},\ssF}_{_{\zb}}\lpar{#2}\rpar}
\newcommand{\Szzd}[2]{\Sigma'^{#1}_{_{\zb\zb}}\lpar{#2}\rpar}
\newcommand{\Swwd}[2]{\Sigma'^{#1}_{_{\wb\wb}}\lpar{#2}\rpar}
\newcommand{\Shhd}[2]{\Sigma'^{#1}_{_{\hb\hb}}\lpar{#2}\rpar}
\newcommand{\ZFren}[2]{{\cal{Z}}^{#1}\lpar{#2}\rpar}
\newcommand{\WFren}[2]{{\cal{W}}^{#1}\lpar{#2}\rpar}
\newcommand{\HFren}[2]{{\cal{H}}^{#1}\lpar{#2}\rpar}
\newcommand{\WI}{\cal{W}}
%
%
\newcommand{\cf}{c_f}
\newcommand{\Cf}{C_{_F}}
\newcommand{\Nf}{N_f}
\newcommand{\Nc}{N_c}
\newcommand{\Ncs}{N^2_c}
\newcommand{\nf }{n_f}
\newcommand{\nfs}{n^2_f}
\newcommand{\nfc}{n^3_f}
\newcommand{\MSB}{\overline{MS}}
\newcommand{\NMSB}{\overline{NMS}}
\newcommand{\LMSB}{\Lambda_{\overline{\mathrm{MS}}}}
\newcommand{\LMSBp}{\Lambda'_{\overline{\mathrm{MS}}}}
\newcommand{\LMSBS}{\Lambda^2_{\overline{\mathrm{MS}}}}
\newcommand{\LMSBv }{\mbox{$\Lambda^{(5)}_{\overline{\mathrm{MS}}}$}}
\newcommand{\LMSBvS}{\mbox{$\left(\Lambda^{(5)}_{\overline{\mathrm{MS}}}\right)^2$}}
\newcommand{\LMSBt }{\mbox{$\Lambda^{(3)}_{\overline{\mathrm{MS}}}$}}
\newcommand{\LMSBtS}{\mbox{$\left(\Lambda^{(3)}_{\overline{\mathrm{MS}}}\right)^2$}}
\newcommand{\LMSBf }{\mbox{$\Lambda^{(4)}_{\overline{\mathrm{MS}}}$}}
\newcommand{\LMSBfS}{\mbox{$\left(\Lambda^{(4)}_{\overline{\mathrm{MS}}}\right)^2$}}
\newcommand{\LMSBn }{\mbox{$\Lambda^{(\nf)}_{\overline{\mathrm{MS}}}$}}
\newcommand{\LMSBnS}{\mbox{$\left(\Lambda^{(\nf)}_{\overline{\mathrm{MS}}}\right)^2$}}
\newcommand{\LMSBnml }{\mbox{$\Lambda^{(\nf-1)}_{\overline{\mathrm{MS}}}$}}
\newcommand{\LMSBnmlS}{\mbox{$\left(\Lambda^{(\nf-1)}_{\overline{\mathrm{MS}}}\right)^2$}}
\newcommand{\Bnf}{\lpar\nf \rpar}
\newcommand{\Bnfm}{\lpar\nf-1 \rpar}
\newcommand{\LuM}{L_{_M}}
\newcommand{\bef}{\beta_{\ff}}
\newcommand{\befs}{\beta^2_{\ff}}
\newcommand{\befc}{\beta^3_{f}}
\newcommand{\alsp}{\alpha'_{_S}}
\newcommand{\api}{\displaystyle \frac{\als(s)}{\pi}}
\newcommand{\alss}{\alpha^2_{_S}}
\newcommand{\ztwo}{\zeta(2)}
\newcommand{\ztri}{\zeta(3)}
\newcommand{\zfor}{\zeta(4)}
\newcommand{\zfiv}{\zeta(5)}
\newcommand{\bi}[1]{b_{#1}}
\newcommand{\ci}[1]{c_{#1}}
\newcommand{\Ci}[1]{C_{#1}}
\newcommand{\bip}[1]{b'_{#1}}
\newcommand{\cip}[1]{c'_{#1}}
%
%
\newcommand{\osps}{16\,\pi^2}
\newcommand{\srt}{\sqrt{2}}
\newcommand{\ospsi}{\displaystyle{\frac{i}{16\,\pi^2}}}
%
%
\newcommand{\tfpromu}{\mbox{$e^+e^-\to \mu^+\mu^-$}}
\newcommand{\tfprotau}{\mbox{$e^+e^-\to \tau^+\tau^-$}}
\newcommand{\tfproe}{\mbox{$e^+e^-\to e^+e^-$}}
\newcommand{\tfpronu}{\mbox{$e^+e^-\to \barnu\nu$}}
\newcommand{\tfproqq}{\mbox{$e^+e^-\to \barq q$}}
\newcommand{\tfprohad}{\mbox{$e^+e^-\to\,$} hadrons}
%
%
\newcommand{\bpromu}{\mbox{$e^+e^-\to \mu^+\mu^-\ph$}}
\newcommand{\bprotau}{\mbox{$e^+e^-\to \tau^+\tau^-\ph$}}
\newcommand{\bproe}{\mbox{$e^+e^-\to e^+e^-\ph$}}
\newcommand{\bpronu}{\mbox{$e^+e^-\to \barnu\nu\ph$}}
\newcommand{\bproqq}{\mbox{$e^+e^-\to \barq q \ph$}}
%
%
\newcommand{\tbprow} {\mbox{$e^+e^-\to \wbp \wbm $}}
\newcommand{\tbproz} {\mbox{$e^+e^-\to \zb  \zb  $}}
\newcommand{\tbproh} {\mbox{$e^+e^-\to \zb  \hb  $}}
\newcommand{\tbprozg}{\mbox{$e^+e^-\to \zb  \ph  $}}
\newcommand{\tbprog} {\mbox{$e^+e^-\to \ph  \ph  $}}
%
%
\newcommand{\Fermionline}[1]{
\vcenter{\hbox{
  \begin{picture}(60,20)(0,{#1})
  \SetScale{2.}
    \ArrowLine(0,5)(30,5)
  \end{picture}}}
}
\newcommand{\AntiFermionline}[1]{
\vcenter{\hbox{
  \begin{picture}(60,20)(0,{#1})
  \SetScale{2.}
    \ArrowLine(30,5)(0,5)
  \end{picture}}}
}
\newcommand{\Photonline}[1]{
\vcenter{\hbox{
  \begin{picture}(60,20)(0,{#1})
  \SetScale{2.}
    \Photon(0,5)(30,5){2}{6.5}
  \end{picture}}}
}
\newcommand{\Gluonline}[1]{
\vcenter{\hbox{
  \begin{picture}(60,20)(0,{#1})
  \SetScale{2.}
    \Gluon(0,5)(30,5){2}{6.5}
  \end{picture}}}
}
\newcommand{\Wbosline}[1]{
\vcenter{\hbox{
  \begin{picture}(60,20)(0,{#1})
  \SetScale{2.}
    \Photon(0,5)(30,5){2}{4}
    \ArrowLine(13.3,3.1)(16.9,7.2)
  \end{picture}}}
}
\newcommand{\Zbosline}[1]{
\vcenter{\hbox{
  \begin{picture}(60,20)(0,{#1})
  \SetScale{2.}
    \Photon(0,5)(30,5){2}{4}
  \end{picture}}}
}
\newcommand{\Philine}[1]{
\vcenter{\hbox{
  \begin{picture}(60,20)(0,{#1})
  \SetScale{2.}
    \DashLine(0,5)(30,5){2}
  \end{picture}}}
}
\newcommand{\Phicline}[1]{
\vcenter{\hbox{
  \begin{picture}(60,20)(0,{#1})
  \SetScale{2.}
    \DashLine(0,5)(30,5){2}
    \ArrowLine(14,5)(16,5)
  \end{picture}}}
}
\newcommand{\Ghostline}[1]{
\vcenter{\hbox{
  \begin{picture}(60,20)(0,{#1})
  \SetScale{2.}
    \DashLine(0,5)(30,5){.5}
    \ArrowLine(14,5)(16,5)
  \end{picture}}}
}
%
%
\newcommand{\gauge}{g}
\newcommand{\gpar}{\xi}
\newcommand{\gpari}[1]{\gpar_{#1}}
\newcommand{\gparis}[1]{\gpar^2_{#1}}
\newcommand{\gpariq}[1]{\gpar^4_{#1}}
\newcommand{\gpars}{\xi^2}
\newcommand{\dgpar}{\Delta\gpar}
\newcommand{\dgparA}{\Delta\gparA}
\newcommand{\dgparZ}{\Delta\gparZ}
\newcommand{\gparq}{\xi^4}
\newcommand{\gparAs}{\xi^2_{_A}}
\newcommand{\gparAq}{\xi^4_{_A}}
\newcommand{\gparZs}{\xi^2_{_Z}}
\newcommand{\gparZq}{\xi^4_{_Z}}
\newcommand{\Rxi}{R_{\gpar}}
\newcommand{\UG}{U}
\newcommand{\UGi}{\ssU}
\newcommand{\hxi}{\chi}
%
%
\newcommand{\LSM}{{\cal{L}}_{_{\rm{SM}}}}
\newcommand{\LSMr}{{\cal{L}}^{\rm{\ssR}}_{_{\rm{SM}}}}
\newcommand{\LYM}{{\cal{L}}_{_{\rm{YM}}}}
\newcommand{\Lzer}{{\cal{L}}_{0}}
\newcommand{\Lone}{{\cal{L}}^{{\bos},{\rm{I}}}}
\newcommand{\Lpro}{{\cal{L}}_{\rm{prop}}}
\newcommand{\Ls  }{{\cal{L}}_{_{\rm{S}}}}
\newcommand{\Lsi }{{\cal{L}}^{\rm{I}}_{_{\rm{S}}}}
\newcommand{\Lgf }{{\cal{L}}_{\rm{gf}}}
\newcommand{\Lgfi}{{\cal{L}}^{\rm{I}}_{\rm{gf}}}
\newcommand{\Lf  }{{\cal{L}}^{{\fer},{\rm{I}}}_{\ssV}}
\newcommand{\LHf }{{\cal{L}}^{\fer}_{\ssS}}
\newcommand{\LHfi}{{\cal{L}}^{{\fer},{\rm{I}}}_{\ssS}}
\newcommand{\Lren}{{\cal{L}}_{\rm{\ssR}}}
\newcommand{\Lct}{{\cal{L}}_{\rm{ct}}}
\newcommand{\Lcti}[1]{{\cal{L}}^{#1}_{\rm{ct}}}
\newcommand{\LctI}{{\cal{L}}^{(2)}_{\rm{ct}}}
\newcommand{\Llone}{{\cal{L}}}
\newcommand{\LQED}{{\cal{L}}_{_{\rm{QED}}}}
\newcommand{\LQEDr}{{\cal{L}}^{\rm{\ssR}}_{_{\rm{QED}}}}
\newcommand{\Greenf}{G}
\newcommand{\Greenfa}[1]{G\lpar{#1}\rpar}
\newcommand{\Greenft}[2]{G\lpar{#1,#2}\rpar}
\newcommand{\FST}[3]{F_{#1#2}^{#3}}
\newcommand{\cD}[1]{D_{#1}}
\newcommand{\pd}[1]{\partial_{#1}}
\newcommand{\tgen}[1]{\tau^{#1}}
\newcommand{\gbl}{g_1}
\newcommand{\lctt}[3]{\varepsilon_{#1#2#3}}
\newcommand{\lctf}[4]{\varepsilon_{#1#2#3#4}}
\newcommand{\lctfb}[4]{\varepsilon\lpar{#1#2#3#4}\rpar}
\newcommand{\slct}{\varepsilon}
\newcommand{\cgfi}[1]{{\cal{C}}^{#1}}
\newcommand{\cgfZ}{{\cal{C}}_{_Z}}
\newcommand{\cgfA}{{\cal{C}}_{_A}}
\newcommand{\cgfZs}{{\cal{C}}^2_{_Z}}
\newcommand{\cgfAs}{{\cal{C}}^2_{_A}}
\newcommand{\hpms}{\mu^2}
\newcommand{\hpal}{\alpha_{_H}}
\newcommand{\hpals}{\alpha^2_{_H}}
\newcommand{\hpbe}{\beta_{_H}}
\newcommand{\hpbep}{\beta^{'}_{_H}}
\newcommand{\hpla}{\lambda}
\newcommand{\hpalf}{\alpha_{f}}
\newcommand{\hpbef}{\beta_{f}}
\newcommand{\tpar}[1]{\Lambda^{#1}}
\newcommand{\Mop}[2]{{\rm{M}}^{#1#2}}
\newcommand{\Lop}[2]{{\rm{L}}^{#1#2}}
\newcommand{\Lgen}[1]{T^{#1}}
\newcommand{\Rgen}[1]{t^{#1}}
\newcommand{\fpari}[1]{\lambda_{#1}}
\newcommand{\fQ}[1]{Q_{#1}}
\newcommand{\unm}{I}
\newcommand{\cDsla}{/\!\!\!\!D}
%
%
\newcommand{\saff}[1]{A_{#1}}                    
\newcommand{\aff}[2]{A_{#1}\lpar #2\rpar}                   
\newcommand{\sbff}[1]{B_{#1}}                    
\newcommand{\sfbff}[1]{B^{\ssF}_{#1}}
\newcommand{\bff}[4]{B_{#1}\lpar #2;#3,#4\rpar}             
\newcommand{\bfft}[3]{B_{#1}\lpar #2,#3\rpar}             
\newcommand{\fbff}[4]{B^{\ssF}_{#1}\lpar #2;#3,#4\rpar}        
\newcommand{\cdbff}[4]{\Delta B_{#1}\lpar #2;#3,#4\rpar}             
\newcommand{\sdbff}[4]{\delta B_{#1}\lpar #2;#3,#4\rpar}             
\newcommand{\cdbfft}[3]{\Delta B_{#1}\lpar #2,#3\rpar}             
\newcommand{\sdbfft}[3]{\delta B_{#1}\lpar #2,#3\rpar}             
\newcommand{\scff}[1]{C_{#1}}                    
\newcommand{\scffo}[2]{C_{#1}\lpar{#2}\rpar}                
\newcommand{\cff}[7]{C_{#1}\lpar #2,#3,#4;#5,#6,#7\rpar}    
\newcommand{\sccff}[5]{c_{#1}\lpar #2;#3,#4,#5\rpar} 
\newcommand{\sdff}[1]{D_{#1}}                    
\newcommand{\dffp}[7]{D_{#1}\lpar #2,#3,#4,#5,#6,#7;}       
\newcommand{\dffm}[4]{#1,#2,#3,#4\rpar}                     
\newcommand{\bzfa}[2]{B^{\ssF}_{_{#2}}\lpar{#1}\rpar}
\newcommand{\bzfaa}[3]{B^{\ssF}_{_{#2#3}}\lpar{#1}\rpar}
\newcommand{\shcff}[4]{C_{_{#2#3#4}}\lpar{#1}\rpar}
\newcommand{\shdff}[6]{D_{_{#3#4#5#6}}\lpar{#1,#2}\rpar}
\newcommand{\scdff}[3]{d_{#1}\lpar #2,#3\rpar} 
\newcommand{\scaldff}[1]{{\cal{D}}^{#1}}
\newcommand{\caldff}[2]{{\cal{D}}^{#1}\lpar{#2}\rpar}
\newcommand{\caldfft}[3]{{\cal{D}}_{#1}^{#2}\lpar{#3}\rpar}
%
%
\newcommand{\slaff}[1]{a_{#1}}                        
\newcommand{\slbff}[1]{b_{#1}}                        
\newcommand{\slbffh}[1]{{\hat{b}}_{#1}}    
\newcommand{\ssldff}[1]{d_{#1}}                        
\newcommand{\sslcff}[1]{c_{#1}}                        
\newcommand{\slcff}[2]{c_{#1}^{(#2)}}                        
\newcommand{\sldff}[2]{d_{#1}^{(#2)}}                        
\newcommand{\lbff}[3]{b_{#1}\lpar #2;#3\rpar}         
\newcommand{\lbffh}[2]{{\hat{b}}_{#1}\lpar #2\rpar}   
\newcommand{\lcff}[8]{c_{#1}^{(#2)}\lpar  #3,#4,#5;#6,#7,#8\rpar}         
\newcommand{\ldffp}[8]{d_{#1}^{(#2)}\lpar #3,#4,#5,#6,#7,#8;}
\newcommand{\ldffm}[4]{#1,#2,#3,#4\rpar}                   
%
%
\newcommand{\Iff}[4]{I_{#1}\lpar #2;#3,#4 \rpar}
\newcommand{\Jff}[4]{J_{#1}\lpar #2;#3,#4 \rpar}
\newcommand{\Jds}[5]{{\bar{J}}_{#1}\lpar #2,#3;#4,#5 \rpar}
\newcommand{\sJds}[1]{{\bar{J}}_{#1}}
%
\newcommand{\nhmt}{\frac{n}{2}-2}
\newcommand{\nhmts}{{n}/{2}-2}
\newcommand{\omnh}{1-\frac{n}{2}}
\newcommand{\nhmo}{\frac{n}{2}-1}
\newcommand{\fmon}{4-n}
\newcommand{\lpi}{\ln\pi}
\newcommand{\lmass}[1]{\ln #1}
\newcommand{\egnh}{\egam{\frac{n}{2}}}
\newcommand{\egomnh}{\egam{1-\frac{n}{2}}}
\newcommand{\egtmnh}{\egam{2-\frac{n}{2}}}
\newcommand{\Ddr}{{\ds\frac{1}{{\bar{\varepsilon}}}}}
\newcommand{\Ddrs}{{\ds\frac{1}{{\bar{\varepsilon}^2}}}}
\newcommand{\Ddrd}{{\bar{\varepsilon}}}
\newcommand{\ept}{\hat\varepsilon}
\newcommand{\Ddrh}{{\ds\frac{1}{\hat{\varepsilon}}}}
\newcommand{\Ddrp}{{\ds\frac{1}{\varepsilon'}}}
\newcommand{\Ddrps}{\lpar{\ds{\frac{1}{\varepsilon'}}}\rpar^2}
\newcommand{\dre}{\varepsilon}
\newcommand{\drei}[1]{\varepsilon_{#1}}
\newcommand{\epp}{\varepsilon'}
\newcommand{\eps}{\varepsilon^*}
\newcommand{\ep}{\epsilon}
\newcommand{\propbt}[6]{{{#1_{#2}#1_{#3}}\over{\lpar #1^2 + #4 
-\ib\ep\rpar\lpar\lpar #5\rpar^2 + #6 -\ib\ep\rpar}}}
\newcommand{\propbo}[5]{{{#1_{#2}}\over{\lpar #1^2 + #3 - \ib\ep\rpar
\lpar\lpar #4\rpar^2 + #5 -\ib\ep\rpar}}}
\newcommand{\propc}[6]{{1\over{\lpar #1^2 + #2 - \ib\ep\rpar
\lpar\lpar #3\rpar^2 + #4 -\ib\ep\rpar
\lpar\lpar #5\rpar^2 + #6 -\ib\ep\rpar}}}
\newcommand{\propa}[2]{{1\over {#1^2 + #2^2 - \ib\ep}}}
\newcommand{\propb}[4]{{1\over {\lpar #1^2 + #2 - \ib\ep\rpar
\lpar\lpar #3\rpar^2 + #4 -\ib\ep\rpar}}}
\newcommand{\propbs}[4]{{1\over {\lpar\lpar #1\rpar^2 + #2 - \ib\ep\rpar
\lpar\lpar #3\rpar^2 + #4 -\ib\ep\rpar}}}
\newcommand{\propat}[4]{{#3_{#1}#3_{#2}\over {#3^2 + #4^2 - \ib\ep}}}
\newcommand{\propaf}[6]{{#5_{#1}#5_{#2}#5_{#3}#5_{#4}\over 
{#5^2 + #6^2 -\ib\ep}}}
\newcommand{\momeps}[1]{#1^2 - \ib\ep}
\newcommand{\mopeps}[1]{#1^2 + \ib\ep}
\newcommand{\propz}[1]{{1\over{#1^2 + \mzs - \ib\ep}}}
\newcommand{\propw}[1]{{1\over{#1^2 + \mws - \ib\ep}}}
\newcommand{\proph}[1]{{1\over{#1^2 + \mhs - \ib\ep}}}
\newcommand{\propf}[2]{{1\over{#1^2 + #2}}}
\newcommand{\propzrg}[3]{{{\delta_{#1#2}}\over{{#3}^2 + \mzs - \ib\ep}}}
\newcommand{\propwrg}[3]{{{\delta_{#1#2}}\over{{#3}^2 + \mws - \ib\ep}}}
\newcommand{\propzug}[3]{{
      {\delta_{#1#2} + \displaystyle{{{#3}^{#1}{#3}^{#2}}\over{\mzs}}}
                         \over{{#3}^2 + \mzs - \ib\ep}}}
\newcommand{\propwug}[3]{{
      {\delta_{#1#2} + \displaystyle{{{#3}^{#1}{#3}^{#2}}\over{\mws}}}
                        \over{{#3}^2 + \mws - \ib\ep}}}
\newcommand{\thf}[1]{\theta\lpar #1\rpar}
\newcommand{\epf}[1]{\varepsilon\lpar #1\rpar}
\newcommand{\singp}{\stackrel{\rm{sing}}{\rightarrow}}
\newcommand{\aint}[3]{\int_{#1}^{#2}\,d #3}
\newcommand{\aroot}[1]{\sqrt{#1}}
\newcommand{\gramc}{\Delta_3}
\newcommand{\gramd}{\Delta_4}
\newcommand{\pinch}[2]{P^{(#1)}\lpar #2\rpar}
\newcommand{\pinchc}[2]{C^{(#1)}_{#2}}
\newcommand{\pinchd}[2]{D^{(#1)}_{#2}}
\newcommand{\loarg}[1]{\ln\lpar #1\rpar}
\newcommand{\loargr}[1]{\ln\lrbr #1\rrbr}
\newcommand{\lsoarg}[1]{\ln^2\lpar #1\rpar}
\newcommand{\ltarg}[2]{\ln\lpar #1\rpar\lpar #2\rpar}
\newcommand{\rfun}[2]{R\lpar #1,#2\rpar}
\newcommand{\pinchb}[3]{B_{#1}\lpar #2,#3\rpar}
\newcommand{\lga}{\ph}
\newcommand{\lzga}{\ssZ\ph}
%
%
\newcommand{\afa}[5]{A_{#1}^{#2}\lpar #3;#4,#5\rpar}
\newcommand{\bfa}[5]{B_{#1}^{#2}\lpar #3;#4,#5\rpar} 
\newcommand{\hfa}[5]{H_{#1}^{#2}\lpar #3;#4,#5\rpar}
\newcommand{\rfa}[5]{R_{#1}^{#2}\lpar #3;#4,#5\rpar}
\newcommand{\afao}[3]{A_{#1}^{#2}\lpar #3\rpar}
\newcommand{\bfao}[3]{B_{#1}^{#2}\lpar #3\rpar}
\newcommand{\hfao}[3]{H_{#1}^{#2}\lpar #3\rpar}
\newcommand{\rfao}[3]{R_{#1}^{#2}\lpar #3\rpar}
\newcommand{\afax}[6]{A_{#1}^{#2}\lpar #3;#4,#5,#6\rpar}
\newcommand{\afas}[2]{A_{#1}^{#2}}
\newcommand{\bfas}[2]{B_{#1}^{#2}}
\newcommand{\hfas}[2]{H_{#1}^{#2}}
\newcommand{\rfas}[2]{R_{#1}^{#2}}
\newcommand{\tfas}[2]{T_{#1}^{#2}}
\newcommand{\afaR}[6]{A_{#1}^{\gpar}\lpar #2;#3,#4,#5,#6 \rpar}
\newcommand{\bfaR}[6]{B_{#1}^{\gpar}\lpar #2;#3,#4,#5,#6 \rpar}
\newcommand{\hfaR}[6]{H_{#1}^{\gpar}\lpar #2;#3,#4,#5,#6 \rpar}
\newcommand{\shfaR}[1]{H_{#1}^{\gpar}}
\newcommand{\rfaR}[6]{R_{#1}^{\gpar}\lpar #2;#3,#4,#5,#6 \rpar}
\newcommand{\srfaR}[1]{R_{#1}^{\gpar}}
\newcommand{\afaRg}[5]{A_{#1 \lga}^{\gpar}\lpar #2;#3,#4,#5 \rpar}
\newcommand{\bfaRg}[5]{B_{#1 \lga}^{\gpar}\lpar #2;#3,#4,#5 \rpar}
\newcommand{\hfaRg}[5]{H_{#1 \lga}^{\gpar}\lpar #2;#3,#4,#5 \rpar}
\newcommand{\shfaRg}[1]{H_{#1\lga}^{\gpar}}
\newcommand{\rfaRg}[5]{R_{#1 \lga}^{\gpar}\lpar #2;#3,#4,#5 \rpar}
\newcommand{\srfaRg}[1]{R_{#1\lga}^{\gpar}}
\newcommand{\afaRt}[3]{A_{#1}^{\gpar}\lpar #2,#3 \rpar}
\newcommand{\hfaRt}[3]{H_{#1}^{\gpar}\lpar #2,#3 \rpar}
\newcommand{\hfaRf}[4]{H_{#1}^{\gpar}\lpar #2,#3,#4 \rpar}
\newcommand{\afasm}[4]{A_{#1}^{\lpar #2,#3,#4 \rpar}}
\newcommand{\bfasm}[4]{B_{#1}^{\lpar #2,#3,#4 \rpar}}
\newcommand{\htf}[2]{H_2\lpar #1,#2\rpar}
\newcommand{\rof}[2]{R_1\lpar #1,#2\rpar}
\newcommand{\rtf}[2]{R_3\lpar #1,#2\rpar}
\newcommand{\rtrans}[2]{R_{#1}^{#2}}
\newcommand{\momf}[2]{#1^2_{#2}}
\newcommand{\Scalvert}[8][70]{
  \vcenter{\hbox{
  \SetScale{0.8}
  \begin{picture}(#1,50)(15,15)
    \Line(25,25)(50,50)      \Text(34,20)[lc]{#6} \Text(11,20)[lc]{#3}
    \Line(50,50)(25,75)      \Text(34,60)[lc]{#7} \Text(11,60)[lc]{#4}
    \Line(50,50)(90,50)      \Text(11,40)[lc]{#2} \Text(55,33)[lc]{#8}
    \GCirc(50,50){10}{1}          \Text(60,48)[lc]{#5} 
  \end{picture}}}
  }
%
%
\newcommand{\tHs}{\mu}
\newcommand{\tHsz}{\mu_{_0}}
\newcommand{\tHss}{\mu^2}
\newcommand{\Reb}{{\rm{Re}}}
\newcommand{\Imb}{{\rm{Im}}}
%
%
\newcommand{\spd}{\partial}
\newcommand{\fun}[1]{f\lpar{#1}\rpar}
\newcommand{\ffun}[2]{F_{#1}\lpar #2\rpar}
\newcommand{\gfun}[2]{G_{#1}\lpar #2\rpar}
\newcommand{\sffun}[1]{F_{#1}}
\newcommand{\csffun}[1]{{\cal{F}}_{#1}}
\newcommand{\sgfun}[1]{G_{#1}}
\newcommand{\tpfi}{\lpar 2\pi\rpar^4\ib}
\newcommand{\ffv}{F_{_V}}
\newcommand{\fga}{G_{_A}}
\newcommand{\ffm}{F_{_M}}
\newcommand{\ffs}{F_{_S}}
\newcommand{\fgp}{G_{_P}}
\newcommand{\fge}{G_{_E}}
\newcommand{\ffa}{F_{_A}}
\newcommand{\ffps}{F_{_P}}
\newcommand{\ffe}{F_{_E}}
\newcommand{\gacom}[2]{\lpar #1 + #2\gfd\rpar}
\newcommand{\mft}{m_{\tilde f}}
\newcommand{\qft}{Q_{f'}}
\newcommand{\vft}{v_{\tilde f}}
\newcommand{\subb}[2]{b_{#1}\lpar #2 \rpar}
\newcommand{\fwfr}[5]{\Sigma\lpar #1,#2,#3;#4,#5 \rpar}
\newcommand{\slim}[2]{\lim_{#1 \to #2}}
\newcommand{\sprop}[3]{
{#1\over {\lpar q^2\rpar^2\lpar \lpar q+ #2\rpar^2+#3^2\rpar }}}
%
%
\newcommand{\xroot}[1]{x_{#1}}
\newcommand{\yroot}[1]{y_{#1}}
\newcommand{\zroot}[1]{z_{#1}}
\newcommand{\lvar}{l}
\newcommand{\rvar}{r}
\newcommand{\tvar}{t}
\newcommand{\uvar}{u}
\newcommand{\vvar}{v}
\newcommand{\xvar}{x}
\newcommand{\yvar}{y}
\newcommand{\zvar}{z}
\newcommand{\yvarp}{y'}
\newcommand{\rvars}{r^2}
\newcommand{\vvars}{v^2}
\newcommand{\xvars}{x^2}
\newcommand{\yvars}{y^2}
\newcommand{\zvars}{z^2}
\newcommand{\rvarc}{r^3}
\newcommand{\xvarc}{x^3}
\newcommand{\yvarc}{y^3}
\newcommand{\zvarc}{z^3}
\newcommand{\rvarq}{r^4}
\newcommand{\xvarq}{x^4}
\newcommand{\yvarq}{y^4}
\newcommand{\zvarq}{z^4}
\newcommand{\avar}{a}
\newcommand{\avars}{a^2}
\newcommand{\avarc}{a^3}
\newcommand{\avari}[1]{a_{#1}}
\newcommand{\avart}[2]{a_{#1}^{#2}}
\newcommand{\delvari}[1]{\delta_{#1}}
\newcommand{\rvari}[1]{r_{#1}}
\newcommand{\xvari}[1]{x_{#1}}
\newcommand{\yvari}[1]{y_{#1}}
\newcommand{\zvari}[1]{z_{#1}}
\newcommand{\rvart}[2]{r_{#1}^{#2}}
\newcommand{\xvart}[2]{x_{#1}^{#2}}
\newcommand{\yvart}[2]{y_{#1}^{#2}}
\newcommand{\zvart}[2]{z_{#1}^{#2}}
\newcommand{\rvaris}[1]{r^2_{#1}}
\newcommand{\xvaris}[1]{x^2_{#1}}
\newcommand{\yvaris}[1]{y^2_{#1}}
\newcommand{\zvaris}[1]{z^2_{#1}}
\newcommand{\Xvar}{X}
\newcommand{\Xvars}{X^2}
\newcommand{\Xvari}[1]{X_{#1}}
\newcommand{\Xvaris}[1]{X^2_{#1}}
\newcommand{\Yvar}{Y}
\newcommand{\Yvars}{Y^2}
\newcommand{\Yvari}[1]{Y_{#1}}
\newcommand{\Yvaris}[1]{Y^2_{#1}}
\newcommand{\lnx}{\ln\xvar}
\newcommand{\lnz}{\ln\zvar}
\newcommand{\lnsx}{\ln^2\xvar}
\newcommand{\lnsz}{\ln^2\zvar}
\newcommand{\lncz}{\ln^3\zvar}
\newcommand{\lnomz}{\ln\lpar 1-\zvar\rpar}
\newcommand{\lnsomz}{\ln^2\lpar 1-\zvar\rpar}
\newcommand{\ccoefi}[1]{c_{#1}}
\newcommand{\ccoeft}[2]{c^{#1}_{#2}}
%
%
\newcommand{\Smat}{{\cal{S}}}
\newcommand{\Mmat}{{\cal{M}}}
\newcommand{\Xmat}[1]{X_{#1}}
\newcommand{\XmatI}[1]{X^{-1}_{#1}}
\newcommand{\unitmat}{I}
\newcommand{\Kmat}{{C}}
\newcommand{\Kmatc}{{C}^{\dagger}}
\newcommand{\Kmati}[1]{{C}_{#1}}
\newcommand{\Kmatci}[1]{{C}^{\dagger}_{#1}}
\newcommand{\ffac}[2]{f_{#1}^{#2}}
\newcommand{\Ffac}[1]{F_{#1}}
\newcommand{\Rvec}[2]{R^{(#1)}_{#2}}
\newcommand{\momfl}[2]{#1_{#2}}
\newcommand{\momfs}[2]{#1^2_{#2}}
\newcommand{\fpseZ}{A^{\ssF\ssP,\ssZ}}
\newcommand{\fpseA}{A^{\ssF\ssP,\ssA}}
\newcommand{\fptZ}{T^{\ssF\ssP,\ssZ}}
\newcommand{\fptA}{T^{\ssF\ssP,\ssA}}
\newcommand{\dprop}{\overline\Delta}
\newcommand{\dpropi}[1]{d_{#1}}
\newcommand{\dpropic}[1]{d^{c}_{#1}}
\newcommand{\dpropii}[2]{d_{#1}\lpar #2\rpar}
\newcommand{\dpropis}[1]{d^2_{#1}}
\newcommand{\dproppi}[1]{d'_{#1}}
\newcommand{\psf}[4]{P\lpar #1;#2,#3,#4\rpar}
\newcommand{\ssf}[5]{S^{(#1)}\lpar #2;#3,#4,#5\rpar}
\newcommand{\csf}[5]{C_{_S}^{(#1)}\lpar #2;#3,#4,#5\rpar}
%
%
\newcommand{\lvec}{l}
\newcommand{\lvecs}{l^2}
\newcommand{\lveci}[1]{l_{#1}}
\newcommand{\mvec}{m}
\newcommand{\mvecs}{m^2}
\newcommand{\mveci}[1]{m_{#1}}
\newcommand{\nvec}{n}
\newcommand{\nvecs}{n^2}
\newcommand{\nveci}[1]{n_{#1}}
\newcommand{\epi}[1]{\epsilon_{#1}}
\newcommand{\phep}[1]{\ep_{#1}}
\newcommand{\php}[3]{\ep^{#1}_{#2}\lpar #3 \rpar}
\newcommand{\sphep}{\ep}
\newcommand{\vbep}[1]{e_{#1}}
\newcommand{\vbepp}[1]{e^{+}_{#1}}
\newcommand{\vbepm}[1]{e^{-}_{#1}}
\newcommand{\svbep}{e}
%
%
\newcommand{\lpol}{\lambda}
\newcommand{\spol}{\sigma}
\newcommand{\rpol}{\rho  }
\newcommand{\kpol}{\kappa}
\newcommand{\lpols}{\lambda^2}
\newcommand{\spols}{\sigma^2}
\newcommand{\rpols}{\rho^2}
\newcommand{\kpols}{\kappa^2}
\newcommand{\lpoli}[1]{\lambda_{#1}}
\newcommand{\spoli}[1]{\sigma_{#1}}
\newcommand{\rpoli}[1]{\rho_{#1}}
\newcommand{\kpoli}[1]{\kappa_{#1}}
%
%
\newcommand{\uvec}{u}
\newcommand{\uveci}[1]{u_{#1}}
%
%
\newcommand{\imom}{q}
\newcommand{\imomi}[1]{q_{#1}}
\newcommand{\imoms}{q^2}
\newcommand{\pmom}{p}
\newcommand{\pmomp}{p'}
\newcommand{\pmoms}{p^2}
\newcommand{\pmomq}{p^4}
\newcommand{\pmomx}{p^6}
\newcommand{\pmomi}[1]{p_{#1}}
\newcommand{\pmomis}[1]{p^2_{#1}}
\newcommand{\Pmom}{P}
\newcommand{\Pmoms}{P^2}
\newcommand{\Pmomi}[1]{P_{#1}}
\newcommand{\Pmomis}[1]{P^2_{#1}}
\newcommand{\Kmom}{K}
\newcommand{\Kmoms}{K^2}
\newcommand{\Kmomi}[1]{K_{#1}}
\newcommand{\Kmomis}[1]{K^2_{#1}}
\newcommand{\kmom}{k}
\newcommand{\kmoms}{k^2}
\newcommand{\kmomi}[1]{k_{#1}}
\newcommand{\lmom}{l}
\newcommand{\lmoms}{l^2}
\newcommand{\lmomi}[1]{l_{#1}}
\newcommand{\qmom}{q}
\newcommand{\qmoms}{q^2}
\newcommand{\qmomi}[1]{q_{#1}}
\newcommand{\qmomis}[1]{q^2_{#1}}
\newcommand{\smom}{s}
\newcommand{\smoms}{s^2}
\newcommand{\smomi}[1]{s_{#1}}
\newcommand{\tmom}{t}
\newcommand{\tmoms}{t^2}
\newcommand{\tmomi}[1]{t_{#1}}
\newcommand{\Trmom}{Q}
\newcommand{\Prmom}{P}
\newcommand{\gmv}{Q^2}
\newcommand{\Trmoms}{Q^2}
\newcommand{\Prmoms}{P^2}
\newcommand{\Ptmoms}{T^2}
\newcommand{\Pumoms}{U^2}
\newcommand{\Trmomq}{Q^4}
\newcommand{\Prmomq}{P^4}
\newcommand{\Ptmomq}{T^4}
\newcommand{\Pumomq}{U^4}
\newcommand{\Trmomx}{Q^6}
\newcommand{\Trmomi}[1]{Q_{#1}}
\newcommand{\Trmomis}[1]{Q^2_{#1}}
\newcommand{\Prmomi}[1]{P_{#1}}
\newcommand{\pone}{p_1}
\newcommand{\ptwo}{p_2}
\newcommand{\ptre}{p_3}
\newcommand{\pfor}{p_4}
\newcommand{\pones}{p_1^2}
\newcommand{\ptwos}{p_2^2}
\newcommand{\ptres}{p_3^2}
\newcommand{\pfors}{p_4^2}
\newcommand{\poneq}{p_1^4}
\newcommand{\ptwoq}{p_2^4}
\newcommand{\ptreq}{p_3^4}
\newcommand{\pforq}{p_4^4}
\newcommand{\modmom}[1]{\mid{\vec{#1}}\mid}
\newcommand{\modmomi}[2]{\mid{\vec{#1}}_{#2}\mid}
\newcommand{\vect}[1]{{\vec{#1}}}
\newcommand{\Energ}{E}
\newcommand{\Energp}{E'}
\newcommand{\Energpp}{E''}
\newcommand{\Energs}{E^2}
\newcommand{\Energc}{E^3}
\newcommand{\Energf}{E^4}
\newcommand{\Energv}{E^5}
\newcommand{\Energx}{E^6}
\newcommand{\Energi}[1]{E_{#1}}
\newcommand{\Energt}[2]{E_{#1}^{#2}}
\newcommand{\Energis}[1]{E^2_{#1}}
\newcommand{\energ}{e}
\newcommand{\energp}{e'}
\newcommand{\energpp}{e''}
\newcommand{\energs}{e^2}
\newcommand{\energi}[1]{e_{#1}}
\newcommand{\energt}[2]{e_{#1}^{#2}}
\newcommand{\energis}[1]{e^2_{#1}}
\newcommand{\wenerg}{w}
\newcommand{\wenergs}{w^2}
\newcommand{\wenergi}[1]{w_{#1}}
\newcommand{\wenergp}{w'}
\newcommand{\wenergpp}{w''}
%
%
\newcommand{\ecut}{e}
\newcommand{\ecuts}{e^2}
\newcommand{\ecuti}[1]{e^{#1}}
\newcommand{\ccut}{c_m}
\newcommand{\ccuti}[1]{c_{#1}}
\newcommand{\ccuts}{c^2_m}
\newcommand{\scuts}{s^2_m}
\newcommand{\ccutis}[1]{c^2_{#1}}
\newcommand{\ccutic}[1]{c^3_{#1}}
\newcommand{\ccutc}{c^3_m}
\newcommand{\rcut}{\varrho}
\newcommand{\rcuts}{\varrho^2}
\newcommand{\rcuti}[1]{\varrho_{#1}}
\newcommand{\rcutu}[1]{\varrho^{#1}}
\newcommand{\Dcut}{\Delta}
%
\newcommand{\dwf}{\delta_{_{\rm{WF}}}}
\newcommand{\gbar}{\overline g}
\newcommand{\PP}{\mbox{PP}}
\newcommand{\mv}{m_{_V}}
\newcommand{\bGv}{{\overline\Gamma}_{_V}}
\newcommand{\Umuv}{\hat{\mu}_\ssV}
\newcommand{\Svv}{{\Sigma}_\ssV}
\newcommand{\muv}{p_\ssV}
\newcommand{\muvb}{\mu_{\ssV_{0}}}
\newcommand{\URPvv}{{P}_\ssV}
\newcommand{\RPvv}{{P}_\ssV}
\newcommand{\Svvrem}{{\Sigma}_\ssV^{\mathrm{rem}}}
\newcommand{\USvvrem}{\hat{\Sigma}_\ssV^{\mathrm{rem}}}
\newcommand{\Gv}{\Gamma_{_V}}
%
%
\newcommand{\param}{p}
\newcommand{\parami}[1]{p^{#1}}
\newcommand{\paramb}{p_{0}}
\newcommand{\Zcon}{Z}
\newcommand{\Zconi}[1]{Z_{#1}}
\newcommand{\zconi}[1]{z_{#1}}
\newcommand{\Zconim}[1]{{Z^-_{#1}}}
\newcommand{\zconim}[1]{{z^-_{#1}}}
\newcommand{\Zcont}[2]{Z_{#1}^{#2}}
\newcommand{\zcont}[2]{z_{#1}^{#2}}
\newcommand{\zcontm}[2]{z_{#1}^{{#2}-}}
\newcommand{\sZconi}[2]{\sqrt{Z_{#1}}^{\;#2}}
\newcommand{\gacome}[1]{\lpar #1 - \gfd\rpar}
\newcommand{\sPj}[2]{\Lambda^{#1}_{#2}}
\newcommand{\sPjs}[2]{\Lambda_{#1,#2}}
\newcommand{\amos}{\mbox{$M^2_{_1}$}}
\newcommand{\amts}{\mbox{$M^2_{_2}$}}
\newcommand{\er}{e_{_{R}}}
\newcommand{\epr}{e'_{_{R}}}
\newcommand{\ers}{e^2_{_{R}}}
\newcommand{\erc}{e^3_{_{R}}}
\newcommand{\erq}{e^4_{_{R}}}
\newcommand{\erf}{e^5_{_{R}}}
\newcommand{\sour}{J}
\newcommand{\sourb}{\overline J}
\newcommand{\lrm}{M_{_R}}
%
%
\newcommand{\vlami}[1]{\lambda_{#1}}
\newcommand{\vlamis}[1]{\lambda^2_{#1}}
\newcommand{\Vvert}{V}
\newcommand{\Avert}{A}
\newcommand{\Svert}{S}
\newcommand{\Pvert}{P}
\newcommand{\vvert}{F}
\newcommand{\Cvert}{\cal{V}}
\newcommand{\Bvert}{\cal{B}}
\newcommand{\Vveri}[2]{V_{#1}^{#2}}
\newcommand{\Fveri}[1]{{\cal{F}}^{#1}}
\newcommand{\Cveri}[1]{{\cal{V}}\lpar{#1}\rpar}
\newcommand{\Bveri}[1]{{\cal{B}}\lpar{#1}\rpar}
\newcommand{\Vverti}[3]{V_{#1}^{#2}\lpar{#3}\rpar}
\newcommand{\Averti}[3]{A_{#1}^{#2}\lpar{#3}\rpar}
\newcommand{\Gverti}[3]{G_{#1}^{#2}\lpar{#3}\rpar}
\newcommand{\Zverti}[3]{Z_{#1}^{#2}\lpar{#3}\rpar}
\newcommand{\Hverti}[2]{H^{#1}\lpar{#2}\rpar}
\newcommand{\Wverti}[3]{W_{#1}^{#2}\lpar{#3}\rpar}
\newcommand{\Cverti}[2]{{\cal{V}}_{#1}^{#2}}
\newcommand{\vverti}[3]{F^{#1}_{#2}\lpar{#3}\rpar}
\newcommand{\averti}[3]{{\overline{F}}^{#1}_{#2}\lpar{#3}\rpar}
\newcommand{\fveone}[1]{f_{#1}}
\newcommand{\fvetri}[3]{f^{#1}_{#2}\lpar{#3}\rpar}
\newcommand{\gvetri}[3]{g^{#1}_{#2}\lpar{#3}\rpar}
\newcommand{\cvetri}[3]{{\cal{F}}^{#1}_{#2}\lpar{#3}\rpar}
\newcommand{\hvetri}[3]{{\hat{\cal{F}}}^{#1}_{#2}\lpar{#3}\rpar}
\newcommand{\avetri}[3]{{\overline{\cal{F}}}^{#1}_{#2}\lpar{#3}\rpar}
\newcommand{\fverti}[2]{F^{#1}_{#2}}
\newcommand{\cverti}[2]{{\cal{F}}_{#1}^{#2}}
\newcommand{\fV}{f_{_{\Vvert}}}
\newcommand{\gA}{g_{_{\Avert}}}
\newcommand{\fVi}[1]{f^{#1}_{_{\Vvert}}}
\newcommand{\seai}[1]{a_{#1}}
\newcommand{\seapi}[1]{a'_{#1}}
\newcommand{\seAi}[2]{A_{#1}^{#2}}
\newcommand{\sewi}[1]{w_{#1}}
\newcommand{\seWi}[1]{W_{#1}}
\newcommand{\seWsi}[1]{W^{*}_{#1}}
\newcommand{\seWti}[2]{W_{#1}^{#2}}
\newcommand{\sewti}[2]{w_{#1}^{#2}}
\newcommand{\seSig}[1]{\Sigma_{#1}\lpar\sla{\pmom}\rpar}
\newcommand{\ww}{{\rm{w}}}
\newcommand{\ew}{{\rm{ew}}}
\newcommand{\nonSE}{\rm{non-SE}}
\newcommand{\leading}{\rm{\ssL}}
%
%
\newcommand{\bbff}[1]{{\overline B}_{#1}}
\newcommand{\sW}{p_{_W}}
\newcommand{\sZ}{p_{_Z}}
\newcommand{\ssp}{s_p}
\newcommand{\fW}{f_{_W}}
\newcommand{\fZ}{f_{_Z}}
\newcommand{\subMSB}[1]{{#1}_{\mbox{$\overline{\scriptscriptstyle MS}$}}}
\newcommand{\supMSB}[1]{{#1}^{\mbox{$\overline{\scriptscriptstyle MS}$}}}
\newcommand{\redMSB}{{\mbox{$\overline{\scriptscriptstyle MS}$}}}
\newcommand{\gpbb}{g'_{0}}
\newcommand{\Zconip}[1]{Z'_{#1}}
\newcommand{\bpff}[4]{B'_{#1}\lpar #2;#3,#4\rpar}             
\newcommand{\xidf}{\xi^2-1}
\newcommand{\tDdr}{1/{\bar{\varepsilon}}}
\newcommand{\cRz}{{\cal R}_{_Z}}
\newcommand{\cRg}{{\cal R}_{\gamma}}
\newcommand{\Sz}{\Sigma_{_Z}}
\newcommand{\alh}{{\hat\alpha}}
\newcommand{\alhz}{\alpha_{\ssZ}}
\newcommand{\Phzg}{{\hat\Pi}_{_{\zb\ab}}}
\newcommand{\fvvert}{F^{\rm vert}_{_V}}
\newcommand{\gavert}{G^{\rm vert}_{_A}}
\newcommand{\bmv}{{\overline m}_{_V}}
\newcommand{\Sgn}{\Sigma_{\gamma\hkn}}
\newcommand{\rmboxd}{{\rm Box}_d\lpar s,t,u;M_1,M_2,M_3,M_4\rpar}
\newcommand{\rmboxc}{{\rm Box}_c\lpar s,t,u;M_1,M_2,M_3,M_4\rpar}
%
%
\newcommand{\Afaci}[1]{A_{#1}}
\newcommand{\Afacis}[1]{A^2_{#1}}
\newcommand{\upar}[1]{u}
\newcommand{\upari}[1]{u_{#1}}
\newcommand{\vpari}[1]{v_{#1}}
\newcommand{\lpari}[1]{l_{#1}}
\newcommand{\Lpari}[1]{l_{#1}}
\newcommand{\Nff}[2]{N^{(#1)}_{#2}}
\newcommand{\Sff}[2]{S^{(#1)}_{#2}}
\newcommand{\sSff}{S}
\newcommand{\FQED}[2]{F_{#1#2}}
\newcommand{\fbpsif}{{\overline{\psi}_f}}
\newcommand{\fpsif}{\psi_f}
\newcommand{\etafd}[2]{\eta_d\lpar#1,#2\rpar}
\newcommand{\sigdu}[2]{\sigma_{#1#2}}
\newcommand{\scalc}[4]{c_{0}\lpar #1;#2,#3,#4\rpar}
\newcommand{\scald}[2]{d_{0}\lpar #1,#2\rpar}
\newcommand{\scaldi}[3]{d_{0}^{#1}\lpar #2,#3\rpar}
\newcommand{\pir}[1]{\Pi^{\rm{\ssR}}\lpar #1\rpar}
\newcommand{\sigh}{\sigma_{\rm had}}
\newcommand{\dah}{\Delta\alpha^{(5)}_{\rm had}}
\newcommand{\dat}{\Delta\alpha_{\rm top}}
\newcommand{\Vqed}[3]{V_1^{\rm sub}\lpar#1;#2,#3\rpar}
\newcommand{\thetah}{{\hat\theta}}
\newcommand{\smlon}{\frac{\mlones}{s}}
\newcommand{\lntwo}{\ln 2}
\newcommand{\wmin}{w_{\rm min}}
\newcommand{\kmin}{k_{\rm min}}
\newcommand{\mdls}{\Big|}
\newcommand{\smf}{\frac{\mfs}{s}}
\newcommand{\bint}{\beta_{\rm int}}
\newcommand{\IRv}{V_{_{\rm IR}}}
\newcommand{\IRr}{R_{_{\rm IR}}}
\newcommand{\fssts}{\frac{s^2}{t^2}}
\newcommand{\fssus}{\frac{s^2}{u^2}}
\newcommand{\optM}{1+\frac{t}{M^2}}
\newcommand{\opuM}{1+\frac{u}{M^2}}
\newcommand{\ftM}{\lpar -\frac{t}{M^2}\rpar}
\newcommand{\fuM}{\lpar -\frac{u}{M^2}\rpar}
\newcommand{\omsM}{1-\frac{s}{M^2}}
\newcommand{\xsf}{\sigma_{_{\rm F}}}
\newcommand{\xsb}{\sigma_{_{\rm B}}}
\newcommand{\afb}{A_{_{\rm FB}}}
\newcommand{\rsoft}{\rm soft}
\newcommand{\rms}{\rm s}
\newcommand{\rsmx}{\sqrt{s_{\rm max}}}
\newcommand{\rspm}{\sqrt{s_{\pm}}}
\newcommand{\rsp}{\sqrt{s_{+}}}
\newcommand{\rsm}{\sqrt{s_{-}}}
\newcommand{\sigmx}{\sigma_{\rm max}}
\newcommand{\gG}[2]{G_{#1}^{#2}}
\newcommand{\gacomm}[2]{\lpar #1 - #2\gfd\rpar}
\newcommand{\fcsx}{\frac{1}{\ctwsix}}
\newcommand{\fcq}{\frac{1}{\ctwf}}
\newcommand{\fcs}{\frac{1}{\ctws}}
\newcommand{\affs}[2]{{\cal A}_{#1}\lpar #2\rpar}                   
\newcommand{\stwei}{s_{\theta}^8}
\def\mdan{\vspace{1mm}\mpar{\hfil$\downarrow$new\hfil}\vspace{-1mm}
          \ignorespaces}
\def\muan{\vspace{-1mm}\mpar{\hfil$\uparrow$new\hfil}\vspace{1mm}\ignorespaces}
\def\mlan{\vspace{-1mm}\mpar{\hfil$\rightarrow$new\hfil}\vspace{1mm}\ignorespaces}
\def\mnnew{\mpar{\hfil NEWNEW \hfil}\ignorespaces}
%
%
\newcommand{\boxc}[2]{{\cal{B}}_{#1}^{#2}}
\newcommand{\boxct}[3]{{\cal{B}}_{#1}^{#2}\lpar{#3}\rpar}
\newcommand{\hboxc}[3]{\hat{{\cal{B}}}_{#1}^{#2}\lpar{#3}\rpar}
\newcommand{\vev}{\langle v \rangle}
\newcommand{\vevi}[1]{\langle v_{#1}\rangle}
\newcommand{\vevs}{\langle v^2   \rangle}
\newcommand{\fwfrV}[5]{\Sigma_{_V}\lpar #1,#2,#3;#4,#5 \rpar}
\newcommand{\fwfrS}[7]{\Sigma_{_S}\lpar #1,#2,#3;#4,#5;#6,#7 \rpar}
\newcommand{\fSi}[1]{f^{#1}_{_{\Svert}}}
\newcommand{\fPi}[1]{f^{#1}_{_{\Pvert}}}
\newcommand{\mXs}{m_{_X}}
\newcommand{\mXss}{m^2_{_X}}
\newcommand{\mYs}{M^2_{_Y}}
\newcommand{\xik}{\xi_k}
\newcommand{\xiks}{\xi^2_k}
\newcommand{\mpls}{m^2_+}
\newcommand{\mmis}{m^2_-}
%
\newcommand{\SN}{\Sigma_{_N}}
\newcommand{\SC}{\Sigma_{_C}}
\newcommand{\SPN}{\Sigma'_{_N}}
\newcommand{\PFf}{\Pi^{\fer}_{_F}}
\newcommand{\PFb}{\Pi^{\bos}_{_F}}
\newcommand{\dPZ}{\Delta{\hat\Pi}_{_Z}}
\newcommand{\Sfin}{\Sigma_{_F}}
\newcommand{\Sfir}{\Sigma_{_R}}
\newcommand{\Sfinh}{{\hat\Sigma}_{_F}}
\newcommand{\Sfinf}{\Sigma^{\fer}_{_F}}
\newcommand{\Sfinbh}{\Sigma^{\bos}_{_F}}
\newcommand{\alf}{\alpha^{\fer}}
\newcommand{\alhfz}{\alpha^{\fer}\lpar{\mzs}\rpar}
\newcommand{\alhfs}{\alpha^{\fer}\lpar{\sman}\rpar}
\newcommand{\gfQ}{g^f_{_{Q}}}
\newcommand{\gfL}{g^f_{_{L}}}
\newcommand{\ccf}{\frac{\gbs}{16\,\pi^2}}
\newcommand{\chq}{{\hat c}^4}
\newcommand{\muuq}{m_{u'}}
\newcommand{\muus}{m^2_{u'}}
\newcommand{\mdd}{m_{d'}}
\newcommand{\clf}[2]{\mathrm{Cli}_{_#1}\lpar\displaystyle{#2}\rpar}
\def\stes{\sin^2\theta}
\def\acal{\cal A}
\def\alr{A_{_{\rm{LR}}}}
\newcommand{\barQ}{\overline Q}
\newcommand{\Sptg}{\Sigma'_{_{3Q}}}
\newcommand{\Sptt}{\Sigma'_{_{33}}}
\newcommand{\Ppgg}{\Pi'_{\ph\ph}}
\newcommand{\Pww}{\Pi_{_{\wb\wb}}}
\newcommand{\capV}[2]{{\cal F}^{#2}_{_{#1}}}
\newcommand{\bt}{\beta_t}
\newcommand{\mhsix}{M^6_{_H}}
\newcommand{\topt}{{\cal T}_{33}}
\newcommand{\topq}{{\cal T}_4}
\newcommand{\Phzgf}{{\hat\Pi}^{\fer}_{_{\zb\ab}}}
\newcommand{\Phzgb}{{\hat\Pi}^{\bos}_{_{\zb\ab}}}
\newcommand{\Sfirh}{{\hat\Sigma}_{_R}}
\newcommand{\Szgh}{{\hat\Sigma}_{_{\zb\ab}}}
\newcommand{\Szghb}{{\hat\Sigma}^{\bos}_{_{\zb\ab}}}
\newcommand{\Szghf}{{\hat\Sigma}^{\fer}_{_{\zb\ab}}}
\newcommand{\Szgb}{\Sigma^{\bos}_{_{\zb\ab}}}
\newcommand{\Szgf}{\Sigma^{\fer}_{_{\zb\ab}}}
\newcommand{\chig}{\chi_{\ph}}
\newcommand{\chiz}{\chi_{\ssZ}}
\newcommand{\Sfih}{{\hat\Sigma}}
\newcommand{\Szzh}{\hat{\Sigma}_{_{\zb\zb}}}
\newcommand{\dPZf}{\Delta{\hat\Pi}^f_{_{\zb}}}
\newcommand{\khZdf}[1]{{\hat\kappa}^{#1}_{_{\zb}}}
\newcommand{\chf}{{\hat c}^4}
\newcommand{\amp}[2]{{\cal{A}}_{_{#1}}^{\rm{#2}}}
\newcommand{\hatvm}[1]{{\hat v}^-_{#1}}
\newcommand{\hatvp}[1]{{\hat v}^+_{#1}}
\newcommand{\hatvpm}[1]{{\hat v}^{\pm}_{#1}}
\newcommand{\kvz}[1]{\kappa^{\zb #1}_{_V}}
\newcommand{\barp}{\overline p}                
\newcommand{\delw}{\Delta_{_{\wb}}}
\newcommand{\bdelw}{{\bar{\Delta}}_{_{\wb}}}
\newcommand{\bdelf}{{\bar{\Delta}}_{\ff}}
\newcommand{\delz}{\Delta_{_\zb}}
\newcommand{\deli}[1]{\Delta\lpar{#1}\rpar}
\newcommand{\hdeli}[1]{{\hat{\Delta}}\lpar{#1}\rpar}
\newcommand{\chizb}{\chi_{_\zb}}
\newcommand{\Swwp}{\Sigma'_{_{\wb\wb}}}
\newcommand{\epph}{\varepsilon'/2}
\newcommand{\sbffp}[1]{B'_{#1}}                    
\newcommand{\epss}{\varepsilon^*}
\newcommand{\Ddrhs}{{\ds\frac{1}{\hat{\varepsilon}^2}}}
\newcommand{\lnmsb}{L_{_\wb}}
\newcommand{\lnsmsb}{L^2_{_\wb}}
\newcommand{\tpni}{\lpar 2\pi\rpar^n\ib}
\newcommand{\tpn}{2^n\,\pi^{n-2}}
\newcommand{\cmf}{M_f}
\newcommand{\cmfs}{M^2_f}
\newcommand{\toDdr}{{\ds\frac{2}{{\bar{\varepsilon}}}}}
\newcommand{\troDdr}{{\ds\frac{3}{{\bar{\varepsilon}}}}}
\newcommand{\totDdr}{{\ds\frac{3}{{2\,\bar{\varepsilon}}}}}
\newcommand{\foDdr}{{\ds\frac{4}{{\bar{\varepsilon}}}}}
\newcommand{\smh}{m_{_H}}
\newcommand{\smhs}{m^2_{_H}}
\newcommand{\Ph}{\Pi_{_\hb}}
\newcommand{\Sphh}{\Sigma'_{_{\hb\hb}}}
\newcommand{\bh}{\beta}
\newcommand{\alsn}{\alpha^{(n_f)}_{_S}}
\newcommand{\smq}{m_q}
\newcommand{\smqp}{m_{q'}}
\newcommand{\shb}{h}
\newcommand{\hab}{A}
\newcommand{\hbpm}{H^{\pm}}
\newcommand{\hbp}{H^{+}}
\newcommand{\hbm}{H^{-}}
\newcommand{\msh}{M_h}
\newcommand{\mha}{M_{_A}}
\newcommand{\mshs}{M^2_h}
\newcommand{\mhss}{M^2_{\ssH}}
\newcommand{\mhas}{M^2_{_A}}
\newcommand{\barfp}{\overline{f'}}                
\newcommand{\chiii}{{\hat c}^3}
\newcommand{\chiv}{{\hat c}^4}
\newcommand{\chv}{{\hat c}^5}
\newcommand{\chvi}{{\hat c}^6}
\newcommand{\alsvi}{\alpha^{6}_{_S}}
\newcommand{\tww}{t_{_W}}
\newcommand{\ti}{t_{1}}
\newcommand{\tii}{t_{2}}
\newcommand{\tiii}{t_{3}}
\newcommand{\tiv}{t_{4}}
\newcommand{\psla}{\hbox{\rlap/p}}
\newcommand{\qsla}{\hbox{\rlap/q}}
\newcommand{\nsla}{\hbox{\rlap/n}}
\newcommand{\lsla}{\hbox{\rlap/l}}
\newcommand{\msla}{\hbox{\rlap/m}}
\newcommand{\cnsla}{\hbox{\rlap/N}}
\newcommand{\clsla}{\hbox{\rlap/L}}
\newcommand{\cmsla}{\hbox{\rlap/M}}
\newcommand{\blmt}{\lrbr - 3\rrbr}
\newcommand{\blfo}{\lrbr 4 1\rrbr}
\newcommand{\bltp}{\lrbr 2 +\rrbr}
\newcommand{\clitwo}[1]{{\rm{Li}}_{2}\lpar{#1}\rpar}
\newcommand{\clitri}[1]{{\rm{Li}}_{3}\lpar{#1}\rpar}
\newcommand{\xt}{x_{\ft}}
\newcommand{\zt}{z_{\ft}}
\newcommand{\Ht}{h_{\ft}}
\newcommand{\xts}{x^2_{\ft}}
\newcommand{\zts}{z^2_{\ft}}
\newcommand{\Hts}{h^2_{\ft}}
\newcommand{\ztc}{z^3_{\ft}}
\newcommand{\Htc}{h^3_{\ft}}
\newcommand{\ztq}{z^4_{\ft}}
\newcommand{\Htq}{h^4_{\ft}}
\newcommand{\ztv}{z^5_{\ft}}
\newcommand{\Htv}{h^5_{\ft}}
\newcommand{\ztx}{z^6_{\ft}}
\newcommand{\Htx}{h^6_{\ft}}
\newcommand{\ztz}{z^7_{\ft}}
\newcommand{\Htz}{h^7_{\ft}}
\newcommand{\sht}{\sqrt{\Ht}}
\newcommand{\atan}[1]{{\rm{arctan}}\lpar{#1}\rpar}
\newcommand{\dbff}[3]{{\hat{B}}_{_{{#2}{#3}}}\lpar{#1}\rpar}
\newcommand{\ztbs}{{\bar{z}}^{2}_{\ft}}
\newcommand{\ztb}{{\bar{z}}_{\ft}}
\newcommand{\Htbs}{{\bar{h}}^{2}_{\ft}}
\newcommand{\Htb}{{\bar{h}}_{\ft}}
\newcommand{\Hztb}{{\bar{hz}}_{\ft}}
\newcommand{\Ln}[1]{{\rm{Ln}}\lpar{#1}\rpar}
\newcommand{\Lns}[1]{{\rm{Ln}}^2\lpar{#1}\rpar}
\newcommand{\wt}{w_{\ft}}
\newcommand{\wts}{w^2_{\ft}}
\newcommand{\wtb}{\overline{w}}
\newcommand{\fra}{\frac{1}{2}}
\newcommand{\frb}{\frac{1}{4}}
\newcommand{\frc}{\frac{3}{2}}
\newcommand{\frd}{\frac{3}{4}}
\newcommand{\fre}{\frac{9}{2}}
\newcommand{\frf}{\frac{9}{4}}
\newcommand{\frg}{\frac{5}{4}}
\newcommand{\frh}{\frac{5}{2}}
\newcommand{\fri}{\frac{1}{8}}
\newcommand{\frj}{\frac{7}{4}}
\newcommand{\frl}{\frac{7}{8}}
\newcommand{\Spzzh}{\hat{\Sigma}'_{_{\zb\zb}}}
\newcommand{\sss}{s\sqrt{s}}
\newcommand{\sqs}{\sqrt{s}}
\newcommand{\Rtg}{R_{_{3Q}}}
\newcommand{\Rtt}{R_{_{33}}}
\newcommand{\Rww}{R_{_{\wb\wb}}}
\newcommand{\ssA}{{\scriptscriptstyle{A}}}
\newcommand{\ssB}{{\scriptscriptstyle{B}}}
\newcommand{\ssC}{{\scriptscriptstyle{C}}}
\newcommand{\ssD}{{\scriptscriptstyle{D}}}
\newcommand{\ssE}{{\scriptscriptstyle{E}}}
\newcommand{\ssF}{{\scriptscriptstyle{F}}}
\newcommand{\ssG}{{\scriptscriptstyle{G}}}
\newcommand{\ssH}{{\scriptscriptstyle{H}}}
\newcommand{\ssI}{{\scriptscriptstyle{I}}}
\newcommand{\ssJ}{{\scriptscriptstyle{J}}}
\newcommand{\ssK}{{\scriptscriptstyle{K}}}
\newcommand{\ssL}{{\scriptscriptstyle{L}}}
\newcommand{\ssM}{{\scriptscriptstyle{M}}}
\newcommand{\ssN}{{\scriptscriptstyle{N}}}
\newcommand{\ssO}{{\scriptscriptstyle{O}}}
\newcommand{\ssP}{{\scriptscriptstyle{P}}}
\newcommand{\ssQ}{{\scriptscriptstyle{Q}}}
\newcommand{\ssR}{{\scriptscriptstyle{R}}}
\newcommand{\ssS}{{\scriptscriptstyle{S}}}
\newcommand{\ssT}{{\scriptscriptstyle{T}}}
\newcommand{\ssU}{{\scriptscriptstyle{U}}}
\newcommand{\ssV}{{\scriptscriptstyle{V}}}
\newcommand{\ssW}{{\scriptscriptstyle{W}}}
\newcommand{\ssX}{{\scriptscriptstyle{X}}}
\newcommand{\ssY}{{\scriptscriptstyle{Y}}}
\newcommand{\ssZ}{{\scriptscriptstyle{Z}}}
\newcommand{\ssWF}{\rm{\scriptscriptstyle{WF}}}
\newcommand{\OLA}{\rm{\scriptscriptstyle{OLA}}}
\newcommand{\QED}{\rm{\scriptscriptstyle{QED}}}
\newcommand{\QCD}{\rm{\scriptscriptstyle{QCD}}}
\newcommand{\EW}{\rm{\scriptscriptstyle{EW}}}
\newcommand{\SM}{\rm{\scriptscriptstyle{SM}}}
\newcommand{\UV}{\rm{\scriptscriptstyle{UV}}}
\newcommand{\IR}{\rm{\scriptscriptstyle{IR}}}
\newcommand{\OMS}{\rm{\scriptscriptstyle{OMS}}}
\newcommand{\GMS}{\rm{\scriptscriptstyle{GMS}}}
\newcommand{\WF}{\rm{\scriptscriptstyle{WF}}}
\newcommand{\NLO}{\rm{\scriptscriptstyle{NLO}}}
\newcommand{\LO}{\rm{\scriptscriptstyle{LO}}}
\newcommand{\CMRP}{\rm{\scriptscriptstyle{CMRP}}}
\newcommand{\CMCP}{\rm{\scriptscriptstyle{CMCP}}}
\newcommand{\DiagramFermionToBosonFullWithMomenta}[8][70]{
  \vcenter{\hbox{
  \SetScale{0.8}
  \begin{picture}(#1,50)(15,15)
    \put(27,22){$\nearrow$}      
    \put(27,54){$\searrow$}    
    \put(59,29){$\to$}    
    \ArrowLine(25,25)(50,50)      \Text(34,20)[lc]{#6} \Text(11,20)[lc]{#3}
    \ArrowLine(50,50)(25,75)      \Text(34,60)[lc]{#7} \Text(11,60)[lc]{#4}
    \Photon(50,50)(90,50){2}{8}   \Text(80,40)[lc]{#2} \Text(55,33)[ct]{#8}
    \Vertex(50,50){2,5}          \Text(60,48)[cb]{#5} 
    \Vertex(90,50){2}
  \end{picture}}}
  }
\newcommand{\DiagramFermionToBosonPropagator}[4][85]{
  \vcenter{\hbox{
  \SetScale{0.8}
  \begin{picture}(#1,50)(15,15)
    \ArrowLine(25,25)(50,50)
    \ArrowLine(50,50)(25,75)
    \Photon(50,50)(105,50){2}{8}   \Text(90,40)[lc]{#2}
    \Vertex(50,50){0.5}         \Text(80,48)[cb]{#3}
    \GCirc(82,50){8}{1}            \Text(55,48)[cb]{#4}
    \Vertex(105,50){2}
  \end{picture}}}
  }
\newcommand{\DiagramFermionToBosonEffective}[3][70]{
  \vcenter{\hbox{
  \SetScale{0.8}
  \begin{picture}(#1,50)(15,15)
    \ArrowLine(25,25)(50,50)
    \ArrowLine(50,50)(25,75)
    \Photon(50,50)(90,50){2}{8}   \Text(80,40)[lc]{#2}
    \BBoxc(50,50)(5,5)            \Text(55,48)[cb]{#3}
    \Vertex(90,50){2}
  \end{picture}}}
  }
\newcommand{\DiagramFermionToBosonFull}[3][70]{
  \vcenter{\hbox{
  \SetScale{0.8}
  \begin{picture}(#1,50)(15,15)
    \ArrowLine(25,25)(50,50)
    \ArrowLine(50,50)(25,75)
    \Photon(50,50)(90,50){2}{8}   \Text(80,40)[lc]{#2}
    \Vertex(50,50){2.5}          \Text(60,48)[cb]{#3}
    \Vertex(90,50){2}
  \end{picture}}}
  }
\newcommand{\expgw}{\frac{\gf\mws}{2\srt\,\pi^2}}
\newcommand{\expgz}{\frac{\gf\mzs}{2\srt\,\pi^2}}
\newcommand{\Spww}{\Sigma'_{_{\wb\wb}}}
\newcommand{\shf}{{\hat s}^4}
\newcommand{\acz}{\scff{0}}
\newcommand{\acoo}{\scff{11}}
\newcommand{\acod}{\scff{12}}
\newcommand{\acdo}{\scff{21}}
\newcommand{\acdd}{\scff{22}}
\newcommand{\acdt}{\scff{23}}
\newcommand{\acdq}{\scff{24}}
\newcommand{\acto}{\scff{31}}
\newcommand{\actd}{\scff{32}}
\newcommand{\actt}{\scff{33}}
\newcommand{\actq}{\scff{34}}
\newcommand{\actc}{\scff{35}}
\newcommand{\acts}{\scff{36}}
\newcommand{\acoA}{\scff{1A}}
\newcommand{\acdA}{\scff{2A}}
\newcommand{\acdB}{\scff{2B}}
\newcommand{\acdC}{\scff{2C}}
\newcommand{\acdD}{\scff{2D}}
\newcommand{\actA}{\scff{3A}}
\newcommand{\actB}{\scff{3B}}
\newcommand{\actC}{\scff{3C}}
\newcommand{\adb}{\sdff{11}}
\newcommand{\adc}{\sdff{12}}
\newcommand{\add}{\sdff{13}}
\newcommand{\ade}{\sdff{21}}
\newcommand{\adf}{\sdff{22}}
\newcommand{\adg}{\sdff{23}}
\newcommand{\adh}{\sdff{24}}
\newcommand{\adi}{\sdff{25}}
\newcommand{\adj}{\sdff{26}}
\newcommand{\adl}{\sdff{27}}
\newcommand{\adm}{\sdff{31}}
\newcommand{\adn}{\sdff{32}}
\newcommand{\ado}{\sdff{33}}
\newcommand{\adp}{\sdff{34}}
\newcommand{\adq}{\sdff{35}}
\newcommand{\adr}{\sdff{36}}
\newcommand{\ads}{\sdff{37}}
\newcommand{\adt}{\sdff{38}}
\newcommand{\adu}{\sdff{39}}
\newcommand{\adw}{\sdff{310}}
\newcommand{\adv}{\sdff{311}}
\newcommand{\ady}{\sdff{312}}
\newcommand{\adz}{\sdff{313}}
\newcommand{\admt}{\frac{\tman}{\sman}}
\newcommand{\admu}{\frac{\uman}{\sman}}
\newcommand{\frm}{\frac{3}{8}}
\newcommand{\frn}{\frac{5}{8}}
\newcommand{\fro}{\frac{15}{8}}
\newcommand{\frp}{\frac{3}{16}}
\newcommand{\frr}{\frac{1}{16}}
\newcommand{\frs}{\frac{7}{2}}
\newcommand{\frt}{\frac{7}{16}}
\newcommand{\fru}{\frac{1}{3}}
\newcommand{\frw}{\frac{2}{3}}
\newcommand{\frz}{\frac{4}{3}}
\newcommand{\fry}{\frac{13}{3}}
\newcommand{\fraa}{\frac{11}{4}}
\newcommand{\bee}{\beta_{e}}
\newcommand{\beW}{\beta_{_\wb}}
\newcommand{\beWs}{\beta^2_{_\wb}}
\newcommand{\etaW}{\eta_{_\wb}}
\newcommand{\toDdrh}{{\ds\frac{2}{{\hat{\varepsilon}}}}}
\newcommand{\bqas}{\begin{eqnarray*}}
\newcommand{\eqas}{\end{eqnarray*}}
\newcommand{\mhcub}{M^3_{_H}}
\newcommand{\adComA}{\sdff{A}}
\newcommand{\adComB}{\sdff{B}}
\newcommand{\adComC}{\sdff{C}}
\newcommand{\adComD}{\sdff{D}}
\newcommand{\adComE}{\sdff{E}}
\newcommand{\adComF}{\sdff{F}}
\newcommand{\adComG}{\sdff{G}}
\newcommand{\adComH}{\sdff{H}}
\newcommand{\adComI}{\sdff{I}}
\newcommand{\adComJ}{\sdff{J}}
\newcommand{\adComL}{\sdff{L}}
\newcommand{\adComM}{\sdff{M}}
\newcommand{\adComN}{\sdff{N}}
\newcommand{\adComO}{\sdff{O}}
\newcommand{\adComP}{\sdff{P}}
\newcommand{\adComQ}{\sdff{Q}}
\newcommand{\adComR}{\sdff{R}}
\newcommand{\adComS}{\sdff{S}}
\newcommand{\adComT}{\sdff{T}}
\newcommand{\adComU}{\sdff{U}}
\newcommand{\adComAc}{\sdff{A}^c}
\newcommand{\adComBc}{\sdff{B}^c}
\newcommand{\adComCc}{\sdff{C}^c}
\newcommand{\adComDc}{\sdff{D}^c}
\newcommand{\adComEc}{\sdff{E}^c}
\newcommand{\adComFc}{\sdff{F}^c}
\newcommand{\adComGc}{\sdff{G}^c}
\newcommand{\adComHc}{\sdff{H}^c}
\newcommand{\adComIc}{\sdff{I}^c}
\newcommand{\adComJc}{\sdff{J}^c}
\newcommand{\adComLc}{\sdff{L}^c}
\newcommand{\adComMc}{\sdff{M}^c}
\newcommand{\adComNc}{\sdff{N}^c}
\newcommand{\adComOc}{\sdff{O}^c}
\newcommand{\adComPc}{\sdff{P}^c}
\newcommand{\adComQc}{\sdff{Q}^c}
\newcommand{\adComRc}{\sdff{R}^c}
\newcommand{\adComSc}{\sdff{S}^c}
\newcommand{\adComTc}{\sdff{T}^c}
\newcommand{\adComUc}{\sdff{U}^c}
\newcommand{\adComAf}{\sdff{A}^f}
\newcommand{\adComBf}{\sdff{B}^f}
\newcommand{\adComCf}{\sdff{F}^f}
\newcommand{\adComDf}{\sdff{D}^f}
\newcommand{\adComEf}{\sdff{E}^f}
\newcommand{\adComFf}{\sdff{F}^f}
\newcommand{\adComGf}{\sdff{G}^f}
\newcommand{\adComHf}{\sdff{H}^f}
\newcommand{\adComIf}{\sdff{I}^f}
\newcommand{\adComJf}{\sdff{J}^f}
\newcommand{\adComLf}{\sdff{L}^f}
\newcommand{\adComMf}{\sdff{M}^f}
\newcommand{\adComNf}{\sdff{N}^f}
\newcommand{\adComOf}{\sdff{O}^f}
\newcommand{\adComPf}{\sdff{P}^f}
\newcommand{\adComQf}{\sdff{Q}^f}
\newcommand{\adComRf}{\sdff{R}^f}
\newcommand{\adComSf}{\sdff{S}^f}
\newcommand{\adComTf}{\sdff{T}^f}
\newcommand{\adComUf}{\sdff{U}^f}
\newcommand{\adComBfc}{\sdff{B}^{fc}} 
\newcommand{\adComCfco}{\sdff{C}^{fc1}}
\newcommand{\adComCfcd}{\sdff{C}^{fc2}} 
\newcommand{\adComCfct}{\sdff{C}^{fc3}} 
\newcommand{\adComDfc}{\sdff{D}^{fc}}
\newcommand{\adComEfc}{\sdff{E}^{fc}}
\newcommand{\adComFfc}{\sdff{F}^{fc}}
\newcommand{\adComGfc}{\sdff{G}^{fc}}
\newcommand{\adComHfc}{\sdff{H}^{fc}}
\newcommand{\adComLfc}{\sdff{L}^{fc}}
\newcommand{\afba}[1]{A^{#1}_{_{\rm FB}}}
\newcommand{\alra}[1]{A^{#1}_{_{\rm LR}}}
\newcommand{\adComAt}{\sdff{A}^t}
\newcommand{\adComBt}{\sdff{B}^t}
\newcommand{\adComCt}{\sdff{T}^t}
\newcommand{\adComDt}{\sdff{D}^t}
\newcommand{\adComEt}{\sdff{E}^t}
\newcommand{\adComFt}{\sdff{T}^t}
\newcommand{\adComGt}{\sdff{G}^t}
\newcommand{\adComHt}{\sdff{H}^t}
\newcommand{\adComIt}{\sdff{I}^t}
\newcommand{\adComJt}{\sdff{J}^t}
\newcommand{\adComLt}{\sdff{L}^t}
\newcommand{\adComMt}{\sdff{M}^t}
\newcommand{\adComNt}{\sdff{N}^t}
\newcommand{\adComOt}{\sdff{O}^t}
\newcommand{\adComPt}{\sdff{P}^t}
\newcommand{\adComQt}{\sdff{Q}^t}
\newcommand{\adComRt}{\sdff{R}^t}
\newcommand{\adComSt}{\sdff{S}^t}
\newcommand{\adComTt}{\sdff{T}^t}
\newcommand{\adComUt}{\sdff{U}^t}
\newcommand{\adComAtt}{\sdff{A}^{\tau}}
\newcommand{\adComBtt}{\sdff{B}^{\tau}}
\newcommand{\adComCtt}{\sdff{T}^{\tau}}
\newcommand{\adComDtt}{\sdff{D}^{\tau}}
\newcommand{\adComEtt}{\sdff{E}^{\tau}}
\newcommand{\adComFtt}{\sdff{T}^{\tau}}
\newcommand{\adComGtt}{\sdff{G}^{\tau}}
\newcommand{\adComHtt}{\sdff{H}^{\tau}}
\newcommand{\adComItt}{\sdff{I}^{\tau}}
\newcommand{\adComJtt}{\sdff{J}^{\tau}}
\newcommand{\adComLtt}{\sdff{L}^{\tau}}
\newcommand{\adComMtt}{\sdff{M}^{\tau}}
\newcommand{\adComNtt}{\sdff{N}^{\tau}}
\newcommand{\adComOtt}{\sdff{O}^{\tau}}
\newcommand{\adComPtt}{\sdff{P}^{\tau}}
\newcommand{\adComQtt}{\sdff{Q}^{\tau}}
\newcommand{\adComRtt}{\sdff{R}^{\tau}}
\newcommand{\adComStt}{\sdff{S}^{\tau}}
\newcommand{\adComTtt}{\sdff{T}^{\tau}}
\newcommand{\adComUtt}{\sdff{U}^{\tau}}
\newcommand{\etavz}[1]{\eta^{\zb #1}_{_V}}
\newcommand{\phanst}{$\hphantom{\sigma^{s+t}\ }$}
\newcommand{\phanat}{$\hphantom{A_{FB}^{s+t}\ }$}
\newcommand{\phanss}{$\hphantom{\sigma^{s}\ }$}
\newcommand{\phanas}{$\hphantom{A_{FB}^{s}\ }$} 
\newcommand{\pbb}{\,\mbox{\bf pb}}
\newcommand{\pe}{\,\%\:}
\newcommand{\pc}{\,\%}
\newcommand{\temiv}{10^{-4}}
\newcommand{\temv}{10^{-5}}
\newcommand{\temvi}{10^{-6}}
\newcommand{\di}[1]{d_{#1}}
\newcommand{\delip}[1]{\Delta_+\lpar{#1}\rpar}
\newcommand{\propbb}[5]{{{#1}\over {\lpar #2^2 + #3 - \ib\varepsilon\rpar
\lpar\lpar #4\rpar^2 + #5 -\ib\varepsilon\rpar}}}
\newcommand{\cfft}[5]{C_{#1}\lpar #2;#3,#4,#5\rpar}    
\newcommand{\ppl}[1]{p_{+{#1}}}
\newcommand{\pmi}[1]{p_{-{#1}}}
\newcommand{\bpox}{\beta^2_{\xi}}
\newcommand{\bffdiff}[5]{B_{\rm d}\lpar #1;#2,#3;#4,#5\rpar}             
\newcommand{\cffdiff}[7]{C_{\rm d}\lpar #1;#2,#3,#4;#5,#6,#7\rpar}    
\newcommand{\affdiff}[2]{A_{\rm d}\lpar #1;#2\rpar}             
\newcommand{\Dqf}{\Delta\qf}
\newcommand{\bposx}{\beta^4_{\xi}}
\newcommand{\svverti}[3]{f^{#1}_{#2}\lpar{#3}\rpar}
\newcommand{\Mods}{\mbox{$M^2_{12}$}}
\newcommand{\Mots}{\mbox{$M^2_{13}$}}
\newcommand{\Motq}{\mbox{$M^4_{13}$}}
\newcommand{\Mdts}{\mbox{$M^2_{23}$}}
\newcommand{\Mdos}{\mbox{$M^2_{21}$}}
\newcommand{\Mtds}{\mbox{$M^2_{32}$}}
\newcommand{\dffpt}[3]{D_{#1}\lpar #2,#3;}           
\newcommand{\quu}{Q_{uu}}
\newcommand{\qdd}{Q_{dd}}
\newcommand{\qud}{Q_{ud}}
\newcommand{\qdu}{Q_{du}}
\newcommand{\msPj}[6]{\Lambda^{#1#2#3}_{#4#5#6}}
\newcommand{\bdiff}[4]{B_{\rm d}\lpar #1,#2;#3,#4\rpar}             
\newcommand{\bdifff}[7]{B_{\rm d}\lpar #1;#2;#3;#4,#5;#6,#7\rpar}             
\newcommand{\adiff}[3]{A_{\rm d}\lpar #1;#2;#3\rpar}  
\newcommand{\aw}{a_{_\wb}}
\newcommand{\az}{a_{_\zb}}
\newcommand{\sct}[1]{sect.~\ref{#1}}
\newcommand{\dreim}[1]{\varepsilon^{\rm M}_{#1}}
\newcommand{\drem}{\varepsilon^{\rm M}}
\newcommand{\hcapV}[2]{{\hat{\cal F}}^{#2}_{_{#1}}}
\newcommand{\swww}{{\scriptscriptstyle \wb\wb\wb}}
\newcommand{\szhz}{{\scriptscriptstyle \zb\hb\zb}}
\newcommand{\shzh}{{\scriptscriptstyle \hb\zb\hb}}
\newcommand{\bwith}[3]{\beta^{#3}_{#1}\lpar #2\rpar}
\newcommand{\Shhh}{{\hat\Sigma}_{_{\hb\hb}}}
\newcommand{\Sphhh}{{\hat\Sigma}'_{_{\hb\hb}}}
\newcommand{\seWilc}[1]{w_{#1}}
\newcommand{\seWtilc}[2]{w_{#1}^{#2}}
\newcommand{\eilc}{\gamma}
\newcommand{\eilcs}{\gamma^2}
\newcommand{\eilcc}{\gamma^3}
\newcommand{\eilcb}{{\overline{\gamma}}}
\newcommand{\eilcbs}{{\overline{\gamma}^2}}
\newcommand{\Sttww}{\Sigma_{_{33;\wb\wb}}}
\newcommand{\bSttww}{{\overline\Sigma}_{_{33;\wb\wb}}}
\newcommand{\Pggtg}{\Pi_{\ph\ph;3Q}}
\def\rmL{{\rm L}}
\def\rmT{{\rm T}}
\def\rmM{{\rm M}}
\newcommand{\hsm}{\hspace{-0.4mm}}
\newcommand{\hsmm}{\hspace{-0.2mm}}
\def\negs{\hspace{-0.26in}}
\def\negss{\hspace{-0.18in}}
\def\negsss{\hspace{-0.09in}}
\def\scee{{\mbox{\lowercase{$\fep\fem$}}}}
\def\scffb{{\mbox{\lowercase{$\ff\barf$}}}}
\def\scqq{{\mbox{\lowercase{$\fq\barq$}}}}
\def\scln{{\mbox{\lowercase{$\fl\fnu$}}}}
\def\app#1#2 {{\it Acta. Phys. Pol.} {\bf#1},#2}
\def\cpc#1#2 {{\it Computer Phys. Comm.} {\bf#1},#2}
\def\np#1#2 {{\it Nucl. Phys.} {\bf#1},#2}
\def\pl#1#2 {{\it Phys. Lett.} {\bf#1},#2}
\def\prep#1#2 {{\it Phys. Rep.} {\bf#1},#2}
\def\prev#1#2 {{\it Phys. Rev.} {\bf#1},#2}
\def\prl#1#2 {{\it Phys. Rev. Lett.} {\bf#1},#2}
\def\zp#1#2 {{\it Zeit. Phys.} {\bf#1},#2}
\def\sptp#1#2 {{\it Suppl. Prog. Theor. Phys.} {\bf#1},#2}
\def\mpl#1#2 {{\it Modern Phys. Lett.} {\bf#1},#2}
\def\jetp#1#2 {{\it Sov. Phys. JETP} {\bf#1},#2}
\def\fpj#1#2 {{\it Fortschr. Phys.} {\bf#1},#2}
\def\afp#1#2 {{\it Acta.Phys. Polon.} {\bf#1},#2}
\def\err#1#2 {{\it Erratum} {\bf#1},#2}
\def\ijmp#1#2 {{\it Int. J. Mod. Phys} {\bf#1},#2}
\def\nc#1#2 {{\it Nuovo Cimento} {\bf#1},#2}
\def\ap#1#2 {{\it Ann. Phys.} {\bf#1},#2}
\def\cmp#1#2 {{\it Comm. Math. Phys.} {\bf#1},#2}
\def\el#1#2 {{\it Europhys. Lett.} {\bf#1},#2}
\def\hpa#1#2 {{\it Helv. Phys. Acta} {\bf#1},#2}
\def\yf#1#2 {{\it Yad. Fiz.} {\bf#1},#2}
\def\nim#1#2 {{\it Nucl. Instrum. Meth.} {\bf#1},#2}
\def\spz#1#2 {{\it Sov. Pisma Zhetf} {\bf#1},#2}
\def\jetpl#1#2 {{\it JETP Lett.} {\bf#1},#2}
\def\sjnp#1#2 {{\it Sov. J. Nucl. Phys.} {\bf#1},#2}
\def\ptp#1#2 {{\it Progr. Theor. Phys. (Kyoto)} {\bf#1},#2}
\def\rmp#1#2  {{\it Rev. Mod. Phys.} {\bf#1},#2}
\def\zhetf#1#2 {{\it ZhETF} {\bf#1},#2}
\def\prs#1#2 {{\it Proc. Roy. Soc.} {\bf#1},#2}
\def\phys#1#2 {{\it Physica} {\bf#1},#2}
\def\itetal{{\it et al.}}
\newcommand{\btp}{\beta'_t}
\newcommand{\kbar}{\overline k}
\newcommand{\qmomit}[2]{q_{#1#2}}
\newcommand{\bDelta}{{\bar\Delta}}
\newcommand{\tDelta}{{\tilde\Delta}}
\newcommand{\tcft}[1]{C_{#1}\lpar t\rpar}
\newcommand{\tcftt}[1]{C_{#1}\lpar tt\rpar}
\newcommand{\tcfp}[1]{C^+_{#1}}
\newcommand{\tcfm}[1]{C^-_{#1}}
\newcommand{\rcn}{{\rm cn}}
\newcommand{\rdn}{{\rm dn}}
\newcommand{\rsn}{{\rm sn}}
\newcommand{\ram}{{\rm am}}
\newcommand{\epsi}[1]{\psi\lpar#1\rpar}               
\newcommand{\egams}[1]{\Gamma^2\lpar#1\rpar}               
\newcommand{\itpf}{\frac{1}{\lpar 2\pi\rpar^4}}
\newcommand{\intfxx}[2]{\int_{\scriptstyle 0}^{\scriptstyle 1}\,d#1\,
                        \int_{\scriptstyle 0}^{\scriptstyle 1}\,d#2}
\newcommand{\lbpa}{\Bigl(}                            
\newcommand{\rbpa}{\Bigl)}                            
\def\bfi{\begin{figure}}
\def\efi{\end{figure}}
\newcommand{\intmomsii}[3]{\int\,d^{#1}#2\,d^{#1}#3}
\newcommand{\intfz}{\int_{\scriptstyle 0}^y\,dz}
\newcommand{\intfzp}{\int_{\scriptstyle 0}^{1-y}\,dz}
\newcommand{\intfy}{\int_z^{\scriptstyle 1}\,dy}
\newcommand{\intfzs}{\int_y^{\scriptstyle 1}\,dz}
\newcommand{\hyper}[4]{{}_2F_1(#1\,,\,#2\,;\,#3\,;\,#4)}
\newcommand{\intfyy}[4]{\int_{\scriptstyle #1}^{\scriptstyle #2}\,dy_1\,
                        \int_{\scriptstyle #3}^{\scriptstyle #4}\,dy_2}
\newcommand{\intfyyy}[6]{\int_{\scriptstyle #1}^{\scriptstyle #2}\,dy_1\,
                         \int_{\scriptstyle #3}^{\scriptstyle #4}\,dy_2\,
                         \int_{\scriptstyle #5}^{\scriptstyle #6}\,dy_3}
\newcommand{\intfyyyp}[6]{\int_{\scriptstyle #1}^{\scriptstyle #2}\,dy_3\,
                          \int_{\scriptstyle #3}^{\scriptstyle #4}\,dy_1\,
                          \int_{\scriptstyle #5}^{\scriptstyle #6}\,dy_2}
\newcommand{\yoij}{y_{\scriptstyle 1ij}}
\newcommand{\ytij}{y_{\scriptstyle 2ij}}
\newcommand{\intfg}[1]{\int\,d[#1]}
\newcommand{\sst}{\scriptstyle}
\newcommand{\hdel}{{\hat\delta}}
\newcommand{\bdel}{{\bar\delta}}
\newcommand{\hgam}{{\hat\gamma}}
\newcommand{\bgam}{{\bar\gamma}}
\newcommand{\hmu}{{\hat{\mu}}}
\newcommand{\hnu}{{\hat{\nu}}}
\newcommand{\bmu}{{\bar{\mu}}}
\newcommand{\htp}{{\hat{p}}}
\newcommand{\brp}{{\bar{p}}}
\newcommand{\NS}[2]{\mathrm{S}_{#1}\lpar\displaystyle{#2}\rpar} 
\newcommand{\intfxxx}[3]{\int_{\scriptstyle 0}^{\scriptstyle 1}\,d#1\,
                        \int_{\scriptstyle 0}^{\scriptstyle 1}\,d#2\,
                        \int_{\scriptstyle 0}^{\scriptstyle 1}\,d#3}
\newcommand{\muva}[1]{p_{\ssV_{#1}}}
\newcommand{\bGva}[1]{{\overline\Gamma}_{\ssV{#1}}}
\newcommand{\bmva}[1]{{\overline m}_{\ssV{#1}}}
\newcommand{\mGam}[2]{\Gamma\Bigl[\ba{c} #1 \\ #2\ea\Bigr]}
\newcommand{\dsimp}[1]{\int\,dS_{#1}}
\newcommand{\dcub}[1]{\int\,dC_{#1}}
\newcommand{\dsimpd}{\int\,dS^d_2}
\newcommand{\dcubs}[2]{\int\,dCS\lpar #1\,;\,#2 \rpar}
\newcommand{\aba}{\ssE}
\newcommand{\aban}[1]{#1;\ssE}
\newcommand{\aca}{\ssI}
\newcommand{\acan}[1]{#1;\ssI}
\newcommand{\ada}{\ssM}
\newcommand{\adan}[1]{#1;\ssM}
\newcommand{\babn}[1]{{_{\scriptstyle{#1;212}}}}
\newcommand{\babb}{{_{\scriptstyle{212}|b}}}
\newcommand{\bba}{\ssG}
\newcommand{\bban}[1]{#1;\ssG}
\newcommand{\bbab}{\ssG|b}
\newcommand{\bbabn}[1]{\ssG|b|#1}
\newcommand{\bca}{\ssK}
\newcommand{\bcan}[1]{#1;\ssK}
\newcommand{\bbb}{\ssH}
\newcommand{\bbbn}[1]{#1;\ssH}
\newcommand{\bto}{\beta_{1t}}
\newcommand{\btt}{\beta_{2t}}
\newcommand{\hPi}{{\hat\Pi}}
\newcommand{\stwvi}{s_{\theta}^6}
\newcommand{\stwviii}{s_{\theta}^8}
\newcommand{\bX}{{\overline X}}
\newcommand{\balpha}{{\overline\alpha}}
\newcommand{\ox}{{\overline x}}
\newcommand{\oy}{{\overline y}}
\newcommand{\bchi}{{\overline \chi}}
\newcommand{\bbeta}{{\overline \beta}}
\newcommand{\bxi}{{\overline \xi}}
\newcommand{\chiu}[1]{\chi_{_{#1}}}
\newcommand{\TS}{{\cal T}op{\cal S}ideF}
\newcommand{\LB}{{\cal L}oop{\cal B}ack}
\newcommand{\GS}{{\cal G}raph{\cal S}hot}
\newcommand{\barX}{{\bar X}}
\newcommand{\intsx}[1]{\int_{\scriptstyle 0}^{\scriptstyle 1}\!\!\!d#1}
\newcommand{\intsxy}[2]{\int_{\scriptstyle 0}^{\scriptstyle 1}\!\!\!d#1
                        \int_{\scriptstyle 0}^{\scriptstyle #1}\!\!\!d#2}
\newcommand{\intsxx}[2]{\int_{\scriptstyle 0}^{\scriptstyle 1}\!\!\!d#1
                        \int_{\scriptstyle 0}^{\scriptstyle 1}\!\!\!d#2}
\newcommand{\intsxyz}[3]{\int_{\scriptstyle 0}^{\scriptstyle 1}\!\!\!d#1
                         \int_{\scriptstyle 0}^{\scriptstyle #1}\!\!\!d#2
                         \int_{\scriptstyle 0}^{\scriptstyle #2}\!\!\!d#3}
\newcommand{\Bbt}{B_{_{\scriptstyle\chi}}}
\newcommand{\Xbt}{X_{_{\scriptstyle\chi}}}
\newcommand{\bXbt}{\bX_{_{\scriptstyle\chi}}}
\newcommand{\bmid}{\Bigr|}
\newcommand{\spliti}[6]{\int_{\scriptstyle #1,#2}^{\scriptstyle #2,#3}\,d#4\,#5_{\scriptstyle 1\oplus 2\,;\,#6}}
\newcommand{\triagi}[3]{\int_{\scriptstyle ( #1\,,\,#2\,,\,#3)}}
\newcommand{\oDUV}{{\overline\Delta}_{\ssU\ssV}}
\newcommand{\DUV}{{\Delta}_{\ssU\ssV}}
\newcommand{\ghat}{{\hat g}}
\newcommand{\Mhat}{{\hat M}}
\newcommand{\shq}{{\hat s}^4}
\newcommand{\shsix}{{\hat s}^6}
\newcommand{\cpw}{s_{\ssW}}
\newcommand{\cpz}{s_{\ssZ}}
\newcommand{\cpv}{s_{\ssV}}
\newcommand{\ext}{\,;\,{\rm ext}}
\newcommand{\rpw}{\mu^2_{\ssW}}
\newcommand{\rpz}{\mu^2_{\ssZ}}
\newcommand{\uv}{\,;\,\ssU\ssV}
\newcommand{\Fin}{\,;\,\ssF}
\newcommand{\LFo}{\scriptscriptstyle{\Phi_1,\dots,\Phi_N\,;\,1}}
\newcommand{\LFt}{\scriptscriptstyle{\Phi_1,\dots,\Phi_N\,;\,2}}
\newcommand{\gpAA}{\xi_{\ssA\ssA}}
\newcommand{\gpAZ}{\xi_{\ssA\ssZ}}
\newcommand{\gpA}{\xi_{\ssA}}
\newcommand{\gpZ}{\xi_{\ssZ}}
\newcommand{\gpW}{\xi_{\ssW}}
\newcommand{\Gred}{G_{\rm red}}
\newcommand{\Girr}{G_{\rm irr}}
\newcommand{\rre}{\,;\,{\rm red}}
\newcommand{\irr}{\,;\,{\rm irr}}
\newcommand{\ren}{\,;\,{\rm ren}}
\newcommand{\ct}{\,;\,{\rm ct}}
\newcommand{\cph}{s_{\ssH}}
\newcommand{\Cph}{S_{\ssH}}
\newcommand{\rph}{\mu^2_{\ssH}}
\newcommand{\srph}{\mu_{\ssH}}
\newcommand{\lgh}{\gamma_{\ssH}}
\newcommand{\brph}{{\overline\mu}^2_{\ssH}}
\newcommand{\sbrph}{{\overline\mu}_{\ssH}}
\newcommand{\blh}{{\Large{$\hookleftarrow$}}}
\newcommand{\brh}{{\Large{$\hookrightarrow$}}}
\newcommand{\xWs}{x^2_{\ssW}}
\newcommand{\xWc}{x^3_{\ssW}}
\newcommand{\xWq}{x^4_{\ssW}}
\newcommand{\xHs}{x^2_{\ssH}}
\newcommand{\xHc}{x^3_{\ssH}}
\newcommand{\xHq}{x^4_{\ssH}}
\newcommand{\xLs}{x^2_{\ssL}}
\newcommand{\xBs}{x^2_{\ssB}}
\newcommand{\xTs}{x^2_{\ssT}}
\newcommand{\xTc}{x^3_{\ssT}}
\newcommand{\xTq}{x^4_{\ssT}}
\newcommand{\xD}{x_{\ssD}}
\newcommand{\xU}{x_{\ssU}}
\newcommand{\xW}{x_{\ssW}}
\newcommand{\xH}{x_{\ssH}}
\newcommand{\xL}{x_{\ssL}}
\newcommand{\xB}{x_{\ssB}}
\newcommand{\xT}{x_{\ssT}}
\newcommand{\xM}{x_{\ssM}}
\newcommand{\xZ}{x_{\ssZ}}
\newcommand{\bxLs}{X^2_{\ssL}}
\newcommand{\bxDs}{X^2_{\ssD}}
\newcommand{\bxUs}{X^2_{\ssU}}
\newcommand{\bxL}{X_{\ssL}}
\newcommand{\bxD}{X_{\ssD}}
\newcommand{\bxU}{X_{\ssU}}
\newcommand{\cL}{{\cal L}}
\newcommand{\cM}{{\cal M}}
\newcommand{\cC}{{\cal C}}
\newcommand{\WEAK}{\rm{\scriptscriptstyle{WEAK}}}
\newcommand{\FT}{\rm{\scriptscriptstyle{FT}}}
\newcommand{\REST}{\rm{\scriptscriptstyle{REST}}}
\newcommand{\BOX}{\rm{\scriptscriptstyle{BOX}}}
\newcommand{\RED}{\rm{\scriptscriptstyle{RED}}}
\newcommand{\WFR}{\rm{\scriptscriptstyle{WFR}}}
\newcommand{\VERT}{\rm{\scriptscriptstyle{VERT}}}
\newcommand{\CTT}{\rm{\scriptscriptstyle{CT2}}}
\newcommand{\ssQQ}{{\scriptscriptstyle{Q}\scriptscriptstyle{Q}}}
\newcommand{\ssVV}{{\scriptscriptstyle{V}\scriptscriptstyle{V}}}
\newcommand{\ssWW}{{\scriptscriptstyle{W}\scriptscriptstyle{W}}}
\newcommand{\ssHH}{{\scriptscriptstyle{H}\scriptscriptstyle{H}}}
\newcommand{\ssAZ}{{\scriptscriptstyle{AZ}}}
\newcommand{\ssZA}{{\scriptscriptstyle{ZA}}}
\newcommand{\ssAA}{{\scriptscriptstyle{AA}}}
\newcommand{\ssZZ}{{\scriptscriptstyle{ZZ}}}
\newcommand{\OS}{{\scriptscriptstyle{OS}}}
\newcommand{\oSigma}{{\overline \Sigma}}
\newcommand{\oF}{{\overline F}}
\newcommand{\tF}{{\tilde F}}
\newcommand{\oD}{{\overline \Delta}}
\newcommand{\Fsc}{\mu^2_{\ssF}}
\newcommand{\Rsc}{\mu^2_{\ssR}}
\newcommand{ \bb}{\beta_t}
\newcommand{ \Ab}{\overline{A}}
\newcommand{ \Zb}{\overline{Z}}
\newcommand{ \Xb}{\overline{X}}
\newcommand{ \mysmall}[1]{\scriptscriptstyle #1} 
\newcommand{ \Mgb}{\bar{g}}
\newcommand{ \ccb}{\bar{c}_\theta}
\newcommand{ \ssb}{\bar{s}_\theta}
\newcommand{ \MMb}{\bar{M}}
\newcommand{\pww}[1]{\Pi^{#1}_{\mu \nu, \mysmall{W} \mysmall{W} }}
\newcommand{\pzz}[1]{\Pi^{#1}_{\mu \nu, \mysmall{Z} \mysmall{Z} }}
\newcommand{\paa}[1]{\Pi^{#1}_{\mu \nu, \mysmall{A} \mysmall{A} }}
\newcommand{\paz}[1]{\Pi^{#1}_{\mu \nu, \mysmall{A} \mysmall{Z} }}
\newcommand{\pza}[1]{\Pi^{#1}_{\mu \nu, \mysmall{Z} \mysmall{A} }}
\newcommand{\pwp}[1]{\Pi^{#1}_{\mu, \mysmall{W \phi} }}
\newcommand{\ppw}[1]{\Pi^{#1}_{\nu, \mysmall{\phi W} }}
\newcommand{\pzp}[1]{\Pi^{#1}_{\mu, \mysmall{Z \phi_o} }}
\newcommand{\ppz}[1]{\Pi^{#1}_{\nu, \mysmall{\phi_o Z} }}
\newcommand{\pap}[1]{\Pi^{#1}_{\mu, \mysmall{A \phi_o}} }
\newcommand{\ppa}[1]{\Pi^{#1}_{\nu, \mysmall{\phi_o A} }}
\newcommand{\ppp}[1]{\Pi^{#1}_{\mysmall{\phi \phi}} }
\newcommand{\ppopo}[1]{\Pi^{#1}_{\mysmall{\phi_o \phi_o}} }
\newcommand{\ppzpz}[1]{\Pi^{#1}_{\mysmall{\phi_o \phi_o}} }
\newcommand{\dd}[2]{D^{(1)}_{\mysmall{{#1} {#2}}}}
\newcommand{\pp}[2]{P^{(1)}_{\mysmall{{#1} {#2}}}}
\newcommand{\GG}[2]{G^{(1)}_{\mysmall{{#1} {#2}}}}
\newcommand{\rr}[2]{R^{(1)}_{\mysmall{{#1} {#2}}}}
\newcommand{\WW}{\mysmall{WW}}
\newcommand{\II}{\mysmall{I}}
\newcommand{\QQ}{\mysmall{QQ}}
\newcommand{\MLL}{\mysmall{LL}}
\newcommand{\OO}{\mysmall{OO}}
\newcommand{\QL}{\mysmall{QL}}
\newcommand{\smallzz}{\mysmall{ZZ}}
\newcommand{\smallaz}{\mysmall{AZ}}
\newcommand{\smallaa}{\mysmall{AA}}
\newcommand{\ST}{\rm{\scriptscriptstyle{ST}}}
\newcommand{\UST}{\rm{\scriptscriptstyle{UST}}}




\subsection{NLO Electroweak for $gg \to H$}
\begin{figure}
\begin{center}
\includegraphics[bb=0 0 567 384,height=6.cm,width=12.0cm]{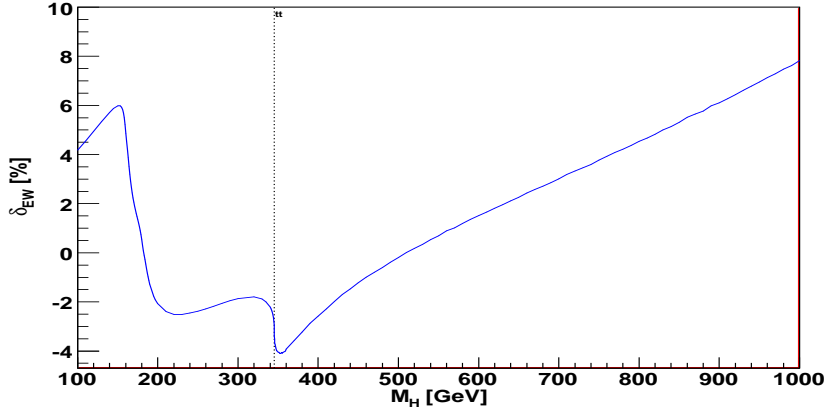}
  \caption[]{NLO electroweak percentage
    corrections to the partonic cross section
    $\sigma( g \, g \to H)$.}
     \label{fig:deltaEW}
\end{center}
\end{figure}

Gluon fusion is the main production channel for the Standard Model 
Higgs boson at hadron colliders. Unsurprisingly, radiative corrections 
have been thoroughly investigated in the past years; in particular, 
since next-to-leading order (NLO) QCD corrections increase the inclusive 
cross section for Higgs production at the LHC by a factor of about $1.5$ to 
$1.7$ with respect to the leading order (LO) term~\cite{Spira:1995rr}, 
there was a flurry of activity on higher order QCD effects. Recent 
reviews on the subject can be found in 
Refs.~\cite{Anastasiou:2008tj,Anastasiou:2005pd,Anastasiou:2004xq,Boughezal:2009py},
Refs.~\cite{Grazzini:2008zz,deFlorian:2009hc,Grazzini:2008fs}
We have recently completed the evaluation of all NLO electroweak corrections 
to the gluon-fusion Higgs production cross section at the partonic 
level in Refs.~\cite{Actis:2008ts,Actis:2008uh,Actis:2008ug,Passarino:2007fp}.
The inclusive cross section for the production of the Standard Model
Higgs boson in hadronic collisions can be written as
\bqa 
  \sigma \lpar s,\mhs \rpar &=&
  \sum_{i,j} \, \int_0^1 \! dx_1  \int_0^1 \! dx_2 \,\,
  f_{i / h_1}\lpar x_1,\Fsc \rpar \, 
  f_{j / h_2}\lpar x_2,\Fsc \rpar \,
\times \nl {}&\times&
  \int_0^1 \! dz \, \delta \lpar z -\frac{\mhs}{s\, x_1 x_2} \rpar 
  \,z\, \sigma^{(0)}\, 
  G_{ij}\lpar z;\alpha_{\ssS}(\Rsc),\mhs/\Rsc; \mhs/ \Fsc \rpar,
\label{eq:CShad}
\eqa
where $\sqrt{s}$ is the center-of-mass energy and $\mu_{\ssF}$ and 
$\mu_{\ssR}$ stand for factorization and renormalization scales.
In \eqn{eq:CShad} the partonic cross section for the sub-process $ij\to H+X$, 
with $i(j) = g, q_f, {\bar q}_f$, has been convoluted with the parton 
densities $f_{a/h_b}$ for the colliding hadrons $h_1$ and $h_2$. The Born 
factor is $\sigma^{(0)}$.
The coefficient functions $G_{ij}$ can be computed in QCD through a 
perturbative expansion in the strong-coupling constant $\alpha_\ssS$,
\bq
  G_{ij} \lpar z ; \alpha_{\ssS}(\Rsc) , \mhs/\Rsc ; \mhs/\Fsc \rpar =
  \alpha_{\ssS}^2(\Rsc) 
  \sum_{n=0}^{\infty} \lpar \frac{\alpha_{\ssS}(\Rsc)}{\pi}\rpar^n
  G_{ij}^{(n)}\lpar z;\mhs/\Rsc;\mhs/\Fsc \rpar,
\eq
with a scale-independent LO contribution given by
  $
    G^{(0)}_{ij}(z) = \delta_{ig}\,\delta_{jg}\,\delta\lpar 1 - z\rpar.
  $
The inclusion of higher order electroweak corrections in \eqn{eq:CShad} 
requires to define a factorization scheme. 
Originally, we introduced two options for replacing the purely QCD-corrected 
partonic cross section in \eqn{eq:CShad} with the expression including NLO 
EW corrections, complete factorization:
$
\sigma^{(0)}\, G_{ij} \to \sigma^{(0)}\,\lpar 1 + \delta_{\EW}\rpar\,G_{ij}$;
and partial factorization:
$
\sigma^{(0)}\,G_{ij} \to \sigma^{(0)}\,\Bigl[ G_{ij} + \alpha_\ssS^2(\Rsc) 
\delta_{\EW}\,G^{(0)}_{ij}\Bigr]$,
where $\delta_{\EW}$ embeds all NLO electroweak corrections to the partonic
cross section $\sigma(gg\to H)$, 
\bq
\sigma_{\EW}=\alpha_{\ssS}^2(\Rsc)\sigma^{(0)}(1+\delta_{\EW}),
\label{eq:deltaPART}
\eq
The CF option amounts to an overall re-scaling of the QCD result, 
dressed at all orders with the NLO electroweak correction factor $\delta_{\EW}$; 
the PF option is equivalent to add electroweak corrections to QCD ones. 
Note that a calculation of the mixed QCD - EW corrections has been performed in
Ref.~\cite{Anastasiou:2008tj}; a significant numerical difference from the
prediction of the complete factorization hypothesis has not been observed.
Results for NLO EW corrections are given in \fig{fig:deltaEW}.
\subsection{Higgs Pseudo-Observables}
The Higgs boson, as well as the $W$ or $Z$ bosons, are unstable 
particles; as such they should be removed from in/out bases in the
Hilbert space, without changing the unitarity of the theory. 
Concepts as the production of an unstable particle or its partial decay widths, 
not having a precise meaning, are only an approximation of a more complete 
description.
The inconsistencies associated with the on-shell LSZ formulation of an
unstable external particles become particularly severe starting from two-loops,
as described in Ref.~\cite{Actis:2008uh}.
At the parton level the $S\,$-matrix for the process $i \to f$ can be written 
as
\bq
S_{fi} = V_i(s)\,\Delta_{\ssH}(s)\,V_f(s) + B_{if}(s),
\label{Smat}
\eq
where $V_i$ is the production vertex $i \to H$ (e.g. $gg \to H$), $V_f$ 
is the decay vertex $H \to f$ (e.g. $H \to \gamma \gamma$), $\Delta_{\ssH}$ 
is the Dyson re-summed Higgs propagator and $B_{if}$ is the non - resonant 
background (e.g. $gg \to \gamma \gamma$ boxes). In the next section we 
will introduce the notion of complex pole. 
A vertex is defined by the following decomposition
\bq
V_f(s) = \sum_a\,V^a_f\lpar s\,,\,\{S\}\rpar\,F^a_f\lpar \{p_f\}\rpar
\eq
where $s = - P_{\ssH}^2$ (with $P_{\ssH} = \sum_f p_f$), $s\,\oplus\,\{S\}$ 
is the set of Mandelstam invariants that characterize the process 
$H \to f$, $V^a_f$ are scalar form factors and the $F^a_f$ contain spinors, 
polarization vectors etc.  
Although a modification of the LSZ reduction formulas has been proposed long
ago for unstable particles we prefer an alternative approach where one considers 
extracting informations on the Higgs boson directly from
\bq
<\,f\;{\rm out}\,|\,H\,>\,<\,H\,|\,i\;{\rm in}\,> +
\sum_{n\,\not=\,H}\,
<\,f\;{\rm out}\,|\,n\,>\,<\,n\,|\,i\;{\rm in}\,>,
\eq
for some initial state $i$ and some final state $f$ and where 
$\{n\}\,\oplus\,H$ is a complete set of states (not as in the in/out bases). 
The price to be paied is the necessity of moving into the complex plane. 
Define $\cph$ and $\Pi_{\ssHH}(s)$ as
\bq
\cph - m^2_{\ssH} + \Sigma_{\ssHH}(\cph) = 0,
\qquad
\Pi_{\ssHH}(s) = \frac{\Sigma_{\ssHH}(s) - \Sigma_{\ssHH}(\cph)}{s-\cph},
\eq
where $\Sigma_{\ssHH}$ is the Higgs self-energy; then the, Dyson re-summed, 
Higgs propagator becomes
\bq
\Delta_{\ssHH}(s)= (s - \cph)^{-1}\,\Bigl[ 1 + \Pi_{\ssHH}(s)\Bigr]^{-1},
\qquad
Z_{\ssH} = 1 + \Pi_{\ssHH}.
\label{tres}
\eq
Using \eqn{tres} we can write \eqn{Smat} as
\bq
S_{fi} = \Bigl[ Z^{-1/2}_{\ssH}(s)\,V_i(s)\Bigr]\,
            \frac{1}{s - \cph}\,
         \Bigl[ Z^{-1/2}_{\ssH}(s)\,V_f(s)\Bigr] + B_{if}(s).
\eq
From the $S\,$-matrix element for a physical process $i \to f$ we extract the 
relevant pseudo - observable,
\bq
S\lpar H_c \to f\rpar = Z^{-1/2}_{\ssH}(\cph)\,V_f(\cph),
\quad
S_{fi} = \frac{S\lpar i \to H_c \rpar\,S\lpar H_c \to f \rpar}{s - \cph} +
\hbox{non resonant terms}.
\label{PO}
\eq
which is gauge parameter independent by construction.
The partial decay width is further defined as
\bq
M_{\ssH}\,\Gamma\lpar H_c \to f\rpar = \frac{(2\,\pi)^4}{2}\,\int\,
d\Phi_f\lpar P_{\ssH}\,,\,\{p_f\}\rpar\,
\sum_{\rm spins}\,\bmid S\lpar H_c \to f\rpar \bmid^2,
\label{GPO}
\eq
where the integration is over the phase space spanned by $| f >$, with the
constraint $P_{\ssH} = \sum\,p_f$. One should not confuse phase space and
the real value of $s= -P^2_{\ssH}$ where the realistic observable is measured
with the complex value for $s$ where to compute gauge invariant loop 
corrections.
The choice of $P^2_{\ssH}$ (phase space) where to define the 
pseudo - observable is conventional and we will use $M^2_{\ssH}= \mid \cph\mid$, 
with $\cph = \rph - i\,\srph\,\lgh$. 

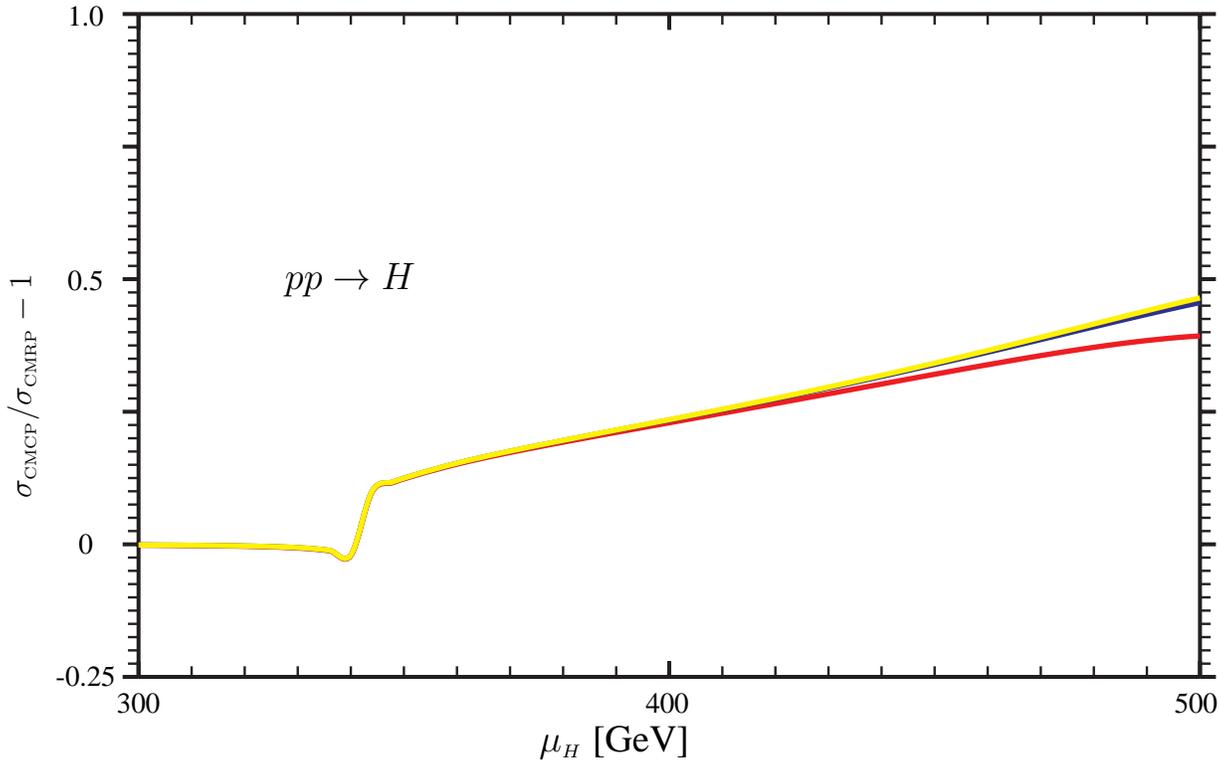
\begin{figure}[ht]
\vspace{-5.cm}
\begin{center} \begin{picture}(360,440)(0,-250)
\SetOffset(-30,-150)
\LinAxis(0,-50)(400,-50)(2,10,5,0,1.5)
\LinAxis(0,200)(400,200)(2,10,-5,0,1.5)
\LinAxis(0,-50)(0,200)(5,10,5,0,1.5)
\LinAxis(400,-50)(400,200)(5,10,-5,0,1.5)
\SetScale{200.} \SetWidth{0.01}
\SetColor{Red}
\Curve{(0.0000,-0.00157)
(0.0400,-0.00182)
(0.0800,-0.00213)
(0.1200,-0.00252)
(0.1600,-0.00302)
(0.2000,-0.00369)
(0.2400,-0.00463)
(0.2800,-0.00601)
(0.3200,-0.00827)
(0.3600,-0.01238)
(0.4000,-0.02126)
(0.4400,0.09887)
(0.4800,0.11664)
(0.5200,0.12944)
(0.5600,0.14076)
(0.6000,0.15113)
(0.6400,0.16045)
(0.6800,0.16907)
(0.7200,0.17722)
(0.7600,0.18504)
(0.8000,0.19264)
(0.8400,0.20009)
(0.8800,0.20743)
(0.9200,0.21471)
(0.9600,0.22195)
(1.0000,0.22917)
(1.0400,0.23640)
(1.0800,0.24362)
(1.1200,0.25087)
(1.1600,0.25814)
(1.2000,0.26543)
(1.2400,0.27275)
(1.2800,0.28010)
(1.3200,0.28747)
(1.3600,0.29486)
(1.4000,0.30223)
(1.4400,0.30961)
(1.4800,0.31696)
(1.5200,0.32427)
(1.5600,0.33158)
(1.6000,0.33868)
(1.6400,0.34569)
(1.6800,0.35253)
(1.7200,0.35913)
(1.7600,0.36546)
(1.8000,0.37144)
(1.8400,0.37698)
(1.8800,0.38201)
(1.9200,0.38641)
(1.9600,0.39011)
(2.0000,0.39293)}
\SetColor{Blue}
\Curve{(0.0000,-0.00123)
(0.0400,-0.00145)
(0.0800,-0.00172)
(0.1200,-0.00207)
(0.1600,-0.00252)
(0.2000,-0.00314)
(0.2400,-0.00403)
(0.2800,-0.00535)
(0.3200,-0.00755)
(0.3600,-0.01160)
(0.4000,-0.02041)
(0.4400,0.09992)
(0.4800,0.11781)
(0.5200,0.13073)
(0.5600,0.14221)
(0.6000,0.15277)
(0.6400,0.16229)
(0.6800,0.17115)
(0.7200,0.17956)
(0.7600,0.18767)
(0.8000,0.19559)
(0.8400,0.20339)
(0.8800,0.21113)
(0.9200,0.21884)
(0.9600,0.22656)
(1.0000,0.23430)
(1.0400,0.24210)
(1.0800,0.24996)
(1.1200,0.25790)
(1.1600,0.26595)
(1.2000,0.27409)
(1.2400,0.28235)
(1.2800,0.29073)
(1.3200,0.29923)
(1.3600,0.30782)
(1.4000,0.31661)
(1.4400,0.32549)
(1.4800,0.33451)
(1.5200,0.34365)
(1.5600,0.35291)
(1.6000,0.36228)
(1.6400,0.37174)
(1.6800,0.38127)
(1.7200,0.39087)
(1.7600,0.40048)
(1.8000,0.41008)
(1.8400,0.41961)
(1.8800,0.42903)
(1.9200,0.43827)
(1.9600,0.44726)
(2.0000,0.45591)}
\SetColor{Yellow}
\Curve{(0.0000,-0.00117)
(0.0400,-0.00139)
(0.0800,-0.00166)
(0.1200,-0.00199)
(0.1600,-0.00244)
(0.2000,-0.00306)
(0.2400,-0.00393)
(0.2800,-0.00525)
(0.3200,-0.00744)
(0.3600,-0.01148)
(0.4000,-0.02028)
(0.4400,0.10008)
(0.4800,0.11798)
(0.5200,0.13092)
(0.5600,0.14243)
(0.6000,0.15301)
(0.6400,0.16257)
(0.6800,0.17146)
(0.7200,0.17990)
(0.7600,0.18806)
(0.8000,0.19603)
(0.8400,0.20388)
(0.8800,0.21167)
(0.9200,0.21944)
(0.9600,0.22723)
(1.0000,0.23505)
(1.0400,0.24293)
(1.0800,0.25088)
(1.1200,0.25893)
(1.1600,0.26708)
(1.2000,0.27534)
(1.2400,0.28373)
(1.2800,0.29225)
(1.3200,0.30091)
(1.3600,0.30969)
(1.4000,0.31865)
(1.4400,0.32775)
(1.4800,0.33700)
(1.5200,0.34640)
(1.5600,0.35594)
(1.6000,0.36562)
(1.6400,0.37541)
(1.6800,0.38532)
(1.7200,0.39534)
(1.7600,0.40538)
(1.8000,0.41548)
(1.8400,0.42556)
(1.8800,0.43559)
(1.9200,0.44551)
(1.9600,0.45523)
(2.0000,0.46469)}
\SetColor{Black}
\Text(0,-60)[]{300}
\Text(200,-60)[]{400}
\Text(400,-60)[]{500}
\Text(-20,-50)[]{-0.25}
\Text(-20,0)[]{0}
\Text(-20,100)[]{0.5}
\Text(-20,200)[]{1.0}
\Text(180,-75)[]{\Large $\srph\;$[GeV]}
\Text(80,100)[]{\Large $pp \to H$}
\rText(-45,60)[][l]{\large $\sigma_{\CMCP}/\sigma_{\CMRP} - 1$}
\end{picture}
\end{center}
\vspace{-0.5cm}
\caption[]{The ratio $\sigma_{\CMCP}/\sigma_{\CMRP}$ for the production
cross-section $pp \to H$, as a function of $\srph$, for
different energies, $\sqrt{s} = 3\,$TeV (red), $\sqrt{s} = 10\,$TeV (blue)
and $\sqrt{s} = 14\,$TeV (yellow). Cross-sections are computed with
MSTW2008 LO PDFs.}
\label{R_ppH}
\end{figure}

We define two different schemes and compare their results:
the CMRP scheme~\cite{Actis:2008uh}, the complex mass scheme with complex 
internal $W$ and $Z$ poles;
the CMCP scheme, the (complete) complex mass scheme with complex,
external, Higgs ($W, Z$ etc.) where the LSZ procedure is carried out at the 
Higgs complex pole (on the second Riemann sheet).
We present the ratio of $\sigma(pp \to H)$ in the two schemes, using MSTW 2008 LO 
partondistribution functions (PDF)~\cite{Martin:2009iq}.
The ratio is given in \fig{R_ppH}.
\subsection*{ACKNOWLEDGEMENTS}
I gratefully acknowledge the unvaluable assistance of Christian Sturm and
Sandro Uccirati in all steps of this project. 




%% file: boughezal/boughezal.tex




%
%

\subsection{Introduction}
The search for the Higgs boson
is a primary goal of the CERN Large Hadron Collider
(LHC), and is a central part of Fermilab's Tevatron program. 
Recently, the Tevatron collaborations reported a 95\% confidence level
exclusion of a Standard Model Higgs boson with a mass in the range
$163-166 \,{\rm GeV}$~\cite{Collaboration:2009je}.  
\\
The dominant production mode at both colliders, gluon fusion
through top-quark loops, receives important QCD radiative
corrections~\cite{Dawson:1990zj,Djouadi:1991tka,Spira:1995rr}.
The inclusive result increases by a factor of 2 at the
LHC and 3.5 at the Tevatron
when perturbative QCD effects through next-to-next-to-leading order (NNLO), 
in the infinite top quark mass limit, are taken into 
account~\cite{Anastasiou:2002yz,Harlander:2002wh}. A review of
the latest theoretical developments in the search for the Higgs boson within 
the Standard Model can be found in~\cite{Boughezal:2009fw} .
The theoretical uncertainty from effects beyond NNLO is estimated  to be
about $\pm 10\%$ by varying renormalization and factorization scales.
At this level of precision, electroweak corrections to the Higgs signal
become important.  In Ref.~\cite{Aglietti:2004nj,Aglietti:2006yd}, the authors
pointed out the importance of a subset of diagrams where the Higgs couples
to the W and Z gauge bosons which subsequently couple to light quarks.   
A careful study of the full 2-loop electroweak effects was performed 
in Ref.~\cite{Actis:2008ug}.  They increase the leading-order cross section 
by up to $5-6\%$ for relevant Higgs masses.
An important question is whether these light-quark contributions receive
the same QCD enhancement as the top quark loops. If they do, then the full
NNLO QCD result is shifted by $+5-6\%$ from these electroweak corrections.
As this effect on the central value of the production cross section
and therefore on the exclusion limits and future measurements is
non-negligible, it is important to quantify it. 
The exact computation
of the mixed electroweak/QCD effects needed to do so requires 3-loop
diagrams with many kinematic scales, and 2-loop diagrams with four external
legs for the real-radiation terms.  Such a computation is prohibitively
difficult with current computational techniques.\\
In Ref.~\cite{Anastasiou:2008tj}, the QCD corrections to the light-quark terms
in the Higgs production cross section via gluon fusion were computed using
an effective theory approach.  This approach was rigorously justified
by applying a hard-mass expansion procedure to the full 3-loop corrections.
In addition to that,  the most up-to-date QCD prediction
for the Higgs boson production cross section in this channel was provided
for use in setting Tevatron exclusion limits. This includes 
the MSTW2008 PDFs, the exact NLO K-factors for the top, 
top-bottom interference and bottom quark contributions, NLO effects arising
from W and Z gauge bosons~\cite{Keung:2009bs} and all the 
new theoretical results.

\subsection{The mixed QCD-electroweak effects}

The cross section for Higgs boson production in hadronic collisions can be written as
\begin{eqnarray}
\sigma(s,M_H^2) &=& \sum_{i,j} \int_{0}^1 dx_1  \int_{0}^1 dx_2 \,f_{i/h_1}(x_1,\mu_F^2)  f_{j/h_2}(x_2,\mu_F^2) \int_{0}^1 dz\,\delta\left(z-\frac{M_H^2}{x_1 x_2 s}\right) \nonumber \\ &\times& z \,\hat{\sigma}_{ij}\left(z; \alpha_s(\mu_R^2), \alpha_{EW},M_H^2/\mu_R^2; M_H^2/\mu_F^2\right).
\end{eqnarray}
Here, $\sqrt{s}$ is the center-of-mass energy of the hadronic collision, $\mu_R$ and $\mu_F$ respectively denote the renormalization and factorization scales, and the $f_{i/h}$ denote the parton densities.  The quantity $z \hat{\sigma}$ is the partonic cross section for the process $ij \to H+X$ with $i,j = g,q,\bar{q}$.  As indicated, it admits a joint perturbative expansion in the strong and electroweak couplings.
Considering QCD and electroweak corrections and suppressing the scale
dependence for simplicity, the partonic cross section can be written as:

\begin{equation}
\hat{\sigma}_{ij} = \sigma^{(0)}_{\rm EW} \,G^{(0)}_{ij}\left(z\right)+\sigma^{(0)} \sum_{n=1}^{\infty} \left(\frac{\alpha_s}{\pi}\right)^n G^{(n)}_{ij}(z).
\label{pcsec2}
\end{equation}
\\
The QCD corrections to the one-loop diagrams coupling the Higgs boson
to gluons via a top-quark loop are given by
\begin{equation} 
G_{ij}(z;\alpha_s) = \sum_{n=1}^{\infty} \left(\frac{\alpha_s}{\pi}\right)^n G^{(n)}_{ij}(z).
\end{equation}
The cross section in Eq.~(\ref{pcsec2}) includes corrections to the
leading-order result valid through ${\cal O}(\alpha)$ in the electroweak
couplings and to ${\cal O}(\alpha_s^2)$ in the QCD coupling constant in the
large top-mass limit upon inclusion of the known results for $G^{(1,2)}_{ij}$.
Since the perturbative corrections to the leading-order result are large, it
is important to quantify the effect of the QCD corrections on the light-quark
electroweak contributions.  This would require knowledge of the mixed
${\cal O}(\alpha\alpha_s)$. In lieu of such a calculation, the authors of
Ref.~\cite{Actis:2008ug} studied two assumptions for the effect of QCD
corrections on the 2-loop light-quark diagrams.
\vspace{3mm}
\begin{itemize}

\item {\it Partial factorization}: no QCD corrections to the light-quark
  electroweak diagrams are included. With this assumption, electroweak
diagrams contribute only a $+1-2\%$ increase to the
Higgs boson production cross section.

\vspace{1mm}

\item{\it Complete factorization}: the QCD corrections to the electroweak
  contributions are assumed to be identical to those affecting the heavy-quark
  diagrams.
\end{itemize}
\vspace{3mm}
\noindent
In this case the light-quark diagrams increase the full NNLO QCD production
cross section by $+5-6\%$.  The last assumption was used in earlier exclusions
of a SM Higgs boson by the Tevatron collaborations.
The calculation of the ${\cal O}(\alpha\alpha_s)$, which allows one to check
these assumptions, can be done in the framework of an effective field theory
where the W-boson is integrated out:
\begin{equation}
{\cal L}_{eff} = -\alpha_s\frac{C_1}{4v} H G_{\mu\nu}^a G^{a\mu\nu}.
\label{lag}
\end{equation}
The Wilson coefficient $C_1$ arising from integrating out the heavy quark
and the W-boson is defined through
\begin{eqnarray}
C_1 &=& -\frac{1}{3\pi}\left\{1+\lambda_{EW}\left[1+a_s C_{1w}+a_s^2 C_{2w}\right]+a_sC_{1q} + a_s^2 C_{2q}  \right\}, \nonumber \\
C_{1q} &=& \frac{11}{4},\;\;\; C_{2q} = \frac{2777}{288} +\frac{19}{16}L_t+N_F
\left(-\frac{67}{96}+\frac{1}{3}L_t\right), \nonumber
\\
\lambda_{EW} &=& \frac{3\alpha}{16\pi
  s_W^2}\left\{\frac{2}{c_W^2}\left[\frac{5}{4}-\frac{7}{3}s_W^2+\frac{22}{9}s_W^4\right]+4\right\}, \nonumber
\end{eqnarray}
where $a_s = \alpha_s/\pi$, $N_F=5$ is the number of active quark flavors,
$L_t = {\rm ln}(\mu_R^2/m_t^2)$ and $s_W,c_W$ are respectively the sine and
cosine of the weak-mixing angle.
The Wilson coefficient obtained from using the complete factorization
assumption is given by
\\
\[ C_1^{fac} = -\frac{1}{3\pi}\left(1+\lambda_{EW}\right) \left\{1+a_sC_{1q} + a_s^2 C_{2q}\right\} .\]
\\
Factorization holds if $C_{1w}=C_{1q}$ and $C_{2w}=C_{2q}$. To test
this assumption, the $C_{1W}$ coefficient was calculated
in~\cite{Anastasiou:2008tj} by expanding the 3-loop
QCD corrections to the light-quark electroweak diagrams, keeping the leading
term. The numerical effect of various choices for $C_{2w}$ was
also studied.
\\
After a computation following the approach outlined above, we obtain the
following  result for $C_{1w}$:
\begin{equation}
C_{1w} = \frac{7}{6}.
\end{equation}
Two points should be noted regarding the comparison of this with
the factorization hypothesis $C_{1w}^{fac} = C_{1q} = 11/4$.
First, there is a fairly large
violation of the factorization result:~$(C_{1q}-C_{1w})/C_{1w} \approx 1.4$.
However, QCD corrections to the Higgs-gluon-gluon matrix elements are much 
larger than this difference, and a large deviation from the $+5-6\%$ shift 
found before does not occur.

\subsection{The updated integrated Higgs cross section}

In tables~(\ref{tab:TEVxsec08}) and~(\ref{tab:LHCxsec08}), 
the numerical results for the new predictions of the gluon fusion cross
section including all currently computed perturbative effects on the cross
section, are shown for both colliders, the Tevatron and the LHC. 
These are: the NNLO $K$-factor
computed in the large-$m_t$ limit and normalized to the exact $m_t$-dependent LO
result, the full light-quark electroweak correction and the ${\cal O}(\alpha_s)
$ correction to this encoded in $C_{1w}$, the bottom-quark contributions
using their NLO K-factors with the exact dependence on the bottom and top
quark masses, NLO effects arising form W and Z gauge 
bosons~\cite{Keung:2009bs} and finally the newest MSTW PDFs from
$2008$~\cite{Martin:2009iq,Martin:2009bu}. 
The uncertainty on the total cross section is estimated by accounting 
for the scale and the PDF uncertainties.
In the fourth and fifth columns we show the PDF uncertainty
alone as compared to the uncertainty obtained by accounting for the PDF 
and $\alpha_s$ uncertainties simultaneously as described 
in~\cite{Martin:2009bu}. The scale uncertainty due to missing higher
order corrections is estimated by varying the scale in the range $[M_H/4,M_H]$,
which is a factor of two around the central value $\mu_R=\mu_F=\mu=M_H/2$.
We note that the new numerical values are $4-6\%$ lower than the numbers in
Ref.~\cite{Catani:2003zt} used in an earlier Tevatron exclusion of a SM
Higgs boson.

{\renewcommand{\arraystretch}{1.2} \renewcommand{\tabcolsep}{.5mm}}
\begin{table}[t]
\begin{minipage}{0.51\textwidth}
\begin{center}
\begin{tabular}{|c|c|c|c|c|}
\hline 
      $m_{H}$[GeV] &$\sigma^{best}$[pb]& scale & PDF only & PDF+$\alpha_s$\\
        \hline \hline
110&  1.429 & $_{- 11.76}^ {+  8.79}$ & $_{-  6.16}^{ +  5.75}$ & $_{-
                        10.99}^{ + 11.46}$
\\[1mm]
\hline

115&  1.252 & $_{- 11.77}^{ +  8.65}$ & $ _{-  6.39}^{ +  5.97}$ & $_{ -
                        11.14}^{+ 11.58}$
\\[1mm]
\hline

120&  1.102 & $_{- 11.71}^{ +  8.53}$ & $_{-  6.62}^{ +  6.20}$ & $_{-
                        11.29}^{ + 11.77}$
\\[1mm]
\hline

125&  0.974 & $_{- 11.65}^{ +  8.42}$ & $_{-  6.85}^{ +  6.42}$ & $_{-
                       11.45}^{ + 12.02}$
\\[1mm]
\hline

130&  0.863 & $_{- 11.65}^{ +  8.34}$ & $_{ -  7.08}^{ +  6.64}$ & $_{ -
                       11.61}^{ + 12.19}$
\\[1mm]
\hline

135&  0.768 & $_{- 11.71}^{ +  8.26}$ & $_{-  7.31}^{ +  6.87}$ & $_{-
                       11.79}^{ + 12.38}$
\\[1mm]
\hline

140&  0.685 & $_{- 11.74}^{+  8.16}$ & $_{-  7.53}^{ +  7.09} $ & $_{-
                       11.95}^{ + 12.51}$
\\[1mm]
\hline

145&  0.613 & $_{- 11.75}^{ +  8.09}$ & $_{ -  7.76}^{ +  7.31}$ & $_{ -
                       12.08}^{ + 12.70}$
\\[1mm]
\hline

150&  0.549 & $_{- 11.75}^{ +  8.11}$ & $_{ -  7.99}^{ +  7.53}$ & $_{ -
                       12.23}^{ + 12.89}$
\\[1mm]
\hline

155&  0.494 & $_{- 11.75}^{+  8.06}$ & $_{ -  8.22}^{ +  7.75}$ & $_{ -
                       12.40}^{ + 13.10}$
\\\hline

\end{tabular}
\end{center}
\end{minipage}
\begin{minipage}{0.31\textwidth}
\begin{center}
    \begin{tabular}{|c|c|c|c|c|}
      \hline 
      $m_{H}$[GeV] &$\sigma^{best}$[fb]& scale & PDF only & PDF+$\alpha_s$\\
        \hline \hline
160&  o.442 & $_{- 11.75}^{ +  8.01}$ & $_{ -  8.45}^{ +  7.98} $ & $_{ -
                       12.56}^{ + 13.32}$
\\[1mm]
\hline

165&  0.389 & $_{- 11.78}^{ +  8.00}$ & $_{ -  8.68}^{ +  8.20}$ & $_{ -
                       12.74}^{ + 13.53}$
\\[1mm]
\hline

170&  0.347 & $_{- 11.79}^{ +  8.03}$ & $_{ -  8.90}^{ +  8.42}$ & $_{ -
                       12.86}^{ + 13.81}$
\\[1mm]
\hline

175&  0.313 & $_{- 11.81}^{ +  8.02}$ & $_{ -  9.13}^{+  8.64}$ & $_{ -
                       13.07}^{ + 14.04}$
\\[1mm]
\hline

180&  0.283 & $_{- 11.84}^{ +  8.04}$ & $_{ -  9.35}^{ +  8.85}$ & $_{ -
                       13.22}^{ + 14.28}$
\\[1mm]
\hline

185&  0.253 & $_{- 11.85}^{ +  8.06}$ & $_{ -  9.57}^{ +  9.07}$ & $_{ -
                       13.38}^{ + 14.52}$
\\[1mm]
\hline

190&  0.229 & $_{- 11.88}^{+  8.14}$ & $_{-  9.79}^{+  9.28}$ & $_{ -
                       13.56}^{+ 14.81}$
\\[1mm]
\hline

195&  0.208 & $_{- 11.96}^{ +  8.16}$ & $_{ -  9.99}^{ +  9.48}$ & $_{ -
                       13.73}^{ + 15.09}$
\\[1mm]  
\hline

200&  0.189 & $_{- 12.00}^{+  8.19}$ & $_{- 10.21}^{ +  9.69}$ & $_{ - 13.88}^{ + 15.36}$
\\[1mm]  
\hline
$-$& $-$& $-$& $-$& $-$
\\\hline

\end{tabular}
\end{center}
\end{minipage}
  \caption{{Higgs production cross
   section (MSTW08) at Tevatron for $\sqrt{s}=1.96$~TeV, with
   $\mu= \mu_R =\mu_F=M_H/2 $ and 
   $\sigma^{best} =
   \sigma_{QCD}^{NNLO}+\sigma_{EW}^{NNLO}$. The scale uncertainty, 
    PDF uncertainty without and accounting for $\alpha_s$ error   
    as described in~\cite{Martin:2009bu} are shown in the third, fourth and 
    fifth colum respectively, as a function of the Higgs boson mass. 
    All the errors are percent ones.}
    \label{tab:TEVxsec08} }
\end{table}
%
%

{\renewcommand{\arraystretch}{1.2} \renewcommand{\tabcolsep}{.5mm}}
\begin{table}[t]
\begin{minipage}{0.51\textwidth}
\begin{center}
\begin{tabular}{|c|c|c|c|c|}
\hline 
      $m_{H}$[GeV] &$\sigma^{best}$[fb]& scale & PDF only & PDF+$\alpha_s$\\
        \hline \hline
110&  37.973  &$_{-8.82}^{+  8.74}$ &$_{-3.16}^{+  2.46}$ &$_{-7.38}^{+
  7.74}$
\\[1mm]
\hline
115&  34.977  &$_{-8.82}^{+  8.48}$ &$_{-3.15}^{+  2.45}$ &$_{-7.35}^{+
  7.66}$
\\[1mm]
\hline
120&  32.301  &$_{-8.74}^{+  8.28}$ &$_{-3.13}^{+  2.44}$ &$_{-7.32}^{+
  7.63}$
\\[1mm]
\hline
125&  29.918  &$_{-8.67}^{+  8.10}$ &$_{-3.12}^{+  2.44}$ &$_{-7.29}^{+
  7.60}$
\\[1mm]
\hline
130&  27.794  &$_{-8.63}^{+  7.90}$ &$_{-3.12}^{+  2.44}$ &$_{-7.27}^{+
  7.57}$
\\[1mm]
\hline
135&  25.879  &$_{-8.55}^{+  7.77}$ &$_{-3.11}^{+  2.45}$ &$_{-7.25}^{+
  7.50}$
\\[1mm]
\hline
140&  24.151  &$_{-8.50}^{+  7.69}$ &$_{-3.11}^{+  2.46}$ &$_{-7.24}^{+
  7.52}$
\\[1mm]
\hline
145&  22.606  &$_{-8.49}^{+  7.53}$ &$_{-3.12}^{+  2.48}$ &$_{-7.23}^{+
  7.50}$
\\[1mm]
\hline
150&  21.204  &$_{-8.47}^{+  7.40}$ &$_{-3.12}^{+  2.49}$ &$_{-7.21}^{+
  7.49}$
\\[1mm]
\hline
155&  19.919  &$_{-8.43}^{+  7.31}$ &$_{-3.13}^{+  2.51}$ &$_{-7.21}^{+
  7.48}$

\\\hline

\end{tabular}
\end{center}
\end{minipage}
\begin{minipage}{0.31\textwidth}
\begin{center}
    \begin{tabular}{|c|c|c|c|c|}
      \hline 
      $m_{H}$[GeV] &$\sigma^{best}$[fb]& scale & PDF only & PDF+$\alpha_s$\\
        \hline \hline
160&  18.619  &$_{-8.40}^{+  7.24}$ &$_{-3.14}^{+  2.54}$ &$_{-7.21}^{+
  7.47}$
\\[1mm]
\hline
165&  17.080  &$_{-8.38}^{+  7.18}$ &$_{-3.15}^{+  2.56}$ &$_{-7.21}^{+
  7.49}$
\\[1mm]
\hline
170&  15.936  &$_{-8.34}^{+  7.11}$ &$_{-3.16}^{+  2.59}$ &$_{-7.20}^{+
  7.48}$
\\[1mm]
\hline
175&  14.979  &$_{-8.33}^{+  7.03}$ &$_{-3.18}^{+  2.62}$ &$_{-7.21}^{+
  7.51}$
\\[1mm]
\hline
180&  14.118  &$_{-8.31}^{+  6.90}$ &$_{-3.19}^{+  2.65}$ &$_{-7.21}^{+
  7.47}$
\\[1mm]
\hline
185&  13.198  &$_{-8.30}^{+  6.84}$ &$_{-3.21}^{+  2.69}$ &$_{-7.22}^{+
  7.43}$
\\[1mm]
\hline
190&  12.408  &$_{-8.26}^{+  6.82}$ &$_{-3.23}^{+  2.72}$ &$_{-7.22}^{+
  7.47}$
\\[1mm]
\hline
195&  11.743  &$_{-8.24}^{+  6.75}$ &$_{-3.25}^{+  2.75}$ &$_{-7.23}^{+
  7.48}$
\\[1mm]
\hline
200&  11.153  &$_{-8.25}^{+  6.69}$ &$_{-3.27}^{+  2.79}$ &$_{-7.24}^{+  7.49}$
\\[1mm]  
\hline
$-$& $-$& $-$& $-$& $-$
\\\hline

\end{tabular}
\end{center}
\end{minipage}
  \caption{{Higgs production cross
   section (MSTW08) at LHC for $\sqrt{s}=10$~TeV, with
   $\mu= \mu_R =\mu_F=M_H/2 $ and
   $\sigma^{best} =
   \sigma_{QCD}^{NNLO}+\sigma_{EW}^{NNLO}$.  The scale uncertainty, 
    PDF uncertainty without and accounting for $\alpha_s$ error   
    as described in~\cite{Martin:2009bu} are shown in the third, fourth and 
    fifth colum respectively, as a function of the Higgs boson mass.
    All the errors are percent ones.}
    \label{tab:LHCxsec08} }
\end{table}

\subsection*{Acknowledgments}

I am grateful to C.~Anastasiou and F. Petriello for collaboration on the topic
discussed in this contribution. This work is supported by the Swiss National 
Science Foundation under contract 200020-116756/2.

%

%% file: binoth/binoth.tex
  Many highly developed Monte Carlo tools for the evaluation of cross
  sections based on tree matrix elements exist and are used by experimental
  collaborations in high energy physics.  As the evaluation of one-loop
  matrix elements has recently been undergoing enormous progress, the
  combination of one-loop matrix elements with existing Monte Carlo tools is
  on the horizon. This would lead to phenomenological predictions at the
  next-to-leading order level.  A complete proposal  (called
  {\em Binoth Les Houches Accord}) for a standard interface between 
  Monte Carlo tools and one-loop matrix element
  programs can be found in~\cite{Binoth:2010xt}. In this Section, we collect
  a few examples of the procedure.

{\em This Section is Dedicated to the memory of, and in tribute to, Thomas Binoth, who led the effort to develop this proposal for Les Houches 2009.
Thomas led the discussions, set up the subgroups, collected the contributions, 
and wrote and edited the proposal.}

%% file: huber/huber.tex







\subsection{INTRODUCTION}

In this section we describe an implementation of an interface between
a Monte Carlo program (MC), {\sc SHERPA} \cite{Gleisberg:2008ta}, and
a one-loop program, {\sc RADY} (RAdiative corrections for Drell-Yan
processes) \cite{Dittmaier:2009cr,Brensing:2007qm,Dittmaier:2001ay},
for the calculation of electroweak next-to-leading order (NLO)
corrections to the neutral-current (NC) Drell--Yan process. Compared
to an MC/OLP interface for NLO QCD corrections this is more involved,
since the treatment of unstable particles and the choice of
electroweak couplings requires the exchange of additional
information. Furthermore, in the case of electroweak corrections the
use of mass regularization for soft and collinear divergences can be
relevant when one considers processes with isolated leptons in the
final state.

As described in~\cite{Binoth:2010xt}, such an
interface should work in two phases, first the \emph{initialization
  phase}, where all the main information is exchanged between the MC
and the OLP, and second the \emph{run-time phase}, where the MC calls
the OLP via the interface. In our case, the interface for calling the
OLP {\sc RADY} (a FORTRAN program) from {\sc SHERPA} (a C++ program)
is using a number of C wrapper functions, which allow linking all
together. The core of the interface functions is written in FORTRAN,
which simplifies the setup and use of internal parameters and
functions of {\sc RADY}.

Much of the technology used in {\sc SHERPA}'s electroweak dipole
subtraction is based on the implementation for the pure QCD case
\cite{Gleisberg:2007md}, using the matrix element generator {\sc
  AMEGIC} \cite{Krauss:2001iv}.

\subsection{INITIALIZATION PHASE}
\label{sec:initialization-phase}

During this phase, the MC requests a particular process and NLO
options from the OLP via an order file, and the OLP in turn creates a
contract file. In the contract file, the OLP confirms or rejects the
individual orders of the MC. In our case, the MC can request the OLP to
create a contract file using the C function
\begin{verbatim}
  void Order(const char* order_file);
\end{verbatim}
The order file contains the setup as wanted from the MC. Everything
after a \verb|#| is treated as a comment. Options are specified by the
name of the option (flag) and the setting for this option separated by
white space. All orders are case insensitive. An example order file
would look like:
\begin{verbatim}
## example order.dat
## OLP settings
CorrectionType                EW
MatrixElementSquareType       CH_SUMMED
CKMInLoops                    Unity
ResonanceTreatment            ComplexMassScheme
# IRRegularisation              DimReg
IRRegularisation              MassReg
IRRegulatorMasses             MU, MC, MD, MS, MB, ME, MMU, MTAU 
IRsubtraction                 None
EWRenScheme                   OLPdefined
Power_Alpha                   2
Power_Alphas                  0
## numerical input parameters (Model file + additional ew. input) 
ModelFile                     model_sm.slha
IN_alpha0                     0.0072973525678993
## processes
2 -> 2 1 -1 13 -13
2 -> 2 2 -2 13 -13
\end{verbatim}
Our interface currently supports the following options:
\begin{itemize}
\item \verb|CorrectionType|:\\[-.5em]
  \begin{itemize}
  \item \verb|EWincluded|: QCD and EW corrections 
  \item \verb|QCDonly| or \verb|QCD|: QCD corrections
  \item \verb|EW|: EW corrections (QED + weak)
  \item \verb|QED|: photonic corrections
  \item \verb|Weak|: genuinely weak corrections 
  \item \verb|BornOnly|: LO only \\[-.5em]
  \end{itemize}
\item \verb|MatrixElementSquareTypeType|: \verb+CH_SUMMED+
\item \verb|CKMInLoops|: \verb|Unity|
\item \verb|ModelFile|: Model file in SLHA format
\item \verb|IRRegularisation|:\\[-.5em]
  \begin{itemize}
  \item \verb|DimReg|
  \item \verb|MassReg|: In this case also the option
    \verb|IRRegulatorMasses| followed by a list of masses to be
    treated as small is necessary\\[-.5em]
  \end{itemize}
\item \verb|IRSubtraction|:\\[-.5em]
  \begin{itemize}
  \item \verb|DipoleSubtraction|: add endpoint contributions to
    virtual corrections
  \item \verb|None|: just virtual corrections\\[-.5em]
  \end{itemize}
\item \verb|ResonanceTreatment|: \\[-.5em]
  \begin{itemize}
  \item \verb|ComplexMassScheme|
  \item \verb|FactorizationScheme|
  \item \verb|PoleScheme|\\[-.5em]
  \end{itemize}
\item \verb|EWRenScheme|: \\[-.5em]
  \begin{itemize}
  \item \verb|alpha0|: $\alpha(0)$ everywhere
  \item \verb|alphaMZ|: $\alpha(M_\mathrm{Z})$ everywhere
  \item \verb|alphaGF|: $\alpha_{G_\mu}$ everywhere
  \item \verb|OLPdefined|: $\alpha(0)$ for photon radiation, i.e.\ the
    photonic NLO correction to the cross section scales with
    $\alpha(0)\alpha_{G_\mu}^2$, whereas the genuinely weak NLO
    correction to the cross section is proportional to
    $\alpha_{G_\mu}^3$. For $\gamma\gamma$ initial states
    $\alpha(0)$ is used everywhere.\\[-.5em]
  \end{itemize}
\end{itemize}
In addition to these options, the order file is also used to exchange
additional parameters, which are not part of the SUSY Les Houches
Accord (SLHA) \cite{Skands:2003cj}. In our case the value of
$\alpha(0)$ is not deduced from $\alpha(M_\mathrm{Z})$, but is set
explicitly via the option \verb|IN_alpha0|. All other model parameters
are passed through a SLHA model file. For this purpose {\sc RADY} uses
the {\sc SLHALib} \cite{Hahn:2004bc}. If an essential option is
missing, the default for this option is set and added to the contract
file.

For the example order file shown above the corresponding contract file
returned by the OLP looks like:
\begin{verbatim}
correctiontype                qcdonly     | 1     # qqcdew =  1,\
 qvirt =  1, qewho =  0, qewhel =  0, qsusy =  0
matrixelementsquaretype       ch_summed     | 1
ckminloops                    unity     | 1
resonancetreatment            complexmassscheme     | 1   \
  # qwidth =  1
irregularisation              massreg      | 1     # qregscheme =  1
irregulatormasses             mu, mc, md, ms, mb, me, mmu, mtau   |\
 1    # Small masses are: MU, MC, MD, MS, MB, ME, MMU, MTAU. 
irsubtraction                 none     | 1     # qbrem =  0
ewrenscheme                   olpdefined     | 1   # qalp =  2,\
 qoptimalscheme =  1
power_alpha                   2     | 1
power_alphas                  0     | 1
modelfile                     model_sm.slha     | 1
in_alpha0                     0.0072973525678993     | 1
2 -> 2 1 -1 13 -13     | 1 2     # proc_label = 2
2 -> 2 2 -2 13 -13     | 1 1     # proc_label = 1
# paremeters used by OLP
# NOTE: EWRenScheme = OLPdefined
# alpha is used for LO q\bar{q} \to l^+l^-, i.e. proc_scheme=1,2,4,
# and alphaIR=alpha0 is used for LO \ga\ga \to l^+l^-,\
 i.e. proc_scheme=8
OUT_alpha       0.7547514055936910E-02
OUT_alphaIR     0.7297352567899300E-02
OUT_MZ          91.15348059999999    
OUT_GZ          2.494266380000000    
OUT_MW          80.37450950000000    
OUT_GW          2.140241340000000    
OUT_CW          0.8817404089366329    ,        0.3238080343356995E-03
OUT_SW          0.4717354368914205    ,       -0.6052431220634457E-03
\end{verbatim}
where a \verb|\| at the end of the line indicates a line break in the
contract file. The answers of the OLP follow after \verb+|+. If the
OLP can provide the order, then it returns a \verb|1|, followed by the
internal options of the OLP. Otherwise, if the option is not supported
a \verb|-1| is returned. For valid subprocesses the OLP returns
\verb|1|, plus a process label to identify the subprocess during the
run-time phase. In addition to the confirmation of the order, the OLP
returns all parameters needed to calculate the LO cross section, for
the case when OLP and MC do not support the same options, in
particular for the treatment of unstable particles. These parameters
are labeled with \verb|OUT_| followed by the name of the
parameter. The first number after the parameter is the real part, and,
if present, the second is the imaginary part. In our case this
additional information was not used, since both {\sc SHERPA} and {\sc
  RADY} support the complex mass scheme. Furthermore, \verb|OUT_alpha|
and \verb|OUT_alphaIR| are returned , these are the couplings used by
the OLP for the calculation of the LO cross section and genuinely weak
corrections, and the photonic corrections respectively. Note that
these couplings are equal in all but the \verb|OLPdefined| scheme.

\subsection{RUN-TIME PHASE}
\label{sec:run-time-phase}

During the run-time phase the MC initializes the OLP for the run
calling the function
\begin{verbatim}
void StartOLP(const char* contract_file);
\end{verbatim}
where the passed string contains the name of the contract file. After
this initialization the MC calls 
\begin{verbatim}
void EvalSubprocess( int proc_label, double* momenta, 
 double ren_scale, double alpha_s, double alpha_ew, double* result );
\end{verbatim}

The first argument, \verb|proc_label|, is an integer label which
encodes bitwise the information of the subprocess to be calculated,
i.e.\ bit 0 for $\mathrm{u}\bar{\mathrm{u}}$, bit 1 for
$\mathrm{d}\bar{\mathrm{d}}$, bit 2 for $\mathrm{b}\bar{\mathrm{b}}$,
bit 3 for $\gamma\gamma$ initial state. The second argument is an
array for passing the momenta with dimension $ 4 \times \# particles
$, ordered in such a way that the first 4 entries correspond to $( E,
p_x, p_y, p_z)$ of the first particles and so on. The third argument,
\verb|ren_scale|, is the renormalization scale. In the case of
dimensional regularization the infrared scale is identified with the
renormalization scale. The values of $\alpha_{\mathrm{s}}$ and
$\alpha_{\mathrm{ew}}$ are passed as \verb|alpha_s| and
\verb|alpha_ew| respectively. Here the electroweak coupling is only
present for a possible implementation of a running coupling. In our
case we did not implement this and our couplings are completely fixed
during the initialization phase, according to the specified input
parameter scheme.

The array \verb|result| contains the information
\begin{verbatim}
  result[0] = PoleCoeff2;
  result[1] = PoleCoeff1;
  result[2] = PoleCoeff0;
  result[3] = BornSq;
  result[4] = alpha_IR;
\end{verbatim}
where \verb|PoleCoeff2|, \ldots, \verb|BornSq| are the results
obtained by the OLP, and \verb|alpha_IR| is the coupling to be used in
the IR modules. Using the result for the Born matrix element squared,
MC then evaluates an effective coupling, defined by
$\alpha_{\mathrm{eff}}^P=\alpha(0)^P
|\mathcal{A}^\mathrm{OLP}_\mathrm{LO}|^2/|\mathcal{A}^\mathrm{MC}_\mathrm{LO}|^2$,
where $P=\mathtt{PowerAlpha}$. This allows to deduce the used values
of $\alpha$ during run-time.

\subsection{COMPARISON OF RESULTS}
\label{sec:comparison-results}

\begin{figure}
  \centering
  \includegraphics[clip,width=0.6\linewidth]{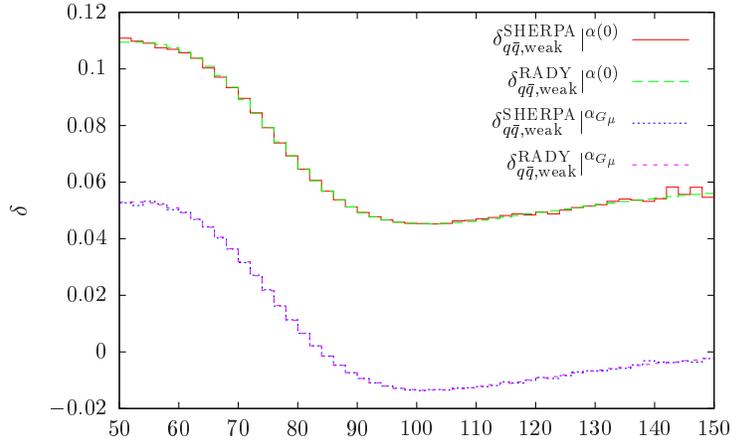}
  \caption{Comparison of weak correction factor in the $\alpha(0)$ and
    the $\alpha_{G_\mu}$-schemes to the di-lepton invariant mass
    distribution obtained by {\sc SHERPA} and {\sc RADY}}
  \label{fig:sherpa-rady-weak-mll}
\end{figure}

\begin{figure}
  \centering
  \includegraphics[clip,width=0.6\linewidth]{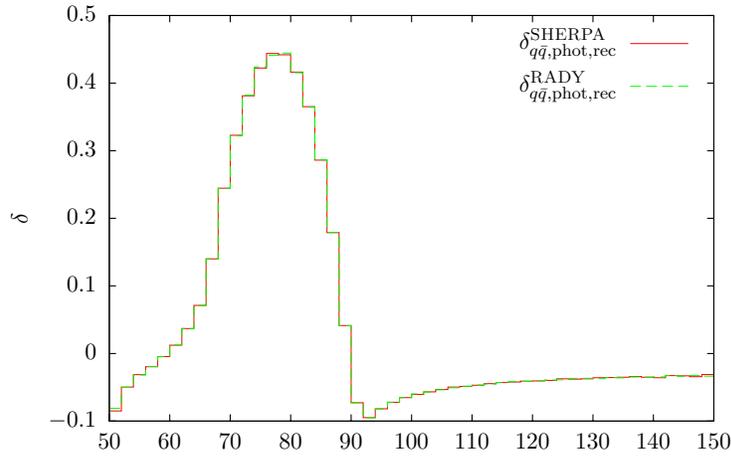}
  \caption{Comparison of photonic correction factor to the di-lepton
    invariant mass distribution with photon recombination in the
    \texttt|OLPdefined| scheme obtained by {\sc SHERPA} and {\sc RADY}.}
  \label{fig:sherpa-rady-mll-qed}
\end{figure}

\begin{figure}
  \centering
  \includegraphics[clip,width=0.6\linewidth]{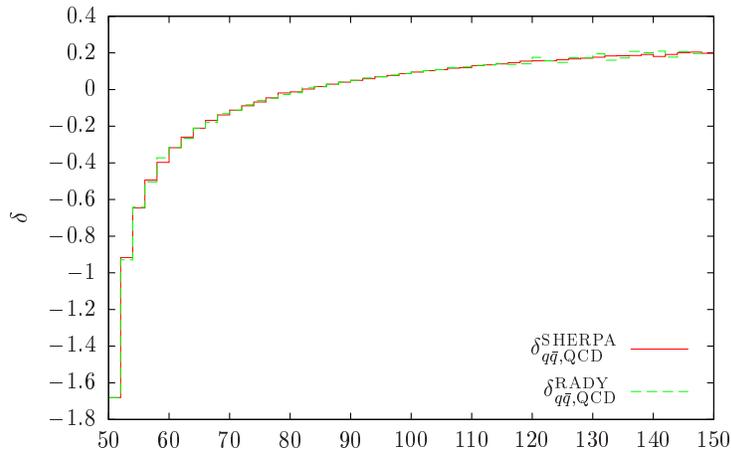}
  \caption{Comparison of QCD correction factor to the di-lepton
    invariant mass distribution obtained by {\sc SHERPA} and {\sc
      RADY}}
  \label{fig:sherpa-rady-mll-qcd}
\end{figure}

As a consistency check we calculated various types of corrections
using different options, with both {\sc SHERPA} using the interface to
{\sc RADY}, and running {\sc RADY} as a standalone program. In
Figs. \ref{fig:sherpa-rady-weak-mll},~\ref{fig:sherpa-rady-mll-qed} and
~\ref{fig:sherpa-rady-mll-qcd} we show the correction factor
\begin{equation}
  \label{eq:1}
  \delta = \frac{\mathrm{d}\sigma_\mathrm{NLO}/ \mathrm{d}M_{ll}}{\mathrm{d}\sigma_\mathrm{LO}/ \mathrm{d}M_{ll}} - 1 \; ,
\end{equation}
for the di-lepton invariant mass distribution. These agree on the
permille level, demonstrating that the interface works.

%% file: frederix/frederix.tex






\subsection{Introduction}
In the last few years much progress has been made in the computation
of next-to-leading (NLO) computations~\cite{thisproc}.  The high
energy physics community has reached the point where many of these
calculations can be done almost automatically.
In many cases two separate codes are needed for a full NLO generator:
\begin{itemize}
\item the One-Loop Program (OLP) which calculates the virtual
  contributions to a process for a given phase-space point,
\item the Monte Carlo (MC) tool which takes care of the real
  emission, the subtraction terms and the phase-space integration.
\end{itemize}
Only together, the OLP and MC codes can provide total cross sections and
distributions at NLO accuracy.

To facilitate the usage of the OLP together with any MC an interface
has been proposed during the Les Houches 2009
workshop~\cite{Binoth:2010xt}.
In this contribution we show how this interface has been implemented
in the MC code
\texttt{MadFKS}~\cite{Frederix:2009yq}~and the OLP codes
\texttt{BlackHat}~\cite{Berger:2006cz,Berger:2008sj,Berger:2008ag}~and
\texttt{Rocket}~\cite{Melnikov:2009dn,KeithEllis:2009bu,Melnikov:2009wh}~and
how it works in practice in the process $e^+e^- \to n$ jets with
$n=2,3,4$. 

\subsection{Code snippets}
Before discussing the details of the interfaces between \texttt{MadFKS}~and
\texttt{BlackHat}~or \texttt{Rocket}, we shortly remind the reader about the proposed
LHA interface. More details about the proposal can be found
in~\cite{Binoth:2010xt}.

\subsubsection{The proposal for the interface in short}
The first stage of using a OLP code with a MC code is the agreement
over the process and all related settings via a so-called order file, written
by the MC and read by the OLP, and the corresponding contract file
written by the OLP. These settings include all relevant input
parameters, possible approximations (e.g. the leading color
approximation) and the treatment of the helicity (e.g. sum of or Monte
Carlo over helicities).
Because this initialization stage does not involve linking the codes
together, it will not be discussed further in this note.

After an agreement has been established on the process that will be
computed and all relevant settings and parameters the actual run can
start. During run-time there are two more stages in the
interface. First the OLP will read the contract file and will set the
corresponding input parameters. The momenta and dynamical input
parameters (such as the renormalization scale and the strong coupling)
are passed from the MC code to the OLP for each phase-space point. The
OLP performs the calculation of the renormalized virtual matrix
element squared and returns the value for the loop corrections as an
expansion in $1/\epsilon$. The MC code then combines this result with
the real radiation term and performs the phase-space integration.

In practice, when interfacing \texttt{MadFKS}~with either \texttt{BlackHat}~or
\texttt{Rocket}, a library from the OLP code has been created. This library
can then be linked to \texttt{MadFKS}~and all information between the OLP code
and the MC code can be passed by subroutine calls.
Notice that the various codes can be written in different languages,
for instance \texttt{BlackHat}~is written in
\texttt{C++}, \texttt{Rocket}~in \texttt{Fortran95}~and \texttt{MadFKS}~in
\texttt{Fortran77}. 
We start by first describing the interface between \texttt{Rocket}~and \texttt{MadFKS},
which is slightly simpler. 

\subsubsection{Linking \texttt{Rocket}~to \texttt{MadFKS}}
Because \texttt{Fortran95}~and \texttt{Fortran77}~are quite similar,
no special treatment is needed if a compiler is used that is
compatible with both languages, such as \texttt{gfortran}.
Notice however that because \texttt{Fortran77} cannot handle
\texttt{modules}, subroutines that need to be called by \texttt{MadFKS}~ have
not been placed inside a module in the \texttt{Rocket}.

In practice we use the following function call in \texttt{MadFKS}~to start-up
the process:
\begin{verbatim}
 call OLP_start(filename,status)
\end{verbatim}
where \verb|filename|~is the name of the agreed-upon contract file
and \verb|status|~is set to `\texttt{1}' by the OLP~for a correctly initialized
contract file. During run-time the following call is used to pass the
information about the phase-space point:
\begin{verbatim}
 call Rocket_EvalSubproc(procnum,nexternal,p,hel,mu,alphaS,virt_wgts).
\end{verbatim}
According to the proposed Les Houches Accord,
\verb|procnum|~is the number that \texttt{Rocket}~gives to the process
in the contract file, \verb|p|~is a linear array containing
four-momenta and masses of the external particles, \verb|mu|~is the
renormalization scale,
\verb|alphaS|~is the strong coupling evaluated at \verb|mu|, and
\verb|virt_wgts|~ is a four-component array with, in order, the double
pole, single pole, finite part
of the virtual matrix element squared and the Born squared for the
given phase-space point \verb|p|.
We also include the number of the external particles and their helicities in the call, even
though this is not (yet) part of the LHA proposal. 
The helicities are passed from \texttt{MadFKS}~to
\texttt{Rocket} by the array \verb|hel|~to facilitate a Monte Carlo over the
helicities\footnote{So far only the helicities of \emph{massless} external
particles have been considered in the interface. For massive
particles it is also needed to agree
upon the basis to project the spins.}.
The number of external particles is passed by \verb|nexternal|. Although not
strictly necessary because this information can be deduced from the
process in the contract file, in practice it is simpler for the
declaration of the arrays \verb|p| and \verb|hel| to have their sizes
available as an argument of the subroutine.

In Fig.~\ref{ROCKETcontract} we show an example of a contract file that is
used for one of the subprocesses contributing to the NLO corrections
to $e^{-} e^{+} \to 3$~jets. All the necessary parameters are
specified in this contract, in the right column a brief explanation of
the parameters is given. The string ``\texttt{OK}'' means that the
running option or parameter is allowed, but other than that this
information is for the user and is not used by the OLP or the MC. A
string ``\texttt{ERROR}'' would indicate that the option/parameter
required through the order file by the MC is not allowed and the
initialization stage of the OLP will fail.

\begin{figure}
\begin{center}
\begin{scriptsize}
\begin{verbatim}
            # contract produced by Rocket from 'order.file'
            
            MatrixElementSquareType C_SUMMED  |  OK note: color summed
            IRregularization DR               |  OK note: DR scheme
            CorrectionType QCD                |  OK note: NLO QCD corrections
            Only_Z false                      |  OK note: photon contribution included
            Only_Photon false                 |  OK note: Z contribution included
            Nf_light 5                        |  OK note: nr of light flavours
            IAlpha_EM_MZ 128.802              |  OK note: Inverse of em coupling
            Alpha_S_MZ  0.1190000             |  OK note: Strong coupling
            SinThW2 0.23                      |  OK note: Sin(Th_W)^2
            Mass_Z 91.188                     |  OK note: Z mass
            Width_Z 2.43000                   |  OK note: Z Width
            Color treatment Full Color        |  OK note: Full Color
            Virtual Full NLO                  |  OK note: Full NLO

            2 -> 3 11 -11 1 -1 21             | 1 4 note: process  4

\end{verbatim}
\end{scriptsize}
\caption{An example of a contract file used with \texttt{Rocket}~in $e^{-}
  e^{+} \to d \bar{d} g$ production. \label{ROCKETcontract}}
\end{center}
\end{figure}

\subsubsection{Linking \texttt{BlackHat}~to \texttt{MadFKS}}
When linking a C++ code to a Fortran code there are a couple of issues
that should be taken into account.
\begin{itemize}
\item First, in \texttt{Fortran} arguments of the subroutines are
  passed by reference, which should be reflected in the \texttt{C++} code.
\item Multi-dimensional arrays are build in different orders in
  \texttt{C++} and \texttt{Fortran}. In particular, this means that in
  \texttt{Fortran} one can define a two-dimensional array of the
  momenta of a phase-space point as
\begin{verbatim}
  real*8 p(0:4, nexternal)
\end{verbatim}
  while in \texttt{C++} this should be
\begin{verbatim}
  double p[nexternal][4]
\end{verbatim}
  to get the one-dimensional array in the layout prescribed by the LHA
  proposal.
\item Because \texttt{Fortran} compilers add (multiple) leading and/or
  trailing underscores to variable names, these need to be added in
  the \texttt{C++} libraries. The number of underscores depends on the
  compiler. Also, \texttt{Fortran} is case insensitive, therefore the
  names of the subroutines in the OLP code that the MC calls, should
  be in capitals or completely without capitals, again depending on
  the fortran compiler. In our case, two trailing underscores and
  subroutine names in lower case letters are needed.
\item Furthermore, strings have to end with a
  null character in \texttt{C}, which means that this character has to be
  added when passing the string of the location of the contract file
  from a \texttt{Fortran} Monte Carlo code to a \texttt{C++} OLP code.
\end{itemize}
In practice, this means that during the start-up phase the following
calls are used in the \texttt{MadFKS}~and \texttt{BlackHat}~codes, respectively:
\begin{verbatim}
  call OLP_Start(filename//Char(0),status)
\end{verbatim}
and
\begin{verbatim}
extern "C" {
   void olp_start__(const char* filename,int& status);
}
\end{verbatim}
During runtime 
\begin{verbatim}
  call OLP_EvalSubprocess(procnum,p,mu,alphaS,alphaEW,virt_wgts)
\end{verbatim}
and
\begin{verbatim}
extern "C" {
   void olp_evalsubprocess__(int& Label,double* p,double& mu,
                    double& alpha_s,double& alpha_ew,double* result);
}
\end{verbatim}
are the call and interface of the subroutine that are used to pass the
momenta and all relevant information from
\texttt{MadFKS}~to \texttt{BlackHat}~which returns the virtual matrix element squared. The
\texttt{extern "C"} is needed in the C++ code to prevent the symbol
names to be mangled by the C++ compiler. It is to be noted that the
parameters have to be passed by reference.

\subsection{Sample results}
As a proof of concept we show here some selected results for NLO
predictions to electron--positron collisions to 2, 3 and 4 jets at
$\sqrt{\hat{s}}=M_Z$ and with the renormalization and factorization
scales also equal to the $Z$ boson mass. To define jets we use the
$k_t$-algorithm and recombine momenta according to the $E$-scheme,
i.e.\ we add up the particles four-momenta. 

\begin{figure}[t]\begin{center}
\includegraphics*[scale=0.8]{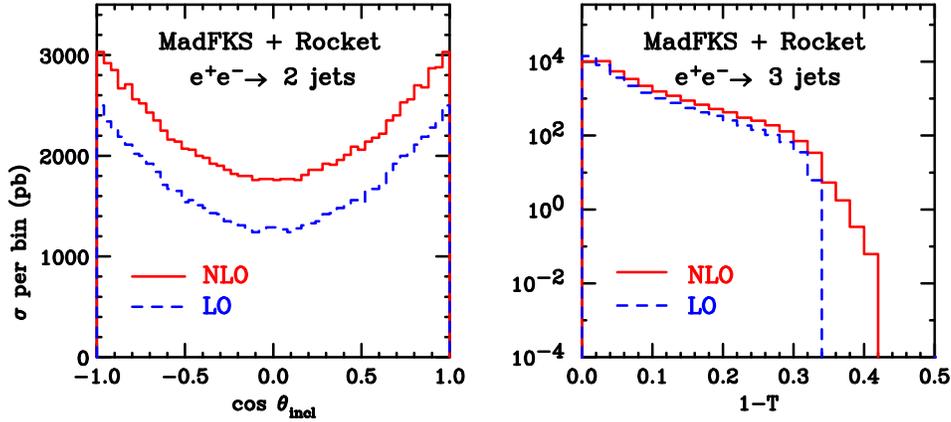}\vspace{-10pt}
\caption{Inclusive $\cos\theta$ for 2 jet production and Thurst
  distribution for 3 jet production at LO (blue dashed) and NLO (red
  solid) using \texttt{MadFKS}~and \texttt{Rocket}. \label{fig:RocketRate}}
\end{center}
\end{figure}

In Fig.~\ref{fig:RocketRate} results for \texttt{Rocket}~linked to \texttt{MadFKS}~are
plotted. On the left hand side are the LO and NLO predictions shown
for the inclusive $\cos\theta$ distribution in 2 jets production. This
distribution is defined as the cosine of the angle between the
incoming electron direction and all of the final state jets, defined
according to the Durham jet algorithm. On the right hand side are the
fixed LO and NLO predictions shown of (one minus) the thrust
distributions, which starts from Born-level 3 parton events and is
therefore shown for $3$-jet events. 

\begin{figure}[t]\begin{center}
\includegraphics*[scale=0.8]{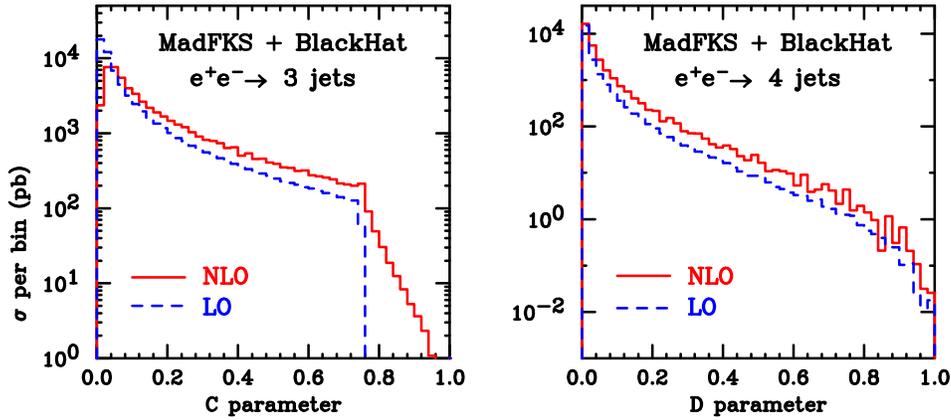}\vspace{-0pt}
\caption{$C$ parameter for 3 jet production and $D$ parameter for 4
  jet production at LO (blue dashed) and NLO (red solid) using
  \texttt{MadFKS}~and \texttt{BlackHat}. \label{fig:BlackHatParam}}
\end{center}
\end{figure}

In Fig.~\ref{fig:BlackHatParam} we show two distributions calculated by
linking \texttt{BlackHat}~code to the \texttt{MadFKS}~MC program. In
the plot on the left hand side, the $C$ parameter is shown in
$e^+e^-\to 3$~jets at LO and NLO, and in the plot of the right hand
side the $D$ parameter in $e^+e^-\to 4$~jets.~\cite{Fox:1978vu,Fox:1978vw}

\subsection{Conclusions}
In this contribution we have shown how the proposed LHA interface
between Monte Carlo tools and one-loop programs works in practice
between \texttt{BlackHat}~or \texttt{Rocket}~together with
\texttt{MadFKS}. The proposal works well even if the codes are written in different
languages. We do not expect that linking other OLP or MC codes using the LHA interface
will lead to any further difficulties.

\subsection{Acknowledgments}
We would like to thank the other members of the \texttt{BlackHat}, \texttt{MadFKS}~and
\texttt{Rocket}~collaborations for their contributions to the development of
these codes.
